 \documentclass[final,3p,times,authoryear]{elsarticle}

\usepackage{amssymb}
\usepackage{hyperref}
\usepackage[tight,footnotesize]{subfigure}
\usepackage{graphicx}
\usepackage{hhline}
\usepackage{booktabs}
\usepackage{xspace}
\usepackage{url}
\usepackage{savesym}
\usepackage{amsmath}
\savesymbol{AND}
\usepackage{algorithmic}
\usepackage{algorithm}
\usepackage{adjustbox}
\usepackage[table]{xcolor}
\usepackage{multirow,makecell}
\usepackage{lscape}
\newcommand*{\eg}{e.g.\@\xspace}
\newcommand*{\ie}{i.e.\@\xspace}
\makeatletter
\newcommand*{\etc}{%
    \@ifnextchar{.}%
        {etc}%
        {etc.\@\xspace}%
}
\makeatother

\usepackage{color}

\journal{Nuclear Physics B}

\begin{document}

\begin{frontmatter}



\title{Community detection algorithm evaluation with ground-truth data}


\author[ub,utm]{Malek Jebabli}
\ead{Malek.Jebabli@u-bourgogne.fr}
\author[ub]{Hocine Cherifi\corref{cor1}}
\ead{hocine.cherifi@u-bourgogne.fr}
\author[ly]{Chantal Cherifi}
\ead{Chantal.BonnerCherifi@univ-lyon2.fr}
\author[utm]{Atef Hamouda}
\ead{atef\char`_hammouda@yahoo.fr}
\cortext[cor1]{Corresponding author}
\address[ub]{University of Burgundy, Esplanade Erasme, 21078, Dijon, FRANCE}
\address[ly]{Laboratoire DISP, IUT Lumi\`{e}re, University of Lyon 2, Lyon, FRANCE}
\address[utm]{University of Tunis El-Manar, El Manar 1, 1068, Tunis, TUNISIA}

\begin{abstract}
    Community structure is of paramount importance for the understanding of complex networks. Consequently, there is a tremendous effort in order to develop efficient community detection algorithms. Unfortunately, the issue of a fair assessment of these algorithms is a thriving open question. If the ground-truth community structure is available, various clustering-based metrics are used in order to compare it versus the one discovered by these algorithms. However, these metrics defined at the node level are fairly insensitive to the variation of the overall community structure. To overcome these limitations, we propose to exploit the topological features of the 'community graphs' (where the nodes are the communities and the links represent their interactions) in order to evaluate the algorithms. To illustrate our methodology, we conduct a comprehensive analysis of overlapping community detection algorithms using a set of real-world networks with known a priori community structure. Results provide a better perception of their relative performance as compared to classical metrics. Moreover, they show that more emphasis should be put on the topology of the community structure. We also investigate the relationship between the topological properties of the community structure and the alternative evaluation measures (quality metrics and clustering metrics). It appears clearly that they present different views of the community structure and that they must be combined in order to evaluate the effectiveness of community detection algorithms.

\end{abstract}


\end{frontmatter}


\section{Introduction}

        In complex network analysis, community detection has attracted increasing attention of researchers in recent years. Several algorithms are introduced almost every day based on a various understanding of what is a community. Usually, it is intuitively recognized as a dense group where members interact with each other more deeply than with those outside the group. This weak structural definition has been approached from many different views, leading to an impressive literature on the subject. The work of \cite{SAMSAM10133} presents an interesting taxonomy of several algorithms proposed in the literature.
        Besides the definition issue, one can also distinguish two types of community structure: non-overlapping communities in which every individual belongs to a single community and overlapping communities in which some entities can belong to several communities.
       Depending on the availability of data with ground truth community structure one is faced with two options in order to evaluate the algorithms. When the ground truth community structure is unknown, the evaluation relies on quality metrics that are supposed to encode what is a 'good' community structure. Several metrics have been introduced \eg \cite{chen2014extension}, \cite{yang2015defining}, \cite{Lancichinetti2008} and \cite{li2008quantitative}. These metrics are of common use to rank the quality of community structures discovered by different community detection algorithms. The most popular and widely used is the modularity \cite{chen2014community}. It reflects the concentration of edges within communities compared with a random model with no community structure. The main drawback of the quality metric approach is that very often they are also used as an optimization criterion in community detection algorithms. Therefore, comparisons can be biased. Furthermore, there is no consensus on desirable properties of a good community. When the ground truth community structure is known, one can evaluate the similarity between the communities discovered by the detection algorithm to the ground truth communities of the network. We can distinguish three main categories of clustering comparison measures used for this purpose \ie \emph{(i)} measures based on pair-counting; \emph{(ii)} set-matching-based measures and \emph{(iii)} information-theoretic-based measures. In measures based on pair-counting, the comparison is based on counting the pairs of points on which two communities agree or disagree. Set-matching-based measures intend to find the largest overlaps between pairs of different communities and then the accuracy of this assignment is measured. Information-theoretic-based measures quantify the mutual information shared by two communities in order to assess their agreement. The main limitation of these measures is that they can be insensitive to the variation of the community structure topology. Indeed, it has been shown, in previous studies, that two community structures very similar according to the clustering based measures can exhibit very different topological properties (embeddedness, average distance, \etc) \cite{orman2012comparative}.
        To overcome this limitation, we propose an alternative evaluation approach based on the topology of the community structure. First of all, we compute the community-graphs for the output of the various community detection algorithms and the ground truth community structure. In these networks, the nodes are the communities and there is a link between two nodes if the two communities interact. Then, the assessment of the algorithms is based on the community-graphs topological properties comparisons. Indeed, we believe that an efficient community detection algorithm should uncover a community structure with similar topological properties as compared to the ground-truth community structure.
        Although the proposed framework is general, in this paper, we restrict our attention to networks with overlapping community structure. Nevertheless, we discuss how it can be applied to networks with non-overlapping community structure.
        To validate our approach, we investigate eleven popular overlapping community detection algorithms on three large-scale networks.
        In a preliminary work, \cite{jebabli2015overlapping}, we conducted a comparative analysis of the topological properties using the AMAZON network. The community structures have been compared at different levels. First of all, we computed basic properties of the community-graphs (average clustering coefficient, average shortest path, diameter, density, and degree correlation). Then we analyzed their various distributions (the distribution of node degree, average clustering coefficient as a function of degree as well as hop distance). Finally, we turned to the original network to compare classical intrinsic features of overlapping communities (community size, overlap size, and membership number distributions). Results showed that the topological properties of the Ground Truth community-graphs and the communities networks based on the community detection algorithms are quite different.
        In this paper, we extend the analysis in various ways. First of all, PGP and aNobii are used in order to check the 'stability of the results'. Indeed, these networks belong to different domains and have different global characteristics as compared to AMAZON \ie range of nodes, edges, communities, \etc. Second, we propose a strategy to rank the algorithms based on the topological properties of their community-graphs. The algorithms are ranked according to each topological properties and the individual rankings are used in a  multiple criteria decision-making approach to obtain a final ranking. Finally, we establish a comparative analysis of the main evaluation approaches (quality metric, clustering measures, and topological properties).

       In this paper,  our main concern is to present and evaluate an efficient alternative methodology as compared to the classical quality and clustering measures. To that end, an extensive empirical comparative evaluation of overlapping community detection algorithms is performed. Our goal is to highlight the importance of the topological characteristics of the community structure to assess the performance of community detection algorithms. We believe that this work provides a promising step towards evaluating community detection algorithms in a more appropriate way.

        The remainder of this paper is organized into four sections. Section 2 discusses related works to the community detection evaluation issue. In section 3, we describe the background on overlapping community detection (the algorithms, the influential quality and clustering measures, the topological properties) and Multiple criteria decision making. Section 4 introduces the data and the methodology to evaluate the community detection algorithms with ground-truth data. In section 5, we report and discuss the results of the topological properties analysis. Section 6 is devoted to the presentation and the discussion of the various rankings of the community detection algorithms. Finally, section 7 summarizes our concluding remarks.

\section{Related works}

        In this section, we survey the most influential related work on comparing, manipulating, and analyzing community structures. We restrict our attention on overlapping community structure. For each study, we mention the data, the measures, and the algorithms used together with the important results. The main characteristics of this works are summarized in Table \ref{table43}

        One of the first comparative studies is reported in \cite{Leskovec:2010:ECA:1772690.1772755}. Four real-world networks with size up to three hundred thousand nodes are used in order to analyze the outputs of five algorithms. In this work, as the ground-truth community structure is not known, eleven quality metrics are investigated. Only two overlapping community detection algorithms are considered. First of all, it appears that the algorithms optimize the quality metrics over a range of size scales. Additionally, many quality metrics favors small clusters. Optimization of the quality metrics in the detection algorithms introduces a systematic bias into the extracted clusters. Indeed, a small variation of the quality scores can lead to great variability in the community structure. This work suggests that the link between quality metrics and the community structure is relatively loose.

        Another widely-recognized analysis is introduced in \cite{journals/csur/XieKS13}. Two real-world networks and synthetic networks with ground-truth community structure are used, together with nine real-world networks with unknown ground-truth community structure. Their size varies from very low (34 nodes) to very high (334863 nodes). Fourteen overlapping community detection algorithms are compared. Their performance is assessed with two version of the overlapping modularity quality metric, four clustering metrics, and two topological properties. Given that the ground truth is not available for most of the real-world networks, performances of the algorithms are assessed only with quality metrics in this case. In the case of synthetic networks, the algorithms are also ranked according to two clustering metrics (NMI and F-score), and two topological properties of the community structure are reported.

        The main lesson of this work is that the clustering metrics (NMI and Omega-Index) are not very sensitive to the overlaps in the community structure. Furthermore, the algorithms can be categorized according to their ability to over-detect or to under-detect the overlapping nodes. Over-detection refers to the case where more overlapping nodes than there exists are claimed, while under-detection refer to the case where only very few overlapping nodes are identified. Experiments on real-world networks show that most of the algorithms belong to the under-detection class. There is a high correlation between the two versions of modularity. Generally, overlapping tend to decrease the modularity scores. The community detection algorithms possess a common feature is that they identify a small fraction of overlapping nodes especially when they are applied to real-world networks. Note that the comparison of the quality and the clustering metrics are not the main issue of this work. Indeed, the authors focus on the ranking of the overlapping community detection algorithms.

        In \cite{almeida2011there}, the authors perform a comparative evaluation of five popular quality metrics (\ie modularity, silhouette index, conductance, coverage, and performance) on seven different real-world networks. Five of them, with size ranging from 12008 to 36682 nodes, are with unknown ground-truth community structure. The remaining are small but with known ground-truth community structure. To compare different metrics, they selected four non-overlapping community detection algorithms from four different, representative categories of clustering algorithms. They conclude that the quality metrics behaves satisfactorily when the communities are well identified. In other words, in the case where the intra-link density value is very high as compared to the inter-link density value. Additionally, they show that the quality metrics have strong biases toward incorrectly awarding good scores to some kinds of clusters, especially seen in larger networks. They indicate that all metrics do not share a common view of what a true clustering should look like and that there is no such a thing as a 'best' quality metric.

        In networks with overlapping community structure, it is commonly admitted that the overlaps are more sparsely connected than the non-overlapping parts. \cite{journals/tist/YangL14}, conducted an extensive analysis of the overlapping community structure. The authors used six real-world networks with explicitly labeled ground-truth communities. They unexpectedly observed that the overlap zones are more densely connected than the non-overlapping ones. Furthermore, the overlaps contain high-degree nodes. As a result, most community detection algorithms identify the overlaps as separate communities. As the network models do not take into account these topological properties, results based on artificial benchmarks are biased. Note that this paper is the first one that clearly points out that functional communities (semantically defined) can be different than structural communities (topologically defined).

        In the same vein as \cite{journals/tist/YangL14}, the work of \cite{hric2014community} presents a comparative study of functional and structural communities. The structural communities discovered by ten community detection algorithms are compared to the ground-truth community structure defined by functional similarity. The authors used fifteen real-world networks with size ranging from 34 to 5189809 nodes. In these networks, the number of functional communities varies greatly (from 2 to 2183754). They also used a medium size synthetic network generated with the LFR algorithm. They conclude that functional communities are not recovered by most of the algorithms. Roughly speaking, there is no simple relation between the functional communities described by the ground-truth and the structural ones recovered by the algorithms.

        Very relevant to our work is that of \cite{WICS:WICS1319}. Five real-world networks\footnote{\url{http://snap.stanford.edu/data/index.html}} with known ground-truth are analyzed. Thirteen community detection methods, including five algorithms that allow overlapping, are compared.
        To evaluate the outputs of the algorithms, quality metrics and clustering measures are used. The results of their experiments show that there is no clear relation between the scores of the quality metrics and the clustering measures. This is in line with recent findings. Indeed, clustering metrics are based on the functional ground-truth community structure while quality metrics describe topological properties linked to cohesiveness.

        Given that there is no universal quality metric, \cite{creusefond2016evaluation} apply a general methodology to identify different contexts,  groups of graphs where the quality functions behave similarly. In these contexts, they identify the most effective quality functions, \ie quality functions whose results are consistent with clustering measures. In other words, a quality function fits a ground-truth if the clusterings that are the closest to the ground-truth are highly ranked with the quality, and conversely. The experiments are performed on ten real-world networks with known ground truth and one synthetic network with size ranging from 115 to 1143395 nodes. Seven non-overlapping community detection algorithms are used. In order to identify contexts, the rankings of the uncovered community structure by the quality functions are compared. Contexts are identified as a set of graphs that are highly correlated. In other words, graphs belong to the same context if the quality functions rank them in the same way. Experiments show that three contexts can be distinguished with their relevant quality functions.
        Table \ref{table43} summarizes the main information about the related works (data, quality metrics and/or clustering measures, community detection algorithms).

        The main lesson learned from all these works is that the community detection evaluation issue is still an open question. First of all, most experiments demonstrate that there is no simple relationship between functional and structural communities. This translates into the fact that quality metrics and clustering measures do not correlate well. Another important aspect is that there is no universal metric. In other words, the efficiency of the metrics is highly dependent on the data. Overall, this suggests that a single feature as computed by a clustering measure or a quality metric is not sufficient to capture the complexity of the community structure evaluation issue. That is the reason why we believe that it must be based on a more detailed analysis of the community structure.

\begin{table}[ph]
  \centering
  
  \scalebox{0.5}{
  
  \setlength{\extrarowheight}{2.5pt}

  \begin{tabular}{|c|c|c|c|c|c|c|c|c|}
    \hline
  \multirow{2}{*}{Papers} & \multicolumn{4}{c}{Data} &\multicolumn{2}{c}{Measures} &\multicolumn{2}{c|}{Algorithms}\\
    
    \hhline{~--------}
  &\multicolumn{1}{c|}{Names}  & \multicolumn{1}{c|}{Ground truth }&\multicolumn{1}{c|}{ Nodes} &\multicolumn{1}{c|} {Edges}& \multicolumn{1}{c|}{Properties}& \multicolumn{1}{c|}{Computed For} &\multicolumn{1}{c|}{ Names} & \multicolumn{1}{c|}{Overlap }\\
    \hline
%
%
   
    \multirow{5}{*}{Lescovec and al. (2010)} & \multicolumn{1}{c|}{DBLP }&\multicolumn{1}{c|}{ No}  & \multicolumn{1}{c|} {317080}& \multicolumn{1}{c|}{1049866}& \multicolumn{1}{c|}{\multirow{5}{*}{\shortstack{Conductance of connected clusters,\\ Average shortest path length,\\ Network community profile,\\ Expansion, Internal density,\\ Cut Ratio, Normalized, Maximum} }}&\multicolumn{1}{c|}{\multirow{5}{*}{All Graphs}}&\multicolumn{1}{c|}{ Local spectral} & \multicolumn{1}{c|}{Yes }\\
               
     & \multicolumn{1}{c|}{Enron email network }&\multicolumn{1}{c|}{ No}  & \multicolumn{1}{c|} {36692}& \multicolumn{1}{c|}{183831}& \multicolumn{1}{p{3cm}|}{}&\multicolumn{1}{c|}{}&\multicolumn{1}{c|}{Metis+MQI} & \multicolumn{1}{c|}{Yes }\\
   
    & \multicolumn{1}{c|}{COAUTH-ASTRO-PH}&\multicolumn{1}{c|}{ No}  & \multicolumn{1}{c|} {18772}& \multicolumn{1}{c|}{198110}& \multicolumn{1}{p{3cm}|}{}&\multicolumn{1}{c|}{}&\multicolumn{1}{c|}{Leighton-Ratio} & \multicolumn{1}{c|}{No }\\

   & \multicolumn{1}{c|}{EPINIONS}&\multicolumn{1}{c|}{ No}  & \multicolumn{1}{c|} {75879}& \multicolumn{1}{c|}{508837}& \multicolumn{1}{p{3cm}|}{}&\multicolumn{1}{c|}{}&\multicolumn{1}{c|}{Graclus} & \multicolumn{1}{c|}{No }\\

   & \multicolumn{1}{c|}{}&\multicolumn{1}{c|}{}  & \multicolumn{1}{c|} {}& \multicolumn{1}{c|}{}& \multicolumn{1}{p{3cm}|}{}&\multicolumn{1}{c|}{}&\multicolumn{1}{c|}{Modulariy} & \multicolumn{1}{c|}{No }\\
   \hhline{---------}

   
    \multirow{14}{*}{Xie and al.(2011a)} & \multicolumn{1}{c|}{LFR }&\multicolumn{1}{c|}{ Yes}  & \multicolumn{1}{c|} { $\ne$ sizes}& \multicolumn{1}{c|}{$\ne$ sizes}& \multicolumn{1}{c|}{Overlapping modularity}&\multicolumn{1}{c|}{LFR, ALL,}&\multicolumn{1}{c|}{ Cfinder} & \multicolumn{1}{c|}{Yes }\\
               
     & \multicolumn{1}{c|}{H.S. friendship }&\multicolumn{1}{c|}{ Yes}  & \multicolumn{1}{c|} {795}& \multicolumn{1}{c|}{795}& \multicolumn{1}{c|}{NMI}&\multicolumn{1}{c|}{LFR, H.S, Friendship}&\multicolumn{1}{c|}{LFM} & \multicolumn{1}{c|}{Yes }\\
   
    & \multicolumn{1}{c|}{Amazon }&\multicolumn{1}{c|}{ Yes}  & \multicolumn{1}{c|} {334863}& \multicolumn{1}{c|}{925872}& \multicolumn{1}{c|}{Omega Index}&\multicolumn{1}{c|}{LFR}&\multicolumn{1}{c|}{EAGLE} & \multicolumn{1}{c|}{Yes }\\
   
    & \multicolumn{1}{c|}{Karate }&\multicolumn{1}{c|}{ No}  & \multicolumn{1}{c|} {34}& \multicolumn{1}{c|}{78}& \multicolumn{1}{c|}{Precision}&\multicolumn{1}{c|}{LFR}&\multicolumn{1}{c|}{CIS} & \multicolumn{1}{c|}{Yes }\\
   
    & \multicolumn{1}{c|}{Football }&\multicolumn{1}{c|}{ No}  & \multicolumn{1}{c|} {115}& \multicolumn{1}{c|}{613}& \multicolumn{1}{c|}{Recall}&\multicolumn{1}{c|}{LFR}&\multicolumn{1}{c|}{GCE} & \multicolumn{1}{c|}{Yes }\\
   
    & \multicolumn{1}{c|}{Lesmis}&\multicolumn{1}{c|}{ No}  & \multicolumn{1}{c|} {77}& \multicolumn{1}{c|}{254}& \multicolumn{1}{c|}{Community Size Distribution}&\multicolumn{1}{c|}{LFR}&\multicolumn{1}{c|}{COPRA} & \multicolumn{1}{c|}{Yes }\\
   
    & \multicolumn{1}{c|}{Dolphins }&\multicolumn{1}{c|}{No}  & \multicolumn{1}{c|} {62}& \multicolumn{1}{c|}{159}& \multicolumn{1}{c|}{Overlapping Density}&\multicolumn{1}{c|}{LFR}&\multicolumn{1}{c|}{Game} & \multicolumn{1}{c|}{Yes }\\
   
    & \multicolumn{1}{c|}{CA-GrQc}&\multicolumn{1}{c|}{ No}  & \multicolumn{1}{c|} {4730}& \multicolumn{1}{c|}{28980}& \multicolumn{1}{c|}{}&\multicolumn{1}{c|}{}&\multicolumn{1}{c|}{NMF} & \multicolumn{1}{c|}{Yes }\\
   
    & \multicolumn{1}{c|}{PGP }&\multicolumn{1}{c|}{ No}  & \multicolumn{1}{c|} {10680}& \multicolumn{1}{c|}{48632}& \multicolumn{1}{c|}{}&\multicolumn{1}{c|}{}&\multicolumn{1}{c|}{MOSES} & \multicolumn{1}{c|}{Yes }\\
   
    & \multicolumn{1}{c|}{Email}&\multicolumn{1}{c|}{ No}  & \multicolumn{1}{c|} {33696}& \multicolumn{1}{c|}{367662}& \multicolumn{1}{c|}{}&\multicolumn{1}{c|}{}&\multicolumn{1}{c|}{Link} & \multicolumn{1}{c|}{Yes }\\
   
    & \multicolumn{1}{c|}{P2P }&\multicolumn{1}{c|}{ No}  & \multicolumn{1}{c|} {62561}& \multicolumn{1}{c|}{295782}& \multicolumn{1}{c|}{}&\multicolumn{1}{c|}{}&\multicolumn{1}{c|}{iLCD} & \multicolumn{1}{c|}{Yes }\\
   
    & \multicolumn{1}{c|}{Epinions}&\multicolumn{1}{c|}{No}  & \multicolumn{1}{c|} {75877}& \multicolumn{1}{c|}{405739}& \multicolumn{1}{c|}{}&\multicolumn{1}{c|}{}&\multicolumn{1}{c|}{UEOC} & \multicolumn{1}{c|}{Yes }\\
   
    & \multicolumn{1}{c|}{}&\multicolumn{1}{c|}{ }  & \multicolumn{1}{c|} {}& \multicolumn{1}{c|}{}& \multicolumn{1}{c|}{}&\multicolumn{1}{c|}{}&\multicolumn{1}{c|}{OSLOM} & \multicolumn{1}{c|}{Yes }\\
   
    & \multicolumn{1}{c|}{}&\multicolumn{1}{c|}{ }  & \multicolumn{1}{c|} {}& \multicolumn{1}{c|}{}& \multicolumn{1}{c|}{}&\multicolumn{1}{c|}{}&\multicolumn{1}{c|}{SLPA} & \multicolumn{1}{c|}{Yes }\\

   \hhline{---------}
   

    \multirow{6}{*}{Almeida and al. (2011)} & \multicolumn{1}{c|}{Karate club }&\multicolumn{1}{c|}{No}  & \multicolumn{1}{c|} { 34}& \multicolumn{1}{c|}{78}& \multicolumn{1}{c|}{Modularity}&\multicolumn{1}{c|}{All Graphs}&\multicolumn{1}{c|}{ Markov Clustering} & \multicolumn{1}{c|}{No }\\
               
     & \multicolumn{1}{c|}{A.C. football}&\multicolumn{1}{c|}{No}  & \multicolumn{1}{c|} {115}& \multicolumn{1}{c|}{615}& \multicolumn{1}{c|}{Silhouette Index}&\multicolumn{1}{c|}{All Graphs}&\multicolumn{1}{c|}{Bisecting K-means} & \multicolumn{1}{c|}{No }\\
   
    & \multicolumn{1}{c|}{Astrophysics}&\multicolumn{1}{c|}{No}  & \multicolumn{1}{c|} {18772}& \multicolumn{1}{c|}{396160}& \multicolumn{1}{c|}{Conductance}&\multicolumn{1}{c|}{All Graphs}&\multicolumn{1}{c|}{Spectral Clustering} & \multicolumn{1}{c|}{No}\\
   
    &\multicolumn{1}{c|}{H.E. Physics }  & \multicolumn{1}{c|} {No}& \multicolumn{1}{c|}{12008}& \multicolumn{1}{c|}{237010}&\multicolumn{1}{c|}{Coverage}& \multicolumn{1}{c|}{All Graphs}&\multicolumn{1}{c|}{Normalized Cut} & \multicolumn{1}{c|}{No }\\
   
    & \multicolumn{1}{c|}{ArXiv}&\multicolumn{1}{c|}{No }  & \multicolumn{1}{c|} {34546}& \multicolumn{1}{c|}{421587}& \multicolumn{1}{c|}{Performance}&\multicolumn{1}{c|}{All Graphs}&\multicolumn{1}{c|}{} & \multicolumn{1}{c|}{}\\

    & \multicolumn{1}{c|}{Gnutella P2P}&\multicolumn{1}{c|}{No}  & \multicolumn{1}{c|} {36682}& \multicolumn{1}{c|}{88328}& \multicolumn{1}{c|}{}&\multicolumn{1}{c|}{}&\multicolumn{1}{c|}{} & \multicolumn{1}{c|}{ }\\

   \hhline{---------}
   

    \multirow{12}{*}{Yang and Leskovec (2014)} & \multicolumn{1}{c|}{LiveJournal }&\multicolumn{1}{c|}{Yes}  & \multicolumn{1}{c|} { 4M}& \multicolumn{1}{c|}{34,9M}& \multicolumn{1}{c|}{Connectivity of communities}&\multicolumn{1}{c|}{LFR, AGM}&\multicolumn{2}{c|}{ No Algorithms}\\

    & \multicolumn{1}{c|}{Friendster }&\multicolumn{1}{c|}{Yes}  & \multicolumn{1}{c|} { 117M}& \multicolumn{1}{c|}{2,586,1M}& \multicolumn{1}{l|}{Edge probability as a function of shared communities}&\multicolumn{1}{c|}{LiveJournal, Friendster, Orkut, DBLP, IMDB, Amazon}&\multicolumn{2}{c|}{ }\\

    & \multicolumn{1}{c|}{Orkut}&\multicolumn{1}{c|}{Yes}  & \multicolumn{1}{c|} { 3M}& \multicolumn{1}{c|}{117,2M}& \multicolumn{1}{c|}{Connector resides in the overlap}&\multicolumn{1}{c|}{LiveJournal}&\multicolumn{2}{c|}{ }\\

    & \multicolumn{1}{c|}{DBLP}&\multicolumn{1}{c|}{Yes}  & \multicolumn{1}{c|} { 0,4M}& \multicolumn{1}{c|}{1,3M}& \multicolumn{1}{c|}{Inside the group}&\multicolumn{1}{c|}{LFR, AGM, LiveJournal}&\multicolumn{2}{c|}{ }\\

    & \multicolumn{1}{c|}{IMDB}&\multicolumn{1}{c|}{Yes}  & \multicolumn{1}{c|} {1,3M}& \multicolumn{1}{c|}{39,8M}& \multicolumn{1}{c|}{Maximal ICDF}&\multicolumn{1}{c|}{LFR, AGM}&\multicolumn{2}{c|}{ }\\

    & \multicolumn{1}{c|}{Amazon }&\multicolumn{1}{c|}{Yes}  & \multicolumn{1}{c|} { 0,3M}& \multicolumn{1}{c|}{0,9M}& \multicolumn{1}{c|}{Community overlaps}&\multicolumn{1}{c|}{LFR, AGM}&\multicolumn{2}{c|}{ }\\

    & \multicolumn{1}{c|}{LFR }&\multicolumn{1}{c|}{Yes}  & \multicolumn{1}{c|} { $\ne$ sizes}& \multicolumn{1}{c|}{ $\ne$ sizes}& \multicolumn{1}{c|}{Degree distribution}&\multicolumn{1}{c|}{All Graphs}&\multicolumn{2}{c|}{ }\\

    & \multicolumn{1}{c|}{AGM}&\multicolumn{1}{c|}{Yes}  & \multicolumn{1}{c|} { $\ne$ sizes}& \multicolumn{1}{c|}{ $\ne$ sizes}& \multicolumn{1}{c|}{Clustering coefficient}&\multicolumn{1}{c|}{All Graphs}&\multicolumn{2}{c|}{ }\\

    & \multicolumn{1}{c|}{}&\multicolumn{1}{c|}{}  & \multicolumn{1}{c|} {}& \multicolumn{1}{c|}{}& \multicolumn{1}{c|}{Hop plot}&\multicolumn{1}{c|}{All Graphs}&\multicolumn{2}{c|}{ }\\
    
   & \multicolumn{1}{c|}{}&\multicolumn{1}{c|}{}  & \multicolumn{1}{c|} {}& \multicolumn{1}{c|}{}& \multicolumn{1}{c|}{Triad participation}&\multicolumn{1}{c|}{All Graphs}&\multicolumn{2}{c|}{ }\\
   
   & \multicolumn{1}{c|}{}&\multicolumn{1}{c|}{}  & \multicolumn{1}{c|} {}& \multicolumn{1}{c|}{}& \multicolumn{1}{c|}{Eigenvalues}&\multicolumn{1}{c|}{All Graphs}&\multicolumn{2}{c|}{ }\\
   
   & \multicolumn{1}{c|}{}&\multicolumn{1}{c|}{}  & \multicolumn{1}{c|} {}& \multicolumn{1}{c|}{}& \multicolumn{1}{c|}{Eigenvector}&\multicolumn{1}{c|}{All Graphs}&\multicolumn{2}{c|}{ }\\
   
   \hhline{---------}
   

    \multirow{16}{*}{Hric and al.} & \multicolumn{1}{c|}{LFR }&\multicolumn{1}{c|}{Yes}  & \multicolumn{1}{c|} {1000}& \multicolumn{1}{c|}{9839}& \multicolumn{1}{c|}{Group sizes,}&\multicolumn{1}{c|}{All Graphs}&\multicolumn{1}{c|}{Louvain} & \multicolumn{1}{c|}{No }\\

      & \multicolumn{1}{c|}{Karate}&\multicolumn{1}{c|}{Yes}  & \multicolumn{1}{c|} {34}& \multicolumn{1}{c|}{78}& \multicolumn{1}{c|}{NMI}&\multicolumn{1}{c|}{All Graphs}&\multicolumn{1}{c|}{infomap} & \multicolumn{1}{c|}{No }\\
   
    & \multicolumn{1}{c|}{Football}&\multicolumn{1}{c|}{Yes}  & \multicolumn{1}{c|} {115}& \multicolumn{1}{c|}{615}& \multicolumn{1}{c|}{Modularity}&\multicolumn{1}{c|}{All Graphs}&\multicolumn{1}{c|}{InfomapSingle} & \multicolumn{1}{c|}{No}\\
   
    &\multicolumn{1}{c|}{Polbooks }  & \multicolumn{1}{c|} {Yes}& \multicolumn{1}{c|}{105}& \multicolumn{1}{c|}{441}&\multicolumn{1}{c|}{Jaccard score}& \multicolumn{1}{c|}{All Graphs}&\multicolumn{1}{c|}{LinkCommunities} & \multicolumn{1}{c|}{Yes }\\
   
    & \multicolumn{1}{c|}{Polblogs}&\multicolumn{1}{c|}{Yes }  & \multicolumn{1}{c|} {1222}& \multicolumn{1}{c|}{16782}& \multicolumn{1}{c|}{Recall score}&\multicolumn{1}{c|}{All Graphs}&\multicolumn{1}{c|}{CliquePerc} & \multicolumn{1}{c|}{Yes}\\

   & \multicolumn{1}{c|}{Dpb}&\multicolumn{1}{c|}{Yes}  & \multicolumn{1}{c|} {35029}& \multicolumn{1}{c|}{161313}& \multicolumn{1}{c|}{Precision score}&\multicolumn{1}{c|}{All Graphs}&\multicolumn{1}{c|}{Conclude} & \multicolumn{1}{c|}{Yes }\\
   
    & \multicolumn{1}{c|}{As-caida}&\multicolumn{1}{c|}{Yes}  & \multicolumn{1}{c|} {46676}& \multicolumn{1}{c|}{262953}& \multicolumn{1}{c|}{}&\multicolumn{1}{c|}{}&\multicolumn{1}{c|}{COPRA} & \multicolumn{1}{c|}{Yes}\\
   
    &\multicolumn{1}{c|}{Fb100 }  & \multicolumn{1}{c|} {Yes}& \multicolumn{1}{c|}{41536}& \multicolumn{1}{c|}{1465654}&\multicolumn{1}{c|}{}& \multicolumn{1}{c|}{}&\multicolumn{1}{c|}{Demon} & \multicolumn{1}{c|}{Yes}\\
   
    & \multicolumn{1}{c|}{PGP}&\multicolumn{1}{c|}{Yes }  & \multicolumn{1}{c|} {81036}& \multicolumn{1}{c|}{190143}& \multicolumn{1}{c|}{}&\multicolumn{1}{c|}{}&\multicolumn{1}{c|}{Ganxis SLPA} & \multicolumn{1}{c|}{Yes}\\

    & \multicolumn{1}{c|}{ANoBII}&\multicolumn{1}{c|}{Yes }  & \multicolumn{1}{c|} {136547}& \multicolumn{1}{c|}{892377}& \multicolumn{1}{c|}{}&\multicolumn{1}{c|}{}&\multicolumn{1}{c|}{GreedyCliqueExp} & \multicolumn{1}{c|}{Yes}\\

    & \multicolumn{1}{c|}{DBLP}&\multicolumn{1}{c|}{Yes }  & \multicolumn{1}{c|} {317080}& \multicolumn{1}{c|}{1049866}& \multicolumn{1}{c|}{}&\multicolumn{1}{c|}{}&\multicolumn{1}{c|}{} & \multicolumn{1}{c|}{}\\

    & \multicolumn{1}{c|}{Amazon}&\multicolumn{1}{c|}{Yes }  & \multicolumn{1}{c|} {366997}& \multicolumn{1}{c|}{1231439}& \multicolumn{1}{c|}{}&\multicolumn{1}{c|}{}&\multicolumn{1}{c|}{} & \multicolumn{1}{c|}{}\\
   
    & \multicolumn{1}{c|}{Flickr}&\multicolumn{1}{c|}{Yes }  & \multicolumn{1}{c|} {1715255}& \multicolumn{1}{c|}{22613981}& \multicolumn{1}{c|}{}&\multicolumn{1}{c|}{}&\multicolumn{1}{c|}{} & \multicolumn{1}{c|}{}\\

    & \multicolumn{1}{c|}{Orkut}&\multicolumn{1}{c|}{Yes }  & \multicolumn{1}{c|} {3072441}& \multicolumn{1}{c|}{117185083}& \multicolumn{1}{c|}{}&\multicolumn{1}{c|}{}&\multicolumn{1}{c|}{} & \multicolumn{1}{c|}{}\\

& \multicolumn{1}{c|}{Lj-backstrom}&\multicolumn{1}{c|}{Yes }  & \multicolumn{1}{c|} {4843953}& \multicolumn{1}{c|}{43362750}& \multicolumn{1}{c|}{}&\multicolumn{1}{c|}{}&\multicolumn{1}{c|}{} & \multicolumn{1}{c|}{}\\

    & \multicolumn{1}{c|}{Lj-mislove}&\multicolumn{1}{c|}{Yes }  & \multicolumn{1}{c|} {5189809}& \multicolumn{1}{c|}{49151786}& \multicolumn{1}{c|}{}&\multicolumn{1}{c|}{}&\multicolumn{1}{c|}{} & \multicolumn{1}{c|}{}\\

   \hhline{---------}


    \multirow{13}{*}{Harenberg and al. (2014)} & \multicolumn{1}{c|}{Amazon}&\multicolumn{1}{c|}{Yes}  & \multicolumn{1}{c|} {8275}& \multicolumn{1}{c|}{22231}& \multicolumn{1}{c|}{Density}&\multicolumn{1}{c|}{All Graphs}&\multicolumn{1}{c|}{SLPA} & \multicolumn{1}{c|}{Yes }\\

      & \multicolumn{1}{c|}{Youtube}&\multicolumn{1}{c|}{Yes}  & \multicolumn{1}{c|} {12091}& \multicolumn{1}{c|}{29775}& \multicolumn{1}{c|}{Clustering coefficient}&\multicolumn{1}{c|}{All Graphs}&\multicolumn{1}{c|}{TopGC} & \multicolumn{1}{c|}{Yes }\\
   
    & \multicolumn{1}{c|}{DBLP}&\multicolumn{1}{c|}{Yes}  & \multicolumn{1}{c|} {26956}& \multicolumn{1}{c|}{88742}& \multicolumn{1}{c|}{Conductance}&\multicolumn{1}{c|}{All Graphs}&\multicolumn{1}{c|}{SVINET} & \multicolumn{1}{c|}{Yes}\\
   
    &\multicolumn{1}{c|}{LiveJournal }  & \multicolumn{1}{c|} {Yes}& \multicolumn{1}{c|}{44093}& \multicolumn{1}{c|}{871409}&\multicolumn{1}{c|}{Triangle participation ratio}& \multicolumn{1}{c|}{All Graphs}&\multicolumn{1}{c|}{MCD} & \multicolumn{1}{c|}{No }\\

   & \multicolumn{1}{c|}{Orkut}&\multicolumn{1}{c|}{Yes}  & \multicolumn{1}{c|} {297691}& \multicolumn{1}{c|}{7747026}& \multicolumn{1}{c|}{Precision}&\multicolumn{1}{c|}{All Graphs}&\multicolumn{1}{c|}{CGGCi-RG} & \multicolumn{1}{c|}{No}\\

   & \multicolumn{1}{c|}{}&\multicolumn{1}{c|}{}  & \multicolumn{1}{c|} {}& \multicolumn{1}{c|}{}& \multicolumn{1}{c|}{Recall}&\multicolumn{1}{c|}{All Graphs}&\multicolumn{1}{c|}{CONCLUDE} & \multicolumn{1}{c|}{No}\\

   & \multicolumn{1}{c|}{}&\multicolumn{1}{c|}{}  & \multicolumn{1}{c|} {}& \multicolumn{1}{c|}{}& \multicolumn{1}{c|}{F-measure}&\multicolumn{1}{c|}{All Graphs}&\multicolumn{1}{c|}{DSE} & \multicolumn{1}{c|}{No}\\

   & \multicolumn{1}{c|}{}&\multicolumn{1}{c|}{}  & \multicolumn{1}{c|} {}& \multicolumn{1}{c|}{}& \multicolumn{1}{c|}{Specificity}&\multicolumn{1}{c|}{All Graphs}&\multicolumn{1}{c|}{SPICi} & \multicolumn{1}{c|}{No}\\

   & \multicolumn{1}{c|}{}&\multicolumn{1}{c|}{}  & \multicolumn{1}{c|} {}& \multicolumn{1}{c|}{}& \multicolumn{1}{c|}{Accuracy}&\multicolumn{1}{c|}{All Graphs}&\multicolumn{1}{c|}{CFinder} & \multicolumn{1}{c|}{Yes}\\

   & \multicolumn{1}{c|}{}&\multicolumn{1}{c|}{}  & \multicolumn{1}{c|} {}& \multicolumn{1}{c|}{}& \multicolumn{1}{c|}{NMI}&\multicolumn{1}{c|}{All Graphs}&\multicolumn{1}{c|}{FastGreedy} & \multicolumn{1}{c|}{Yes}\\

   & \multicolumn{1}{c|}{}&\multicolumn{1}{c|}{}  & \multicolumn{1}{c|} {}& \multicolumn{1}{c|}{}& \multicolumn{1}{c|}{Similarity}&\multicolumn{1}{c|}{All Graphs}&\multicolumn{1}{c|}{LPA} & \multicolumn{1}{c|}{No}\\

   & \multicolumn{1}{c|}{}&\multicolumn{1}{c|}{}  & \multicolumn{1}{c|} {}& \multicolumn{1}{c|}{}& \multicolumn{1}{c|}{}&\multicolumn{1}{c|}{}&\multicolumn{1}{c|}{LE} & \multicolumn{1}{c|}{No}\\

   & \multicolumn{1}{c|}{}&\multicolumn{1}{c|}{}  & \multicolumn{1}{c|} {}& \multicolumn{1}{c|}{}& \multicolumn{1}{c|}{}&\multicolumn{1}{c|}{}&\multicolumn{1}{c|}{Walktrap} & \multicolumn{1}{c|}{No}\\

   \hhline{---------}


    \multirow{12}{*}{Creusefond and al. (2016)} & \multicolumn{1}{c|}{DBLP}&\multicolumn{1}{c|}{Yes}  & \multicolumn{1}{c|} {129981}& \multicolumn{1}{c|}{332595}& \multicolumn{1}{c|}{The Local internal clustering coefficient}&\multicolumn{1}{c|}{All except LFR}&\multicolumn{1}{c|}{Louvain} & \multicolumn{1}{c|}{No}\\

      & \multicolumn{1}{c|}{CS}&\multicolumn{1}{c|}{Yes}  & \multicolumn{1}{c|} {400657}& \multicolumn{1}{c|}{1428030}& \multicolumn{1}{c|}{Performance}&\multicolumn{1}{c|}{All except LFR}&\multicolumn{1}{c|}{Clauset} & \multicolumn{1}{c|}{No }\\
   
    & \multicolumn{1}{c|}{Actors(imdb)}&\multicolumn{1}{c|}{Yes}  & \multicolumn{1}{c|} {124414}& \multicolumn{1}{c|}{20489642}& \multicolumn{1}{c|}{Flak-ODF}&\multicolumn{1}{c|}{All except LFR}&\multicolumn{1}{c|}{MCL} & \multicolumn{1}{c|}{No}\\
   
    &\multicolumn{1}{c|}{Github }  & \multicolumn{1}{c|} {Yes}& \multicolumn{1}{c|}{39845}& \multicolumn{1}{c|}{22277795}&\multicolumn{1}{c|}{Fraction Over Median Degree}& \multicolumn{1}{c|}{All except LFR}&\multicolumn{1}{c|}{Infomap} & \multicolumn{1}{c|}{No }\\

   & \multicolumn{1}{c|}{LiveJournal}&\multicolumn{1}{c|}{Yes}  & \multicolumn{1}{c|} {1143395}& \multicolumn{1}{c|}{16880773}& \multicolumn{1}{c|}{Conductance}&\multicolumn{1}{c|}{All except LFR}&\multicolumn{1}{c|}{LexDFS} & \multicolumn{1}{c|}{No}\\

   & \multicolumn{1}{c|}{Youtube}&\multicolumn{1}{c|}{Yes}  & \multicolumn{1}{c|} {51204}& \multicolumn{1}{c|}{317393}& \multicolumn{1}{c|}{Cut-ratio}&\multicolumn{1}{c|}{All except LFR}&\multicolumn{1}{c|}{3-score} & \multicolumn{1}{c|}{No}\\

   & \multicolumn{1}{c|}{Flickr}&\multicolumn{1}{c|}{Yes}  & \multicolumn{1}{c|} {368285}& \multicolumn{1}{c|}{11915549}& \multicolumn{1}{c|}{Compactness}&\multicolumn{1}{c|}{All except LFR}&\multicolumn{1}{c|}{label propagation} & \multicolumn{1}{c|}{No}\\

   & \multicolumn{1}{c|}{Amazon}&\multicolumn{1}{c|}{Yes}  & \multicolumn{1}{c|} {147510}& \multicolumn{1}{c|}{267135}& \multicolumn{1}{c|}{Modulariy}&\multicolumn{1}{c|}{All except LFR}&\multicolumn{1}{c|}{} & \multicolumn{1}{c|}{}\\

   & \multicolumn{1}{c|}{Football}&\multicolumn{1}{c|}{Yes}  & \multicolumn{1}{c|} {115}& \multicolumn{1}{c|}{613}& \multicolumn{1}{c|}{Surprise}&\multicolumn{1}{c|}{All except LFR}&\multicolumn{1}{c|}{} & \multicolumn{1}{c|}{}\\
   
   & \multicolumn{1}{c|}{Cora}&\multicolumn{1}{c|}{Yes}  & \multicolumn{1}{c|} {23165}& \multicolumn{1}{c|}{89156}& \multicolumn{1}{c|}{Significance}&\multicolumn{1}{c|}{All except LFR}&\multicolumn{1}{c|}{} & \multicolumn{1}{c|}{}\\
   
   & \multicolumn{1}{c|}{LFR}&\multicolumn{1}{c|}{Yes}  & \multicolumn{1}{c|} {$\ne$ sizes}& \multicolumn{1}{c|}{$\ne$ sizes}& \multicolumn{1}{c|}{NMI}&\multicolumn{1}{c|}{All Graphs}&\multicolumn{1}{c|}{} & \multicolumn{1}{c|}{}\\
   
   & \multicolumn{1}{c|}{}&\multicolumn{1}{c|}{}  & \multicolumn{1}{c|} {}& \multicolumn{1}{c|}{}& \multicolumn{1}{c|}{F-BCubed}&\multicolumn{1}{c|}{All Graphs}&\multicolumn{1}{c|}{} & \multicolumn{1}{c|}{}\\

   \hhline{---------}

    \hline
  \end{tabular}
  
  }
  \caption{\label{table43} Main characteristics of related works.}
\end{table}


\section{Background}
        In this section, we present the overlapping community detection algorithms analyzed in our study, together with the quality and clustering metrics designed for the purpose of evaluating community structure. We recall the network topological properties classically computed in the network science literature. As we plan to compare the detection algorithms trough this set of features rather than a single property, we present the most influential multiple criteria decision making algorithms that are used in order to rank the community detection algorithms.

\subsection{Overlapping community detection algorithms}

        There is a great deal of work devoted to the community detection issues. Many solutions based on various definitions are frequently published. In order to get a better understanding on the subject, some recent surveys have proposed taxonomies of the community detection methods \cite{SAMSAM10133, journals/csur/XieKS13}. In this work, ten overlapping community detection methods are evaluated. Our choice is based on various criteria: the availability of their source code, their complexity, and their popularity. Moreover, we selected them such that they belong to various categories according to the classification reported in \cite{journals/csur/XieKS13}.

        Table \ref{table41} reports the complexity and the classification of the considered algorithms.

        \begin{table}[htpb!]
        \renewcommand{\arraystretch}{1.3}
        \caption{Algorithms used for detecting overlapping community structure ranked by year. The classes are Clique Percolation (CP), Local Expansion/Optimization (LE/O), Fuzzy Detection (FD), Line Graph/Link Partitioning. (LG/LP), Label Propagation (LP)}
        \label{table41}
        \centering
        \begin{tabular}{lllll}
        \hline
         Algorithm&Classes&Reference&Complexity
        \\  \hline
                CFINDER&CP& \cite{Palla2005uncovering}&polynomial\\
                LFM&LE/O&\cite{Lancichinetti2008}&$O(n^2)$\\
                GCE&LE/O&\cite{lee2010detecting}&$O(mh)$\\
                OSLOM&LE/O&\cite{lancichinetti2011finding}&$O(n^2)$\\
                LINKC&LG/LP&\cite{Ahn2009}&$O(nk_{max}^2)$\\
                SVINET&LG/LP&\cite{Gopalan03092013}& not explicitly stated\\
                MOSES&FD&\cite{conf/asunam/McDaidH10}&$O(en^2)$\\
                SLPA&LP&\cite{conf/icdm/XieSL11}&$O(tm)$\\
                DEMON&LP&\cite{conf/kdd/CosciaRGP12}&$O(n+m)$\\

        \hline
        \end{tabular}
        \end{table}

        \textbf{Clique Finder}\protect\footnote{\url{http://www.cfinder.org/}} (CFINDER).
        It is the implementation of the Clique Percolation method. It assumes that a community is made of highly connected cliques. Indeed, it is defined as the largest subgraph composed of adjacent k-clique. Note that a k-clique is a subset of $k$ vertices which form a complete subgraph. Two k-clique are adjacent if they share (k-1) links. CFINDER has a polynomial time data complexity.

        \textbf{Lancichinetti Fortunato Method}\protect\footnote{\url{https://github.com/sumnous/LFM_improve}} (LFM). It takes a random seed node and adds nodes to it until a fitness function is locally maximal. After assembling one community, the same process is applied on another seed node not yet assigned to any community in order to grow a new community. The fitness function controls the strength and the size of the communities. The worst-case complexity is $O(n^2)$ where $n$ is the number of nodes.

        \textbf{Greedy Clique Expansion}\protect\footnote{\url{https://sites.google.com/site/greedycliqueexpansion/}} (GCE). It is based on the same principle that LFM. Rather than using a random node as a seed, maximal cliques are the starting elements of a community. These seeds are expanded by greedily optimizing a local fitness function. The time complexity for GCE is $O(mh)$, where $m$ is the number of edges, and $h$ is the number of cliques.

        \textbf{Order Statistics Local Optimization Method}\protect\footnote{\url{http://oslom.org/}} (OSLOM). It starts by detecting seed communities using a non-overlapping community detection algorithm (Infomap or Louvain).  Then, a random node from these seeds is linked with an arbitrary number of neighbors to establish the overlap zones. For each grain, OSLOM applies rules to successively add and remove nodes until reaching a stable state. Its time complexity is $O(n^2)$, where $n$ is the number of nodes.

        \textbf{Link Communities}\protect\footnote{\url{http://barabasilab.neu.edu/projects/linkcommunities/}} (LINKC). It builds a partition of links via hierarchical clustering of edge similarity. It uses the Jaccard similarity coefficient for links with at least one node in common. Then, a classical hierarchical clustering process builds a link dendrogram which is cut at some clustering threshold in order to optimize the partition density. Its time complexity is $O(nk_{max}^2)$ where $n$ is the number of nodes and $k_{max}$ is the maximum node degree in a network.

        \textbf{Stochastic Variational Inference NETwork}\protect\footnote{\url{https://github.com/premgopalan/svinet}}(SVINET). This algorithm considers a probabilistic membership model in order to create overlap zones. It begins by defining a posterior distribution of overlap size that ensures the high density of overlap zones. Then, sub-sampling the network, analyzing the sub-sample, and updating the estimated community structure is done in order to approximate the posterior. Its complexity is not explicitly stated.

       \textbf{Model-Based Overlapping ExpanSion}\protect\footnote{\url{https://sites.google.com/site/aaronmcdaid/moses}} (MOSES). It computes the Fuzzy Detection with a fitness function based on OSBM (Overlapping Stochastic Block Models) proposed by \cite{Latouche2011}. It uses extensive probability for nodes connection in order to take prior community assignments equivalence. As a result, the number of communities possesses a realistic distribution (power law). The computational time complexity is equal to $O(en^2)$ where $n$ is the number of nodes and $e$ is the number of edges to be expanded.

       \textbf{Speaker-listener Label Propagation Algorithm}\protect\footnote{\url{https://sites.google.com/site/communitydetectionslpa/}} (SLPA). It is an extension of the Label Propagation Algorithm (LPA). While in LPA, each node holds only a single label that is iteratively updated by adopting the majority label in the neighborhood, in SLPA each node possesses a memory containing multiple labels. Starting from a node selected as a listener, its neighbors send out a label following certain speaking rules. The listener selects one label according to a listening rule and adds it to its memory. Once all the nodes have been visited, the communities are extracted from the node's memory converted into a probability distribution of labels that defines the membership degree to communities. SLPA has a time complexity equals to $O(tm)$  when $m$ is the total number of edges and t is the memory size.

       \textbf{Democratic Estimate of the Modular Organization of a Network}\protect\footnote{\url{http://www.michelecoscia.com/?page_id=42}} (DEMON). This method tends to affect a node to the most frequent community by the application of a label propagation algorithm on its neighbors sub-graphs. In other words, for each node, their neighbors vote for its community membership. All the votes are then combined to construct the overlapping community structure. Its time complexity equals to $O(n+m)$ where $n$ is the number of nodes and $m$ is the number of edges.

\subsection{Quality metrics}\label{quality}
        The quality metrics tends to answer the question: What is a good community structure? They are usually based on local properties of the communities. The knowledge of the ground-truth community membership is not necessary in this case.

        We use five quality metrics that are reported in \cite{yang2015defining}. According to these authors, the quality metrics can be categorized into four classes (internal connectivity, external connectivity, internal and external connectivity combination, network model). In our study, we restrict our attention to metrics belonging to three classes.

\subsubsection{Scoring functions based on internal connectivity}
\paragraph*{Average degree}
        This measure computes the average internal degree of the members of a community. It is given by $f(S)=\frac{2m_{s}}{n_{s}}$, where S is the community, $m_{s}$ is the number of links of S and $n_{s}$ is the number of nodes of S.
\paragraph*{Internal density}
        The internal density is the edge density of nodes of a community. For a community S, the internal density is given by $ f(S)=\frac{m_{s}}{n_{s}(n_{s}-1)/2}$, where $m_{s}$ is the number of links of S and $n_{s}$ is the number of nodes of S.
        \vspace{.3cm}
\subsubsection{Scoring functions that combine internal and external connectivity}
\paragraph*{Maximum-Out Degree Fraction (Max-ODF)}
        The Max-ODF is the maximum fraction of edges of a node that point outside its community. It is given by $f(S)= max_{u\in S}\frac{|\{(u,v)\in E:v\notin S\}|}{d(u)}$, where d(u) is the degree of node u.
\paragraph*{Average-Out Degree Fraction (Average-ODF)}
        The Average-ODF gives the information of the inter-edges of a community. For a community S, the Average-ODF is given by $f(S)=\frac{1}{n_{s}}\sum_{u\in S}\frac{|\{(u,v)\in E:v\notin S\}|}{d(u)}$, where $n_{s}$ is the number of nodes of S and d(u) is the degree of node u.
\paragraph*{Flake-Out Degree Fraction (Flake-ODF)}
        The Flake-ODF is the fraction of nodes in S that have fewer intra-edges than the inter-edges. It is given by $f(S)=\frac{|\{u: u\in S,|\{(u,v)\in E:v\in S\}|<d(u)/2\}|}{n_{s}}$, where S is a community, E the set of edges of the graph,  d(u) is the degree of node u, and $n_{s}$ is the number of nodes of S.

        Note that these definitions are given for a single community. They must be averaged in order to qualify the overall community structure quality.
\subsubsection{Scoring function based on a network model}
\paragraph*{Overlapping Modularity}
        The modularity was introduced by \cite{newman2004finding} in order to formulate the fact that a subgraph is a community if the number of connections between its nodes is higher than what would be expected if links were randomly assigned. It is described as the proportion of incident edges on a given subgraph minus the number of edges arranged randomly on the same subgraph. High modularity means that connections of nodes within communities are denser than those between nodes in different modules. The 'Newman' definition of modularity is specific for non-overlapping communities. Several extensions to the overlapping case have been proposed in the literature. We use the one recently introduced by \cite{chen2015fuzzy}. It is defined as follows:
        \begin{equation}\label{eq1}
          Q_{ov}=\displaystyle \sum_{c\in C}[\frac{|E_{c}^{in}|}{|E|}-(\frac{2|E_{c}^{in}|+|E_{c}^{out}|}{2|E|})^{2}].
        \end{equation}
        where $|E|$ is the number of edges, $|E_{c}^{in}|$ are the $c$ intra-community edges and $|E_{c}^{out}|$ are the $c$ inter-community edges.

\subsection{Clustering metrics}
        The clustering metrics compare the communities discovered by the algorithms to the ones given by the ground-truth. A lot of metrics have been proposed in the literature. They can be classified into three main categories: measures based on information theory, measures based on pair counting, and set-matching-based measures. Note that they are more or less correlated \cite{labatut:hal-00611319}. Indeed, most of them can be derived from the confusion matrix whose elements are the number of nodes that are common to both partitions.

\subsubsection{Information-theoretic-based measures}
        The metrics of this category are based on the mutual information shared by two partitions. When two partitions are independent, they do not share any information, while when they are identical, the information shared is maximum.

        The normalized mutual information (NMI), defined in order to compare two partitions, is the most famous Information-theoretic-based measure. Its extension to compare overlapping communities is not trivial, and there are several alternatives \cite{Lancichinetti2008, MEILA2007873}. In this work, we use the version proposed by \cite{mcdaid2011normalized}. It is defined by:

        \begin{equation}\label{eq41}
          NMI_{max}=\frac{I(C_1:C_2)}{max(H(C_1),H(C_2))}
        \end{equation}
        where
        \begin{equation}\label{eq42}
        I(C_1 : C_2 )= 1/2*[H(C_1)-H(C_1|C_2)+H(C_2)-H(C_2|C_1)]
        \end{equation}
        and $H(C_1|C_2)$ is the normalized conditional entropy of a cover $C_1$ with respect to $C_2$.

\subsubsection{Pair counting based measures}
        In this category, clustering comparison is based on counting the pairs of points on which two partitions agree or disagree. Rand Index (RI) \cite{Rand:71}  and the Jaccard Index are well-known measures in this class for comparing two partitions. The Omega-Index is the most influential pair counting based measure in the overlapping community detection literature \cite{journals/csur/XieKS13, springerlink:10.1007/978-3-642-01206-8_5, xie2012towards}. It is based on pairs of nodes in agreement in two covers. Here, a pair of nodes is considered to be in agreement if they are clustered in exactly the same number of communities. It is the overlapping extension of Adjusted Rand Index introduced by \cite{hubert1985comparing}. It is given by:
         \begin{equation}\label{eq5}
         {\omega {{(C}}_{1}{{,C}}_{2})=\frac {\omega _{u}{{(C}}_{1}{{,C}}_{2})-\omega _{e}{{(C}}_{1}{{,C}}_{2})}{1-\omega _{e}{{(C}}_{1}{{,C}}_{2})}}.
         \end{equation}
        where
         \begin{equation}\label{eq51}
         \omega _{u}{{(C}}_{1}{{,C}}_{2}) = \frac{1}{M}*\sum^{max(K_{1},K_{2})}_{j=0}|t_j(C_1)\cap t_j(C_2)|
        \end{equation}
        and
        \begin{equation}\label{eq52}
        \omega _{e}{{(C}}_{1}{{,C}}_{2}) = \frac{1}{M^2}*\sum^{max(K_{1},K_{2})}_{j=0}|t_j(C_1)* t_j(C_2)|
        \end{equation}
        $C_1, C_2$  are covers with a number of communities $K_1,K_2$. $M$ equal to $n(n-1)/2$ represents the number of node pairs and $t_j(C)$ is the set of pairs that appear exactly $j$ times in a community C. Its value ranges between 0 (no matching) and 1 (perfect match).

\subsubsection{Set-matching-based measures}
        Based on set cardinality, this class of measures intends to find the largest overlaps between pairs of communities. The proportion of correctly assigned nodes is known as Purity. Each identified community is matched to the one with the maximum overlap in the reference one, and then the accuracy of this assignment is measured by counting the number of correctly assigned nodes. Precision and Recall are the most frequently Set-matching-based used measures.

        Let us consider that instances belong either to a positive class or to a negative class. The entries of a confusion matrix are true positives (TP) (correctly classified positive instances), false positives (FP) (misclassified negatives), true negatives (TN) (correctly classified negatives) and false negatives (FN)  (misclassified positives). In an N classification problem Precision, Recall and F1-score represent the performance of the prediction for only one class. They are defined by:
        \begin{equation}\label{eq3}
          Precision = \frac{TP}{TP+FP}, \qquad
          Recall = \frac{TP}{TP + FN}
        \end{equation}
        where $TP$ is the number of true positives, $FP$ the number of false positives and $FN$ the number of false negatives.

        The F1-score (also known as balanced F-score or F-measure) is defined as the harmonic mean of Precision and Recall. It is given by:
        \begin{equation}\label{eq4}
          F1-score = 2*\frac{precision*recall}{precision + recall}
        \end{equation}
\subsection{Network topological properties} \label{TopoPropCom}
        The topological properties can be categorized into three classes: Basic properties, Microscopic, and Mesoscopic. The basic properties summarize the overall network features. The microscopic properties reflect the features of the nodes. The mesoscopic properties characterize the modular structure of the network.
\subsubsection{Basic properties}
        The distance between two nodes is defined to be the length of the shortest path between them. \emph{The average shortest path} is the average number of edges along the shortest paths between all possible pairs of network nodes. \emph{The diameter} is defined to be the maximum of all possible distances. Most of real-world networks satisfy the small-world property \ie most nodes are just a few edges away on average and the diameter is small.

        \emph{The degree correlation} measures the tendency of nodes to associate with other nodes sharing the same characteristics and especially the same degree values. In assortative networks, the nodes tend to associate with their connectivity peers, and the degree correlation is positive. In disassortative networks, high-degree nodes tend to associate with low-degree ones, and the degree correlation is negative.

        \emph{The global clustering coefficient} reflects the tendency of link formation between neighboring nodes in a network. It is defined as the proportion of triangles in networks. Usually, social networks are characterized by a high clustering coefficient.
\subsubsection{Microscopic properties}
        In order to characterize the microscopic properties of the networks, three distributions are used. One is linked to the degree of nodes, the second one is related to their clustering property and the third one describes the statistics of distance between nodes.
        \emph{The degree distribution} measures the statistical repartition of the network nodes' degrees. For a large number of networks, such distribution can be adequately described as a power-law. It can be written as $(P(k)\sim k^{-\alpha})$, where $\alpha$ is a positive exponent. Related experimental studies show that the exponent value of the power law usually ranges from 2 to 3.

        \emph{The average clustering coefficient as a function of node degree} gives details of a network triangular clustering structure. In order to estimate this distribution,  we first compute the local clustering coefficient for every node in the network. Then, for each set of nodes that has the same degree, we compute the average clustering coefficient. For a large number of networks, this distribution can be adequately represented by a Power-Law \cite{Cheng2009}.

        \emph{The hop plot} represents the distribution of pairwise distances in a network \cite{siganos2003power}. Generally, it can be well estimated by a Gaussian law. It is usually represented as a cumulative distribution in order to extract the diameter (100-percentile), the effective diameter (90-percentile) and the median path length (50-percentile).
\subsubsection{Mesoscopic properties}
        At the mesoscopic level, \cite{Palla2005uncovering} introduces four measures in order to quantify the overlapping community structure of complex networks. Three of them are related to the communities (degree, size, overlap) and one is related to the nodes (membership).

        \emph{The degree of a community} is defined as the number of communities that overlap with it. In other words, it is the degree distribution of the 'community-graph'.

        \emph{The size of a community} is the number of nodes it contains.

        \emph{The overlap size} between two communities is the number of their common nodes.

        \emph{The membership} of a node is the number of communities to which it belongs.

        The distributions of these four basic quantities allow characterizing the community structure of a network.
        Note that for a large number of networks, they can be adequately described by a power-law distribution.

\subsection{Multiple Criteria Decision-Making}
        In order to assess the effectiveness of the community detection algorithms, one cannot rely on a single property. Besides, computing multiple properties can lead to contradictory results. Therefore, a Multiple-Criteria Decision-Making strategy must be implemented in order to find the best compromise.
        To rank the algorithms, we propose to use a two steps process. In the first step, the algorithms are ranked according to each individual topological property. In the second step, all those rankings are combined using a Multiple Criteria Decision-Making strategy in order to reduce the sets of individual rankings into a unique one.
        Many Multi-Criteria Decision Making (MCDM) algorithms have been proposed in order to choose the best alternative from a set of alternatives \cite{aruldoss2013survey}.

        In our analysis, we consider two popular algorithms in the MCDM literature: Kemeny consensus and TOPSIS.

        \textbf{Kemeny consensus} (also known as rank aggregation). In this voting scheme, voters (the topological properties in our case) rank choices (the community detection algorithms) according to their order of preference.
        The Kemeny-score calculation is done in two steps. The first step is to create a matrix that counts pairwise voter preferences. The second step is to test all possible rankings, calculate a score for each  ranking, and compare the scores. Each ranking score equals the sum of the pairwise counts that apply to that ranking. The ranking that has the largest score is identified as the overall ranking \cite{betzler2010partial}.

        \textbf{Technique for Order Preference by Similarity to Ideal Solution (TOPSIS)}.
        It is based on the principle of compromise between the best and the worst solution. In other words, the chosen alternative should have the shortest distance from the positive ideal solution (PIS) and the farthest distance from the negative ideal solution (NIS).
        TOPSIS  assumes that each criterion has a tendency of monotonically increasing or decreasing utility. This allows defining easily the positive and the negative ideal solutions. The final ranking is given by a  series of comparisons of the various alternative relative distances.

\section{Data and Methods}
        This section describes the datasets, the proposed ranking methodology and the construction of the 'community-graph'.
\subsection{Data}
        The choice of a dataset is a quite difficult sensitive problem for several reasons. First of all, the real networks must be provided with a ground-truth community structure. Second, they must contain a large number of overlapping communities in order to build a community-graph with an acceptable size. Indeed, as we plan to compute topological properties of these graphs, they must be enough big so that these statistics are relevant. The last constraint is contradictory with the previous one. The size of the networks must be appropriate to the complexity issues of the topological properties computation and the overlapping community detection algorithms. Among a large number of networks available, three graphs are the best fit for these constraints: American electronic commerce company (AMAZON), Pretty Good Privacy (PGP), and social bookmarking (aNobii). AMAZON is available in the Stanford large network dataset collection (snap). PGP and aNobii have been provided by \cite{hric2014community}.

        \textbf{AMAZON\protect\footnote{\url{http://snap.stanford.edu/}}.} The product co-purchasing site that needs no introduction. At first, 'Amazon.com'\protect\footnote{\url{https://www.amazon.com/}} was specifically designed for the sale of books. After the company goes public, it becomes the first Internet retailer to secure one million customers in the sale of all types of cultural products. This website is a gold mine for the complex networks analysis. It can be represented by a graph where the nodes are the products and the links connect commonly co-purchased products. The product categories provided by AMAZON defines the ground-truth communities. They can be overlapping or hierarchically nested.

        \textbf{PGP.}  Pretty Good Privacy is the world's most widely used email encryption software. In many fields, this software is used for signing, encrypting, and decrypting different forms of data \ie texts, files, emails, \etc. In the PGP network, the nodes represent email addresses and links represent the signature of emails key. In fact, each email address has a unique key. When an individual trust another, he trust his key \cite{weippl2005pgp} with a numerical signature. The ground-truth communities are email domain or sub-domain names. The nodes can belong to multiple groups. In social research, this network has received a lot of attention \cite{kaur2016survey,dar2015survey}.

        \textbf{aNobii.} It is a social bookmarking site created for readers and book fans \cite{aiello2010link}. It is designed to record and share personal libraries and book lists. The users of aNobii give information about their books and reading interests. They can establish typed social ties to other users and belong to groups. In this network, the nodes are the users and links represent their social ties.
        Recently, several studies have been carried out on aNobii \cite{aiello2010link,scholz2010node,li2014survey}.

        \begin{table}[ht!]
        \centering
        \caption{Global properties of used networks. The calculated properties are number of nodes (V), number of edges (E), Density ($\rho$), Diameter ($d$), Average shortest path ($l_{G}$), Average node degree ($\widetilde{deg}$), Max node degree ($\delta(G)$), Assortativity Coefficient ($\tau$), and Clustering Coefficient ($C$)}
        \label{table1}
                \begin{tabular}{lccccccccc}
                \hline
                  &V&E&$\rho$&$d$&$l_{G}$&$\widetilde{deg}$&$\delta(G)$&$\tau$&$C$\\
                \hline
                PGP&81036&190143&$5.79e^{-05}$&24&7.43&4.69&8741&-0.03&0.03 \\
                AMAZON&334863&925872&$8.25e^{-06}$&44&2.78 &5.53&549&-0.06&0.21\\
                aNobii &136547&892377&$9.57e^{-05}$&17&5.21&13.07 &6037&-0.13&0.01\\
                \hline
                \end{tabular}
        \end{table}
        The summary of the basic properties of these networks is reported in Table \ref{table1}. PGP is the one with the smallest size. AMAZON is four times bigger and the size of aNobii is in between.
        PGP has a density in the same range that aNobii while AMAZON's one is around ten times smaller. All of them are small-world networks with an average shortest path ranging from $2.7$ to $7.4$. They are disassortative and except for AMAZON, their clustering coefficient value is very low. The basic properties of these networks are very typical of what is generally observed in many real-world situations.

\subsection{Methodology of the comparative evaluation} \label{Method}
\subsubsection{General Framework}
        \begin{figure}[ht!]
        \centering
        \includegraphics[width=.99\textwidth]{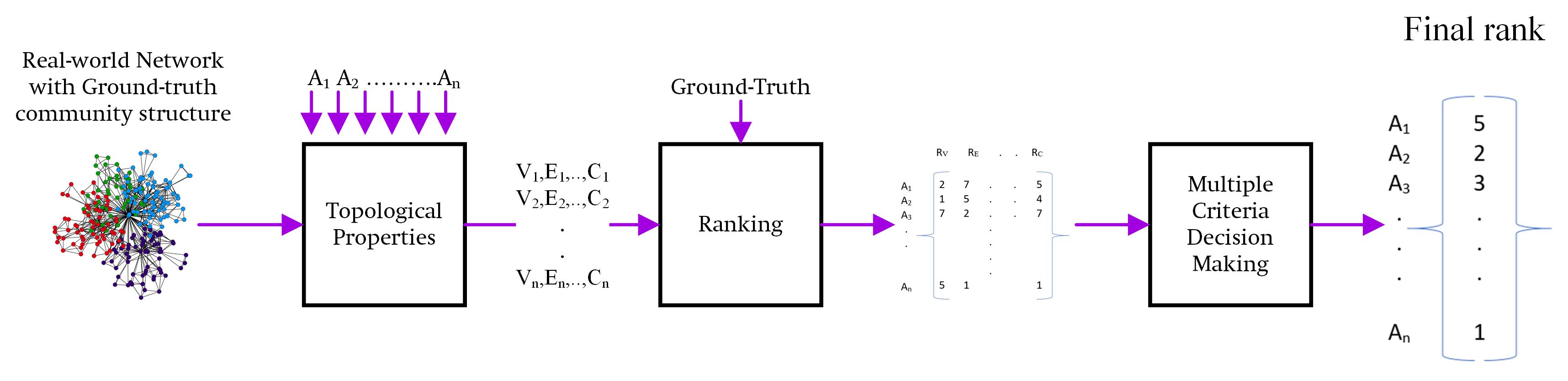}
        \caption{\label{fig03} General Framework. $A_{1},A_{2},...,A_{n}$ are the community detection algorithms. $V, E,...,C$ are the topological properties. $R_{V},R_{E},...,R_{C}$ are the ranking of $V, E,...,C$}
        \subfiguretopcaptrue
        \end{figure}
        Figure \ref{fig03} illustrates the general framework of the proposed approach in order to evaluate overlapping community detection algorithm using data with known community structure. As input, a real-world network with its ground-truth community structure is needed. The $n$ overlapping community detection methods that we want to compare are run on this real-world network in order to uncover its community structure. Then, various topological properties ($V_i, E_i,..., C_i$) are computed on the $n$ resulting community structure. Based on the comparison with those of the ground-truth community structure, a local ranking of algorithms is established for each property. All these local rankings are finally merged on a global ranking by an MCDM. Note that the local ranking strategy depends on the nature of the considered topological property. We distinguish two cases \ie the case where the topological property is a scalar value and the case where it is a probability distribution.

\subsubsection{Evaluation based on scalar properties}

        \begin{figure}[ht!]
        \centering
        \includegraphics[width=.99\textwidth]{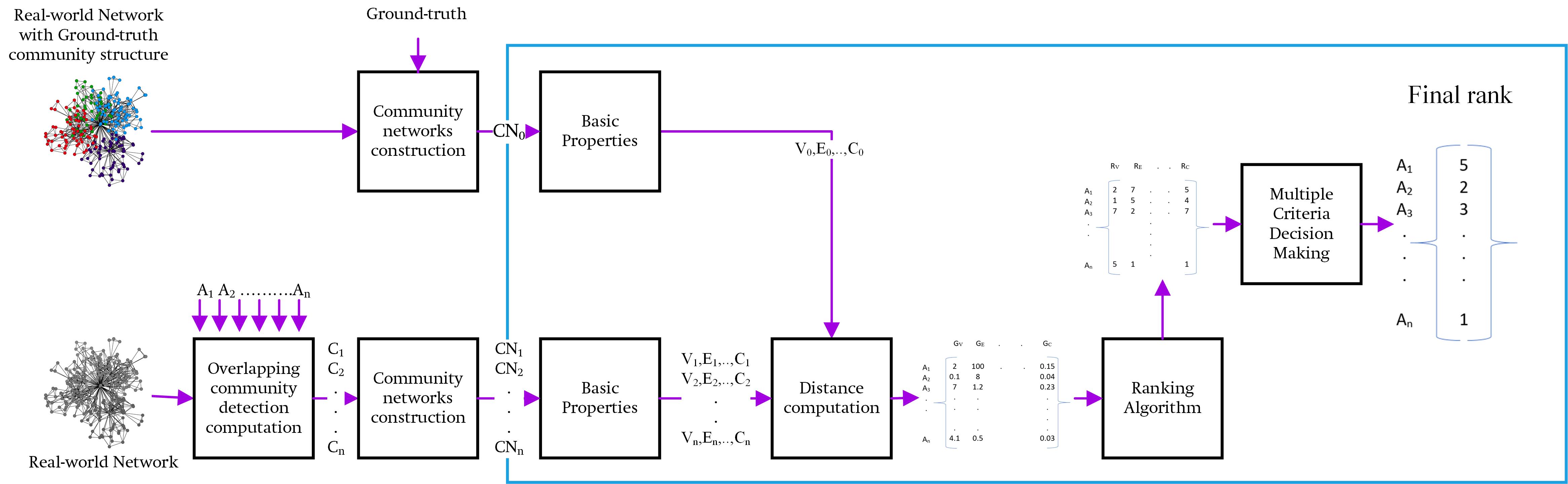}
        \caption{\label{fig031} Evaluation based on scalar properties. $A_{1},A_{2},...,A_{n}$ are the community detection algorithms. $C_{1},C_{2},...C_{n}$ are the unveiled community structure. $CN_{1},CN_{2},...,CN_{n}$ are the community-graphs of $C_{1},C_{2},...C_{n}$. $V, E,...,C$ are the topological properties. $G_{V},G_{E},...,G_{C}$ are the fits of $V, E,...,C$ based on ground-truth best fit. $R_{V},R_{E},...,R_{C}$ are the ranking of $V, E,...,C$}
        \subfiguretopcaptrue
        \end{figure}

     The main steps of the scalar properties evaluation framework are illustrated in figure \ref{fig031}. There are two parallel processes: one is dedicated to the ground-truth community structure, while the second concerns the discovered community structure by the $n$ algorithms under evaluation. In both cases, the 'community-graphs' are computed $(\{CN_0\},\{CN_1..CN_n\})$. More details are given in section \ref{commNetw} about this step. After that, various scalar topological properties $(V_i,E_i,...,C_i)$ are extracted from all these graphs $(i=0...n)$. In the next step, a distance between the ground-truth 'community-graph' topological property value and the ones extracted from the 'community-graphs' built using the unveiled community structure is computed. The algorithms are then sorted in ascending order according to their distance values. Finally, as there is a local ranking for each scalar property, all these local rankings are input in an MCDM method in order to obtain a final ranking. This process is applied on the basic topological properties (number of nodes (V), number of edges (E), Density ($\rho$), Diameter ($d$), Average shortest path ($l_{G}$), Average node degree ($\widetilde{deg}$), Max node degree ($\delta(G)$), Assortativity Coefficient ($\tau$), and Clustering Coefficient(C)). It has been also used to merge the local rankings given by various classical quality and clustering metrics. Note that in this case, these properties are computed on the community structures rather than on the 'community-graphs'.

\subsubsection{Evaluation based on probability distribution properties}
\begin{figure}[ht!]
        \centering
        \includegraphics[width=.99\textwidth]{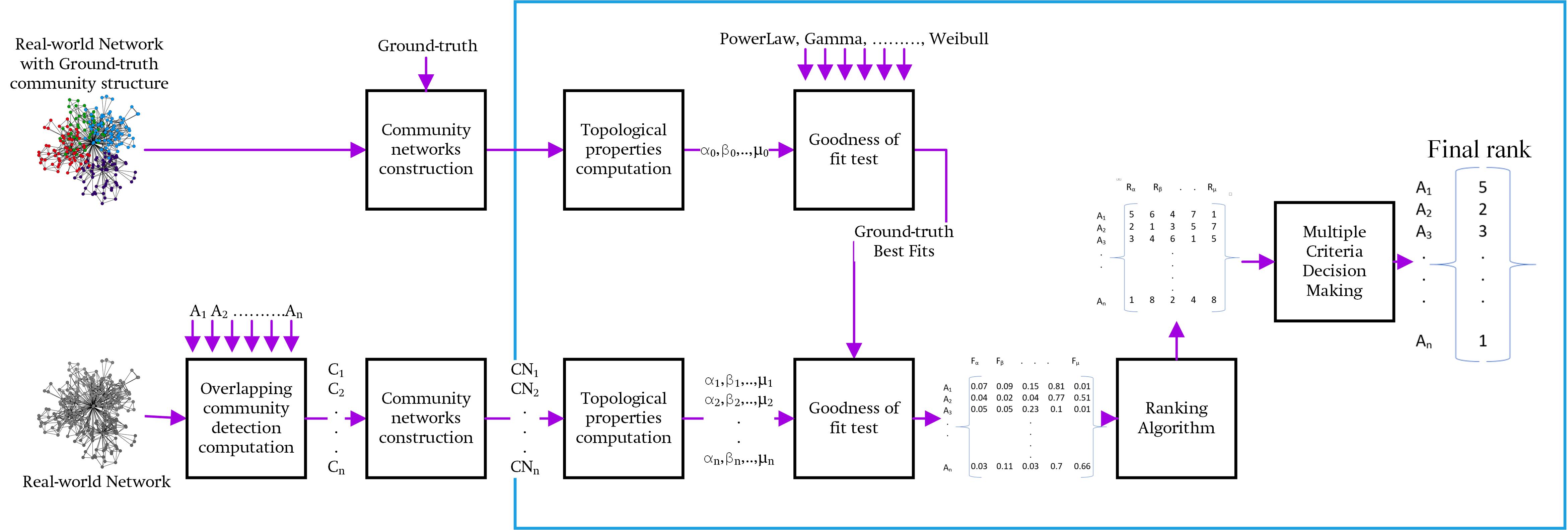}
        \caption{\label{fig00} Evaluation based on probability distribution properties. $A_{1},A_{2},...,A_{n}$ are the community detection algorithms. $C_{1},C_{2},...C_{n}$ are the unveiled community structure. $CN_{1},CN_{2},...,CN_{n}$ are the community-graphs of $C_{1},C_{2},...C_{n}$. $\alpha, \beta,...,\mu$ are the topological properties. $F_{\alpha},F_{\beta},...,F_{\mu}$ are the fits of $\alpha, \beta,...,\mu$ based on ground-truth best fit. $R_{\alpha},R_{\beta},...,R_{\mu}$ are the ranking of $\alpha, \beta,...,\mu$}
        \subfiguretopcaptrue
        \end{figure}

        Figure \ref{fig00} illustrates the main steps for evaluating the community detection algorithms in the case where the topological properties are probability distribution estimates. The overall process is very similar to the previous one \ie 'community-graphs' are build using both the ground-truth community structure and the outputs of the community detection algorithms. The main difference is in the ranking process. Once a topological property based on the ground-truth community structure is computed, a goodness of fit test is applied in order to estimate the underlying distribution. Nine alternative distributions (Beta, Cauchy, Exponential, Gamma, Logistic, Log-Normal, Normal, Uniform, and Weibull) are investigated. The best fit according to the Kolmogorov-Smirnov (KS) test is retained as the true distribution for the topological property under evaluation. It is then used as a reference in order to compute the ranking of the algorithms for this property. Under this hypothesis, the KS distance between the theoretical distribution and the empirical distribution is computed for each algorithm. They are ranked by increasing order of KS distance values for this property. Finally, the MCDM algorithm is used to merge all the individual rankings.

        For example,  let's consider the case where the best fit for the degree distribution of the ground-truth 'community-graph'  is the power-law according to the KS test. In this case, the degree distribution of the 'community-graphs' build from the uncovered community structure by the algorithms are fitted by the power-law. The KS test values between the empirical and the estimated power-law are computed for each algorithm. The detection algorithms are then sorted by increasing value of their  KS distance for this topological property. As there is a ranking for each individual property, the final ranking is the result of the MCDM  process.

        This process is performed to rank the algorithms according to the set of  the microscopic properties (degree distribution, average clustering coefficient as a function of node degree, hop plot). Ranking the algorithms according to their mesoscopic properties is  also based on this process. Indeed, the mesoscopic properties are described by probability distributions (community degree, community size, overlap size, \etc). The main difference is that they are computed on the community structure rather than the 'community-graphs'.

\subsection{Community-graph construction }\label{commNetw}
        To our knowledge, there are two well-known techniques to represent the community structure as a network.
        The first one is reported in \cite{Palla2005uncovering}. In this paper, the so-called 'community-graph' is defined as follows. The nodes refer to communities and a link is drawn if two communities share at least one node. The second representation is described by \cite{yang2012community} with the name 'network communities'. The nodes refer to communities. If two communities share at least one link their representative nodes in the graph are linked.
        In our analysis, we adopt the definition of \cite{Palla2005uncovering}. Indeed, the definition proposed by \cite{yang2012community} does not take into account the overlaps between the communities. It can describe indifferently overlapping and non-overlapping community structure. Furthermore, very often, this definition applied to real-world networks leads to almost complete graphs.

        \begin{figure}[ht!]
        \centering
        \subfigure[Community structure]{\includegraphics[width=.3\textwidth]{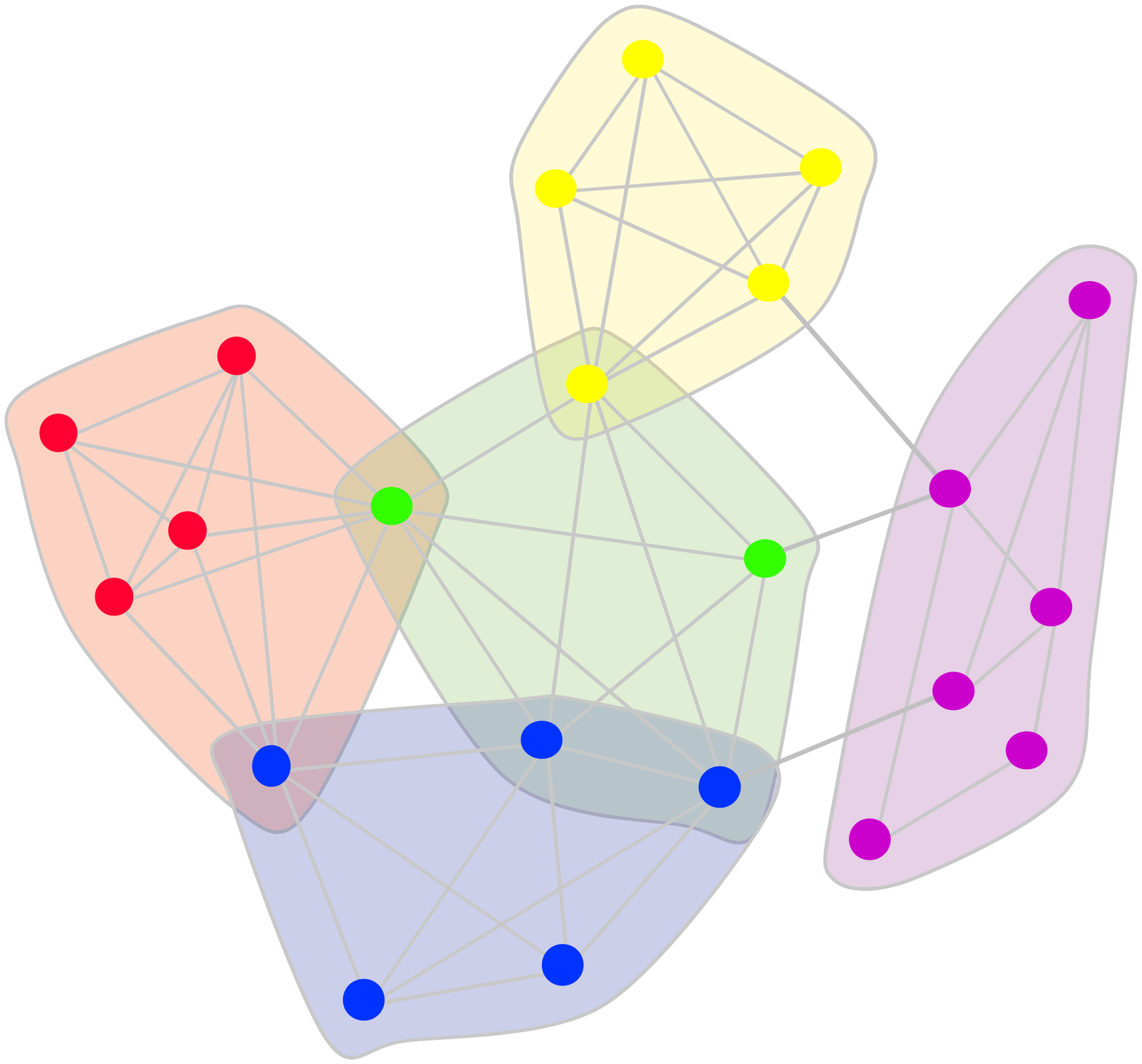}}
        \subfigure[Community-graph]{\includegraphics[width=.3\textwidth]{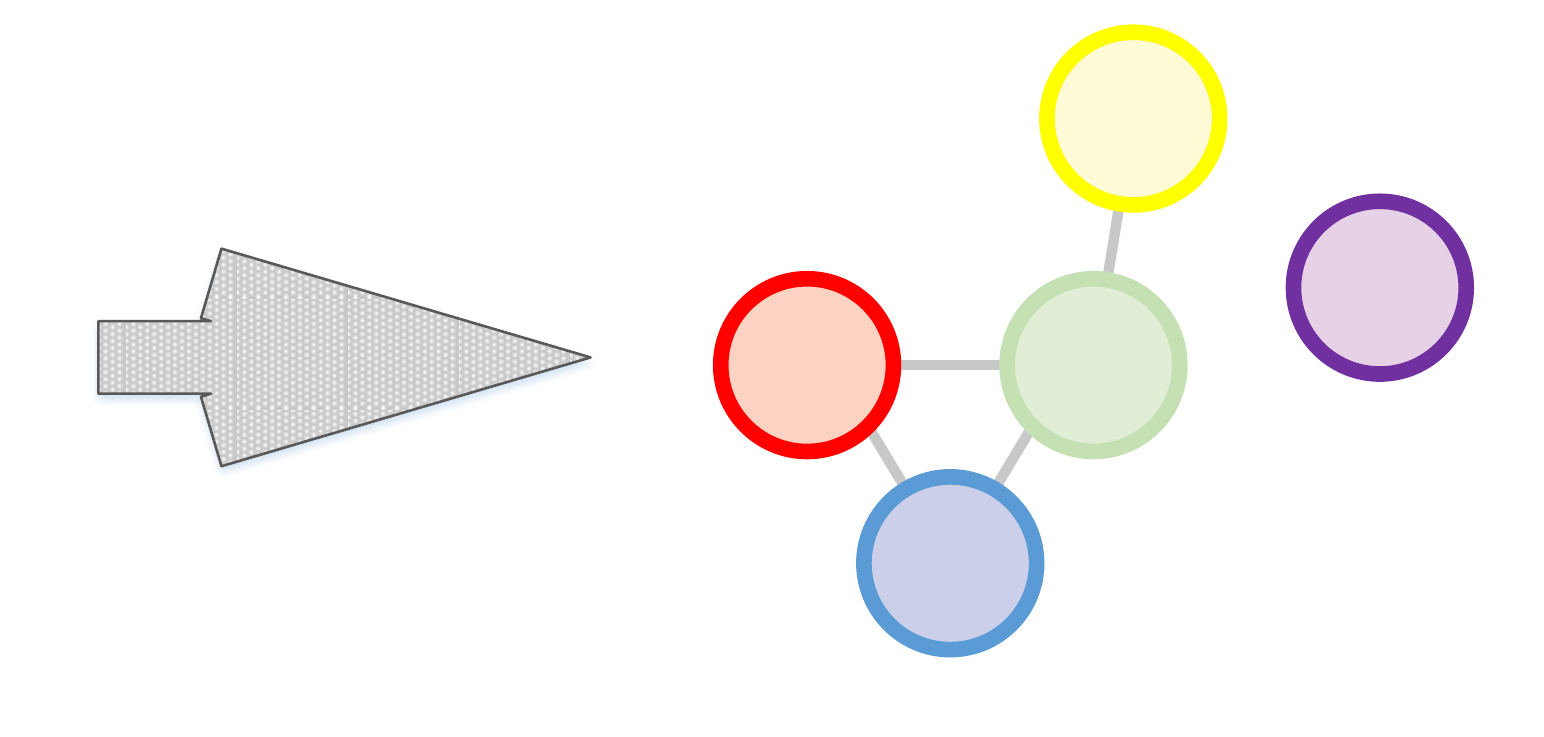}}
        \caption{\label{fig0} (a) A network with overlapping community structure (b) Its 'community-graph'}
        \subfiguretopcaptrue
        \end{figure}

        Figure \ref{fig0} illustrates the 'community-graph' construction. Note that the 'community-graph' is made of a set of connected components. Generally, on real-world networks, one can observe a 'giant'  component and some components of small size. In the following, when we mention the 'community-graph' we refer to its 'giant' component. In other words, the 'homeless' (non-overlapping) communities are ignored. The 'community-graph' is undirected and unweighted.

        The pseudo-code to build the 'community-graph' is reported in Algorithm \ref{algorithm}. The input is the community structure. The output is a 'community-graph'. The algorithm is very basic: for each pair of communities, if there is at least one shared node, then we add these two communities as linked nodes. Once the community-graph is built, we extract its "giant"  component.
        \begin{algorithm}[H]
        \caption{Construction of 'community-graph'}\label{algorithm}
        \begin{algorithmic}
        \REQUIRE Communities
        \ENSURE 'Community-graph'
        \FOR{ $i \leftarrow 1$ \TO numberOfCommunity - 1}
            \FOR{$j \leftarrow 2$ \TO numberOfCommunity}
                \IF{Communities(i).nodes $\bigcap$ Communities(j).nodes $\neq \varnothing$}
                \STATE Community-graph.AddLink(i,j)
                \ENDIF
            \ENDFOR
        \ENDFOR
        \STATE Community-graph.GetGiantConnectedComponent()
        \end{algorithmic}
        \end{algorithm}
        In order to distinguish between the real-world networks from their ground-truth 'community-graphs' we will use the following notation. For simplicity, we use the same name for both of them, and a star is appended for the 'community-graph'. For example, AMAZON is the real-world network and AMAZON* its 'community-graph'. We use the same notation to distinguish the community structure discovered by a detection algorithm with its 'community-graph'. For example, for a given real-world network (PGP, AMAZON, aNobii), SLPA is the community structure uncovered by the SLPA algorithms and SLPA* refers to its 'community-graph'.

        The function 'NetworkOfCommunity.AddLink(i,j)' join the pair of nodes 'i' and 'j' to the edge-list file and the function 'Community-graph.GetGiantConnectedComponent()' removes the small connected components and keeps the 'giant' connected component.

\section{Data Analysis and Discussion}
        In order to perform the analysis of the overlapping community structure, we build the 'community-graph' of the ground-truth and the 'community-graphs' of the unveiled community structure for PGP, AMAZON, and aNobii.

        For the sake of clarity, we cannot report all the figures and tables related to the three datasets (PGP, AMAZON, and aNobii). Therefore, we choose to provide in this section the results for PGP. AMAZON and aNobii figures and tables are available in the appendix section. Nevertheless, even if we concentrate on PGP, the conclusions are based on the analysis of all the datasets.

        Note that some community detection algorithm does not run to completion on the largest datasets in a reasonable time. In this case, they are excluded from the analysis.

\subsection{Basic properties}

        \begin{table}[ht!]
        \centering
        \caption{Global properties of PGP* and 'community-graph' of the overlapping community detection algorithms. The calculated properties are Number of nodes (V), Number of edges (E), Density ($\rho$), Diameter ($d$), Average shortest path ($l_{G}$), Average node degree ($\widetilde{deg}$), Max node degree ($\delta(G)$), Assortativity Coefficient ($\tau$), and Clustering Coefficient ($C$)}
        \label{table2}
        \begin{tabular}{lccccccccc}
        \hline
          &V&E&$\rho$&$d$&$l_{G}$&$\widetilde{deg}$&$\delta(G)$&$\tau$&$C$\\
        \hline
        PGP* & 11074 & 23091 & 3.77E-04 & 15 & 7.43 & 4.17 & 4292 & -0.12 & 0.01 \\
        LFM* & 43558 & 146969 & 1.55E-04 & 26 & 9.12 & 6.75 & 234 & 0.15 & 0.61 \\
        GCE* & 741 & 2840 & 1.04E-02 & 10 & 5.77 & 7.67 & 126 & -0.02 & 0.2 \\
        OSLOM* & 1972 & 22778 & 1.17E-02 & 10 & 4.1 & 23.1 & 348 & 0.21 & 0.64 \\
        LINKC* & 42443 & 664348 & 7.38E-04 & 24 & 8.14 & 31.31 & 8186 & 0.08 & 0.75 \\
        SVINET* & 3325 & 9177 & 1.6E-03 & 14 & 5.7 & 5.52 & 941 & -0.15 & 0.04 \\
        SLPA* & 2666 & 5111 & 1.44E-03 & 13 & 5.5 & 3.8 & 468 & -0.15 & 0.05 \\
        DEMON* & 369 & 5537 & 8.16E-02 & 5 & 3.75 & 30.01 & 192 & -0.32 & 0.47 \\
        \hline
        \end{tabular}
        \end{table}
        Table \ref{table2} describes the global features of PGP* as well as the 'community-graphs' related to the community detection algorithms. The first impression given by the results reported in this table is that there is a great variability of the basic topological properties. If we look at the number of nodes (V) and links (E), we note that the algorithms can be grouped into two classes. The first class contains DEMON*, GCE*, OSLOM*, SLPA* and SVINET*, while the second one contains LFM* and LINKC*. In the first class, both the number of communities and the overlaps are under estimated while in the second class they are over estimated. Whatever the case, the values are far from the reference (PGP*). Let's check the other properties, LFM* and LINKC* have very close density ($\rho$) values to that of PGP*, and LFM* performs well in regards to 'average node degree' ($\widetilde{deg}$) value. Results reported for SLPA* and SVINET* concerning the Diameter ($d$), Assortativity Coefficient ($\tau$), and Clustering Coefficient ($C$) are not far from the reference. LFM*, LINKC*, and OSLOM* are assortative while the reference is disassortative. Furthermore, their clustering coefficient values are very high as compared to the reference.

        We note a relative similarity for the results of the community detection algorithms on the two real graphs AMAZON and aNobii according to the tables \ref{table3}, \ref{table4}. Indeed, the community detection algorithms underestimate the number of communities ('community-graphs' nodes) and the number of overlaps ('community-graphs' links). SVINET* for AMAZON and GCE*, OSLOM*, SLPA* for aNobii are the 'community-graphs' that have a comparable density to those of AMAZON* and aNobii* respectively. All 'community-graphs' built from the unveiled community structure have a comparable diameter and average node degree as compared to those of the references (AMAZON* and aNobii*). For the average shortest path, DEMON* and MOSES* have €‹similar values than those of AMAZON* and aNobii*. Similarly to the references (AMAZON* and aNobii*), DEMON*, GCE*, MOSES*, SLPA* and CFINDER* are disassortative. In most cases, the clustering coefficient of the 'community-graphs' is higher than the reference. This suggests that even if the number of communities and overlaps are globally under-estimated, the uncovered ones are highly overlapping.

\subsection{Microscopic properties}
        Fitting distributions to data consist in choosing a probability distribution modeling the random variable, as well as finding parameter estimates for that distribution. Usually, it is done in an iterative process of distribution choice, parameter estimation, and quality of fit assessment. In this work, we use the R package fitdistrplus (\cite{delignette2015fitdistrplus}). It implements several methods for fitting univariate parametric distributions using various estimation methods (maximum likelihood estimation (MLE), moment matching estimation (MME), \etc). In order to measure the distance between the fitted parametric distribution and the empirical distribution, different goodness-of-fit statistics are proposed (Cramer-von Mises, Kolmogorov-Smirnov and Anderson-Darling). We retained the Kolmogorov-Smirnov statistic in our work. The fit of ten distributions (Power-Law (PL), Beta (BE), Cauchy (CA), Exponential (E), Gamma (GM), Logistic (LO), Log-Normal (LN), Normal (N), Uniform (U), and Weibull (WB)) has been investigated. This has been done systematically for every distribution and for each 'community-graph€' under evaluation. For clarity, in the following, we only report the goodness-of-fit of the reference 'community-graph€' (ground-truth).
\subsubsection{Degree Distribution}

        The result of the goodness-of-fit test are reported in Table \ref{table5} for PGP*. It appears clearly that the Power-Law is the best fit for the degree distribution. The estimate of the exponent $\alpha = 2.33$ is in the same range as usually reported for real-world complex networks.

        Except for DEMON* where the KS-value of the Power-Law and the Log-Normal are identical, the former is the best fit for all the 'community-graphs' built from the unveiled community structures. The low values of the KS distance reported in Table \ref{table511} corroborate these findings. Note that the estimated exponent values are globally satisfactory.

\begin{figure}[ht!]
        \subfigure[PGP*]{\includegraphics[width=.121\textwidth]{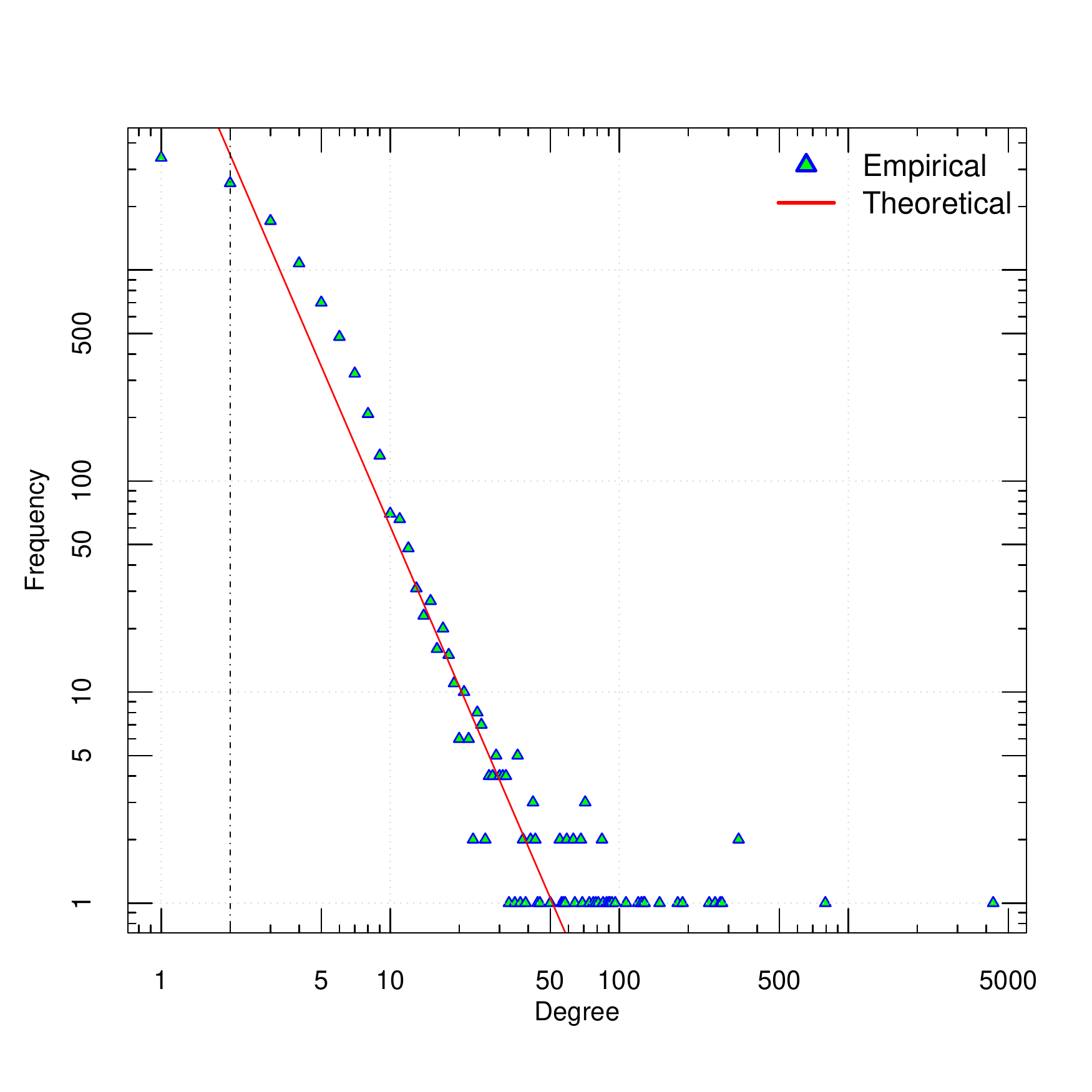}}
        \subfigure[LFM*]{\includegraphics[width=.121\textwidth]{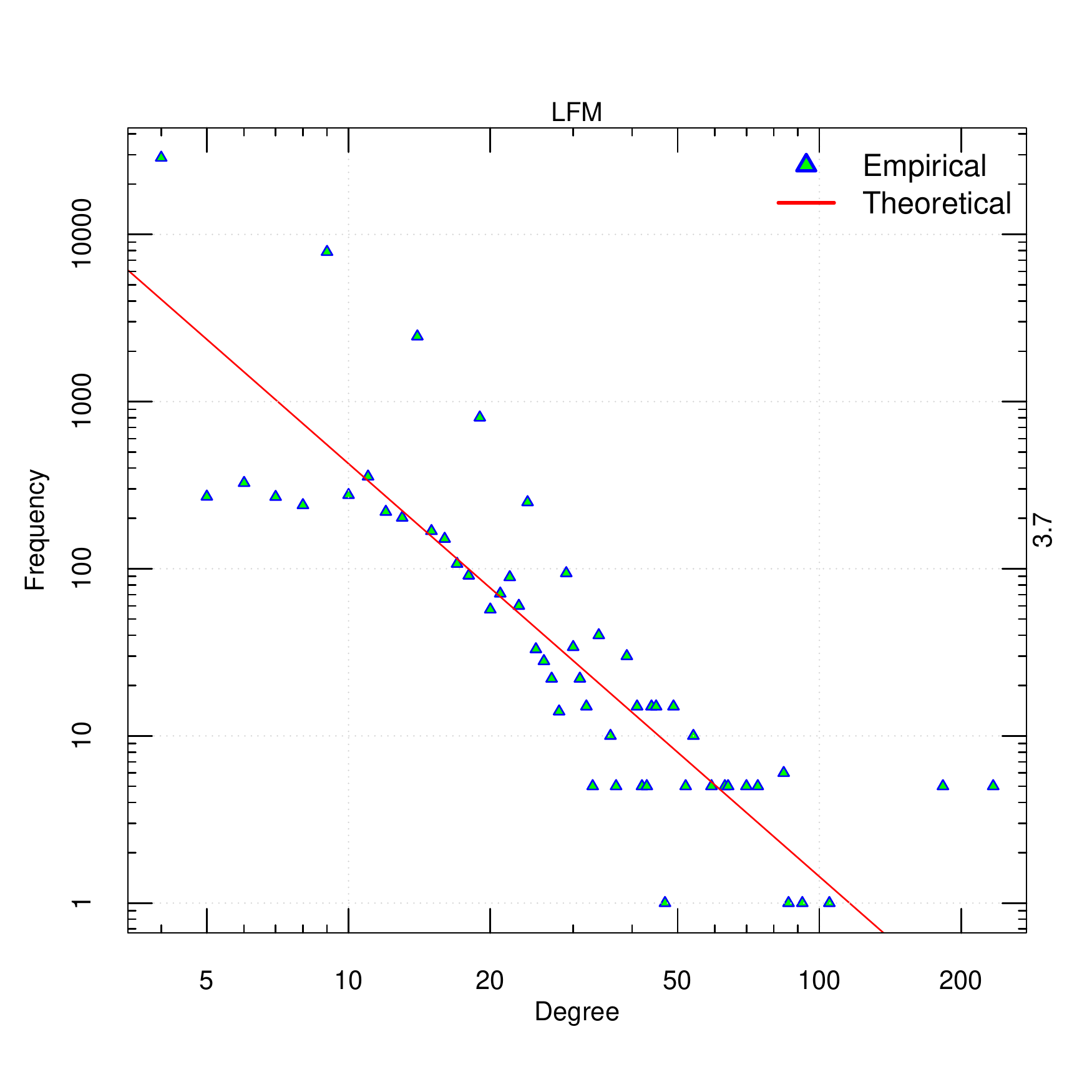}}
        \subfigure[GCE*]{\includegraphics[width=.121\textwidth]{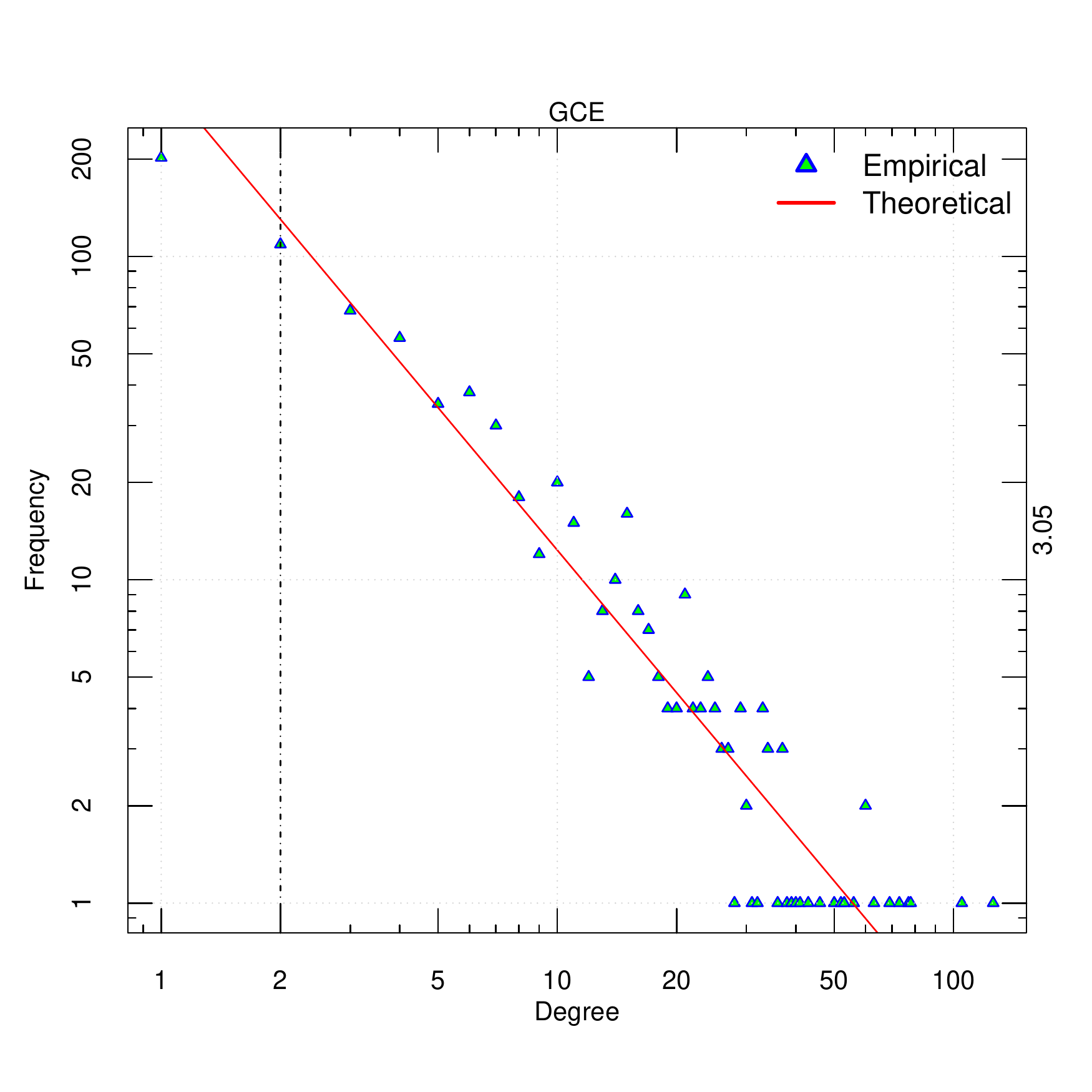}}
        \subfigure[OSLOM*]{\includegraphics[width=.121\textwidth]{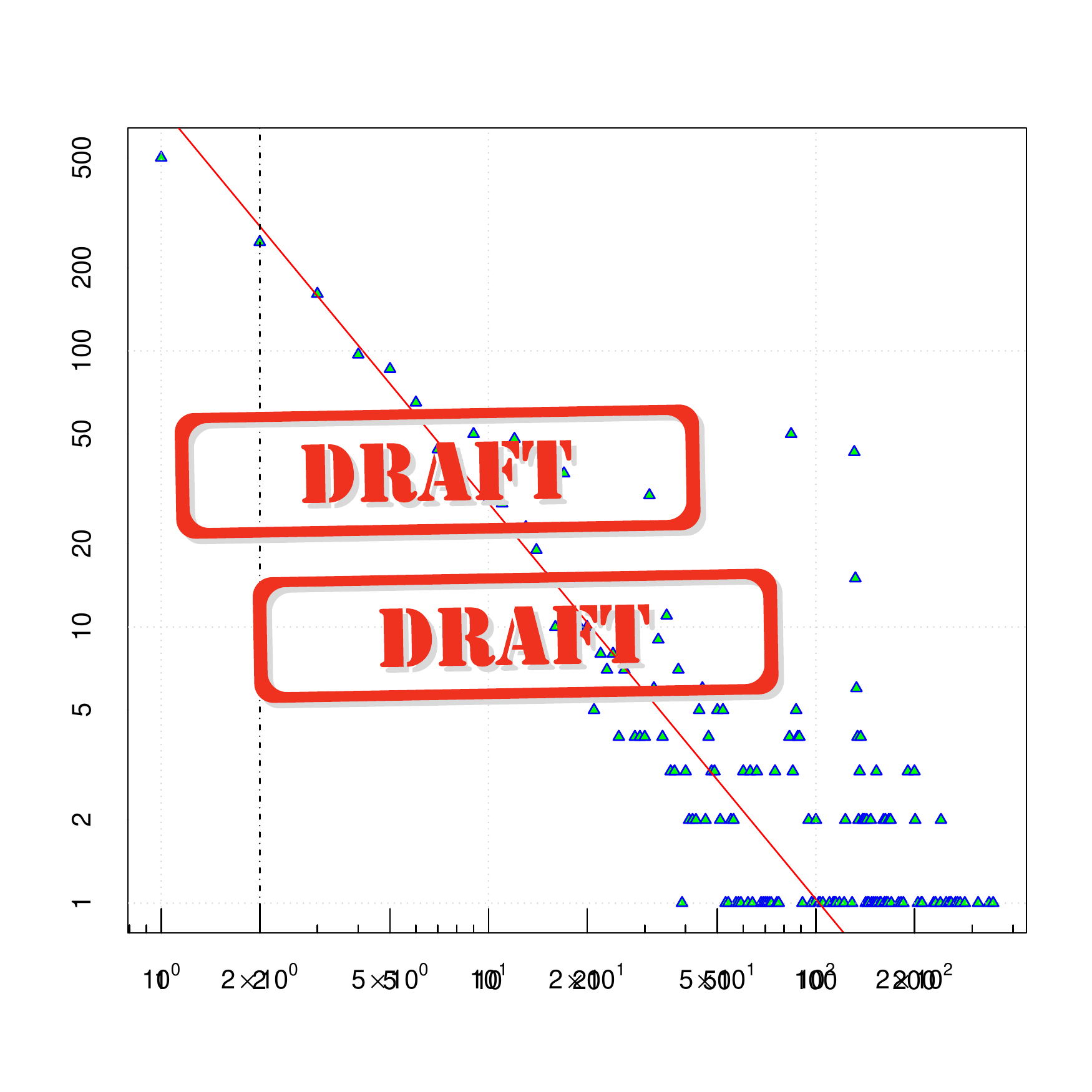}}
        \subfigure[LINKC*]{\includegraphics[width=.121\textwidth]{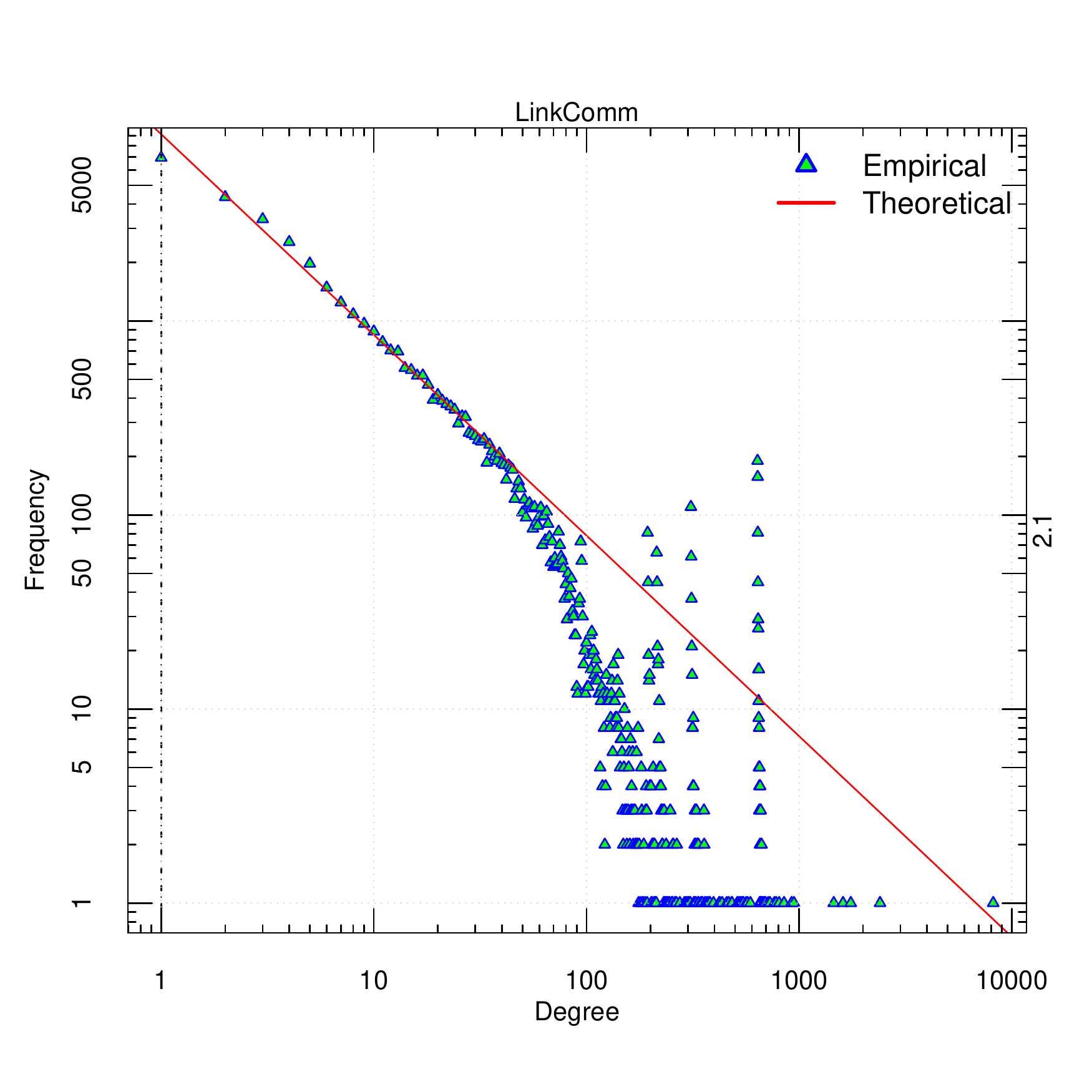}}
        \subfigure[SVINET*]{\includegraphics[width=.121\textwidth]{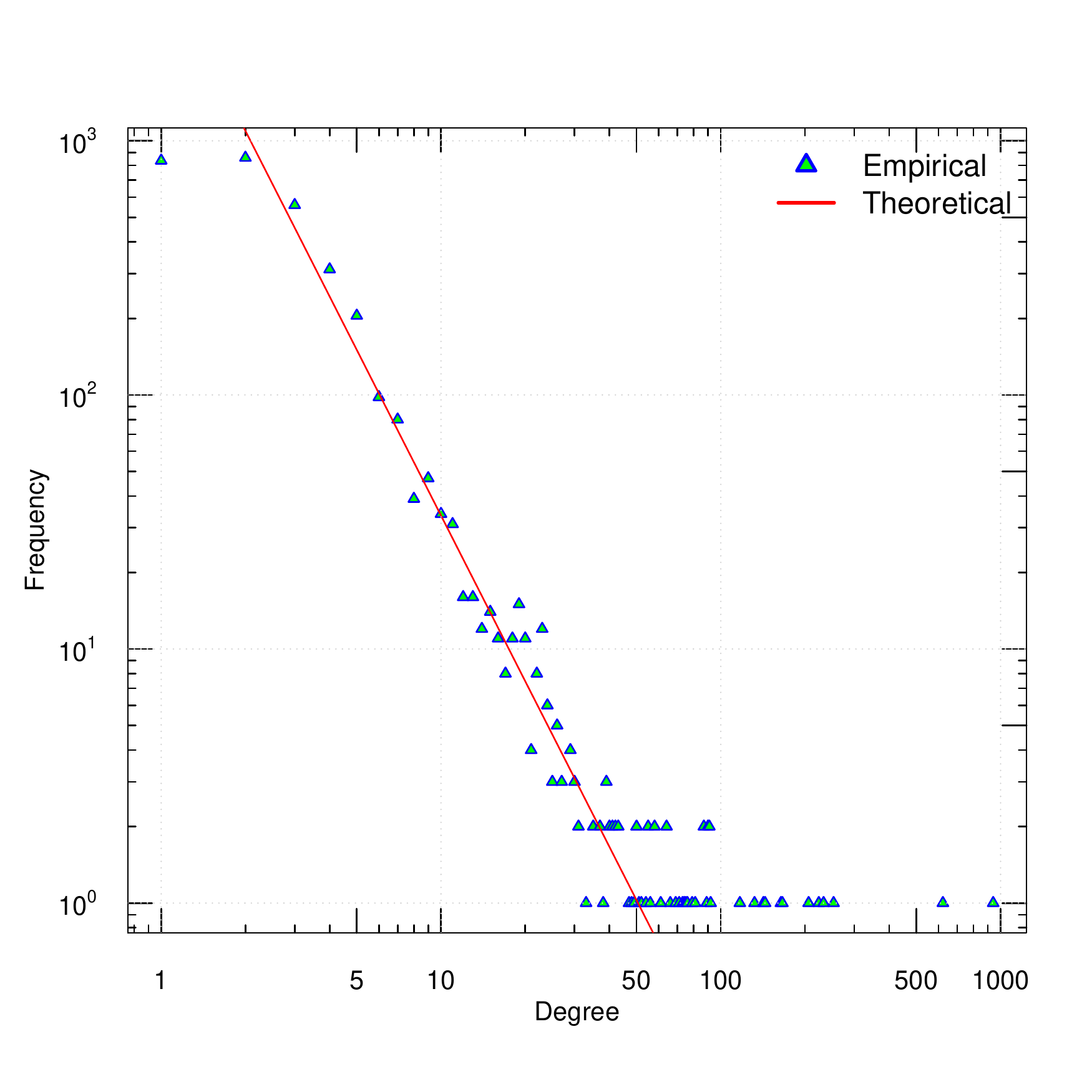}}
        \subfigure[SLPA*]{\includegraphics[width=.121\textwidth]{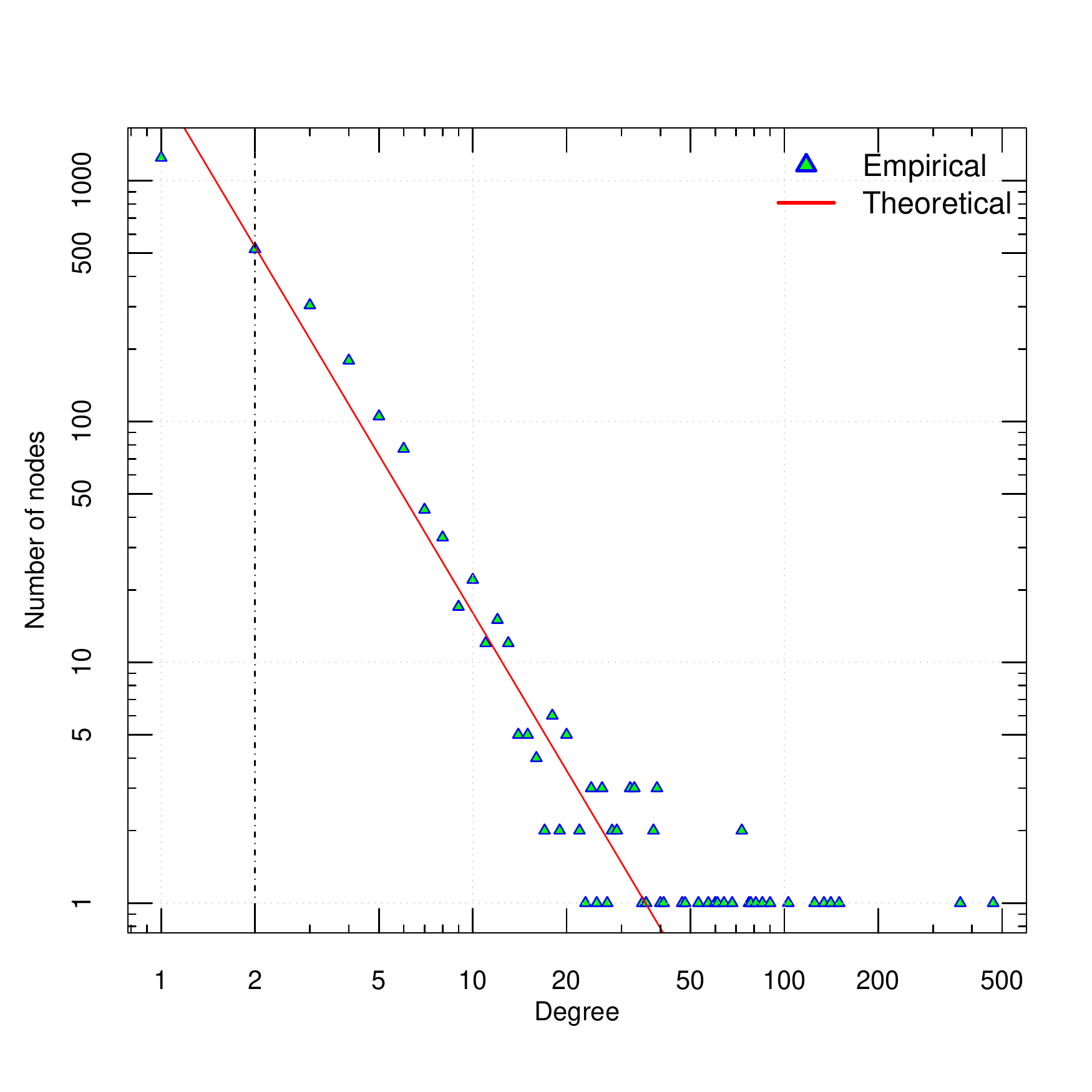}}
        \subfigure[DEMON*]{\includegraphics[width=.121\textwidth]{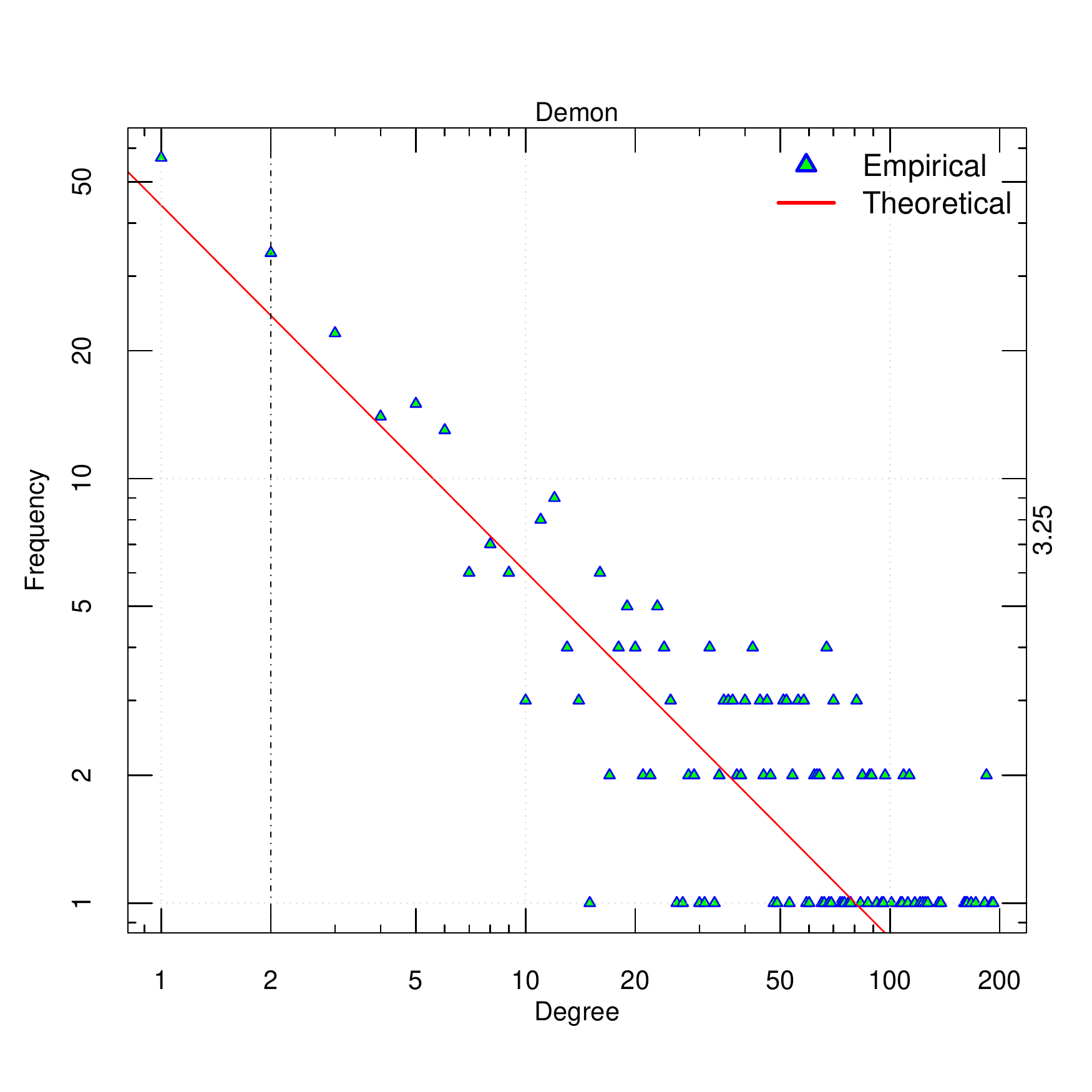}}
        \caption{\label{fig1}Log-log empirical degree distribution (dot) and Power-Law estimates (line) for PGP* (a), LFM* (b), GCE* (c),  OSLOM* (d), LINKC* (e), SVINET* (f), SLPA* (g) and  DEMON*(h)}
        \subfiguretopcaptrue
        \end{figure}

        Figure \ref{fig1} reports the empirical degree distribution of the 'community-graphs' together with their estimated distribution under the power-law hypothesis. These results go in the same direction as those reported in the previous Table \ref{table511}.

        According to Table \ref{table6} and Table \ref{table7}, the Power-Law is also the best fit for aNobii* and AMAZON*. However, it is not as clear as in the PGP* case. Indeed, the KS-test values of the Log-Normal and the Power-Law distributions are very close. The explanation may be that for low degree values, the empirical distribution is well approximated by the Log-Normal and that the Power-Law is a better fit for the tail. Note that this is not surprising as very similar basic generative models can lead to either Power-Law or Log-Normal distributions.

        For the 'community-graphs' built from the unveiled community structures, results show clearly the good fit of the Power-Law distribution (see Figure \ref{fig2}, Figure \ref{fig3}, Table \ref{table6}, Table \ref{table7}).

        \begin{table}[ht!]
        \centering
        \caption{KS-test values for the degree distribution for PGP*. The distributions under test are the Power-Law (PL), Beta (BE), Cauchy (CA), Exponential (E), Gamma (GM), Logistic (LO), Log-Normal (LN), Normal (N), Uniform (U), and Weibull (WB)}
        \label{table5}
        \begin{tabular}{lcccccccccc}
        \hline
         &  PL & BE & CA & E & GM & LO & LN & N & U & WB \\
        \hline
        KS&0.04&0.66&0.27&0.31&0.66&0.47&0.19&0.47&0.64&0.14\\
        \hline
        \end{tabular}
        \end{table}

        \begin{table}[htbp]
          \centering
          \caption{KS-test values for the degree distribution considering the Power-Law hypothesis for the 'community-graphs'}
          \label{table511}
            \begin{tabular}{lccccccc}
            \hline
              & LFM* & GCE* & OSLOM* & LINKC* & SVINET* & SLPA* & DEMON* \\
            \hline
            KS(Power-Law) & 0.06 & 0.05 & 0.08 & 0.04 & 0.02 & 0.02 & 0.09 \\
            \hline
            \end{tabular}%
        \end{table}%

\subsubsection{Average Clustering Coefficient as a Function of Degree}

        Generally, in the literature, the authors calculate the overall clustering coefficient of the network. Few studies have considered the transitivity through the distribution of 'the clustering coefficient as a function of degree'. We can mention the works of \cite{ahn2007analysis} and  \cite{gulyas2015navigable}. Results of their analysis on real-world networks show that this distribution tends to follow a Power-Law.

\begin{figure}[ht!]
        \subfigure[PGP*]{\includegraphics[width=.121\textwidth]{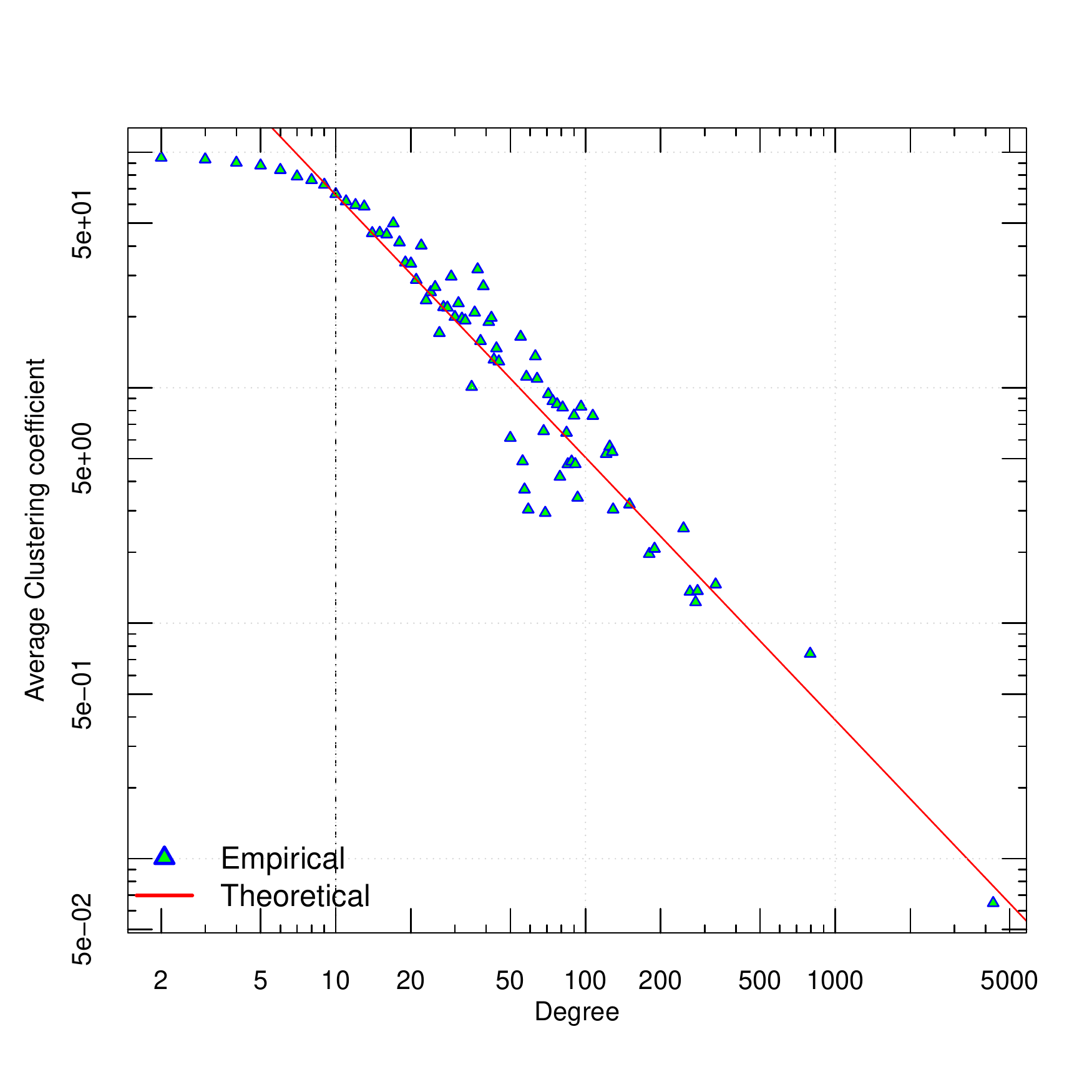}}
        \subfigure[LFM*]{\includegraphics[width=.121\textwidth]{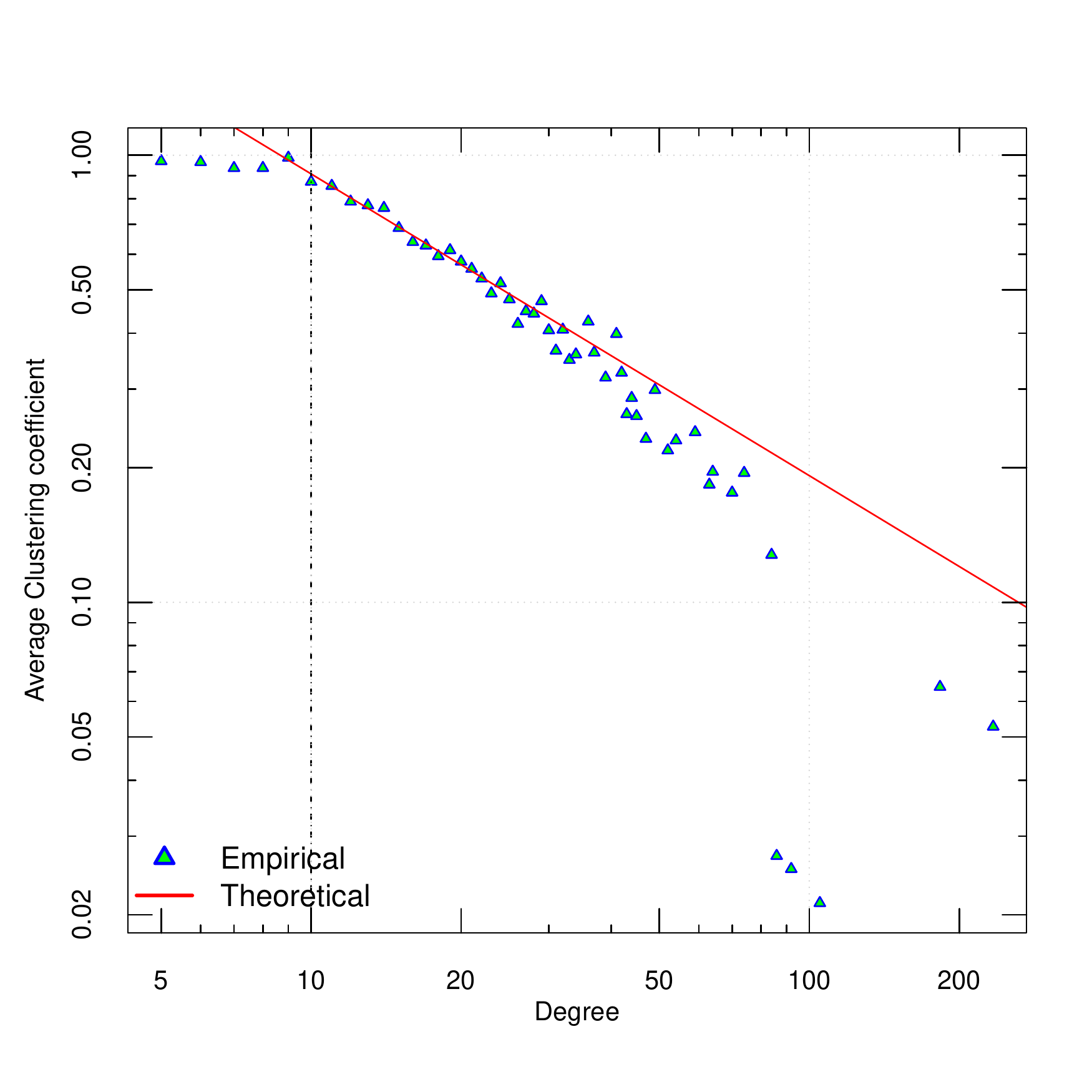}}
        \subfigure[GCE*]{\includegraphics[width=.121\textwidth]{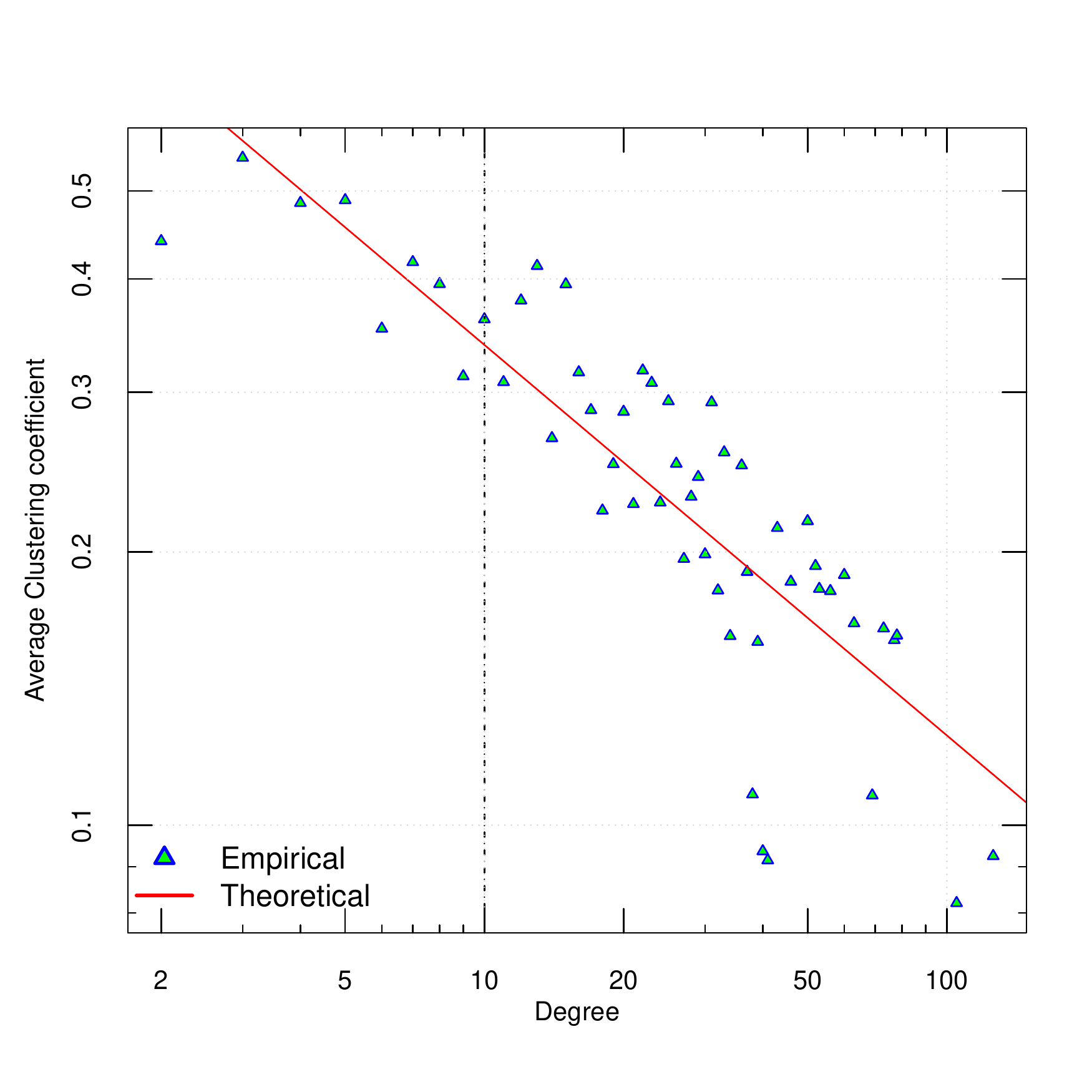}}
        \subfigure[OSLOM*]{\includegraphics[width=.121\textwidth]{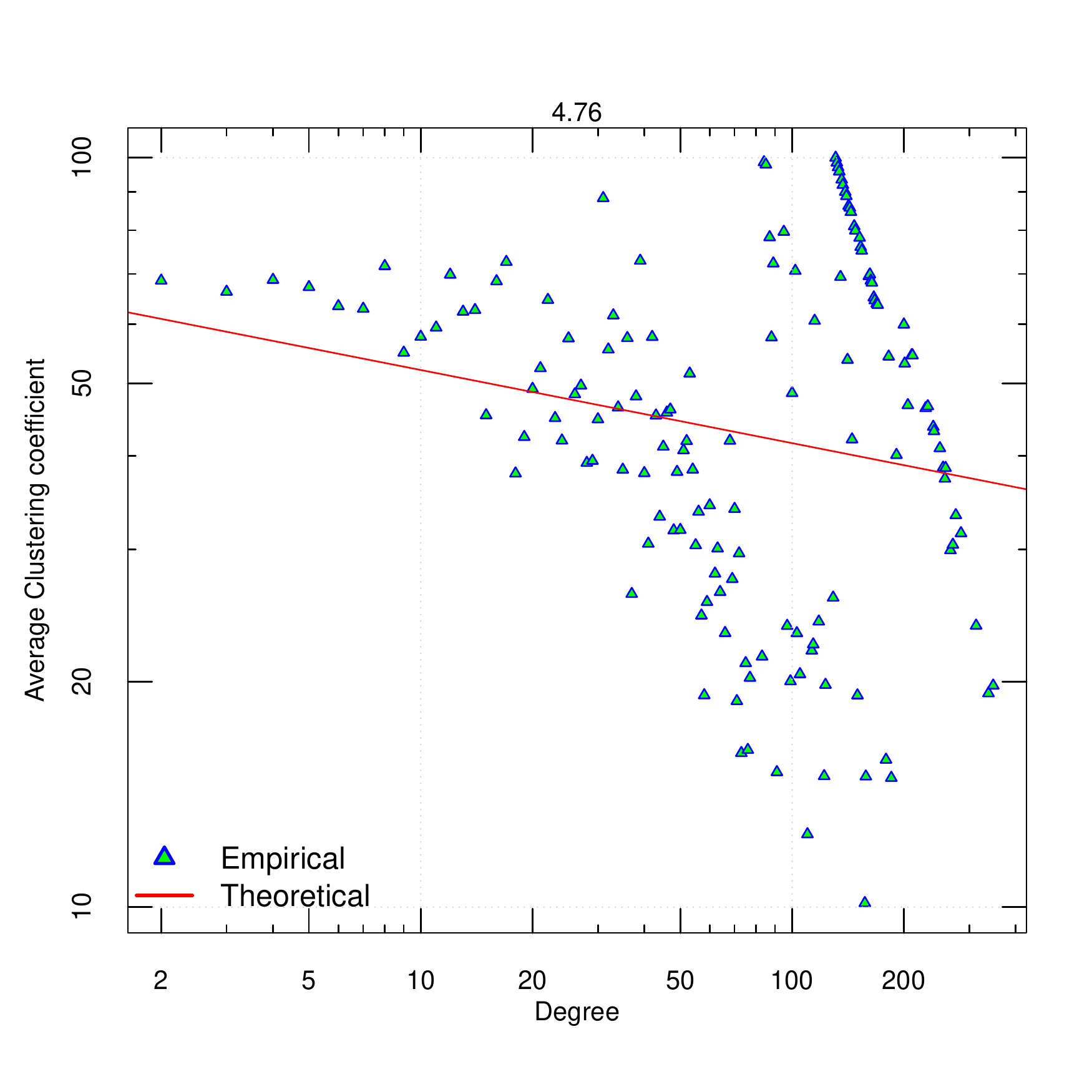}}
        \subfigure[LINKC*]{\includegraphics[width=.121\textwidth]{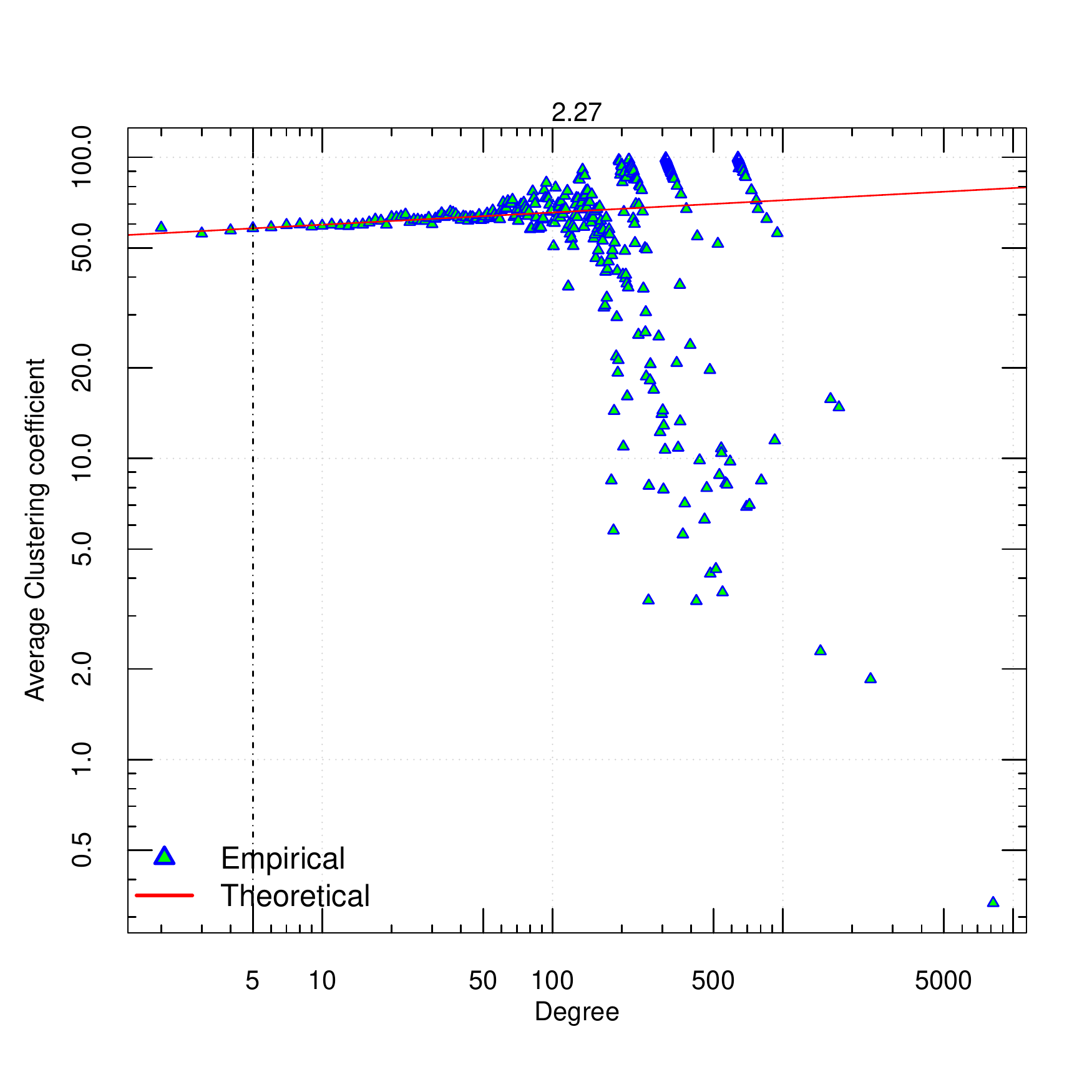}}
        \subfigure[SVINET*]{\includegraphics[width=.121\textwidth]{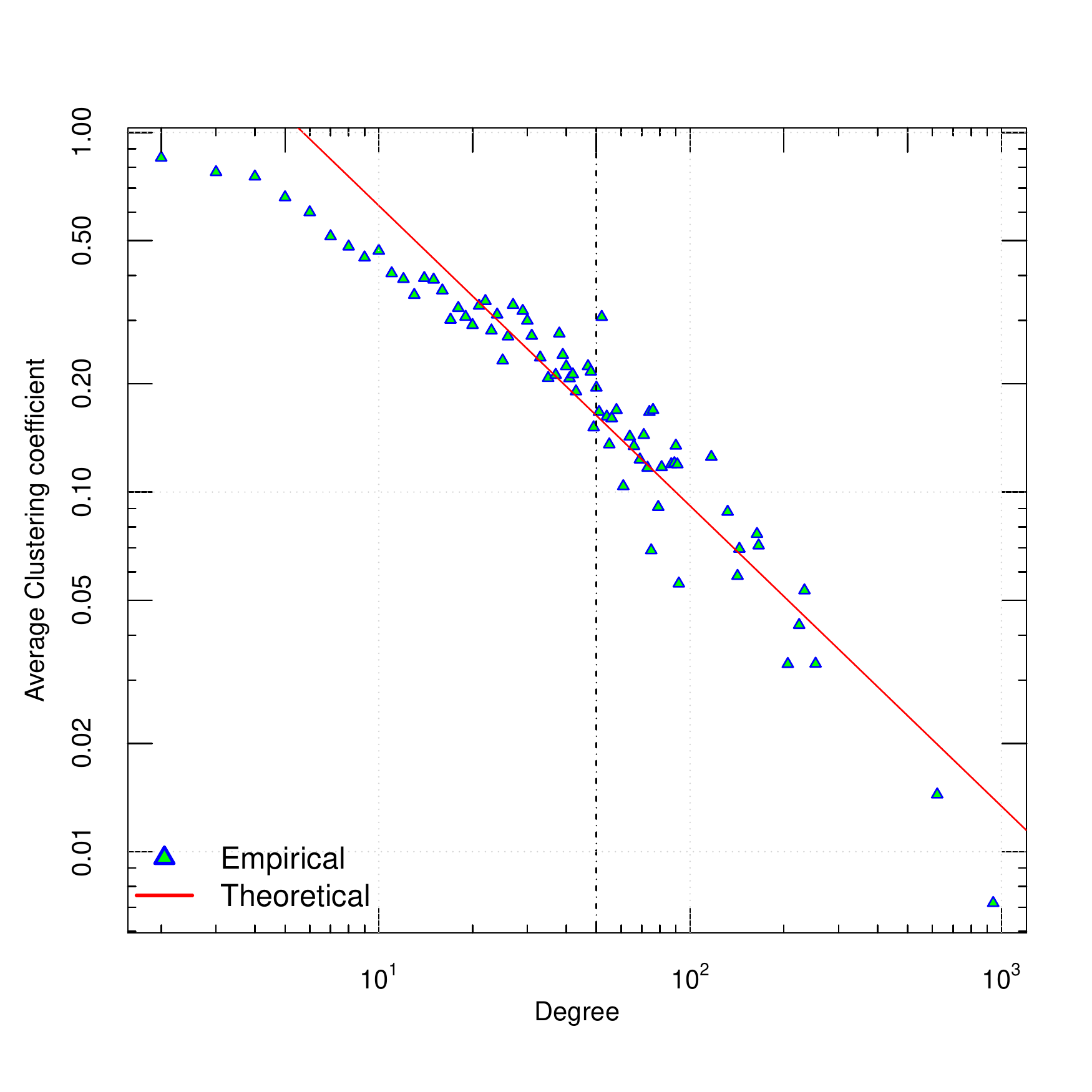}}
        \subfigure[SLPA*]{\includegraphics[width=.121\textwidth]{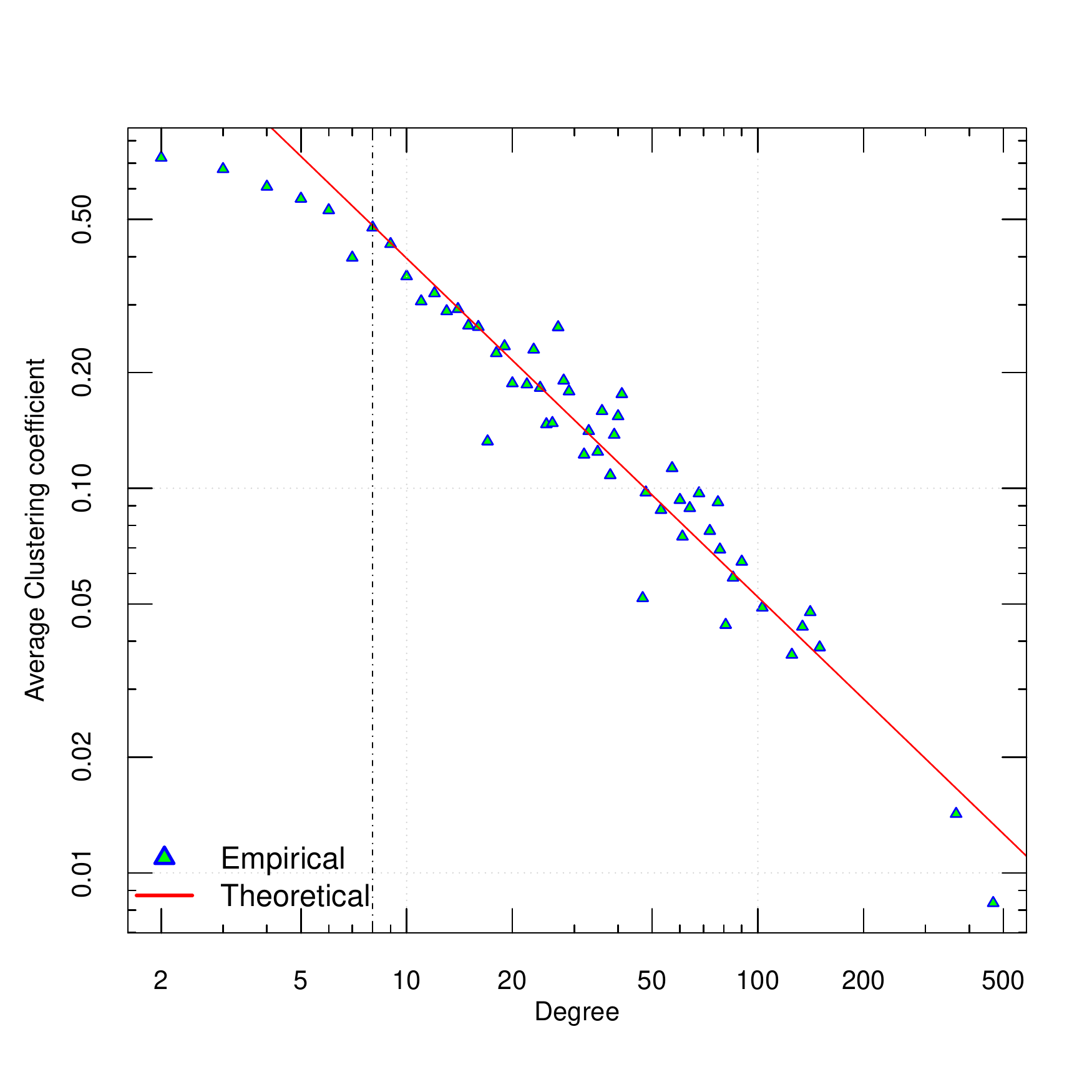}}
        \subfigure[DEMON*]{\includegraphics[width=.121\textwidth]{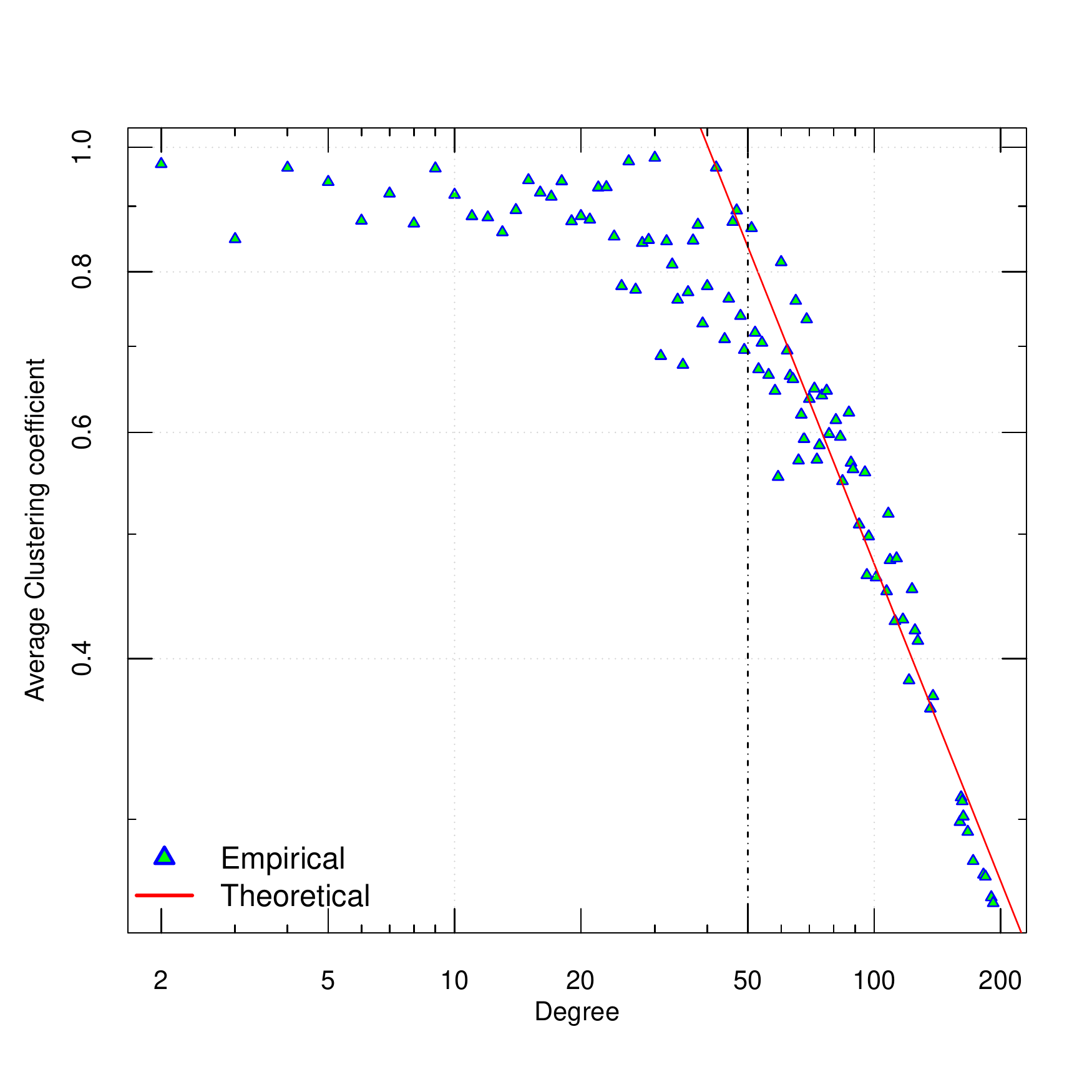}}

        \caption{\label{fig4}Log-log empirical average clustering coefficient distributions as a function of the degree (dots) and Power-Law estimates (line) for PGP* (a), LFM* (b), GCE* (c),  OSLOM* (d), LINKC* (e), SVINET* (f), SLPA* (g) and  DEMON*(h)}
       \end{figure}

        According to the KS-test for PGP*, the Log-Normal distribution is the best fit (See Table \ref{table8} ). It is closely followed by the Power-Law. If we look at Figure \ref{fig4}, it appears that the Power-Law is more appropriate in the tail of the distribution. In any case, both distributions are heavy tailed. Note that the estimated exponent of the Power-Law is slightly high ($\alpha= 3.25$). The Log-Normal is a two parameters distribution (location $\mu=-4.84$ and scale $\sigma = 1.11$). It is, therefore, more flexible to fit empirical data.

        Table \ref{table81} reports the KS-value for the 'community-graphs' under both hypotheses (Power-Law and Log-Normal). Globally it is very difficult to draw a conclusion according to these values. Indeed, when the KS-values are not equal, they are very close. To get a better understanding, one has to look at Figure \ref{fig4}. Globally, it seems that the empirical distributions can be well approximated by a Power-Law in the tails. Additionally, in some cases (OSLOM*, LINKC*, DEMON*) the parameters estimates seems to be of poor quality.

        Analysis of the results for the dataset AMAZON leads to very similar conclusions than those of PGP (See  Table \ref{table9}). For aNobii*, the Power-Law is clearly not the best fit according to the KS-test values reported in Table \ref{table10}. Three  distributions with two parameters (BETA, GAMMA, WEIBULL) are more appropriate. No law emerges particularly for the 'community-graphs' associated with the community detection algorithms (see Figure \ref{fig5} and Figure \ref{fig6}).

        The overall results concerning this property leads us to believe that the underlying distribution is not easy to uncover. Nevertheless, it is with a heavy tail.

        \begin{table}[ht!]
        \centering
        \caption{KS-test values for the average clustering coefficient as a function of degree for PGP*. The distributions under test are the Power-Law (PL), Beta (BE), Cauchy (CA), Exponential (E), Gamma (GM), Logistic (LO), Log-Normal (LN), Normal (N), Uniform (U), and Weibull (WB)}
        \label{table8}
        \begin{tabular}{lcccccccccc}
        \hline
         &  PL & BE & CA & E & GM & LO & LN & N & U & WB \\
        \hline
        KS&0.06&0.72&0.26&0.11&0.7&0.4&0.04&0.41&0.96&0.31\\
        \hline
        \end{tabular}
        \end{table}

        \begin{table}[ht!]
          \centering
          \caption{KS-test values for the average clustering coefficient as a function of degree considering the Power-Law and the Log-Normal hypothesis for the 'community-graphs'}
          \label{table81}
            \begin{tabular}{lccccccc}
            \hline
              & LFM* & GCE* & OSLOM* & LINKC* & SVINET* & SLPA* & DEMON* \\
            \hline
            KS(Power-Law) & 0.07 & 0.09 & 0.06 & 0.14 & 0.07 & 0.05 & 0.08 \\
            KS(Log-Normal) & 0.08 & 0.09 & 0.16 & 0.07 & 0.07 & 0.1 & 0.09 \\

            \hline
            \end{tabular}%
        \end{table}%

\subsubsection{Hop Distance Distribution}

\begin{figure}[ht!]
        \subfigure[PGP*]{\includegraphics[width=.121\textwidth]{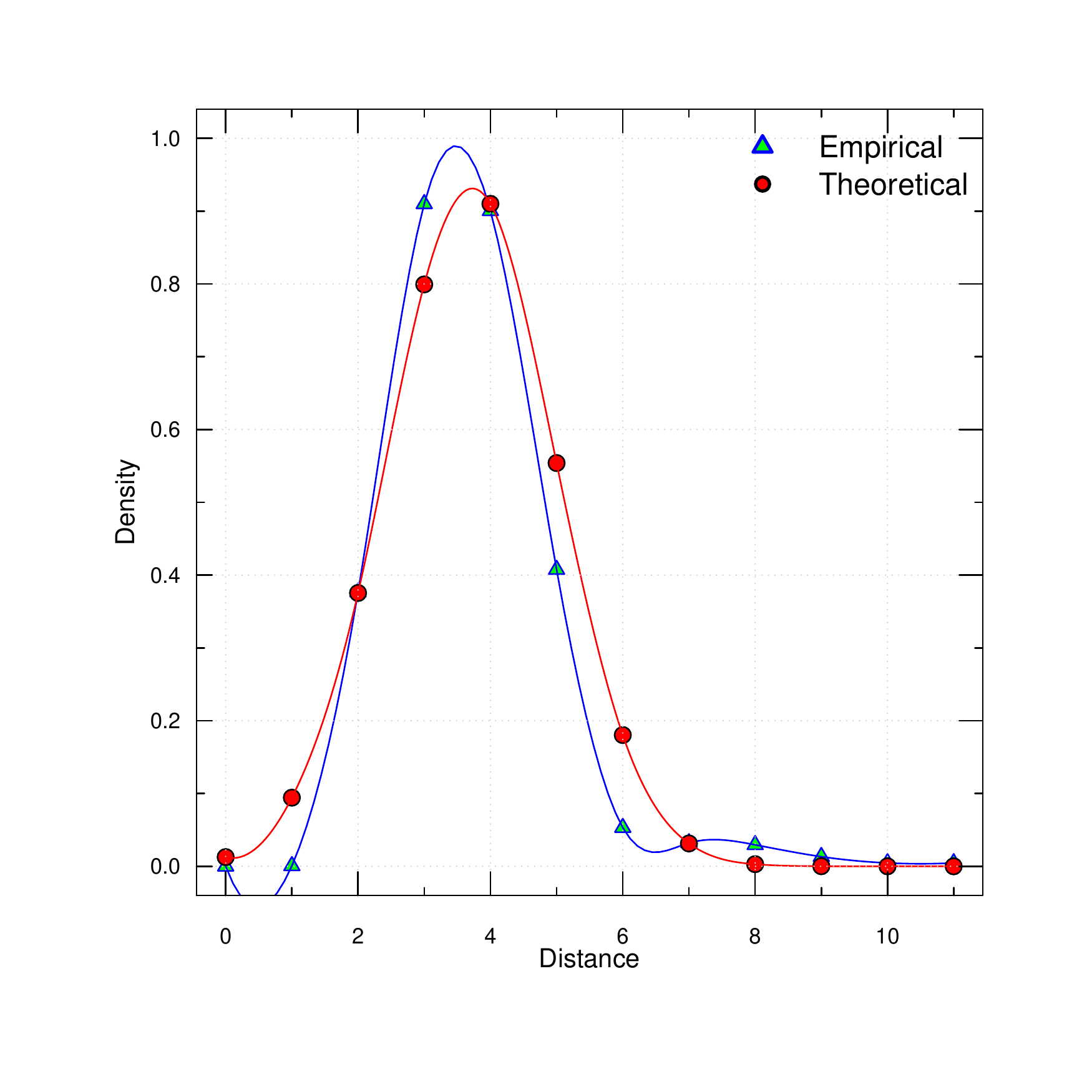}}
        \subfigure[LFM*]{\includegraphics[width=.121\textwidth]{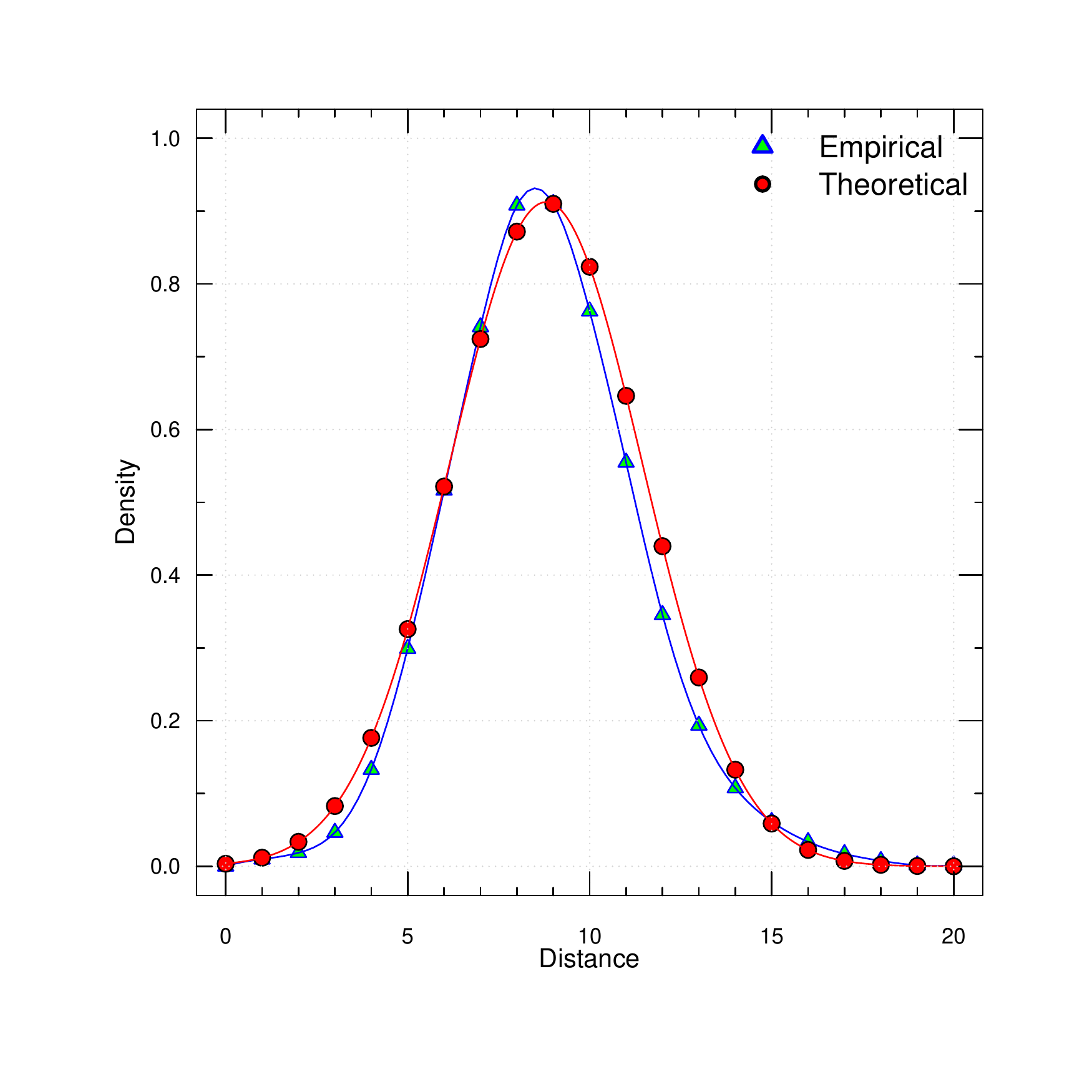}}
        \subfigure[GCE*]{\includegraphics[width=.121\textwidth]{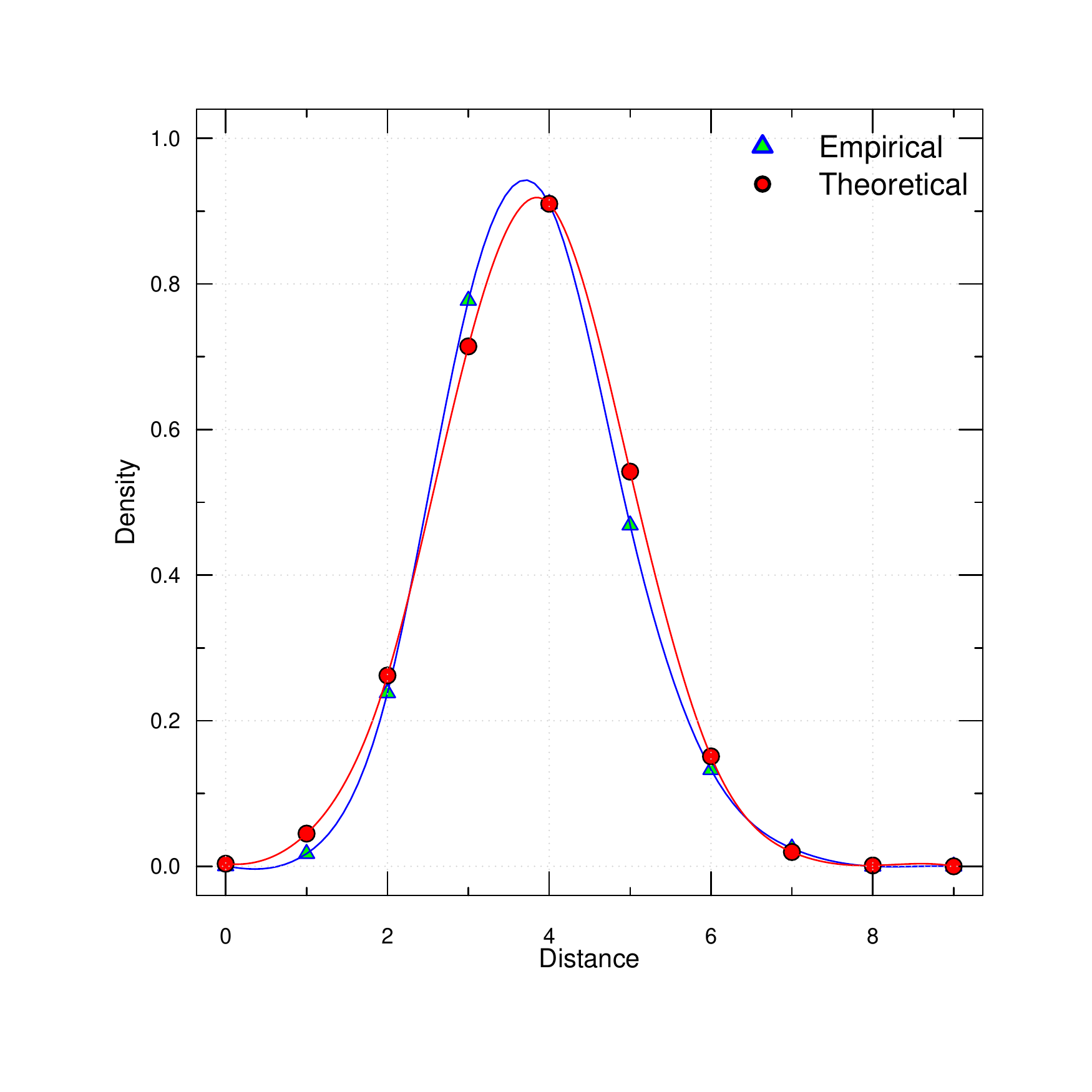}}
        \subfigure[OSLOM*]{\includegraphics[width=.121\textwidth]{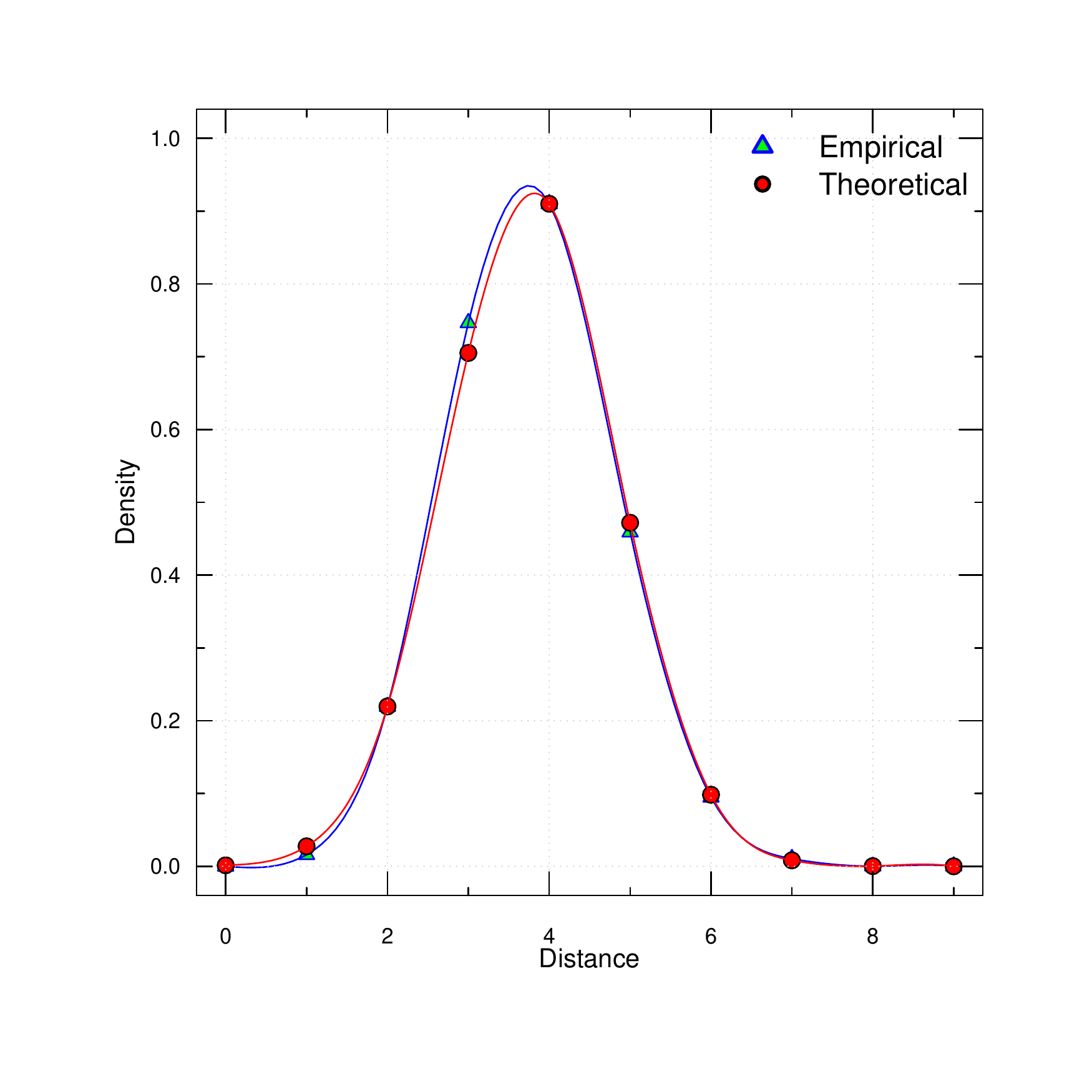}}
        \subfigure[LINKC*]{\includegraphics[width=.121\textwidth]{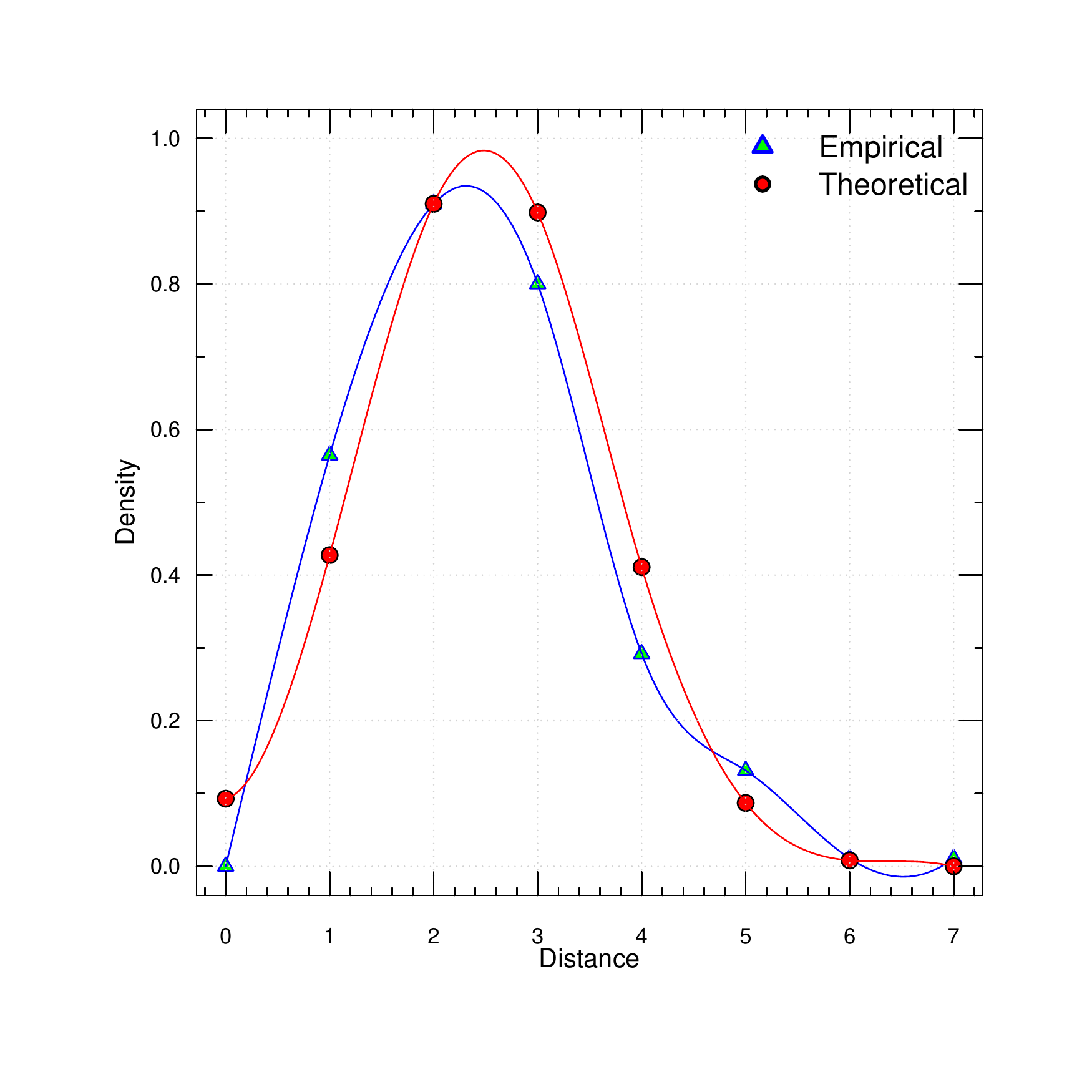}}
        \subfigure[SVINET*]{\includegraphics[width=.121\textwidth]{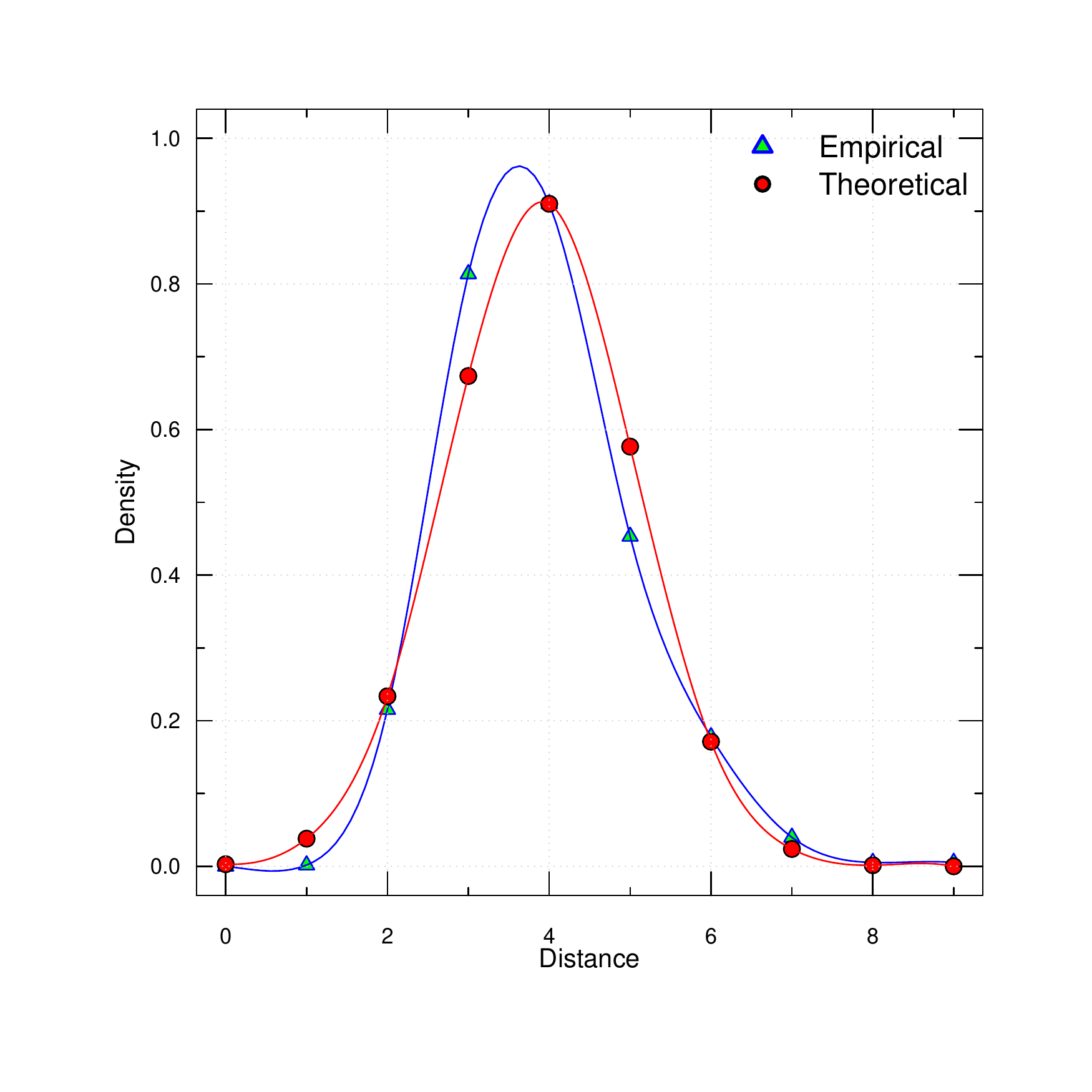}}
        \subfigure[SLPA*]{\includegraphics[width=.121\textwidth]{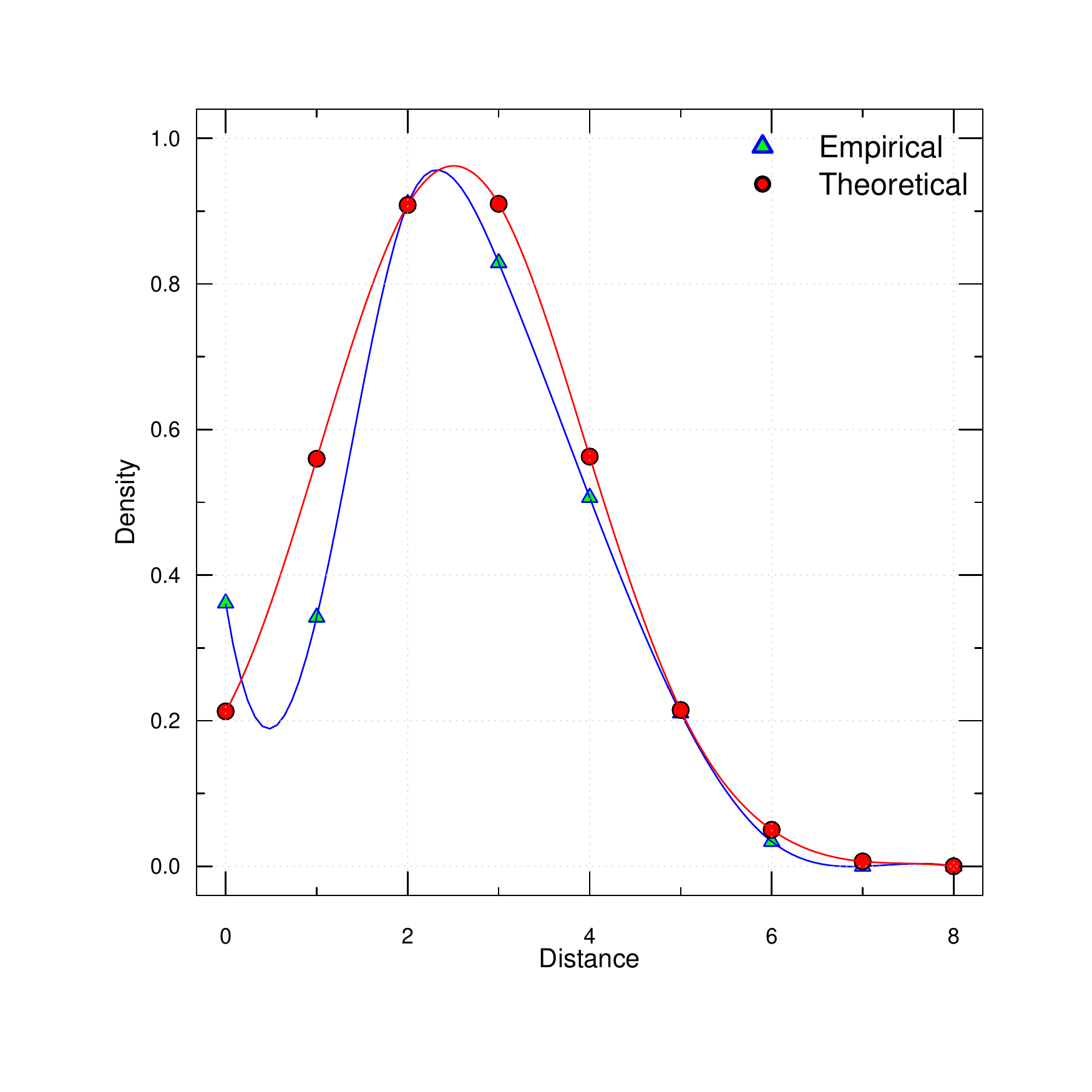}}
        \subfigure[DEMON*]{\includegraphics[width=.121\textwidth]{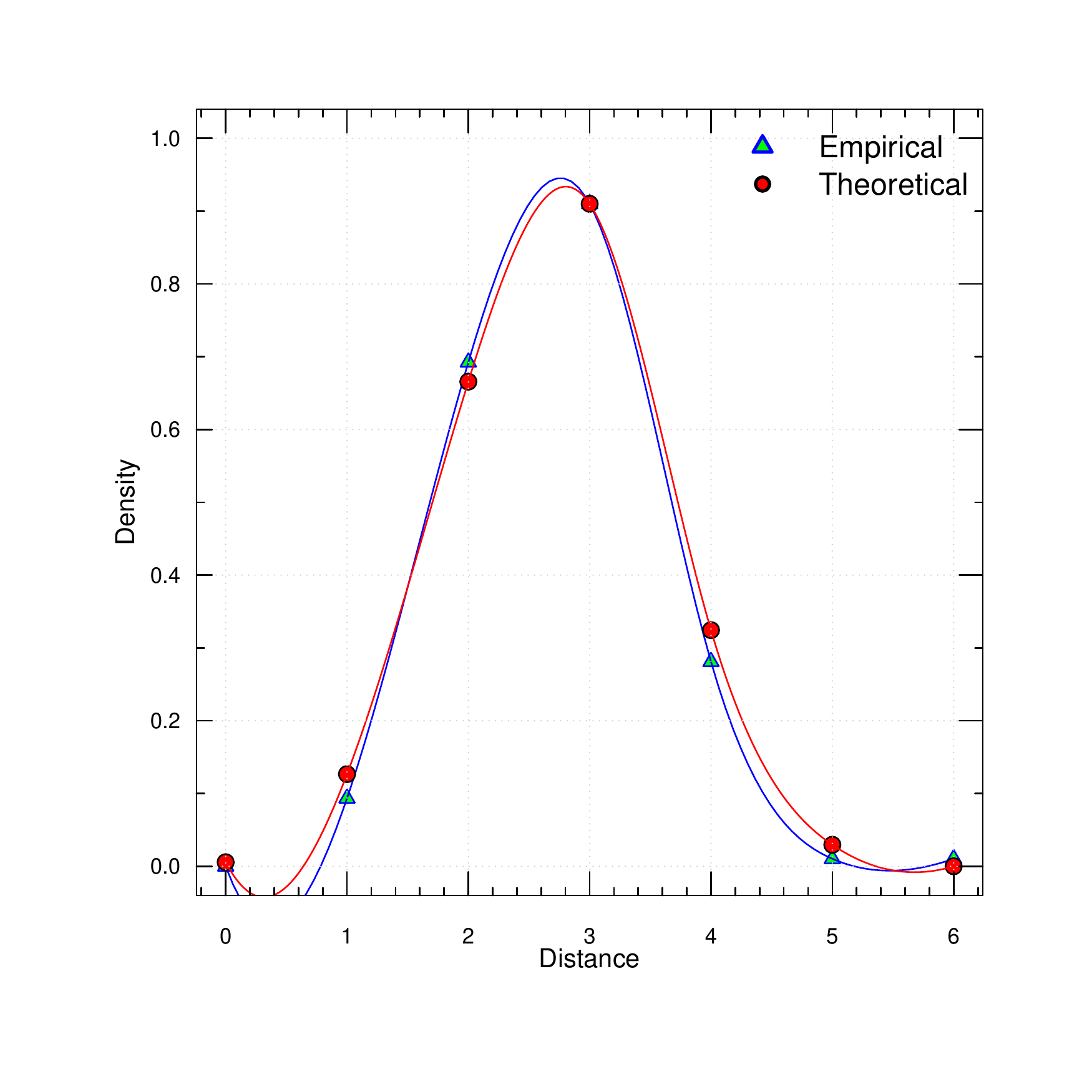}}

        \caption{\label{fig7}Hop Distance distribution forPGP* (a), LFM* (b), GCE* (c),  OSLOM* (d), LINKC* (e), SVINET* (f), SLPA* (g) and  DEMON*(h)}
        \end{figure}

\begin{figure}[ht!]
        \subfigure[PGP*]{\includegraphics[page=2,width=.121\textwidth]{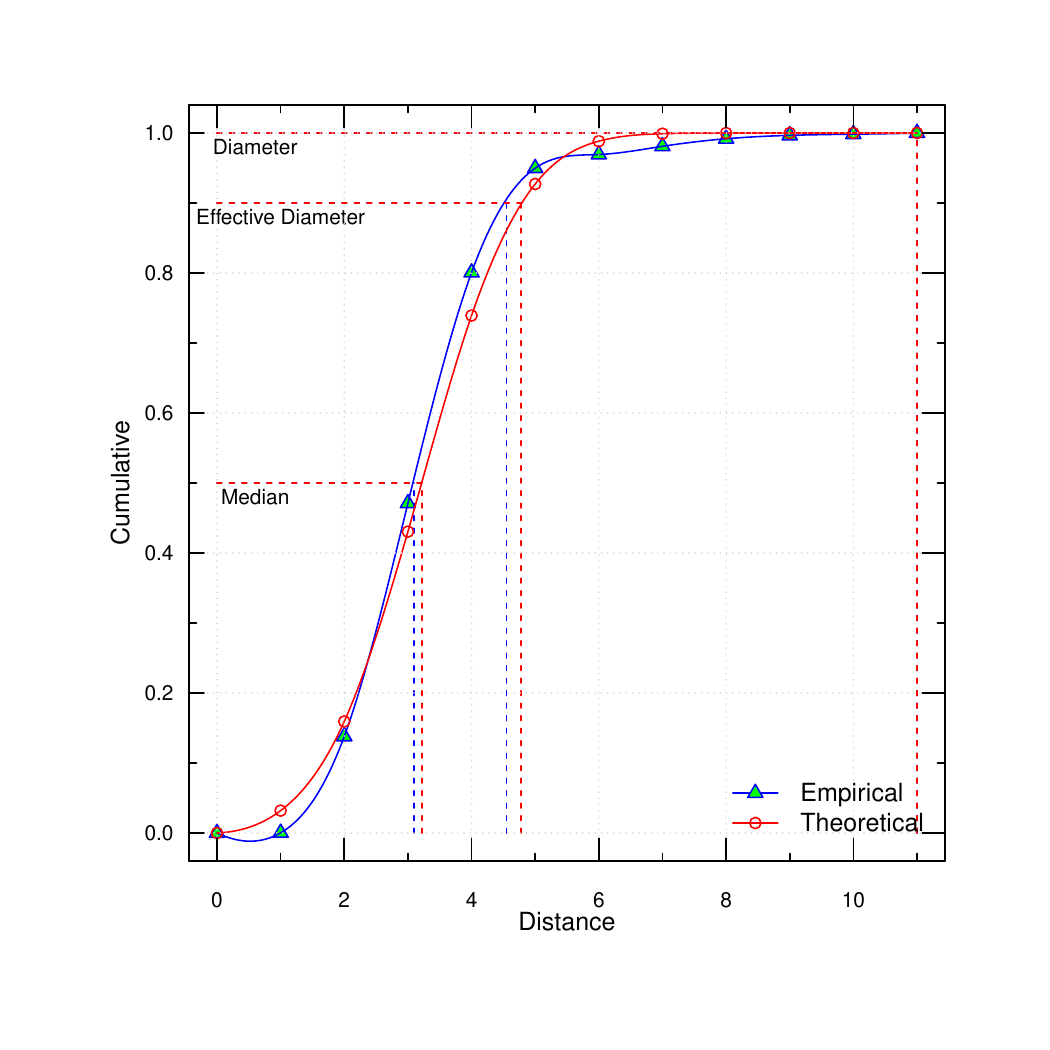}}
        \subfigure[LFM*]{\includegraphics[page=2,width=.121\textwidth]{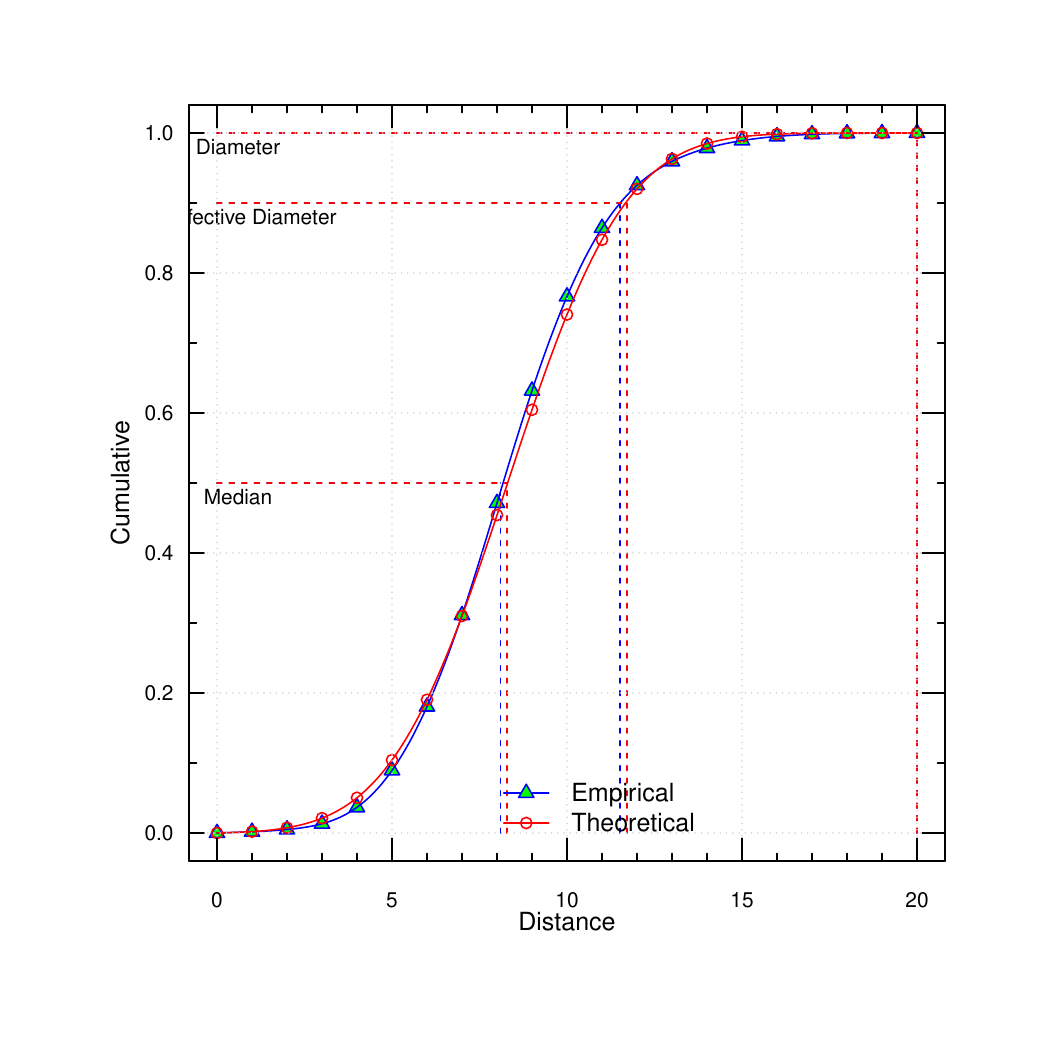}}
        \subfigure[GCE*]{\includegraphics[page=2,width=.121\textwidth]{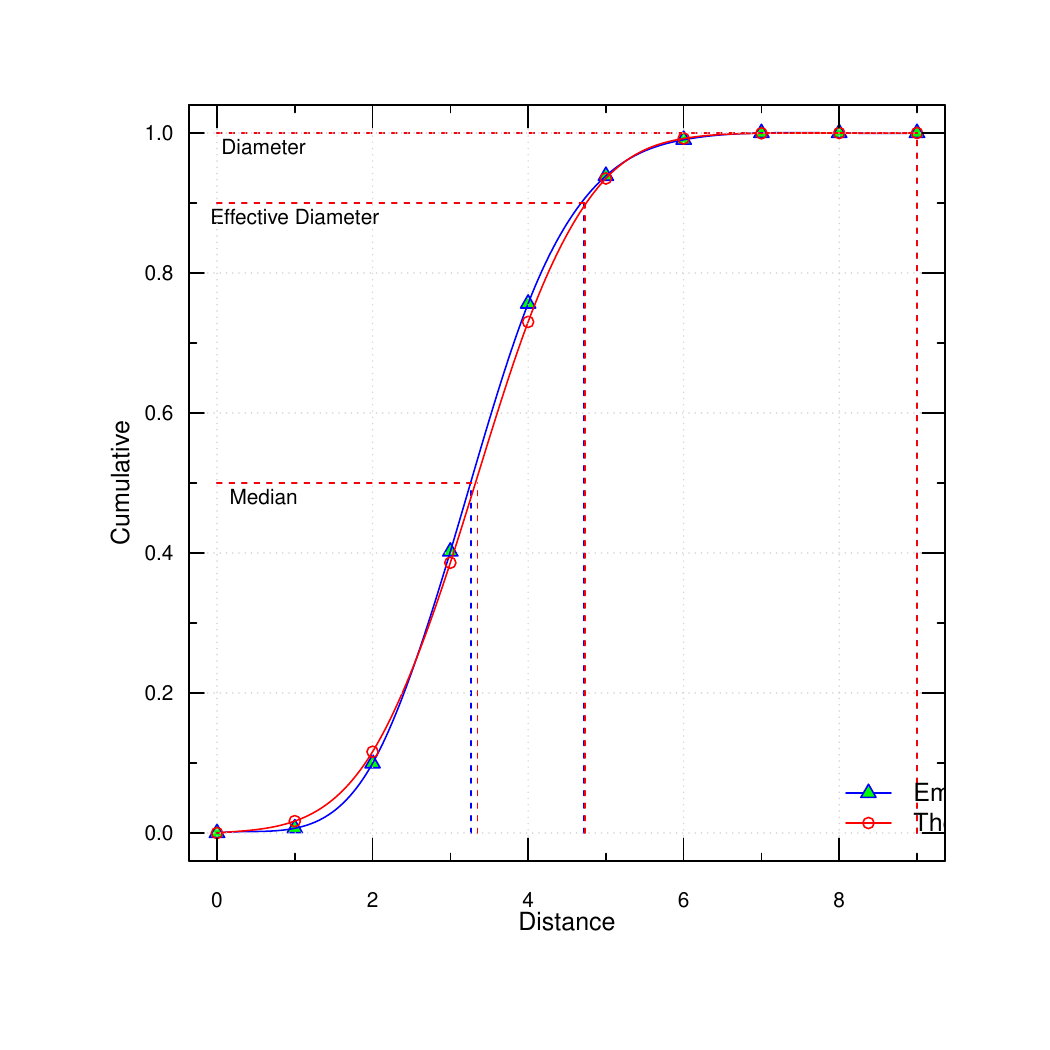}}
        \subfigure[OSLOM*]{\includegraphics[page=2,width=.121\textwidth]{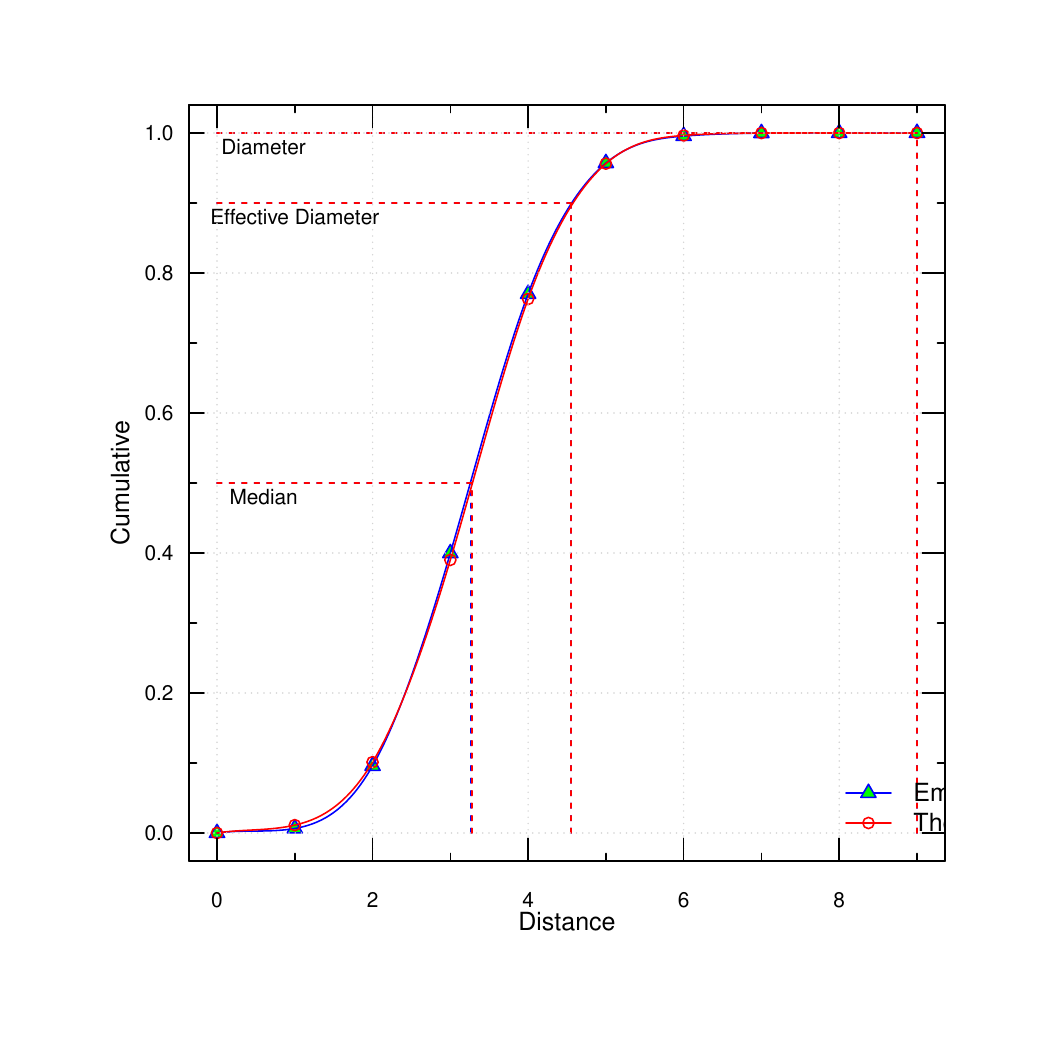}}
        \subfigure[LINKC*]{\includegraphics[page=2,width=.121\textwidth]{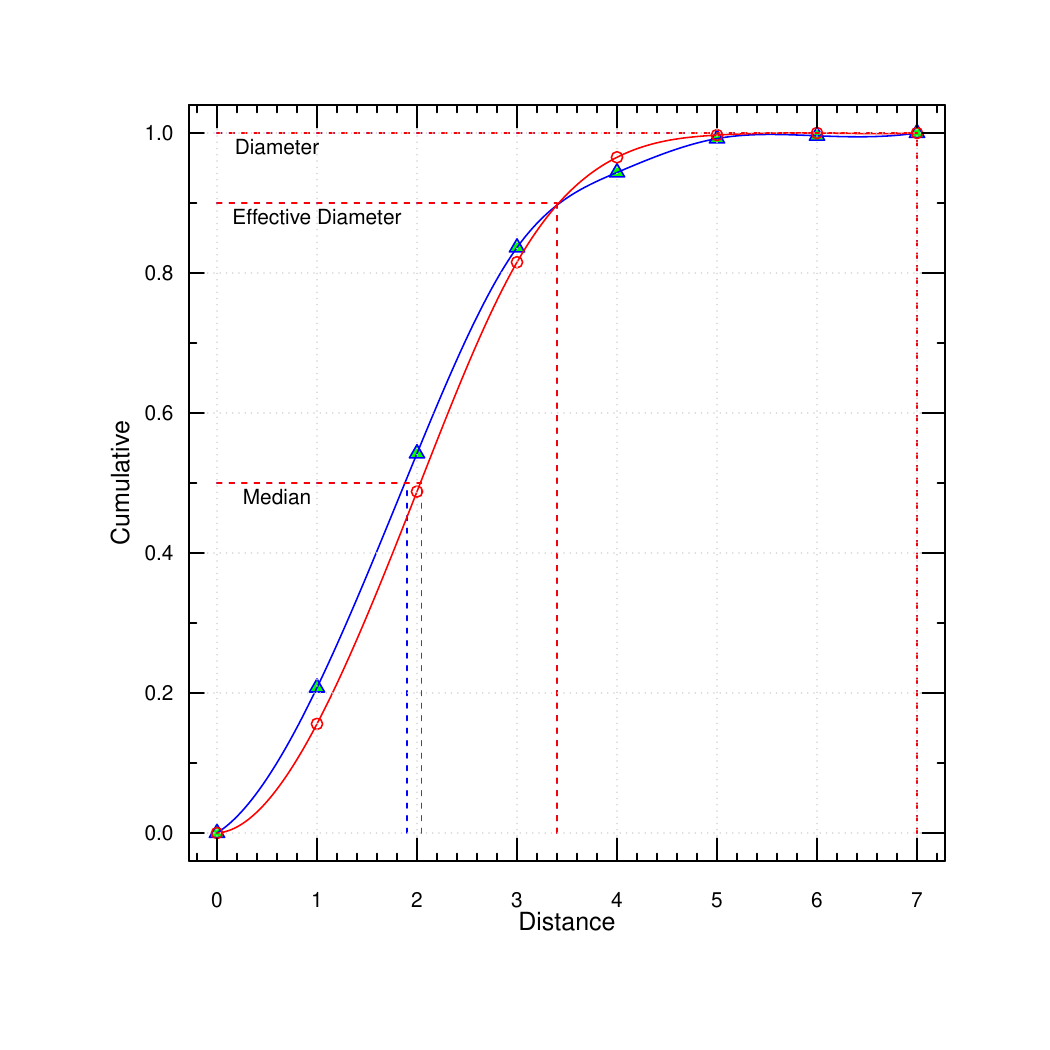}}
        \subfigure[SVINET*]{\includegraphics[page=2,width=.121\textwidth]{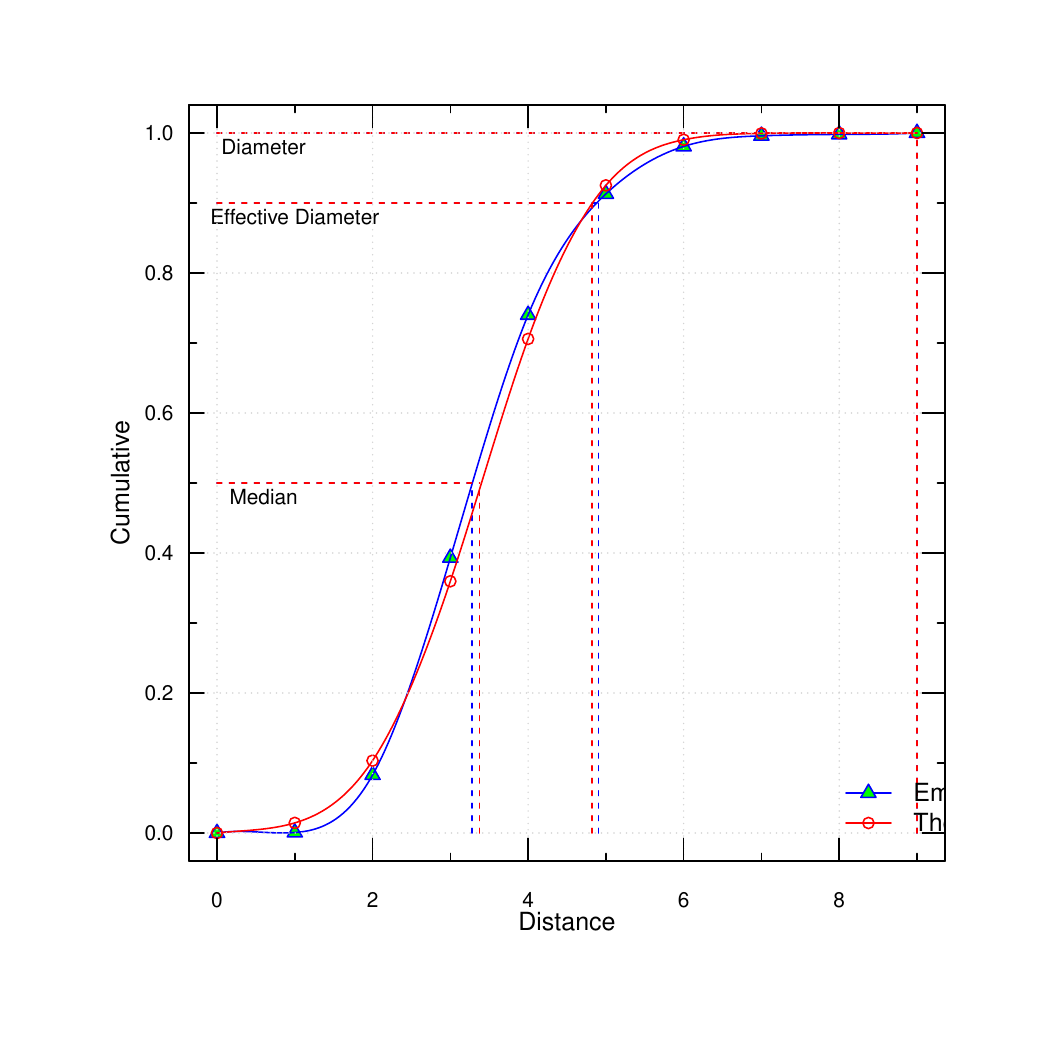}}
        \subfigure[SLPA*]{\includegraphics[page=2,width=.121\textwidth]{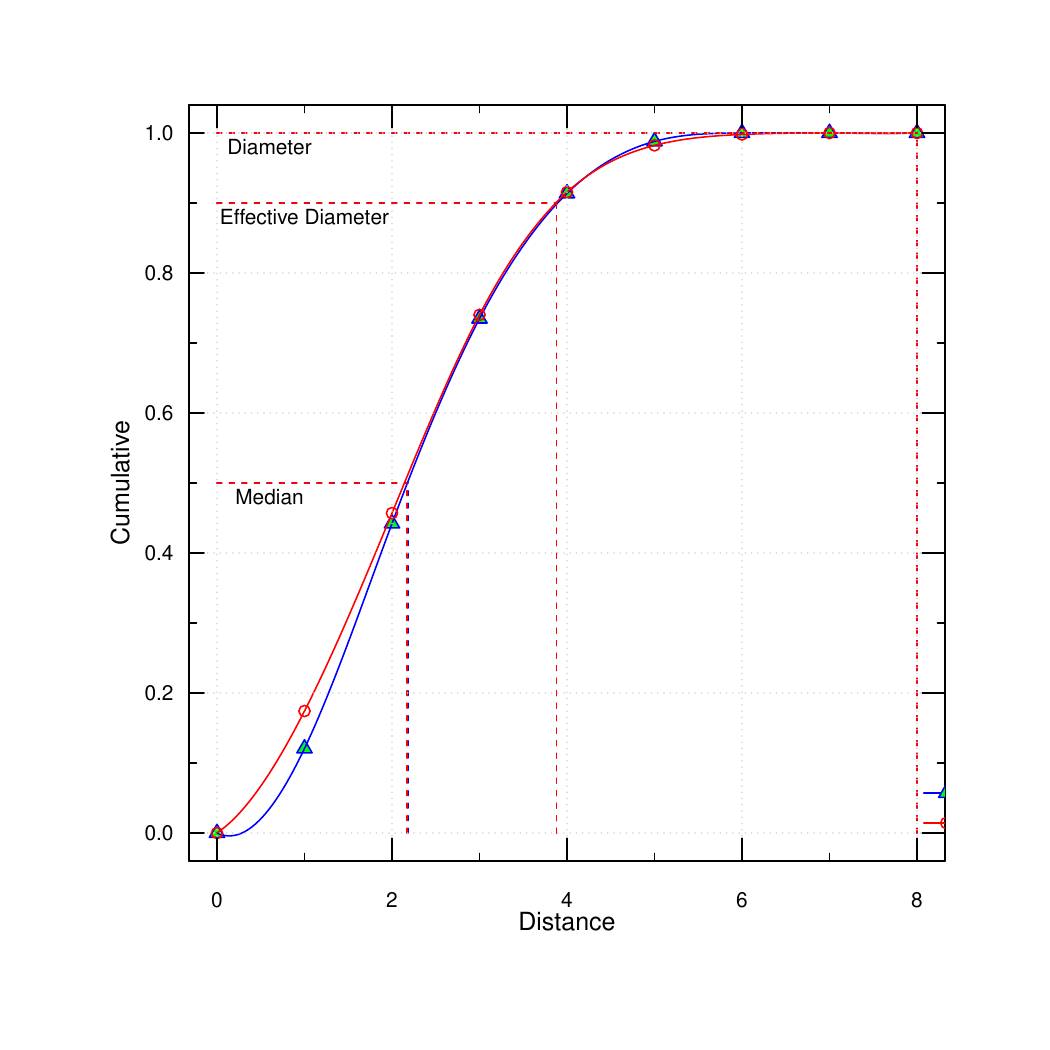}}
        \subfigure[DEMON*]{\includegraphics[page=2,width=.121\textwidth]{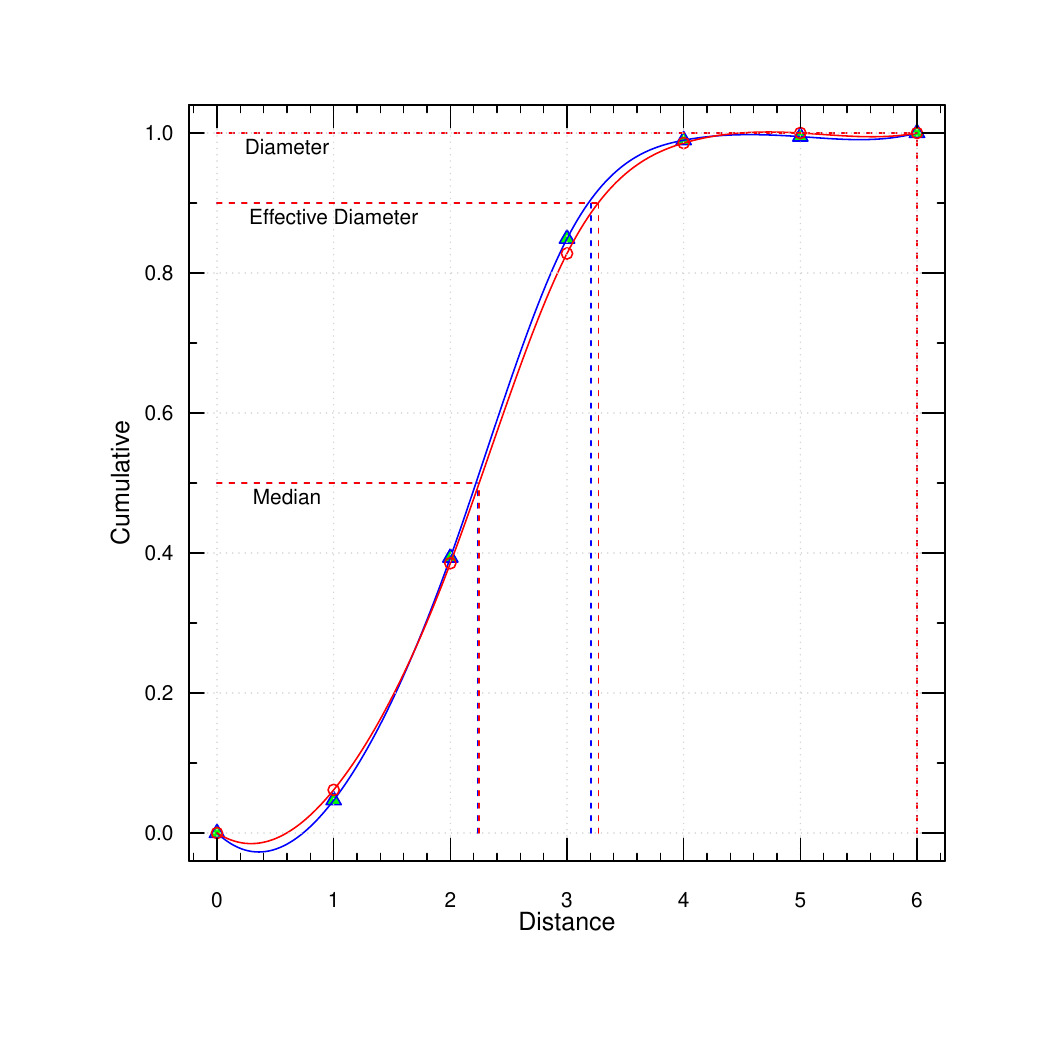}}

        \caption{\label{fig10}Hop distance cumulative distributions for PGP* (a), LFM* (b), GCE* (c),  OSLOM* (d), LINKC* (e), SVINET* (f), SLPA* (g) and  DEMON* (h)}
        \end{figure}

        Table \ref{table13} reports the KS-test values for the various distributions tested on PGP*. According to these results, the Gaussian distribution is clearly the best fit. The goodness-of-fit test results under the hypothesis that the hop distance distribution is Gaussian are shown in Table \ref{table810} for the other 'community-graphs'. The low value of the KS distance supports this hypothesis. Note that for LINKC* and SLPA*, the Exponential distribution is the best fit. Indeed, in this case, the KS distance value is slightly lower ($0.04$ for LINKC* and $0.06$ for SLPA*).
        Figure \ref{fig7} represents the Gaussian estimated density and the empirical distribution for all the 'community-graphs'. It shows that in some cases (PGP*, SLPA* and LINKC*) the empirical distributions are asymmetric. This may explain the better fits of a non-Gaussian distribution. The estimated values of the mean and the standard deviation are displayed in Table \ref{table11}. We note that their values are very close to the reference ones (PGP*) for DEMON*, GCE*, as well as SLPA*. The cumulative distributions are also plotted in Figure \ref{fig10}. Their parameters which are the median path length, the effective diameter, and the diameter are also given in Table \ref{table12}. We mention that OSLOM* and GCE* give very similar values to those of the ground-truth PGP*.

        In the case of AMAZON*, the hop distance distribution follows a Normal law with a KS-test value equal to $0.05$ as shown in Figure \ref{fig8} and Table \ref{table14}. Except for OSLOM* which is heavily asymmetric, the Normal distribution is always the best fit for the hop distance distribution of the 'community-graphs' (See Table \ref{table14}). The parameters of the Normal law for DEMON* and SLPA* are very close  as compared to those of the ground truth 'community-graph' (Table \ref{table11}). The parameters (median path length, effective diameter, diameter) extracted from the cumulative distribution of DEMON* and SVINET* are the nearest to those of AMAZON* (see Figure \ref{fig11} and Table \ref{table12}).

        In the case of the aNobii dataset, the results are very consensual. In any case, the Normal distribution is the best fit (see Table \ref{table15}, Table \ref{table11}, Table \ref{table12}, Figure \ref{fig9} and Figure \ref{fig12}).

        \begin{table}[ht!]
        \centering
        \caption{KS-test values for the Hop distance for PGP*. The distributions under test are the Power-Law (PL), Beta (BE), Cauchy (CA), Exponential (E), Gamma (GM), Logistic (LO), Log-Normal (LN), Normal (N), Uniform (U), and Weibull (WB)}
        \label{table13}
        \begin{tabular}{lcccccccccc}
        \hline
         &  PL & BE & CA & E & GM & LO & LN & N & U & WB \\
        \hline
        KS&0.21&0.31&0.54&0.47&0.06&0.47&0.34&0.03&0.44&0.47\\
         \hline
        \end{tabular}
        \end{table}

        \begin{table}[ht!]
          \centering
          \caption{KS-test values for the Hop distance considering the Normal hypothesis for the 'community-graphs'}
          \label{table810}
            \begin{tabular}{lccccccc}
            \hline
              & LFM* & GCE* & OSLOM* & LINKC* & SVINET* & SLPA* & DEMON* \\
            \hline
            KS(Normal) & 0.01 & 0.03 & 0.06 & 0.07 & 0.03 & 0.09 & 0.04 \\

            \hline
            \end{tabular}%
        \end{table}%

\subsection{Mesoscopic properties}
        In this section, we analyze the distribution of the community size, the membership of nodes, and the overlap size. Previous analysis of \cite{Palla2005uncovering} and \cite{7024681} have shown that they can be adequately described by a Power-Law.
        Note that these properties are related to the internal characteristics of the communities and not to the 'community-graphs'.

\subsubsection{Community Size}

        \begin{figure}[ht!]
        \subfigure[Ground-truth]{\includegraphics[width=.121\textwidth]{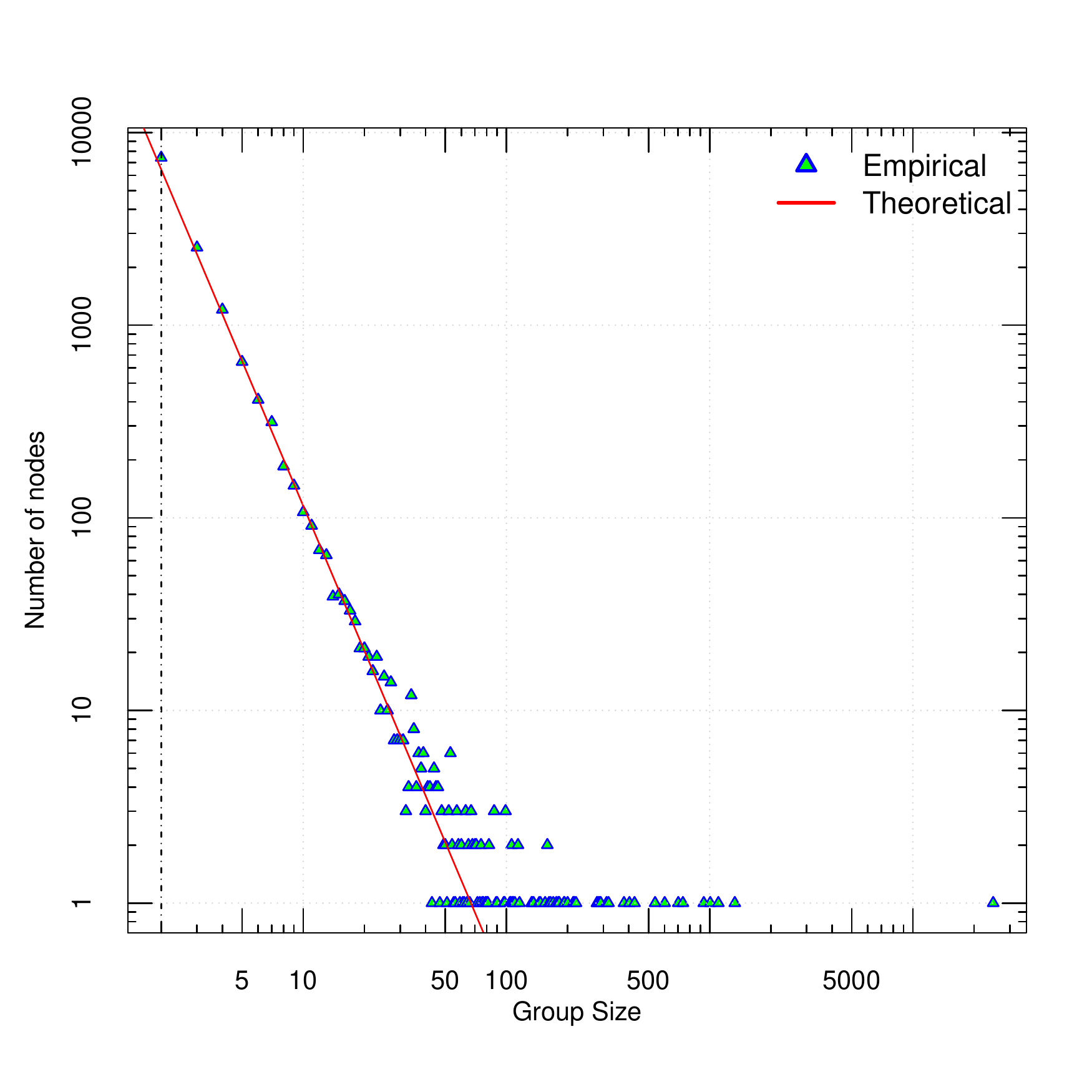}}
        \subfigure[LFM]{\includegraphics[width=.121\textwidth]{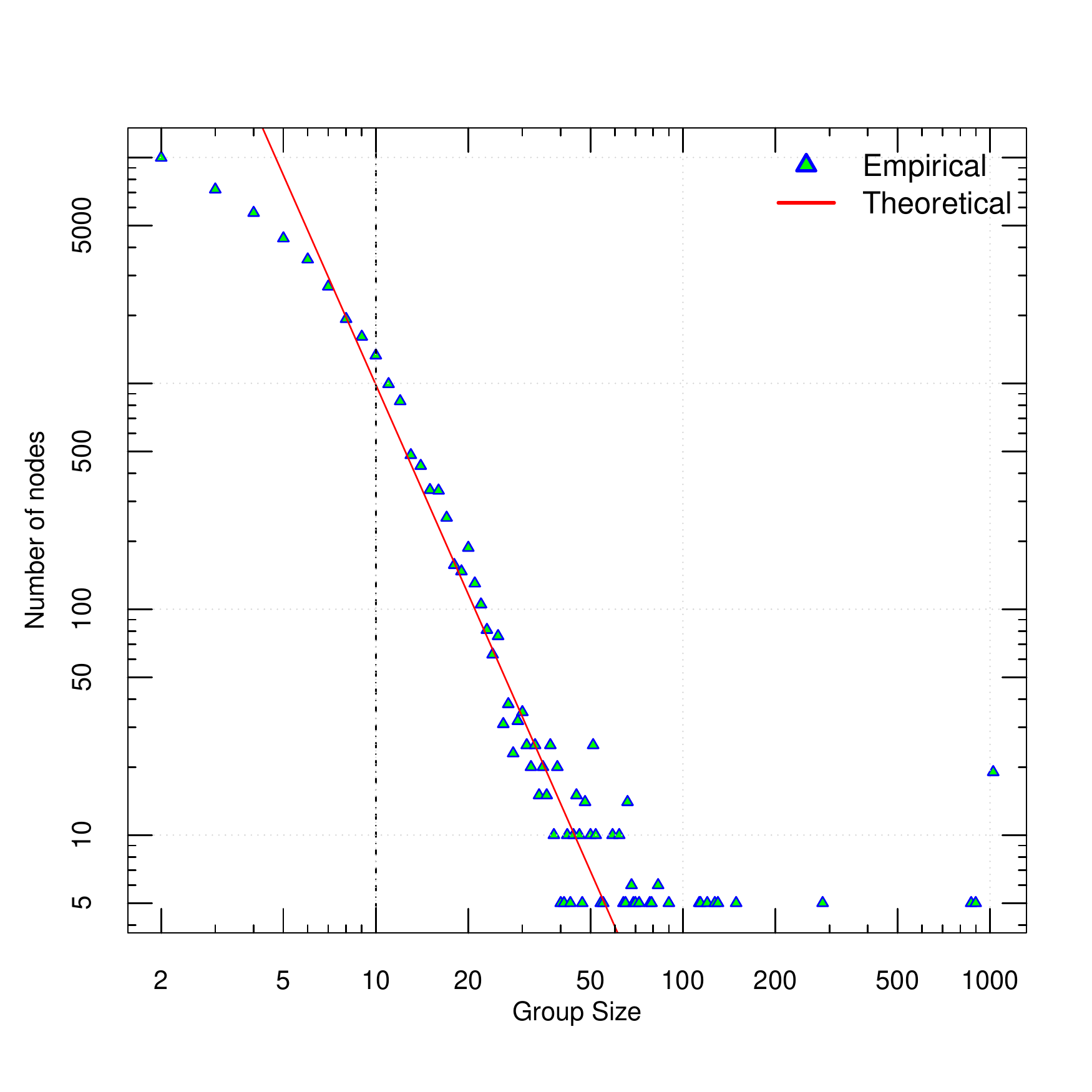}}
        \subfigure[GCE]{\includegraphics[width=.121\textwidth]{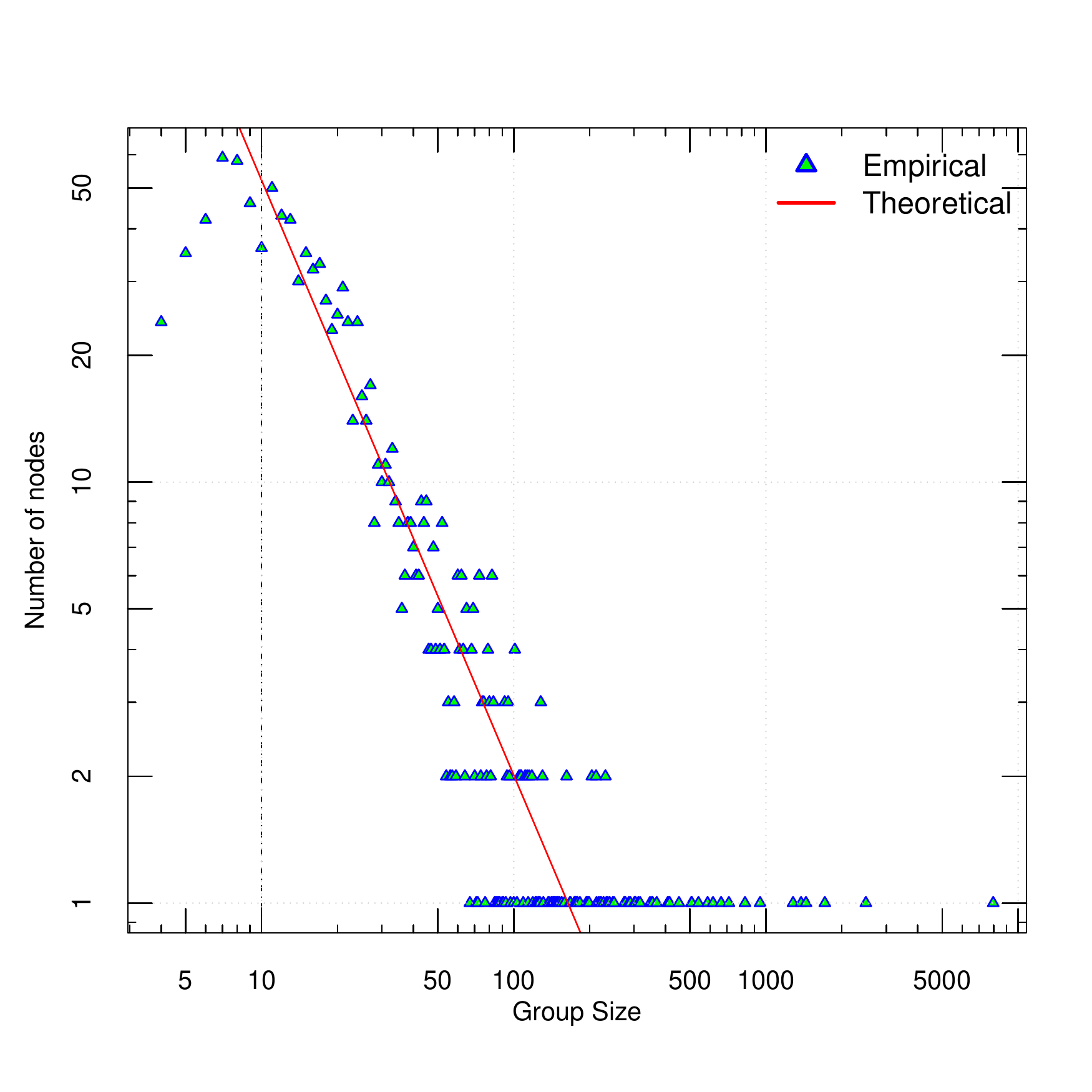}}
        \subfigure[OSLOM]{\includegraphics[width=.121\textwidth]{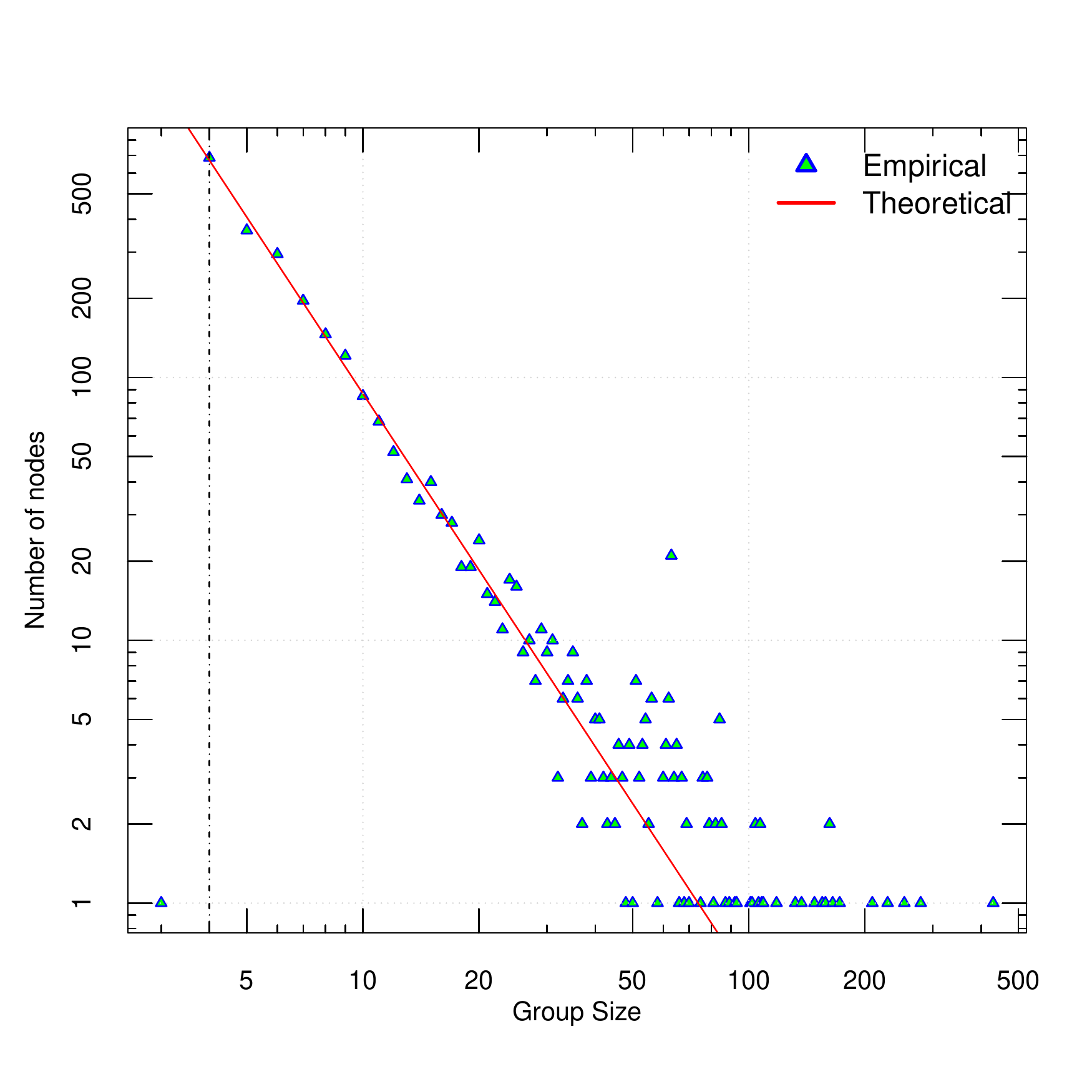}}
        \subfigure[LINKC]{\includegraphics[width=.121\textwidth]{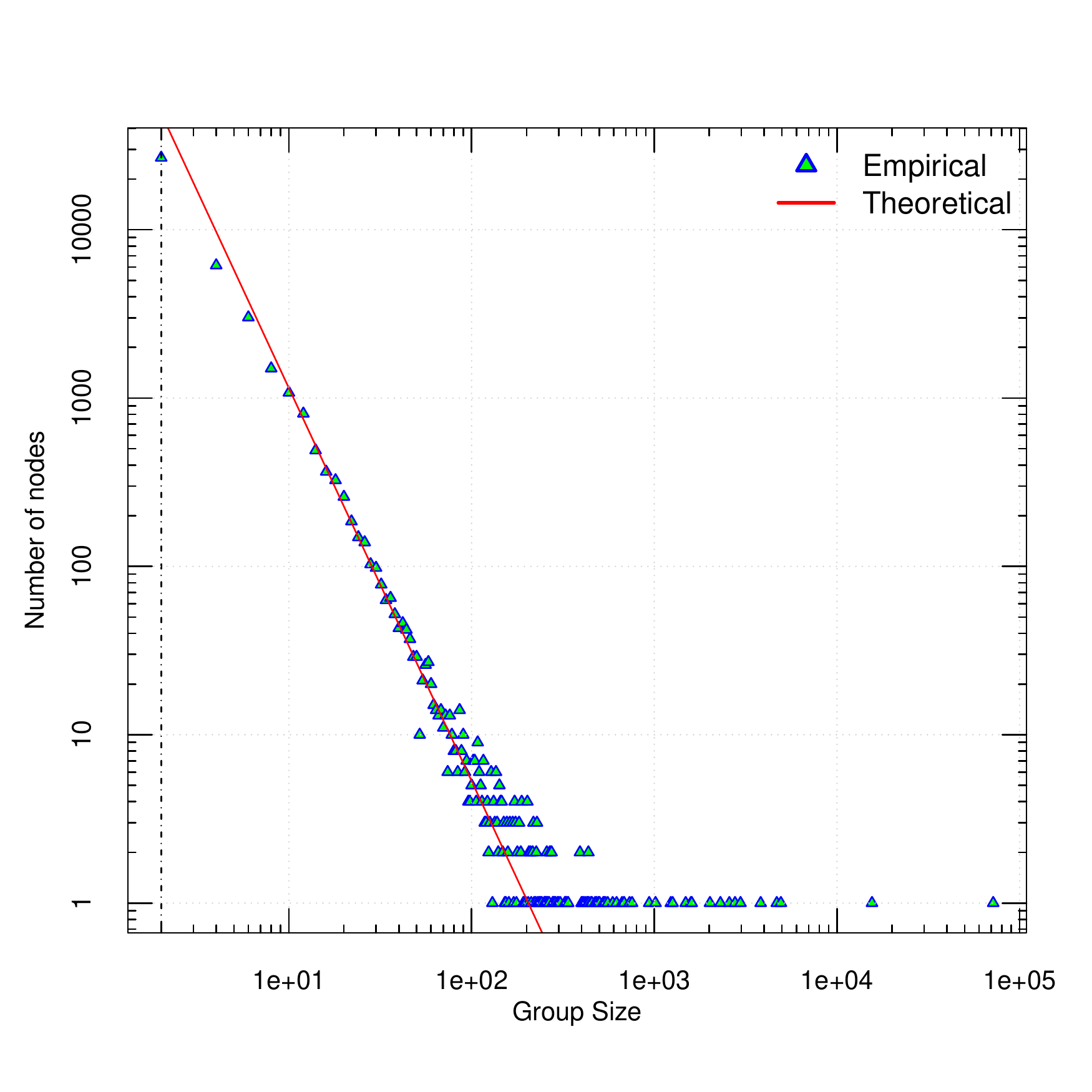}}
        \subfigure[SVINET]{\includegraphics[width=.121\textwidth]{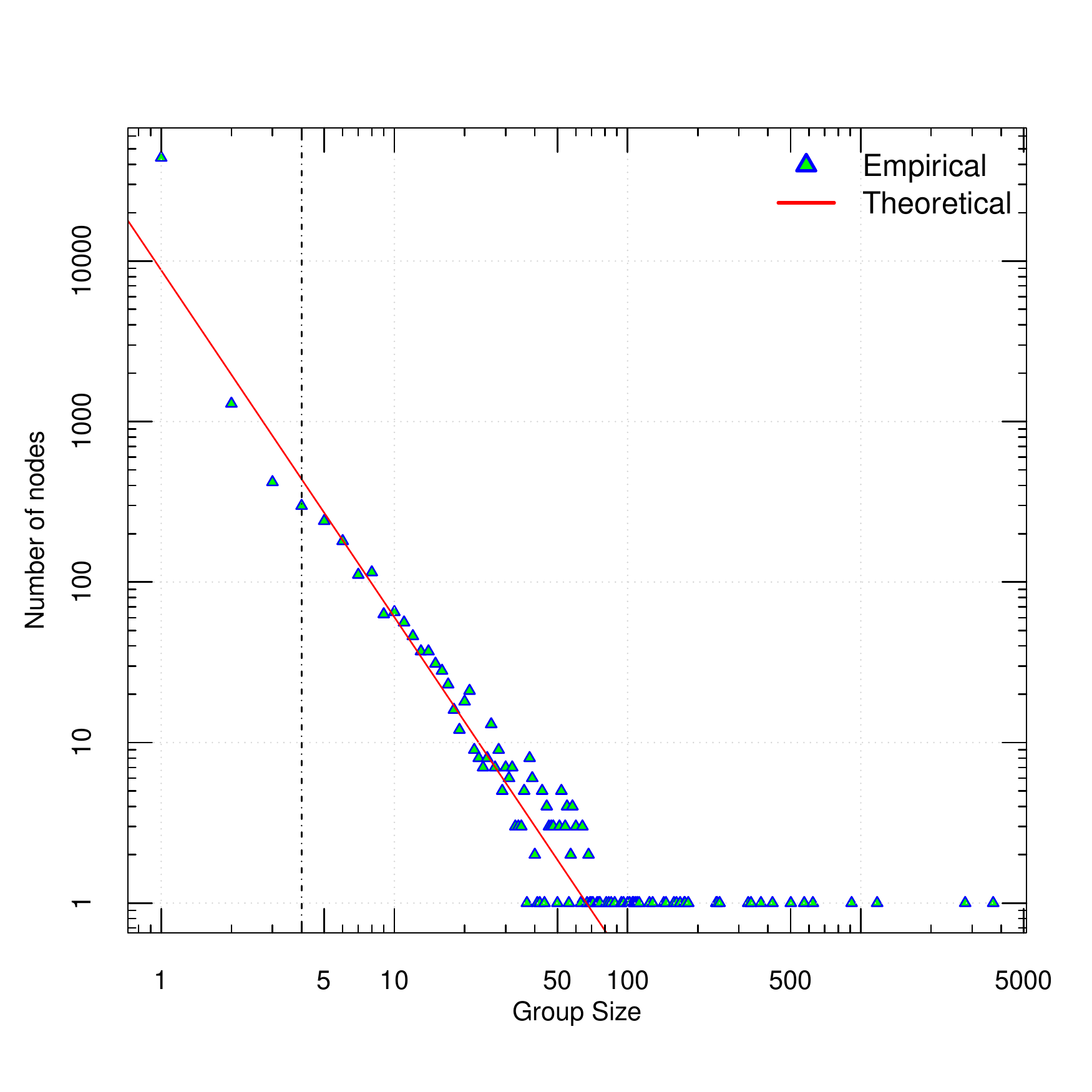}}
        \subfigure[SLPA]{\includegraphics[width=.121\textwidth]{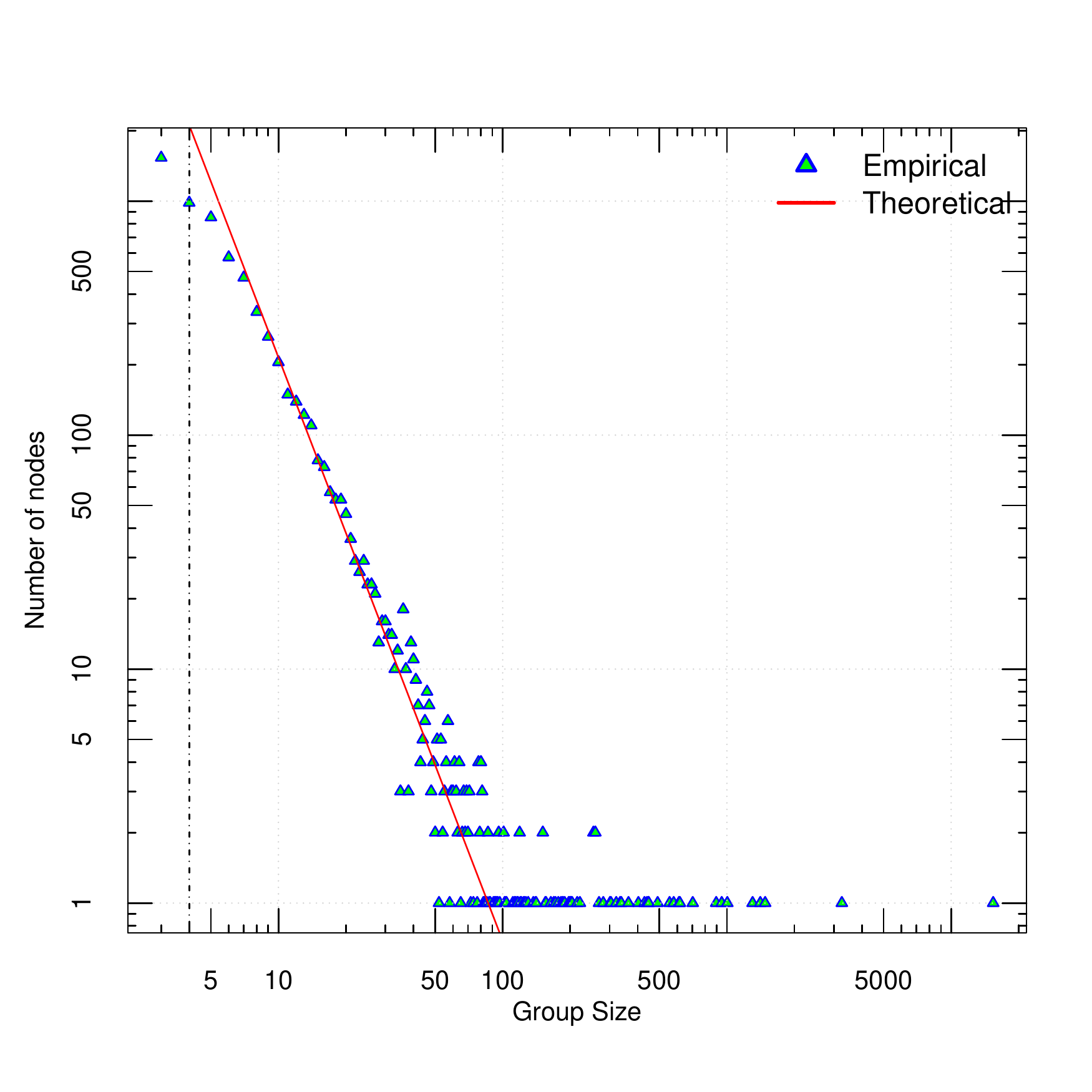}}
        \subfigure[DEMON]{\includegraphics[width=.121\textwidth]{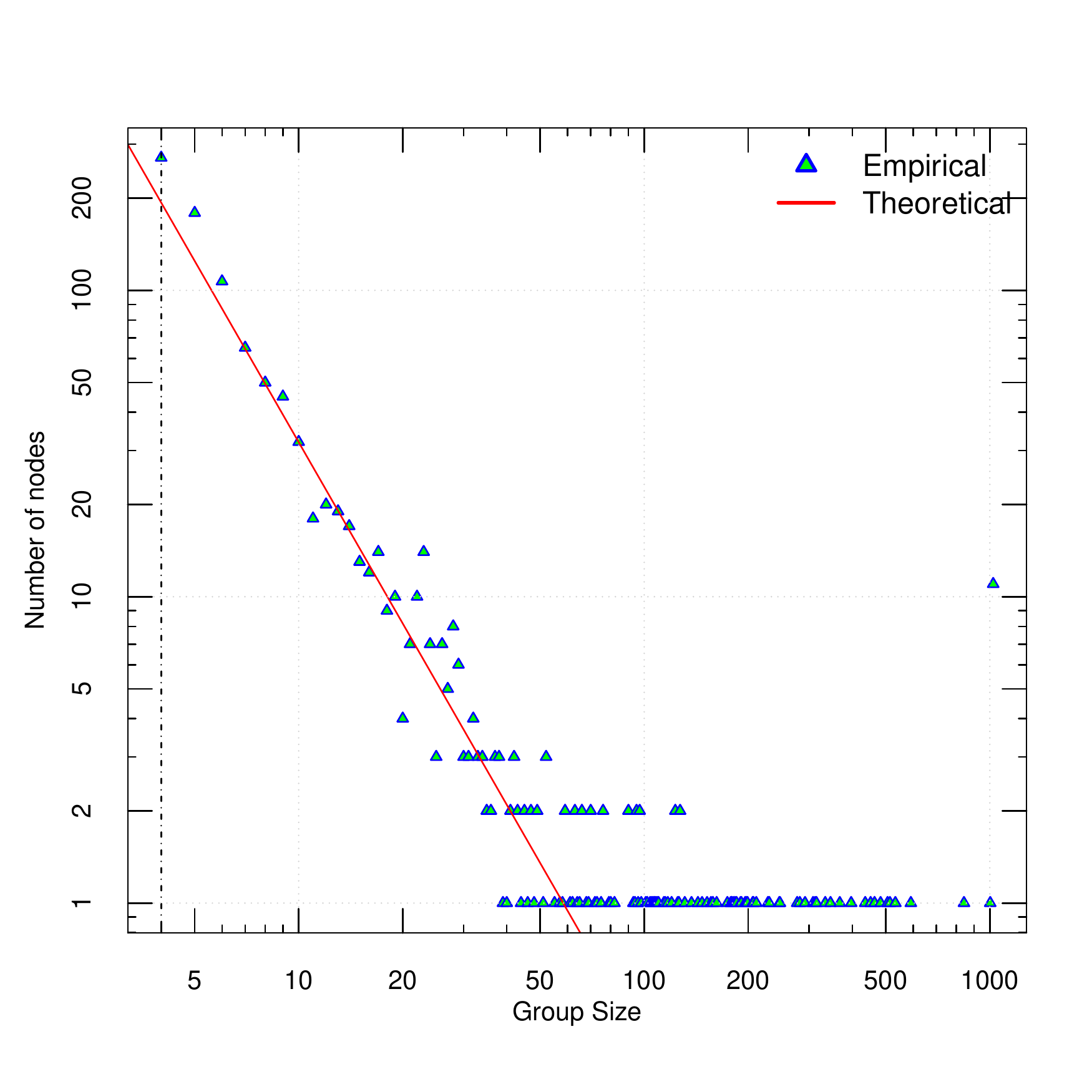}}

        \caption{\label{fig13} Log-log empirical Community size distribution (dots) and Power-Law estimate (line) of PGP Ground-truth (a), LFM (b), GCE (c),  OSLOM (d), LINKC (e), SVINET (f), SLPA (g) and  DEMON (h)}
        \end{figure}

        The community size distributions of the ground-truth community structure of PGP and the unveiled community structure by the algorithms are shown in Figure \ref{fig13}. It is clear that they follow a Power-Law. Results of the KS-test reported in Table \ref{table17} confirm that the Power-Law is the most suitable hypothesis in any case.

        The parameters of the Power-Law (average, maximal community size and the exponent) together with the number of communities are given in Table \ref{table16}. It shows that no algorithm provide a number of communities close to that of the ground-truth community structure. Globally, the Power-Law exponents are in the same range than the reference one. Nevertheless, when we look at maximum and average community size, we observe a great dispersion of the results. It seems that it is difficult for all the algorithms to uncover the biggest communities. They are generally split into smaller ones.

        \begin{table}[ht!]
        \centering
        \caption{KS-test values for the Community size. The distributions under test are the Power-Law (PL), Beta (BE), Cauchy (CA), Exponential (E), Gamma (GM), Logistic (LO), Log-Normal (LN), Normal (N), Uniform (U), and Weibull (WB)}
        \label{table17}
        \begin{tabular}{lcccccccccc}
        \hline
         &  PL & BE & CA & E & GM & LO & LN & N & U & WB \\
        \hline
        Ground-truth&0.01&0.54&0.14&0.54&0.54&0.49&0.21&0.49&0.99&0.18\\
        DEMON&0.03&0.55&0.26&0.5&0.44&0.38&0.13&0.4&0.85&0.16\\
        LFM&0.02&0.65&0.39&0.23&0.64&0.42&0.13&0.43&0.96&0.28\\
        SLPA&0.01&0.75&0.28&0.38&0.74&0.47&0.1&0.48&0.98&0.35\\
        LINKC&0.03&0.63&0.29&0.63&0.63&0.49&0.21&0.49&0.99&0.35\\
        GCE&0.03&0.79&0.18&0.29&0.76&0.41&0.05&0.42&0.94&0.28\\
        OSLOM&0.01&0.51&0.16&0.26&0.48&0.32&0.14&0.34&0.86&0.25\\
        SVINET&0.02&0.45&0.26&0.35&0.71&0.54&0.11&0.51&0.88&0.44\\
        \hline
        \end{tabular}
        \end{table}

        We also analyzed the community size distribution of AMAZON as well as aNobii. Figure \ref{fig14} reports the empirical distributions and the estimated Power-Law for the ground-truth community structure of AMAZON and the outputs of the community detection algorithm. The Power-Law is always a very good fit. The KS-test results reported in Table \ref{table18} confirm this feeling. Indeed, the Power-Law exhibits the smallest KS distance values.

        Note that the Log-Normal is not far behind for most of the algorithms (LFM, MOSES, GCE, OSLOM, DEMON, SLPA and SVINET). Concerning the parameters of the Power-Law, results are very similar than those of the PGP dataset: the exponents of Power-Law values are acceptable, the number of communities and the maximum community size are always under estimated (see Table \ref{table16}).

        In the case of aNobii, the results are summarized in Table \ref{table19}, Table \ref{table16} and Figure \ref{fig15}. Globally in accordance with the previous conclusions. Nevertheless, there are a few differences. Indeed, some algorithms (GCE and SLPA) uncover communities which are bigger than the reference.

\subsubsection{Membership}

        \begin{figure}[ht!]
        \subfigure[Ground-truth]{\includegraphics[width=.121\textwidth]{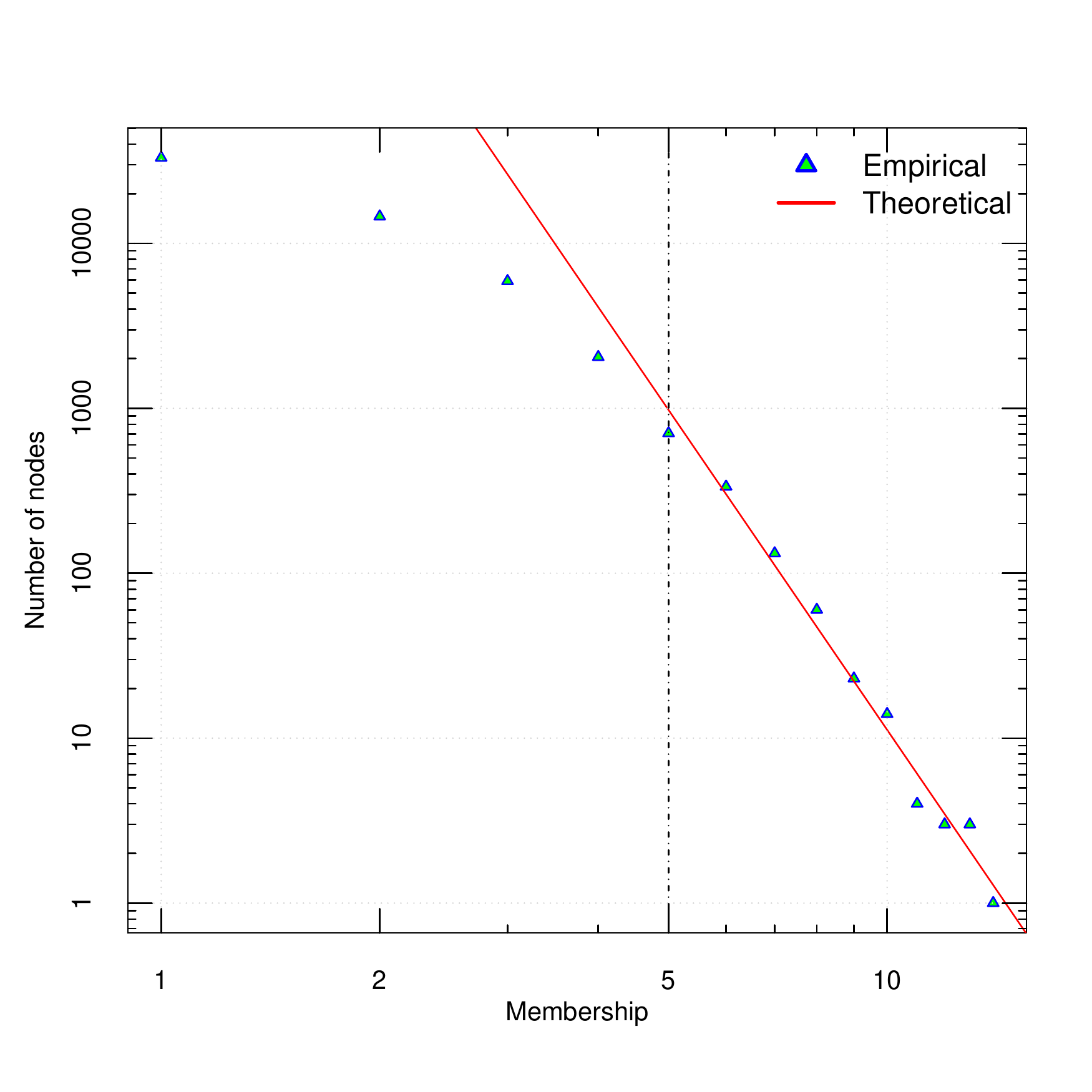}}
        \subfigure[LFM]{\includegraphics[width=.121\textwidth]{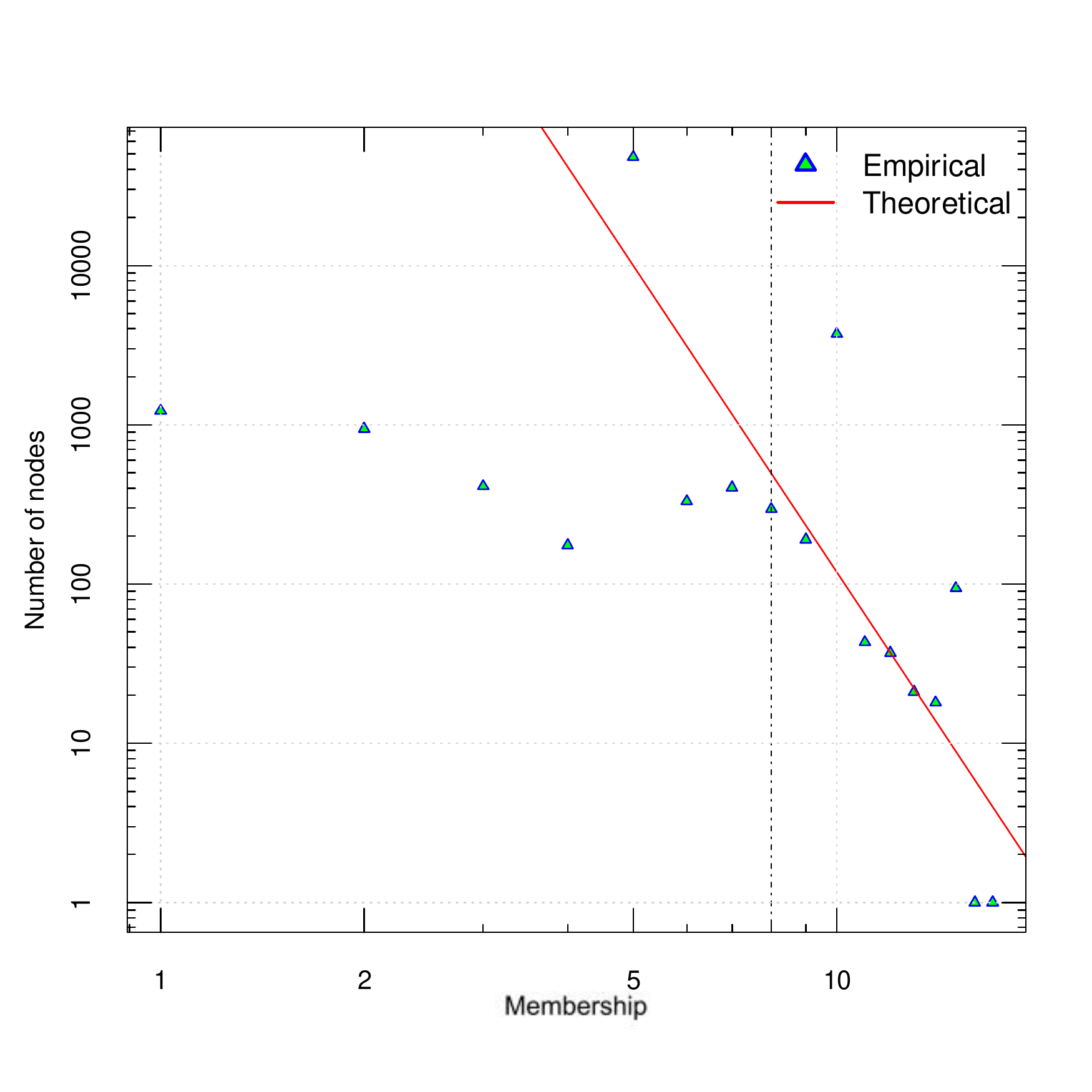}}
        \subfigure[GCE]{\includegraphics[width=.121\textwidth]{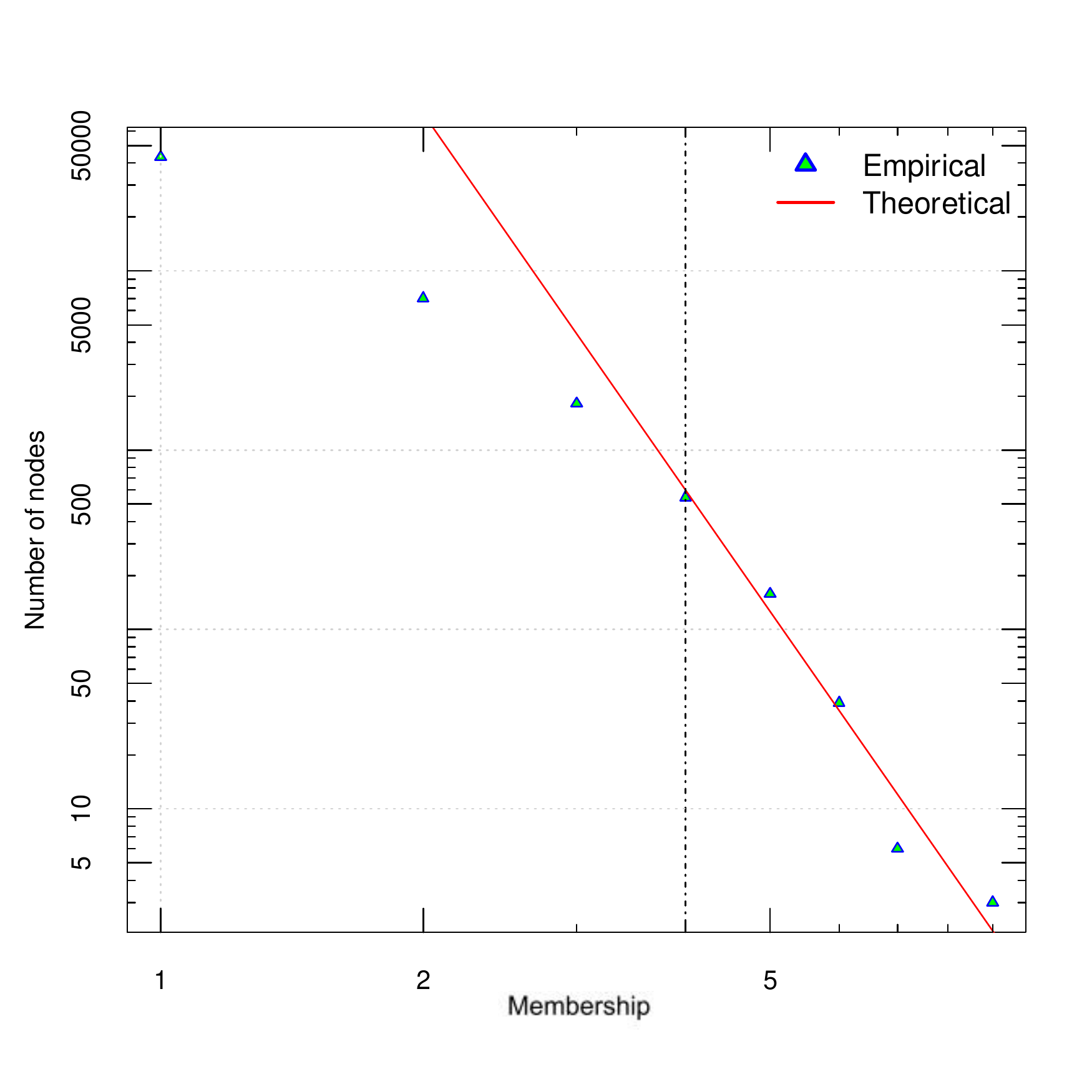}}
        \subfigure[OSLOM]{\includegraphics[width=.121\textwidth]{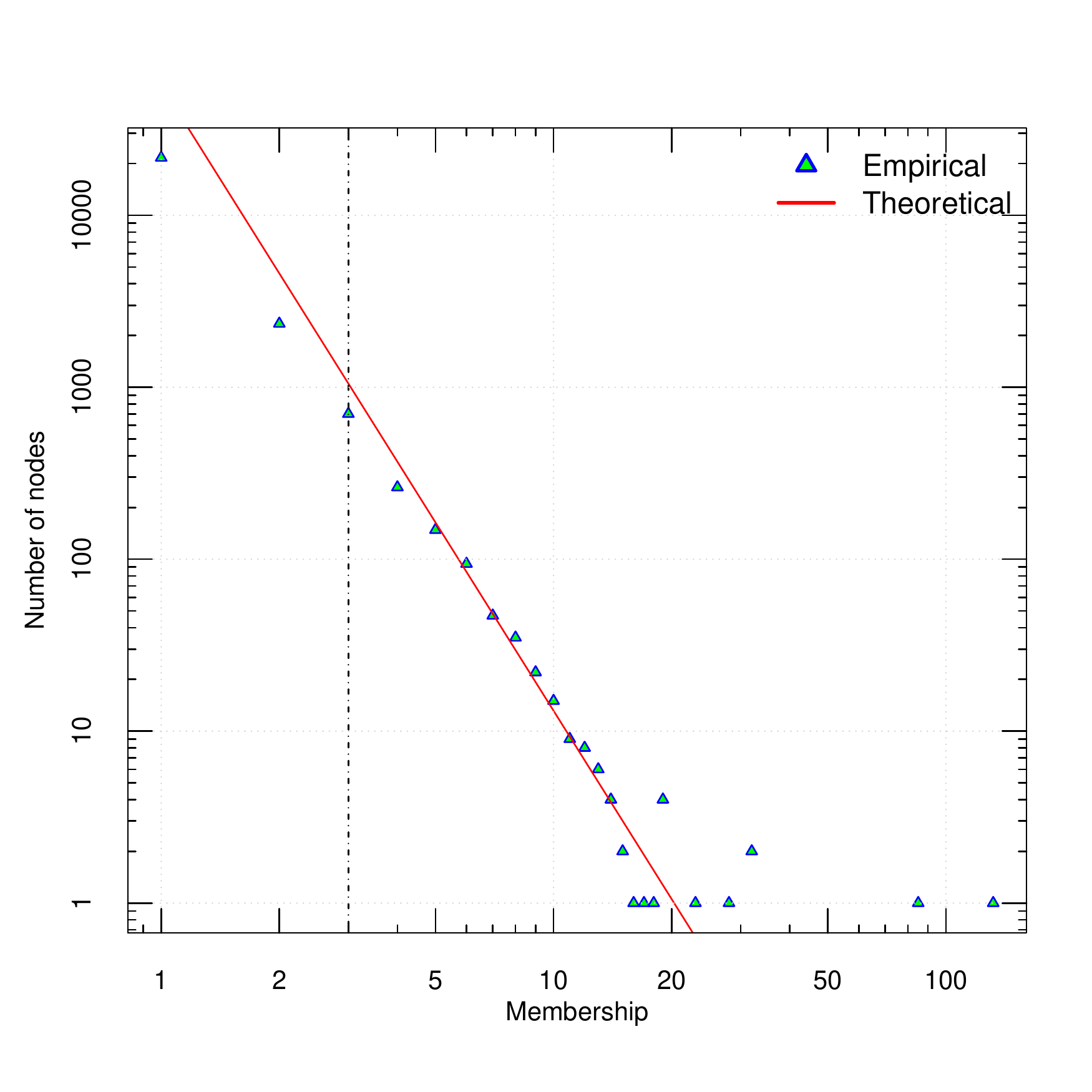}}
        \subfigure[LINKC]{\includegraphics[width=.121\textwidth]{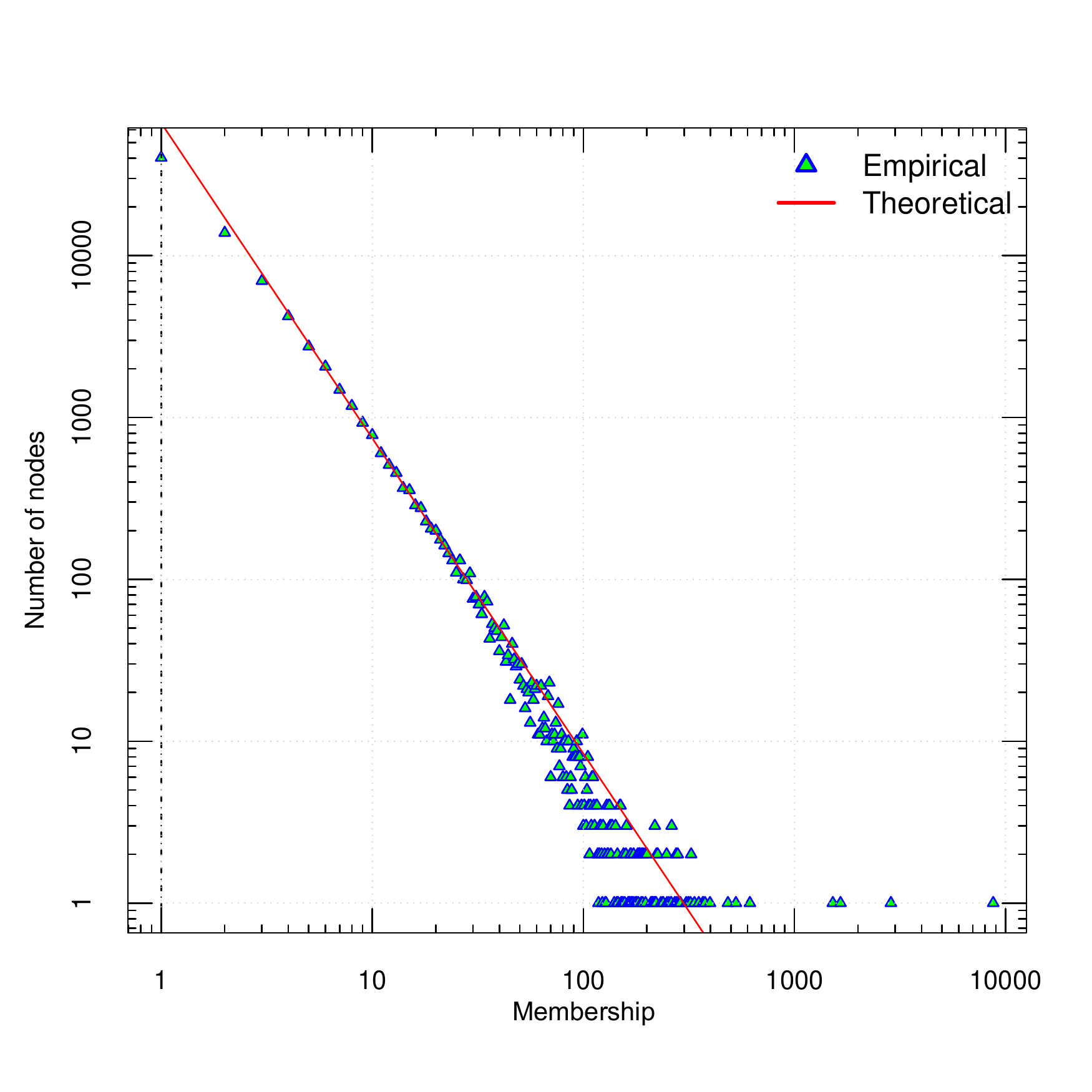}}
        \subfigure[SVINET]{\includegraphics[width=.121\textwidth]{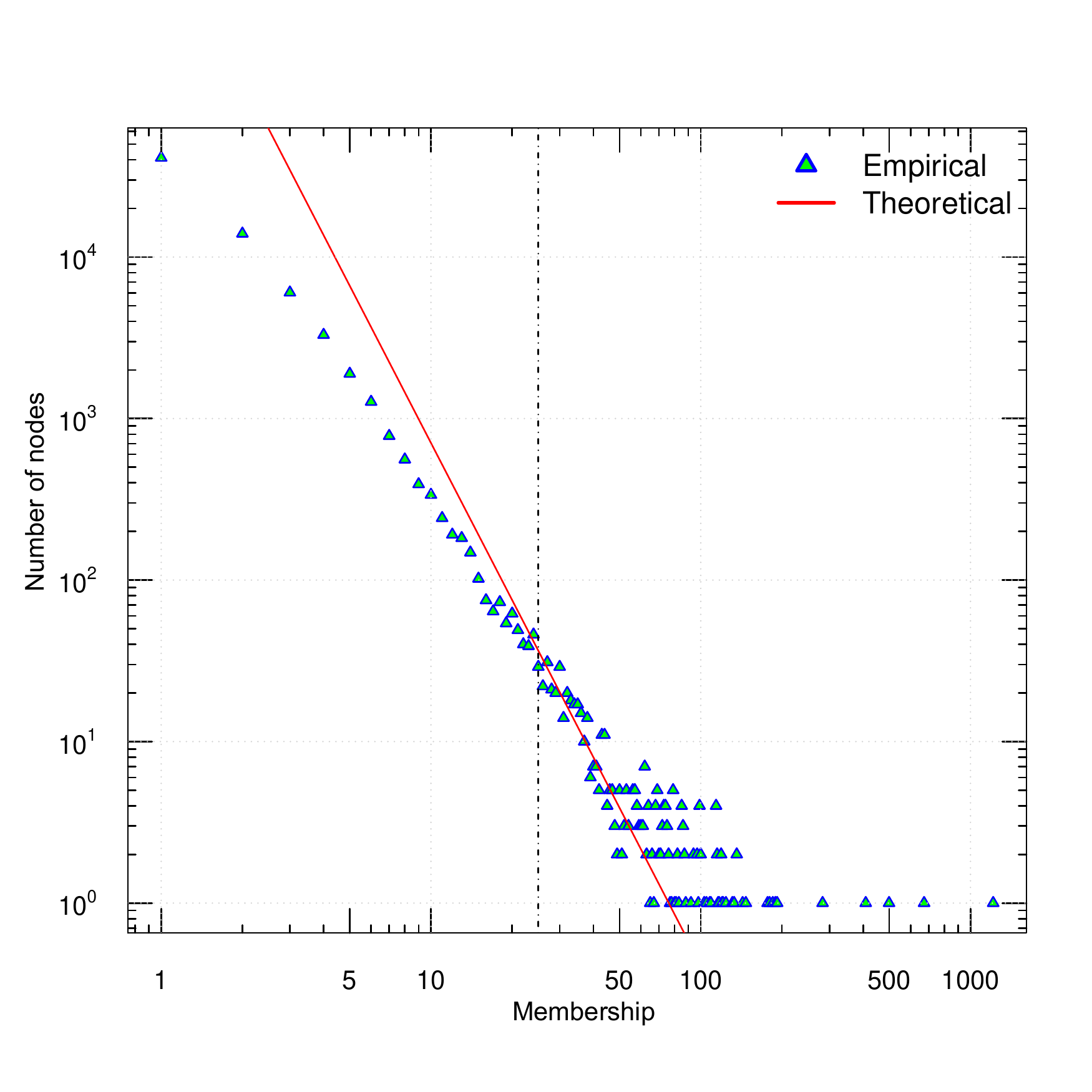}}
        \subfigure[SLPA]{\includegraphics[width=.121\textwidth]{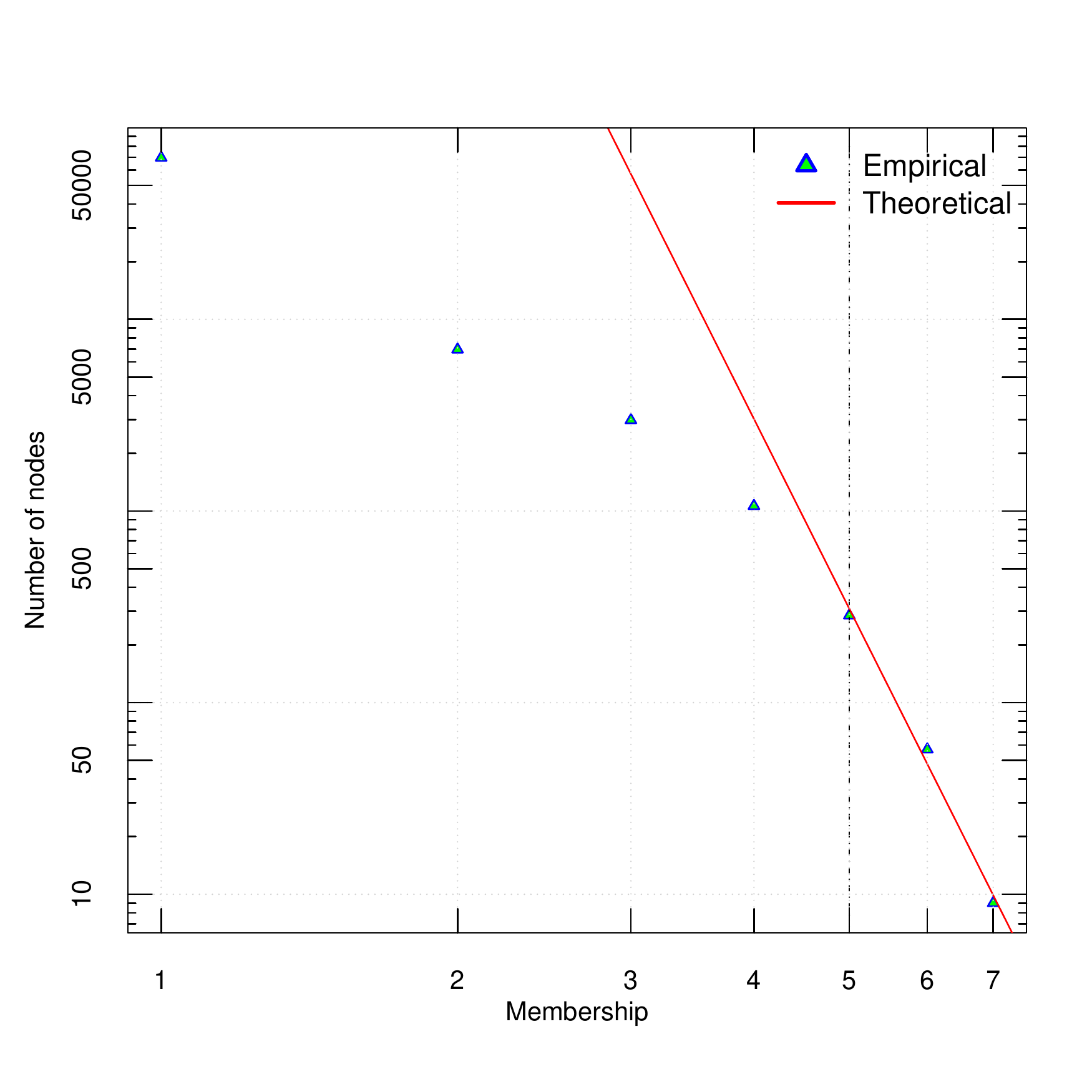}}
        \subfigure[DEMON]{\includegraphics[width=.121\textwidth]{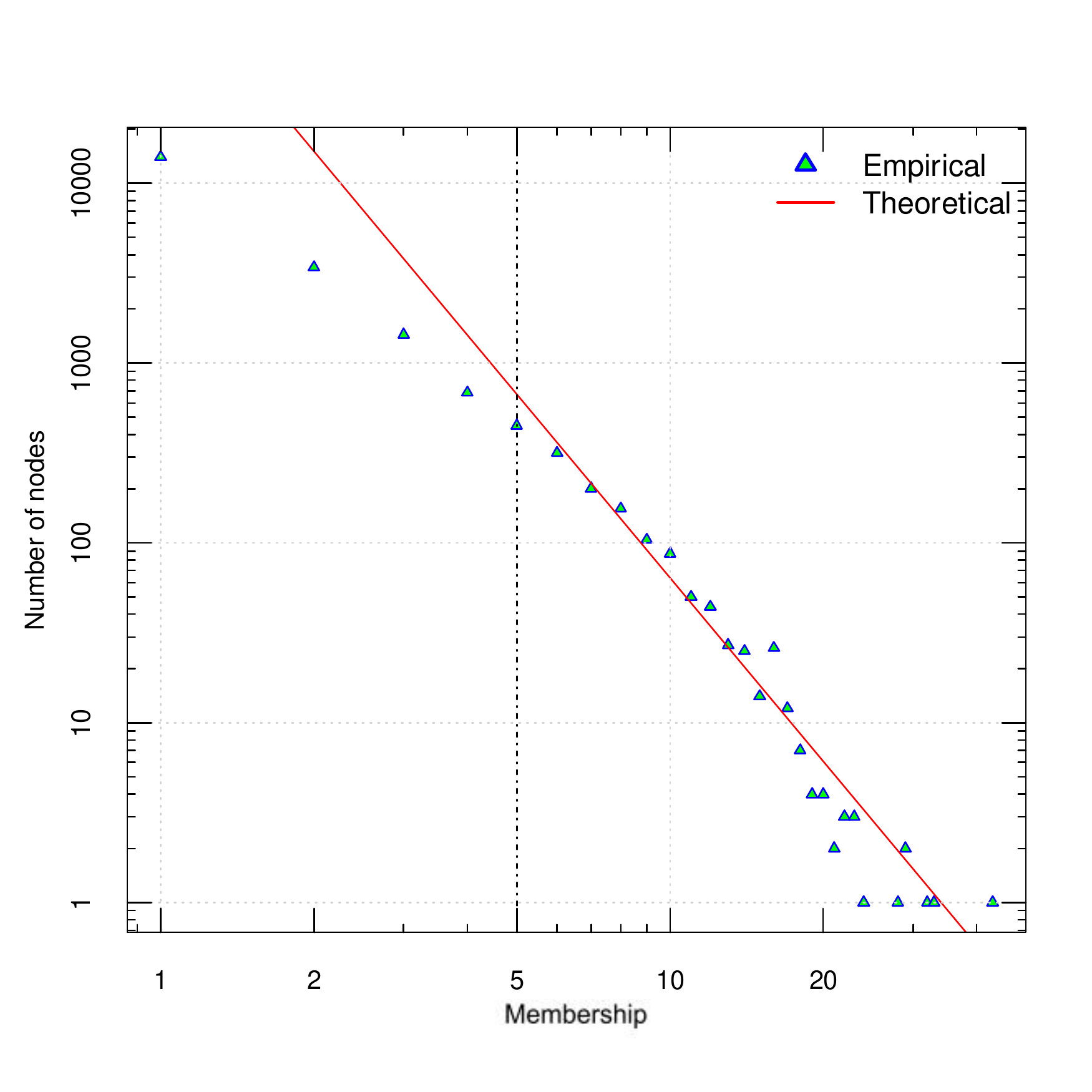}}

        \caption{\label{fig16}Log-log empirical Membership distribution (dots) and Power-Law estimate (line) of PGP Ground-truth (a), LFM (b), GCE (c),  OSLOM (d), LINKC (e), SVINET (f), SLPA (g) and  DEMON(h)}
        \end{figure}

        \begin{table}[ht!]
        \centering
        \caption{KS-test values for the Membership. The distributions under test are the Power-Law (PL), Beta (BE), Cauchy (CA), Exponential (E), Gamma (GM), Logistic (LO), Log-Normal (LN), Normal (N), Uniform (U), and Weibull (WB)}
        \label{table20}
        \begin{tabular}{lcccccccccc}
        \hline
         &  PL & BE & CA & E & GM & LO & LN & N & U & WB \\
        \hline
        Ground-truth&0.02&0.58&0.14&0.58&0.58&0.35&0.38&0.32&0.79&0.32\\
        DEMON&0.03&0.66&0.26&0.66&0.66&0.35&0.29&0.33&0.86&0.25\\
        LFM&0.14&0.44&0.21&0.56&0.43&0.48&0.46&0.47&0.66&0.48\\
        SLPA&0.01&0.86&0.25&0.86&0.86&0.51&0.39&0.5&0.86&0.34\\
        LINKC&0.03&0.5&0.18&0.5&0.5&0.45&0.18&0.46&0.99&0.4\\
        GCE&0.02&0.82&0.17&0.82&0.82&0.49&0.45&0.48&0.83&0.4\\
        OSLOM&0.03&0.57&0.24&0.85&0.85&0.44&0.38&0.43&0.96&0.28\\
        SVINET&0.01&0.77&0.44&0.34&0.71&0.54&0.11&0.49&0.88&0.44\\
        \hline
        \end{tabular}
        \end{table}

        We notice that PGP membership values vary from 1 to 100. This is not the case for the uncovered community structures; Indeed membership can reach 10000 for LINKC and SVINET. Except for LFM, the membership distribution follows a Power-Law (see Table \ref{table20} and Figure \ref{fig16}).

        The membership values for AMAZON are in the same range of those of PGP. The distributions of the membership of AMAZON and the unveiled community structures are shown in Figure \ref{fig17}. The KS-test values, reported in Table \ref{table21}, show that the Power-Law is the best fit for all the unveiled community structures.

        In the case of aNobii, the membership values of the unveiled community structures are much more lower as compared to those of PGP and AMAZON. These values vary from 1 to 500 as shown in Figure \ref{fig18}. Nevertheless, the distributions of the unveiled community structure follow a Power-Law. The KS distance values in Table \ref{table22} confirm this behavior.

\subsubsection{Overlap Size}

        \begin{figure}[ht!]
        \subfigure[Ground-truth]{\includegraphics[width=.121\textwidth]{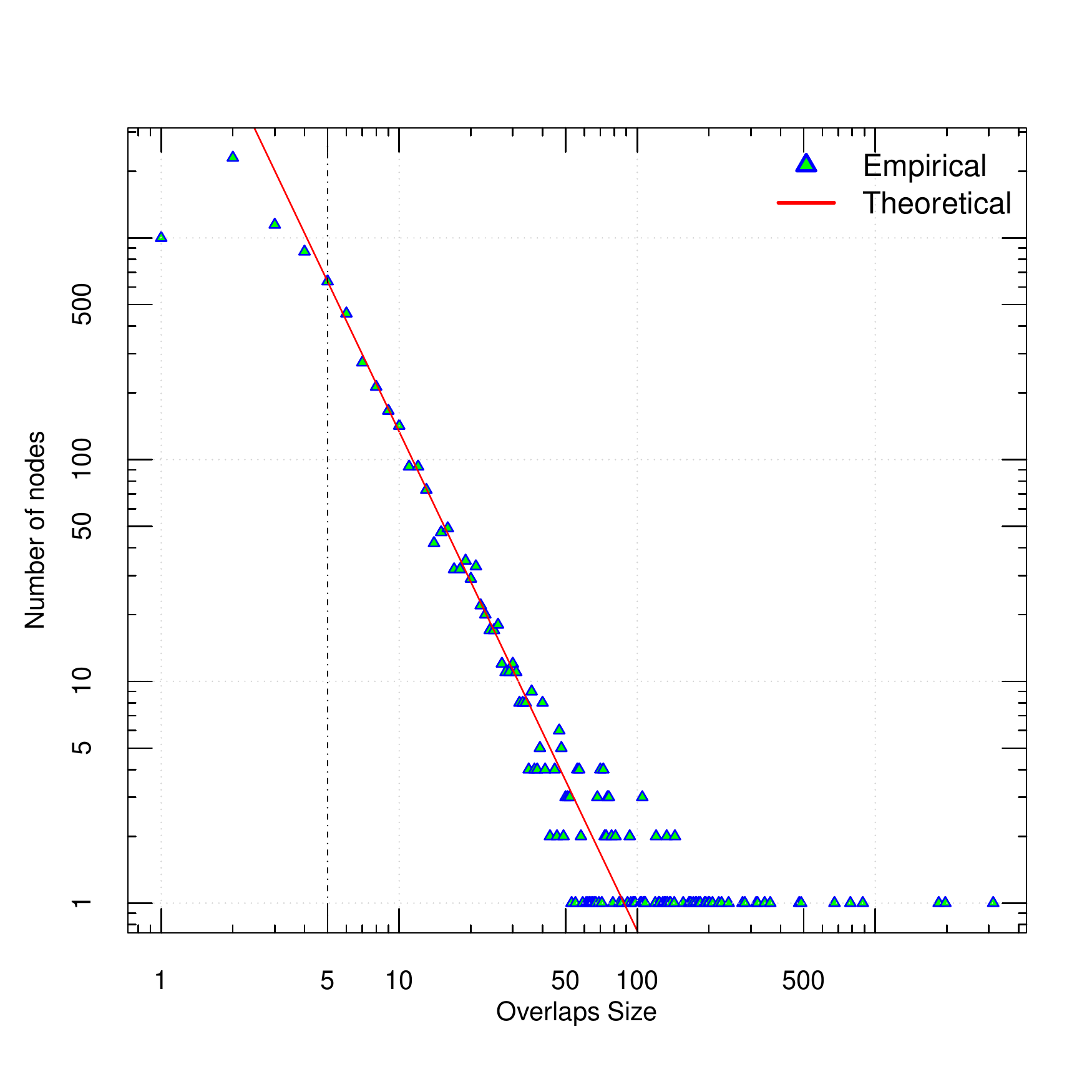}}
        \subfigure[LFM]{\includegraphics[width=.121\textwidth]{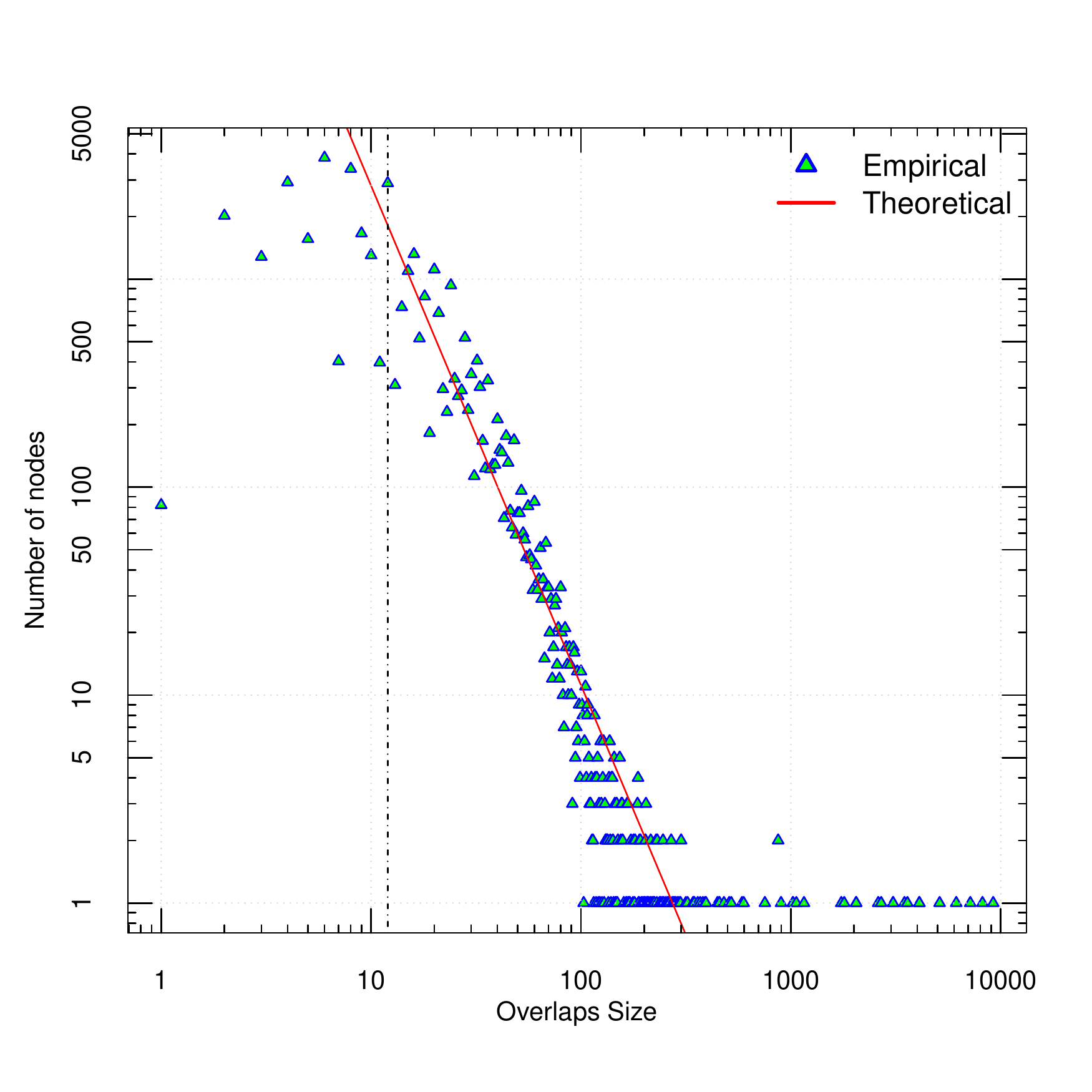}}
        \subfigure[GCE]{\includegraphics[width=.121\textwidth]{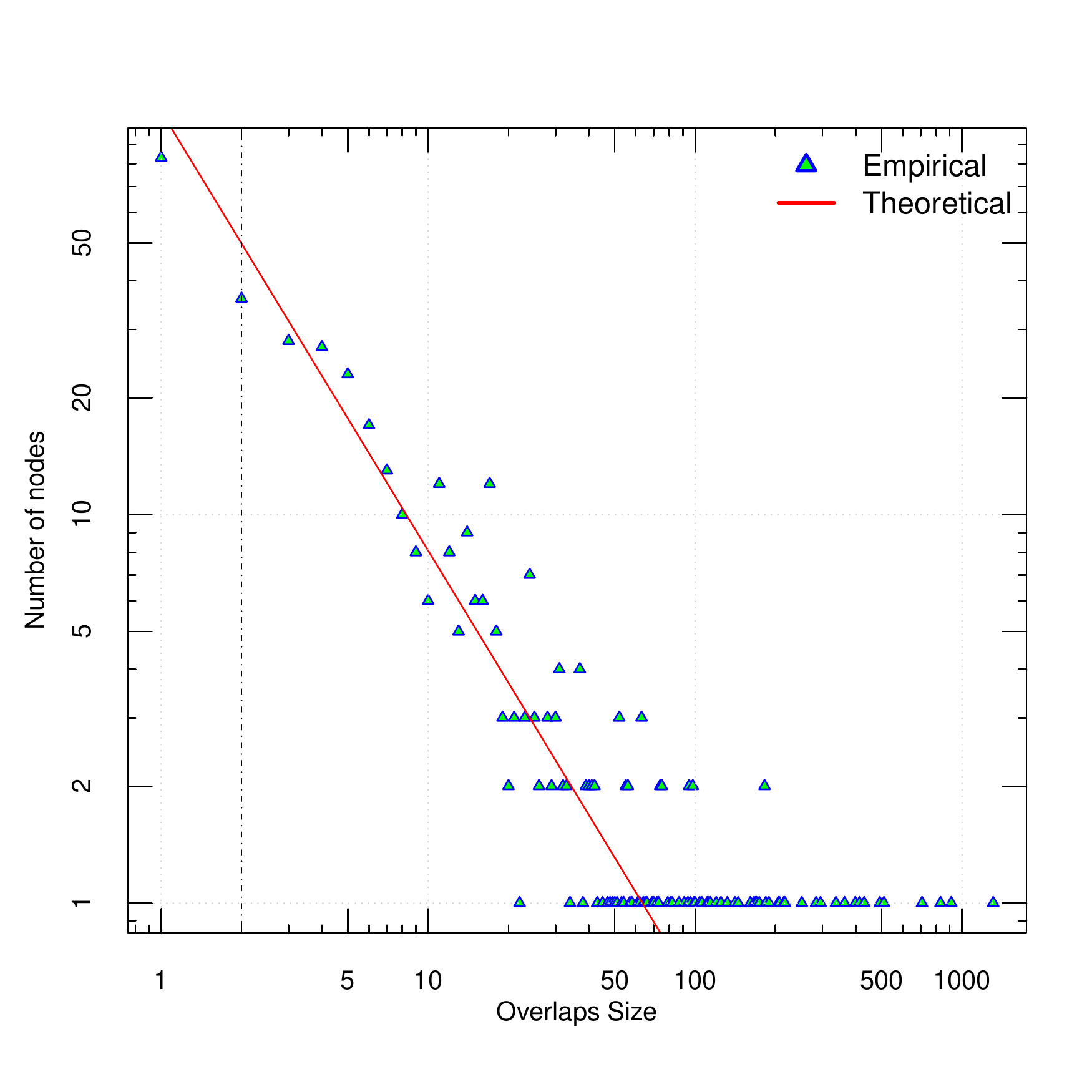}}
        \subfigure[OSLOM]{\includegraphics[width=.121\textwidth]{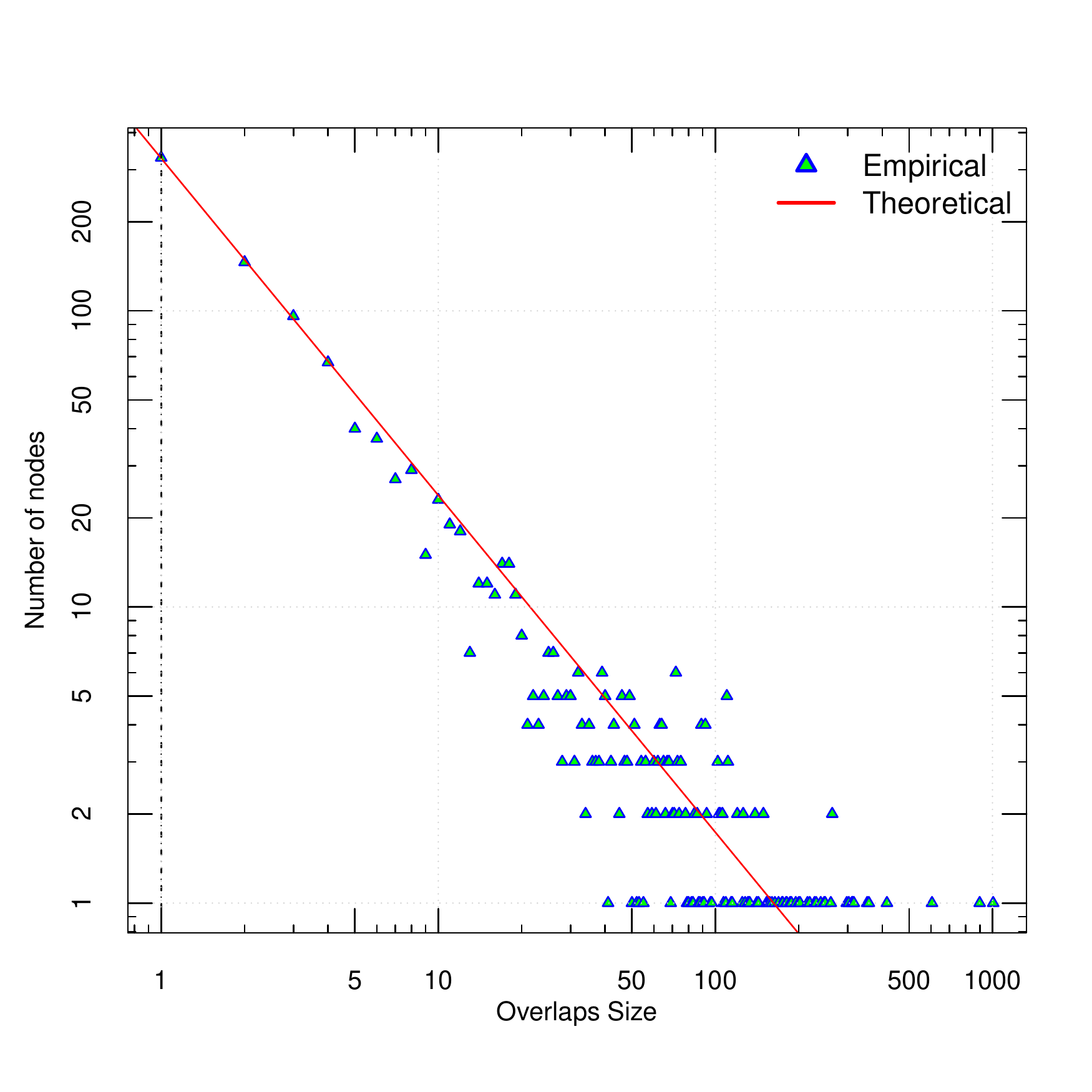}}
        \subfigure[LINKC]{\includegraphics[width=.121\textwidth]{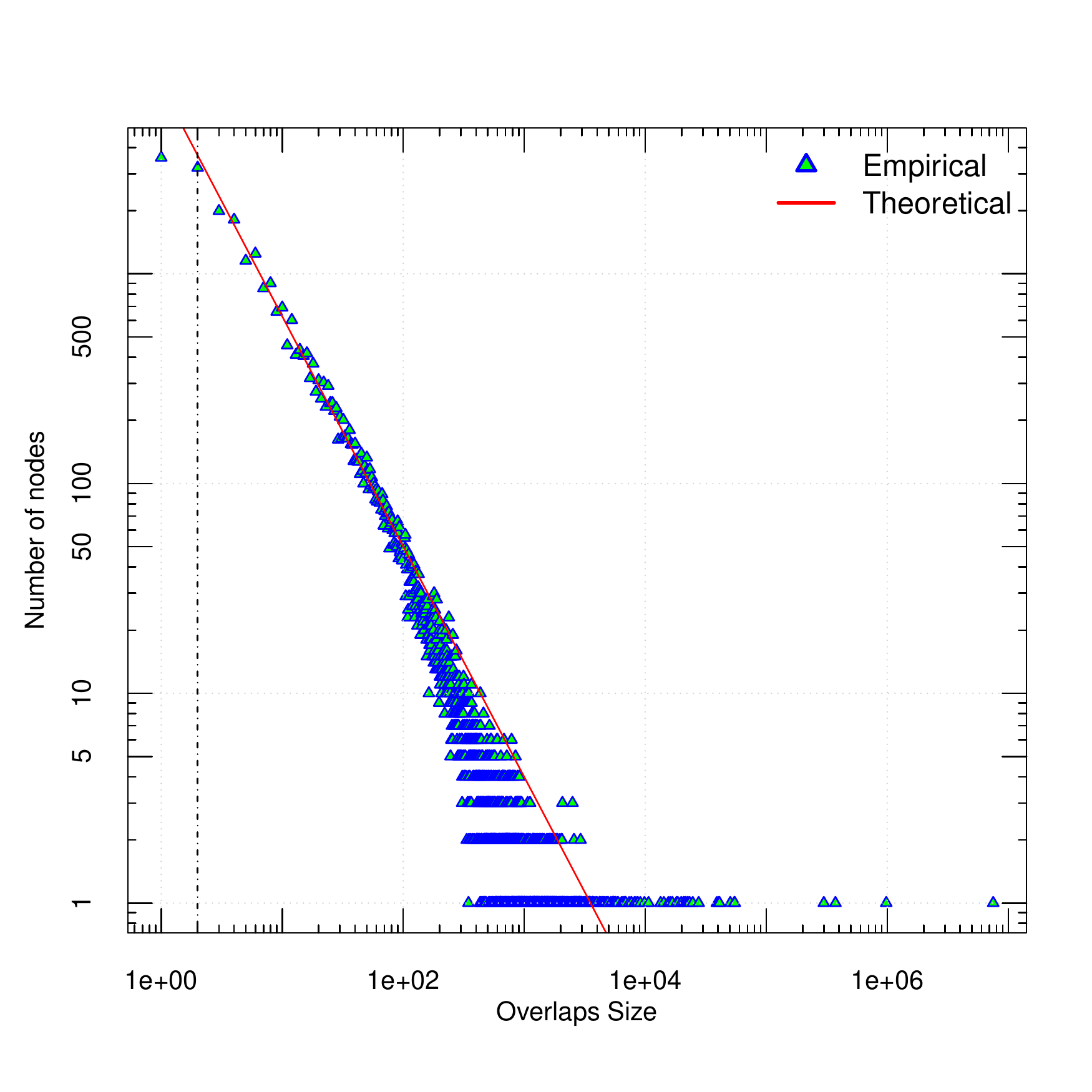}}
        \subfigure[SVINET]{\includegraphics[width=.121\textwidth]{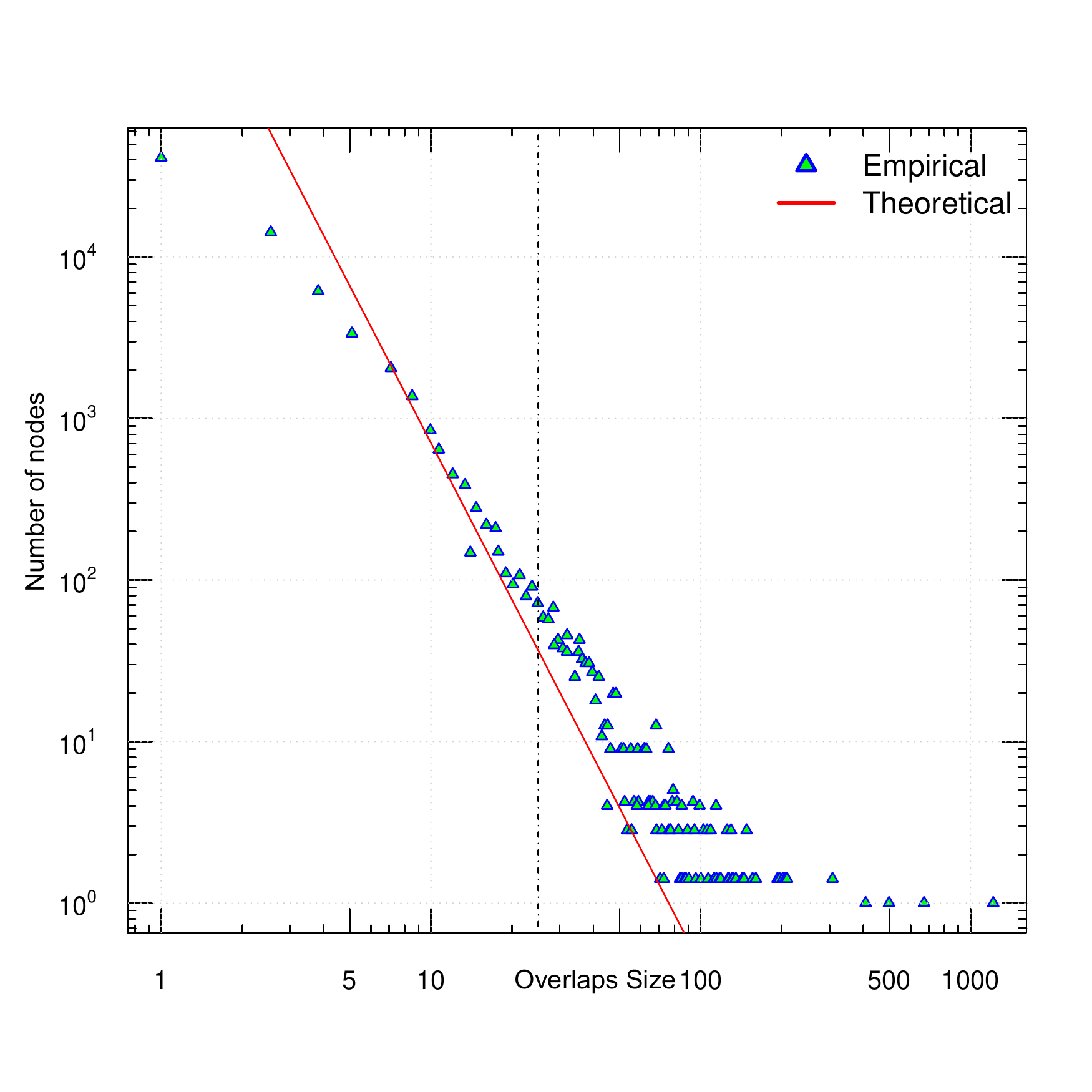}}
        \subfigure[SLPA]{\includegraphics[width=.121\textwidth]{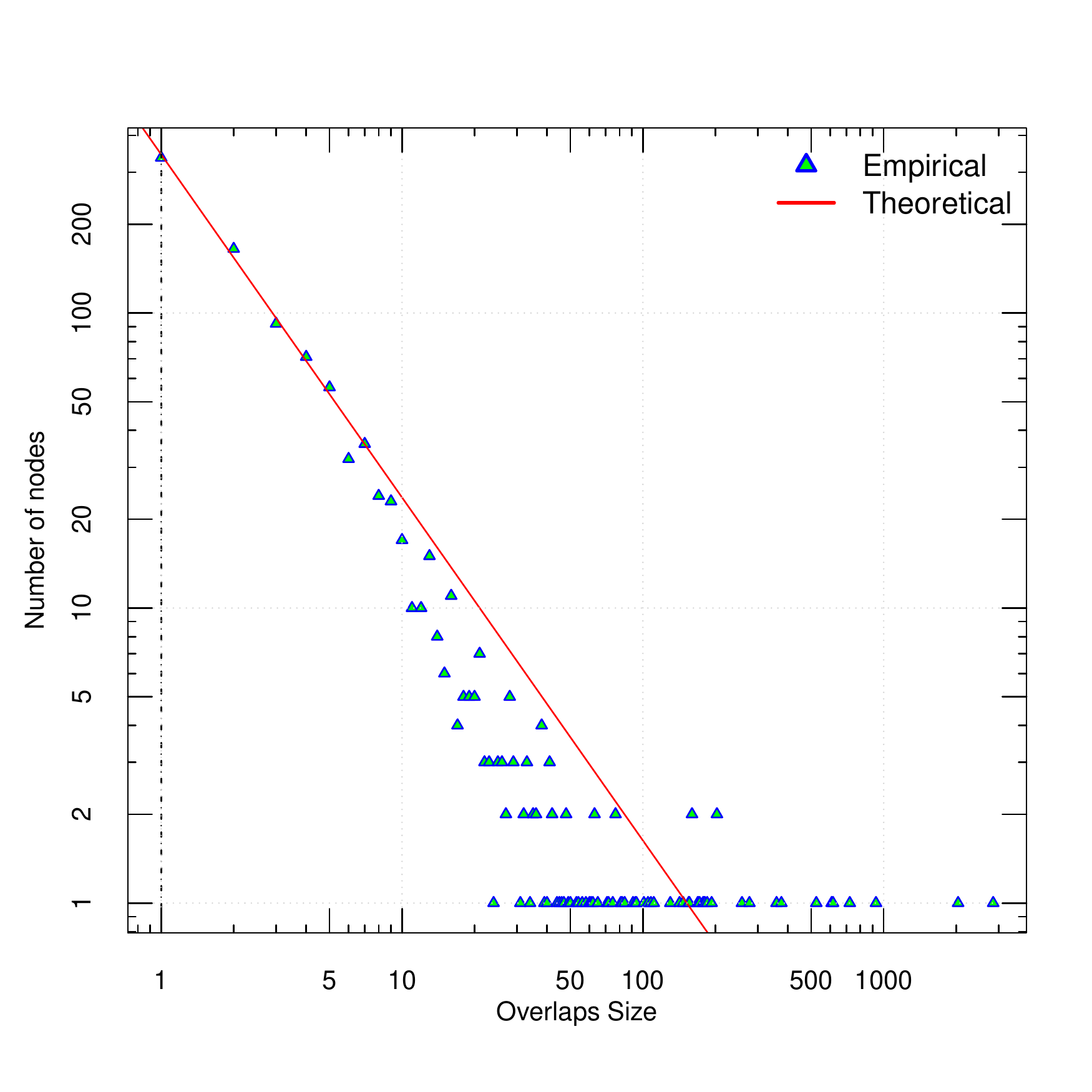}}
        \subfigure[DEMON]{\includegraphics[width=.121\textwidth]{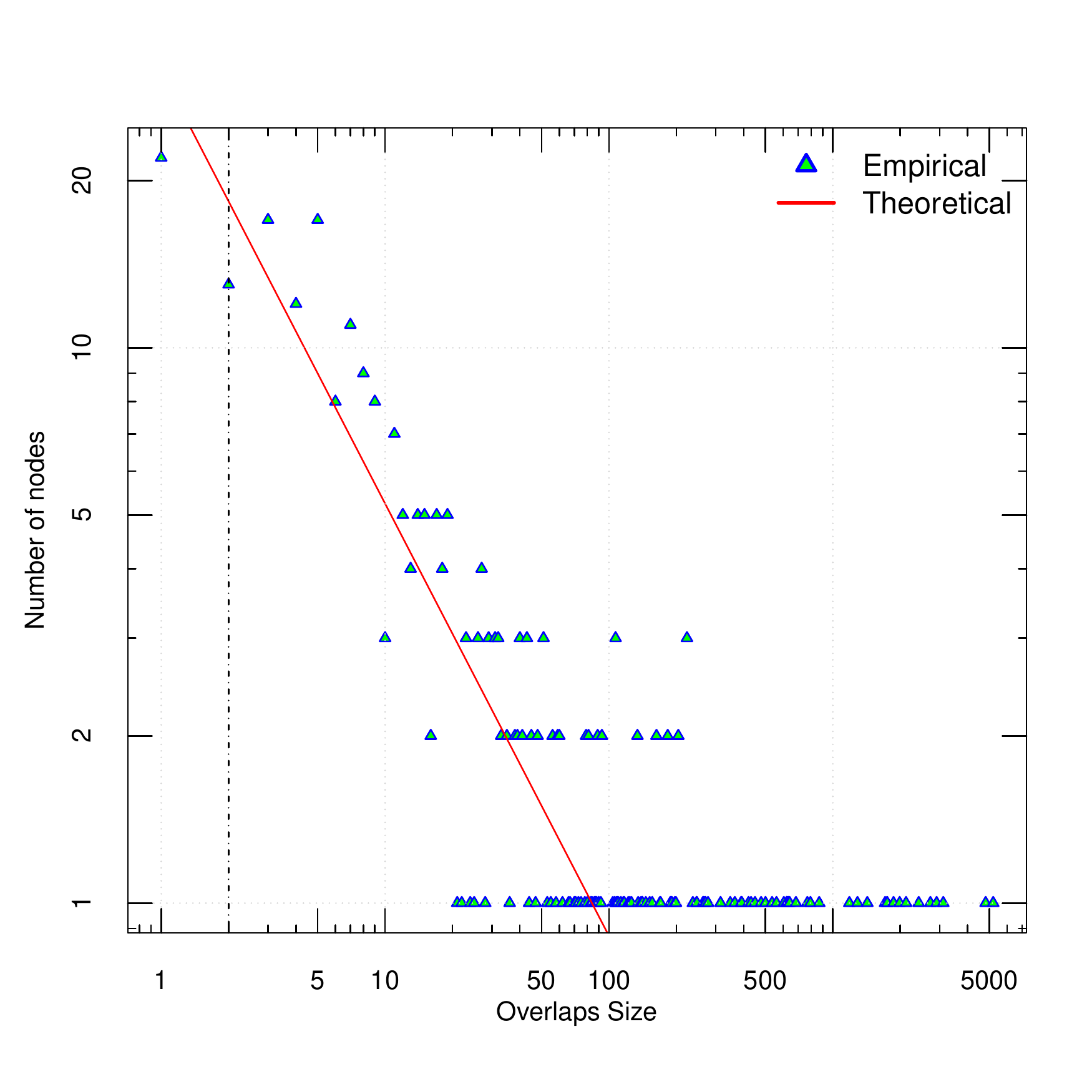}}

        \caption{\label{fig19}Log-log empirical Overlap size distribution (dots) and Power-Law estimate (line) of PGP Ground-truth (a), LFM (b), GCE (c),  OSLOM (d), LINKC (e), SVINET (f), SLPA (g) and  DEMON(h)}
        \end{figure}

        \begin{table}[ht!]
        \centering
        \caption{KS-test values for the overlap size. The distribution under test are the Power-Law (PL), Beta (BE), Cauchy (CA), Exponential (E), Gamma (GM), Logistic (LO), Log-Normal (LN), Normal (N), Uniform (U), and Weibull (WB)}
        \label{table23}
        \begin{tabular}{lcccccccccc}
        \hline
         &  PL & BE & CA & E & GM & LO & LN & N & U & WB \\
        \hline
        Ground-truth&0.01&0.79&0.3&0.3&0.79&0.44&0.17&0.45&0.97&0.2\\
        DEMON&0.06&0.52&0.25&0.42&0.4&0.36&0.06&0.37&0.83&0.15\\
        LFM&0.02&0.85&0.19&0.13&0.84&0.42&0.06&0.43&0.98&0.11\\
        SLPA&0.02&0.58&0.21&0.49&0.55&0.43&0.13&0.44&0.94&0.24\\
        LINKC&0.05&0.9&0.26&0.61&0.9&0.5&0.06&0.5&0.77&0.35\\
        GCE&0.08&0.41&0.23&0.36&0.33&0.34&0.07&0.36&0.84&0.21\\
        OSLOM&0.06&0.3&0.25&0.4&0.27&0.34&0.1&0.35&0.85&0.25\\
        SVINET&0.15&0.32&0.18&0.43&0.35&0.75&0.14&0.41&0.79&0.2\\
        \hline
        \end{tabular}
        \end{table}

        In Figure \ref{fig19}, we present the overlap size distribution for PGP ground-truth and the community structures given by algorithms. Indeed, it is clear that these distributions follow a Power-Law. This is also confirmed by the KS-test values reported in Table \ref{table23}.

        In the case of the aNobii and AMAZON datasets, the results are very similar to those of PGP. In any case, the Power-Law distribution is the best fit (see Figure \ref{fig20}, Figure \ref{fig21}, Table \ref{table24}, and \ref{table25}).

\section{Ranking the detection algorithms}
        In this section, we present the results of the comparison of the detection algorithms according to various types of evaluation measures. The main objective is to investigate the relationships between the topological properties,  the quality metrics, and clustering metrics. First of all,  the topological properties of the uncovered community structures are considered.  Ranking of the algorithms based on the basic properties, microscopic properties, and mesoscopic properties are compared. To do so,  local rankings are calculated for each individual property (see section \ref{Method} for more details about the calculation of local rankings) and merged together into a global ranking for each set of properties using an MCDM strategy (Kconsensus and TOPSIS).  Scalar properties are ranked in ascending order according to the Manhattan distance between the ground-truth and the unveiled 'community-graph' value. For example, to sort the algorithms according to their number of nodes, we compute $|V_0-V_i|$ where $V_0$ is the number of nodes of the ground-truth 'community-graph' and $V_i, (i=1,...,n)$ is the number of nodes of the 'community-graphs' built with the uncovered community structures by the community detection algorithms under study. The algorithms are then ranked in ascending order from smallest to highest distance. Using the same methodology we rank the algorithms according to the sets of quality metrics and clustering metrics. Results are compared to all the topological properties grouped in a single set. Finally, we give the ranking obtained by merging the individual ranks of all the properties.
\subsection{Topological ranking}
\subsubsection{Basic properties}
        Table \ref{table54} presents the local basic properties rankings and the merged one using Kconsensus and TOPSIS for the PGP dataset processed by the various community detection algorithms. Both MCDM strategies agree for the ranking of the SVINET and SLPA algorithms. They are ranked respectively first and second. Indeed the basic properties of their 'community-graphs' are the closest to the ones of the ground-truth 'community-graph'. Note that this is also the case for the AMAZON dataset with Kconsensus as a merging strategy of the individual rankings (See  Table \ref{table55}). If TOPSIS is used, SLPA rank third.
        For the aNobii dataset MOSES rank first and SLPA still rank second whatever merging strategy is used (see Table \ref{table57}). Note that in this case, there is no results for  SVINET because the algorithm did not work on this dataset. Concerning the other algorithms, the global ranking results are very mixed. If we look at the correlation values between the individual properties ranking for the PGP dataset, as reported in  Table \ref{table550} , the clustering coefficient is highly correlated with the average node degree and the assortativity. Note that two rankings are considered correlated if their correlation value is around ($0.8$). In order to check if this result is not an isolated case, we look at Table \ref{table56} and Table \ref{table58} that present the same type of results for  AMAZON and aNobii. According to these results, there is no strong evidence that the observed high correlation values are meaningful, whatever the dataset. Indeed, correlation values vary in large proportions from one dataset to another. In other words, there are no two basic properties that are correlated in any case. Therefore, it is highly recommended to take into account all these properties in order to perform the ranking of the algorithms. In order to compare the MCDM strategies, we computed the correlation between the ranking given by Kconsensus and TOPSIS for each dataset. Except for the PGP dataset which exhibits a very low correlation value ($0.41$), the results indicates that both strategies are very similar. Indeed the correlation is equal to $0.78$ in the case of the AMAZON dataset, and $0.82$ for aNobii.

        \begin{table}[ht!]
        \centering
        \caption{Ranking based on the basic properties of PGP 'community-graphs' built from the unveiled community structure. The calculated properties are number of nodes (V), number of edges (E), Density ($\rho$), Diameter ($d$), Average shortest path ($l_{G}$), Average node degree ($\widetilde{deg}$), Max node degree ($\delta(G)$), Assortativity Coefficient ($\tau$), and Clustering Coefficient ($C$). Kconsensus denotes the final ranking using Kemeny consensus and TOPSIS denotes the final ranking obtained by TOPSIS.}
        \label{table54}
        \begin{tabular}{lccccccccccc}
        \hline
          &V&E&$\rho$&$d$&$l_{G}$&$\widetilde{deg}$&$\delta(G)$&$\tau$&$C$&Kconsensus&TOPSIS\\
        \hline
            LFM & 7 & 6 & 1 & 7 & 4 & 3 & 5 & 6 & 5 & 7 & 5\\
            GCE & 4 & 5 & 5 & 3 & 3 & 4 & 7 & 3 & 3 & 3 & 7\\
            OSLOM & 3 & 1 & 6 & 3 & 7 & 5 & 4 & 7 & 6 & 6 & 4\\
            LINKC & 6 & 7 & 2 & 5 & 1 & 7 & 3 & 4 & 7 & 5 & 3\\
            SVINET & 1 & 2 & 4 & 1 & 5 & 2 & 1 & 1 & 1 & 1 & 1\\
            SLPA & 2 & 4 & 3 & 2 & 6 & 1 & 2 & 2 & 2 & 2 & 2\\
            DEMON & 5 & 3 & 7 & 6 & 2 & 6 & 6 & 4 & 4 & 4 & 6\\

        \hline
        \end{tabular}
        \end{table}

        \begin{table}[ht!]
          \centering
          \caption{Correlation of basic properties rankings. The calculated properties are number of nodes (V), number of edges (E), Density ($\rho$), Diameter ($d$), Average shortest path ($l_{G}$), Average node degree ($\widetilde{deg}$), Max node degree ($\delta(G)$), Assortativity Coefficient ($\tau$), and Clustering Coefficient ($C$)}
          \label{table550}
            \begin{tabular}{lccccccccc}
            \toprule
               &V&E&$\rho$&$d$&$l_{G}$&$\widetilde{deg}$&$\delta(G)$&$\tau$&$C$ \\
            \midrule
            $V$ & 1 &   &   &   &   &   &   &   &  \\
            $E$ &  0.71 & 1 &   &   &   &   &   &   &  \\
            $\rho$ & -0.36 & -0.71 & 1 &   &   &   &   &   &  \\
            $d$ &0.95 & 0.53 & -0.21 & 1 &   &   &   &   &  \\
            $l_{G}$ & -0.64 & -0.68 & 0.11 & -0.56 & 1 &   &   &   &  \\
            $deg$ & 0.57 & 0.21 & 0.29 & 0.53 & -0.64 & 1 &   &   &  \\
            $\widetilde{deg}$ & 0.57 & 0.21 & 0.36 & 0.56 & -0.39 & 0.43 & 1 &   &  \\
            $\delta(G)$ & 0.58 & 0 & 0.04 & 0.61 & 0.11 & 0.47 & 0.44 & 1 &  \\
            $C$ & 0.71 & 0.36 & -0.14 & 0.63 & -0.32 & 0.79 & 0.29 & 0.8 & 1 \\
            \bottomrule
            \end{tabular}%
        \end{table}%

\subsubsection{Microscopic properties}

        Individual rankings according to the three microscopic properties (Degree distribution, Average clustering coefficient as a function of degree and the Hop distance distribution) and the merged rankings using Kconsensus and TOPSIS are reported in Table \ref{table38} for the PGP dataset. SVINET and GCE  are respectively ranked first and second by both MCDM strategies. SLPA has a very bad score. It ranks fourth out of seven according to Kconsensus and sixth using TOPSIS. SLPA and SVINET rank respectively first and second according to Kconsensus and first and third using TOPSIS with the AMAZON dataset. GCE scores very poorly in that case (See Table \ref{table560} ). For the aNobii dataset, SLPA is still one of the highly ranked algorithms together with MOSES (See Table \ref{table580}). When we look at the correlation between the rankings given by each property individually, it clearly appears that there no strong relations between them whatever the dataset (PGP, AMAZON, and aNobii) (See Table \ref{table62}, Table \ref{table63}, Table \ref{table64}). These findings confirm that they provide useful complementary information about the community structure.

        The correlation between the global rankings due to Kconsensus and TOPSIS are still very high for two datasets ( $0.75$ in the case of the PGP dataset and $0.76$ in the case of the AMAZON dataset). However, it is not the case for the aNobii dataset with a correlation value equal to $0.37$.

        \begin{table}[!ht]
          \centering
          \caption{Microscopic properties ranking for PGP. The distributions under test are the degree distribution (DD), the average clustering coefficient as function of degree (Av), the hop distance (HD). Kconsensus denotes the topological microscopic ranking using Kemeny consensus and TOPSIS denotes the final ranking obtained by TOPSIS.}
               \label{table38}
            \begin{tabular}{lccccc}
            \hline
            & \multicolumn{1}{l}{DD} & \multicolumn{1}{l}{Av} & \multicolumn{1}{l}{HD} & \multicolumn{1}{l}{Kconsensus} & \multicolumn{1}{l}{TOPSIS}\\
            \hline
            LFM & 5 & 1 & 4 & 5 &3\\
            GCE & 3 & 5 & 1 & 2 &2\\
            OSLOM & 6 & 6 & 7 & 6 &7\\
            LINKC & 2 & 7 & 8 & 3 &4\\
            SVINET & 1 & 2 & 2 & 1 &1\\
            SLPA & 4 & 4 & 6 & 4 &6\\
            DEMON & 7 & 3 & 5 & 7 &5\\
            \hline
            \end{tabular}%
        \end{table}%

        \begin{table}[!ht]
          \centering
          \caption{Correlation of the rankings of the microscopic properties for PGP (degree distribution (DD), the average clustering coefficient as function of degree (Av), the hop distance (HD))}
          \label{table62}
            \begin{tabular}{lccc}
            \hline
              & \multicolumn{1}{c}{DD} & \multicolumn{1}{c}{Av} & \multicolumn{1}{c}{HD} \\
              \hline
            DD & 1 &   &  \\
            Av & -0.1 & 1 &  \\
            HD & 0.3 & 0.54 & 1 \\
            \hline
            \end{tabular}%
        \end{table}%

\subsubsection{Mesoscopic properties}
        The algorithms are ranked according to the distance between the distributions (the community size, the overlap size and the membership of nodes)  of the unveiled community structures and the one estimated using the ground truth.  For the PGP dataset, SLPA  rank first ( See Table \ref{table380}). It is followed by SVINET and LFM which rank respectively second and third. Note that both merging strategies (TOPSIS and Kconsensus) give the same rankings for theses algorithms. They also agree on the fact that DEMON is the less effective according to the mesoscopic properties distances.  Rankings are very different when we examine the other datasets. Indeed, in the case of AMAZON, CFINDER and GCE exhibit the best scores, while SLPA, SVINET and LFM rank at the bottom (Table \ref{table561}). For the aNobii dataset, MOSES and DEMON are ranking in the top 2 (Table \ref{table581}), while LFM and GCE occupy the last position if one refer to Kconsensus or TOPSIS.  The explanation of this great variability may lie on the fact that all the community graphs are able to reproduce fairly well the power law distribution of the mesoscopic properties. Consequently, the KS values used for the individual rankings are very close and they do not reflect significant differences between the algorithms, while the ranking has a tendency to amplify these differences. Table \ref{table65} reports the rank correlation between the mesoscopic properties for the PGP dataset. It indicates that there is no correlation between these properties, hence all these mesoscopic properties need to be considered. This is also the case for AMAZON (See Table \ref{table66}) and for aNobii (See Table \ref{table67}). Finally, the PGP dataset is the only one for which the correlation between the rankings of Kconsensus and TOPSIS is high ($0.89$). Its value is below $0.6$  for the two other datasets.
        \begin{table}[!ht]
          \centering
          \caption{Mesoscopic properties ranking for PGP. The distribution under test are the community size (CS), the membership (M), the overlap size (OS). Kconsensus denotes the topological mesoscopic ranking using Kemeny consensus and TOPSIS denotes the final ranking obtained by TOPSIS.}
               \label{table380}
            \begin{tabular}{lccccccc}
            \hline
            & \multicolumn{1}{l}{CS} & \multicolumn{1}{l}{M} & \multicolumn{1}{l}{OS} & \multicolumn{1}{l}{Kconsensus} & \multicolumn{1}{l}{TOPSIS}\\
            \hline
            LFM & 6 & 6 & 1 & 3 &3\\
            GCE & 3 & 7 & 6 & 6 &4\\
            OSLOM & 4 & 5 & 5 & 5 &6\\
            LINKC & 7 & 3 & 3 & 4 &5\\
            SVINET & 2 & 2 & 4 & 2 &2\\
            SLPA & 1 & 1 & 2 & 1 &1\\
            DEMON & 5 & 4 & 7 & 7 &7\\
            \hline
            \end{tabular}%
        \end{table}%

        \begin{table}[!ht]
          \centering
          \caption{Correlation of the rankings of the microscopic properties for PGP (the community size (CS), the membership (M), the overlap size (OS))}
          \label{table65}
            \begin{tabular}{lccc}
            \hline
              & \multicolumn{1}{c}{CS} & \multicolumn{1}{c}{M} & \multicolumn{1}{c}{OS} \\
              \hline
            CS & 1 &   &  \\
            M & 0.39 & 1 &  \\
            OS & -0.07 & 0.28 & 1 \\
            \hline
            \end{tabular}%
        \end{table}%

\subsubsection{All topological properties}
        Given that we considered three sets of topological properties (Basic, microscopic, mesoscopic), this raises the question of their correlation. Indeed, if a strong correlation is observed between two sets, we do not need to take into account both sets in order to evaluate the algorithms. Table \ref{table88} reports the correlation matrix for the ranks obtained by the algorithms according to each set of topological property using the PGP dataset. It shows that there is no correlation between these sets. This result is confirmed by the experiments with the AMAZON (See Table \ref{table89}) and aNobii (See Table \ref{table90}) datasets.
        The correlation between the topological properties are shown in Table \ref{table68} for the PGP dataset, in Table \ref{table69} for AMAZON, and in Table \ref{table70} for aNobii. It allows a finer view of the correlation between the topological properties taken individually. For the sake of clarity, the correlation between properties belonging to different sets of topological properties is reported in red. From the analysis of these results, it emerges that there is no strong evidence that one can consider that a strong correlation exists between some couples of topological properties. Indeed, when the correlation value is high for a dataset, it is not the case for the others. A typical example is given by the high correlation value observed($0.86$) between the community size distribution ranking and the one based on the clustering coefficient for the PGP dataset. Its value is ($0.38$) for AMAZON and ($-0.37$) for aNobii. Therefore, if the topological properties rankings are not well correlated, they all have to be considered in order to evaluate the overlapping community detection algorithms. Table \ref{table381} shows the rankings obtained by merging the individual rankings of all the topological properties for the PGP dataset. Both MCDM strategies agree about the extremes rankings: SVINET and SLPA are leading, while OSLOM and DEMON are at the end. Note that SVINET is also ranked first by both strategies for the AMAZON dataset while SLPA is third according to Kconsensus and fourth according to TOPSIS (See Table \ref{table562} ). OSLOM is also at the end of this dataset. If the results are quite consistent for these two datasets, this is not the case for aNobii. In this case, DEMON and MOSES have the best ranks (See Table \ref{table582}). The lack of consensus among the merged strategies is clearly reflected in the observed correlation between Kconsensus and TOPSIS rankings. It goes from $0.85$ for the PGP datasets to $0.52$  for AMAZON and $0.21$ for aNobii.

        \begin{table}[!ht]
          \centering
          \caption{All topological properties ranking for the PGP dataset. The calculated properties are number of nodes (V), number of edges (E), Density ($\rho$), Diameter ($d$), Average shortest path ($l_{G}$), Average node degree ($\widetilde{deg}$), Max node degree ($\delta(G)$), Assortativity Coefficient ($\tau$), and Clustering Coefficient ($C$), the degree distribution (DD), the average clustering coefficient as function of degree (Av), the Hop distance (HD), the community size (CS), the membership (M), the overlap size (OS).}
          \label{table381}
          \small
            \begin{tabular}{lccccccccccccccccc}
            \hline
            & \multicolumn{9}{c}{Basic properties} & \multicolumn{3}{c}{Microscopic} & \multicolumn{3}{c}{Mesoscopic} & \multicolumn{2}{c}{MCDM Ranking}\\
            \hline
              &V&E&$\rho$&$d$&$l_{G}$&$\widetilde{deg}$&$\delta(G)$&$\tau$&$C$& DD & Av & HD & CS & M & OS & Kconsensus & TOPSIS\\
            \hline
            LFM & 7 & 6 & 1 & 7 & 4 & 3 & 5 & 6 & 5 & 5 & 1 & 4 & 6 & 6 & 1 &  5&3\\
            GCE & 4 & 5 & 5 & 3 & 3 & 4 & 7 & 3 & 3 & 3 & 5 & 1 & 3 & 7 & 6 &  3&5\\
            OSLOM & 3 & 1 & 6 & 3 & 7 & 5 & 4 & 7 & 6 & 6 & 6 & 7 & 4 & 5 & 5 & 6& 6\\
            LINKC & 6 & 7 & 2 & 5 & 1 & 7 & 3 & 4 & 7 & 2 & 7 & 8 & 7 & 3 & 3 & 4& 4\\
            SVINET & 1 & 2 & 4 & 1 & 5 & 2 & 1 & 1 & 1 & 1 & 2 & 2 & 2 & 2 & 4 & 1& 1\\
            SLPA & 2 & 4 & 3 & 2 & 6 & 1 & 2 & 2 & 2 & 4 & 4 & 6 & 1 & 1 & 2 & 2& 2\\
            DEMON & 5 & 3 & 7 & 6 & 2 & 6 & 6 & 4 & 4 & 7 & 3 & 5 & 5 & 4 & 7 & 7 &7\\
            \hline
            \end{tabular}%
        \end{table}

        \begin{table}[!ht]
          \centering
          \footnotesize
          \caption{Correlation of ranking of all topological properties for PGP dataset. The calculated properties are Number of nodes (V), Number of edges (E), Density ($\rho$), Diameter ($d$), Average shortest path ($l_{G}$), Average node degree ($\widetilde{deg}$), Max node degree ($\delta(G)$), Assortativity Coefficient ($\tau$), and Clustering Coefficient ($C$), the Degree distribution (DD), the Average clustering coefficient as function of degree (Av), the Hop distance (HD), the Community size (CS), the Membership (M), the Overlap size (OS).}
          \label{table68}
            \begin{tabular}{lccccccccccccccc}
            \hline
            & V  & E  & $\rho$  & $d$  & $l_{G}$  & $\widetilde{deg}$  & $\delta(G)$  & $\tau$  & $C$  & DD  & Av  & HD  & CS  & M  & OS \\
            \hline
            V  & 1 &   &   &   &   &   &   &   &   &   &   &   &   &   &  \\
            E  & 0.71 & 1 &   &   &   &   &   &   &   &   &   &   &   &   &  \\
            $\rho$  & -0.36 & -0.71 & 1 &   &   &   &   &   &   &   &   &   &   &   &  \\
            $d$  & 0.95 & 0.53 & -0.21 & 1 &   &   &   &   &   &   &   &   &   &   &  \\
            $l_{G}$  & -0.64 & -0.68 & 0.11 & -0.56 & 1 &   &   &   &   &   &   &   &   &   &  \\
            $\widetilde{deg}$  & 0.57 & 0.21 & 0.29 & 0.53 & -0.64 & 1 &   &   &   &   &   &   &   &   &  \\
            $\delta(G)$  & 0.57 & 0.21 & 0.36 & 0.56 & -0.39 & 0.43 & 1 &   &   &   &   &   &   &   &  \\
            $\tau$  & 0.58 & 0.01 & 0.04 & 0.61 & 0.11 & 0.47 & 0.44 & 1 &   &   &   &   &   &   &  \\
            $C$  & 0.71 & 0.36 & -0.14 & 0.63 & -0.32 & 0.79 & 0.29 & 0.8 & 1 &   &   &   &   &   &  \\
            DD  & \textcolor[rgb]{ 1,  0,  0}{\textbf{0.32}} & \textcolor[rgb]{ 1,  0,  0}{\textbf{-0.29}} & \textcolor[rgb]{ 1,  0,  0}{\textbf{0.46}} & \textcolor[rgb]{ 1,  0,  0}{\textbf{0.53}} & \textcolor[rgb]{ 1,  0,  0}{\textbf{0.14}} & \textcolor[rgb]{ 1,  0,  0}{\textbf{0.25}} & \textcolor[rgb]{ 1,  0,  0}{\textbf{0.54}} & \textcolor[rgb]{ 1,  0,  0}{\textbf{0.66}} & \textcolor[rgb]{ 1,  0,  0}{\textbf{0.32}} & 1 &   &   &   &   &  \\
            Av  & \textcolor[rgb]{ 1,  0,  0}{\textbf{0.01}} & \textcolor[rgb]{ 1,  0,  0}{\textbf{0.11}} & \textcolor[rgb]{ 1,  0,  0}{\textbf{0.18}} & \textcolor[rgb]{ 1,  0,  0}{\textbf{-0.18}} & \textcolor[rgb]{ 1,  0,  0}{\textbf{-0.14}} & \textcolor[rgb]{ 1,  0,  0}{\textbf{0.57}} & \textcolor[rgb]{ 1,  0,  0}{\textbf{0.04}} & \textcolor[rgb]{ 1,  0,  0}{\textbf{0.18}} & \textcolor[rgb]{ 1,  0,  0}{\textbf{0.54}} & -0.11 & 1 &   &   &   &  \\
            HD  & \textcolor[rgb]{ 1,  0,  0}{\textbf{0.24}} & \textcolor[rgb]{ 1,  0,  0}{\textbf{0.09}} & \textcolor[rgb]{ 1,  0,  0}{\textbf{-0.12}} & \textcolor[rgb]{ 1,  0,  0}{\textbf{0.26}} & \textcolor[rgb]{ 1,  0,  0}{\textbf{0.01}} & \textcolor[rgb]{ 1,  0,  0}{\textbf{0.45}} & \textcolor[rgb]{ 1,  0,  0}{\textbf{-0.27}} & \textcolor[rgb]{ 1,  0,  0}{\textbf{0.45}} & \textcolor[rgb]{ 1,  0,  0}{\textbf{0.69}} & 0.3 & 0.54 & 1 &   &   &  \\
            CS  & \textcolor[rgb]{ 1,  0,  0}{\textbf{0.89}} & \textcolor[rgb]{ 1,  0,  0}{\textbf{0.54}} & \textcolor[rgb]{ 1,  0,  0}{\textbf{-0.25}} & \textcolor[rgb]{ 1,  0,  0}{\textbf{0.84}} & \textcolor[rgb]{ 1,  0,  0}{\textbf{-0.64}} & \textcolor[rgb]{ 1,  0,  0}{\textbf{0.79}} & \textcolor[rgb]{ 1,  0,  0}{\textbf{0.36}} & \textcolor[rgb]{ 1,  0,  0}{\textbf{0.62}} & \textcolor[rgb]{ 1,  0,  0}{\textbf{0.86}} & \textcolor[rgb]{ 1,  0,  0}{\textbf{0.21}} & \textcolor[rgb]{ 1,  0,  0}{\textbf{0.18}} & \textcolor[rgb]{ 1,  0,  0}{\textbf{0.42}} & 1 &   &  \\
            M  & \textcolor[rgb]{ 1,  0,  0}{\textbf{0.54}} & \textcolor[rgb]{ 1,  0,  0}{\textbf{0.18}} & \textcolor[rgb]{ 1,  0,  0}{\textbf{0.14}} & \textcolor[rgb]{ 1,  0,  0}{\textbf{0.46}} & \textcolor[rgb]{ 1,  0,  0}{\textbf{-0.18}} & \textcolor[rgb]{ 1,  0,  0}{\textbf{0.32}} & \textcolor[rgb]{ 1,  0,  0}{\textbf{0.86}} & \textcolor[rgb]{ 1,  0,  0}{\textbf{0.58}} & \textcolor[rgb]{ 1,  0,  0}{\textbf{0.36}} & \textcolor[rgb]{ 1,  0,  0}{\textbf{0.32}} & \textcolor[rgb]{ 1,  0,  0}{\textbf{0.01}} & \textcolor[rgb]{ 1,  0,  0}{\textbf{-0.36}} & 0.39 & 1 &  \\
            OS & \textcolor[rgb]{ 1,  0,  0}{\textbf{-0.18}} & \textcolor[rgb]{ 1,  0,  0}{\textbf{-0.46}} & \textcolor[rgb]{ 1,  0,  0}{\textbf{0.93}} & \textcolor[rgb]{ 1,  0,  0}{\textbf{-0.11}} & \textcolor[rgb]{ 1,  0,  0}{\textbf{-0.21}} & \textcolor[rgb]{ 1,  0,  0}{\textbf{0.46}} & \textcolor[rgb]{ 1,  0,  0}{\textbf{0.5}} & \textcolor[rgb]{ 1,  0,  0}{\textbf{-0.04}} & \textcolor[rgb]{ 1,  0,  0}{\textbf{-0.07}} & \textcolor[rgb]{ 1,  0,  0}{\textbf{0.29}} & \textcolor[rgb]{ 1,  0,  0}{\textbf{0.25}} & \textcolor[rgb]{ 1,  0,  0}{\textbf{-0.24}} & -0.07 & 0.29 & 1 \\
            \hline
            \end{tabular}%
        \end{table}%

        \begin{table}[htbp!]
          \centering
          \caption{Correlation of the basic, microscopic, and mesoscopic rankings.}
          \label{table88}
            \begin{tabular}{lccc}
            \hline
              & \multicolumn{1}{c}{Basic} & \multicolumn{1}{c}{Micro} & \multicolumn{1}{c}{Meso} \\
              \hline
            Basic & 1 &   &  \\
            Micro & 0.61 & 1 &  \\
            Meso & 0.11 & 0.28 & 1 \\
            \hline
            \end{tabular}%
        \end{table}%
\subsection{Classical metrics ranking}
\subsubsection{Quality metrics}
        Here we analyze the six quality measures presented in section \ref{quality}. These metrics are computed for the ground-truth community structure and the outputs of the overlapping community detection algorithms.
        The results of these quality measures in the case of PGP dataset are shown in Table \ref{table42}. The first line of this table referred as PGP contains the computed value for the ground-truth community structure, while the remaining ones concern the quality measure obtained for the community structures uncovered by the various community detection algorithms under test. One can notice that the values of Average Degree, Flake-ODF and Internal Density for the detected community structures are more or less in the same order of magnitude than those of the ground-truth. This is not the case for the Average-ODF and Max-ODF. Indeed, they exhibit a greater variability. Note that, except for SVINET, and LINKC, the overlapping modularity values are relatively low. The quality metrics rankings are presented in Table \ref{table51}. In this case, the results of the ground-truth are considered as a reference in order to compute the distances. LINKC  and SVINET are ranked respectively first and second by Kconsensus, while SVINET is first followed by LFM for the TOPSIS merging strategy. In both cases, SLPA is ranked third and DEMON is the last one. Table \ref{table45} shows the quality metric values computed on AMAZON ground-truth and its uncovered community structures. We remark that in this case, the results of the overlapping community structure are comparable to those of the ground truth for all the quality metrics under test. In Table \ref{table52}, the individual and the final rankings are reported. SVINET is the leading algorithm while DEMON and OSLOM are the less performings according to the merging strategies.  In the case of aNobii, the results of the quality measures are very mixed as shown in Table \ref{table46}. For the Average Degree, all the algorithms have comparable values to those of the ground-truth. MOSES and DEMON have the nearest value of Average ODF while all the other algorithms exhibit quite lower values for this property.  The community structures uncovered by all the algorithms have a lower internal density and overlapping modularity as compared to the ground-truth community structure. We observe a great variability of the values of Max ODF. MOSES and OSLOM are the best algorithms considering the final ranking based on all the quality metrics as shown in Table \ref{table53}. Note that whatever the dataset, DEMON is always ranked at the end.  The correlation between the ranks given by the quality metrics are reported in Table \ref{table82} for PGP,  in Table \ref{table83} for AMAZON and in Table \ref{table84} for aNobii. Again, overall, the measures seems to be fairly uncorrelated. We also observe that the merging strategies lead to quite similar results. Indeed, the correlation between Kconsensus and TOPSIS equals to $0.71$ in the case of the PGP dataset, $0.88$ in the case of the AMAZON dataset, and $0.66$ in the case of the aNobii dataset.

        \begin{table}[!ht]
          \centering

          \caption{Quality metrics values for PGP ground-truth and the uncovered community structure. The calculated properties are Average Degree (AD), Average ODF (AO), Flake ODF (FO), Internal Density (ID), Max ODF (MO), and Overlapping Modularity (OM).}
          \label{table42}
            \begin{tabular}{lcccccc}
            \hline
              & AD & AO & FO & ID & MO & OM \\
            \hline
            PGP&1.45&6.45&1.71&0.79&17.1&0.37\\
            LFM&1.39&2.93&3&0.45&12.75&0.13\\
            GCE&3.88&1.46&2.94&0.27&35.85&0.14\\
            OSLOM&4.18&74.2&3.54&0.62&602.76&0.16\\
            LINKC&3.15&5.14&7.14&0.41&114.1&0.37\\
            SVINET&2.01&7.14&3.15&0.66&18.2&0.41\\
            SLPA&2.19&1.18&1.43&0.47&7.31&0.24\\
            DEMON&4.35&9.67&10.4&0.55&385.75&0.17\\
            \hline
            \end{tabular}%
        \end{table}%

        \begin{table}[!ht]
          \centering

          \caption{Quality metrics ranking for overlapping community detection algorithms applied on PGP. The calculated properties are Average Degree (AD), Average ODF (AO), Flake ODF (FO), Internal Density (ID), Max ODF (MO), and Overlapping Modularity (OM). Kconsensus denotes the quality metrics ranking using Kemeny consensus.and TOPSIS denotes the final ranking obtained by TOPSIS}
          \label{table51}%
            \begin{tabular}{lcccccccc}
            \hline
              & AD & AO & FO & ID & MO & OM & \multicolumn{1}{c}{Kconsensus} & \multicolumn{1}{c}{TOPSIS} \\
            \hline
            LFM & 1 & 4 & 3 & 5 & 2 & 7 & 4 &2\\
            GCE & 5 & 5 & 2 & 7 & 4 & 6 & 6 &5\\
            OSLOM & 6 & 7 & 5 & 2 & 7 & 5 & 5 &6\\
            LINKC & 4 & 2 & 6 & 6 & 5 & 1 & 1 &4\\
            SVINET & 2 & 1 & 4 & 1 & 1 & 2 & 2 &1\\
            SLPA & 3 & 6 & 1 & 4 & 3 & 3 & 3 &3\\
            DEMON & 7 & 3 & 7 & 3 & 6 & 4 & 7 &7\\
            \hline
            \end{tabular}%
        \end{table}%

        \begin{table}[htbp!]
          \centering

                  \caption{Correlation of the quality metrics ranking. The calculated properties are Average Degree (AD), Average ODF (AO), Flake ODF (FO), Internal Density (ID), Max ODF (MO), and Overlapping Modularity (OM). }
          \label{table82}%
            \begin{tabular}{lcccccc}
            \hline
              & AD & AO & FO & ID & MO & OM \\
              \hline
            AD & 1 &   &   &   &   &  \\
            AO & 0.29 & 1 &   &   &   &  \\
            FO & 0.54 & -0.43 & 1 &   &   &  \\
            ID & -0.04 & 0.11 & -0.29 & 1 &   &  \\
            MO & 0.89 & 0.43 & 0.57 & 0.04 & 1 &  \\
            OM & 0 & 0.54 & -0.32 & 0.25 & 0.04 & 1 \\
            \hline
            \end{tabular}%
        \end{table}%

\subsubsection{Clustering metrics}
        Table \ref{table26} reports the clustering metrics values for the PGP dataset. We can notice that these very low values indicate that all the algorithms perform poorly.  This is also true for the other datasets (See  \ref{table27} and  Table \ref{table28} ). Table \ref{table32} gives the individual rankings of the clustering metrics and the merged one using Kconsensus and TOPSIS for the PGP dataset. It shows that SVINET is the best algorithm for both merging strategies. The second algorithm is SLPA  according to Kconsensus while it is LINKC according to TOPSIS. At the other extreme for the two merging strategies DEMON and LFM are considered as the worst algorithms. Indeed they are ranked respectively 6 and 7. Note that the rankings are not homogeneous for the three networks. For the AMAZON dataset, we can see in Table \ref{table33} that DEMON, CFINDER and SLPA rank respectively first, second and third for both merging strategies. LFM and OSLOM are the fewer performings. Indeed they rank respectively 7 and 8. For this dataset, the merging strategies are very consensual, while that it is not the case for the aNobii dataset. In this case,  SLPA is ranked first by both merged strategies as indicated in Table \ref{table34}. It is the only case where both merging strategies agree on the ranks of the algorithms. When we look at the correlation between the rankings of the three clustering metrics (See Table \ref{table85}, Table \ref{table86} and Table \ref{table87}), it appears that globally, the correlation values are very low. We expected a better agreement between these metrics as they are related more or less to the confusion matrix. Therefore, these results indicate that it is better to consider them all rather than relying on one of them in order to evaluate the effectiveness of the community detection algorithms.

        \begin{table}[h!]
        \centering
        \caption{Clustering metrics for PGP ground-truth and the uncovered community structure by overlapping community detection algorithms. The calculated properties are NMI, Omega Index (OI) and F1-score.}
        \label{table26}
		\begin{tabular}{lcccc}
                \hline
				 &  NMI  &  OI  &  F1-score \\
                \hline
				LFM&0.06&0.12&0.37\\
                GCE&0.51&0.16&0.11\\
                OSLOM&0.31&0.2&0.28\\
                LINKC&0.24&0.41&0.66\\
                SVINET&0.64&0.34&0.71\\
                SLPA&0.6&0.25&0.51\\
                DEMON&0.19&0.21&0.16\\
                \hline
		\end{tabular}
	   \end{table}

        \begin{table}[!ht]
          \centering
          \caption{Clustering metrics ranking for overlapping community detection algorithms applied on PGP. The calculated properties are NMI, Omega Index (OI) and F1-score. Kconsensus denotes the clustering metrics ranking using Kemeny consensus and TOPSIS denotes the final ranking obtained by TOPSIS.}
          \label{table32}
            \begin{tabular}{lccccc}
            \hline
            & \multicolumn{1}{l}{NMI} & \multicolumn{1}{l}{OI} & \multicolumn{1}{l}{F1-score} & \multicolumn{1}{l}{Kconsensus} & \multicolumn{1}{l}{TOPSIS} \\
            \hline
            LFM & 7 & 7 & 4 & 7 & 7 \\
            GCE & 3 & 6 & 7 & 3 & 4 \\
            OSLOM & 4 & 5 & 5 & 4 & 5 \\
            LINKC & 5 & 1 & 2 & 5 & 2 \\
            SVINET & 1 & 2 & 1 & 1 & 1 \\
            SLPA & 2 & 3 & 3 & 2 & 3 \\
            DEMON & 6 & 4 & 6 & 6 & 6 \\
            \hline
        \end{tabular}%
       \end{table}%

        \begin{table}[htbp!]
          \centering
          \caption{Correlation of the clustering metrics ranking for overlapping community detection algorithms applied on PGP. The calculated properties are NMI, Omega Index (OI) and F1-score. }
                  \label{table85}
            \begin{tabular}{lccc}
            \hline
              & NMI & OI & F1-score \\
              \hline
            NMI & 1 &   &  \\
            OI & 0,42 & 1 &  \\
            F1-score & 0,35 & 0,71 & 1 \\
            \hline
            \end{tabular}%
        \end{table}%

\subsection{Ranking based on all properties}
        In order to investigate the relationship between the three types of properties that can be used in order to compare the algorithms, we compute the correlation between their rankings for both merging strategies. Table \ref{table97} reports the results for the PGP dataset with the rankings given by Kconsensus. With a correlation value equal to $0.82$, it appears that the topological properties and the clustering ones are well related. To a lesser extent, topological properties and quality metrics exhibit some relations, while the low correlation value ($0.32$) between the rank given by the clustering and the quality metrics suggest clearly that these two types of measures are complementary. Table \ref{table970} reports the same type of results for the TOPSIS merging strategy. It appears that there is no correlation between the quality and the clustering metrics as observed with the alternative strategy. However, this time, the correlation of the rankings obtained by merging the topological properties with the quality metrics ones is more much higher than the correlation value with the clustering metrics. The results with the AMAZON dataset (See Table \ref{table98}) point in the same direction. There is a weak correlation between the clustering and quality metrics rankings for both merging strategies. We also observe a  stronger correlation between the topological rankings and the clustering based ones than between the topological and the quality metrics rankings using the Kconsensus  merging strategy. For the TOPSIS strategy, this is the contrary (see Table \ref{table980}).  With the aNobii dataset (See Table \ref{table99}, and Table \ref{table990}) the results are quite similar. The main differences are that the correlation values are a little bit smaller. To summarize, the results are fairly independent of the datasets. Clustering and Quality metrics rankings of the algorithms appear to be not correlated while the topological properties rankings correlate with either the clustering or the quality metrics ranking depending of the merging strategy. However, the correlation observed between the topological properties and its alternative is not enough strong in order to substitute one to the others. Although the topological properties seem to be more efficient, as they capture part of the information of both the quality and the clustering metrics, it can be interesting to use all the information given by these three types of properties in order to get a more accurate ranking. Table \ref{table59} illustrates the ranking obtained using all the properties (topological, quality and clustering properties) for the PGP dataset. All the individual rankings are merged into a single one using Kconsensus and TOPSIS. According to both strategies, SVINET is the best algorithm. It is followed by SLPA, while DEMON is the less performing algorithm.  SVINET is ranked first by TOPSIS and third by Kconsensus with the AMAZON dataset, while it is MOSES that rank first for Kconsensus (See Table \ref{table60}) and fourth for TOPSIS. SLPA is in the middle range. Overall, the rankings are quite different than with the other dataset.  Indeed, the two merging strategies rank SLPA second for the aNobii dataset, and they do not agree for the first place. Nevertheless, they all rank DEMON six out of six (See Table \ref{table61}). Globally, SVINET and SLPA are very often ranked in the top tier. However, this finding has to be taken with caution. Indeed, the efficiency is greatly dependent of the dataset and the results suggest that there is no universal solution to the community detection problem.

        \begin{table}[ht!]
          \centering
          \caption{Correlation of the topological properties, the quality metrics and the clustering measures rankings using the Kconsensus strategy}
          \label{table97}%
            \begin{tabular}{lccc}
            \hline
              & Topo & Quality & Clustering \\
            \hline
            Topo & 1 &   &  \\
            Quality & 0.6 & 1 &  \\
            Clustering & 0.82 & 0.32 & 1 \\
            \hline
            \end{tabular}%
        \end{table}%

        \begin{table}[ht!]
          \centering
          \caption{Correlation of the topological properties. the quality metrics and the clustering measures rankings using the TOPSIS strategy}
          \label{table970}%
            \begin{tabular}{lccc}
            \hline
              & Topo & Quality & Clustering \\
            \hline
            Topo & 1 &   &  \\
            Quality & 0.96 & 1 &  \\
            Clustering & 0.57 & 0.43 & 1 \\
            \hline
            \end{tabular}%
        \end{table}%

        \begin{table}[!h]
          \centering
          \scriptsize

          \caption{All properties for PGP dataset. The calculated properties are number of nodes (V), number of edges (E), Density ($\rho$), Diameter ($d$), Average shortest path ($l_{G}$), Average node degree ($\widetilde{deg}$), Max node degree ($\delta(G)$), Assortativity Coefficient ($\tau$), and Clustering Coefficient ($C$), the degree distribution (DD), the average clustering coefficient as function of degree (Av), the hop distance (HD), the community size (CS), the membership (M), the overlap size (OS), Average Degree (AD), Average ODF (AO), Flake ODF (FO), Internal Density (ID), Max ODF (MO), and Overlapping Modularity (OM), NMI, Omega Index (OI) and F1-score. Kconsensus denotes the final ranking using Kemeny consensus and TOPSIS is the final ranking using TOPSIS.}
          \label{table59}
            \begin{tabular}{{p{.9cm}|}*{5}{p{.008cm}}*{1}{p{.1cm}}*{1}{p{.18cm}}*{1}{p{.008cm}}*{1}{p{.008cm}|}*{2}{p{.08cm}}*{1}{p{.08cm}|}*{2}{p{.08cm}}*{1}{p{.08cm}|}*{5}{p{.08cm}}*{1}{p{.18cm}|}*{1}{p{.18cm}}*{1}{p{.04cm}}*{1}{p{.9cm}|}*{2}{p{.7cm}}}
            \hline
              & \multicolumn{9}{c|}{Basic properties} & \multicolumn{3}{c|}{Microscopic} & \multicolumn{3}{c|}{Mesoscopic} & \multicolumn{6}{c|}{Clustering} & \multicolumn{3}{c|}{Quality} & \multicolumn{2}{c}{MCDM Ranking} \\
              \hline
              &V&E&$\rho$&$d$&$l_{G}$&$\widetilde{deg}$&$\delta(G)$&$\tau$&$C$& DD & Av & HD & CS & M & OS & AD & AO & FO & ID & MO & OM & NMI & OI & F1-score & Kconsensus & TOPSIS \\
            \hline
            LFM & 7 & 6 & 1 & 7 & 4 & 3 & 5 & 6 & 5 & 5 & 1 & 4 & 6 & 6 & 1 & 1 & 4 & 3 & 5 & 2 & 7 & 7 & 7 & 4 & 5 & 3 \\
            GCE & 4 & 5 & 5 & 3 & 3 & 4 & 7 & 3 & 3 & 3 & 5 & 1 & 3 & 7 & 6 & 5 & 5 & 2 & 7 & 4 & 6 & 3 & 6 & 7 & 7 & 5 \\
            OSLOM & 3 & 1 & 6 & 3 & 7 & 5 & 4 & 7 & 6 & 6 & 6 & 7 & 4 & 5 & 5 & 6 & 7 & 5 & 2 & 7 & 5 & 4 & 5 & 5 & 4 & 6 \\
            LINKC & 6 & 7 & 2 & 5 & 1 & 7 & 3 & 4 & 7 & 2 & 7 & 8 & 7 & 3 & 3 & 4 & 2 & 6 & 6 & 5 & 1 & 5 & 1 & 2 & 3 & 4 \\
            SVINET & 1 & 2 & 4 & 1 & 5 & 2 & 1 & 1 & 1 & 1 & 2 & 2 & 2 & 2 & 4 & 2 & 1 & 4 & 1 & 1 & 2 & 1 & 2 & 1 & 1 & 1 \\
            SLPA & 2 & 4 & 3 & 2 & 6 & 1 & 2 & 2 & 2 & 4 & 4 & 6 & 1 & 1 & 2 & 3 & 6 & 1 & 4 & 3 & 3 & 2 & 3 & 3 & 2 & 2 \\
            DEMON & 5 & 3 & 7 & 6 & 2 & 6 & 6 & 4 & 4 & 7 & 3 & 5 & 5 & 4 & 7 & 7 & 3 & 7 & 3 & 6 & 4 & 6 & 4 & 6 & 6 & 7 \\
            \hline
            \end{tabular}%
        \end{table}%

\section{CONCLUSION}
       In this paper, we propose a methodology in order to evaluate overlapping community detection algorithms with data including a reference ground truth community structure. Our work departs from the classical approach that relies on clustering metrics to assess the efficiency of the community detection algorithms. It is based on the comparison of the ground-truth community structure, which is considered as a reference, with the one uncovered by the algorithm.
        Various basic and microscopic topological properties of the so-called 'community-graph' where the nodes are the communities and the links describe the overlap between two communities are compared. Furthermore, classical mesoscopic properties distributions such as the community size, the overlap size, and the membership of nodes are used to evaluate the differences between the ground truth community structure and the one uncovered by the algorithms. The study has shown that an extensive topological analysis is more appropriate to highlight the deviations between the reference and the discovered community structures. Indeed, clustering metrics may assign the same value for very different situations. Additionally, results show that there is no single metric or topological property that allows a better understanding of the strengths and limitations of each community detection method. Therefore, one recommendation from this study is to combine the multiple views of the community structure carried by the various measures in order to assess the performance of the algorithms.  To do so, the proposed scheme consists in ranking the algorithms according to each individual property and to merge all these local rankings into a global one using an MCDM strategy. The properties have been grouped into three main categories: topological properties, (basic, microscopic and mesoscopic), quality metrics and clustering metrics. For each category, a merged ranking is given using two MCDM strategies. Results reveal that the local rankings are fairly uncorrelated. Consequently, evaluating the overlapping community structure cannot rely on a single evaluation criterion. Comparisons of the global rankings based on the three types of measures give rather clear results. They do not carry the same information about the underlying community structure. Quality and clustering metrics are always uncorrelated, while topological properties are often well correlated either with quality or clustering metrics depending on the MCDM strategy. For this reason, they must be preferred to their alternative. However, using simultaneously all the information given by all the measures from the three categories must be preferred.
        Another important concern brought forth by our results is the impact of variations in data on the community detection performance. The results indicate that there is no method that clearly outperforms all methods in all situations. Future research effort should focus on investigating the possibility to combine a minimal subset of measures that can be computed efficiently and different methods of combining the individual rankings should be explored.
\clearpage
\appendix
\section{AMAZON}
        This section is devoted to present the different distributions for AMAZON, AMAZON*, and all 'algorithms-community-graph'.
        \begin{figure}[!ht]
        \subfigure[AMAZON*]{\includegraphics[width=.121\textwidth]{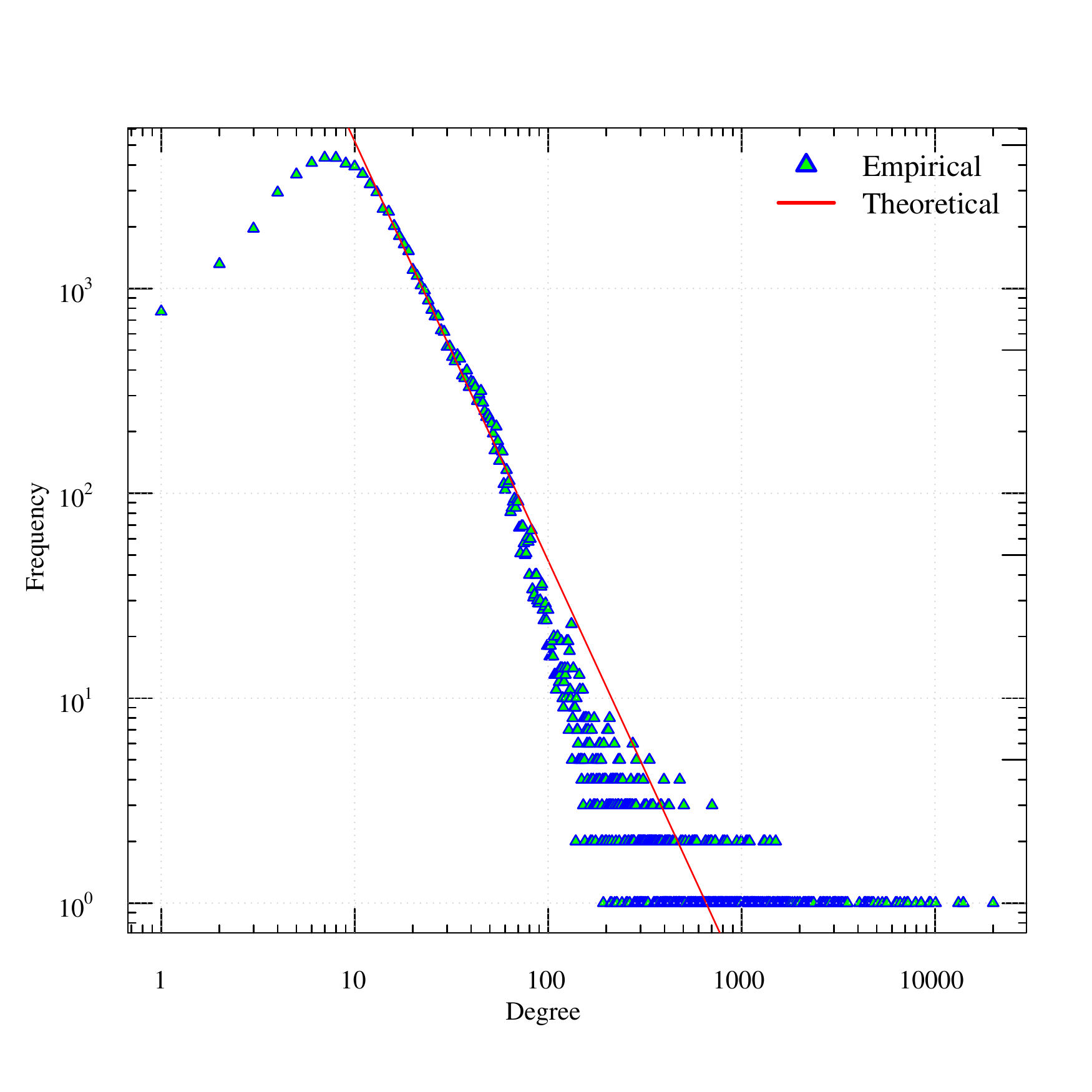}}
        \subfigure[CFINDER*]{\includegraphics[width=.121\textwidth]{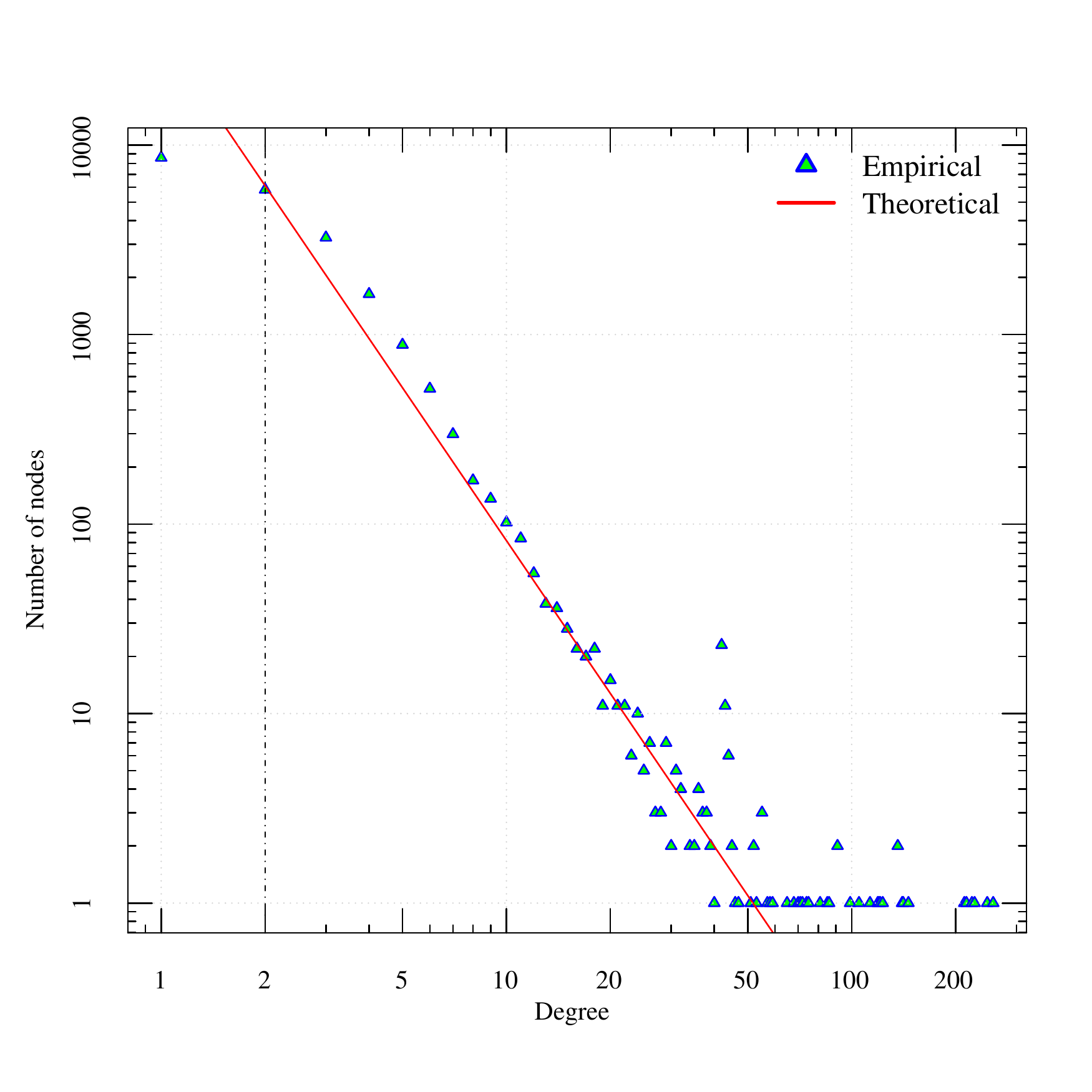}}
        \subfigure[LFM*]{\includegraphics[width=.121\textwidth]{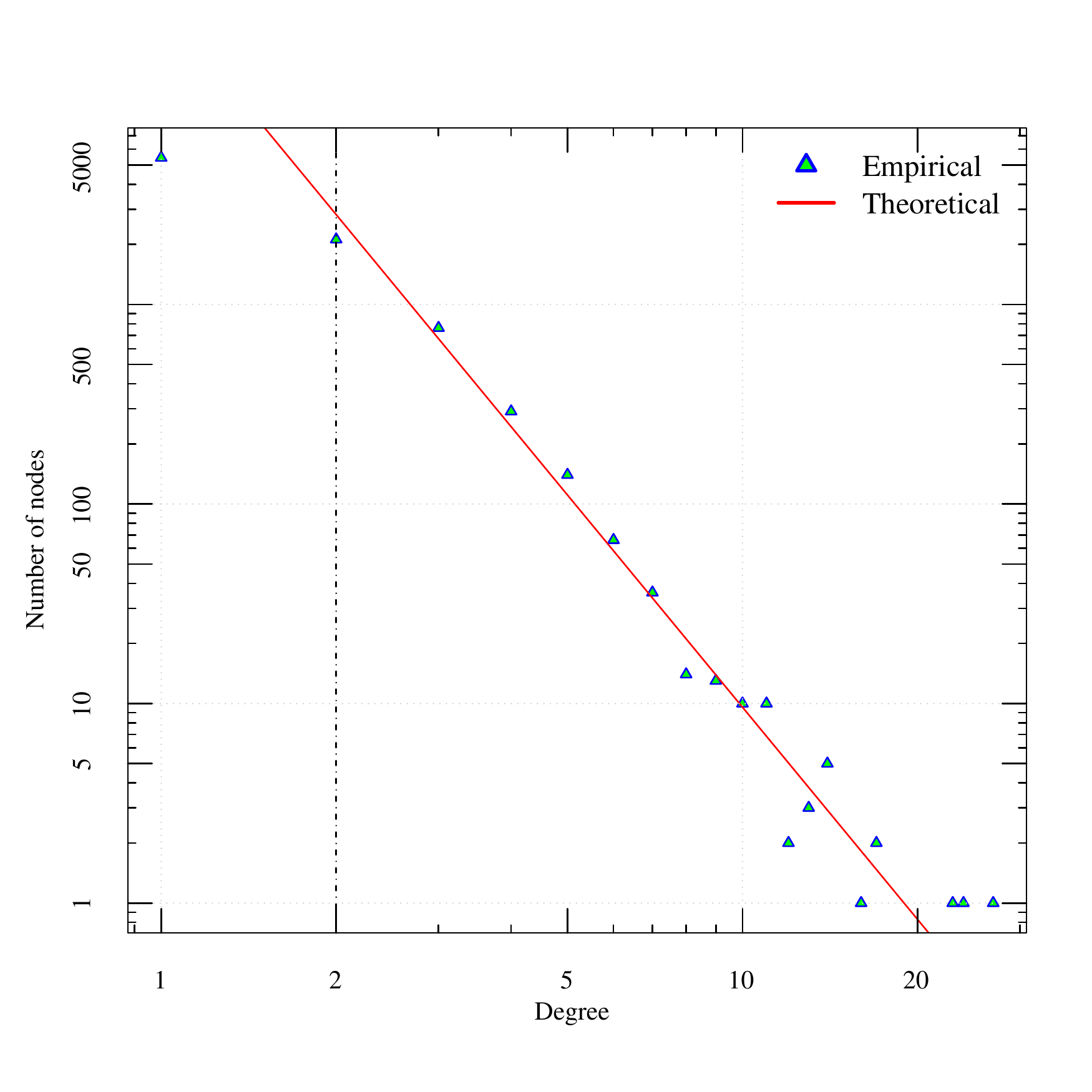}}
        \subfigure[GCE*]{\includegraphics[width=.121\textwidth]{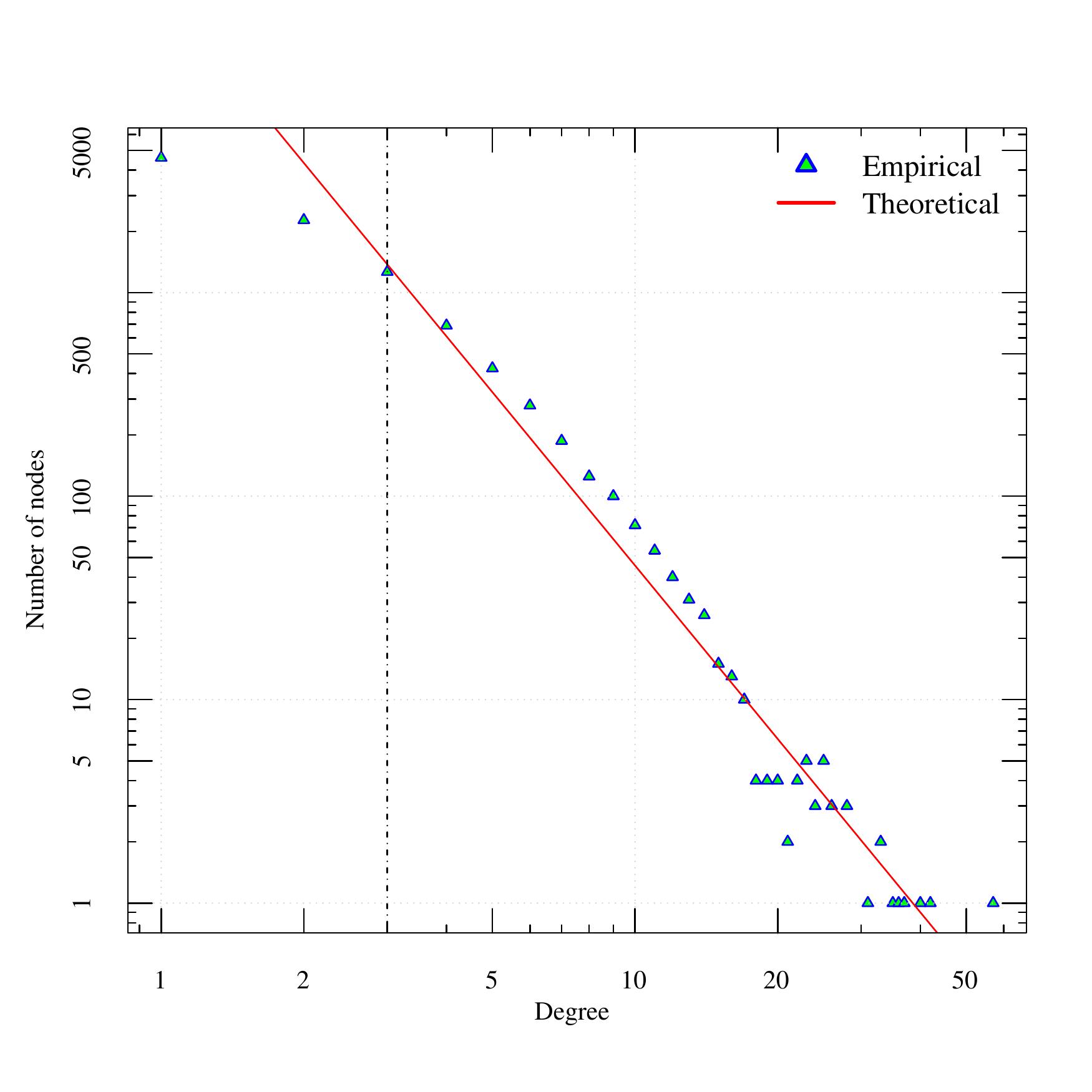}}
        \subfigure[OSLOM*]{\includegraphics[width=.121\textwidth]{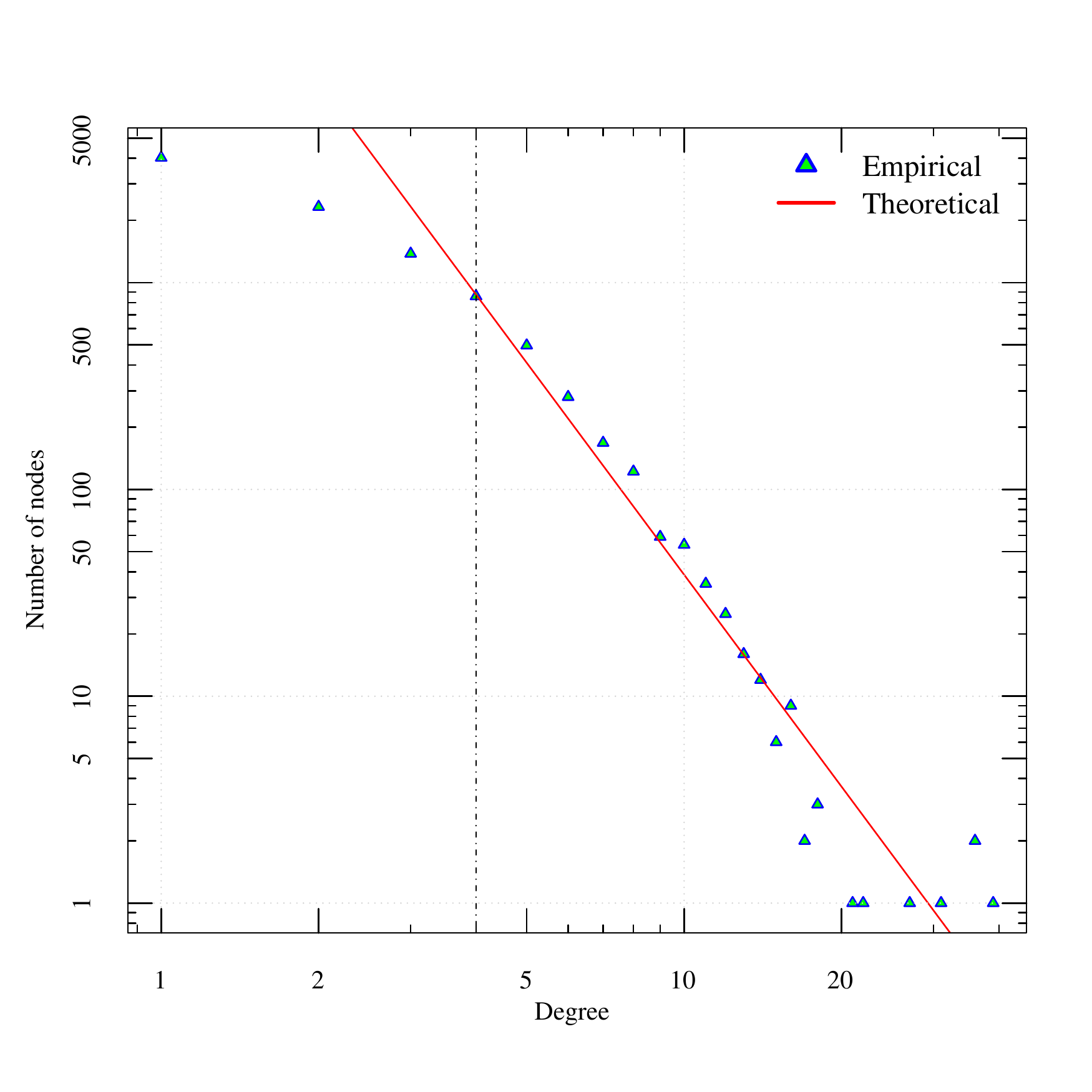}}
        \subfigure[SVINET*]{\includegraphics[width=.121\textwidth]{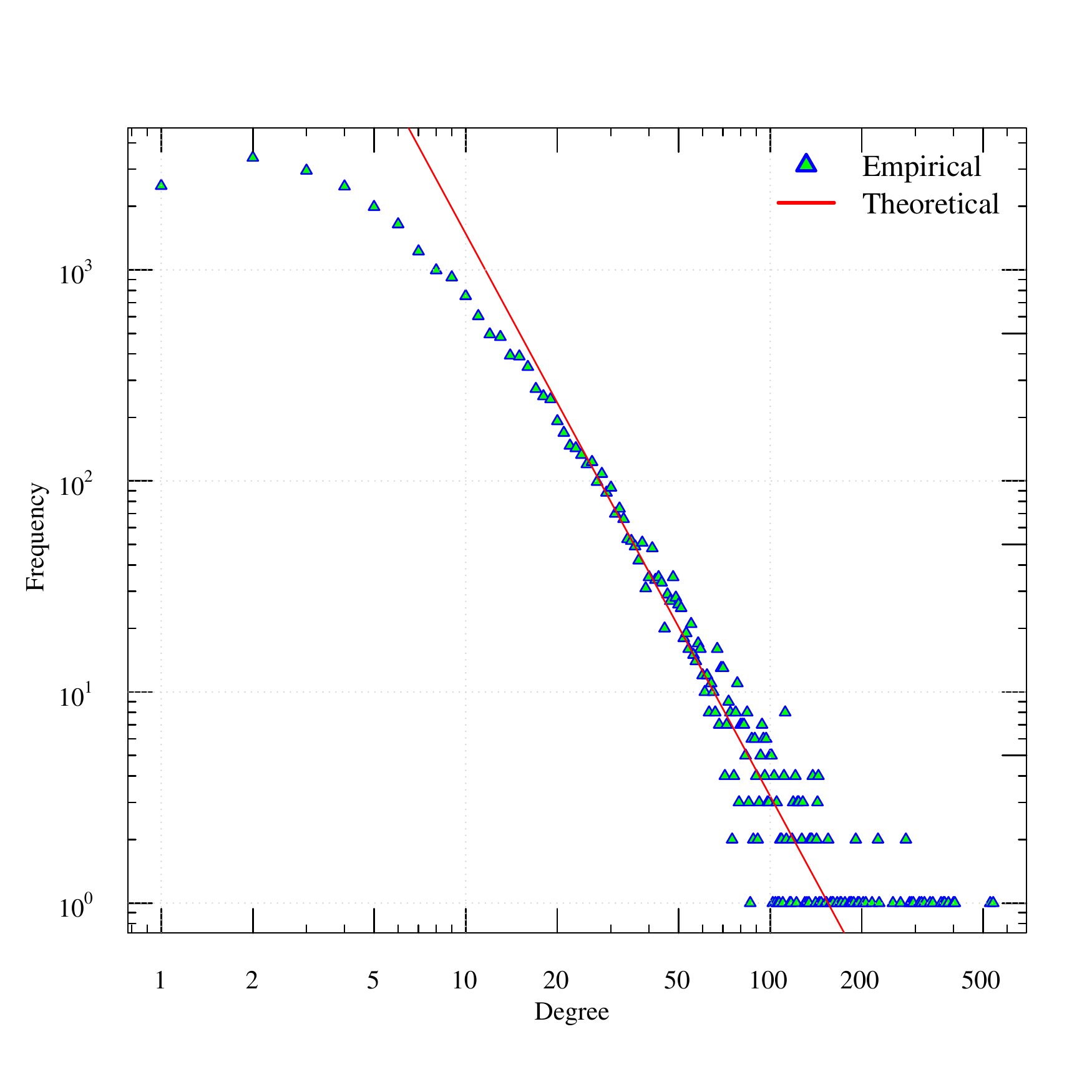}}
        \subfigure[MOSES*]{\includegraphics[width=.121\textwidth]{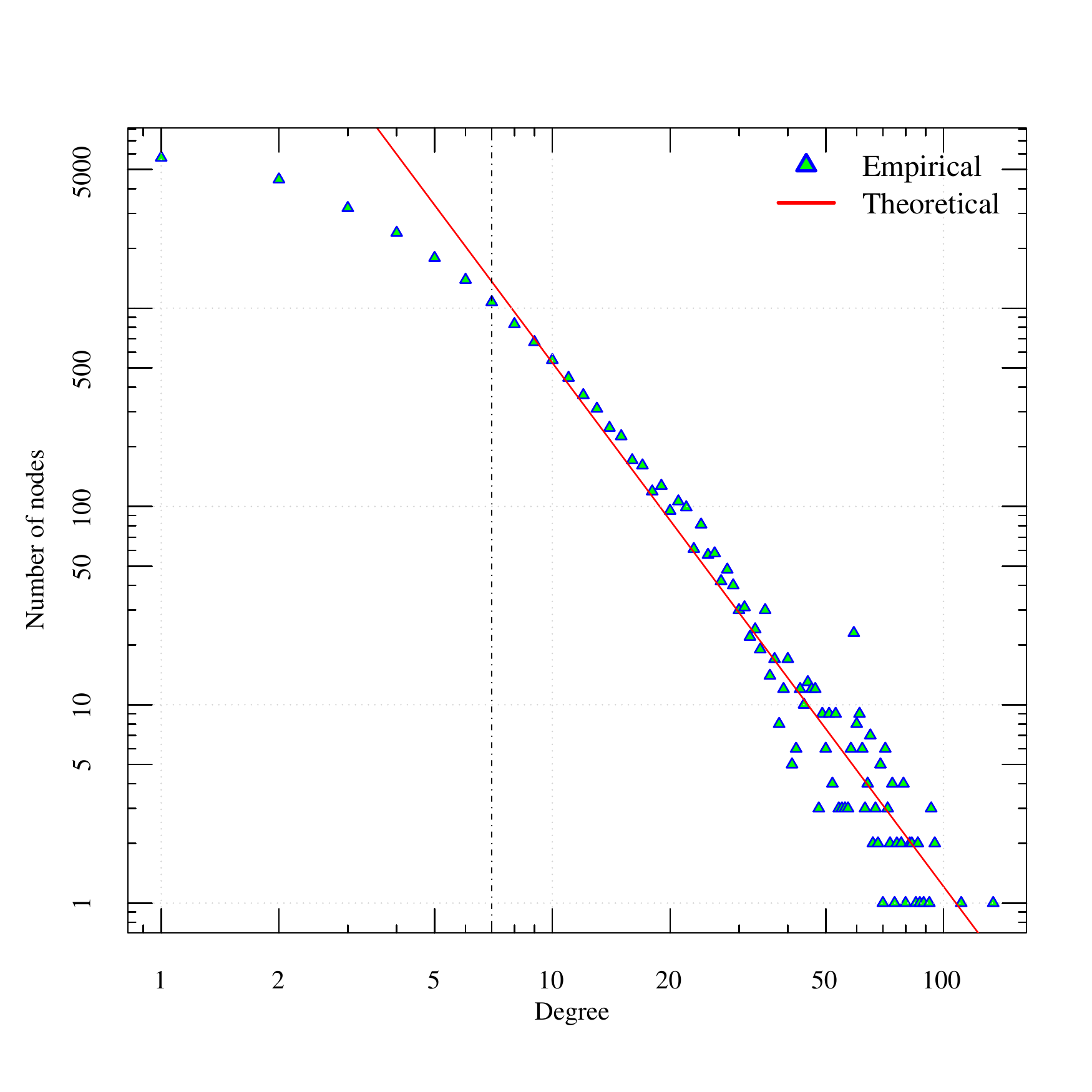}}
        \subfigure[SLPA*]{\includegraphics[width=.121\textwidth]{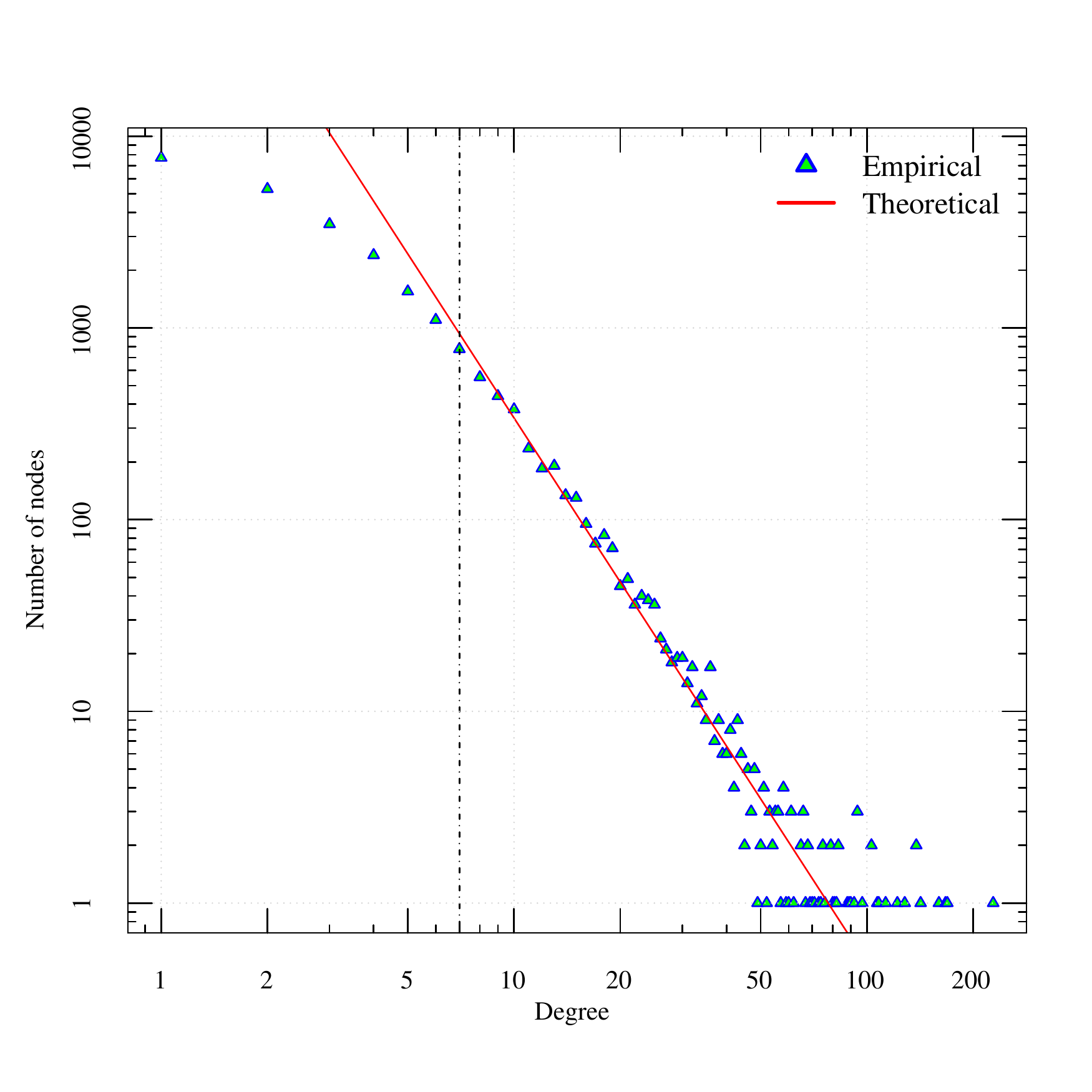}}
        \subfigure[DEMON*]{\includegraphics[width=.121\textwidth]{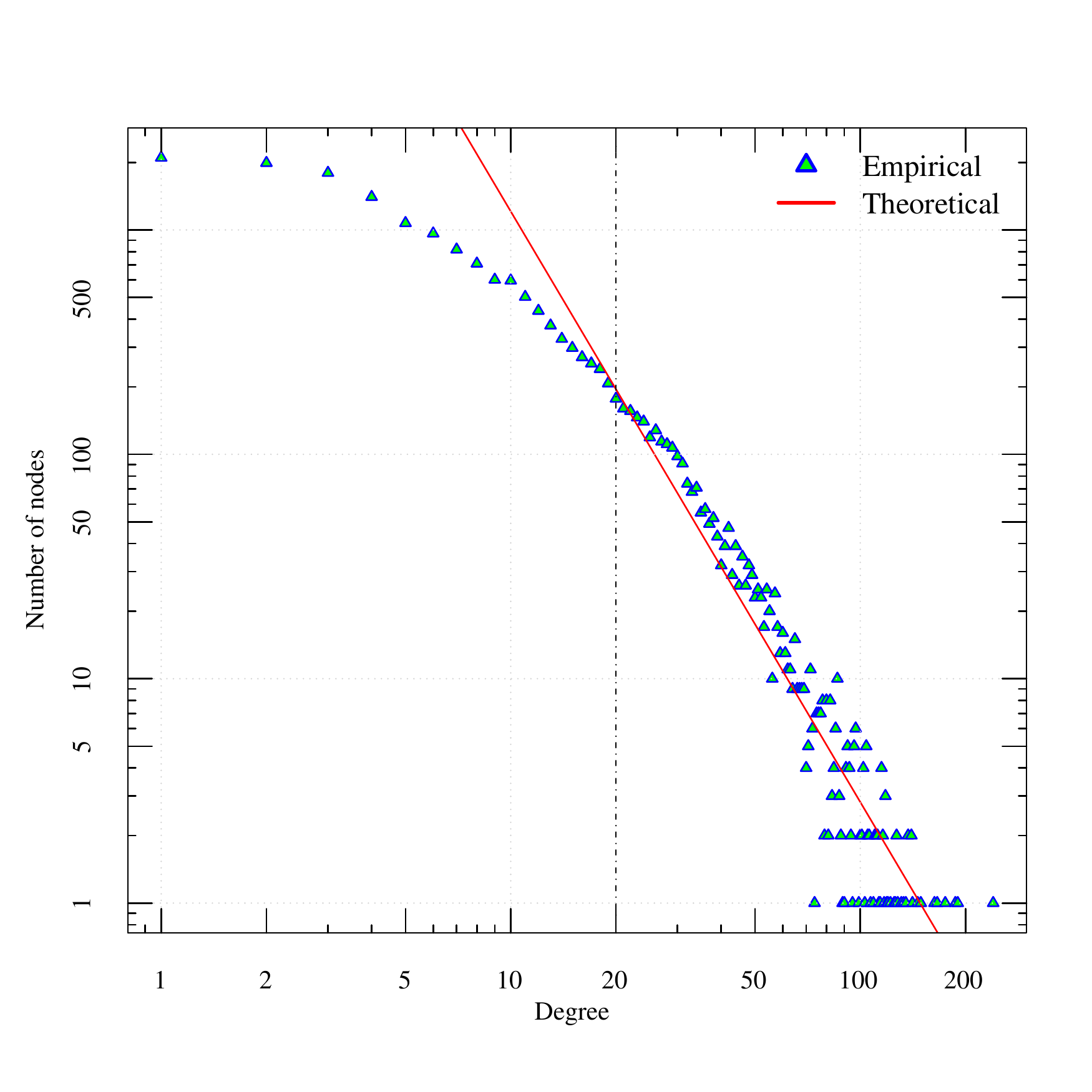}}
        \caption{\label{fig2}Log-log empirical degree distribution (dots) and Power-Law estimate (line) for AMAZON* (a), CFINDER* (b), LFM* (c),  GCE* (d), OSLOM* (e), SVINET** (f), MOSES* (g), SLPA* (h), and DEMON* (i)}
        \end{figure}

        \begin{figure}[!ht]
        \subfigure[AMAZON*]{\includegraphics[width=.121\textwidth]{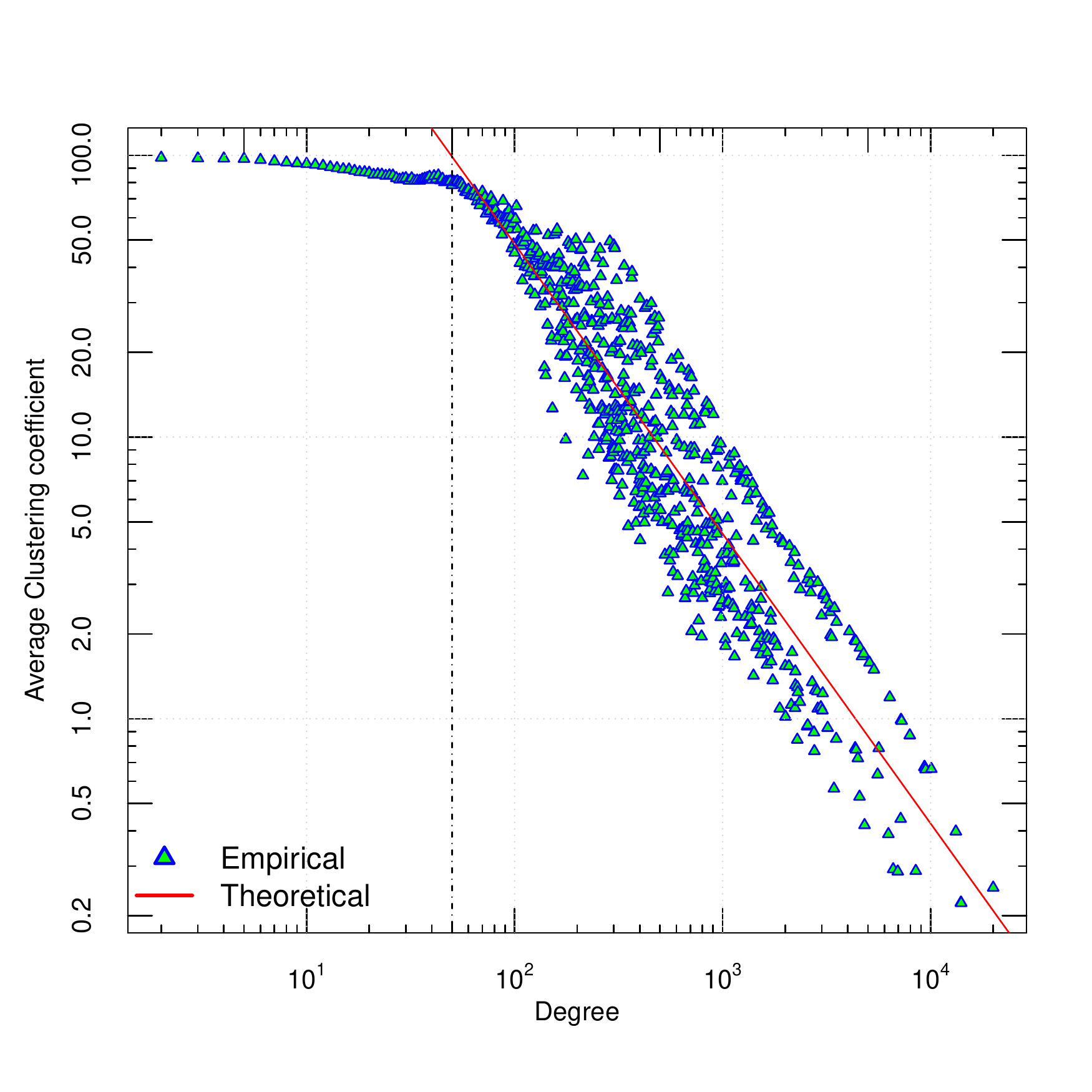}}
        \subfigure[CFINDER*]{\includegraphics[width=.121\textwidth]{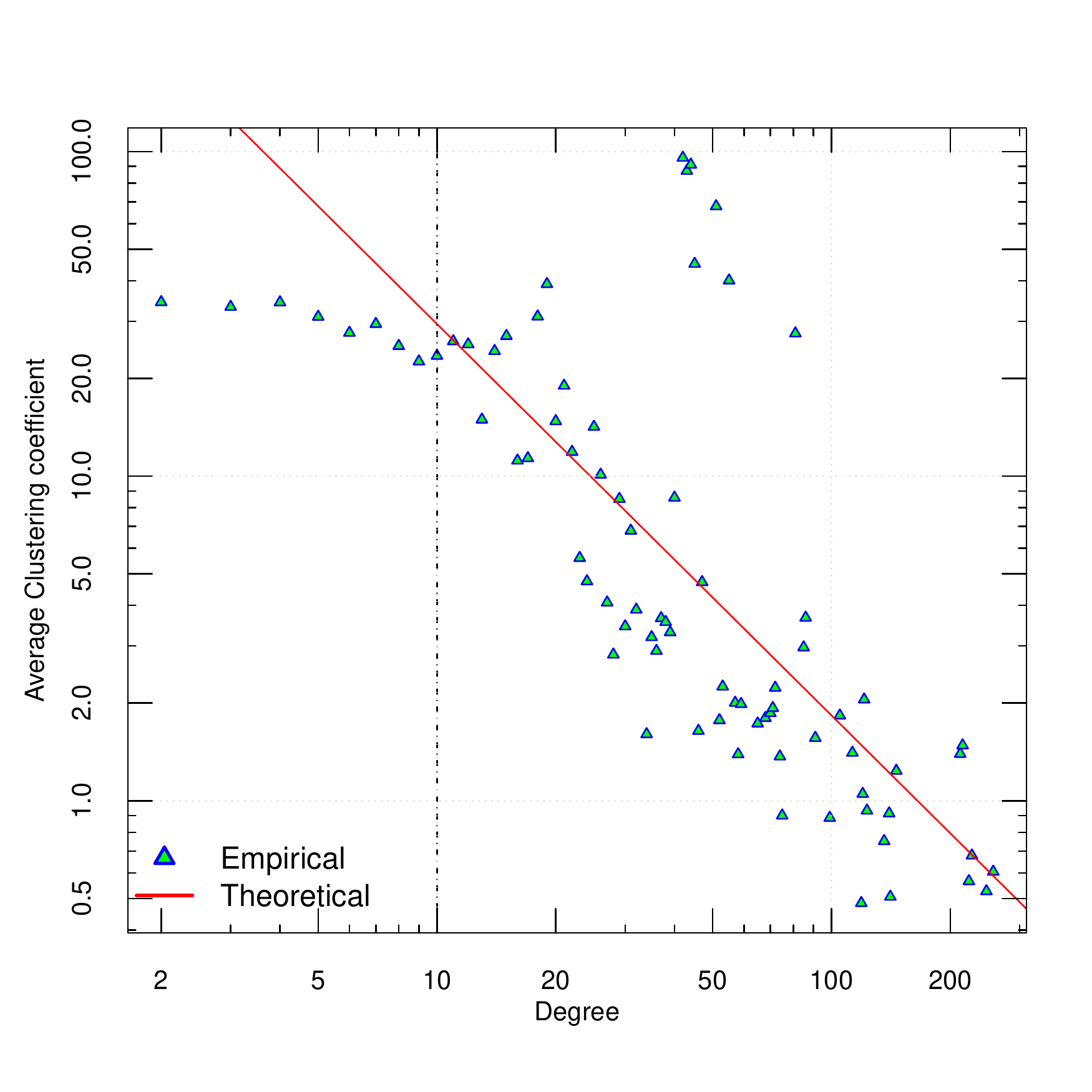}}
        \subfigure[LFM*]{\includegraphics[width=.121\textwidth]{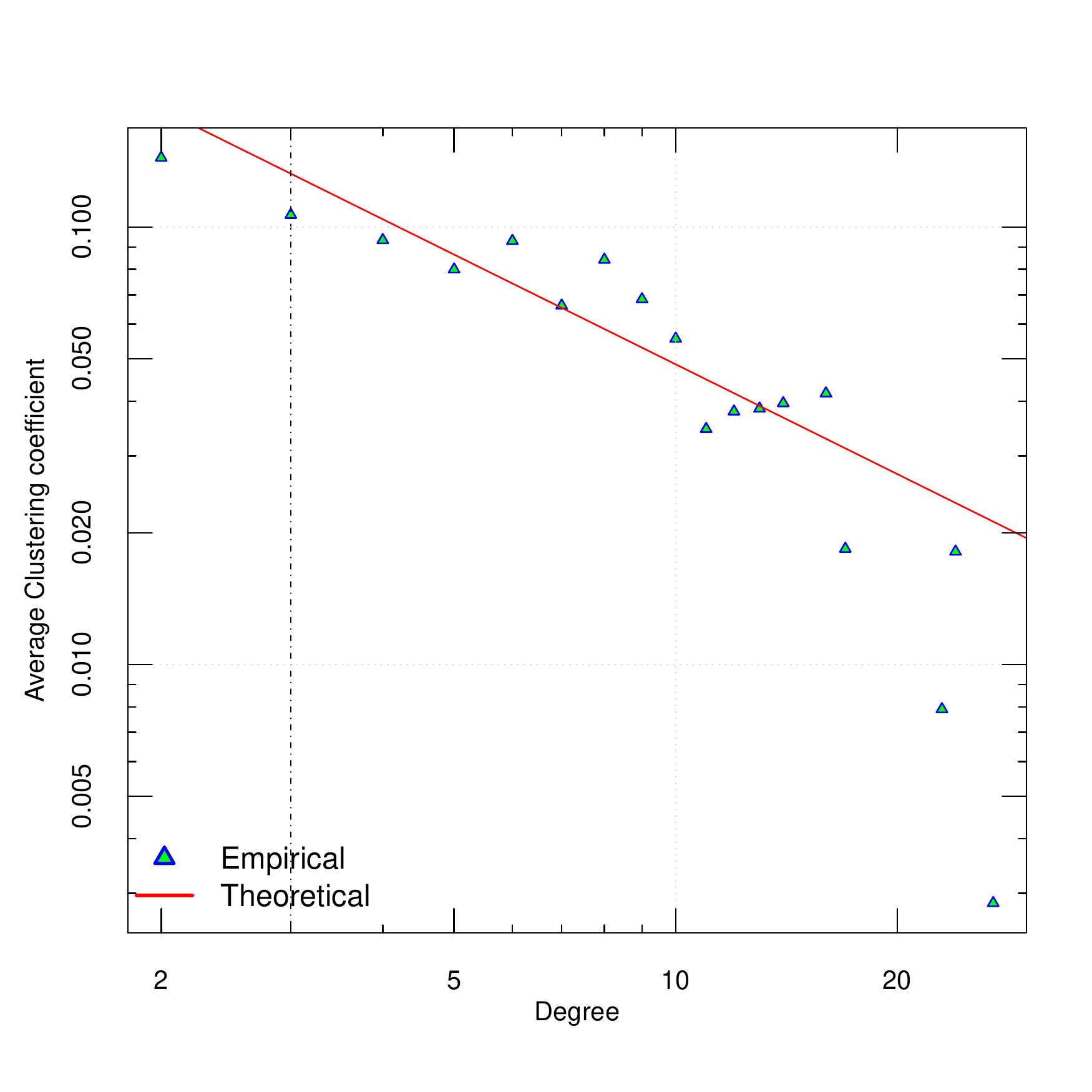}}
        \subfigure[GCE*]{\includegraphics[width=.121\textwidth]{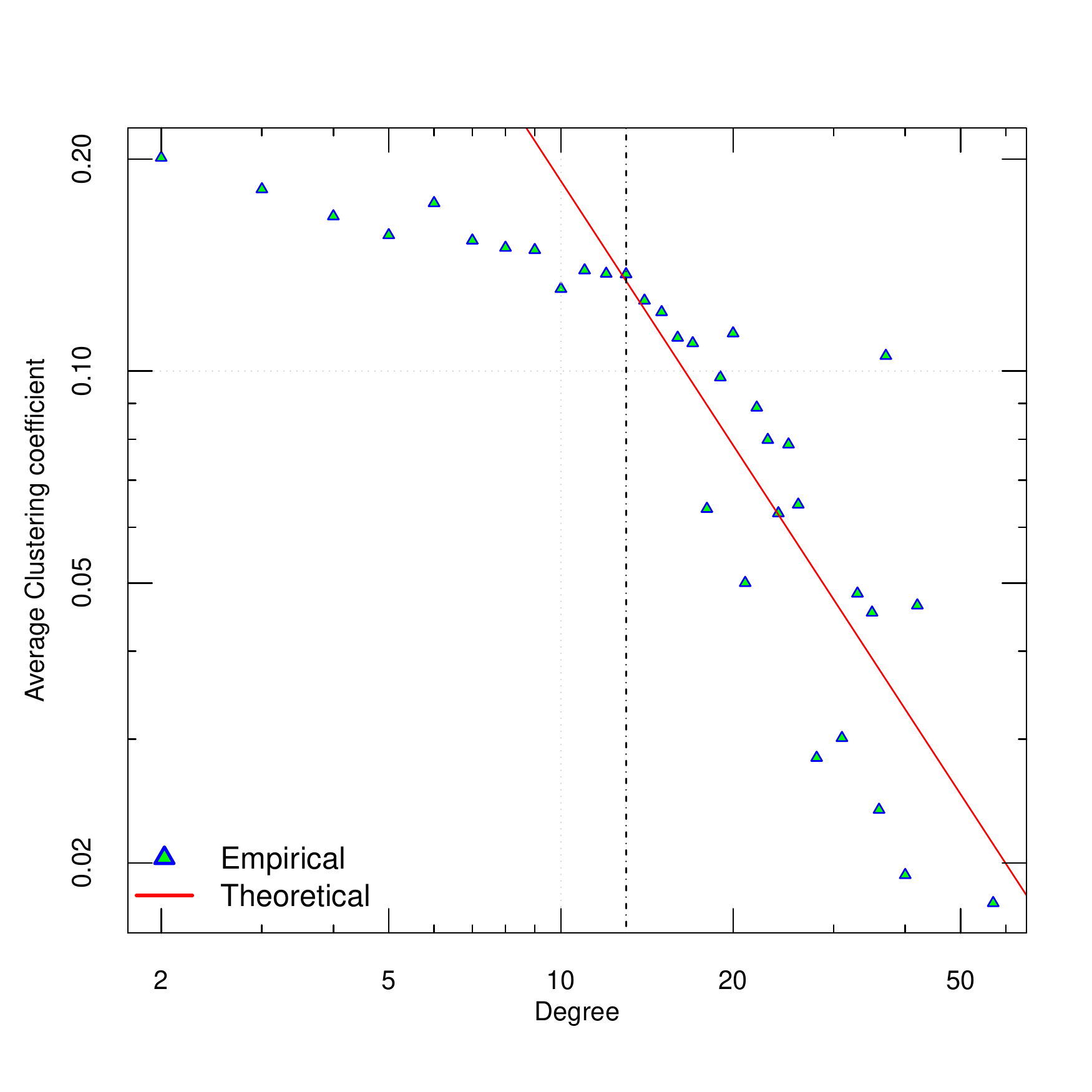}}
        \subfigure[OSLOM*]{\includegraphics[width=.121\textwidth]{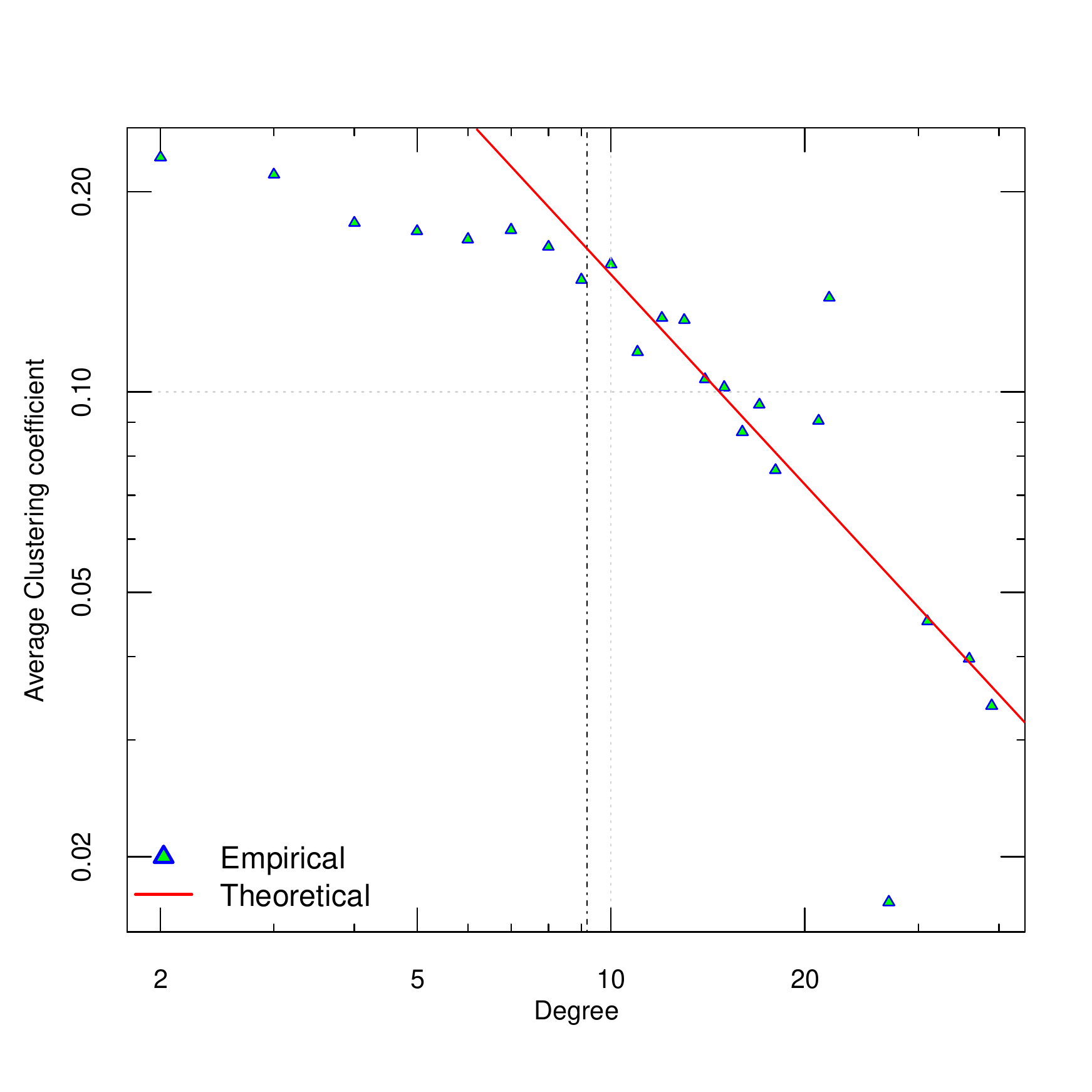}}
        \subfigure[SVINET*]{\includegraphics[width=.121\textwidth]{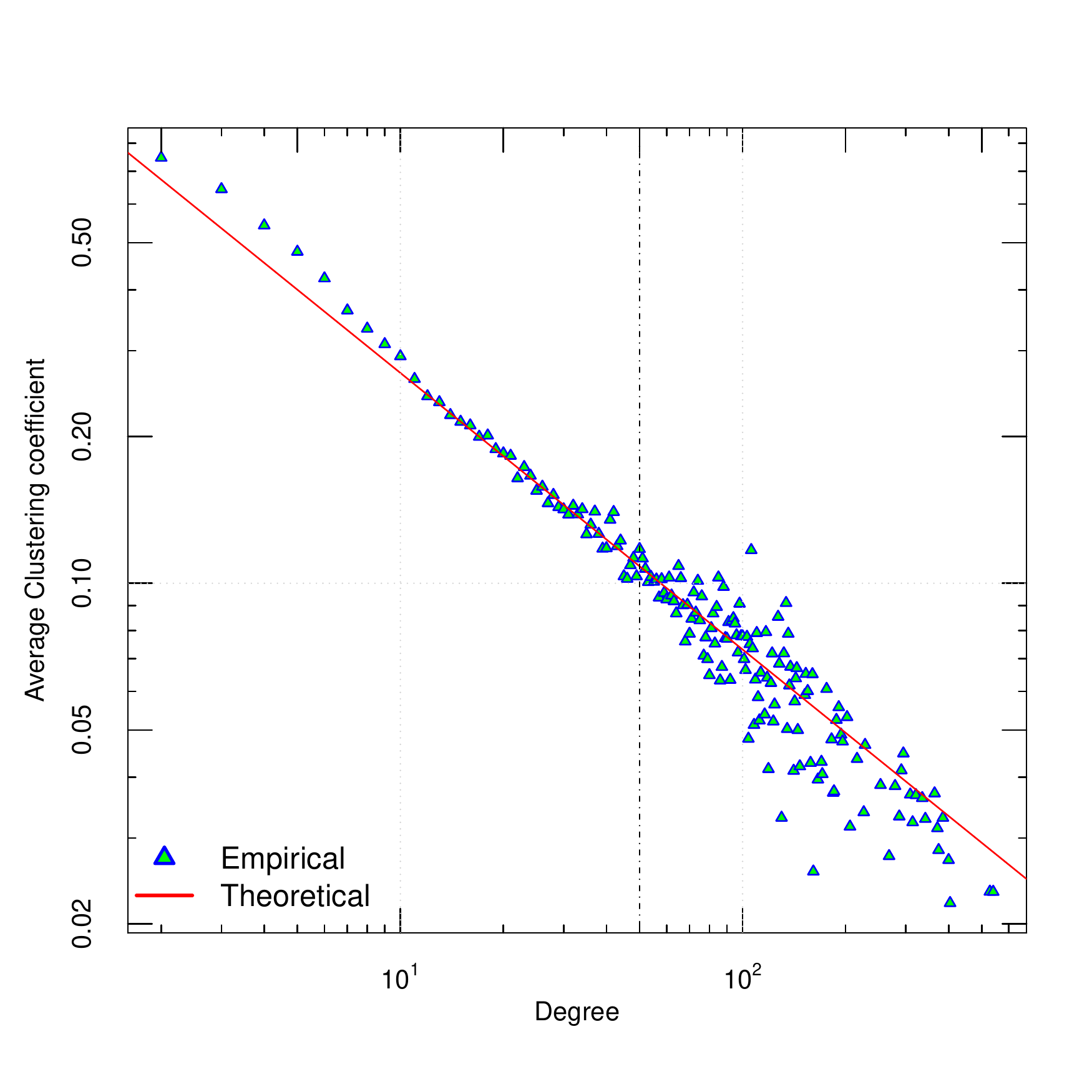}}
        \subfigure[MOSES*]{\includegraphics[width=.121\textwidth]{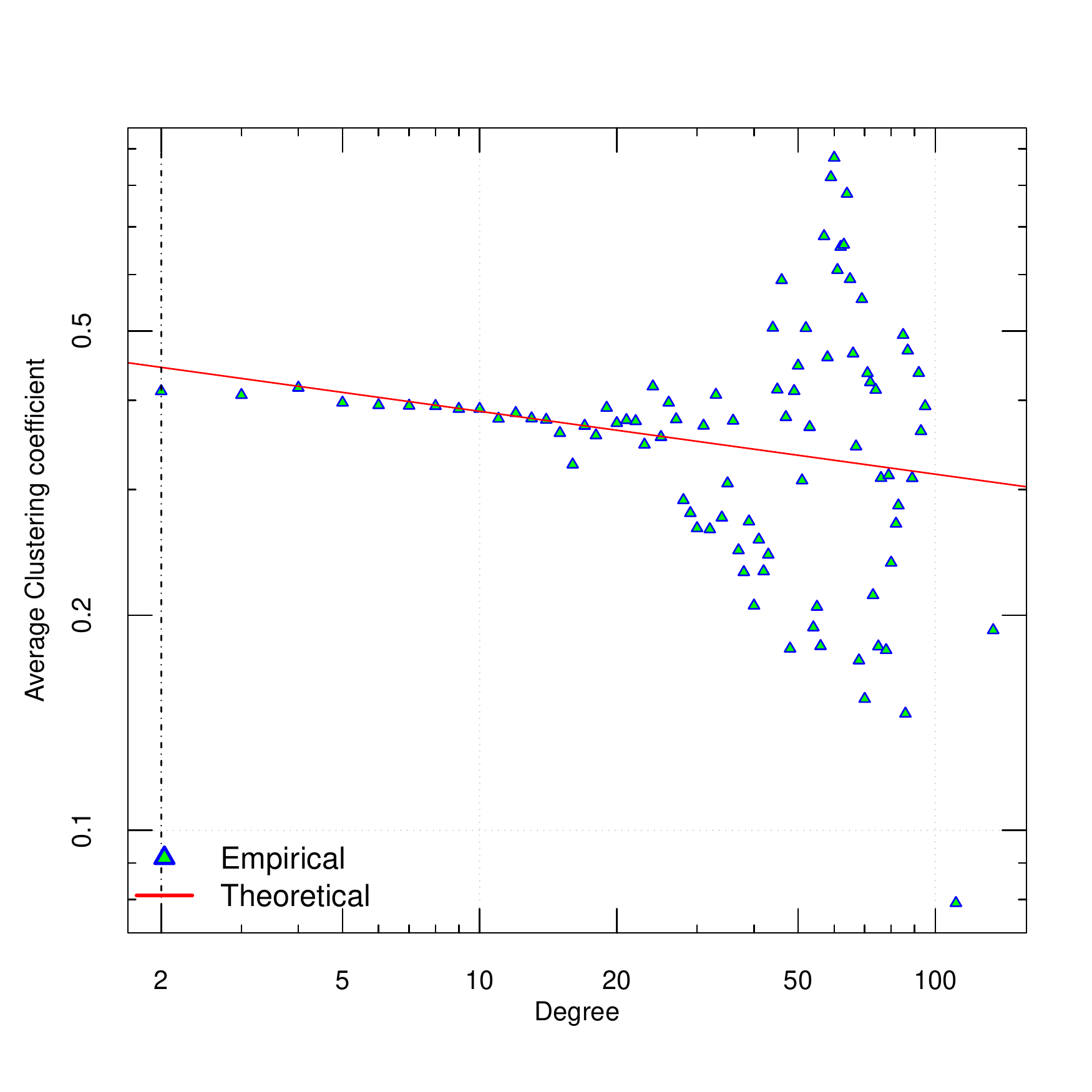}}
        \subfigure[SLPA*]{\includegraphics[width=.121\textwidth]{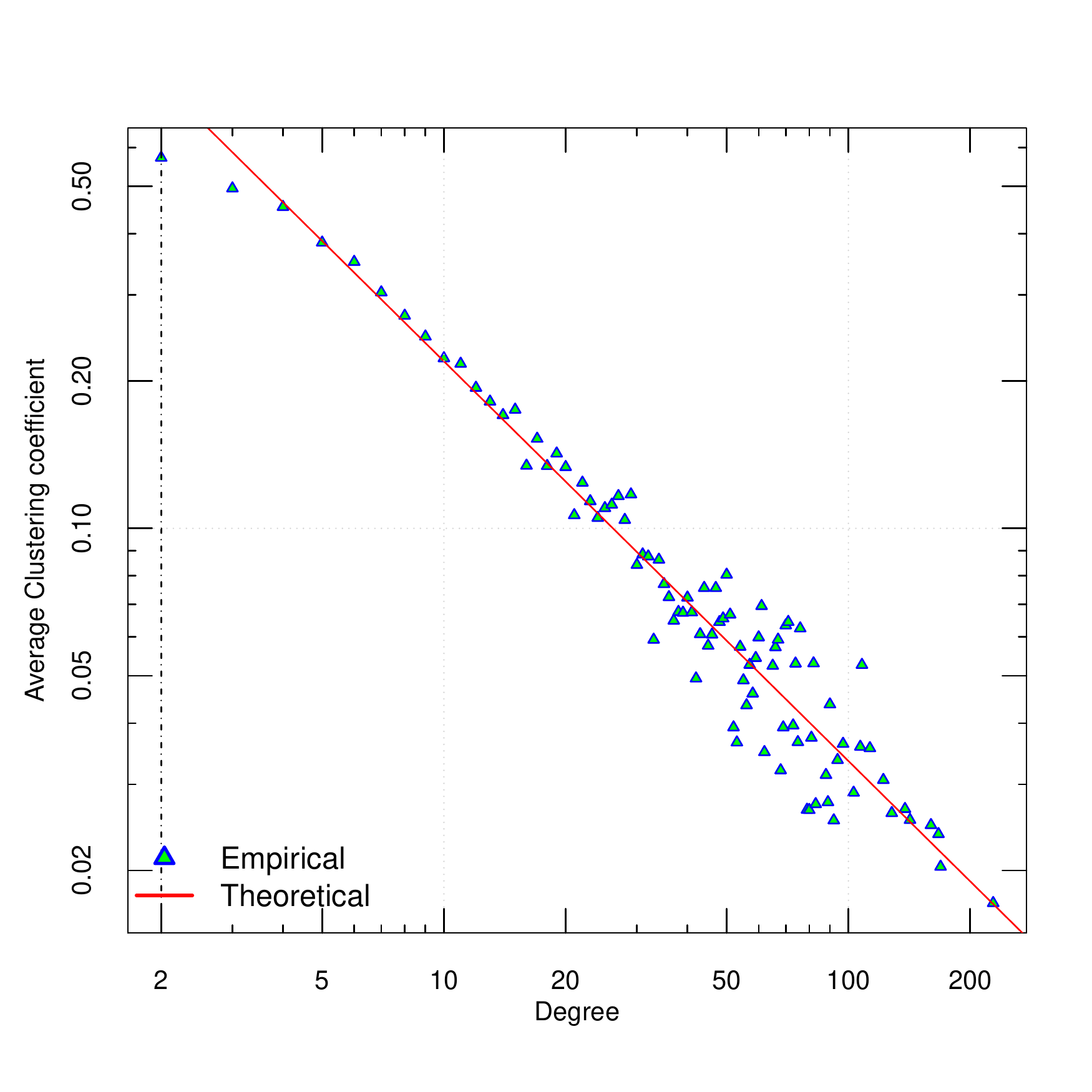}}
        \subfigure[DEMON*]{\includegraphics[width=.121\textwidth]{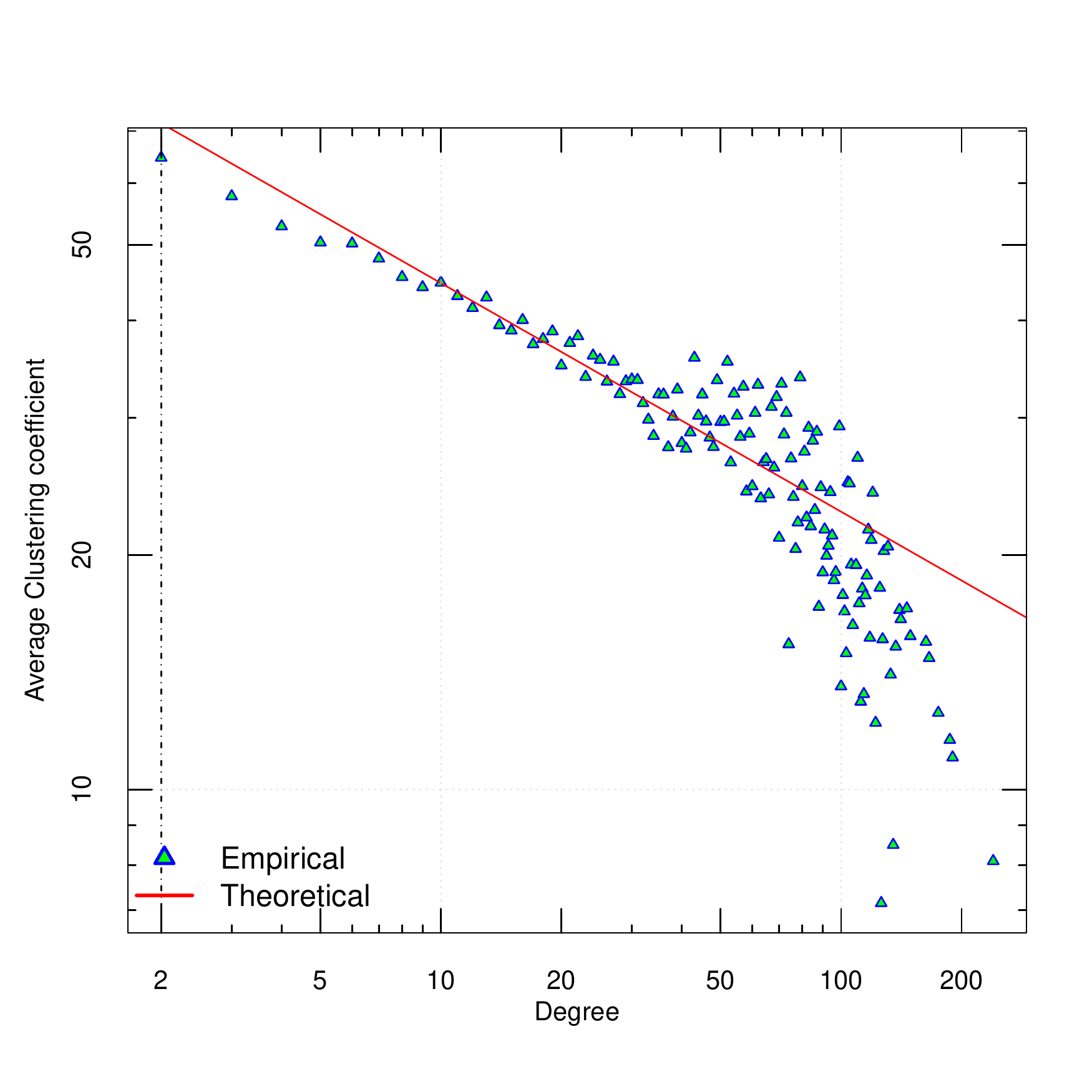}}

        \caption{\label{fig5} Log-log empirical Average clustering coefficient distribution as a function of the degree (dots) and Power-Law estimate (line) for AMAZON* (a), CFINDER* (b), LFM* (c),  GCE* (d), OSLOM* (e), SVINET** (f), MOSES* (g), SLPA* (h), and DEMON* (i)}
        \end{figure}

        \begin{figure}[!ht]
        \subfigure[AMAZON*]{\includegraphics[width=.121\textwidth]{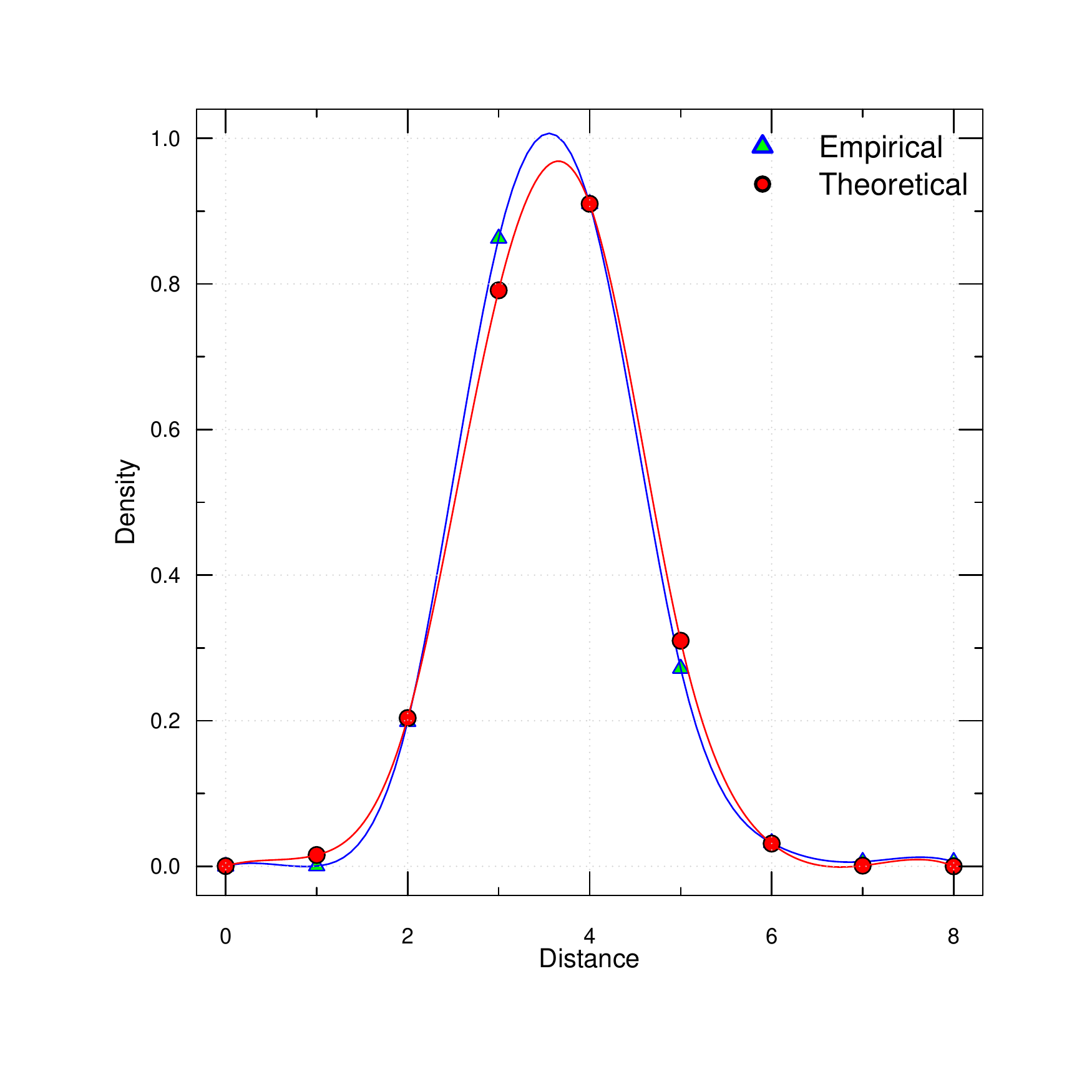}}
        \subfigure[CFINDER*]{\includegraphics[width=.121\textwidth]{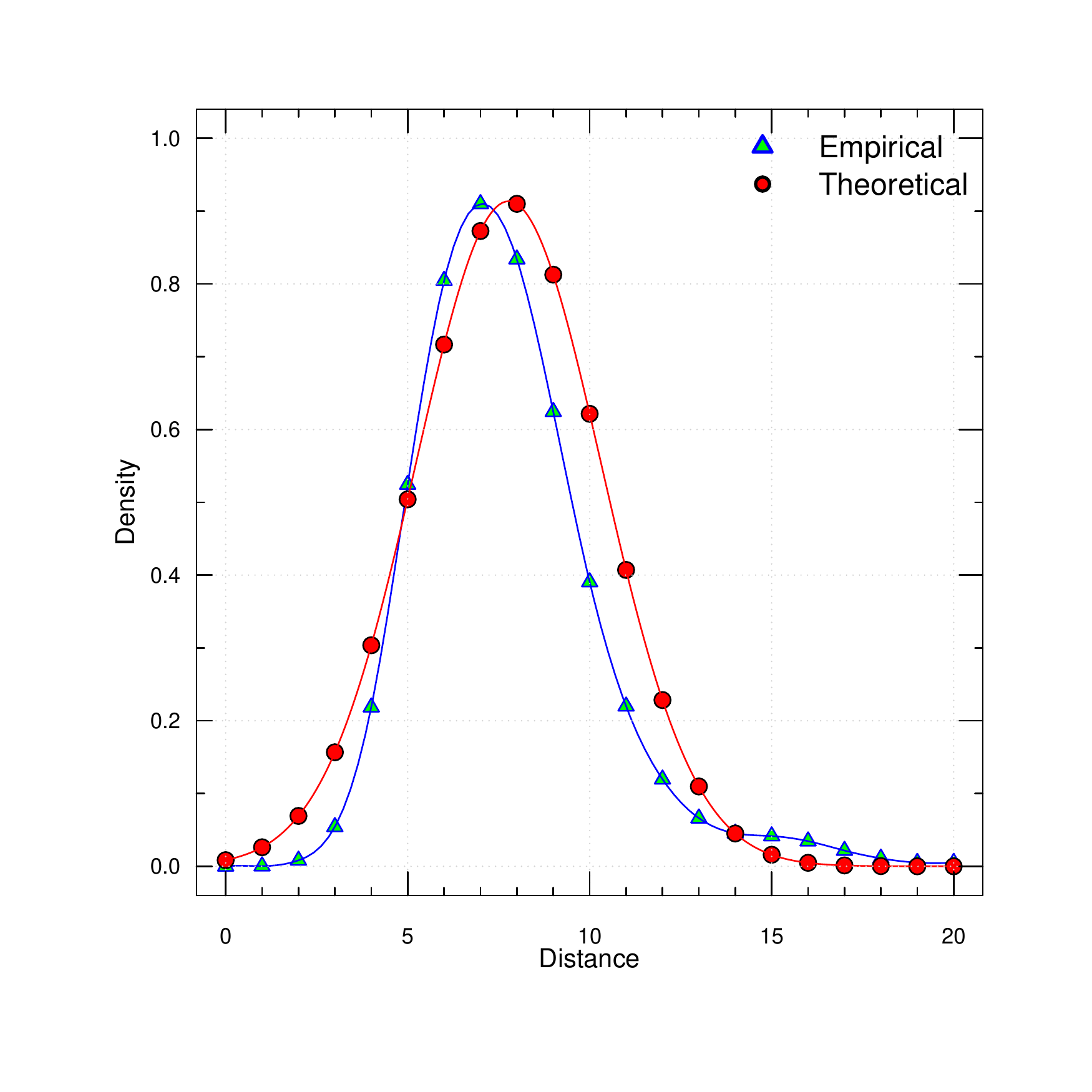}}
        \subfigure[LFM*]{\includegraphics[width=.121\textwidth]{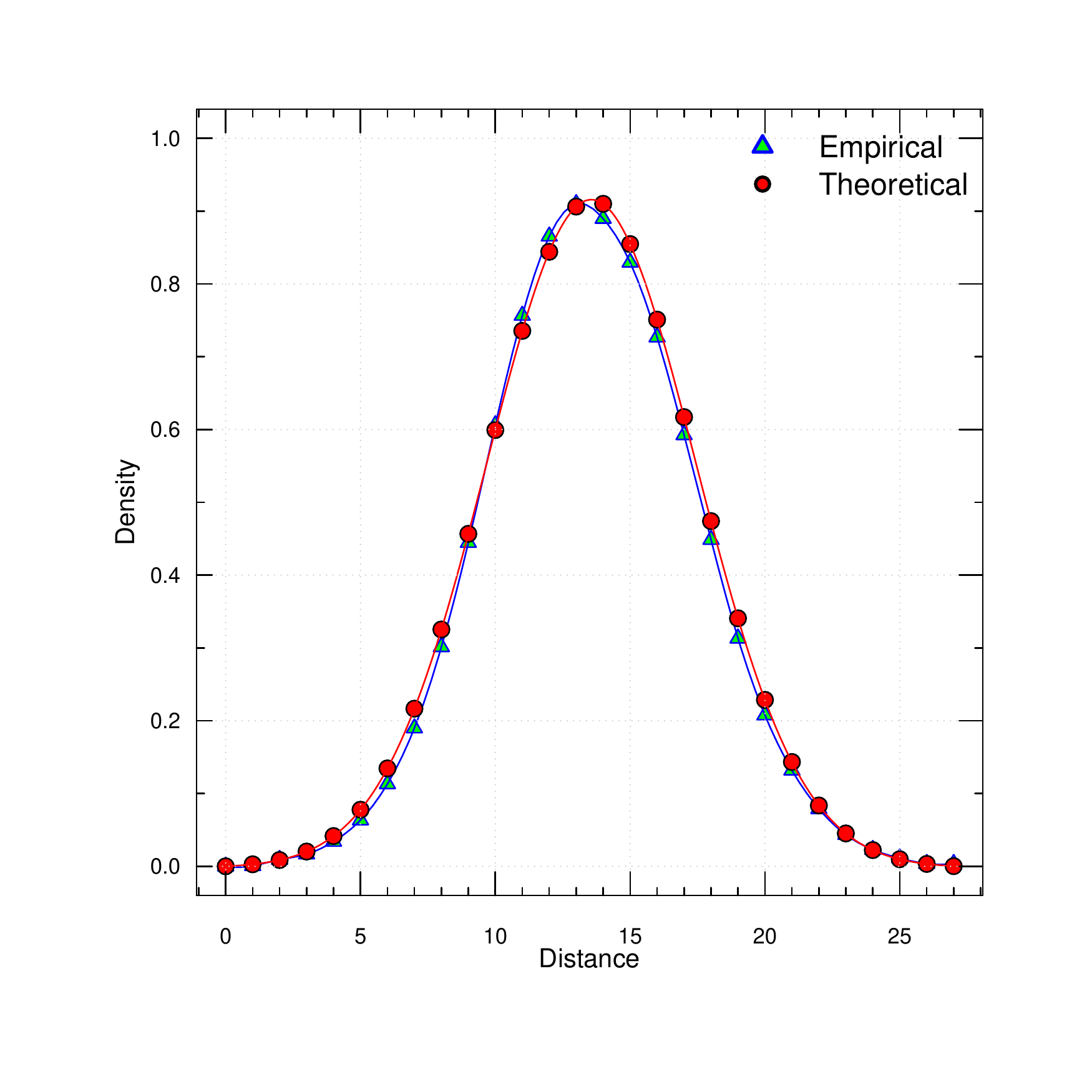}}
        \subfigure[GCE*]{\includegraphics[width=.121\textwidth]{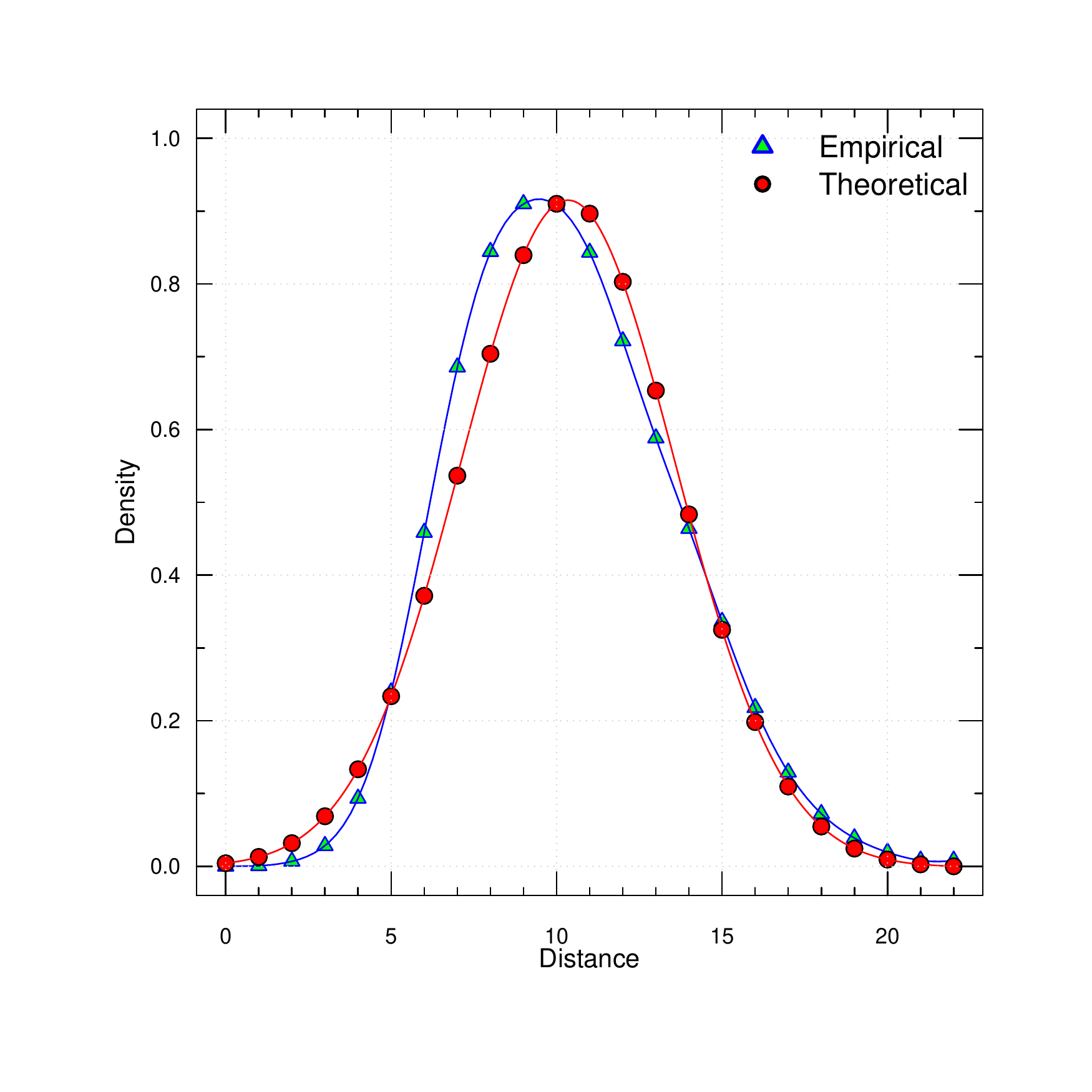}}
        \subfigure[OSLOM*]{\includegraphics[width=.121\textwidth]{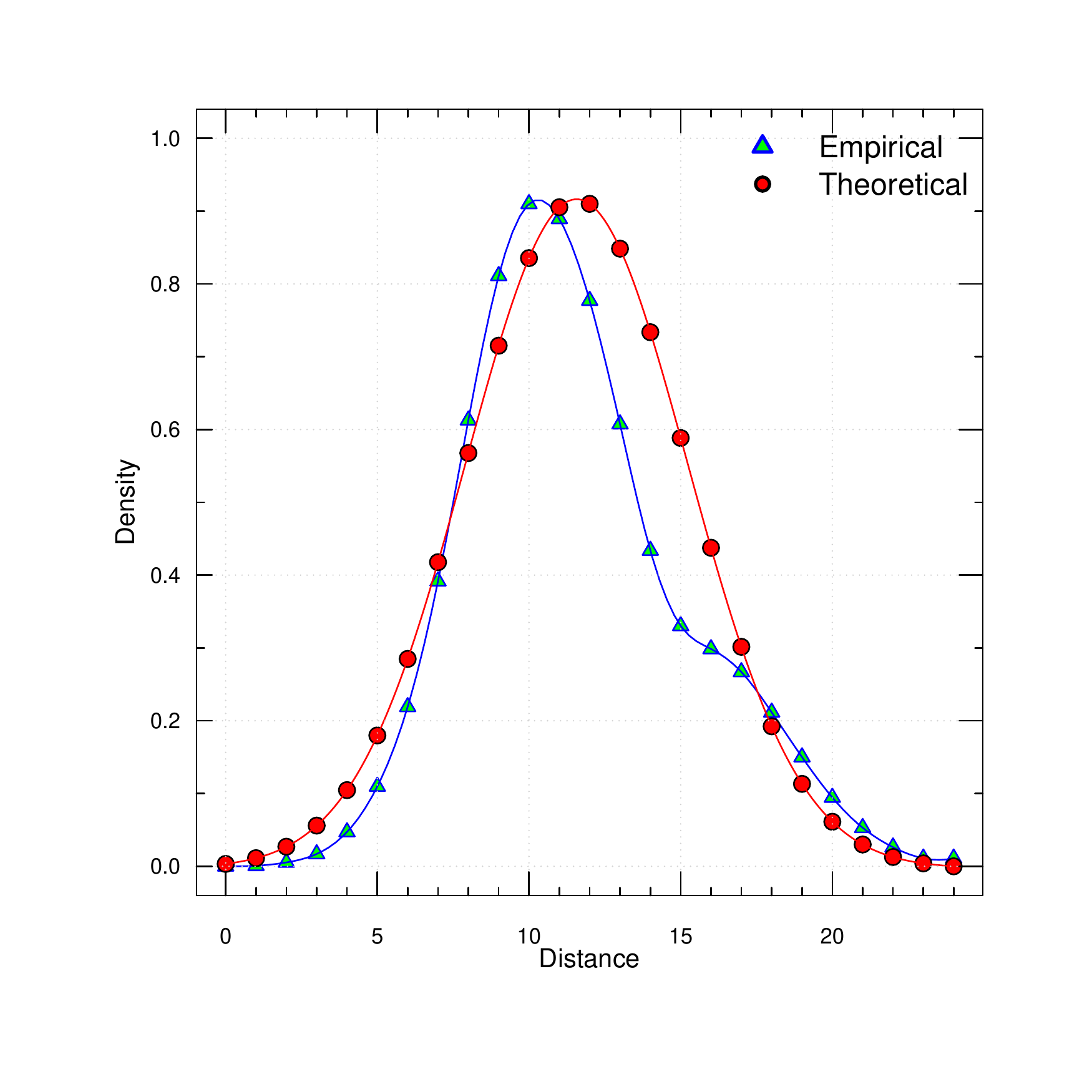}}
        \subfigure[SVINET*]{\includegraphics[width=.121\textwidth]{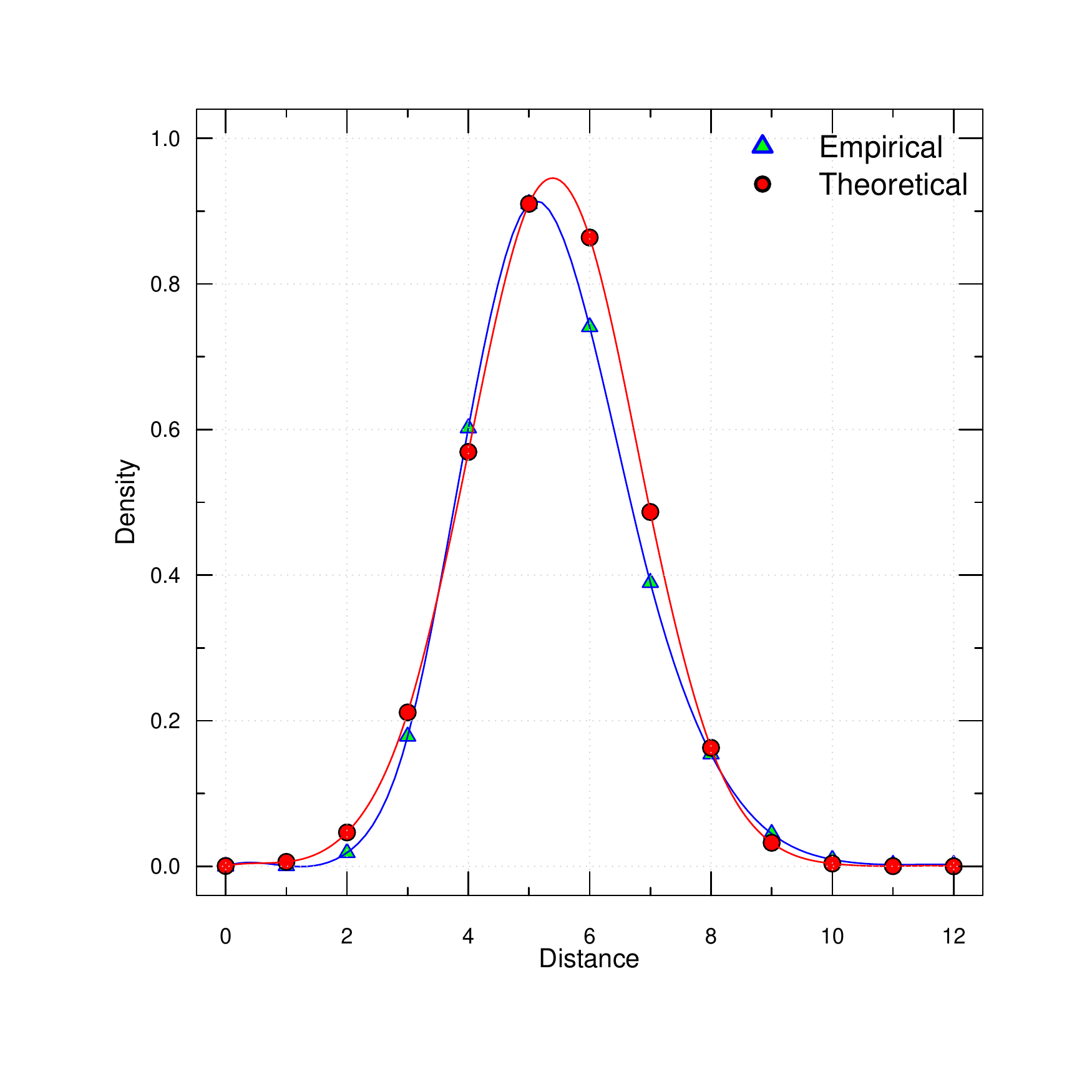}}
        \subfigure[MOSES*]{\includegraphics[width=.121\textwidth]{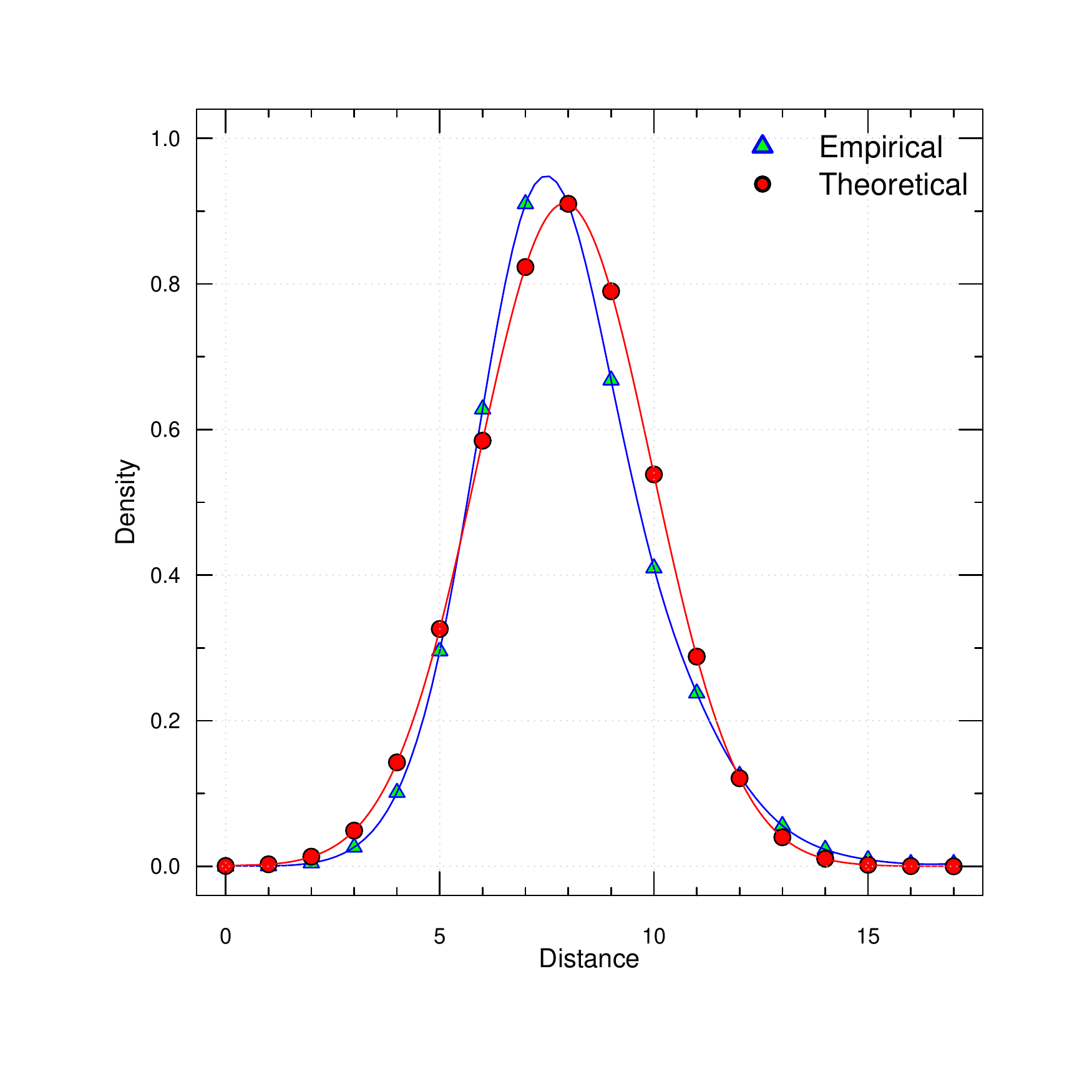}}
        \subfigure[SLPA*]{\includegraphics[width=.121\textwidth]{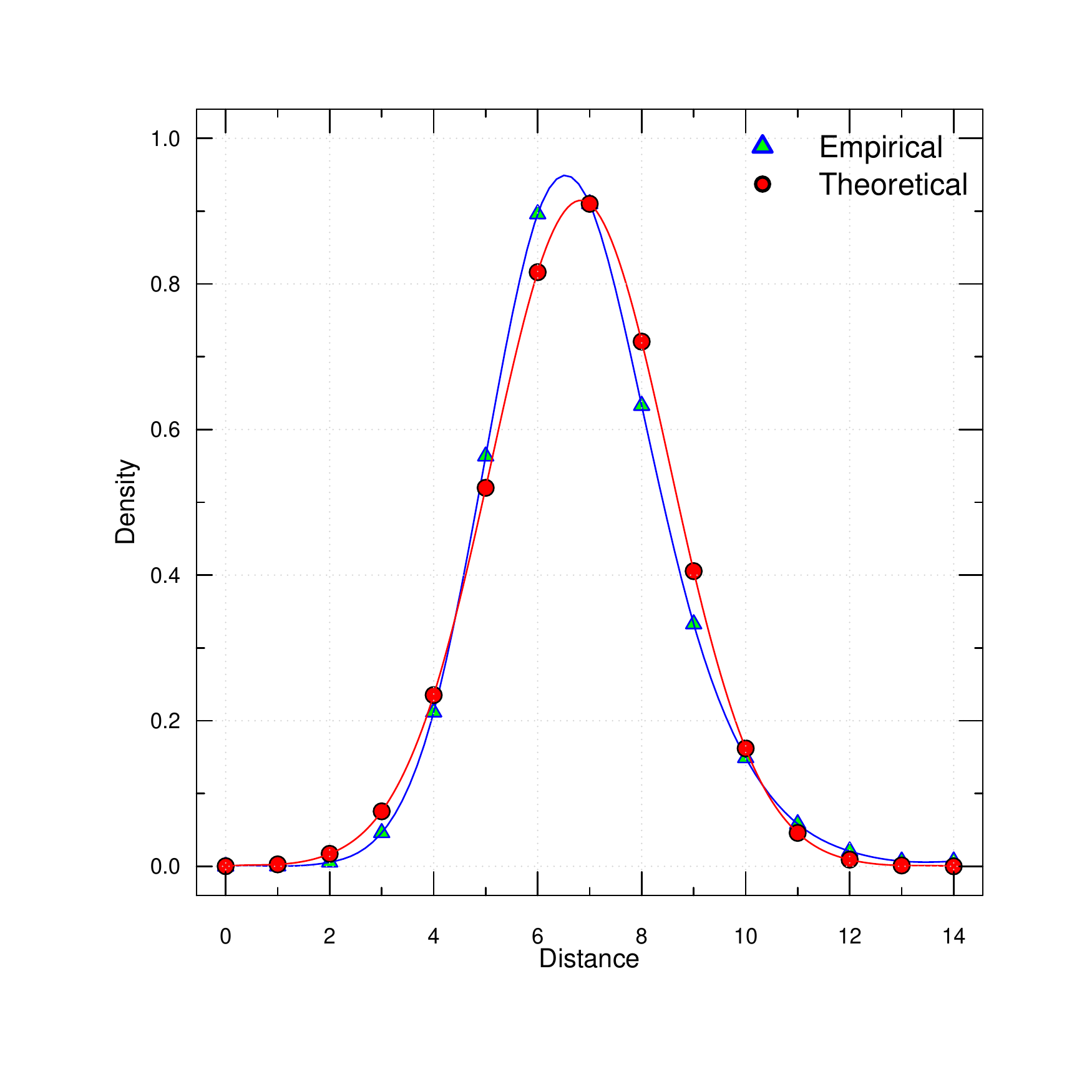}}
        \subfigure[DEMON*]{\includegraphics[width=.121\textwidth]{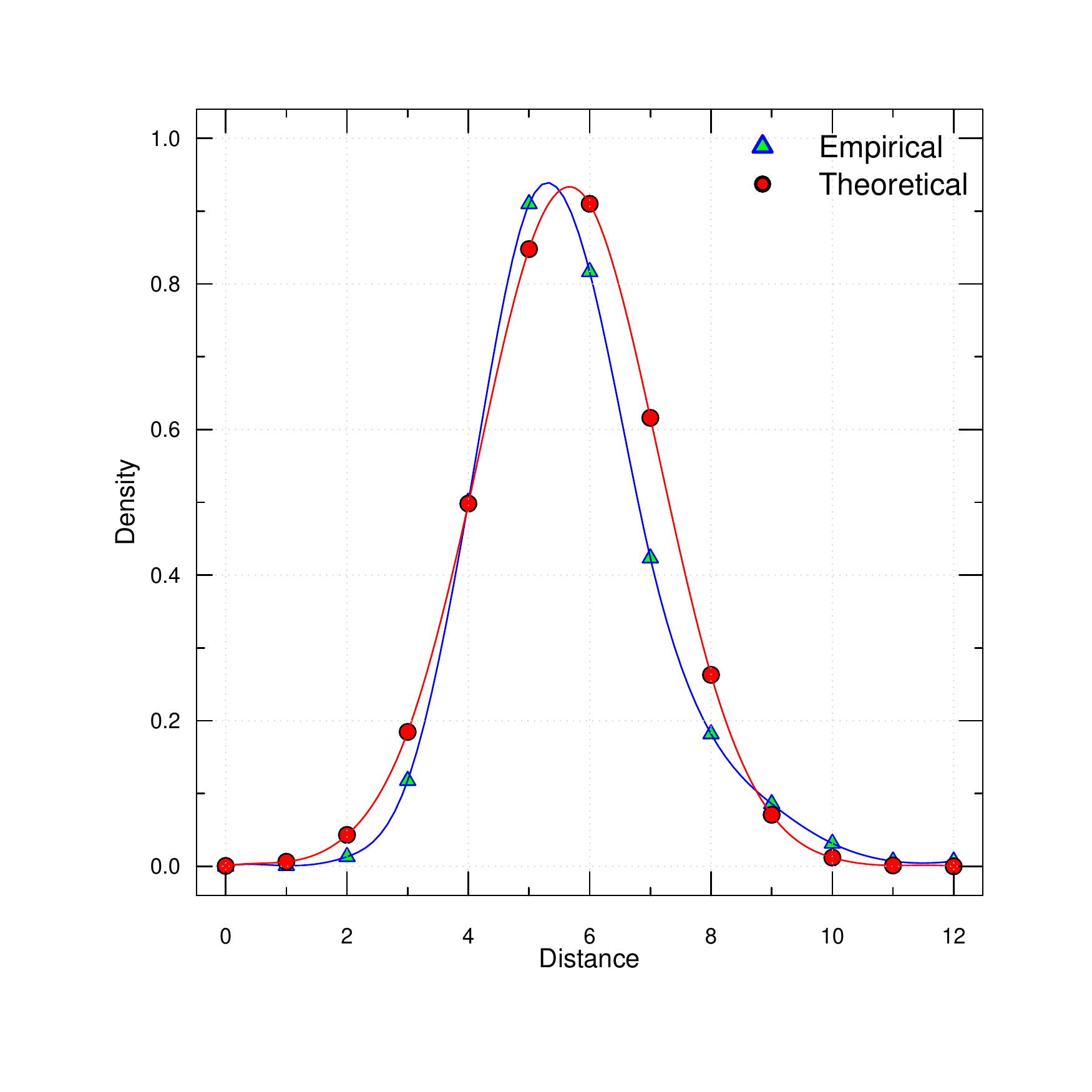}}

        \caption{\label{fig8}Empirical and estimated Hop Distance distribution for AMAZON* (a), CFINDER* (b), LFM* (c),  GCE* (d), OSLOM* (e), SVINET* (f), MOSES* (g), SLPA* (h), and DEMON* (i)}
        \end{figure}

        \begin{figure}[!ht]
        \subfigure[AMAZON*]{\includegraphics[width=.121\textwidth]{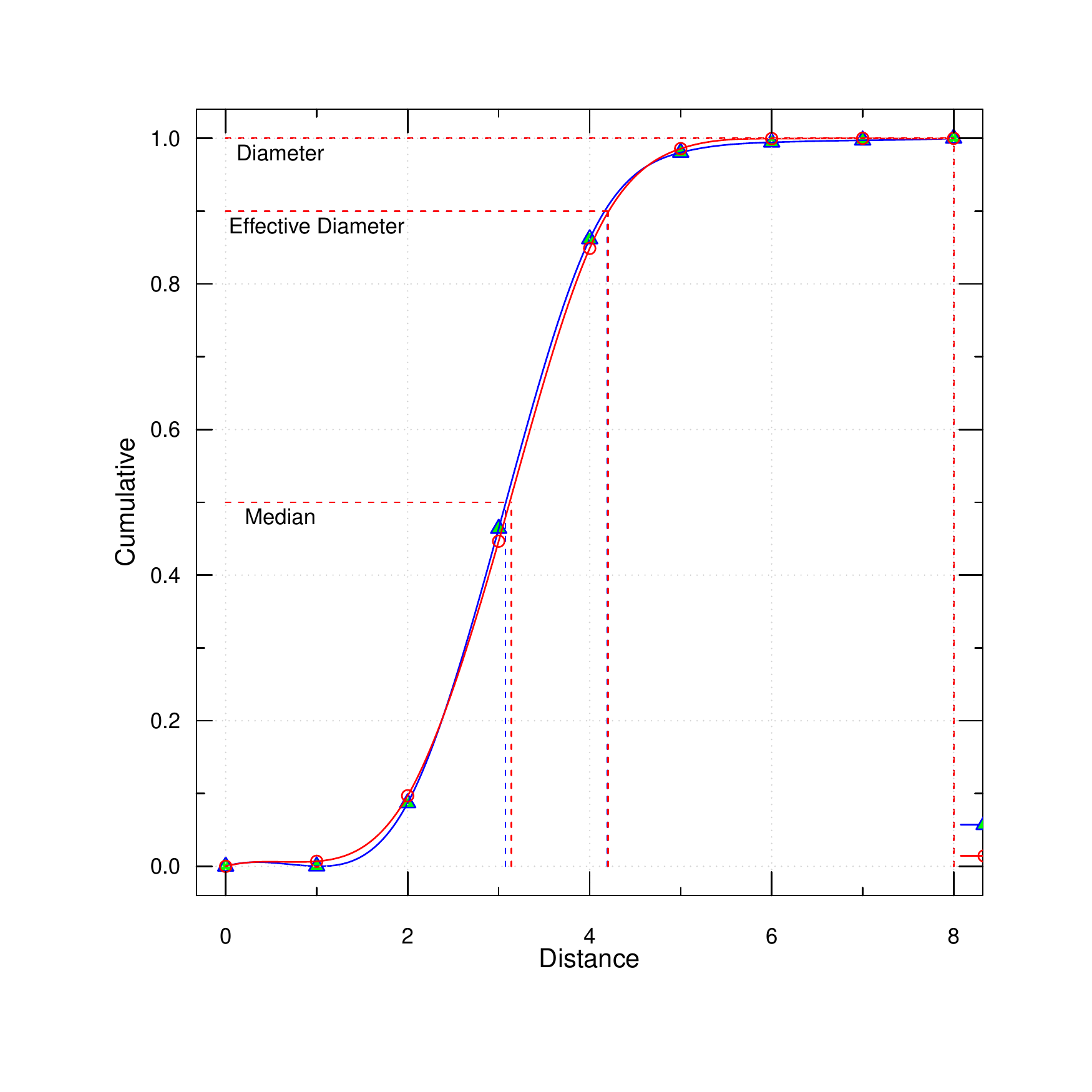}}
        \subfigure[CFINDER*]{\includegraphics[width=.121\textwidth]{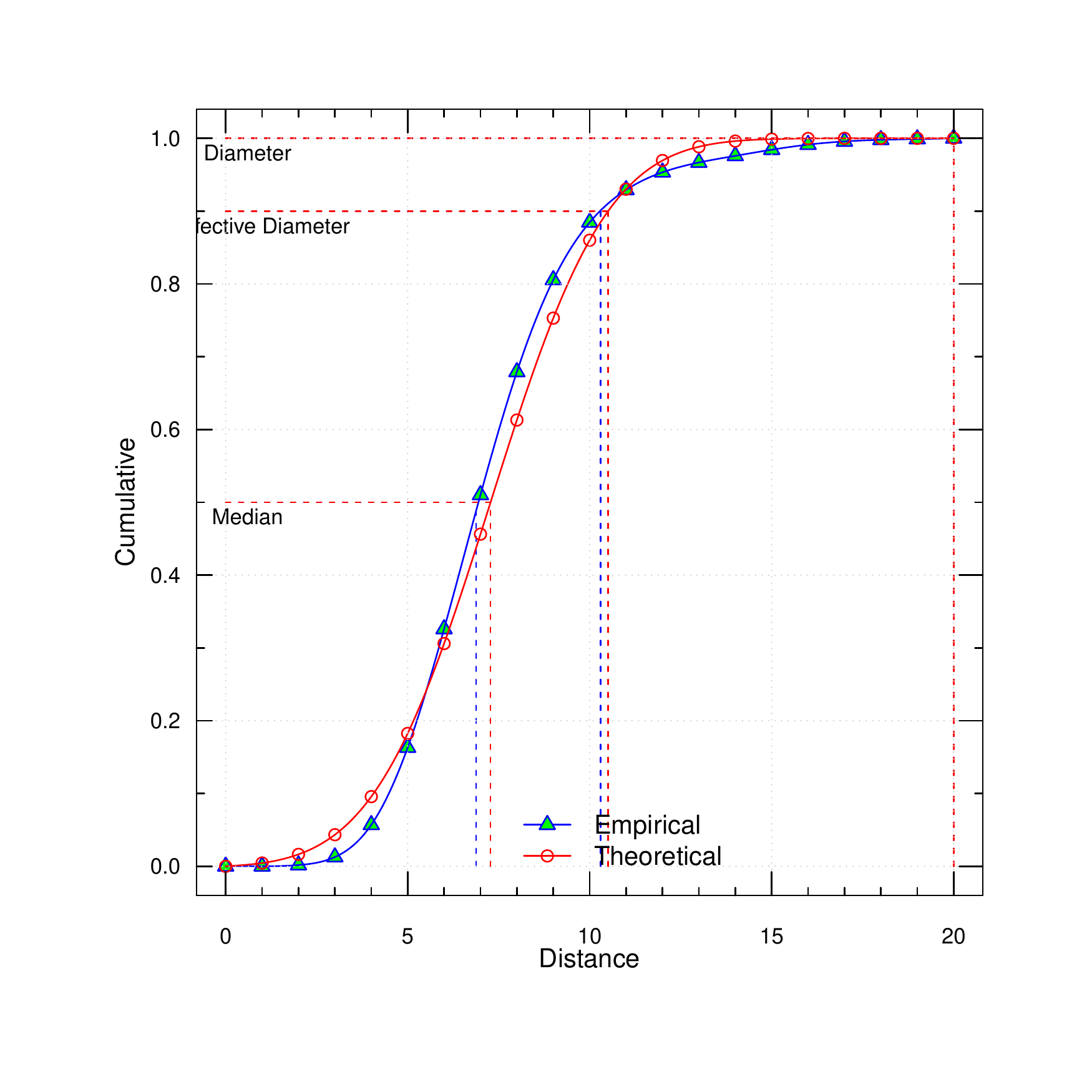}}
        \subfigure[LFM*]{\includegraphics[width=.121\textwidth]{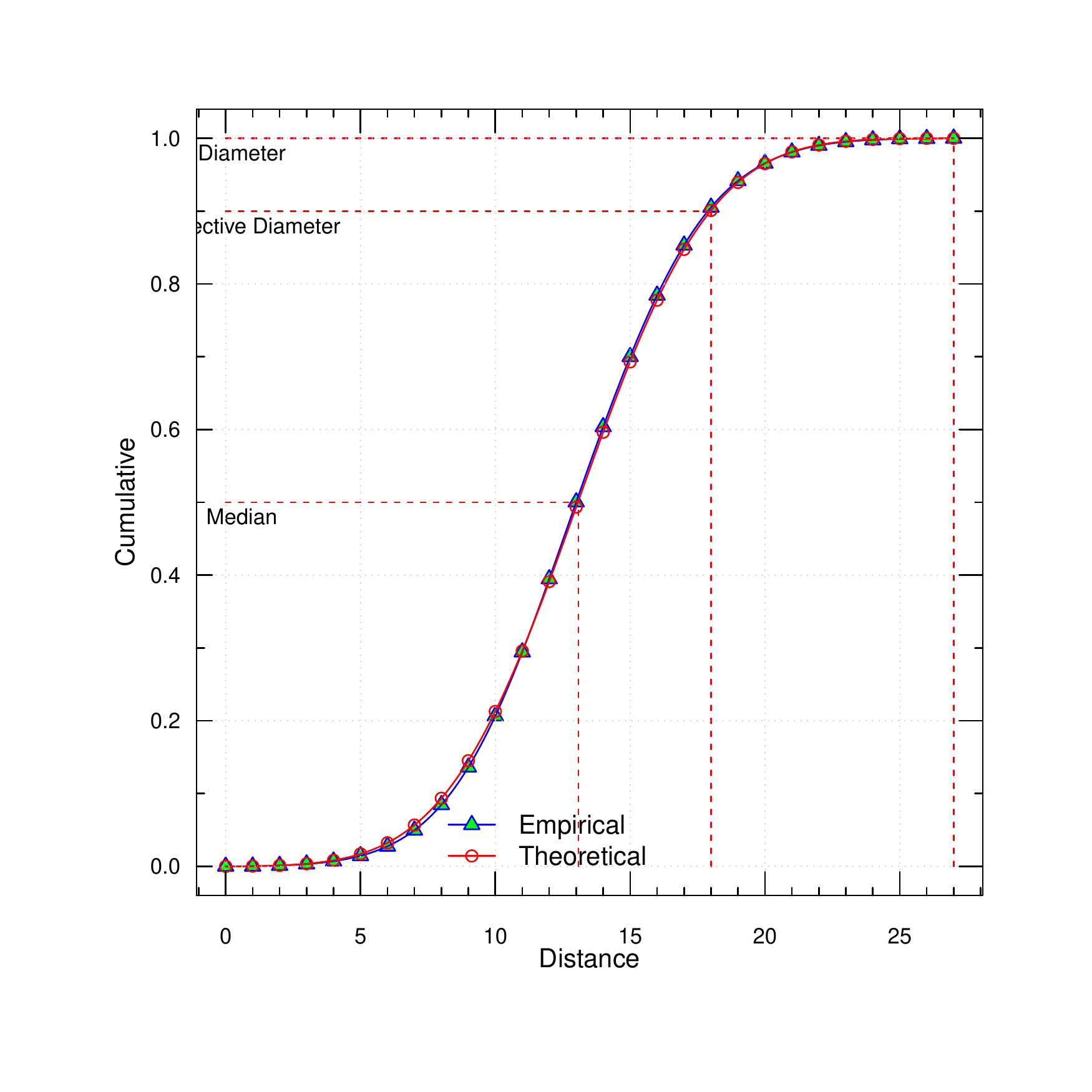}}
        \subfigure[GCE*]{\includegraphics[width=.121\textwidth]{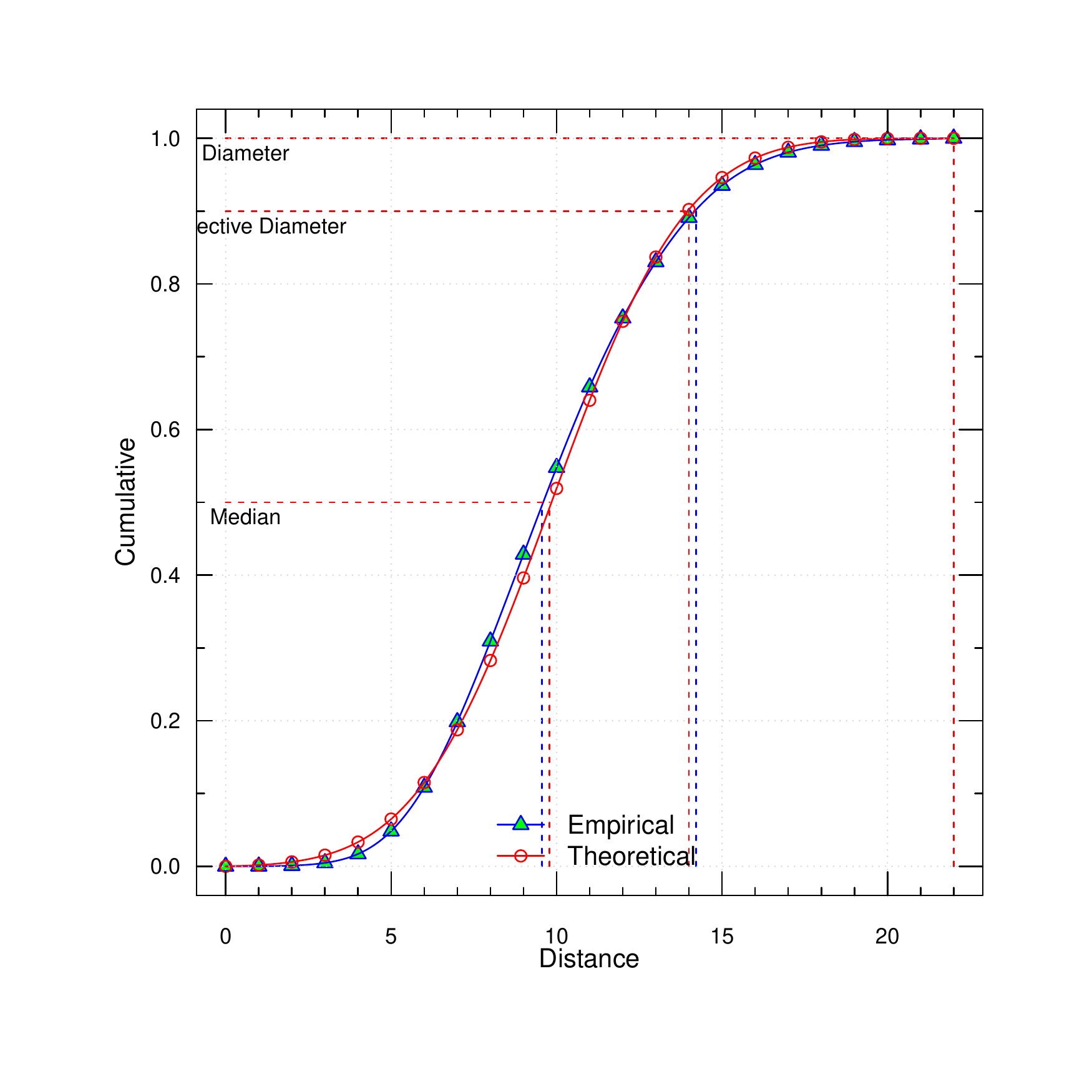}}
        \subfigure[OSLOM*]{\includegraphics[width=.121\textwidth]{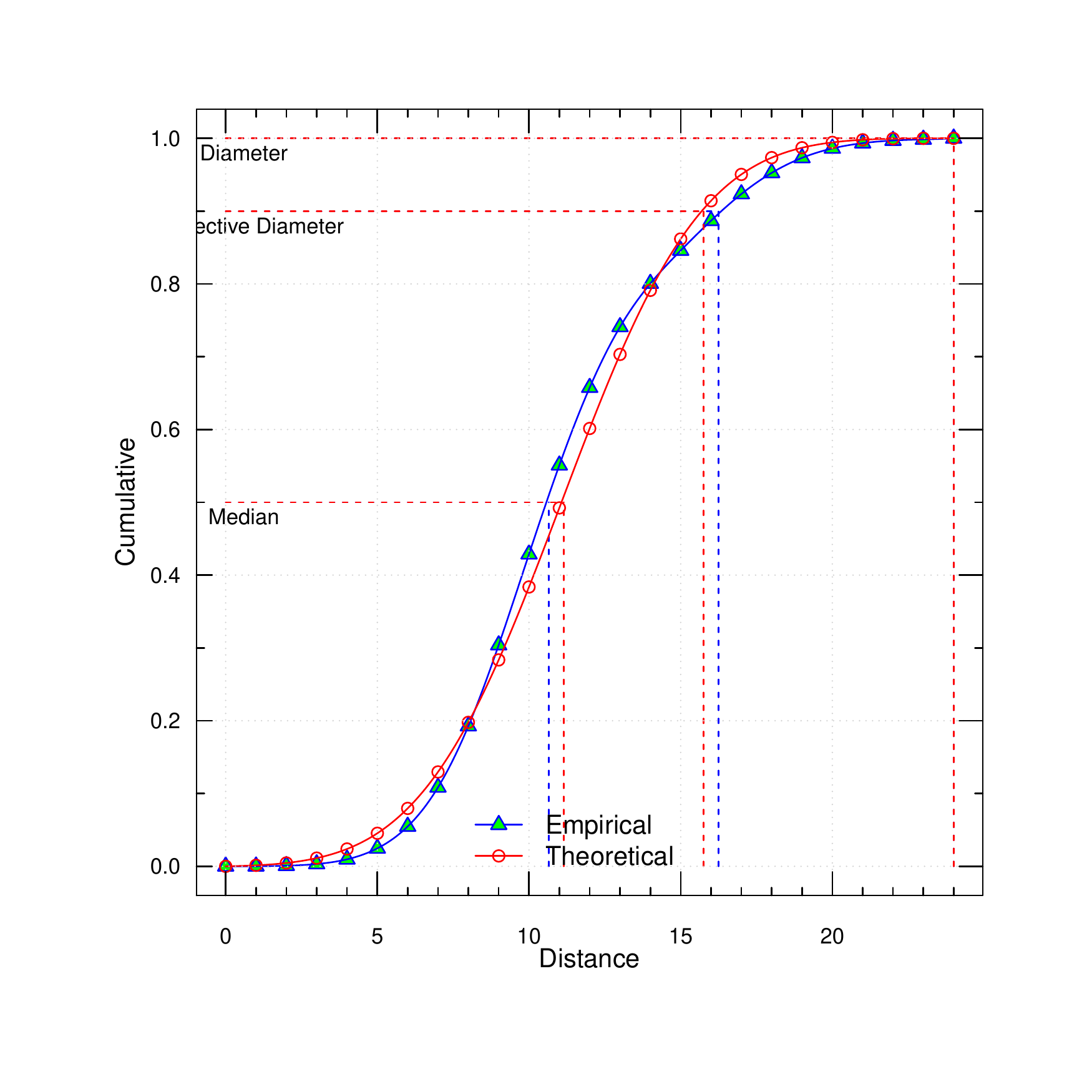}}
        \subfigure[SVINET*]{\includegraphics[width=.121\textwidth]{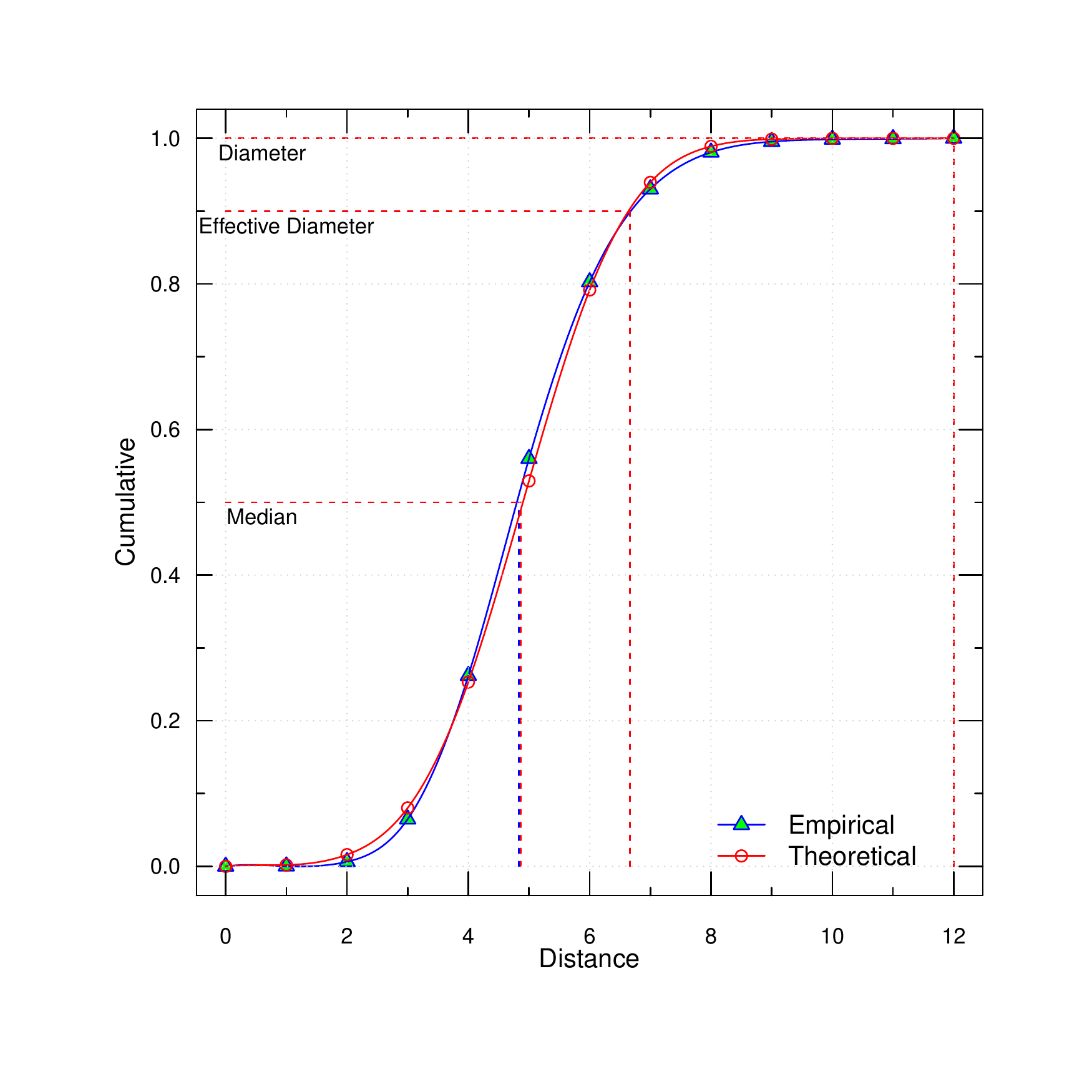}}
        \subfigure[MOSES*]{\includegraphics[width=.121\textwidth]{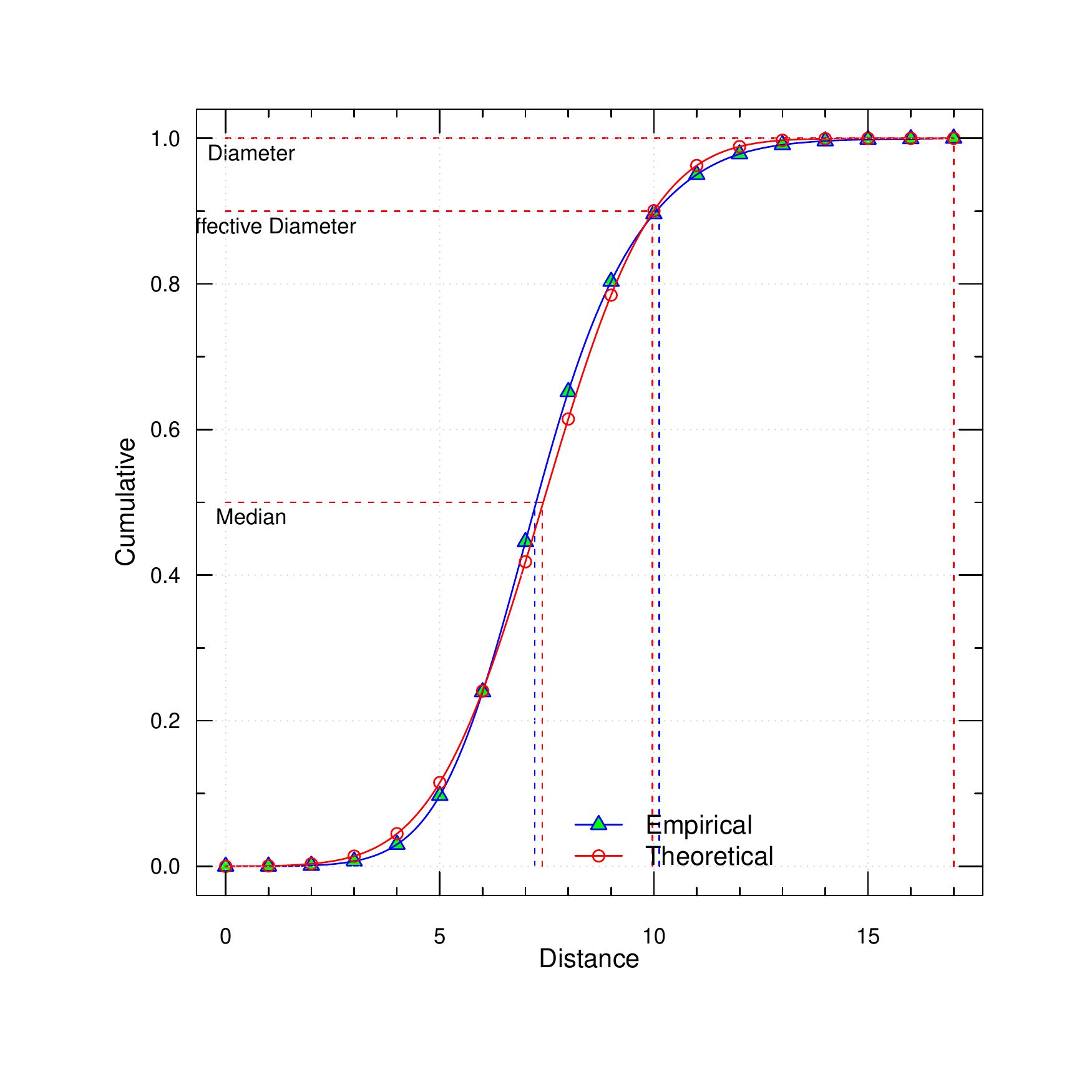}}
        \subfigure[SLPA*]{\includegraphics[width=.121\textwidth]{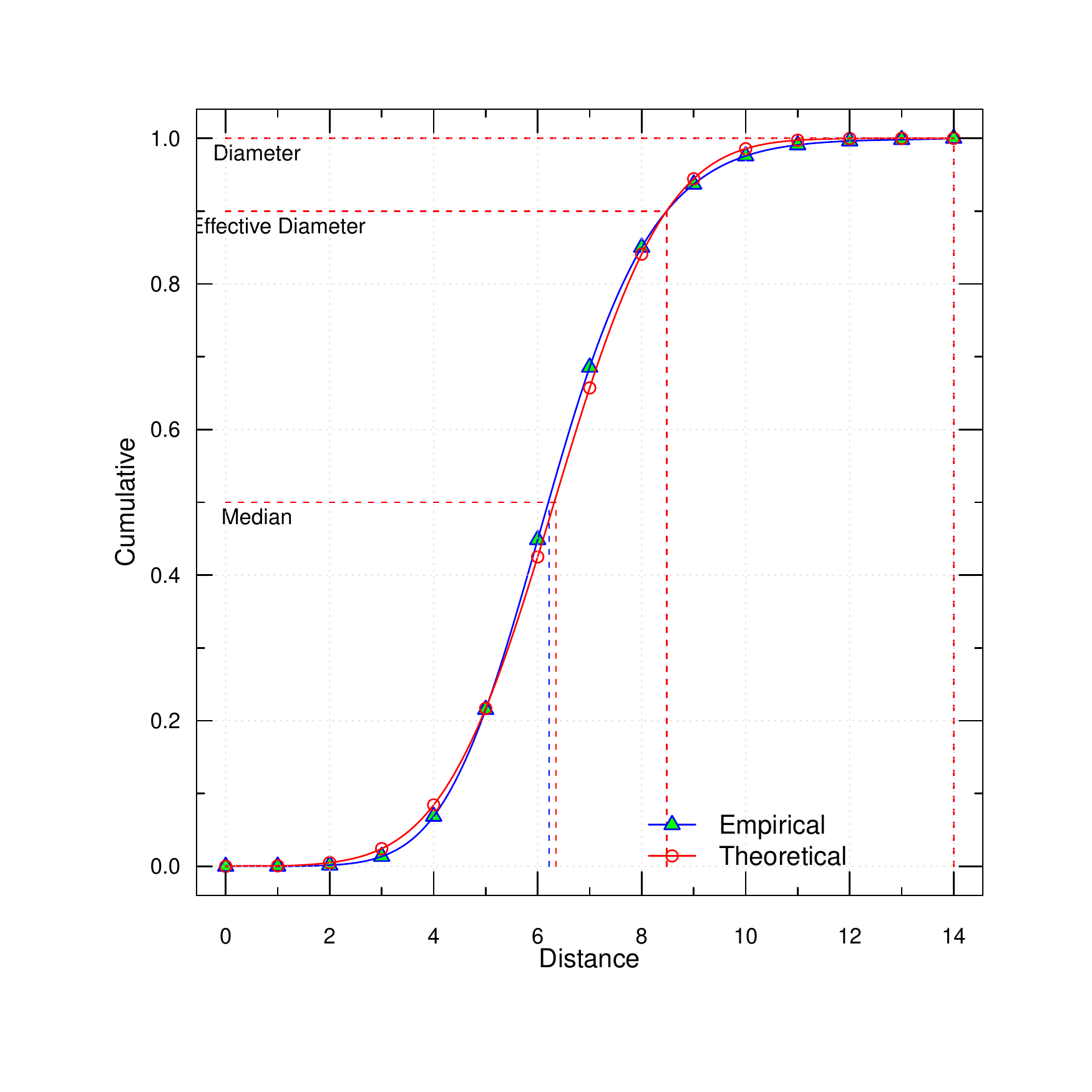}}
        \subfigure[DEMON*]{\includegraphics[width=.121\textwidth]{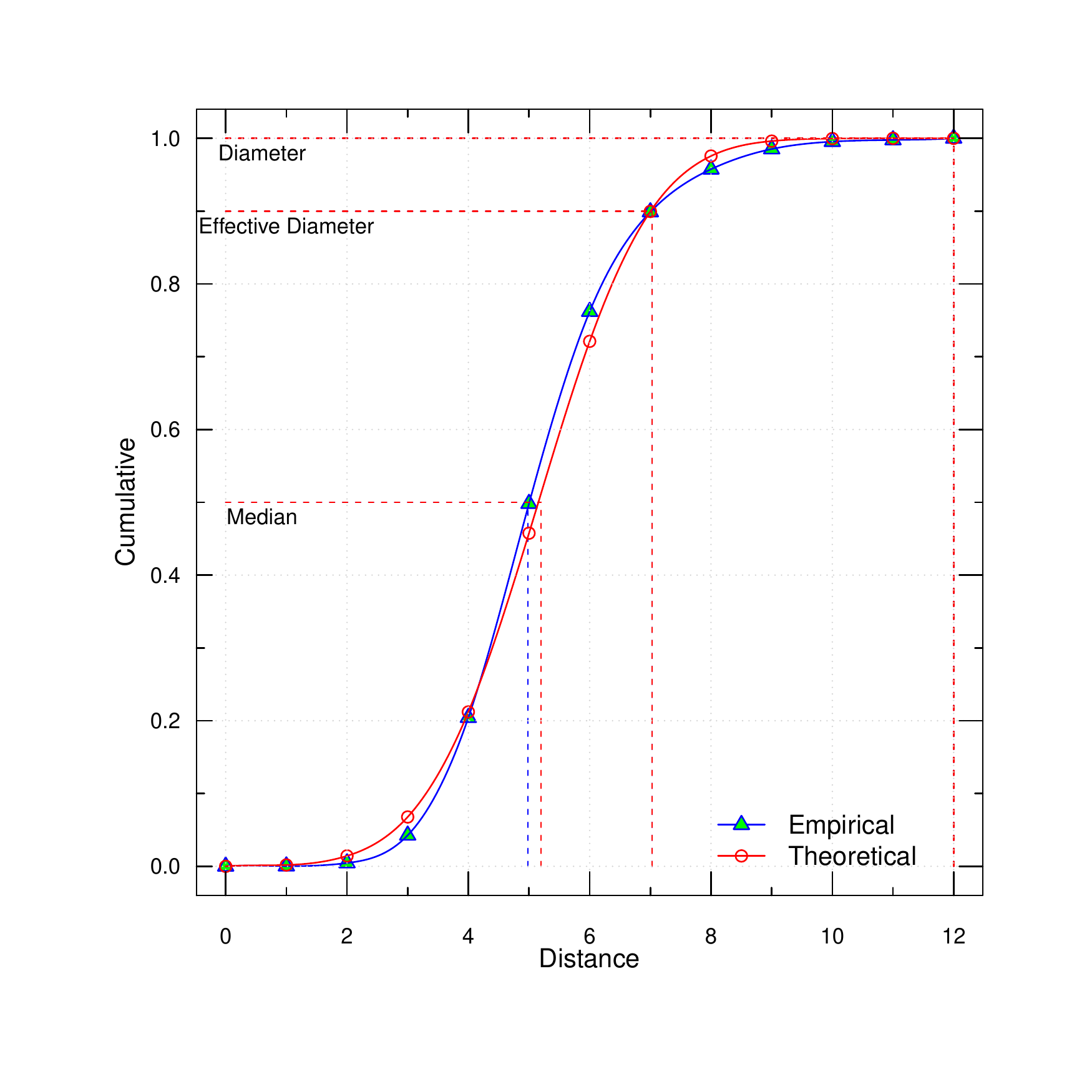}}

        \caption{\label{fig11} Empirical and estimated Hop distance cumulative distributions for AMAZON* (a), CFINDER* (b), LFM* (c),  GCE* (d), OSLOM* (e), SVINET** (f), MOSES* (g), SLPA* (h), and DEMON* (i)}
        \end{figure}

        \begin{figure}[!ht]
        \subfigure[Ground-truth]{\includegraphics[width=.121\textwidth]{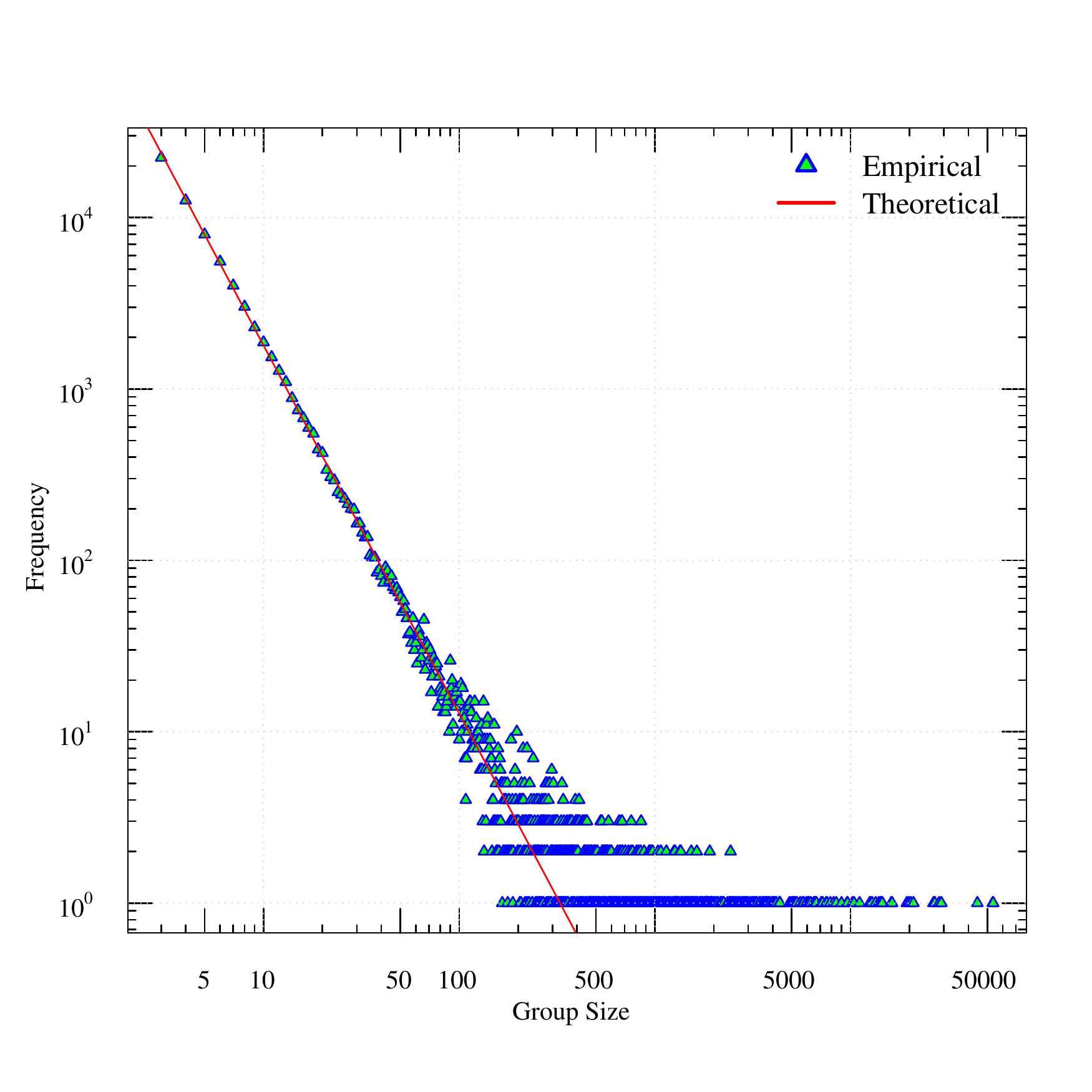}}
        \subfigure[CFINDER]{\includegraphics[width=.121\textwidth]{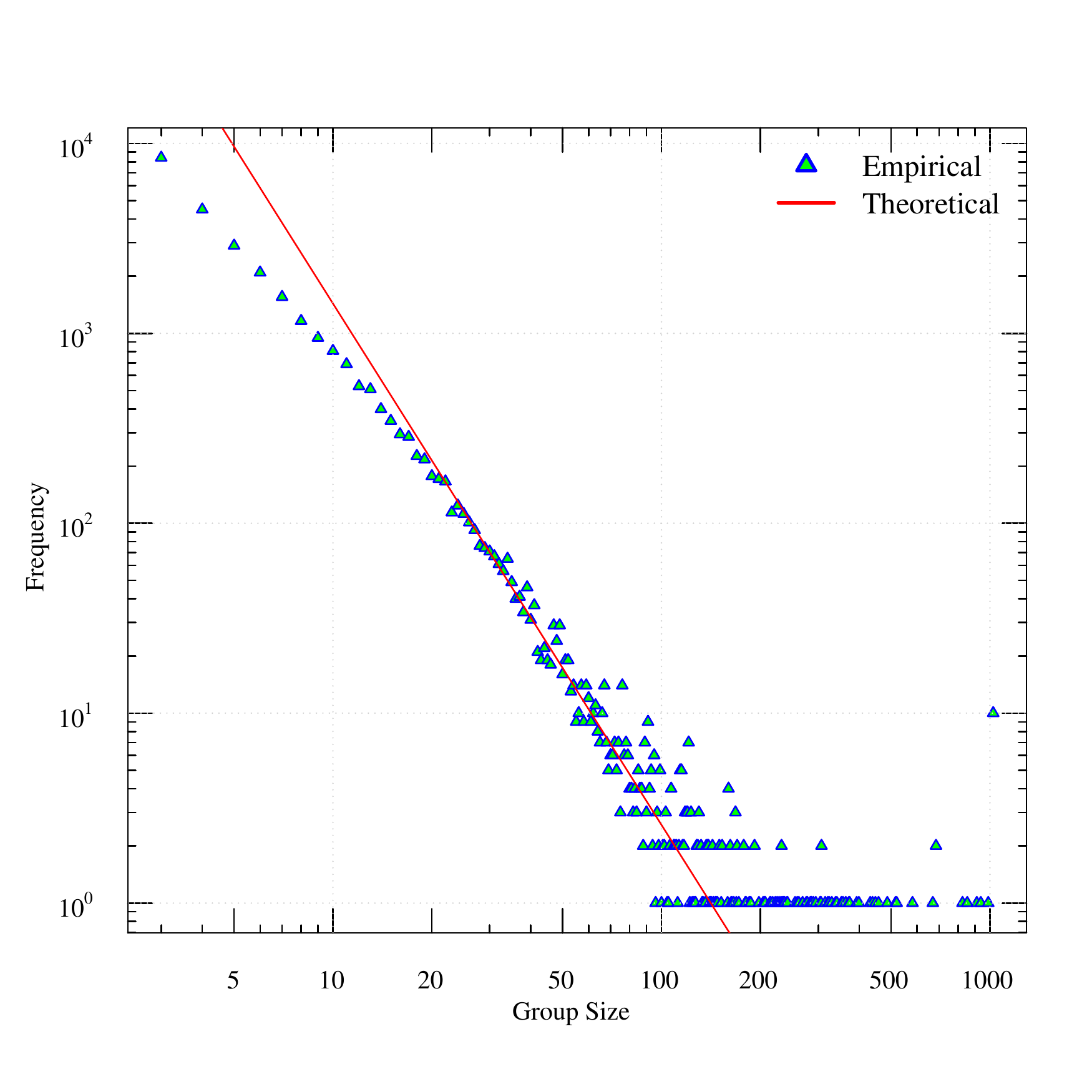}}
        \subfigure[LFM]{\includegraphics[width=.121\textwidth]{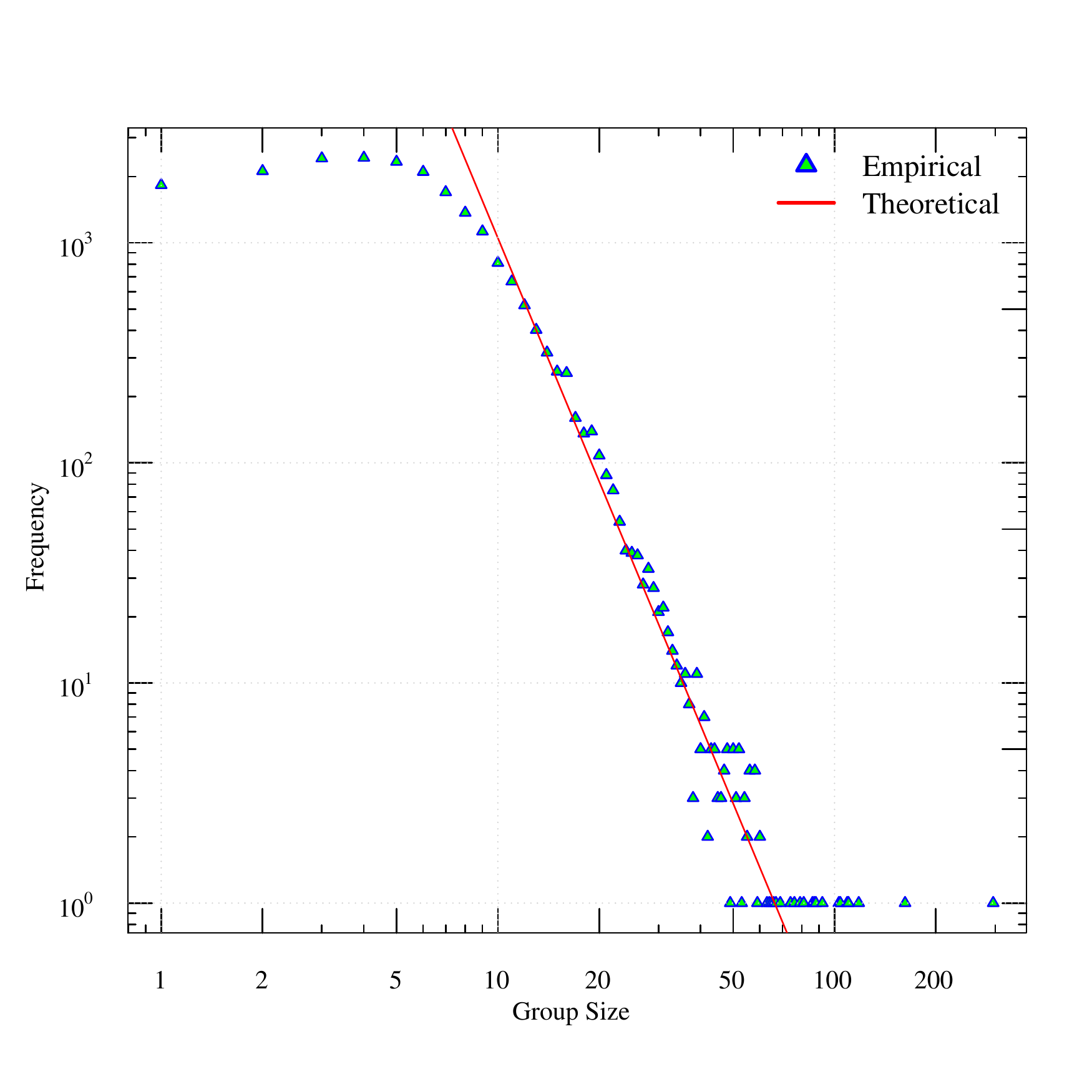}}
        \subfigure[GCE]{\includegraphics[width=.121\textwidth]{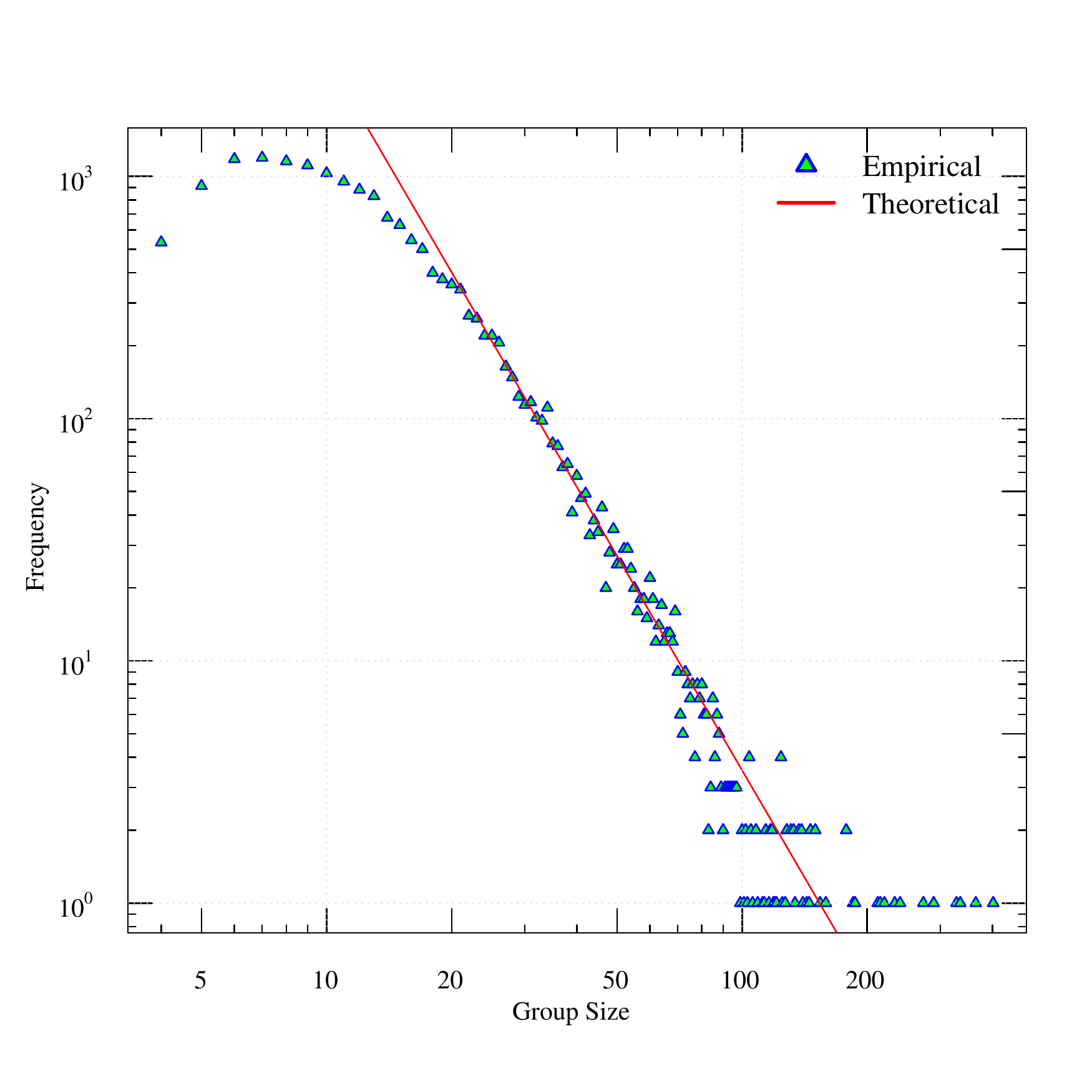}}
        \subfigure[OSLOM]{\includegraphics[width=.121\textwidth]{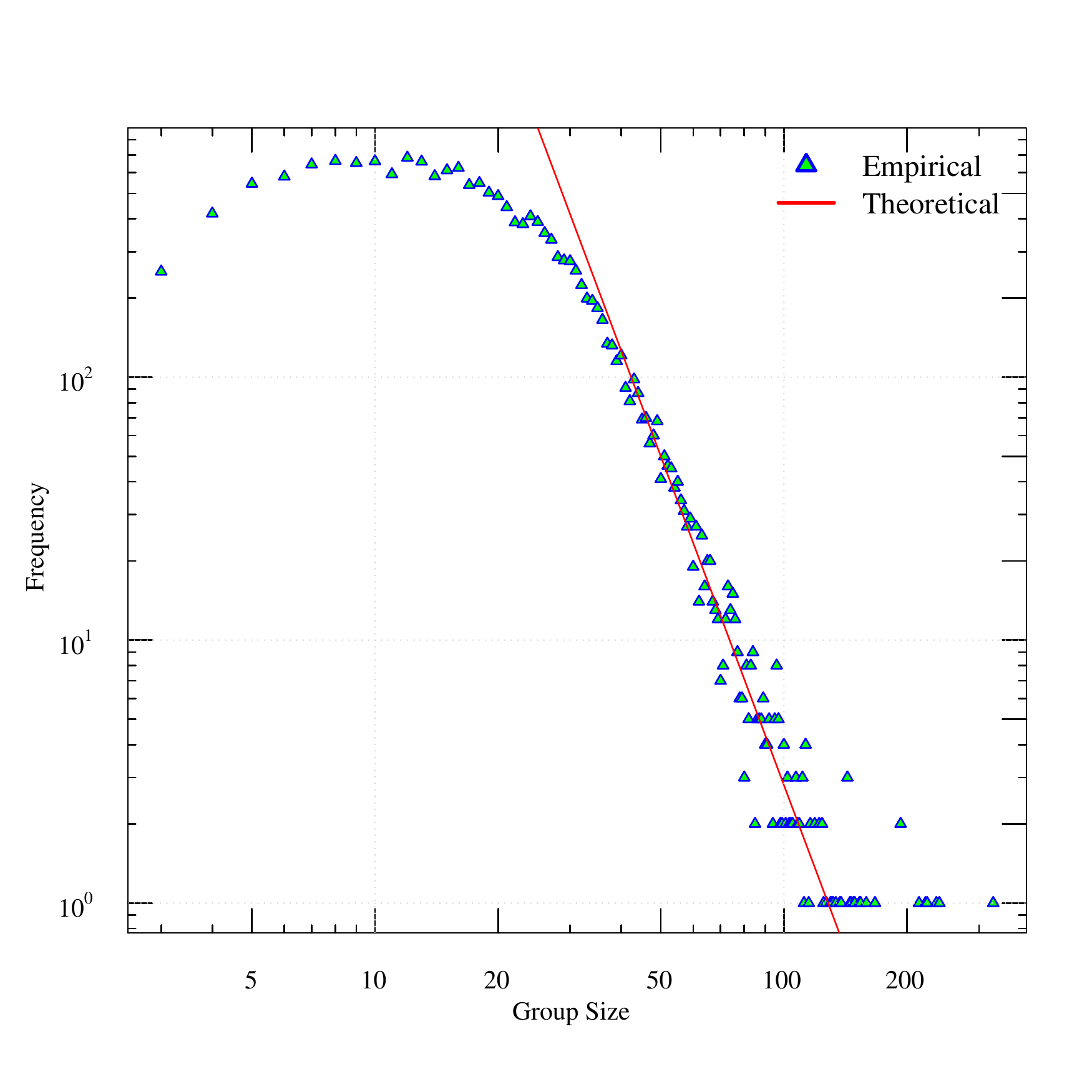}}
        \subfigure[SVINET]{\includegraphics[width=.121\textwidth]{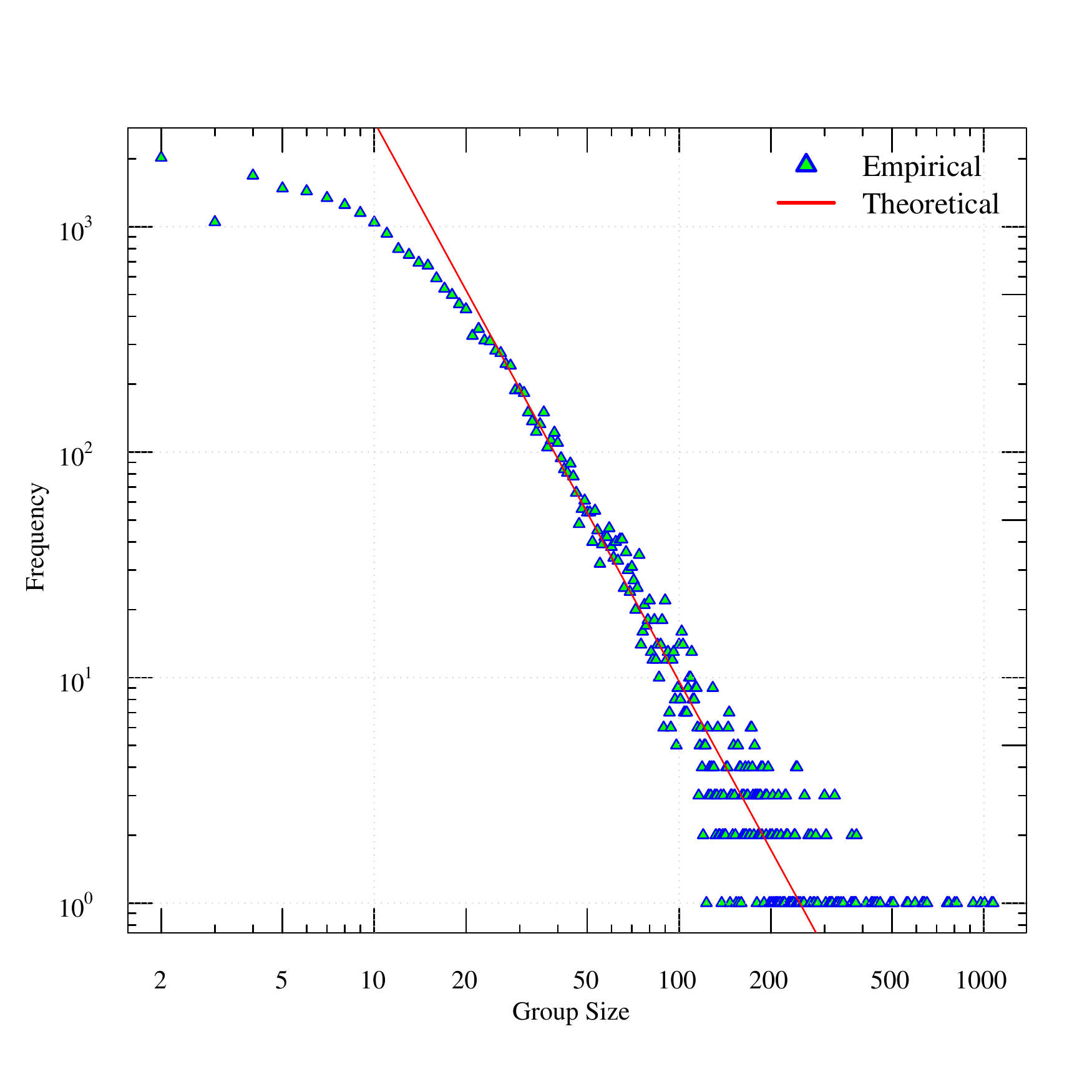}}
        \subfigure[MOSES]{\includegraphics[width=.121\textwidth]{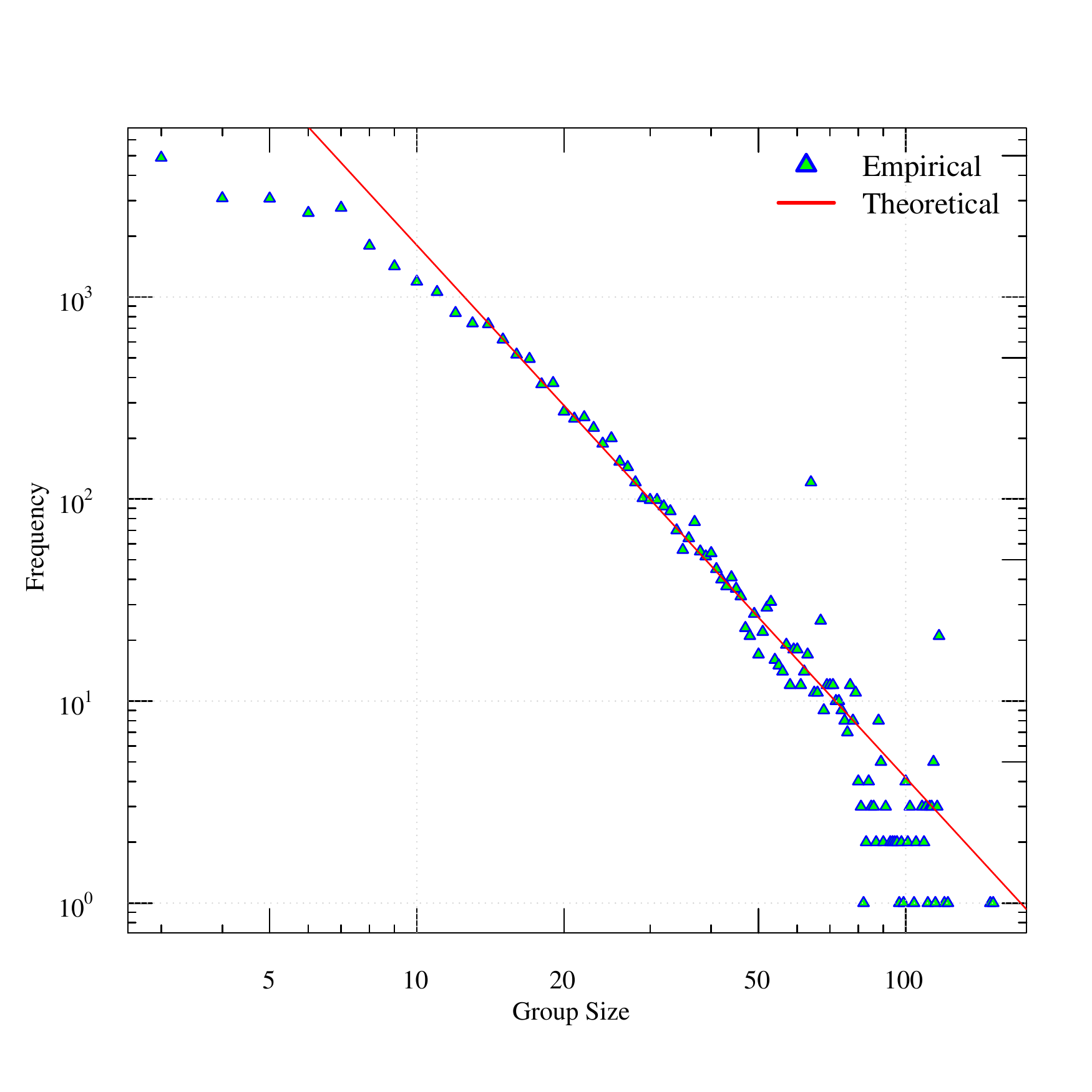}}
        \subfigure[SLPA]{\includegraphics[width=.121\textwidth]{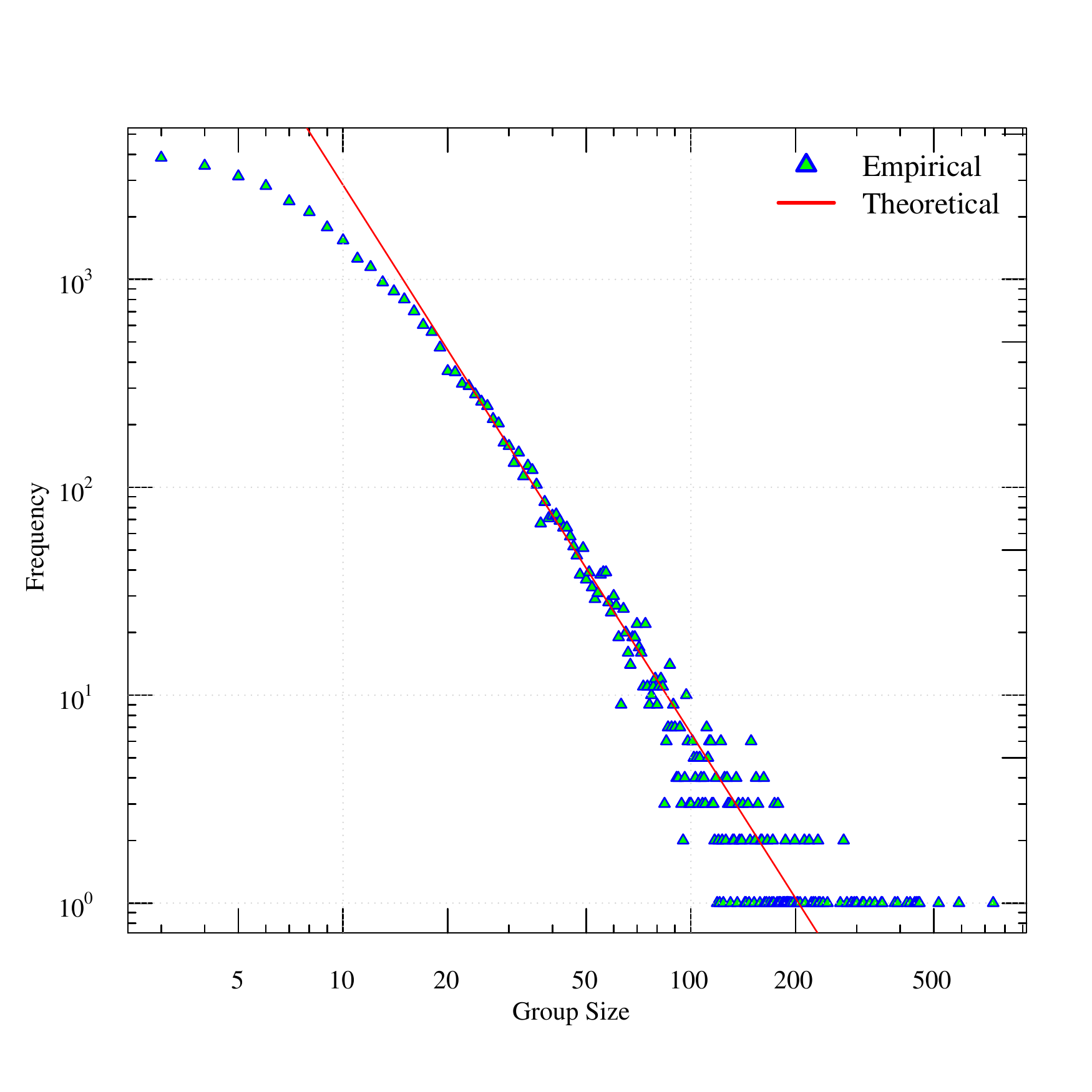}}
        \subfigure[DEMON]{\includegraphics[width=.121\textwidth]{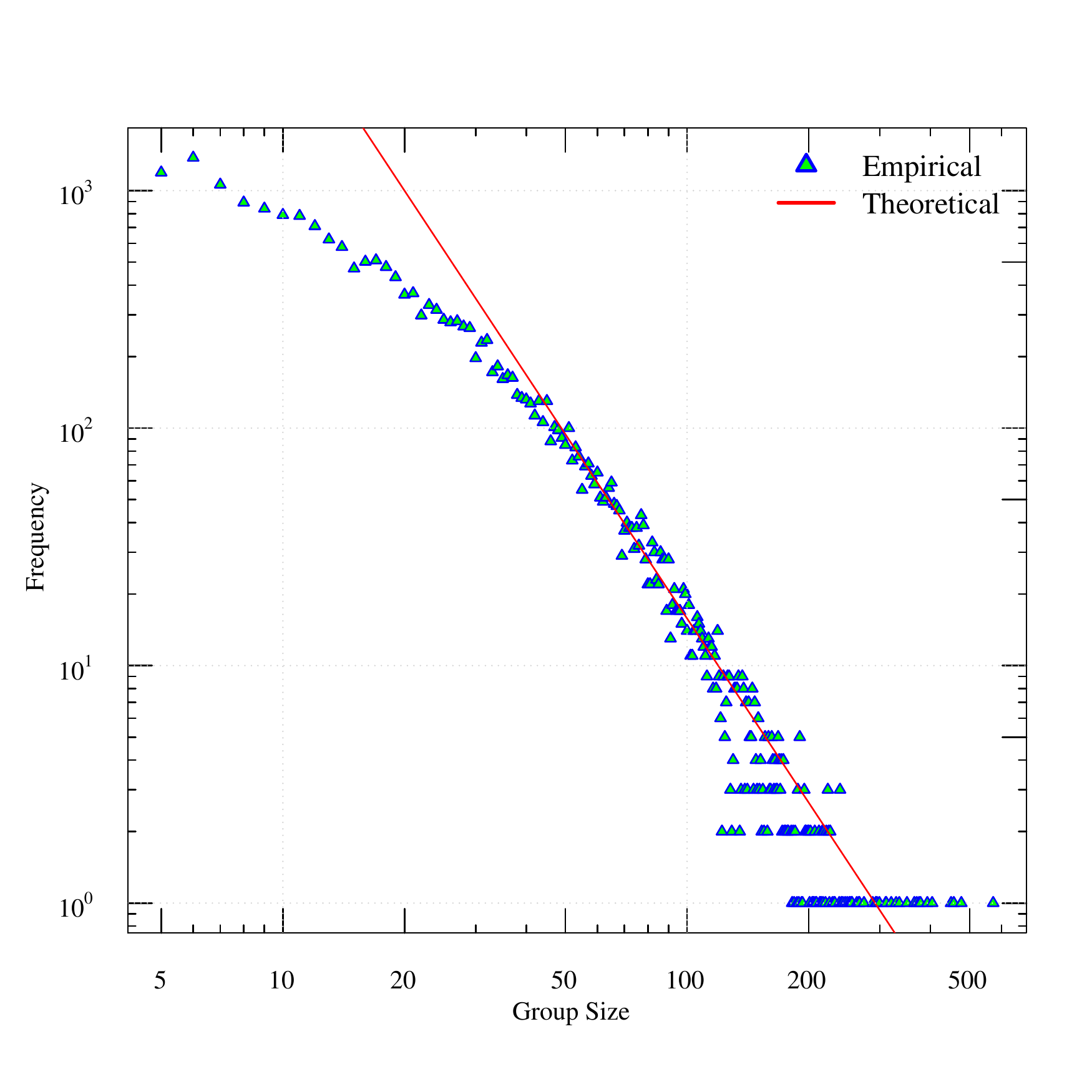}}

        \caption{\label{fig14}Log-log empirical Community size distribution (dots) and Power-Law estimate (line) of AMAZON Ground-truth (a), CFINDER (b), LFM (c),  GCE (d), OSLOM (e), SVINET (f), MOSES (g), SLPA (h), and DEMON (i)}
        \end{figure}

        \begin{figure}[!ht]
        \subfigure[Ground-truth]{\includegraphics[width=.121\textwidth]{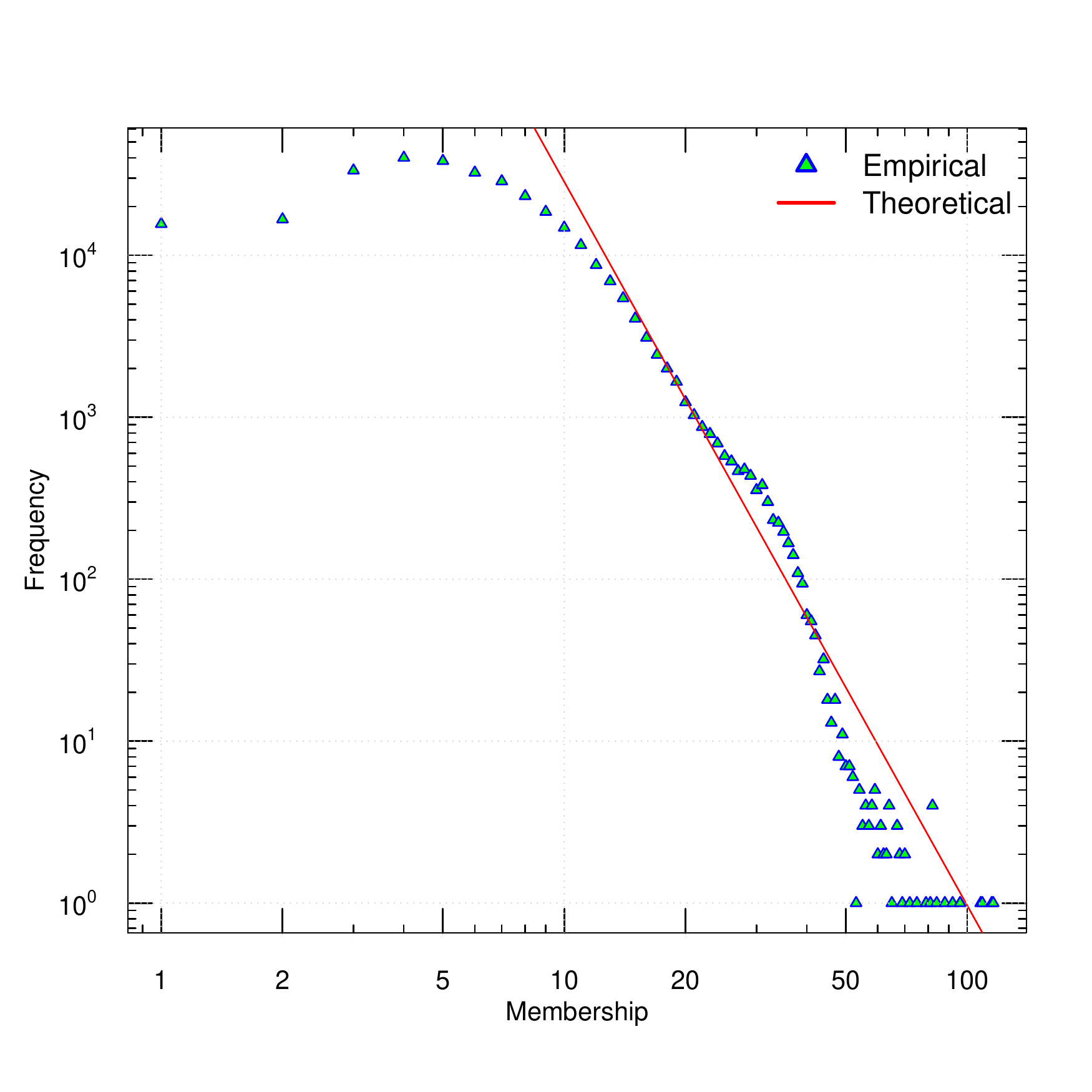}}
        \subfigure[CFINDER]{\includegraphics[width=.121\textwidth]{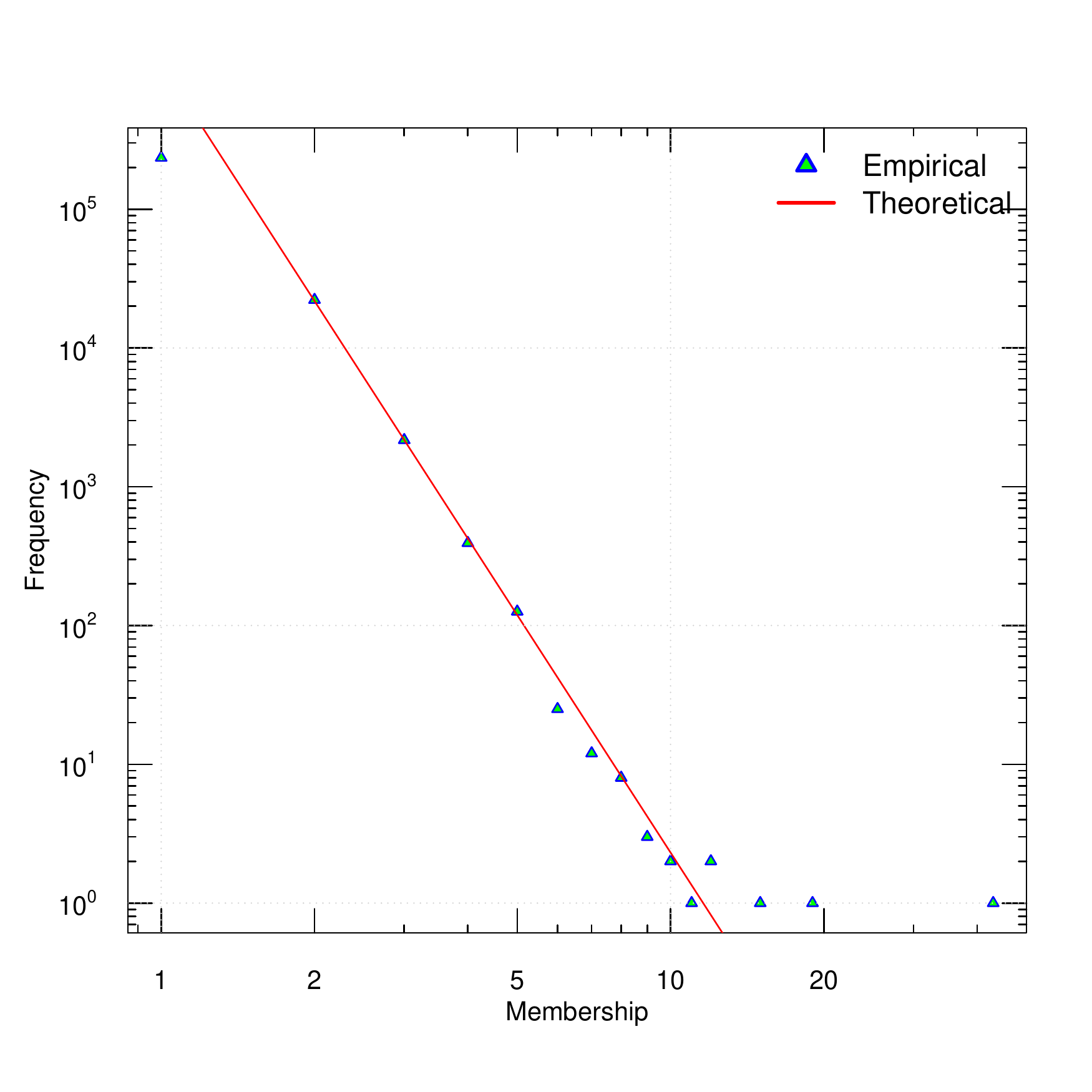}}
        \subfigure[LFM]{\includegraphics[width=.121\textwidth]{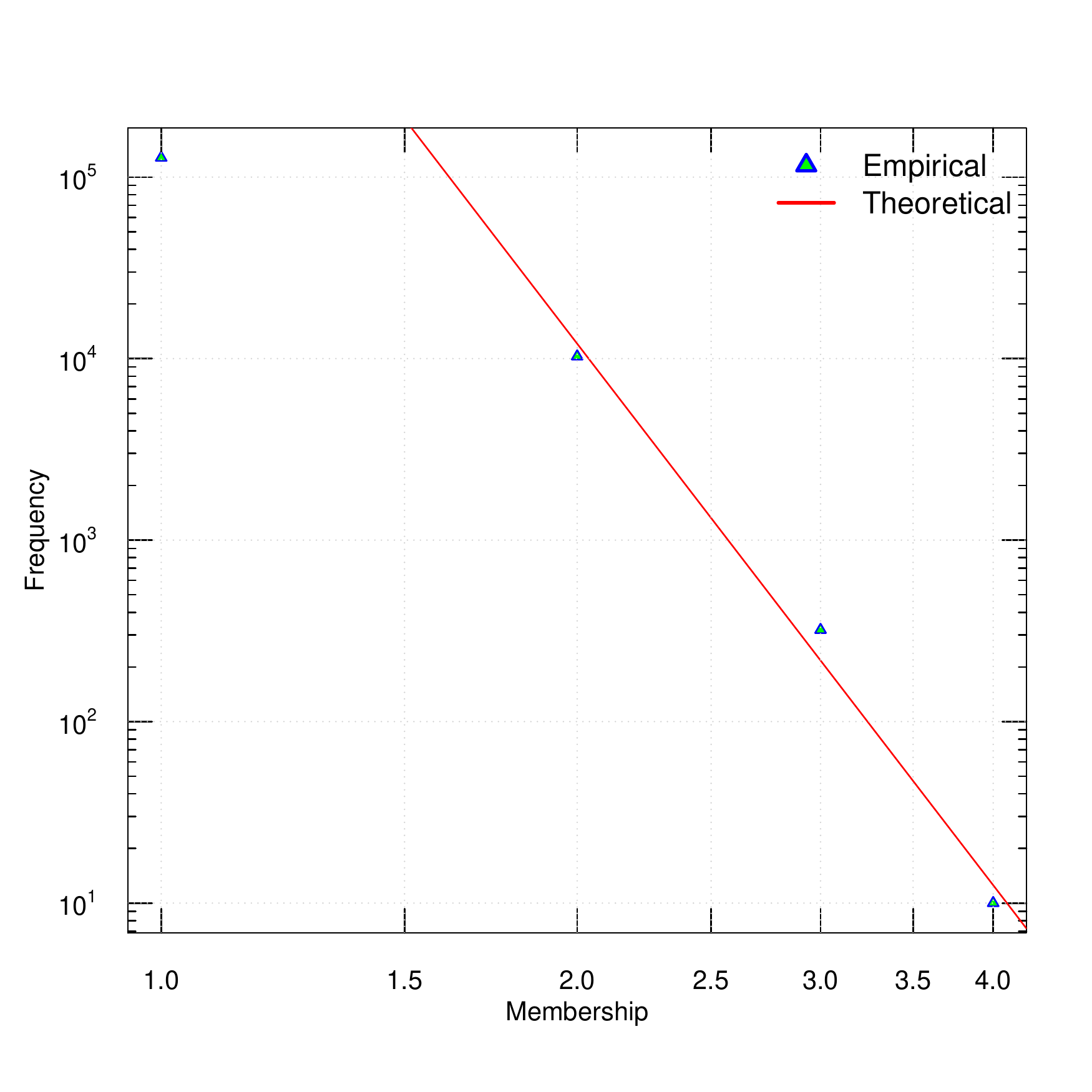}}
        \subfigure[GCE]{\includegraphics[width=.121\textwidth]{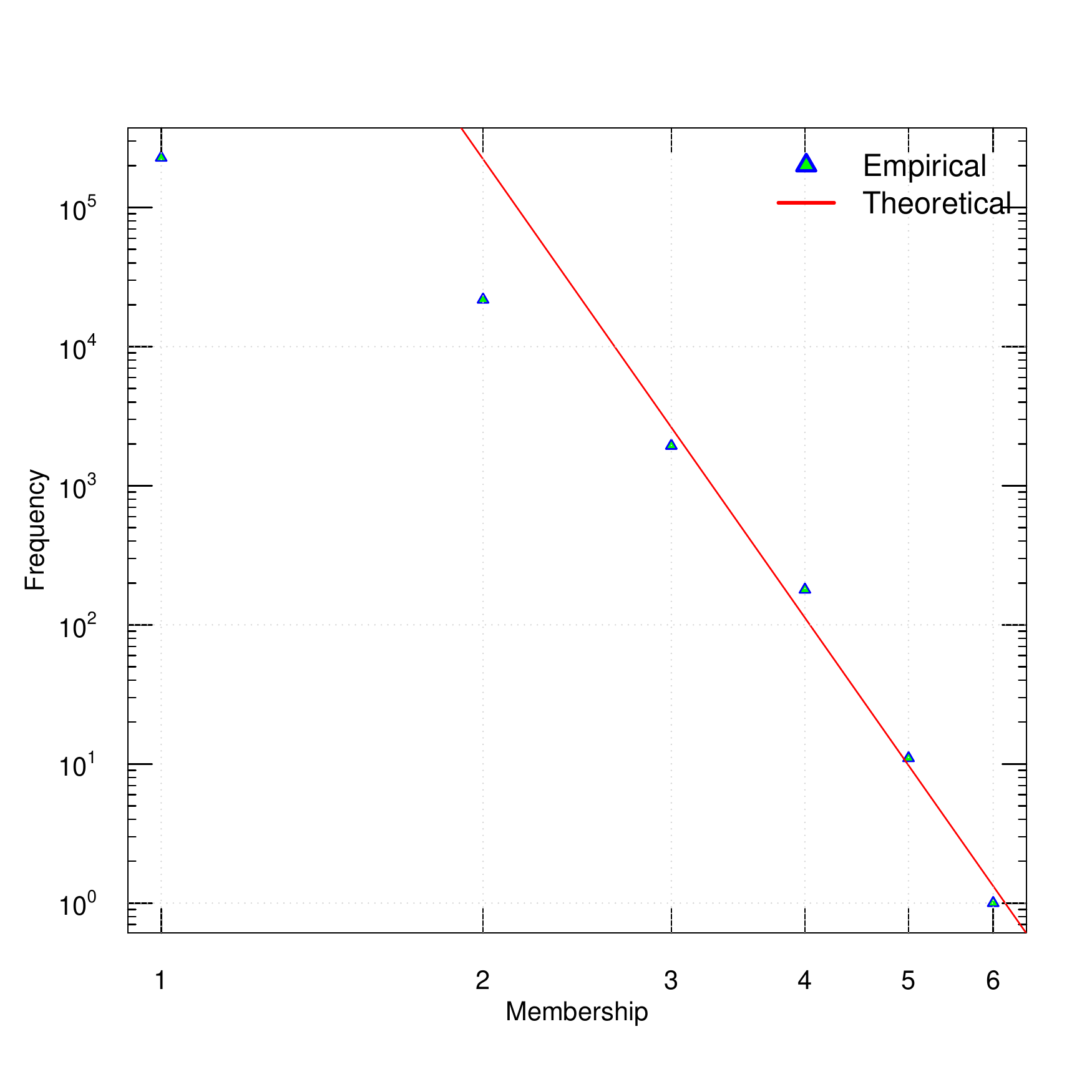}}
        \subfigure[OSLOM]{\includegraphics[width=.121\textwidth]{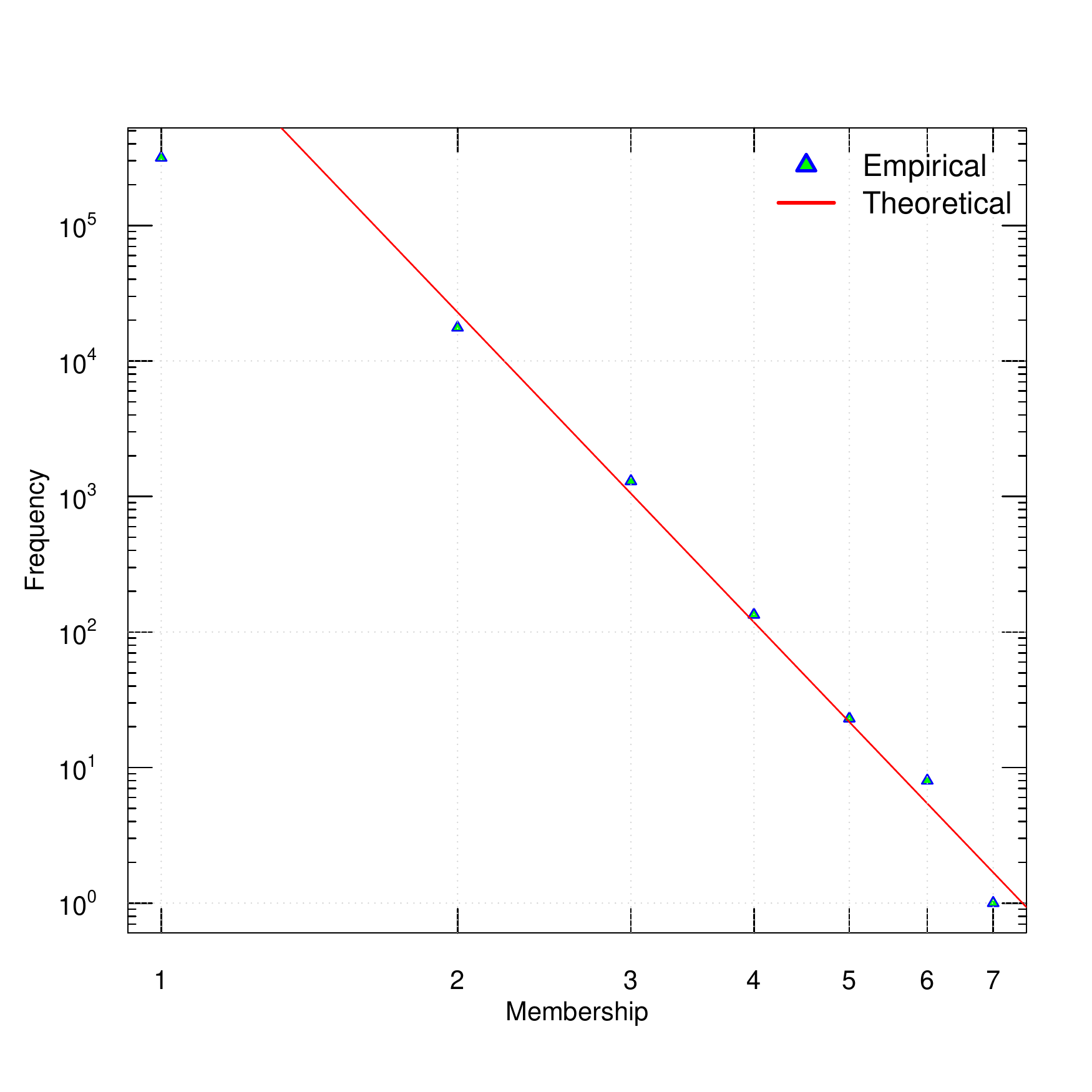}}
        \subfigure[SVINET]{\includegraphics[width=.121\textwidth]{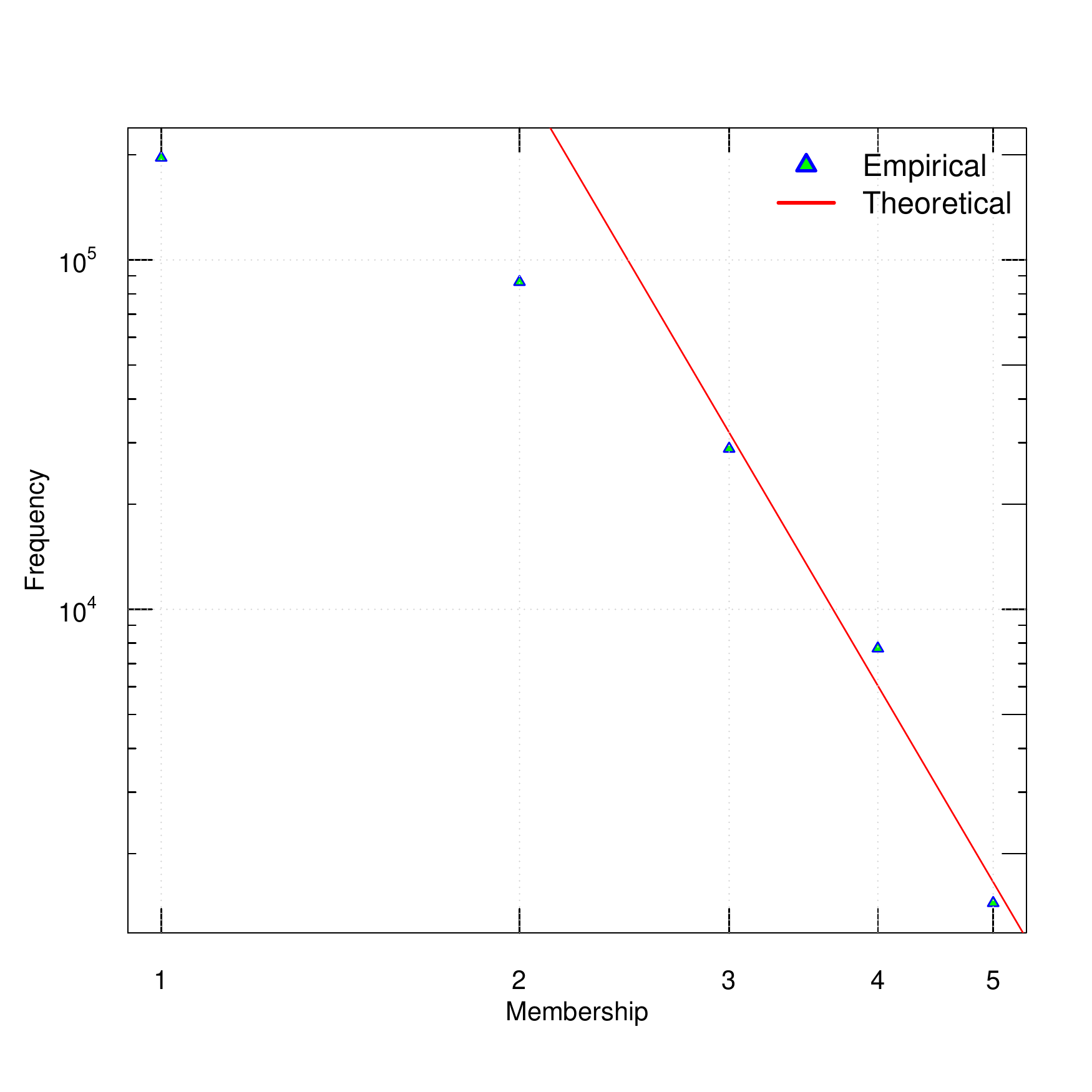}}
        \subfigure[MOSES]{\includegraphics[width=.121\textwidth]{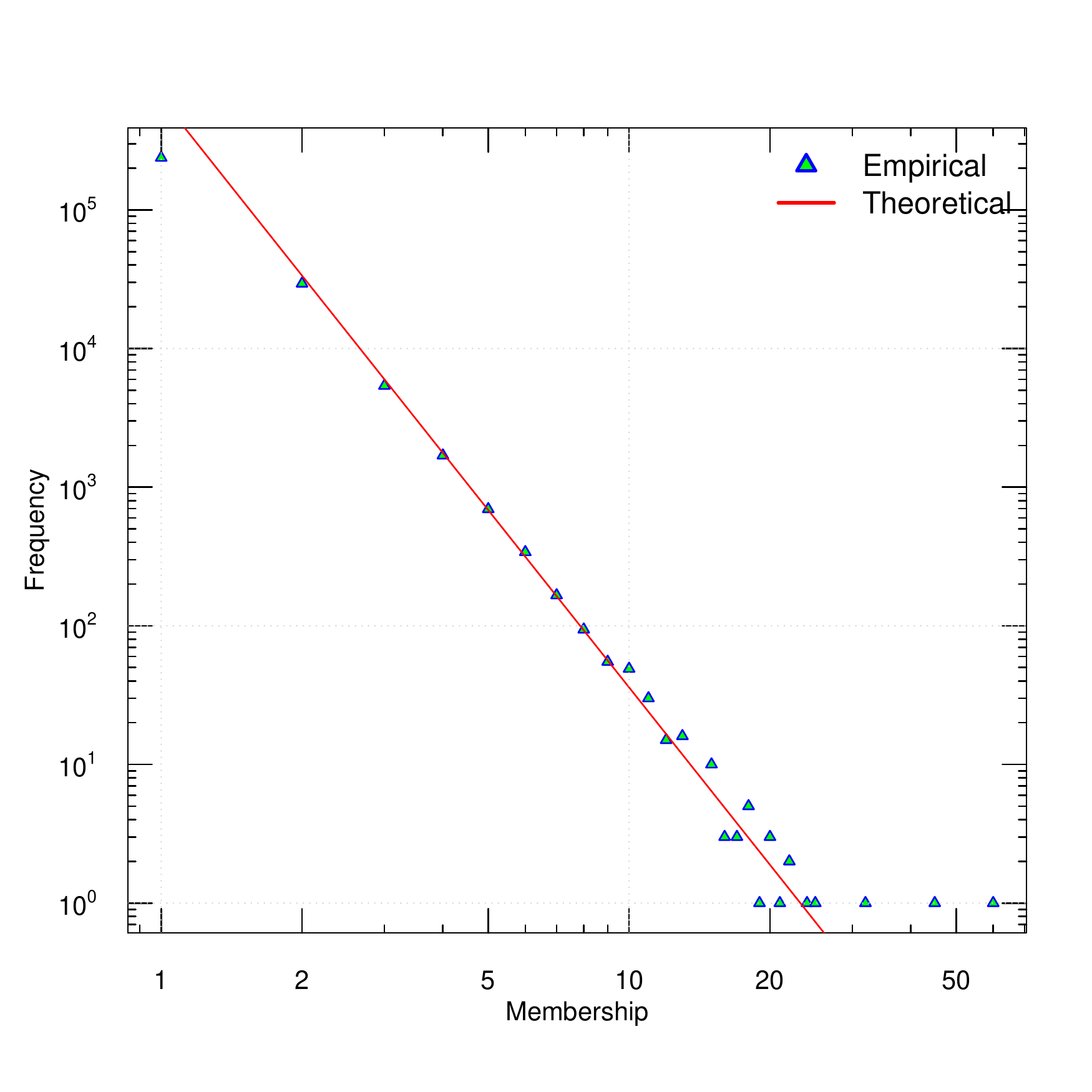}}
        \subfigure[SLPA]{\includegraphics[width=.121\textwidth]{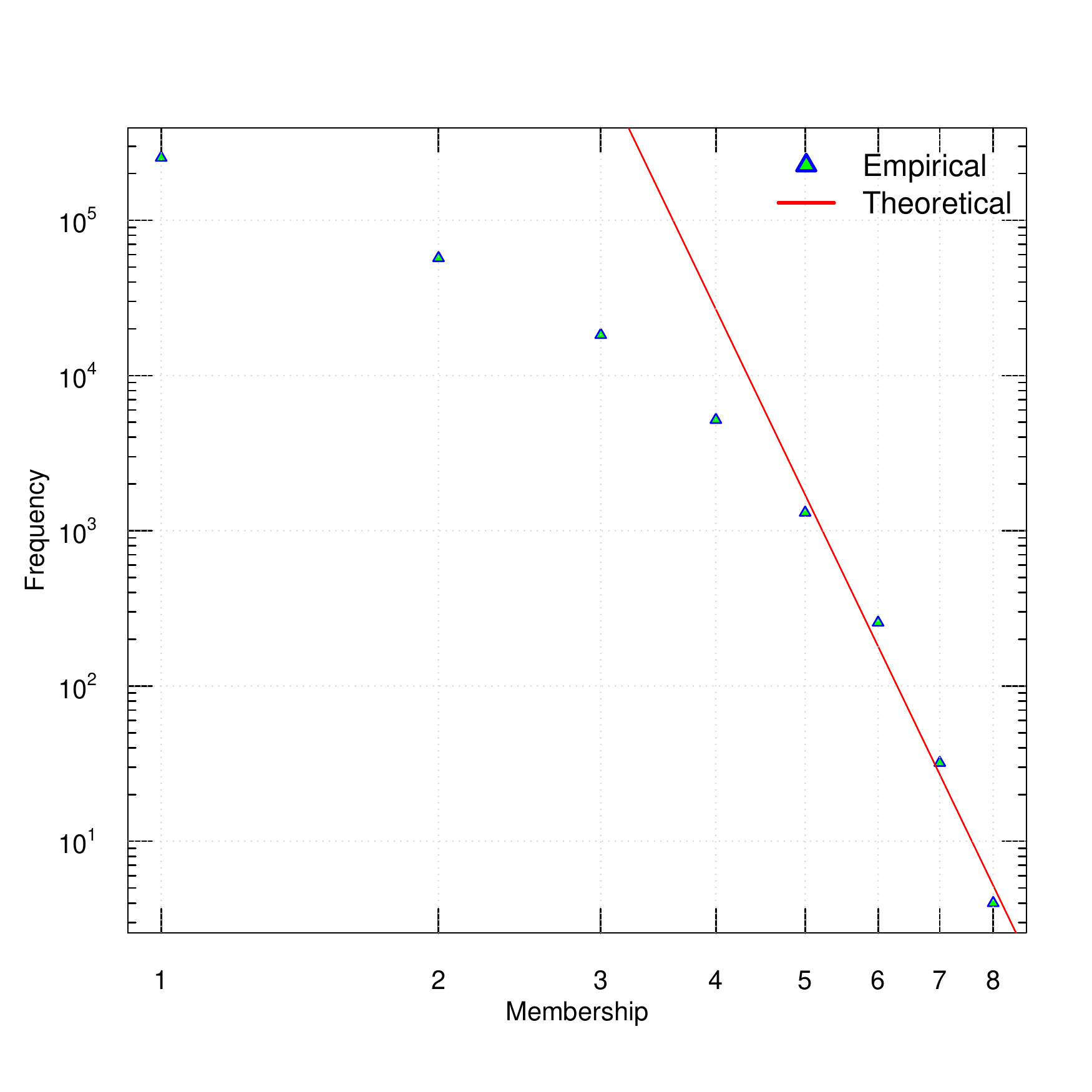}}
        \subfigure[DEMON]{\includegraphics[width=.121\textwidth]{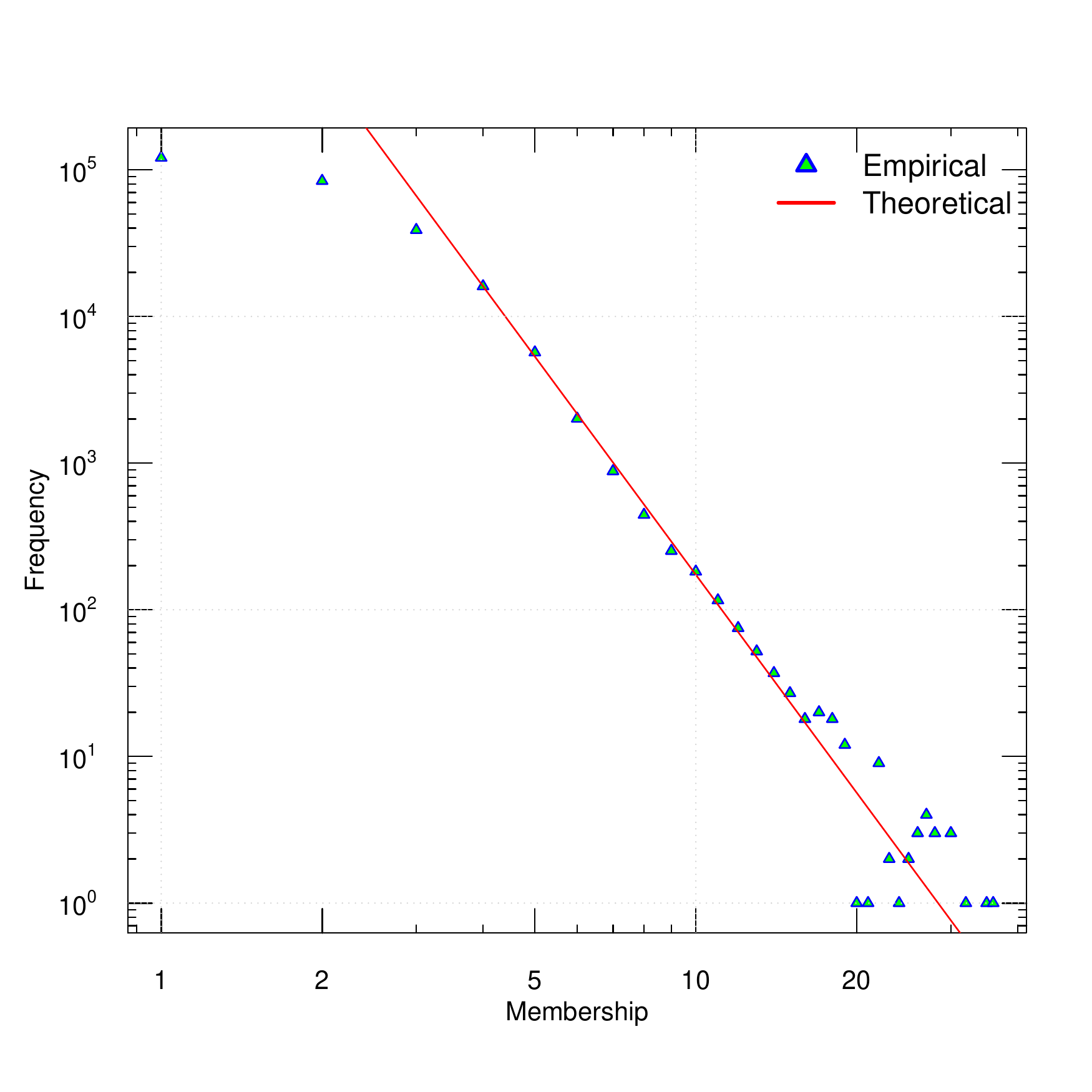}}

        \caption{\label{fig17}Log-log empirical Membership distribution (dots) and Power-Law estimate (line) of AMAZON Ground-truth (a), CFINDER (b), LFM (c),  GCE (d), OSLOM (e), SVINET (f), MOSES (g), SLPA (h), and DEMON (i)}
        \end{figure}

        \begin{figure}[!ht]
        \subfigure[Ground-truth]{\includegraphics[width=.121\textwidth]{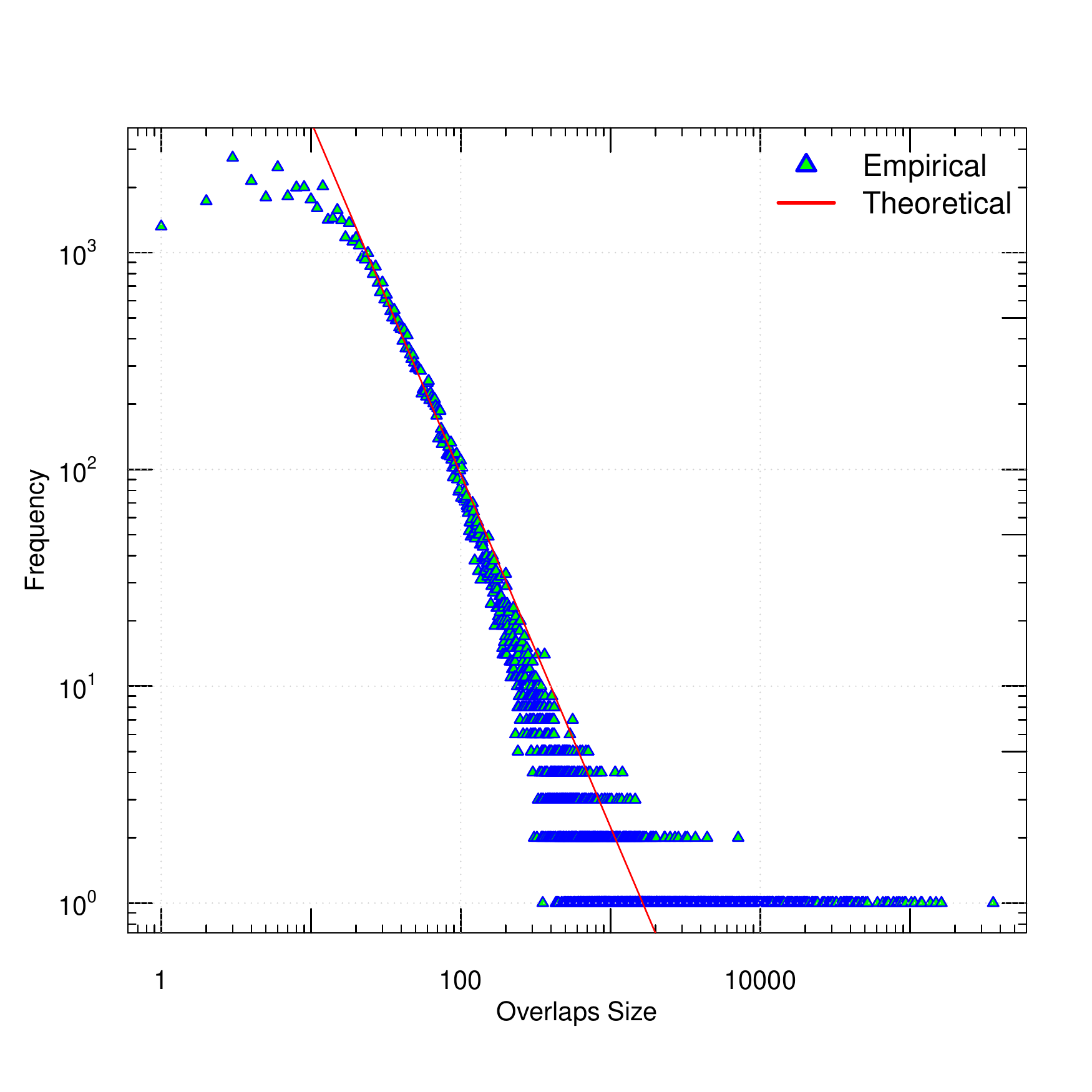}}
        \subfigure[CFINDER]{\includegraphics[width=.121\textwidth]{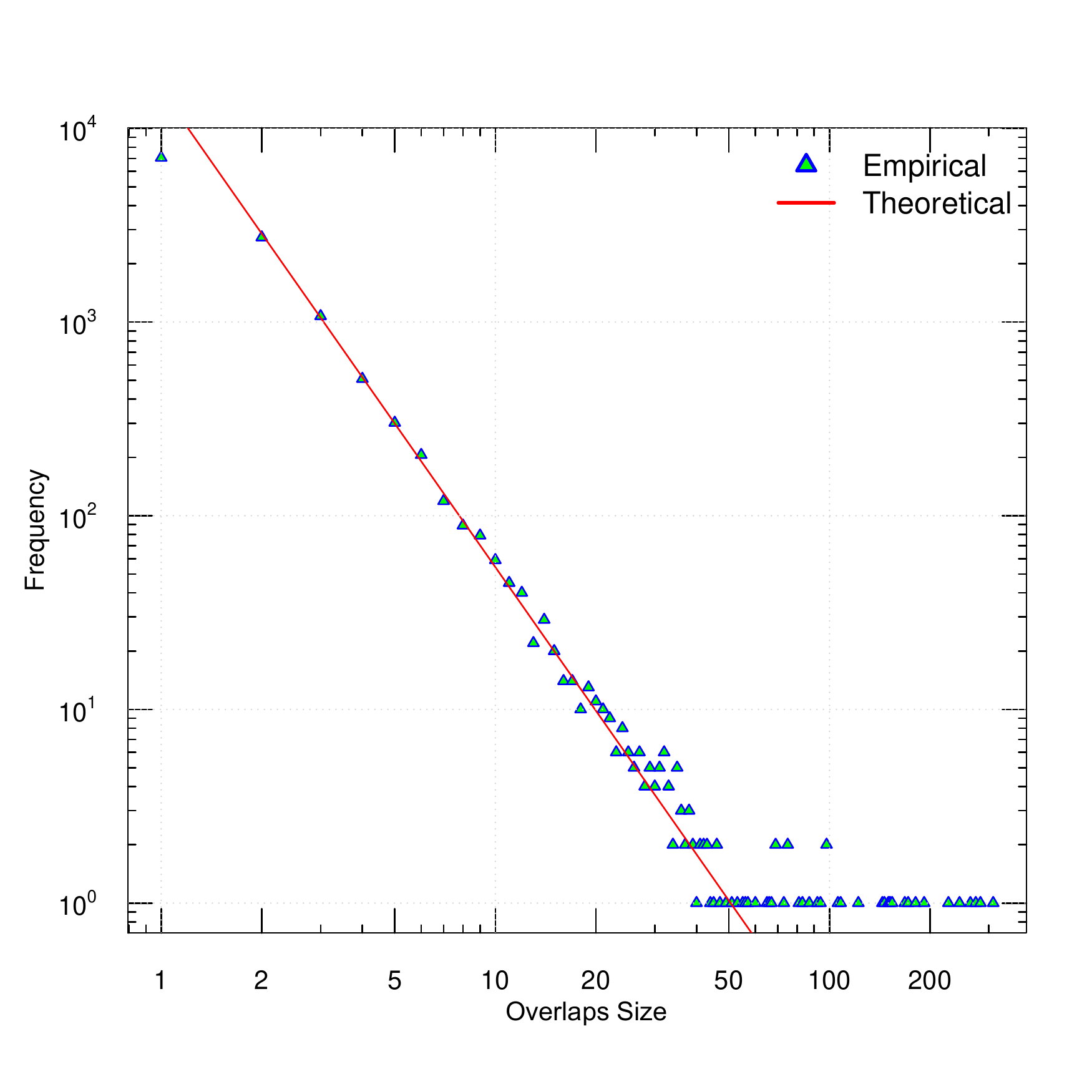}}
        \subfigure[LFM]{\includegraphics[width=.121\textwidth]{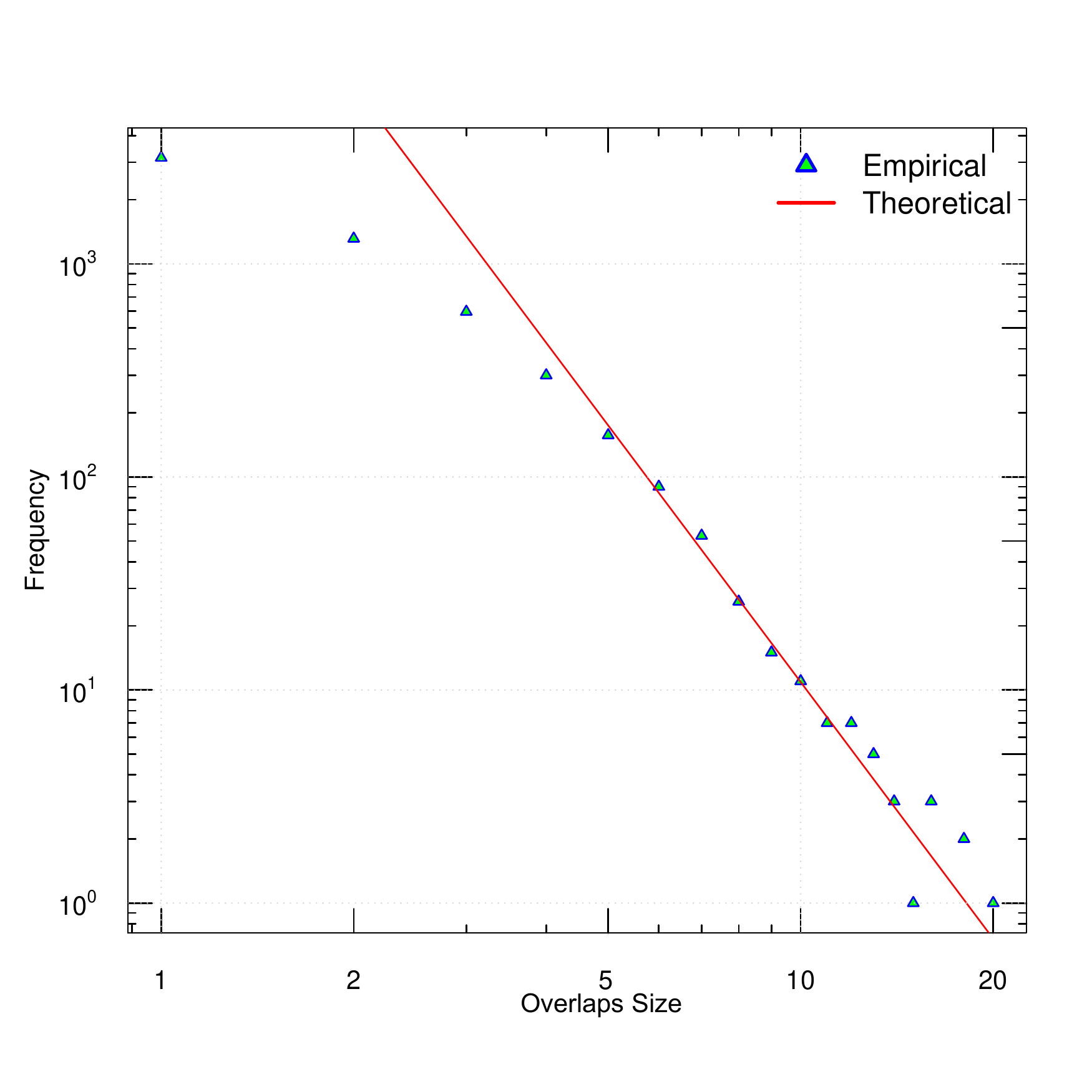}}
        \subfigure[GCE]{\includegraphics[width=.121\textwidth]{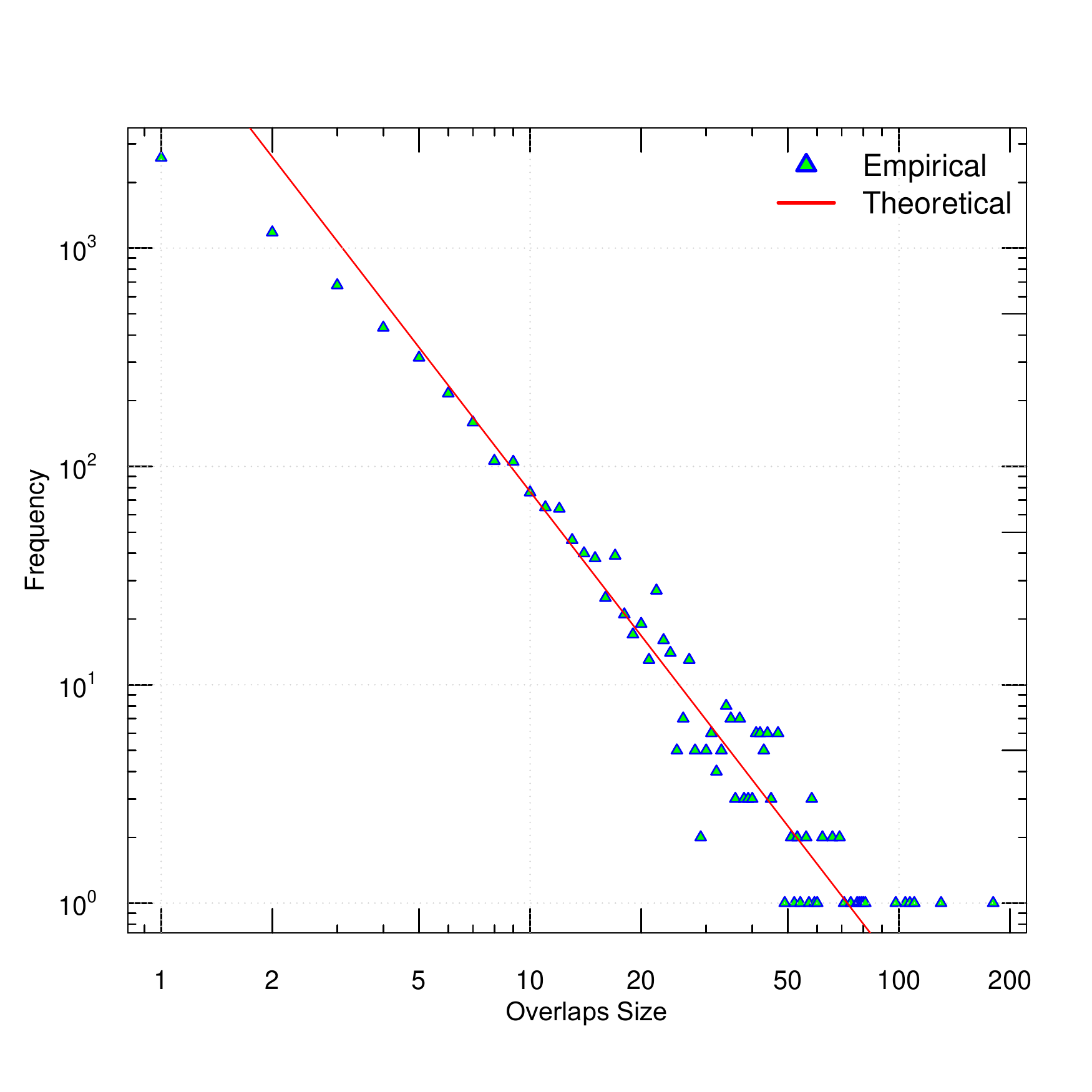}}
        \subfigure[OSLOM]{\includegraphics[width=.121\textwidth]{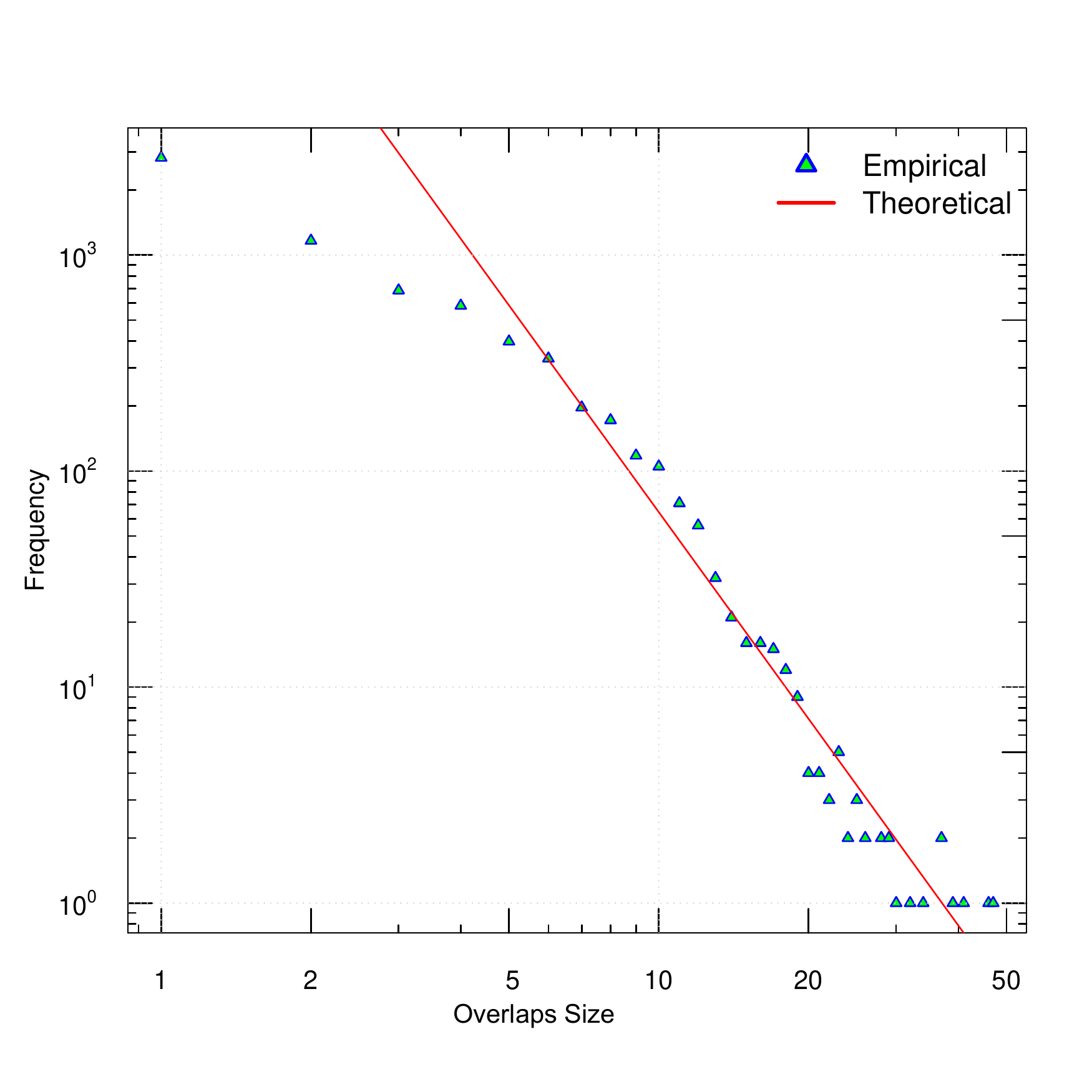}}
        \subfigure[SVINET]{\includegraphics[width=.121\textwidth]{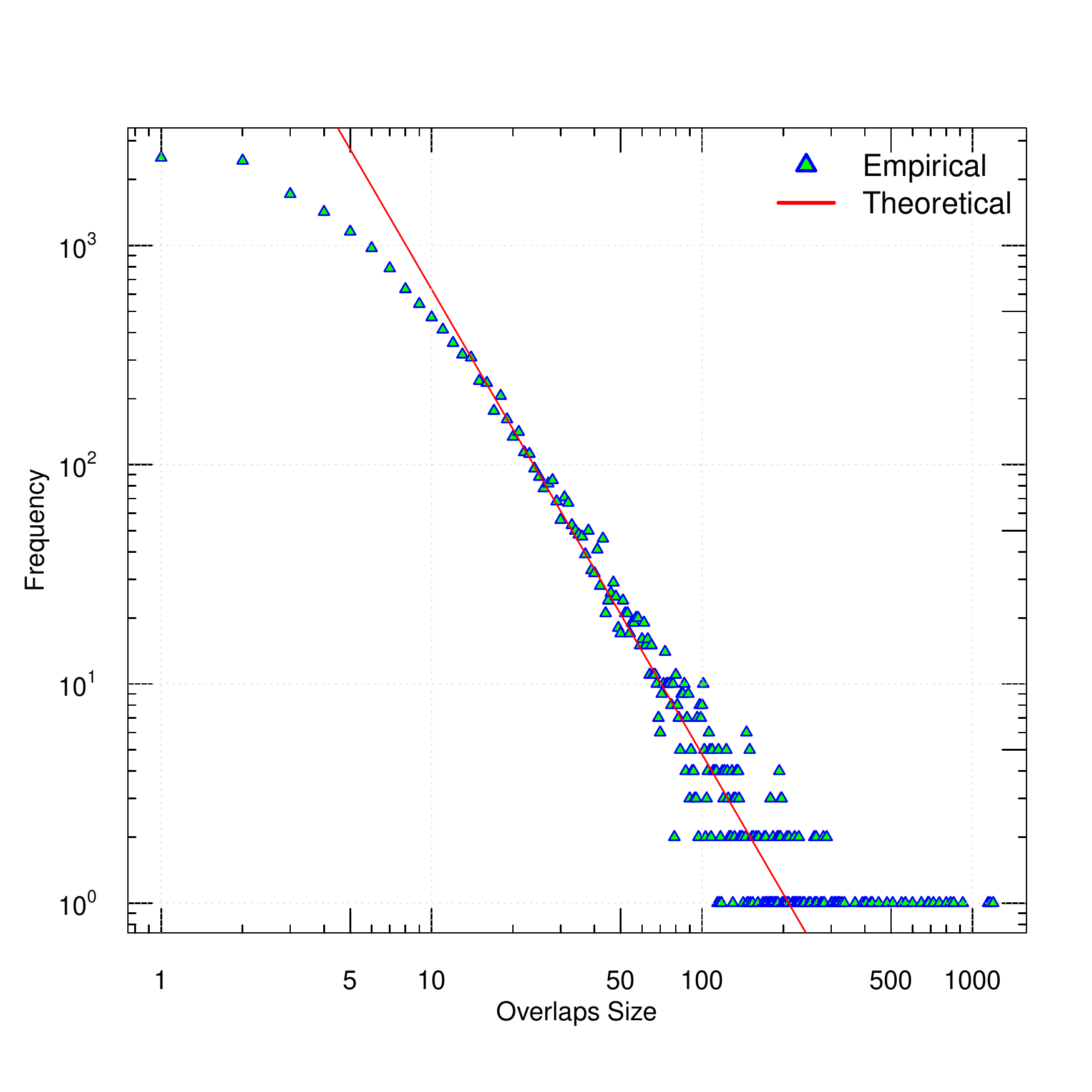}}
        \subfigure[MOSES]{\includegraphics[width=.121\textwidth]{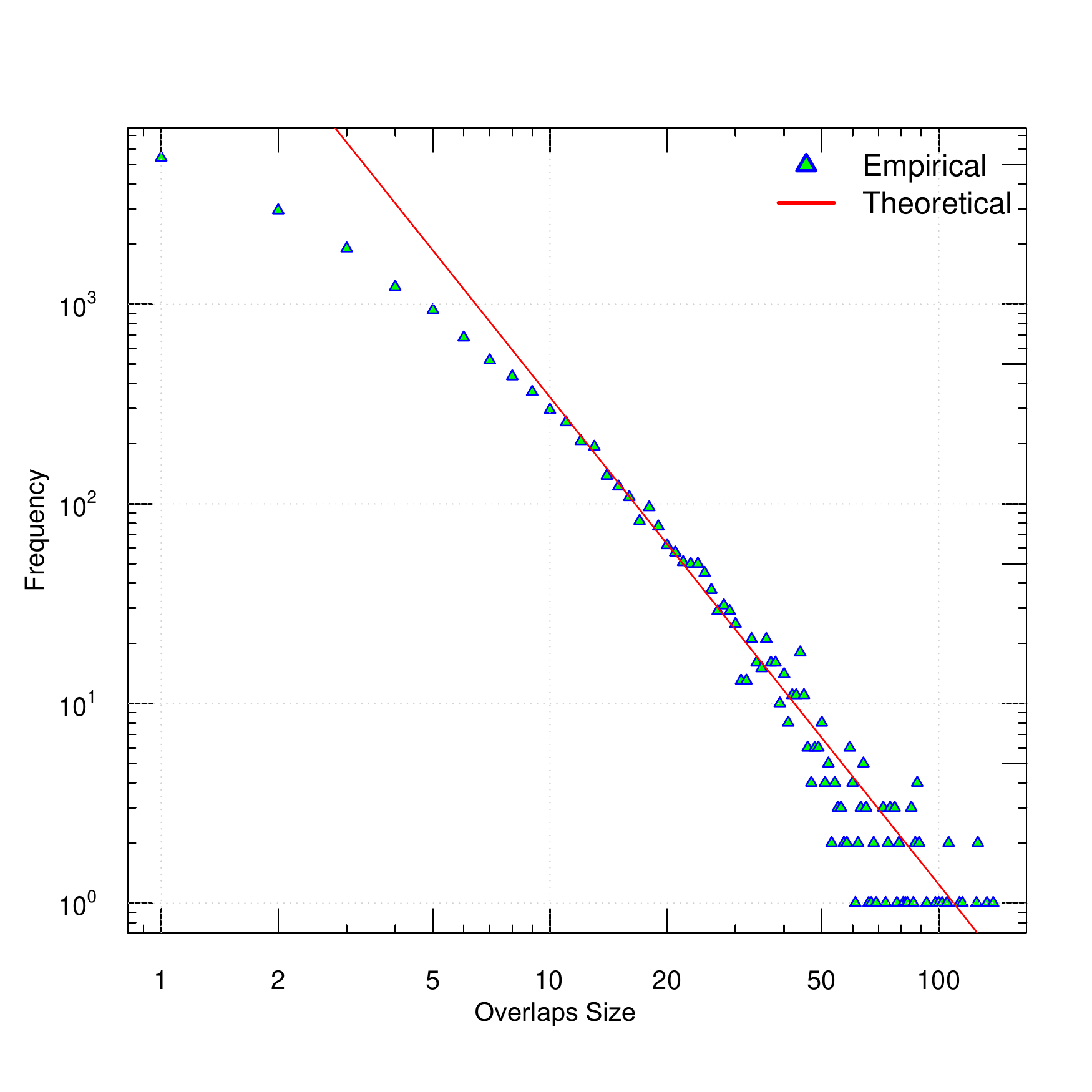}}
        \subfigure[SLPA]{\includegraphics[width=.121\textwidth]{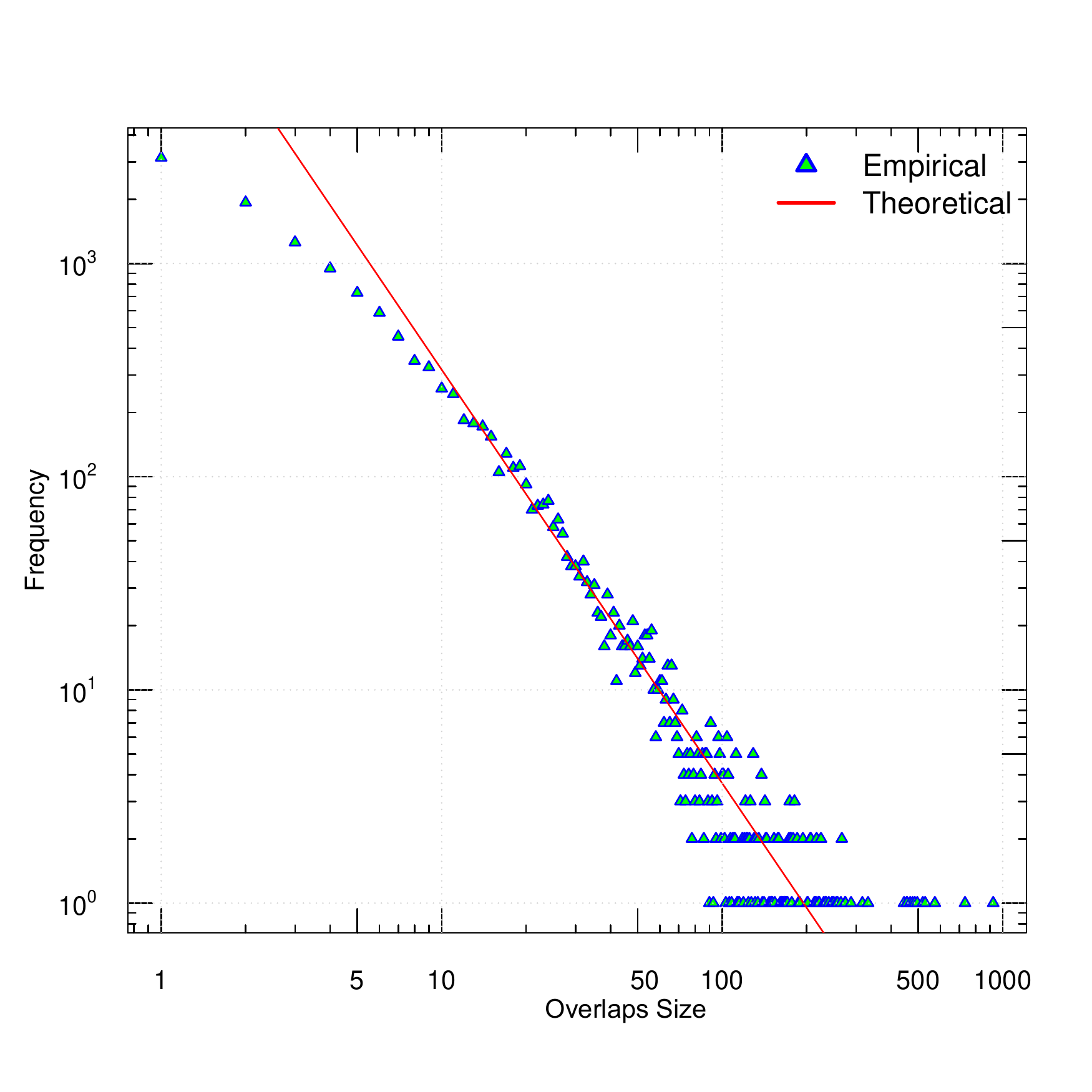}}
        \subfigure[DEMON]{\includegraphics[width=.121\textwidth]{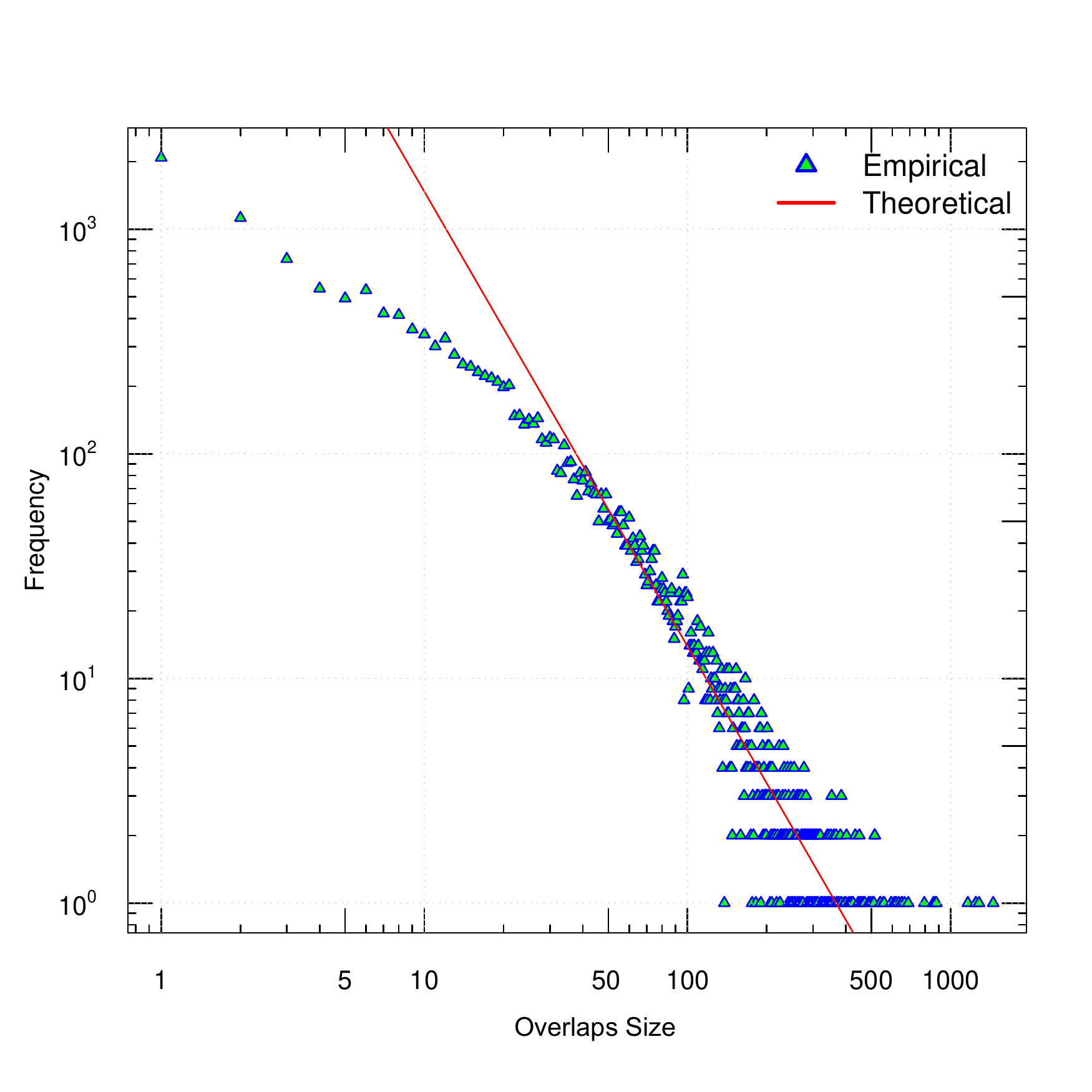}}

        \caption{\label{fig20}Log-log empirical Overlap Size distribution (dots) and Power-Law estimate (line) of AMAZON Ground-truth (a), CFINDER (b), LFM (c),  GCE (d), OSLOM (e), SVINET (f), MOSES (g), SLPA (h), and DEMON (i)}
        \end{figure}

        \begin{table}[!ht]
        \centering
        \caption{Global properties of AMAZON* and 'community-graph' of the overlapping community detection algorithms. The calculated properties are Number of nodes (V), Number of edges (E), Density ($\rho$), Diameter ($d$), Average shortest path ($l_{G}$), Average node degree ($\widetilde{deg}$), Max node degree ($\delta(G)$), Assortativity Coefficient ($\tau$), and Clustering Coefficient ($C$)}
        \label{table3}
        \begin{tabular}{lccccccccc}
        \hline
         &V&E&$\rho$&$d$&$l_{G}$&$\widetilde{deg}$&$\delta(G)$&$\tau$&$C$\\
        \hline
        AMAZON* & 74698 & 1062092 & 3.8E-04 & 27 & 28.43 & 2.13 & 19991 & -0.16 & 0.02 \\
        CFINDER* & 21888 & 31522 & 6.5E-05 & 24 & 2.88 & 2.66 & 257 & -0.02 & 0.15 \\
        LFM* & 8914 & 7585 & 9.5E-05 & 37 & 1.71 & 3.88 & 27 & 0.11 & 0.09 \\
        GCE* & 10256 & 13526 & 1.2E-05 & 31 & 2.63 & 3.51 & 57 & 0.25 & 0.13 \\
        OSLOM* & 9876 & 12613 & 1.2E-05 & 29 & 2.55 & 3.78 & 39 & 0.23 & 0.16 \\
        SVINET* & 25162 & 123947 & 3.9E-04 & 28 & 9.81 & 3.08 & 540 & 0.03 & 0.09 \\
        MOSES* & 25415 & 72499 & 1.1E-05 & 31 & 17.08 & 3.11 & 502 & 0.51 & 0.41 \\
        SLPA* & 25455 & 53442 & 8.2E-05 & 22 & 4.19 & 3.01 & 228 & 0.03 & 0.13 \\
        DEMON* & 17809 & 99293 & 3.1E-05 & 16 & 11.15 & 3.04 & 240 & 0.23 & 0.29 \\
			\hline
        \end{tabular}
        \end{table}

        \begin{table}[!ht]
        \small
        \centering
        \caption{KS-test values for the degree distribution with the AMAZON dataset. The distribution under test are the Power-Law (PL), Beta (BE), Cauchy (CA), Exponential (E), Gamma (GM), Logistic (LO), Log-Normal (LN), Normal (N), Uniform (U), and Weibull (WB)}
        \label{table6}
        \begin{tabular}{lcccccccccc}
        \hline
        &  PL & BE & CA & E & GM & LO & LN & N & U & WB \\
        \hline
        AMAZON*&0.03&0.87&0.23&0.23&0.87&0.44&0.06&0.44&0.98&0.19\\
        CFINDER*&0.02&0.4&0.22&0.39&0.4&0.36&0.25&0.38&0.94&0.25\\
        LFM*&0.02&0.61&0.41&0.61&0.61&0.33&0.37&0.31&0.86&0.33\\
        MOSES*&0.03&0.23&0.23&0.23&0.23&0.25&0.13&0.27&0.83&0.22\\
        GCE*&0.02&0.45&0.25&0.45&0.45&0.27&0.24&0.28&0.84&0.27\\
        OSLOM*&0.03&0.41&0.29&0.41&0.41&0.25&0.24&0.24&0.82&0.29\\
        DEMON*&0.04&0.17&0.22&0.15&0.14&0.22&0.08&0.24&0.8&0.22\\
        SLPA*&0.01&0.3&0.24&0.3&0.3&0.29&0.17&0.31&0.9&0.26\\
        SVINET*&0.03&0.36&0.22&0.17&0.34&0.28&0.08&0.31&0.89&0.26\\
        \hline
        \end{tabular}
        \end{table}

        \begin{table}[!ht]
        \centering
        \caption{KS-test values for the Average clustering coefficient as a function of degree distribution with the AMAZON dataset. The distribution under test are the Power-Law (PL), Beta (BE), Cauchy (CA), Exponential (E), Gamma (GM), Logistic (LO), Log-Normal (LN), Normal (N), Uniform (U), and Weibull (WB)}
        \label{table9}
        \begin{tabular}{lcccccccccc}
        \hline
         &  PL & BE & CA & E & GM & LO & LN & N & U & WB \\
        \hline
        AMAZON*&0.03&0.39&0.21&0.1&0.37&0.31&0.04&0.33&0.93&0.05\\
        CFINDER*&0.16&0.19&0.24&0.2&0.2&0.13&0.21&0.14&0.73&0.31\\
        LFM*&0.07&0.15&0.19&0.15&0.15&0.15&0.1&0.14&0.47&0.2\\
        MOSES*&0.08&0.09&0.11&0.18&0.14&0.1&0.17&0.08&0.31&0.11\\
        GCE*&0.09&0.06&0.18&0.09&0.08&0.12&0.1&0.11&0.47&0.2\\
        OSLOM*&0.09&0.08&0.18&0.09&0.1&0.13&0.1&0.13&0.44&0.24\\
        DEMON*&0.06&0.04&0.16&0.1&0.08&0.1&0.12&0.09&0.43&0.16\\
        SLPA*&0.05&0.06&0.19&0.11&0.06&0.17&0.07&0.19&0.61&0.22\\
        SVINET*&0.05&0.07&0.23&0.08&0.04&0.17&0.09&0.19&0.65&0.04\\
        \hline
        \end{tabular}
        \end{table}

        \begin{table}[!ht]
        \centering
        \caption{KS-test values for the Hop distance distribution with the AMAZON dataset. The distribution under test are the Power-Law (PL), Beta (BE), Cauchy (CA), Exponential (E), Gamma (GM), Logistic (LO), Log-Normal (LN), Normal (N), Uniform (U), and Weibull (WB)}
        \label{table14}
        \begin{tabular}{lcccccccccc}
        \hline
        &  PL & BE & CA & E & GM & LO & LN & N & U & WB \\
        \hline
        AMAZON* & 0.4 & 0.27 & 0.59 & 0.66 & 0.22 & 0.41 & 0.43 & 0.05 & 0.86 & 0.91\\
		CFINDER & 0.26 & 0.27 & 0.1 & 0.31 & 0.34 & 0.29 & 0.51 & 0.03 & 0.18 & 0.48\\
		LFM* & 0.13 & 0.31 & 0.22 & 0.66 & 0.25 & 0.8 & 0.26 & 0.05 & 0.29 & 0.61\\
		MOSES* & 0.22 & 0.21 & 0.14 & 0.8 & 0.55 & 0.6 & 0.13 & 0.04 & 0.49 & 0.78\\
		GCE* & 0.88 & 0.53 & 0.51 & 0.76 & 0.76 & 0.1 & 0.15 & 0.01 & 0.88 & 0.39\\
		OSLOM* & 0.7 & 0.21 & 0.44 & 0.15 & 0.66 & 0.11 & 0.23 & 0.11 & 0.43 & 0.43\\
		DEMON* & 0.43 & 0.41 & 0.74 & 0.8 & 0.19 & 0.46 & 0.63 & 0.01 & 0.09 & 0.82\\
		SLPA* & 0.1 & 0.35 & 0.45 & 0.13 & 0.28 & 0.71 & 0.89 & 0.05 & 0.35 & 0.59\\
		SVINET* & 0.75 & 0.8 & 0.73 & 0.87 & 0.61 & 0.45 & 0.67 & 0.06 & 0.72 & 0.29\\
			\hline
        \end{tabular}
        \end{table}

        \begin{table}[htpb!]
        \centering
        \caption{KS-test values for the Community size distribution for AMAZON dataset. The distribution under test are the Power-Law (PL), Beta (BE), Cauchy (CA), Exponential (E), Gamma (GM), Logistic (LO), Log-Normal (LN), Normal (N), Uniform (U), and Weibull (WB)}
        \label{table18}
        \begin{tabular}{lcccccccccc}
        \hline
        &  PL & BE & CA & E & GM & LO & LN & N & U & WB \\
        \hline
		Ground-truth & 0.01 & 0.68 & 0.27 & 0.57 & 0.68 & 0.47 & 0.14 & 0.48 & 0.98 & 0.2\\
		CFINDER & 0.01 & 0.5 & 0.26 & 0.32 & 0.49 & 0.39 & 0.12 & 0.41 & 0.94 & 0.23\\
		LFM & 0.01 & 0.16 & 0.24 & 0.11 & 0.16 & 0.17 & 0.09 & 0.19 & 0.91 & 0.31\\
		MOSES & 0.03 & 0.19 & 0.24 & 0.16 & 0.16 & 0.23 & 0.07 & 0.25 & 0.78 & 0.2\\
		LINKC & 0.03 & 0.74 & 0.21 & 0.74 & 0.74 & 0.38 & 0.3 & 0.4 & 0.95 & 0.24\\
		GCE & 0.02 & 0.19 & 0.21 & 0.06 & 0.18 & 0.19 & 0.05 & 0.22 & 0.86 & 0.27\\
		OSLOM & 0.02 & 0.07 & 0.22 & 0.1 & 0.06 & 0.13 & 0.07 & 0.13 & 0.81 & 0.27\\
		CLIPERC & 0.02 & 0.55 & 0.14 & 0.55 & 0.55 & 0.36 & 0.43 & 0.34 & 0.61 & 0.4\\
		DEMON & 0.04 & 0.13 & 0.21 & 0.11 & 0.1 & 0.21 & 0.05 & 0.24 & 0.82 & 0.23\\
		SLPA & 0.02 & 0.3 & 0.23 & 0.14 & 0.29 & 0.28 & 0.07 & 0.3 & 0.91 & 0.25\\
		SVINET & 0.02 & 0.35 & 0.21 & 0.13 & 0.33 & 0.29 & 0.04 & 0.31 & 0.9 & 0.24\\
		\hline
        \end{tabular}
        \end{table}

        \begin{table}[htpb!]
        \centering
        \caption{KS-test values for the membership distribution with the AMAZON dataset. The distribution under test are the Power-Law (PL), Beta (BE), Cauchy (CA), Exponential (E), Gamma (GM), Logistic (LO), Log-Normal (LN), Normal (N), Uniform (U), and Weibull (WB)}
        \label{table21}
        \begin{tabular}{lcccccccccc}
        \hline
         &  PL & BE & CA & E & GM & LO & LN & N & U & WB \\
        \hline
		Ground-truth & 0.02 & 0.12 & 0.25 & 0.18 & 0.12 & 0.16 & 0.08 & 0.16 & 0.82 & 0.25\\
		CFINDER & 0.02 & 0.43 & 0.84 & 0.88 & 0.79 & 0.44 & 0.9 & 0.39 & 0.89 & 0.14\\
		LFM & 0.04 & 0.44 & 0.67 & 0.47 & 0.51 & 0.21 & 0.42 & 0.65 & 0.78 & 0.67\\
		MOSES & 0.01 & 0.81 & 0.13 & 0.26 & 0.35 & 0.26 & 0.41 & 0.41 & 0.41 & 0.38\\
		LINKC & 0.03 & 0.26 & 0.34 & 0.77 & 0.24 & 0.35 & 0.79 & 0.4 & 0.87 & 0.25\\
		GCE & 0.01 & 0.82 & 0.35 & 0.64 & 0.56 & 0.62 & 0.3 & 0.86 & 0.44 & 0.13\\
		OSLOM & 0.04 & 0.5 & 0.37 & 0.65 & 0.39 & 0.21 & 0.76 & 0.65 & 0.34 & 0.53\\
		CLIPERC & 0.03 & 0.51 & 0.73 & 0.55 & 0.65 & 0.57 & 0.75 & 0.73 & 0.82 & 0.88\\
		DEMON & 0.01 & 0.76 & 0.17 & 0.85 & 0.24 & 0.65 & 0.18 & 0.25 & 0.88 & 0.66\\
		SLPA & 0.04 & 0.35 & 0.37 & 0.85 & 0.66 & 0.39 & 0.82 & 0.46 & 0.4 & 0.37\\
		SVINET & 0.04 & 0.61 & 0.34 & 0.61 & 0.61 & 0.39 & 0.43 & 0.36 & 0.63 & 0.39\\
			\hline
        \end{tabular}
        \end{table}

        \begin{table}[htpb!]
        \centering
        \caption{KS-test values for the overlap size distribution with the AMAZON dataset. The distribution under test are the Power-Law (PL), Beta (BE), Cauchy (CA), Exponential (E), Gamma (GM), Logistic (LO), Log-Normal (LN), Normal (N), Uniform (U), and Weibull (WB)}
        \label{table24}
        \begin{tabular}{lcccccccccc}
        \hline
         &  PL & BE & CA & E & GM & LO & LN & N & U & WB \\
        \hline
		Ground-truth & 0.02 & 0.95 & 0.28 & 0.5 & 0.94 & 0.47 & 0.04 & 0.47 & 0.99 & 0.22\\
		CFINDER & 0.03 & 0.56 & 0.3 & 0.56 & 0.56 & 0.41 & 0.27 & 0.42 & 0.94 & 0.21\\
		LFM & 0.02 & 0.55 & 0.35 & 0.55 & 0.55 & 0.29 & 0.31 & 0.28 & 0.78 & 0.27\\
		MOSES & 0.04 & 0.32 & 0.23 & 0.32 & 0.32 & 0.28 & 0.14 & 0.3 & 0.83 & 0.22\\
		LINKC & 0.01 & 0.55 & 0.2 & 0.19 & 0.55 & 0.36 & 0.05 & 0.38 & 0.98 & 0.33\\
		GCE & 0.04 & 0.4 & 0.24 & 0.4 & 0.4 & 0.31 & 0.17 & 0.33 & 0.87 & 0.22\\
		OSLOM & 0.02 & 0.41 & 0.25 & 0.41 & 0.41 & 0.24 & 0.18 & 0.26 & 0.77 & 0.25\\
		CLIPERC & 0.02 & 0.26 & 0.23 & 0.12 & 0.23 & 0.25 & 0.04 & 0.27 & 0.85 & 0.22\\
		DEMON & 0.02 & 0.21 & 0.22 & 0.22 & 0.19 & 0.29 & 0.05 & 0.31 & 0.87 & 0.25\\
		SLPA & 0.03 & 0.38 & 0.22 & 0.3 & 0.37 & 0.34 & 0.1 & 0.36 & 0.91 & 0.24\\
		SVINET & 0.03 & 0.49 & 0.23 & 0.24 & 0.47 & 0.35 & 0.08 & 0.37 & 0.92 & 0.25\\
			\hline
        \end{tabular}
        \end{table}

        \begin{table}[!ht]
        \centering
        \caption{Ranking of the algorithms based on basic properties with the the AMAZON dataset. The calculated properties are Number of nodes (V), Number of edges (E), Density ($\rho$), Diameter ($d$), Average shortest path ($l_{G}$), Average node degree ($\widetilde{deg}$), Max node degree ($\delta(G)$), Assortativity Coefficient ($\tau$), and Clustering Coefficient ($C$). Kconsensus  and TOPSIS denotes respectively the final ranking using Kemeny consensus and TOPSIS.}
        \label{table55}
        \begin{tabular}{lccccccccccc}
        \hline
          &V&E&$\rho$&$d$&$l_{G}$&$\widetilde{deg}$&$\delta(G)$&$\tau$&$C$&Kconsensus&TOPSIS\\
        \hline
            CFINDER &4&5&4&3&5&3&1&1&5&4&2\\
            LFM &8&8&2&7&8&8&8&4&2&8&6\\
            GCE &6&6&6&5&6&6&6&7&4&6&8\\
            OSLOM &7&7&6&2&7&7&7&5&6&7&7\\
            SVINET &3&1&1&1&3&1&4&2&1&1&1\\
            MOSES &2&3&8&4&1&2&5&8&8&5&4\\
            SLPA &1&4&3&6&4&5&2&3&3&2&3\\
            DEMON &5&2&5&8&2&4&3&5&7&3&5\\
        \hline
        \end{tabular}
        \end{table}

        \begin{table}[!ht]
          \centering
          \caption{Correlation of basic properties rankings for AMAZON dataset. The calculated properties are Number of nodes (V), Number of edges (E), Density ($\rho$), Diameter ($d$), Average shortest path ($l_{G}$), Average node degree ($\widetilde{deg}$), Max node degree ($\delta(G)$), Assortativity Coefficient ($\tau$), and Clustering Coefficient ($C$)}
          \label{table56}
            \begin{tabular}{lccccccccc}
            \hline
               &V&E&$\rho$&$d$&$l_{G}$&$\widetilde{deg}$&$\delta(G)$&$\tau$&$C$ \\
            \hline
            V & 1 &   &   &   &   &   &   &   &  \\
            E & 0.71 & 1 &   &   &   &   &   &   &  \\
            $\rho$ & -0.01 & 0.09 & 1 &   &   &   &   &   &  \\
            $d$ & 0.17 & 0.14 & 0.04 & 1 &   &   &   &   &  \\
            $l_{G}$  & 0.76 & 0.9 & -0.26 & 0 & 1 &   &   &   &  \\
            $\widetilde{deg}$  & 0.74 & 0.88 & 0.01 & 0.43 & 0.83 & 1 &   &   &  \\
            $\delta(G)$ & 0.71 & 0.6 & 0.14 & 0.02 & 0.55 & 0.62 & 1 &   &  \\
            $\tau$  & 0.13 & 0.11 & 0.79 & 0.26 & -0.18 & 0.18 & 0.53 & 1 &  \\
            $C$ & -0.07 & -0.1 & 0.89 & 0.14 & -0.43 & -0.12 & -0.1 & 0.57 & 1 \\
            \hline
            \end{tabular}%
        \end{table}%

        \begin{table}[!ht]
          \centering
          \caption{Ranking of the algorithms based on microscopic properties with the the  AMAZON dataset. The distribution under test are the degree distribution (DD), the average clustering coefficient as function of degree (Av), the hop distance (HD). Kconsensus  and TOPSIS denotes respectively the final ranking using Kemeny consensus and TOPSIS.}
          \label{table560}
            \begin{tabular}{lccccc}
            \hline
              & \multicolumn{1}{l}{DD} & \multicolumn{1}{l}{Av} & \multicolumn{1}{l}{HD} & \multicolumn{1}{l}{Kconsensus} & \multicolumn{1}{l}{TOPSIS} \\
            \hline
            CFINDER & 4 & 8 & 5 & 8  & 7 \\
            LFM & 6 & 4 & 2 &  4 & 4 \\
            GCE & 5 & 6 & 7 & 6  & 8 \\
            OSLOM & 3 & 7 & 8 & 7  & 6 \\
            SVINET & 2 & 2 & 3 & 2  & 3 \\
            MOSES & 1 & 5 & 4 & 5  & 2 \\
            SLPA & 7 & 1 & 1 & 1  & 1 \\
            DEMON & 8 & 3 & 6 &  3 & 5 \\
            \hline
            \end{tabular}%
        \end{table}%

        \begin{table}[!ht]
          \centering
          \caption{Correlation of the rankings of the microscopic properties for AMAZON (degree distribution (DD), the Average clustering coefficient as function of degree (Av), the Hop distance (HD))}
                  \label{table63}
            \begin{tabular}{lccc}
            \hline
              & DD & Av & HD \\
              \hline
            DD & 1 &   &  \\
            Av & -0.4 & 1 &  \\
            HD & -0.1 & 0.69 & 1 \\
            \hline
            \end{tabular}%
        \end{table}%

        \begin{table}[!ht]
          \centering
          \caption{Ranking of the algorithms based on mesoscopic properties with the the  AMAZON dataset. Mesoscopic properties ranking for AMAZON. The distribution under test are the community size (CS), the membership (M), the overlap size (OS).  Kconsensus  and TOPSIS denotes respectively the final ranking using Kemeny consensus and TOPSIS.}
          \label{table561}
            \begin{tabular}{lccccc}
            \hline
              & \multicolumn{1}{l}{CS} & \multicolumn{1}{l}{MC} & \multicolumn{1}{l}{OS} & \multicolumn{1}{l}{Kconsensus} & \multicolumn{1}{l}{TOPSIS} \\
            \hline
            CFINDER & 1 & 1 & 4 &  1 & 1 \\
            LFM & 2 & 5 & 1 & 5  & 2 \\
            GCE & 3 & 2 & 8 & 2  & 3 \\
            OSLOM & 4 & 8 & 2 &  8 & 4 \\
            SVINET & 6 & 7 & 6 & 7  & 8 \\
            MOSES & 7 & 4 & 7 & 4  & 6 \\
            SLPA & 5 & 6 & 5 & 6  & 7 \\
            DEMON & 8 & 3 & 3 & 3  & 5 \\
            \hline
            \end{tabular}%
        \end{table}%

        \begin{table}[!ht]
          \centering
          \caption{Correlation of the rankings of the microscopic properties for AMAZON (the community size (CS), the membership (M), the overlap size (OS))}
                  \label{table66}
            \begin{tabular}{lccc}
            \hline
              & CS & MC & OS \\
              \hline
            CS & 1 &   &  \\
            MC & 0.26 & 1 &  \\
            OS & 0.24 & -0.29 & 1 \\
            \hline
            \end{tabular}%
        \end{table}%

        \begin{table}[!ht]
          \centering
          \footnotesize
          \caption{Ranking of the Algorithms based on all topological properties with the AMAZON dataset. The calculated properties are Number of nodes (V), Number of edges (E), Density ($\rho$), Diameter ($d$), Average shortest path ($l_{G}$), Average node degree ($\widetilde{deg}$), Max node degree ($\delta(G)$), Assortativity Coefficient ($\tau$),Clustering Coefficient ($C$), the Degree distribution (DD), the Average clustering coefficient as function of degree (Av), the hop distance (HD), the community size (CS), the membership (M), the overlap size (OS).}
          \label{table562}
            \begin{tabular}{lccccccccccccccccc}
            \hline
              & \multicolumn{9}{c}{Basic properties} & \multicolumn{3}{c}{Microscopic properties} & \multicolumn{3}{c}{Mesoscopic} & \multicolumn{2}{c}{MCDM Ranking}  \\
            \hline
              &V&E&$\rho$&$d$&$l_{G}$&$\widetilde{deg}$&$\delta(G)$&$\tau$&$C$& DD & Av & HD & CS & MC & OS & Kconsensus & TOPSIS \\
            \hline
            CFINDER & 4 & 5 & 4 & 3 & 5 & 3 & 1 & 1 & 5 & 4 & 8 & 5 & 1 & 1 & 4 &  5 & 2 \\
            LFM & 8 & 8 & 2 & 7 & 8 & 8 & 8 & 4 & 2 & 6 & 4 & 2 & 2 & 5 & 1 & 8  & 5 \\
            GCE & 6 & 6 & 6 & 5 & 6 & 6 & 6 & 7 & 4 & 5 & 6 & 7 & 3 & 2 & 8 & 6  & 8 \\
            OSLOM & 7 & 7 & 6 & 2 & 7 & 7 & 7 & 5 & 6 & 3 & 7 & 8 & 4 & 8 & 2 &  7 & 7 \\
            SVINET & 3 & 1 & 1 & 1 & 3 & 1 & 4 & 2 & 1 & 2 & 2 & 3 & 6 & 7 & 6 & 1  & 1 \\
            MOSES & 2 & 3 & 8 & 4 & 1 & 2 & 5 & 8 & 8 & 1 & 5 & 4 & 7 & 4 & 7 & 3  & 4 \\
            SLPA & 1 & 4 & 3 & 6 & 4 & 5 & 2 & 3 & 3 & 7 & 1 & 1 & 5 & 6 & 5 & 4  & 3 \\
            DEMON & 5 & 2 & 5 & 8 & 2 & 4 & 3 & 5 & 7 & 8 & 3 & 6 & 8 & 3 & 3 & 2  & 6 \\
            \hline
            \end{tabular}%
        \end{table}%

        \begin{table}[!ht]
          \centering
          \footnotesize
          \caption{Correlation of ranking of all topological properties with the AMAZON dataset. The calculated properties are Number of nodes (V), Number of edges (E), Density ($\rho$), Diameter ($d$), Average shortest path ($l_{G}$), Average node degree ($\widetilde{deg}$), Max node degree ($\delta(G)$), Assortativity Coefficient ($\tau$), Clustering Coefficient ($C$), the Degree distribution (DD), the Average clustering coefficient as function of degree (Av), the Hop distance (HD), the Community size (CS), the Membership (M), the Overlap size (OS).}
          \label{table69}
            \begin{tabular}{lccccccccccccccc}
            \hline
              & V & E & $\rho$ & $d$ & $l_{G}$ & $\widetilde{deg}$ & $\delta(G)$ & $\tau$ & $C$ & DD & Av & HD & CS & M & OS \\
              \hline
            V  & 1 &   &   &   &   &   &   &   &   &   &   &   &   &   &  \\
            E  & 0.71 & 1 &   &   &   &   &   &   &   &   &   &   &   &   &  \\
            $\rho$  & -0.01 & 0.09 & 1 &   &   &   &   &   &   &   &   &   &   &   &  \\
            $d$  & 0.17 & 0.14 & 0.04 & 1 &   &   &   &   &   &   &   &   &   &   &  \\
            $l_{G}$  & 0.76 & 0.9 & -0.26 & 0 & 1 &   &   &   &   &   &   &   &   &   &  \\
            $\widetilde{deg}$  & 0.74 & 0.88 & 0.01 & 0.43 & 0.83 & 1 &   &   &   &   &   &   &   &   &  \\
            $\delta(G)$  & 0.71 & 0.6 & 0.14 & 0.02 & 0.55 & 0.62 & 1 &   &   &   &   &   &   &   &  \\
            $\tau$  & 0.13 & 0.11 & 0.79 & 0.26 & -0.18 & 0.18 & 0.53 & 1 &   &   &   &   &   &   &  \\
            $C$  & -0.07 & -0.1 & 0.89 & 0.14 & -0.43 & -0.12 & -0.1 & 0.57 & 1 &   &   &   &   &   &  \\
            DD  & \textcolor[rgb]{ 1,  0,  0}{\textbf{0.19}} & \textcolor[rgb]{ 1,  0,  0}{\textbf{0.14}} & \textcolor[rgb]{ 1,  0,  0}{\textbf{-0.26}} & \textcolor[rgb]{ 1,  0,  0}{\textbf{0.83}} & \textcolor[rgb]{ 1,  0,  0}{\textbf{0.19}} & \textcolor[rgb]{ 1,  0,  0}{\textbf{0.48}} & \textcolor[rgb]{ 1,  0,  0}{\textbf{-0.19}} & \textcolor[rgb]{ 1,  0,  0}{\textbf{-0.16}} & \textcolor[rgb]{ 1,  0,  0}{\textbf{-0.12}} & 1 &   &   &   &   &  \\
            Av  & \textcolor[rgb]{ 1,  0,  0}{\textbf{0.45}} & \textcolor[rgb]{ 1,  0,  0}{\textbf{0.52}} & \textcolor[rgb]{ 1,  0,  0}{\textbf{0.51}} & \textcolor[rgb]{ 1,  0,  0}{\textbf{-0.3}} & \textcolor[rgb]{ 1,  0,  0}{\textbf{0.38}} & \textcolor[rgb]{ 1,  0,  0}{\textbf{0.19}} & \textcolor[rgb]{ 1,  0,  0}{\textbf{0.17}} & \textcolor[rgb]{ 1,  0,  0}{\textbf{0.13}} & \textcolor[rgb]{ 1,  0,  0}{\textbf{0.43}} & -0.36 & 1 &   &   &   &  \\
            HD  & \textcolor[rgb]{ 1,  0,  0}{\textbf{0.48}} & \textcolor[rgb]{ 1,  0,  0}{\textbf{0.19}} & \textcolor[rgb]{ 1,  0,  0}{\textbf{0.61}} & \textcolor[rgb]{ 1,  0,  0}{\textbf{-0.2}} & \textcolor[rgb]{ 1,  0,  0}{\textbf{0.14}} & \textcolor[rgb]{ 1,  0,  0}{\textbf{0.17}} & \textcolor[rgb]{ 1,  0,  0}{\textbf{0.21}} & \textcolor[rgb]{ 1,  0,  0}{\textbf{0.38}} & \textcolor[rgb]{ 1,  0,  0}{\textbf{0.55}} & -0.14 & 0.69 & 1 &   &   &  \\
            CS  & \textcolor[rgb]{ 1,  0,  0}{\textbf{-0.45}} & \textcolor[rgb]{ 1,  0,  0}{\textbf{-0.76}} & \textcolor[rgb]{ 1,  0,  0}{\textbf{0.24}} & \textcolor[rgb]{ 1,  0,  0}{\textbf{0.14}} & \textcolor[rgb]{ 1,  0,  0}{\textbf{-0.81}} & \textcolor[rgb]{ 1,  0,  0}{\textbf{-0.48}} & \textcolor[rgb]{ 1,  0,  0}{\textbf{-0.14}} & \textcolor[rgb]{ 1,  0,  0}{\textbf{0.38}} & \textcolor[rgb]{ 1,  0,  0}{\textbf{0.38}} & \textcolor[rgb]{ 1,  0,  0}{\textbf{-0.02}} & \textcolor[rgb]{ 1,  0,  0}{\textbf{-0.57}} & \textcolor[rgb]{ 1,  0,  0}{\textbf{-0.02}} & 1 &   &  \\
            M  & \textcolor[rgb]{ 1,  0,  0}{\textbf{0.02}} & \textcolor[rgb]{ 1,  0,  0}{\textbf{0.03}} & \textcolor[rgb]{ 1,  0,  0}{\textbf{-0.29}} & \textcolor[rgb]{ 1,  0,  0}{\textbf{-0.4}} & \textcolor[rgb]{ 1,  0,  0}{\textbf{0.14}} & \textcolor[rgb]{ 1,  0,  0}{\textbf{0.14}} & \textcolor[rgb]{ 1,  0,  0}{\textbf{0.38}} & \textcolor[rgb]{ 1,  0,  0}{\textbf{-0.09}} & \textcolor[rgb]{ 1,  0,  0}{\textbf{-0.31}} & \textcolor[rgb]{ 1,  0,  0}{\textbf{-0.24}} & \textcolor[rgb]{ 1,  0,  0}{\textbf{-0.4}} & \textcolor[rgb]{ 1,  0,  0}{\textbf{-0.19}} & 0.26 & 1 &  \\
            OS  & \textcolor[rgb]{ 1,  0,  0}{\textbf{-0.57}} & \textcolor[rgb]{ 1,  0,  0}{\textbf{-0.45}} & \textcolor[rgb]{ 1,  0,  0}{\textbf{0.31}} & \textcolor[rgb]{ 1,  0,  0}{\textbf{-0.3}} & \textcolor[rgb]{ 1,  0,  0}{\textbf{-0.5}} & \textcolor[rgb]{ 1,  0,  0}{\textbf{-0.55}} & \textcolor[rgb]{ 1,  0,  0}{\textbf{-0.24}} & \textcolor[rgb]{ 1,  0,  0}{\textbf{0.35}} & \textcolor[rgb]{ 1,  0,  0}{\textbf{0.05}} & \textcolor[rgb]{ 1,  0,  0}{\textbf{-0.38}} & \textcolor[rgb]{ 1,  0,  0}{\textbf{-0.07}} & \textcolor[rgb]{ 1,  0,  0}{\textbf{0.02}} & 0.24 & -0.29 & 1 \\
            \hline
            \end{tabular}%
        \end{table}%

        \begin{table}[htbp!]
          \centering
          \caption{Correlation of the basic, microscopic, and mesoscopic rankings for AMAZON dataset.}
          \label{table89}
            \begin{tabular}{lccc}
            \hline
              & Basic & Micro & Meso \\
              \hline
            Basic & 1 &   &  \\
            Micro & 0.16 & 1 &  \\
            Meso & 0.38 & -0.4 & 1 \\
            \hline
            \end{tabular}%
        \end{table}%

        \begin{table}[!ht]
          \centering

          \caption{Quality metrics values for AMAZON ground-truth and the uncovered community structure. The calculated properties are Average Degree (AD), Average ODF (AO), Flake ODF (FO), Internal Density (ID), Max ODF (MO), and Overlapping Modularity (OM).}
          \label{table45}
            \begin{tabular}{lcccccc}
            \hline
              & AD & AO & FO & ID & MO & OM \\
            \hline
            CFINDER&3.44&3.48&1.49&0.73&11.97&0.45\\
            LFM&1.57&3.42&4.23&0.34&8.45&0.32\\
            GCE&4.29&1.18&1.23&0.43&6.85&0.47\\
            OSLOM&4.17&1.31&2.08&0.33&10.3&0.31\\
            SVINET&2.66&3.88&5.71&2.01&12.01&0.46\\
            MOSES&3.73&4.32&2.04&0.61&20.23&0.22\\
            SLPA&3.09&1.97&2.86&0.46&7.29&0.5\\
            DEMON&4.45&2.83&4.58&0.34&22.26&0.4\\
            \hline
            \end{tabular}%
        \end{table}%

        \begin{table}[!ht]
          \centering

          \caption{Quality metrics ranking for overlapping community detection algorithms with the AMAZON dataset. The calculated properties are Average Degree (AD), Average ODF (AO), Flake ODF (FO), Internal Density (ID), Max ODF (MO), and Overlapping Modularity (OM).  Kconsensus  and TOPSIS denotes respectively the final ranking using Kemeny consensus and TOPSIS.}
          \label{table52}
            \begin{tabular}{lcccccccc}
            \hline
              & AD & AO & FO & ID & MO & OM & Kconsensus & TOPSIS  \\
            \hline
            CFINDER&4&3&7&2&1&3&4&3\\
            LFM&3&4&3&6&4&6&3&2\\
            GCE&7&8&8&5&6&1&5&6\\
            OSLOM&6&7&5&8&3&7&8&7\\
            SVINET&1&2&1&1&2&2&1&1\\
            MOSES&5&1&6&3&7&8&6&5\\
            SLPA&2&6&4&4&5&4&2&4\\
            DEMON&8&5&2&6&8&5&7&8\\
            \hline
            \end{tabular}%
    \end{table}%

        \begin{table}[htbp!]
          \centering
          \caption{Correlation of the quality metrics ranking for AMAZON dataset. The calculated properties are Average Degree (AD), Average ODF (AO), Flake ODF (FO), Internal Density (ID), Max ODF (MO), and Overlapping Modularity (OM).}
          \label{table83}%
            \begin{tabular}{lcccccc}
            \hline
              & AD & AO & FO & ID & MO & OM \\
              \hline
            AD & 1 &   &   &   &   &  \\
            AO & 0.45 & 1 &   &   &   &  \\
            FO & 0.38 & 0.29 & 1 &   &   &  \\
            ID & 0.59 & 0.69 & 0.04 & 1 &   &  \\
            MO & 0.6 & 0.17 & 0 & 0.34 & 1 &  \\
            OM & 0.17 & -0.26 & -0.1 & 0.44 & 0.29 & 1 \\
            \hline
            \end{tabular}%
        \end{table}%

        \begin{table}[!ht]
          \centering
          \caption{Clustering metrics for AMAZON ground-truth and the uncovered community structure by overlapping community detection algorithms. The calculated properties are NMI, Omega Index (OI) and F1-score.}
          \label{table27}
            \begin{tabular}{lccc}
            \hline
              &  NMI  &  OI  &  F1-score \\
            \hline
            CFINDER&0.26&0.44&0.14\\
            LFM&0.08&0.12&0.06\\
            GCE&0.19&0.27&0.13\\
            OSLOM&0.13&0.09&0.11\\
            SVINET&0.15&0.17&0.6\\
            MOSES&0.22&0.21&0.21\\
            SLPA&0.23&0.31&0.36\\
            DEMON&0.41&0.15&0.61\\
            \hline
            \end{tabular}%
        \end{table}%

        \begin{table}[!ht]
          \centering
          \caption{Clustering metrics ranking for overlapping community detection algorithms with the AMAZON dataset. The calculated properties are NMI, Omega Index (OI) and F1-score.  Kconsensus  and TOPSIS denotes respectively the final ranking using Kemeny consensus and TOPSIS.}
                \label{table33}
            \begin{tabular}{lccccc}
            \hline
              & NMI & OI & F1-score & Kconsensus & TOPSIS \\
            \hline
            CFINDER & 2 & 1 & 5 & 2 & 2 \\
            LFM & 8 & 7 & 8 & 8 & 8 \\
            GCE & 5 & 3 & 6 & 5 & 6 \\
            OSLOM & 7 & 8 & 7 & 7 & 7 \\
            SVINET & 6 & 5 & 2 & 4 & 4 \\
            MOSES & 4 & 4 & 4 & 6 & 5 \\
            SLPA & 3 & 2 & 3 & 3 & 3 \\
            DEMON & 1 & 6 & 1 & 1 & 1 \\
            \hline
            \end{tabular}%
        \end{table}

        \begin{table}[htbp!]
          \centering
          \caption{Correlation of the clustering metrics ranking for overlapping community detection algorithms with the  AMAZON dataset. The calculated properties are NMI, Omega Index (OI) and F1-score.}
          \label{table86}
            \begin{tabular}{lccc}
            \hline
              & NMI & OI & F1-score \\
              \hline
            NMI & 1 &   &  \\
            OI & 0.59 & 1 &  \\
            F1-score & 0.69 & 0.26& 1 \\
            \hline
            \end{tabular}%

        \end{table}%

        \begin{table}[ht!]
          \centering
          \caption{Correlation of the topological properties, the quality metrics and the clustering measures rankings using the Kconsensus strategy for AMAZON dataset.}
          \label{table98}%
            \begin{tabular}{lccc}
            \hline
              & \multicolumn{1}{l}{Topo} & \multicolumn{1}{l}{Quality} & \multicolumn{1}{l}{Clustering} \\
            \hline
            Topo & 1 &   &  \\
            Quality & 0.21 & 1 &  \\
            clustering & 0.64 & 0.09 & 1 \\
            \hline
            \end{tabular}%
        \end{table}%

        \begin{table}[ht!]
          \centering
          \caption{Correlation of the topological properties, the quality metrics and the clustering measures rankings using the TOPSIS strategy for AMAZON dataset.}
          \label{table980}%
            \begin{tabular}{lccc}
            \hline
              & \multicolumn{1}{l}{Topo} & \multicolumn{1}{l}{Quality} & \multicolumn{1}{l}{Clustering} \\
            \hline
            Topo & 1 &   &  \\
            Quality & 0.76 & 1 &  \\
            Clustering & 0.43 & -0.14 & 1 \\
            \hline
            \end{tabular}%
        \end{table}%

        \begin{table}[!ht]
          \centering
          \caption{Ranking of the algorithms based on all the properties with the AMAZON dataset. The calculated properties are Number of nodes (V), Number of edges (E), Density ($\rho$), Diameter ($d$), Average shortest path ($l_{G}$), Average node degree ($\widetilde{deg}$), Max node degree ($\delta(G)$), Assortativity Coefficient ($\tau$), Clustering Coefficient ($C$), the Degree distribution (DD), the Average clustering coefficient as function of degree (Av), the Hop distance (HD), the Community size (CS), the Membership (M), the Overlap size (OS), Average Degree (AD), Average ODF (AO), Flake ODF (FO), Internal Density (ID), Max ODF (MO), and Overlapping Modularity (OM), NMI, Omega Index (OI) and F1-score. Kconsensus  and TOPSIS denotes respectively the final ranking using Kemeny consensus and TOPSIS.}
          \label{table60}
          \scriptsize
            \begin{tabular}{{p{.9cm}|}*{5}{p{.008cm}}*{1}{p{.1cm}}*{1}{p{.18cm}}*{1}{p{.008cm}}*{1}{p{.008cm}|}*{2}{p{.08cm}}*{1}{p{.08cm}|}*{2}{p{.08cm}}*{1}{p{.08cm}|}*{5}{p{.08cm}}*{1}{p{.18cm}|}*{1}{p{.18cm}}*{1}{p{.04cm}}*{1}{p{.9cm}|}*{2}{p{.7cm}}}
            \hline
              & \multicolumn{9}{c|}{Basic properties} & \multicolumn{3}{c|}{Microscopic} & \multicolumn{3}{c|}{Mesoscopic} & \multicolumn{6}{c|}{Clustering} & \multicolumn{3}{c|}{Quality} & \multicolumn{2}{c}{MCDM Ranking} \\
              \hline
              &V&E&$\rho$&$d$&$l_{G}$&$\widetilde{deg}$&$\delta(G)$&$\tau$&$C$& DD & Av & HD & CS & M & OS & AD & AO & FO & ID & MO & OM & NMI & OI & F1-score & Kconsensus & TOPSIS \\
            \hline
            CFINDER & 4 & 5 & 4 & 3 & 5 & 3 & 1 & 1 & 5 & 4 & 8 & 5 & 1 & 1 & 4 & 4 & 3 & 7 & 2 & 1 & 3 & 2 & 1 & 5 & 5 & 2 \\
            LFM & 8 & 8 & 2 & 7 & 8 & 8 & 8 & 4 & 2 & 6 & 4 & 2 & 2 & 5 & 1 & 3 & 4 & 3 & 6 & 4 & 6 & 8 & 7 & 8 & 8 & 6 \\
            GCE & 6 & 6 & 6 & 5 & 6 & 6 & 6 & 7 & 4 & 5 & 6 & 7 & 3 & 2 & 8 & 7 & 8 & 8 & 5 & 6 & 1 & 5 & 3 & 6 & 6 & 7 \\
            OSLOM & 7 & 7 & 6 & 2 & 7 & 7 & 7 & 5 & 6 & 3 & 7 & 8 & 4 & 8 & 2 & 6 & 7 & 5 & 8 & 3 & 7 & 7 & 8 & 7 & 7 & 8 \\
            SVINET & 3 & 1 & 1 & 1 & 3 & 1 & 4 & 2 & 1 & 2 & 2 & 3 & 6 & 7 & 6 & 1 & 2 & 1 & 1 & 2 & 2 & 6 & 5 & 2 & 3 & 1 \\
            MOSES & 2 & 3 & 8 & 4 & 1 & 2 & 5 & 8 & 8 & 1 & 5 & 4 & 7 & 4 & 7 & 5 & 1 & 6 & 3 & 7 & 8 & 4 & 4 & 4 & 1 & 4 \\
            SLPA & 1 & 4 & 3 & 6 & 4 & 5 & 2 & 3 & 3 & 7 & 1 & 1 & 5 & 6 & 5 & 2 & 6 & 4 & 4 & 5 & 4 & 3 & 2 & 3 & 4 & 3 \\
            DEMON & 5 & 2 & 5 & 8 & 2 & 4 & 3 & 5 & 7 & 8 & 3 & 6 & 8 & 3 & 3 & 8 & 5 & 2 & 6 & 8 & 5 & 1 & 6 & 1 & 2 & 5 \\

            \hline
            \end{tabular}%
        \end{table}%

\clearpage
\section{aNobii}
        \begin{figure}[!ht]
        \subfigure[aNobii*]{\includegraphics[width=.121\textwidth]{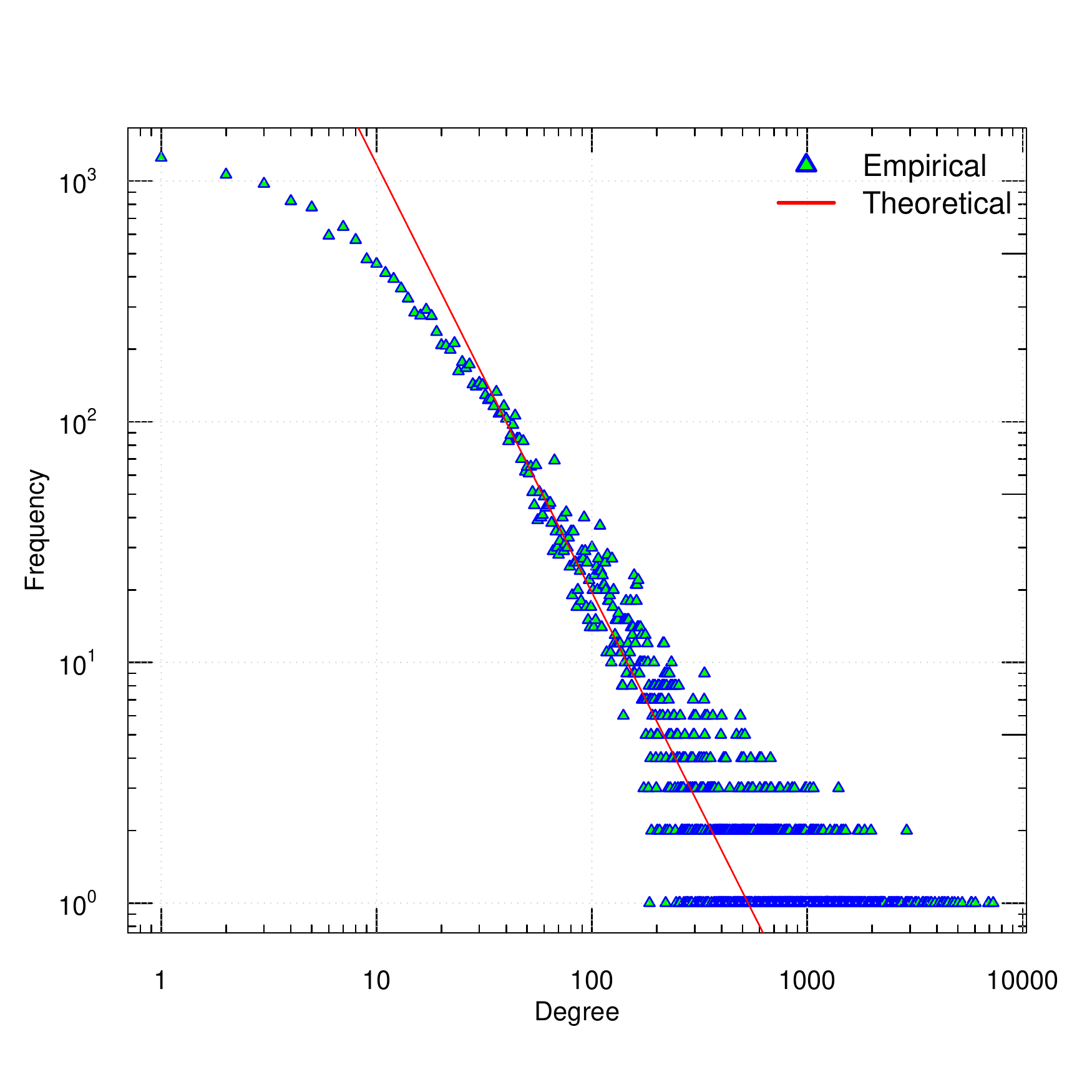}}
        \subfigure[LFM*]{\includegraphics[width=.121\textwidth]{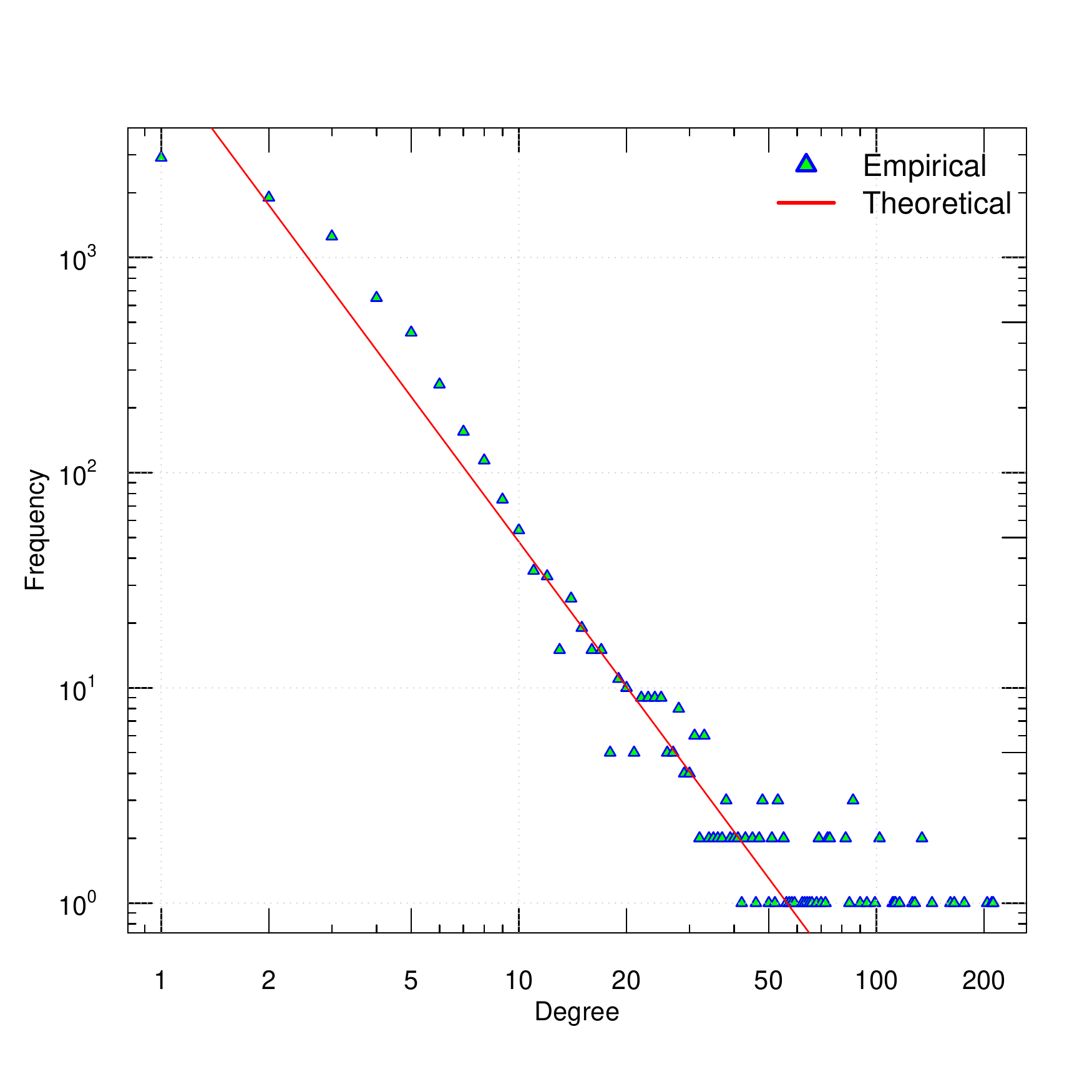}}
        \subfigure[GCE*]{\includegraphics[width=.121\textwidth]{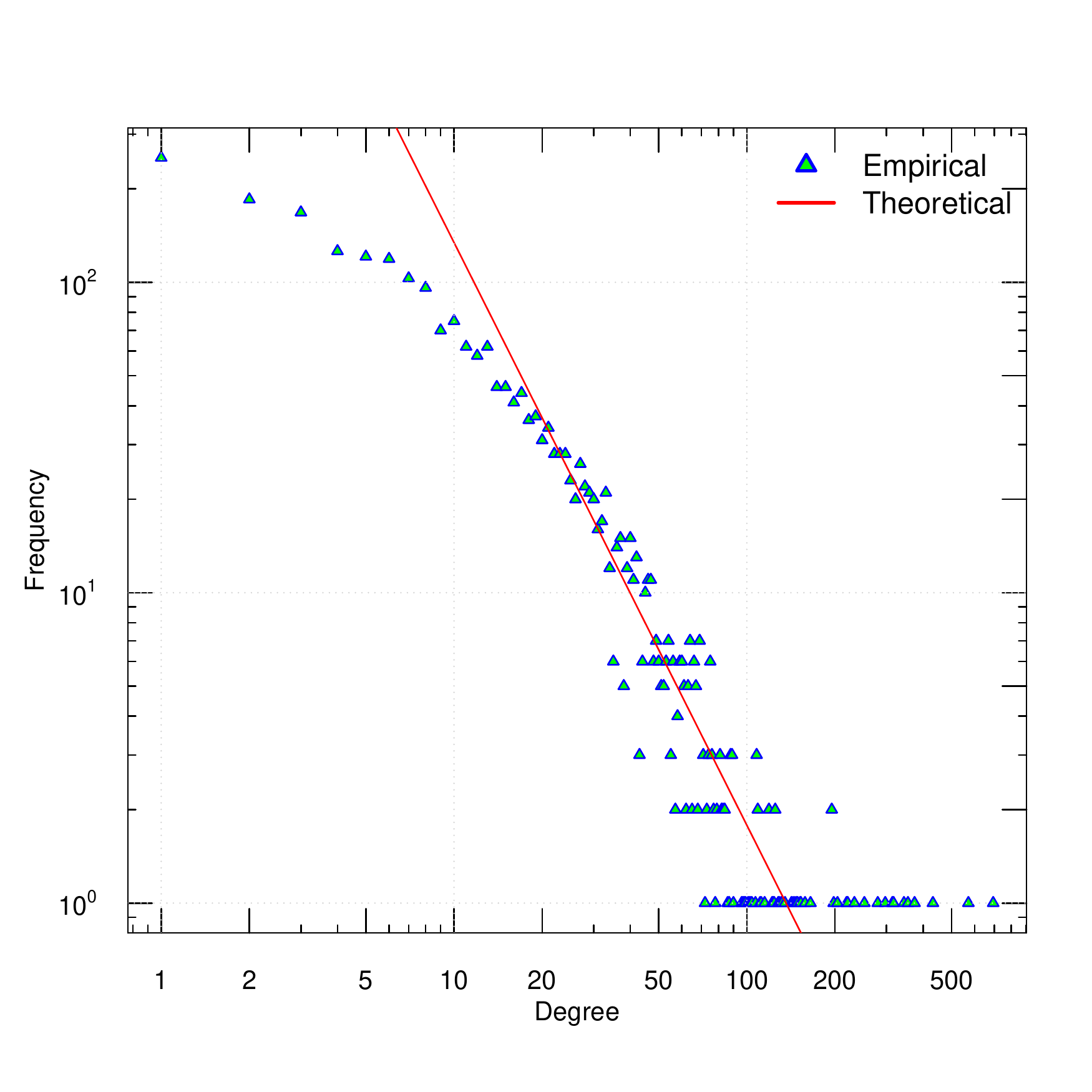}}
        \subfigure[OSLOM*]{\includegraphics[width=.121\textwidth]{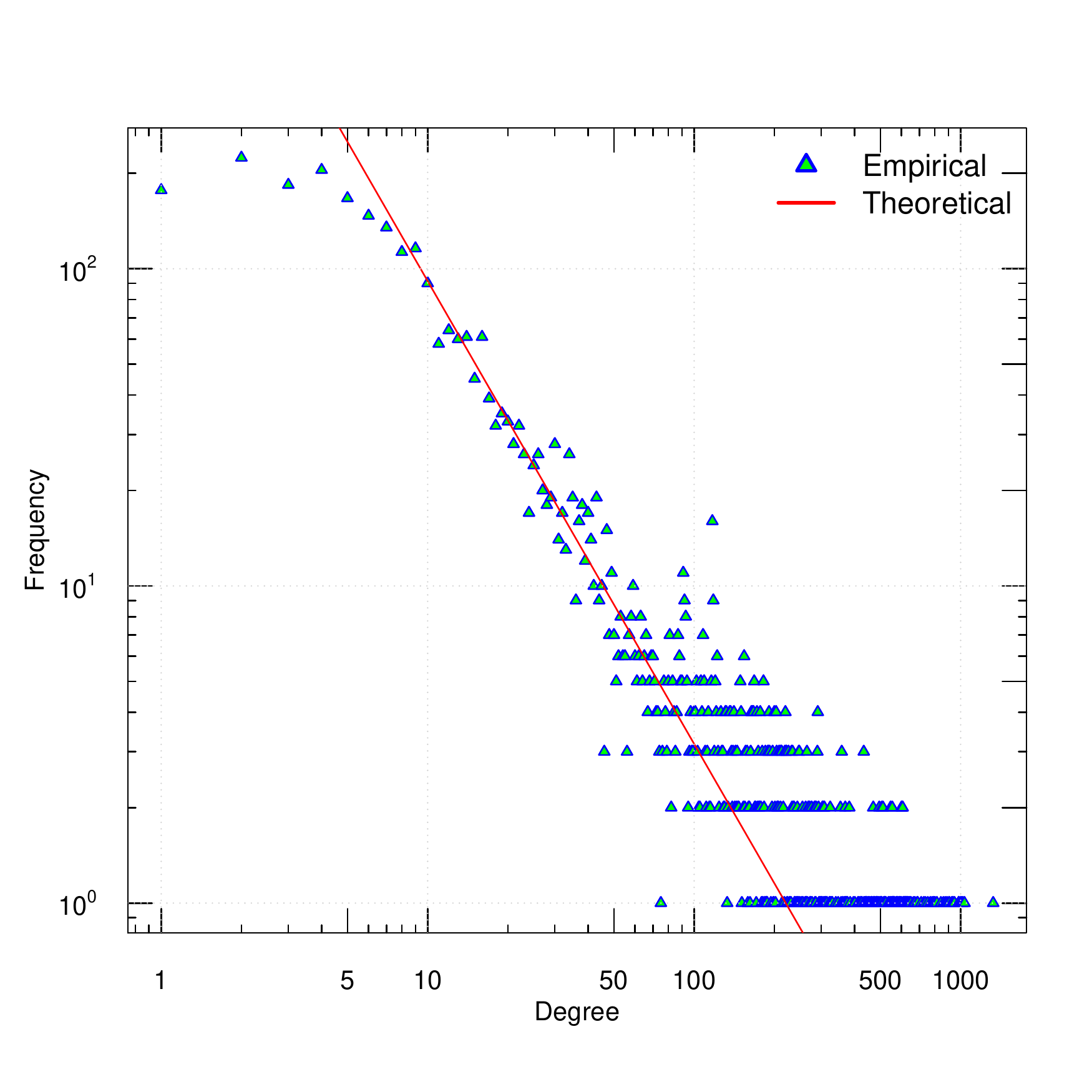}}
        \subfigure[MOSES*]{\includegraphics[width=.121\textwidth]{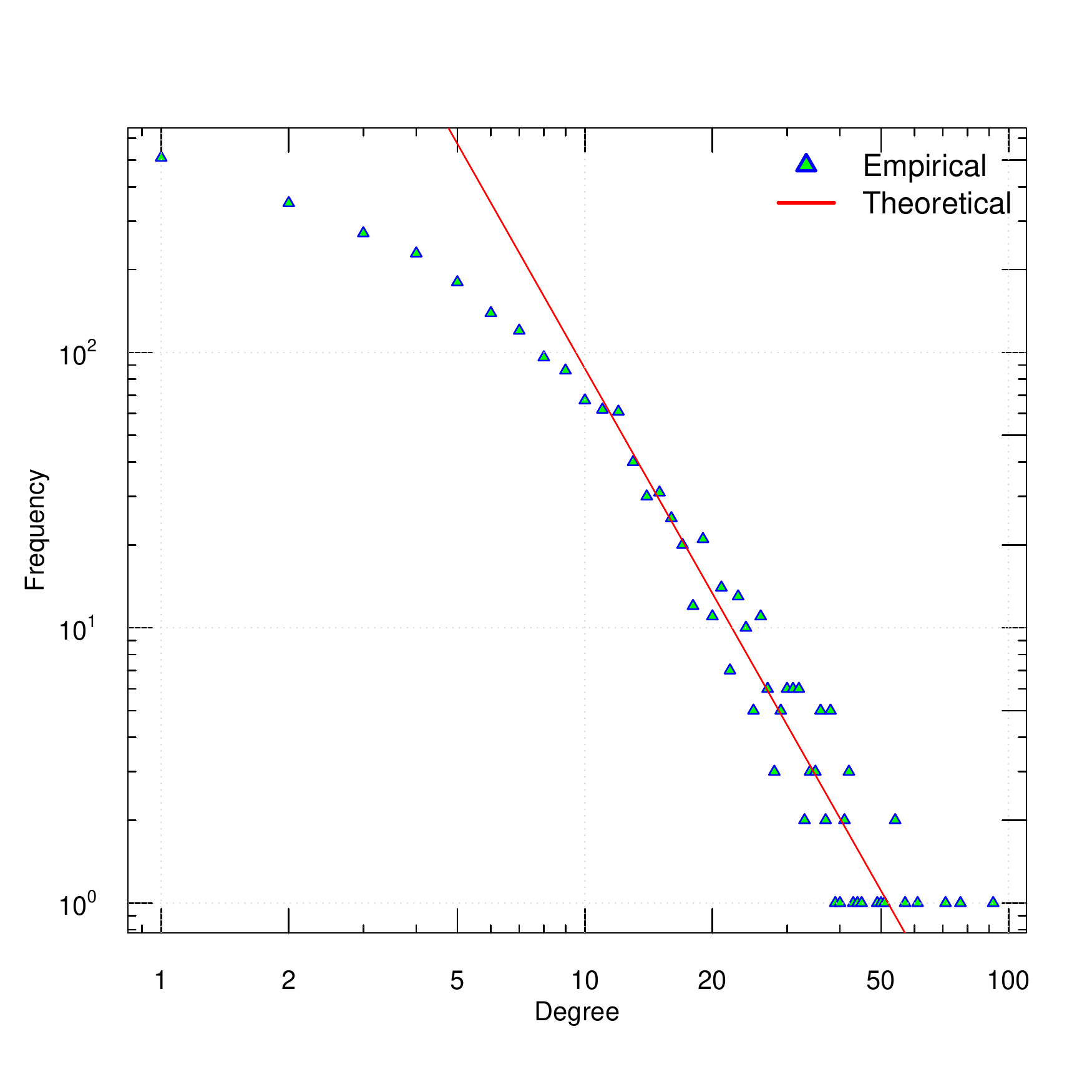}}
        \subfigure[SLPA*]{\includegraphics[width=.121\textwidth]{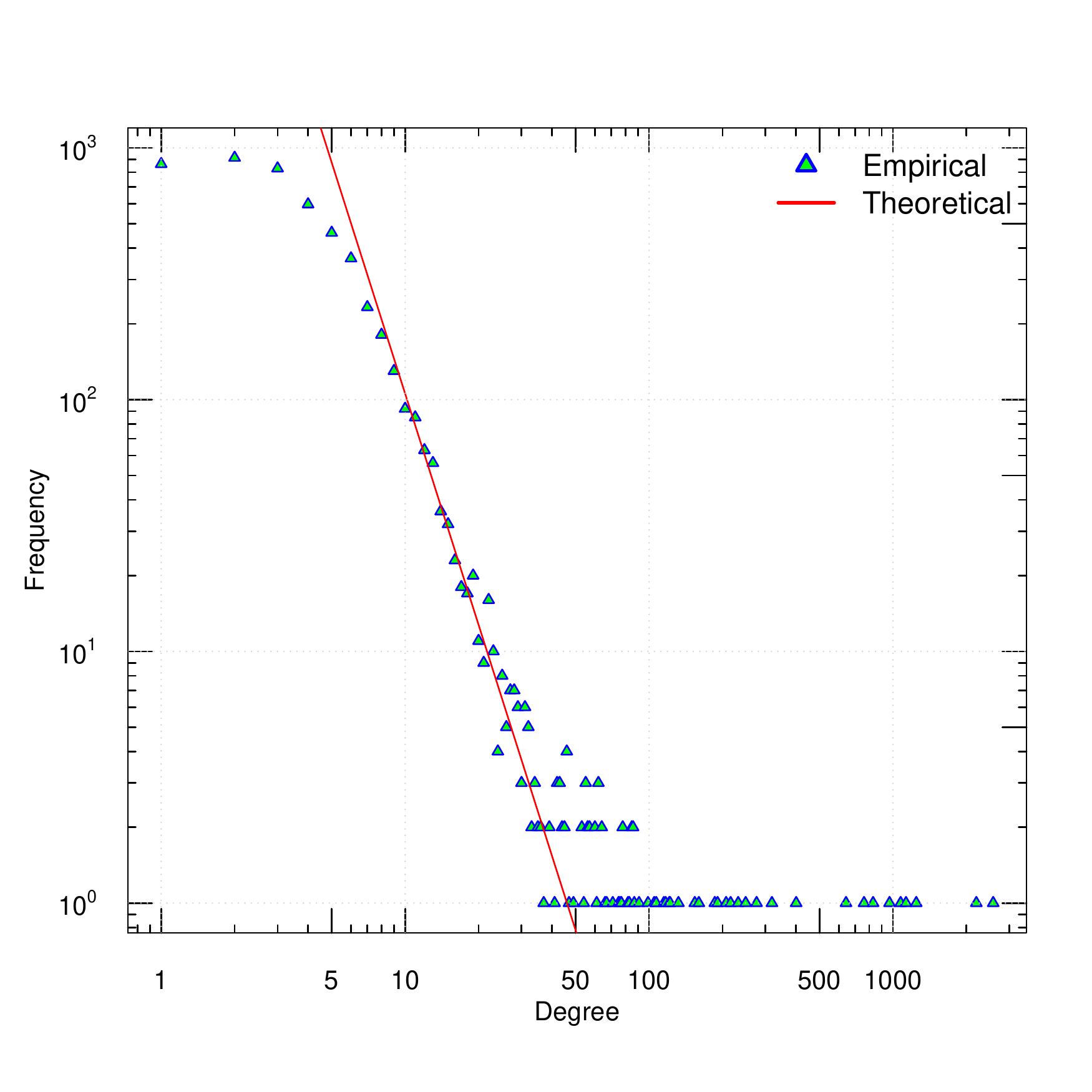}}
        \subfigure[DEMON*]{\includegraphics[width=.121\textwidth]{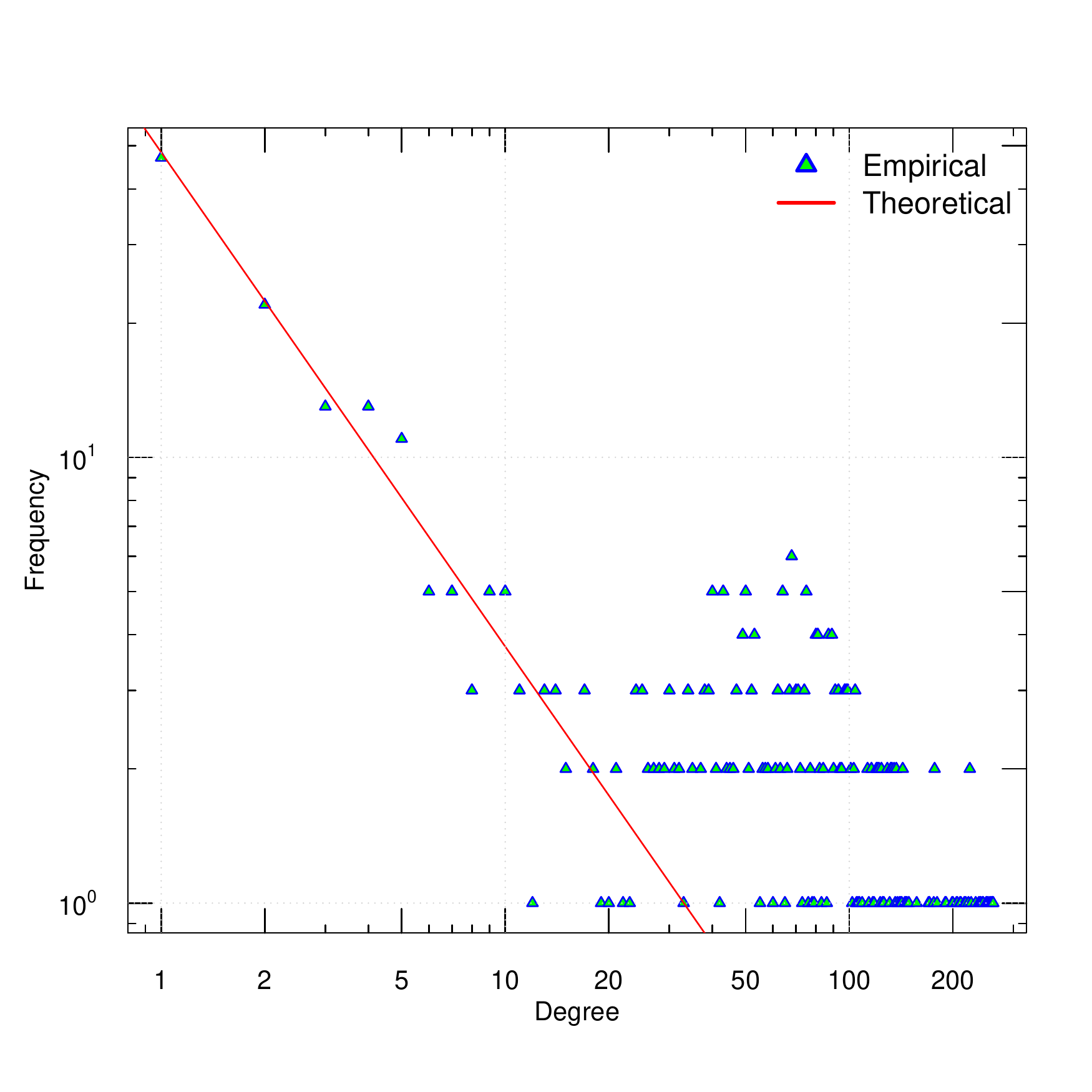}}

        \caption{\label{fig3}Log-log empirical degree distribution (dots) and Power-Law estimate (line) for aNobii* (a), LFM* (b),  GCE* (c), OSLOM* (d), MOSES* (e), SLPA* (f), and DEMON* (g))}
        \end{figure}

        \begin{figure}[!ht]
        \subfigure[aNobii*]{\includegraphics[width=.121\textwidth]{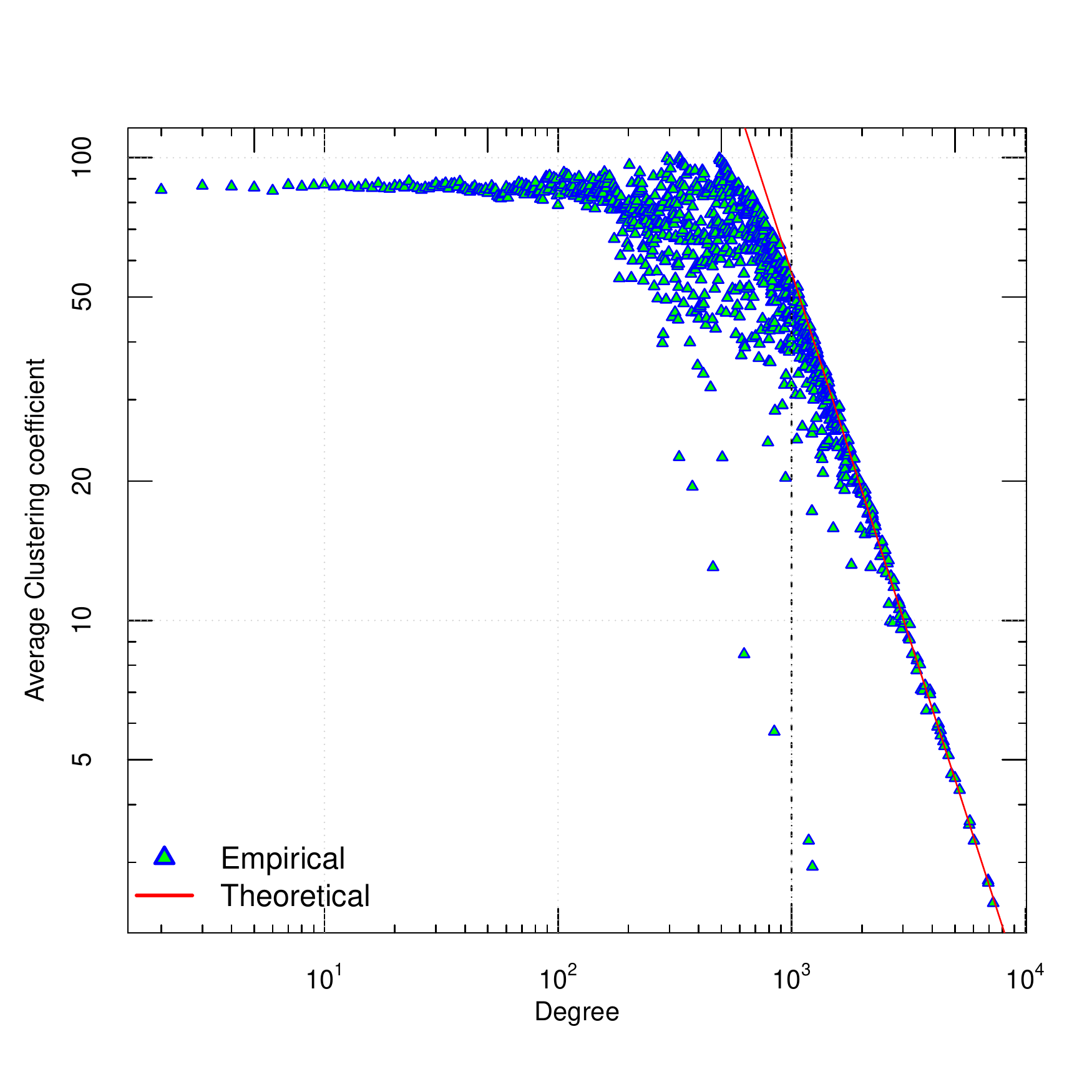}}
        \subfigure[LFM*]{\includegraphics[width=.121\textwidth]{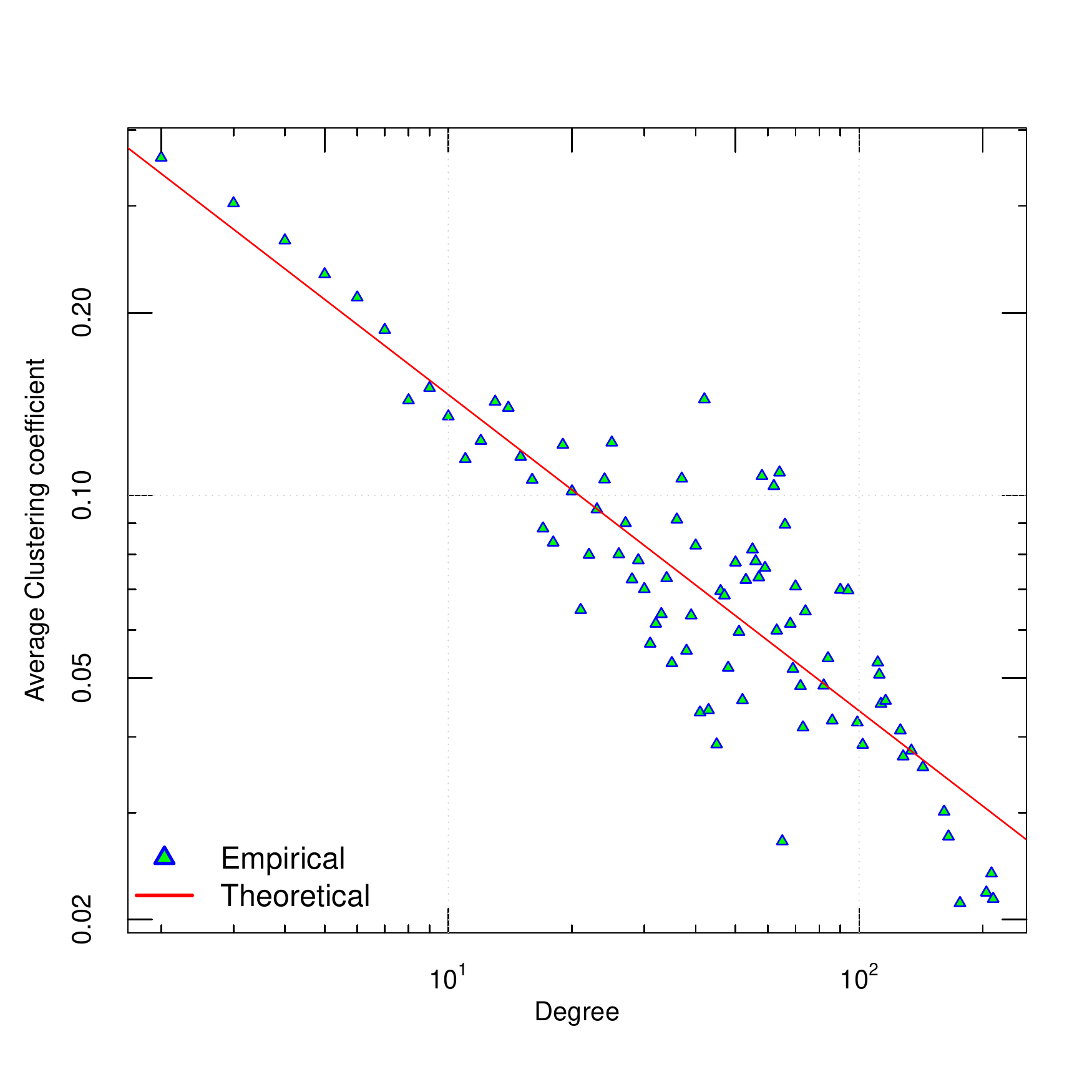}}
        \subfigure[GCE*]{\includegraphics[width=.121\textwidth]{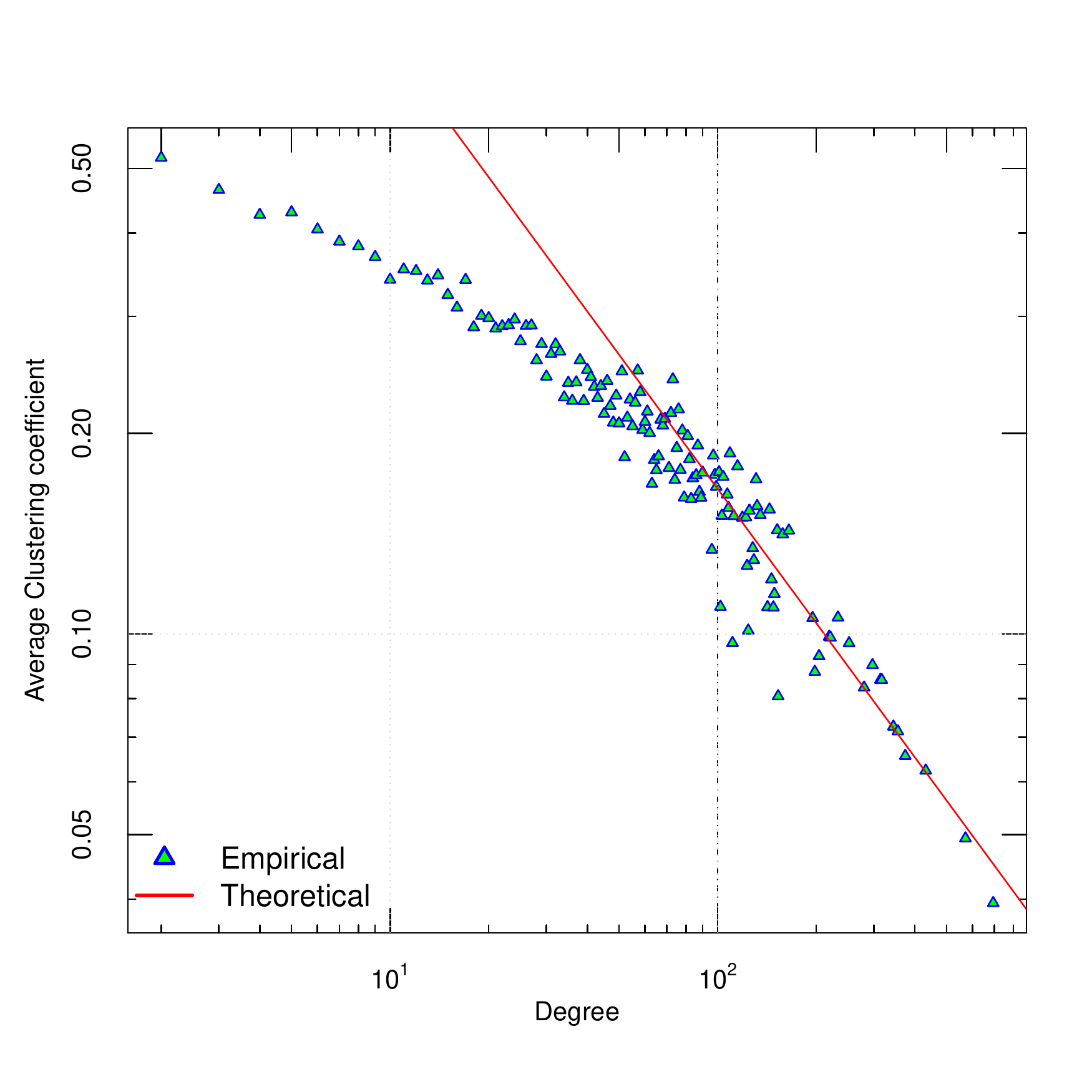}}
        \subfigure[OSLOM*]{\includegraphics[width=.121\textwidth]{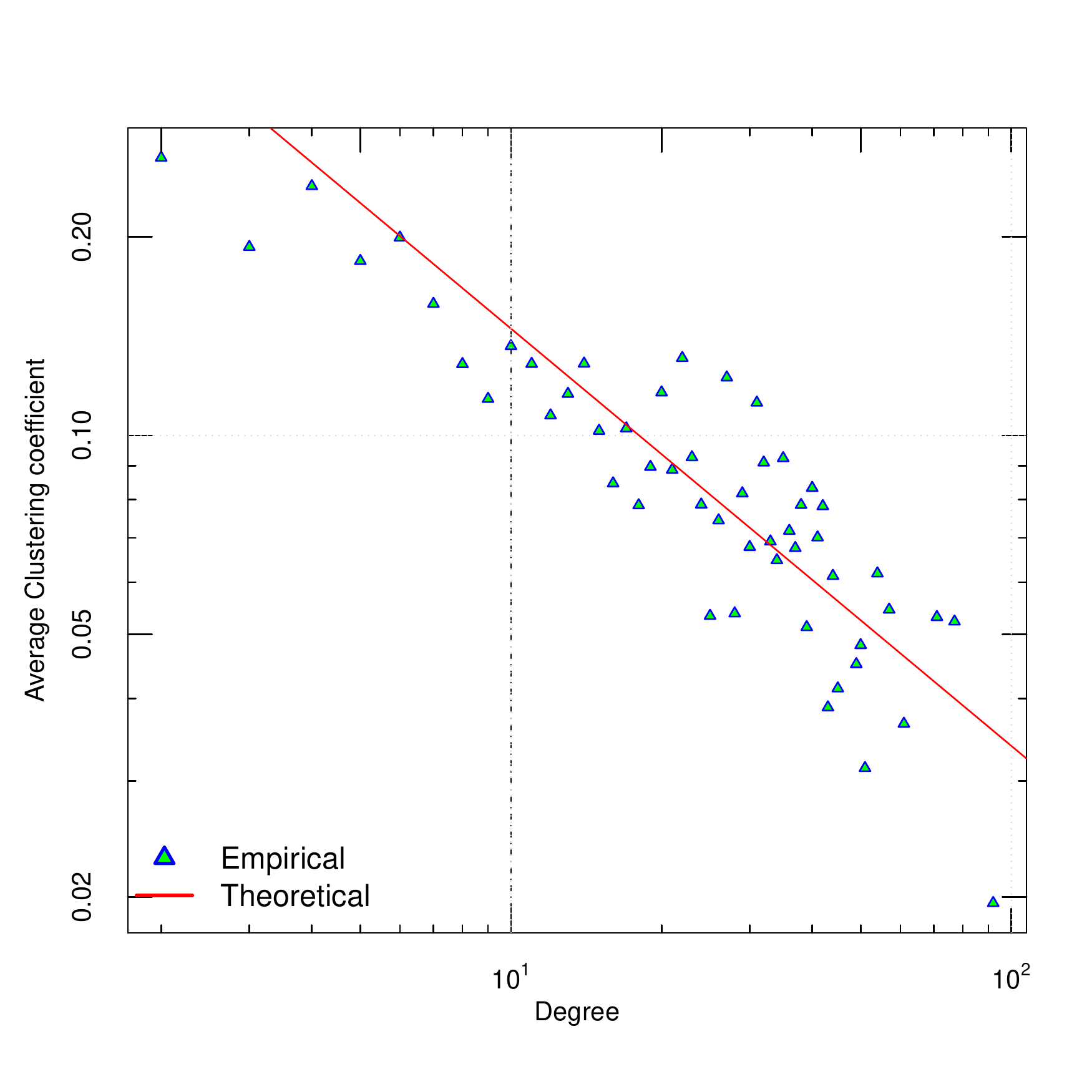}}
        \subfigure[MOSES*]{\includegraphics[width=.121\textwidth]{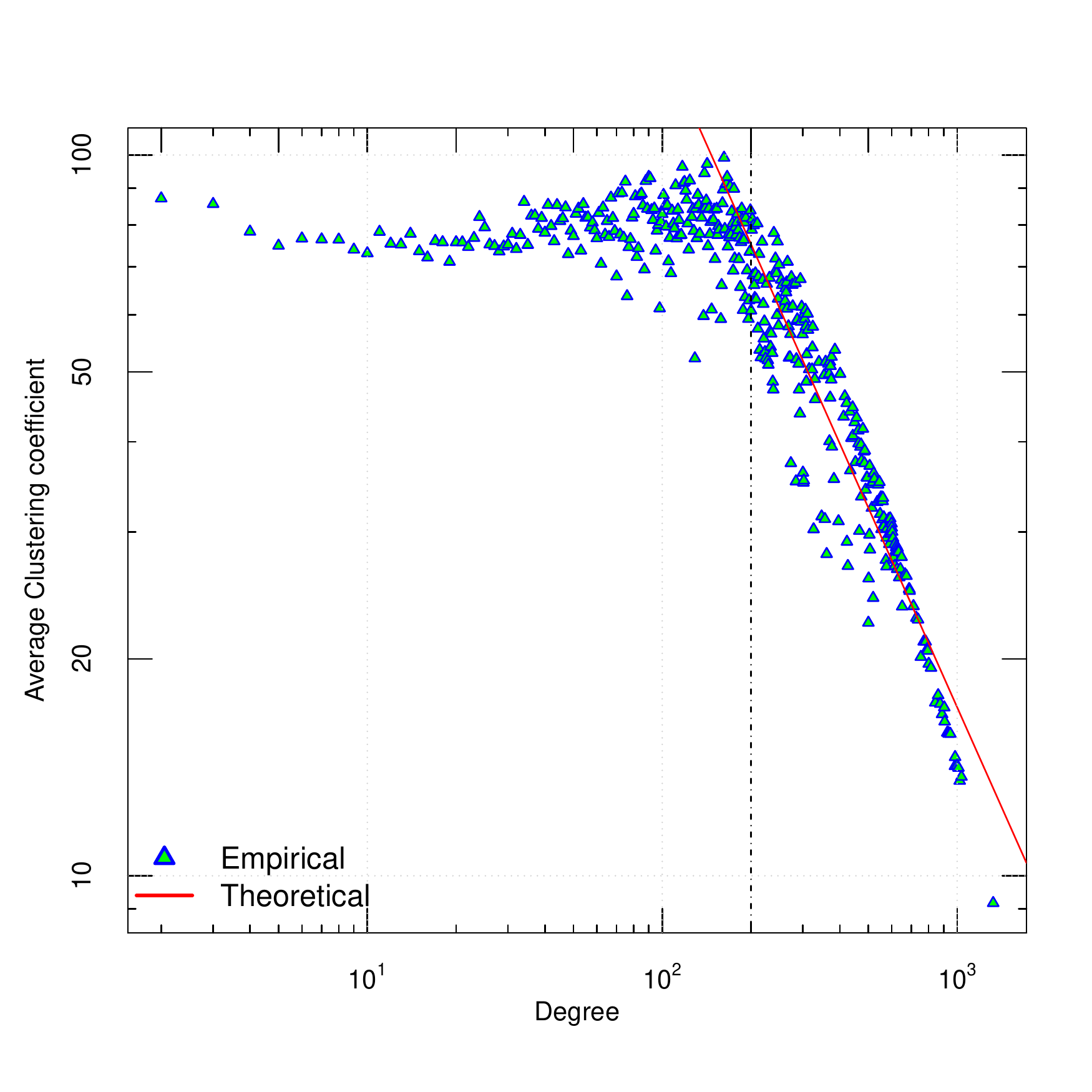}}
        \subfigure[SLPA*]{\includegraphics[width=.121\textwidth]{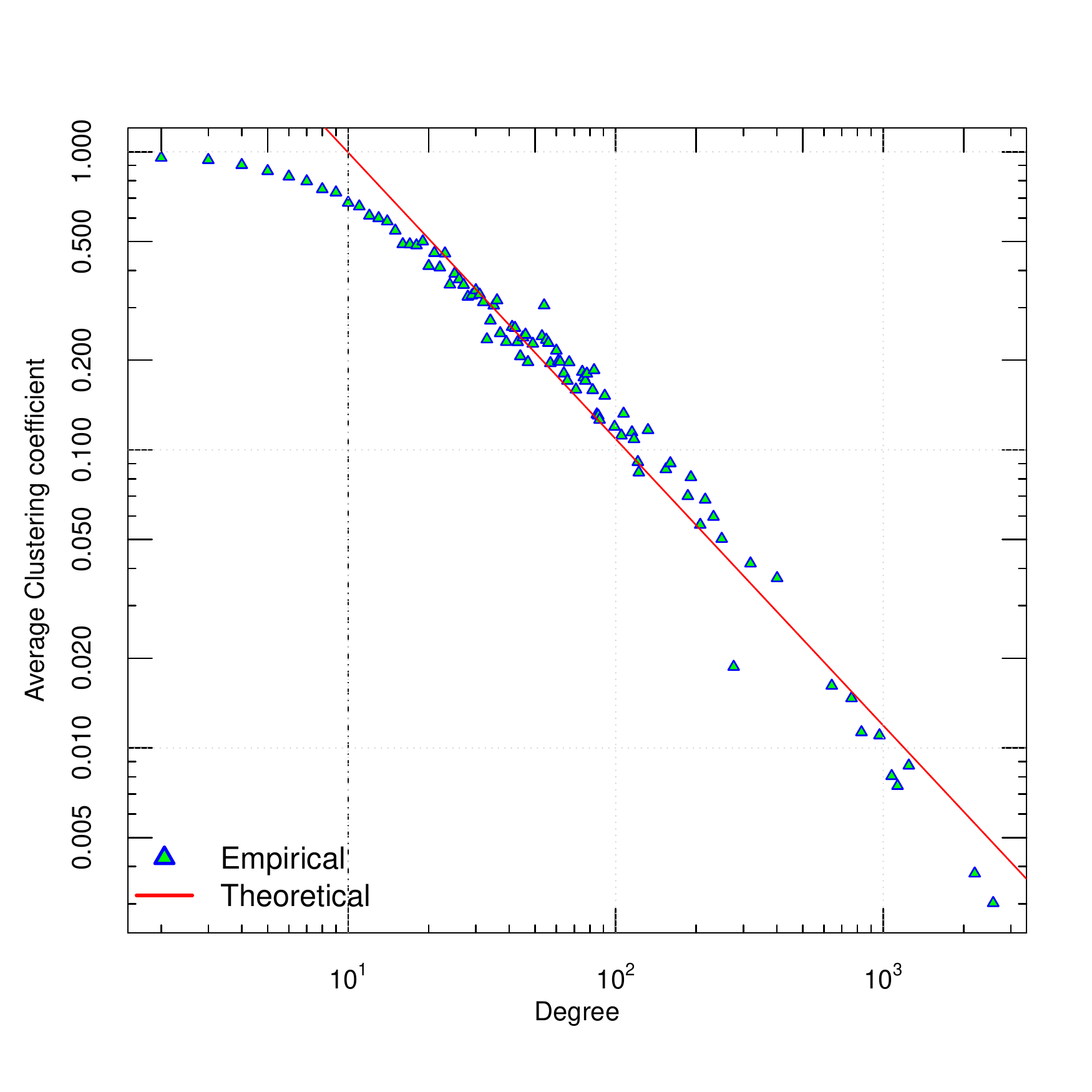}}
        \subfigure[DEMON*]{\includegraphics[width=.121\textwidth]{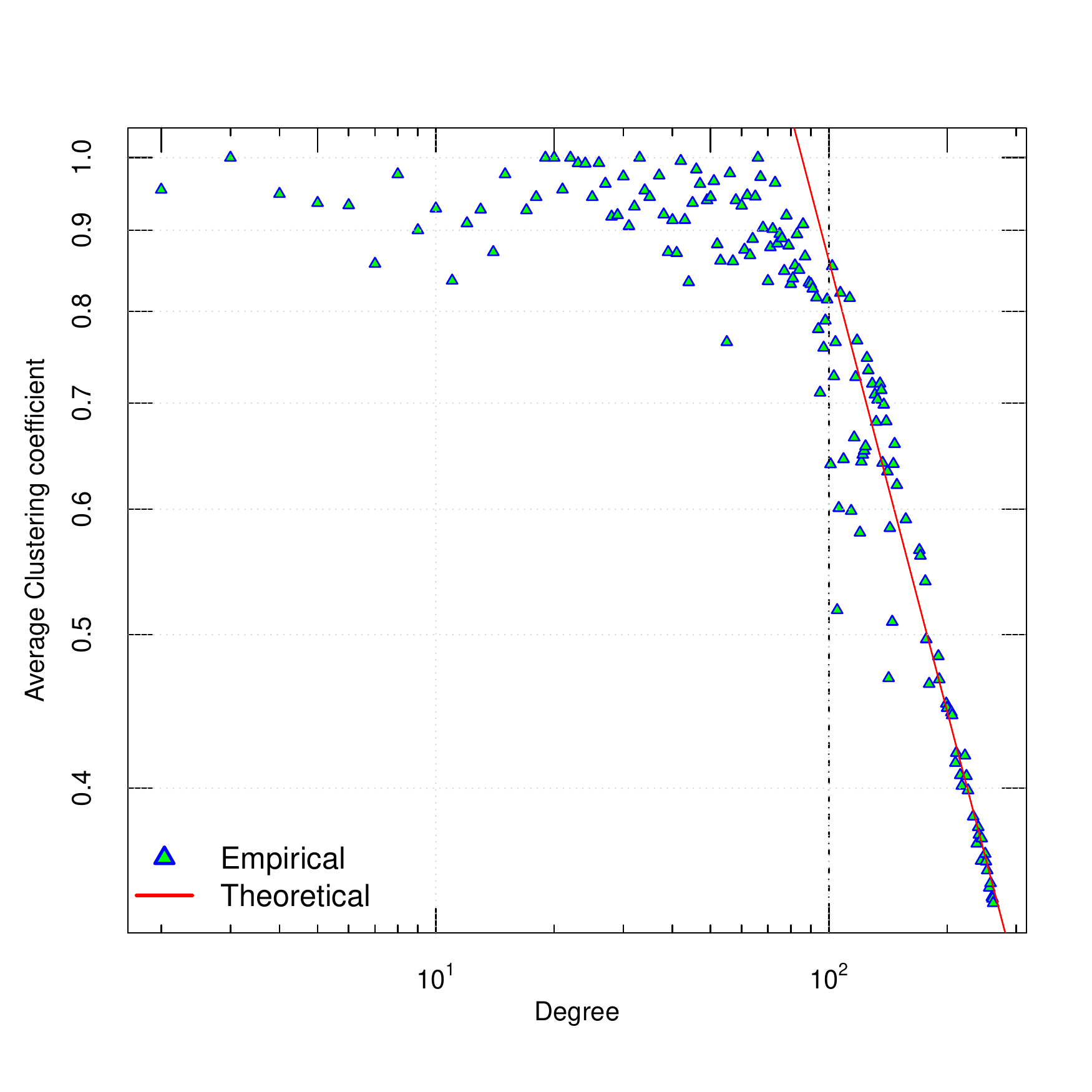}}

        \caption{\label{fig6}Log-log empirical Average clustering coefficient distributions as a function of the degree (dots) and Power-Law estimate (line) for aNobii* (a), LFM* (b),  GCE* (c), OSLOM* (d), MOSES* (e), SLPA* (f), and DEMON* (g)}
        \end{figure}

        \begin{figure}[!ht]
        \subfigure[aNobii*]{\includegraphics[width=.121\textwidth]{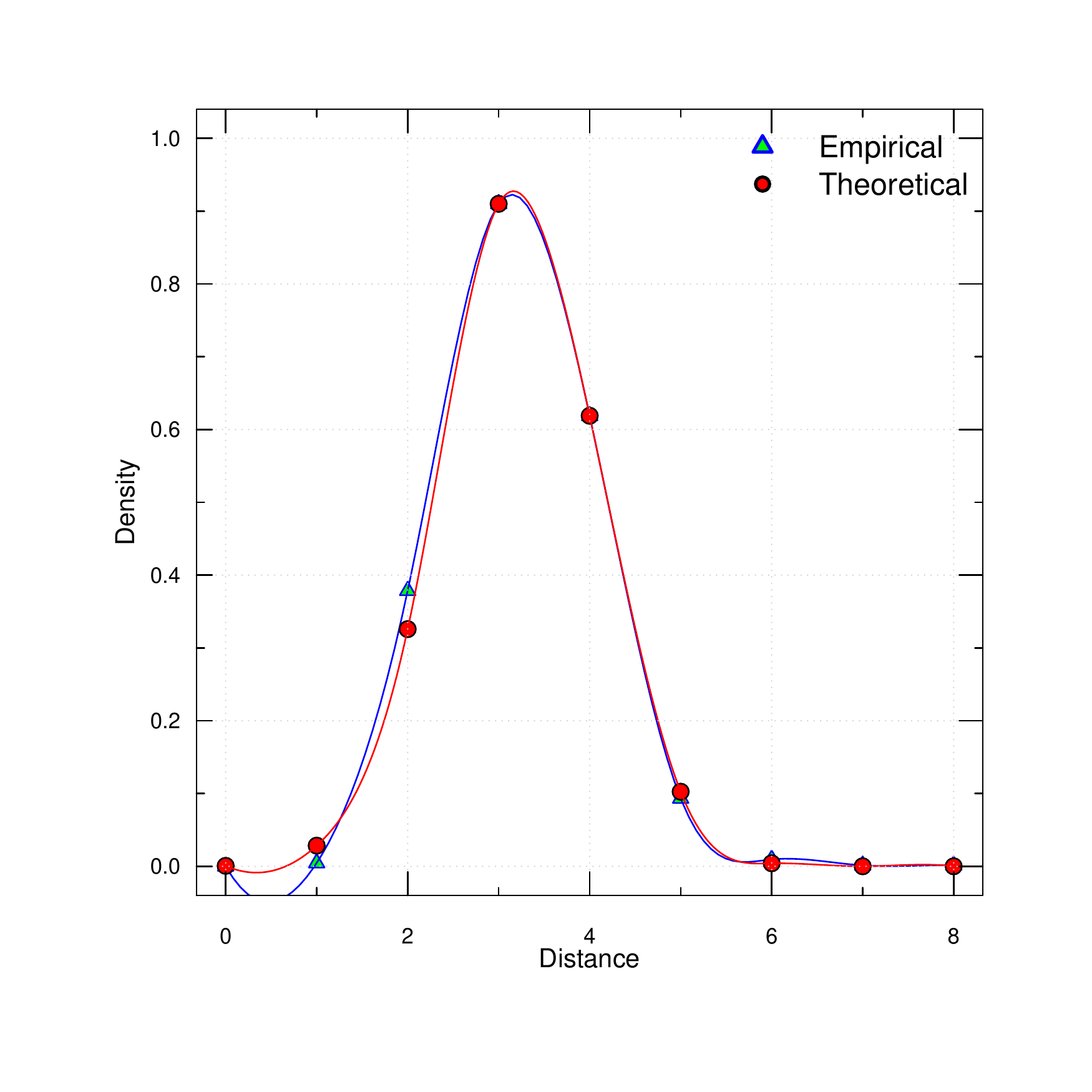}}
        \subfigure[LFM*]{\includegraphics[width=.121\textwidth]{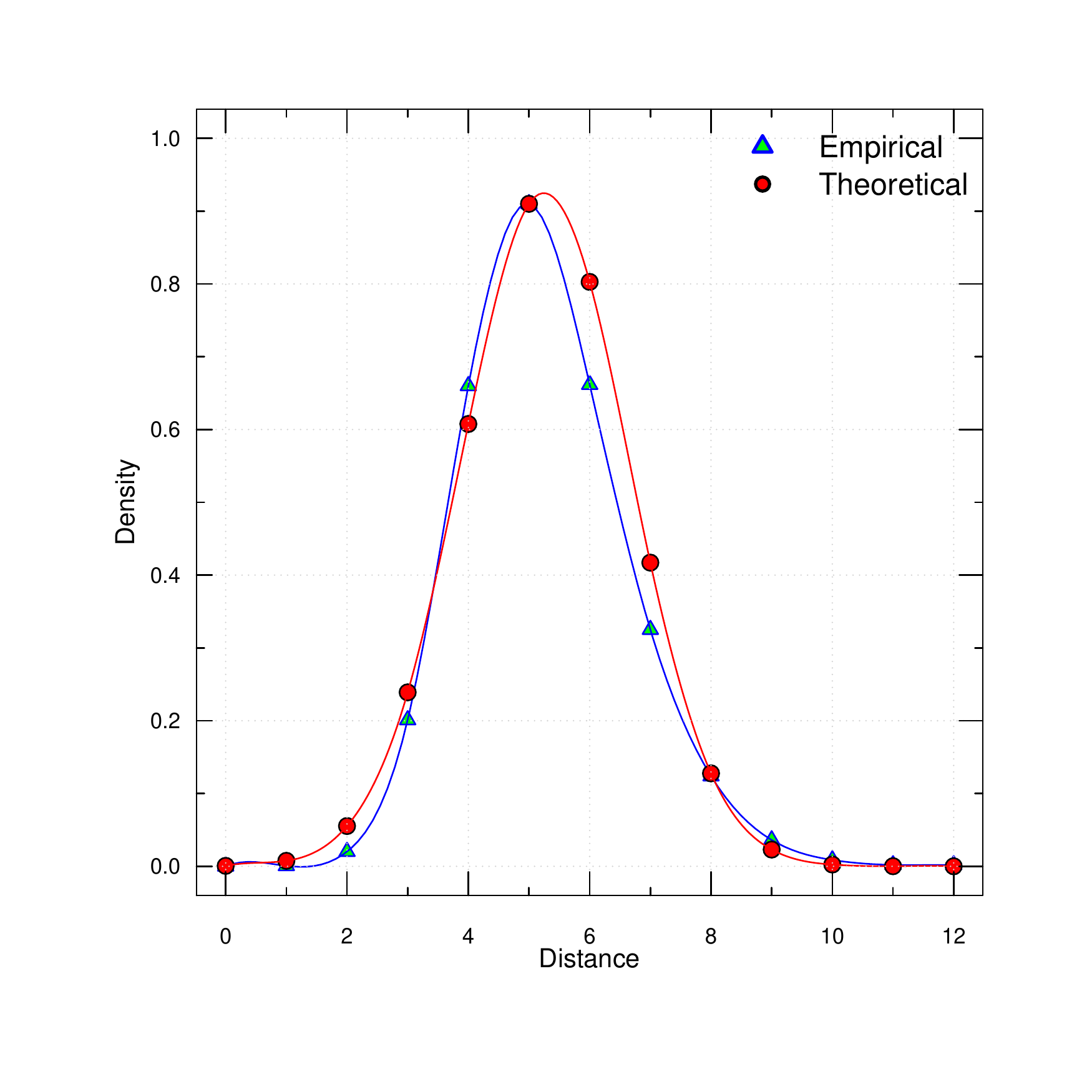}}
        \subfigure[GCE*]{\includegraphics[width=.121\textwidth]{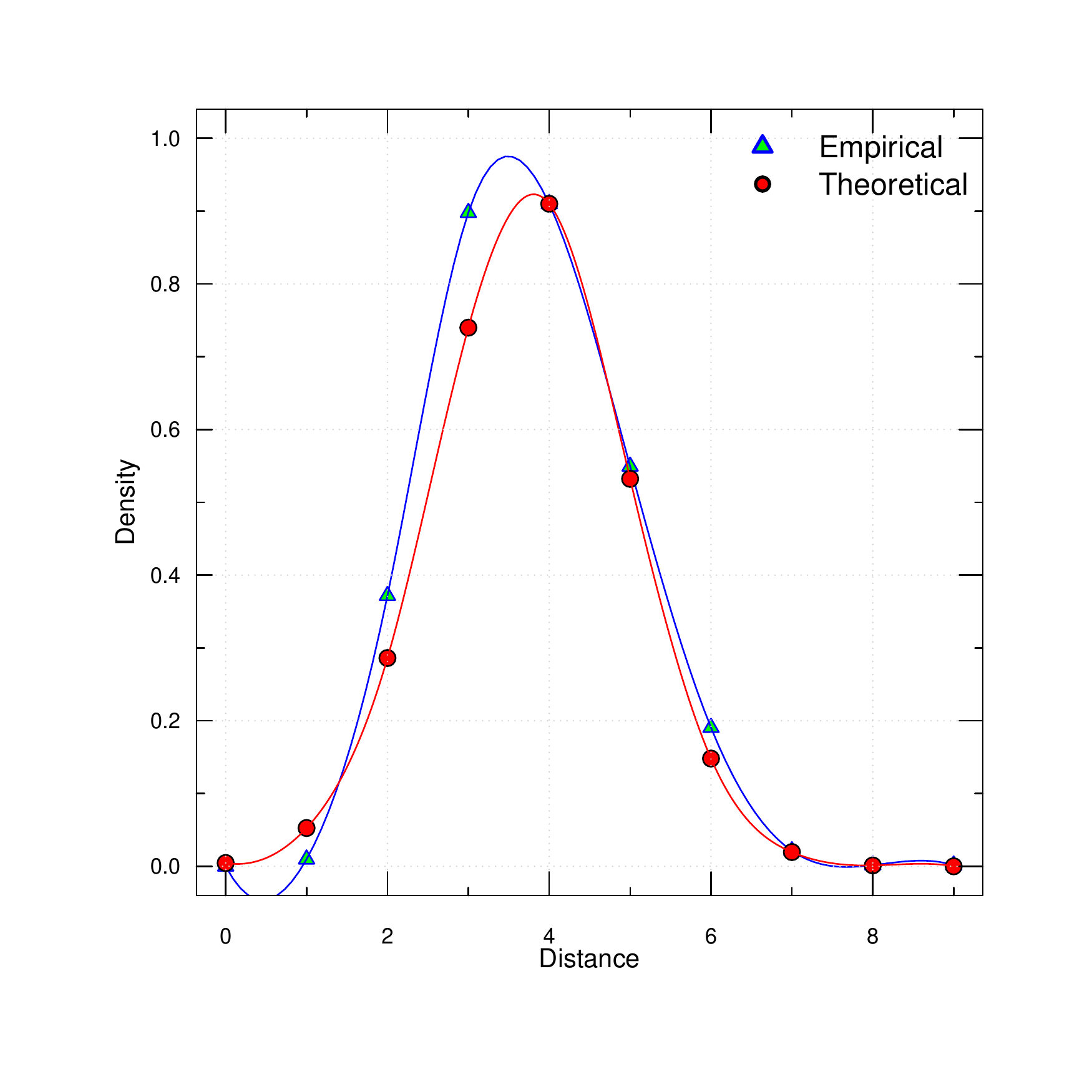}}
        \subfigure[OSLOM*]{\includegraphics[width=.121\textwidth]{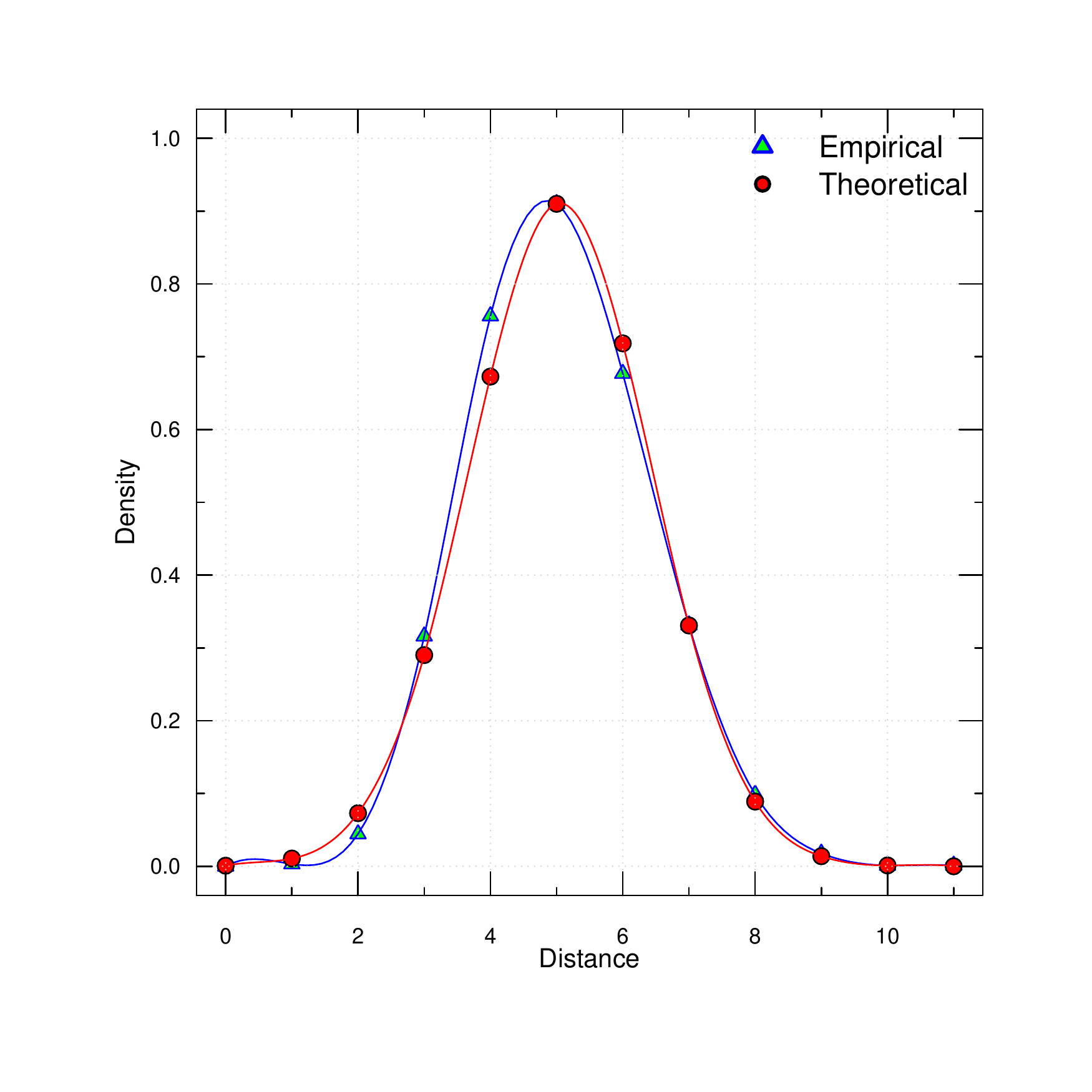}}
        \subfigure[MOSES*]{\includegraphics[width=.121\textwidth]{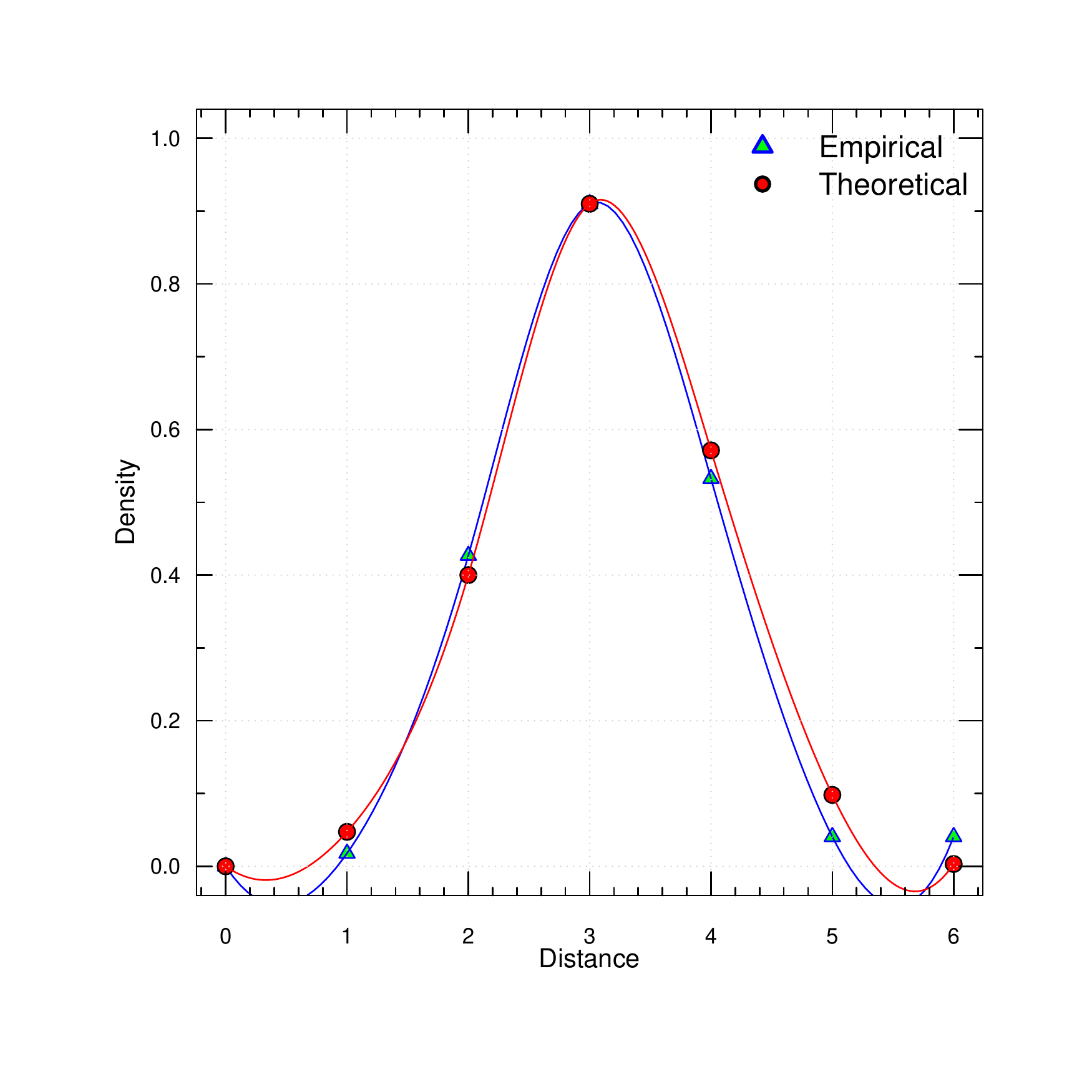}}
        \subfigure[SLPA*]{\includegraphics[width=.121\textwidth]{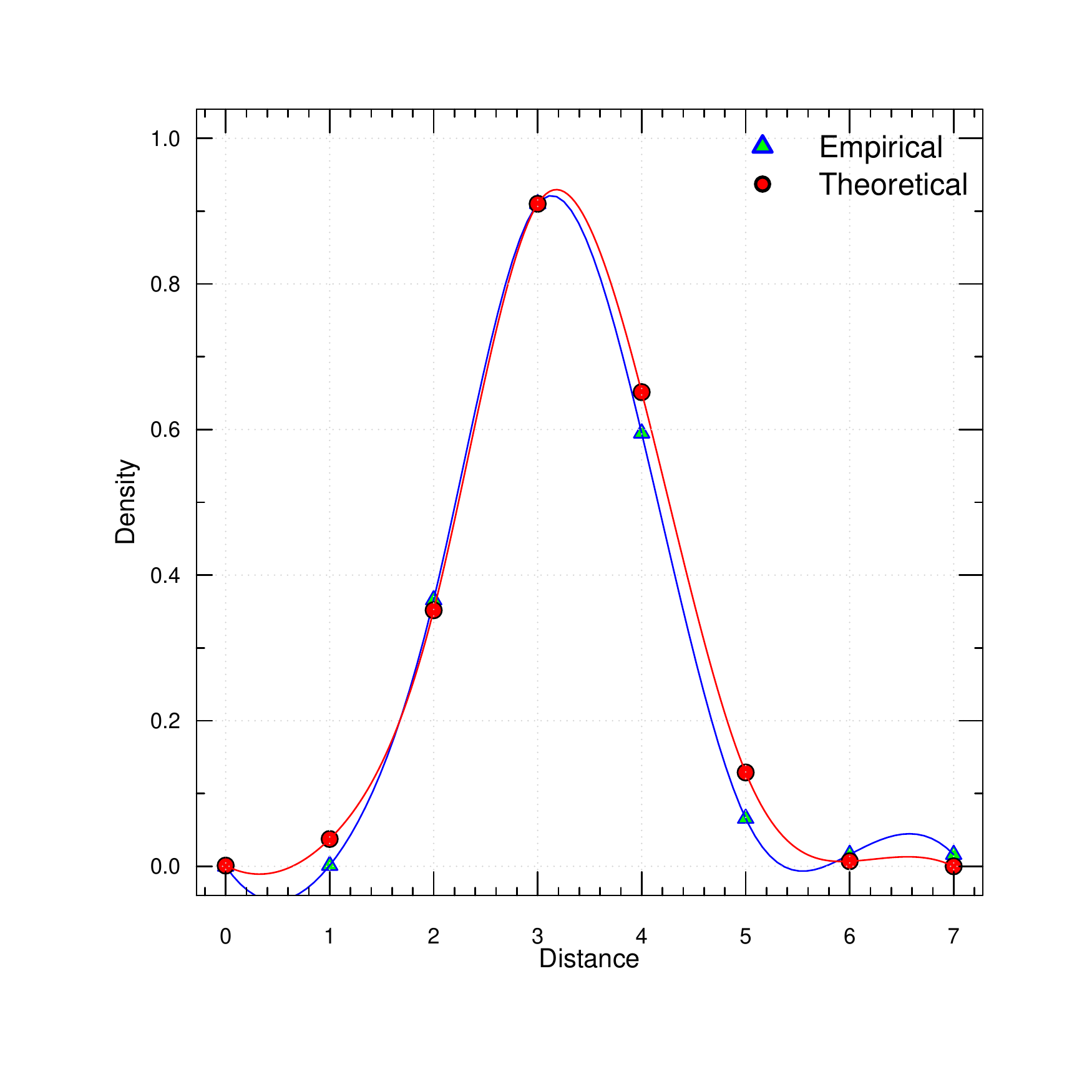}}
        \subfigure[DEMON*]{\includegraphics[width=.121\textwidth]{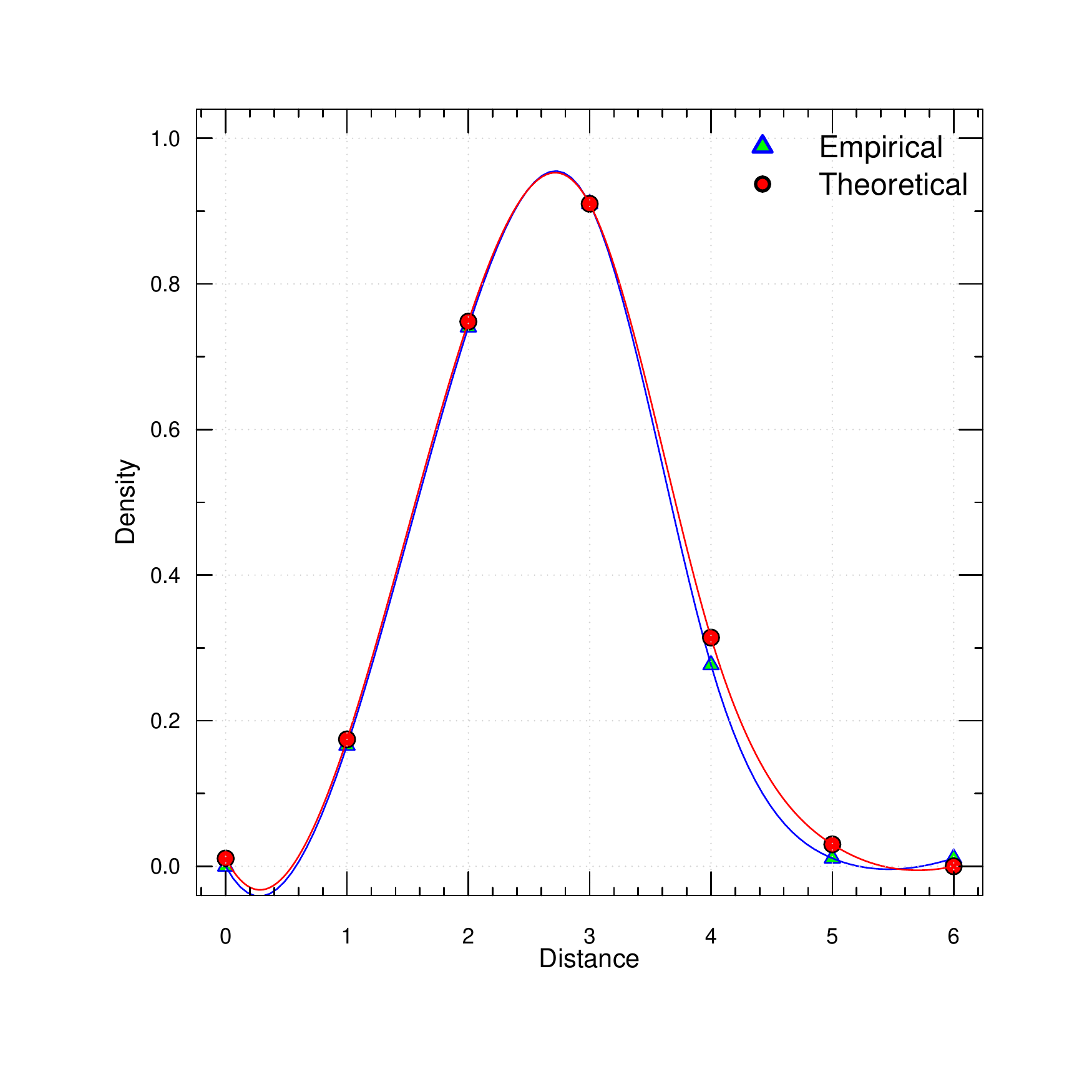}}

        \caption{\label{fig9} Empirical and estimated Hop Distance distribution for aNobii* (a), LFM* (b),  GCE* (c), OSLOM* (d), MOSES* (e), SLPA* (f), and DEMON* (g)}
        \end{figure}

        \begin{figure}[!ht]
        \subfigure[aNobii*]{\includegraphics[page=2,width=.121\textwidth]{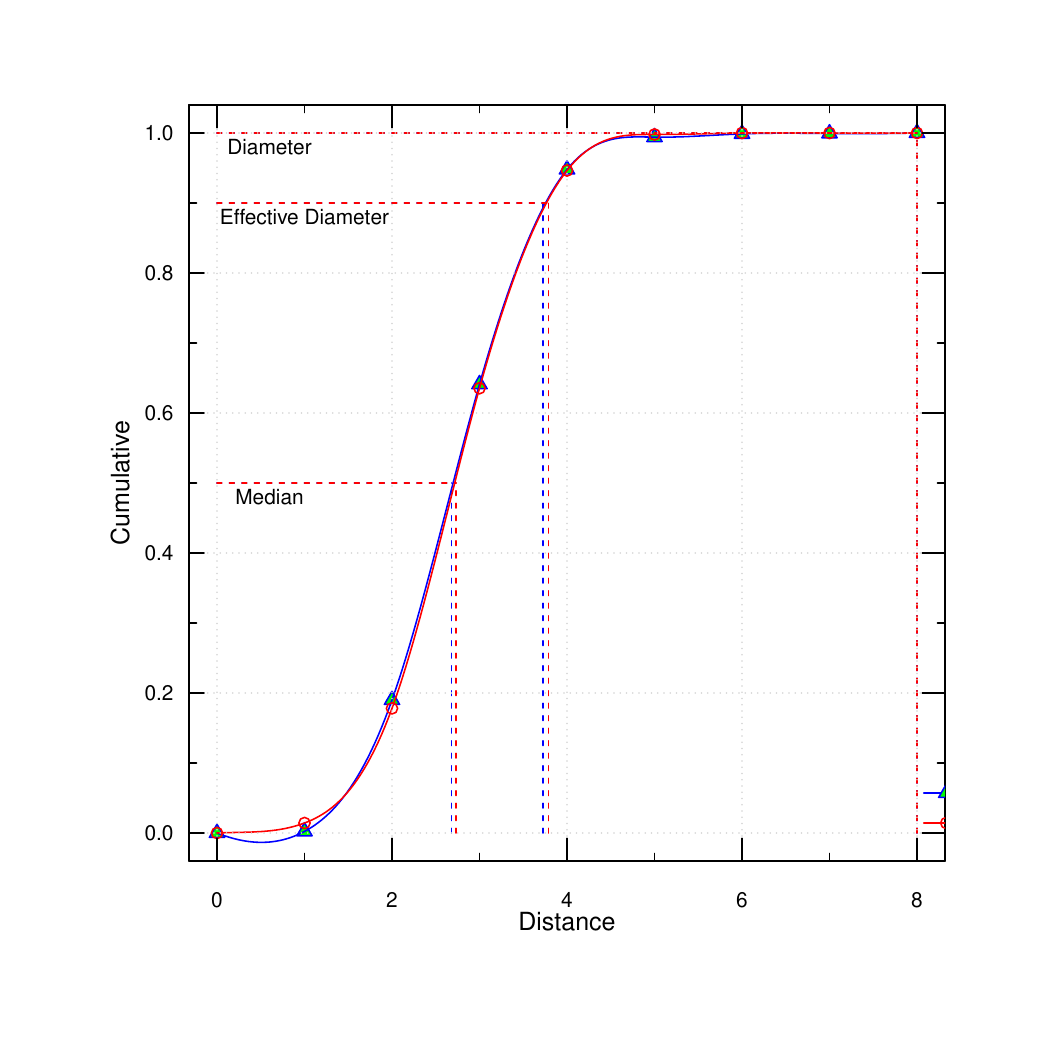}}
        \subfigure[LFM*]{\includegraphics[page=2,width=.121\textwidth]{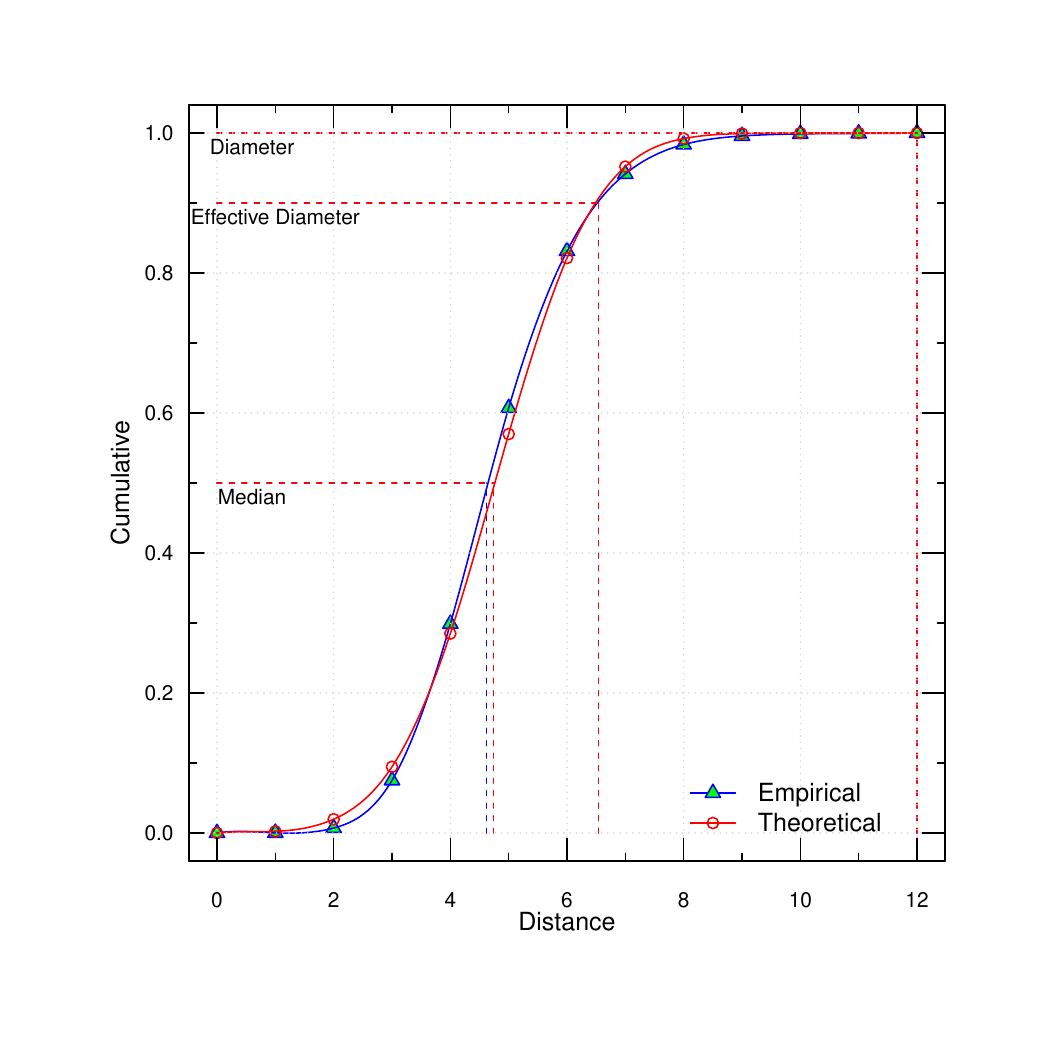}}
        \subfigure[GCE*]{\includegraphics[page=2,width=.121\textwidth]{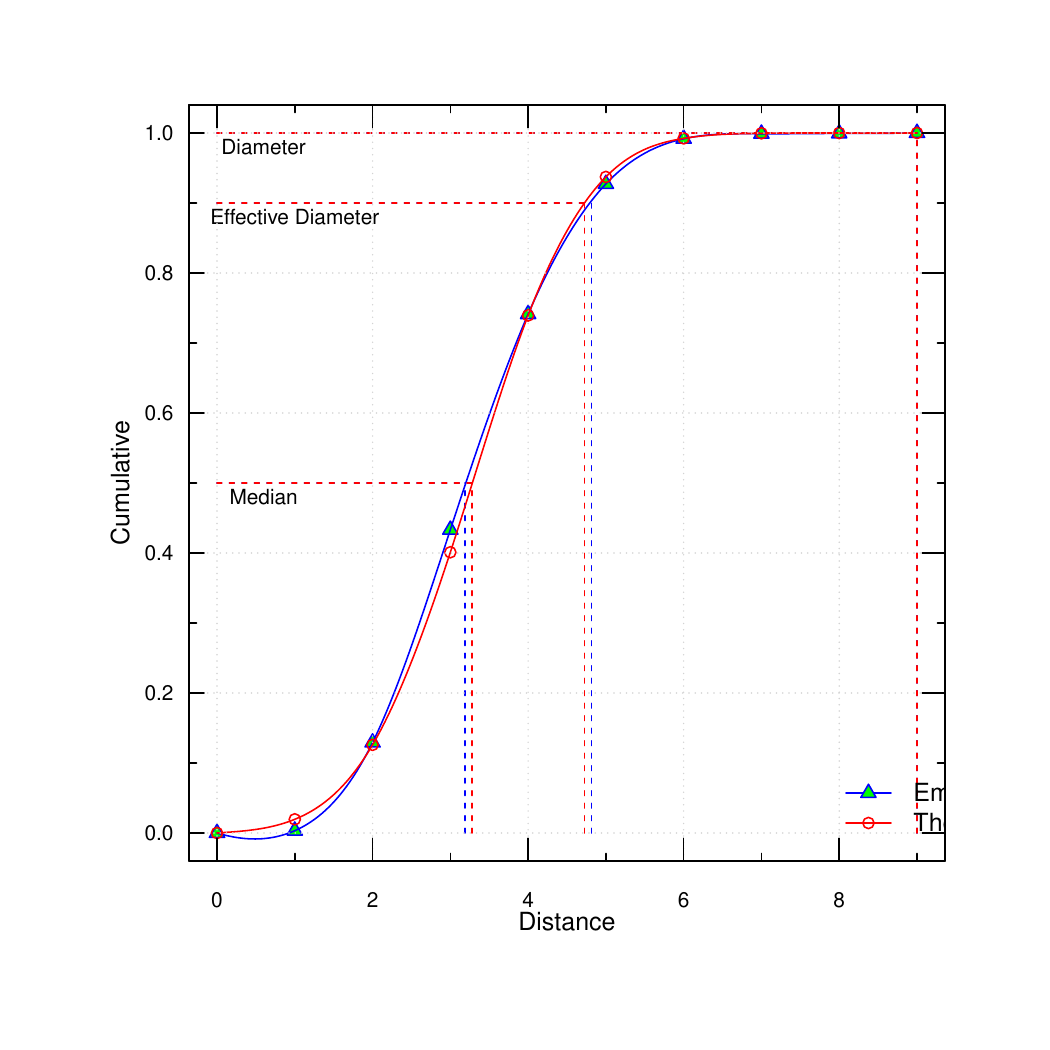}}
        \subfigure[OSLOM*]{\includegraphics[page=2,width=.121\textwidth]{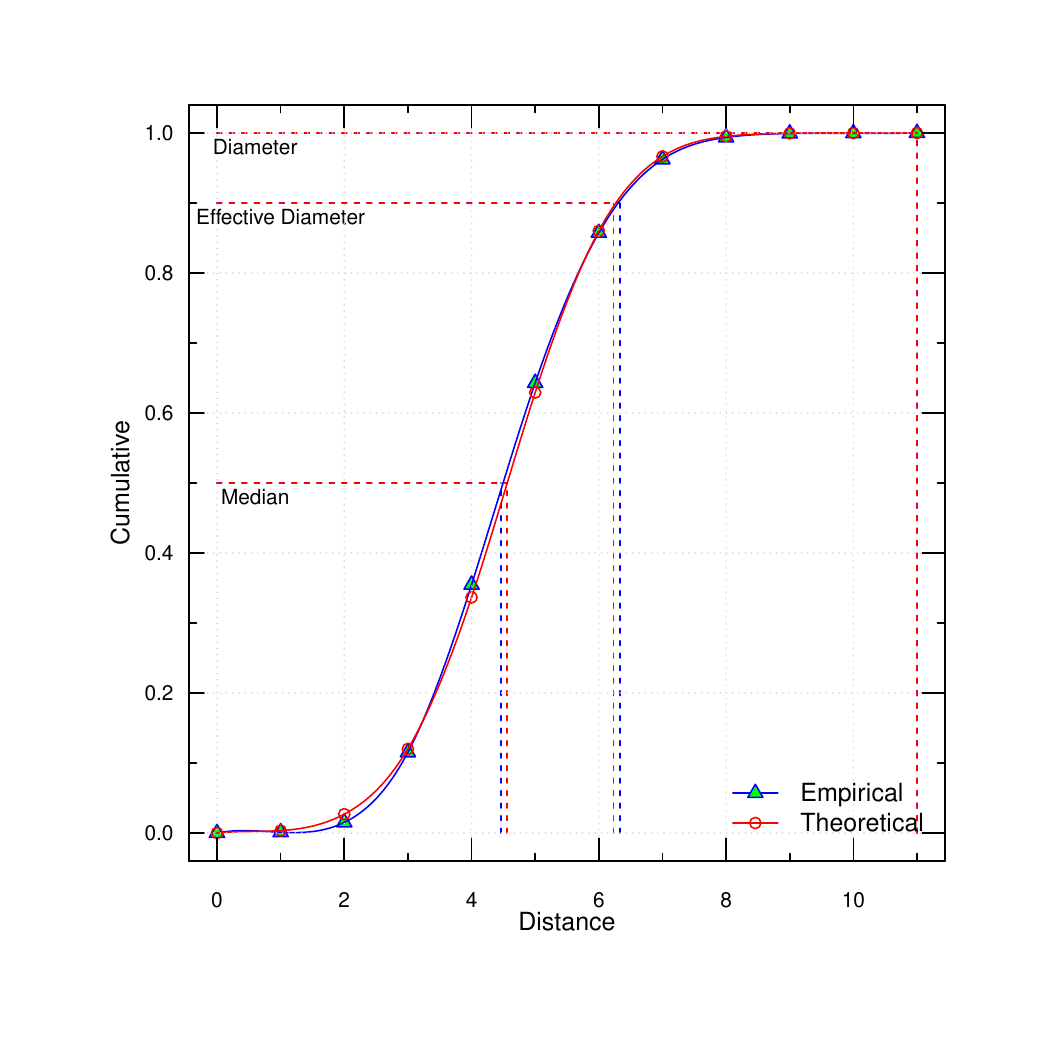}}
        \subfigure[MOSES*]{\includegraphics[page=2,width=.121\textwidth]{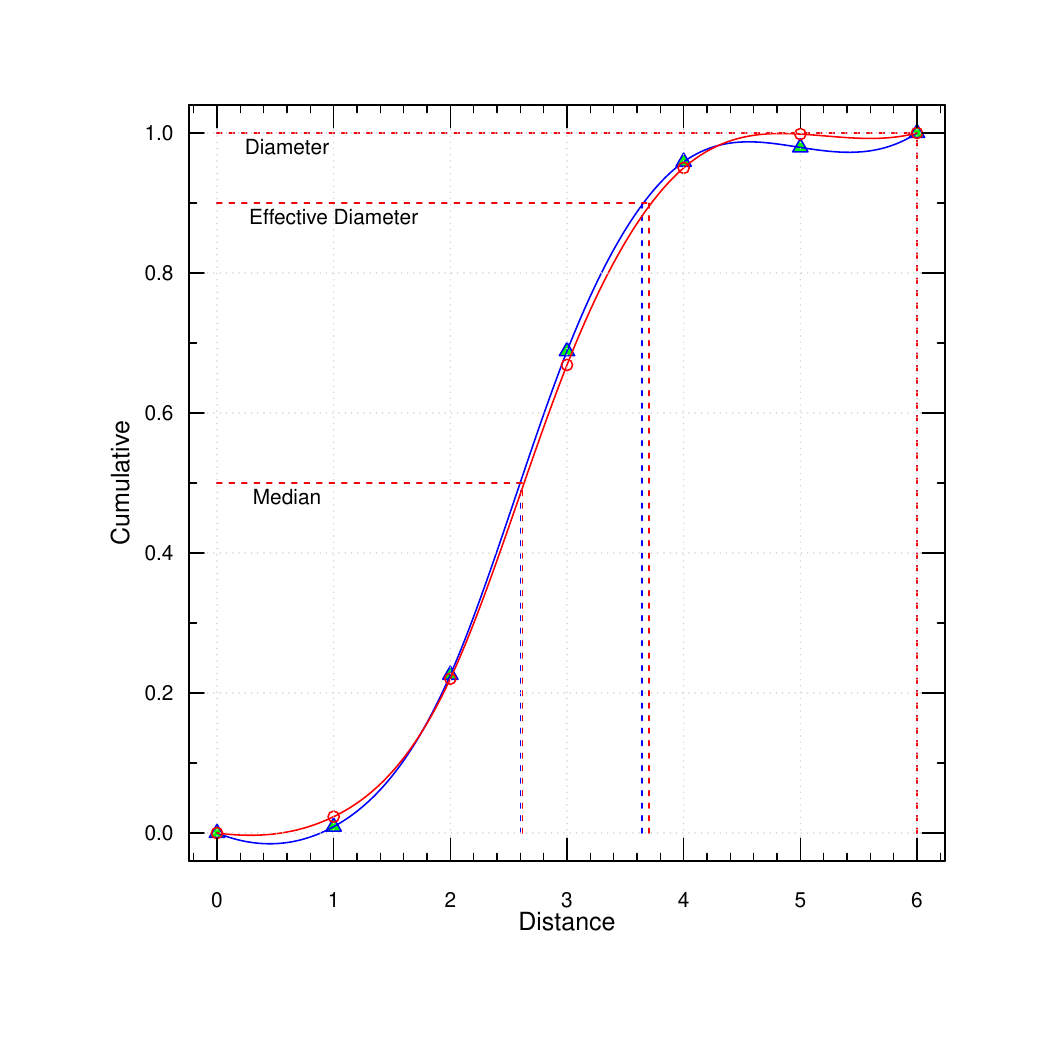}}
        \subfigure[SLPA*]{\includegraphics[page=2,width=.121\textwidth]{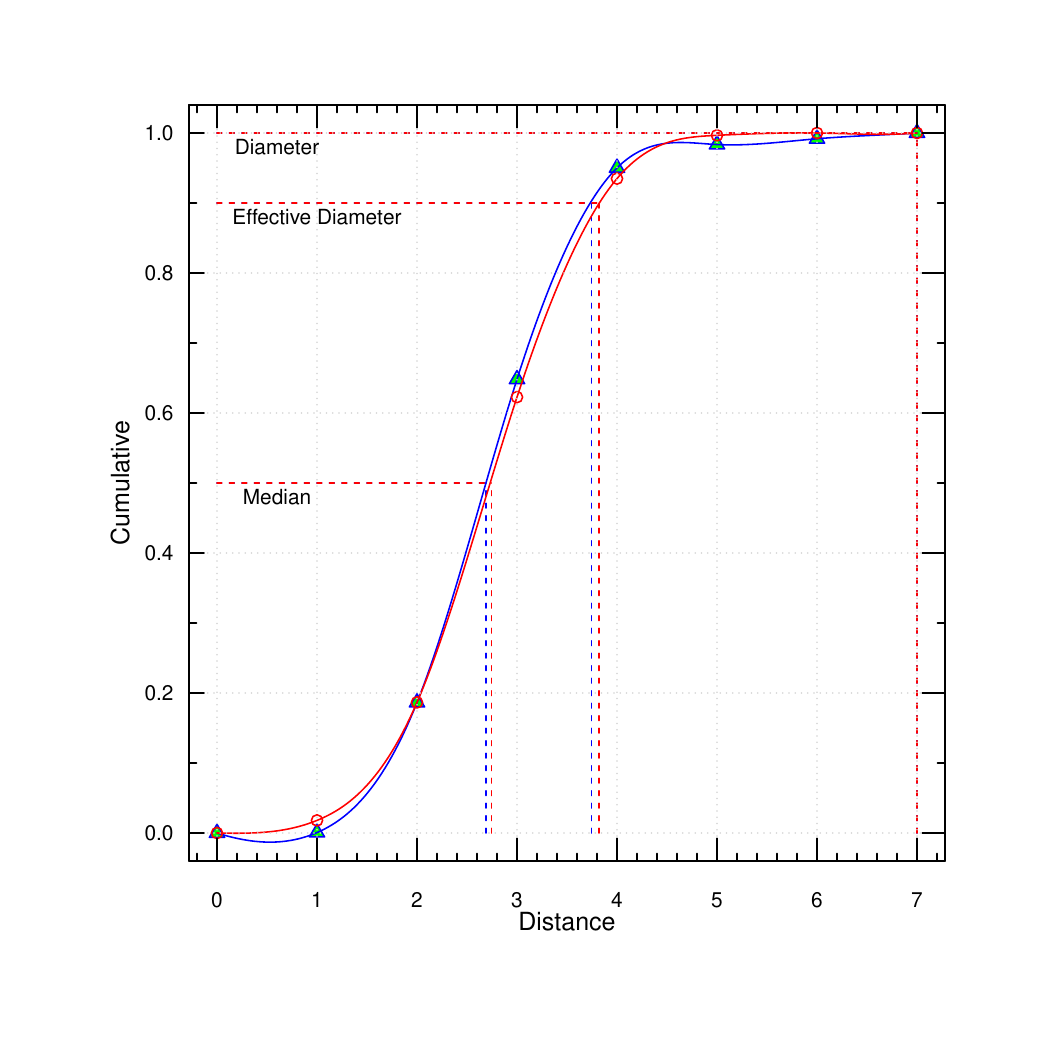}}
        \subfigure[DEMON*]{\includegraphics[page=2,width=.121\textwidth]{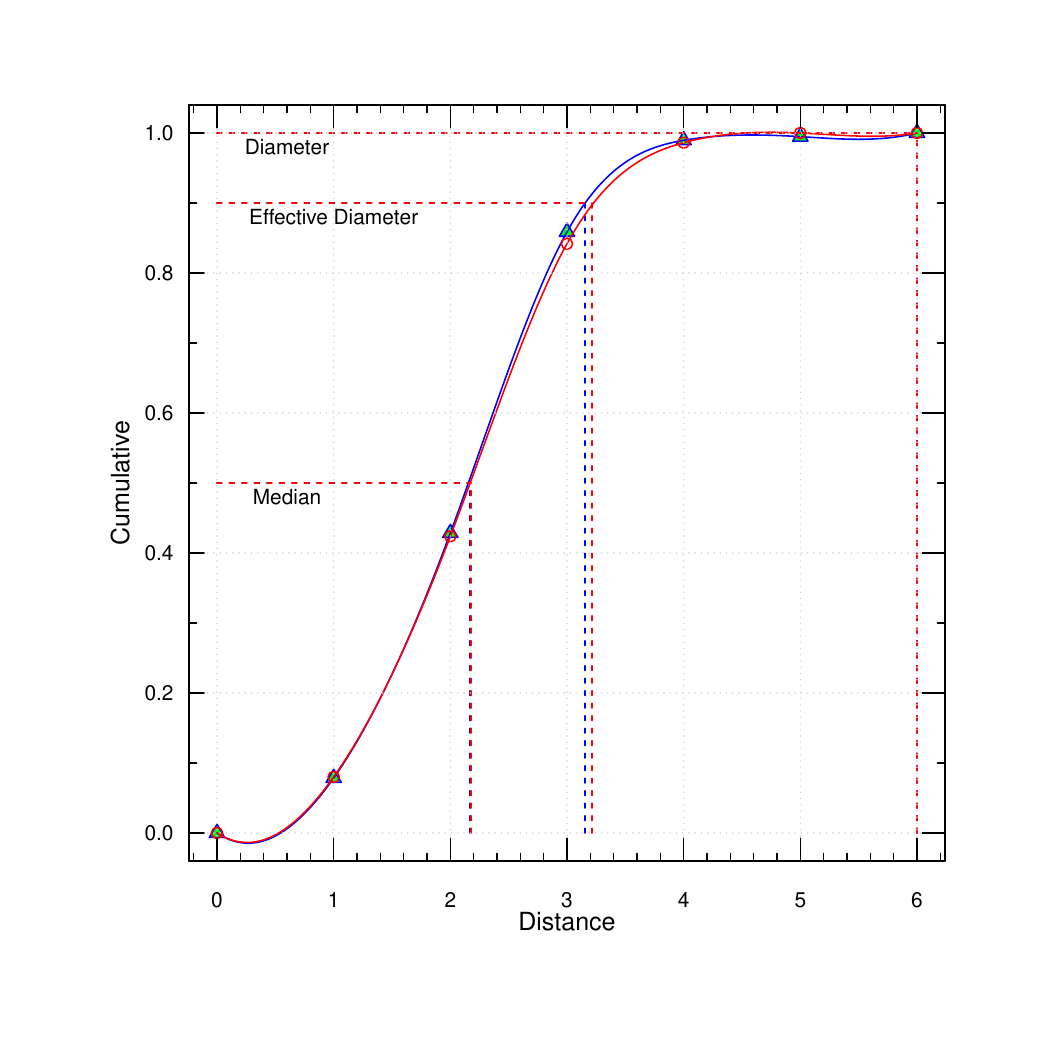}}

        \caption{\label{fig12} Empirical and estimated Hop distance cumulative distributions for aNobii* (a), LFM* (b),  GCE* (c), OSLOM* (d), MOSES* (e), SLPA* (f), and DEMON* (g)}
        \end{figure}

        \begin{figure}[!ht]
        \subfigure[Ground-truth]{\includegraphics[width=.121\textwidth]{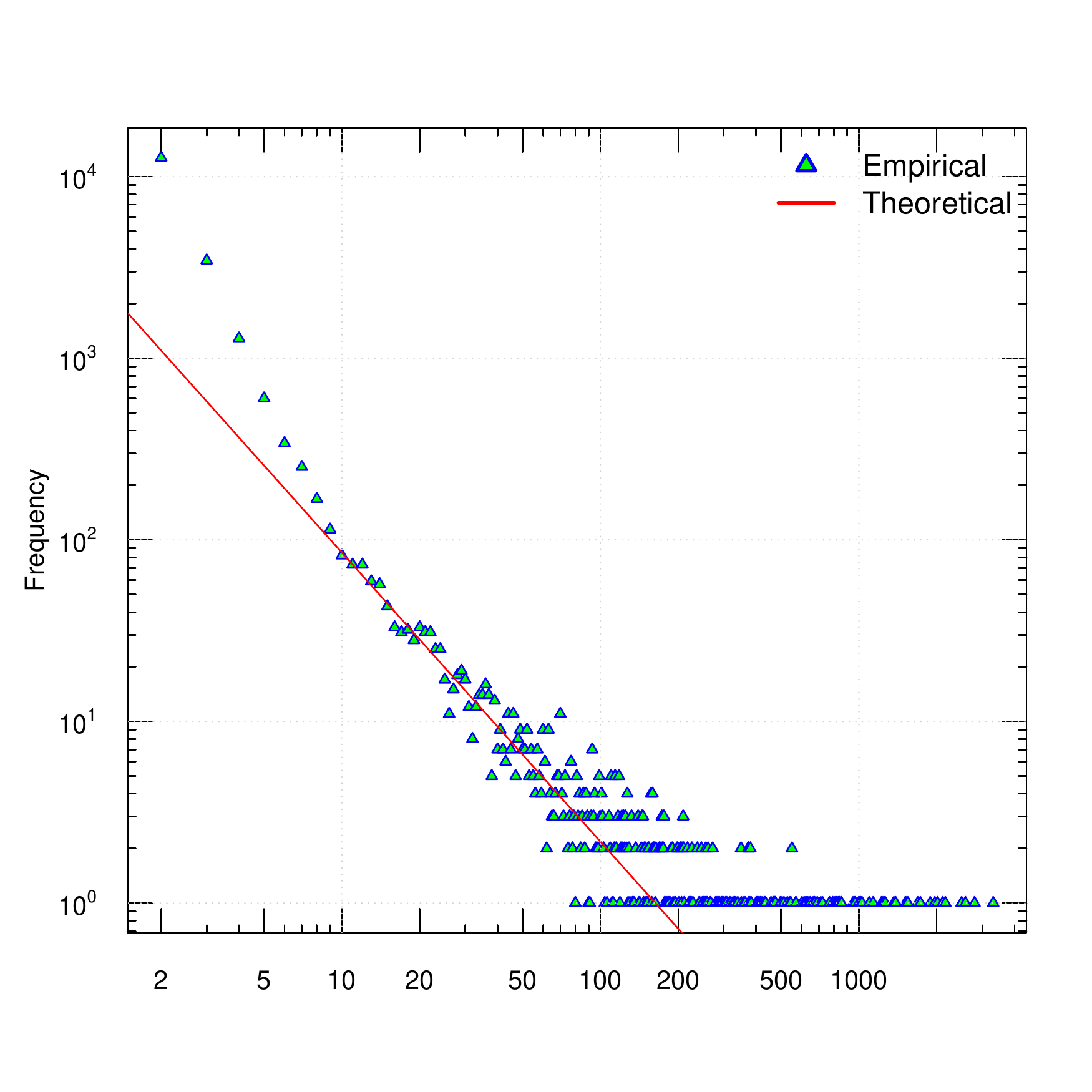}}
        \subfigure[LFM]{\includegraphics[width=.121\textwidth]{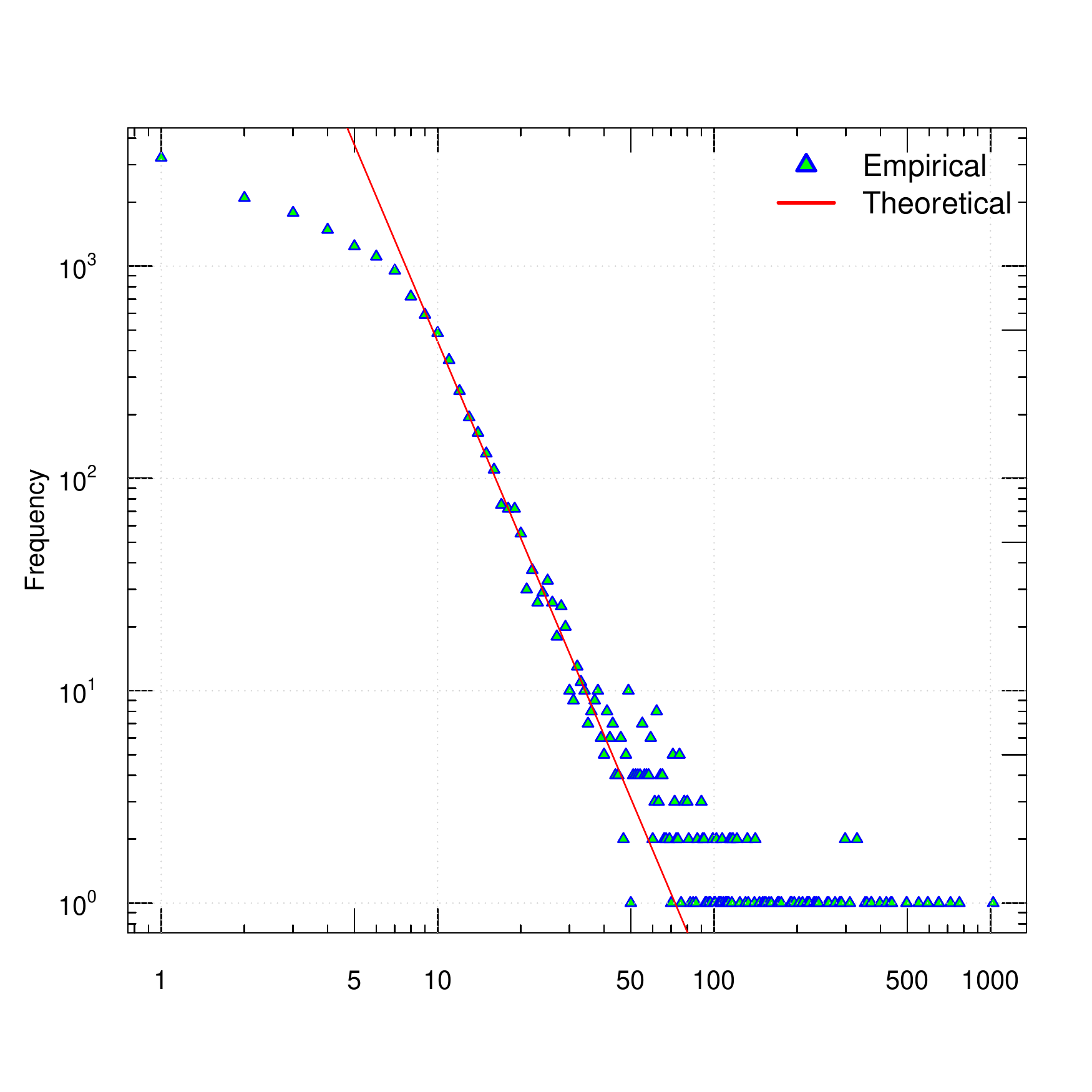}}
        \subfigure[GCE]{\includegraphics[width=.121\textwidth]{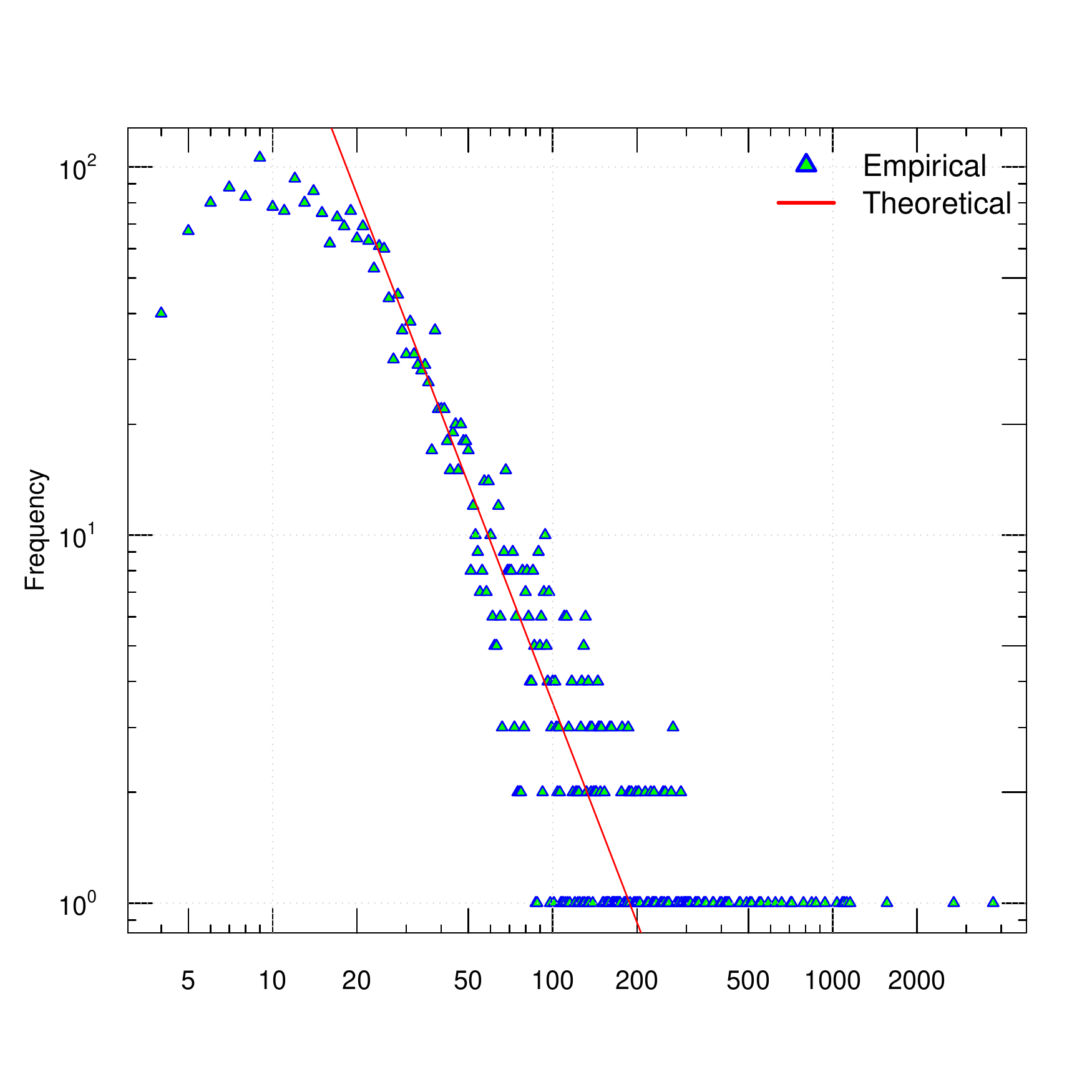}}
        \subfigure[OSLOM]{\includegraphics[width=.121\textwidth]{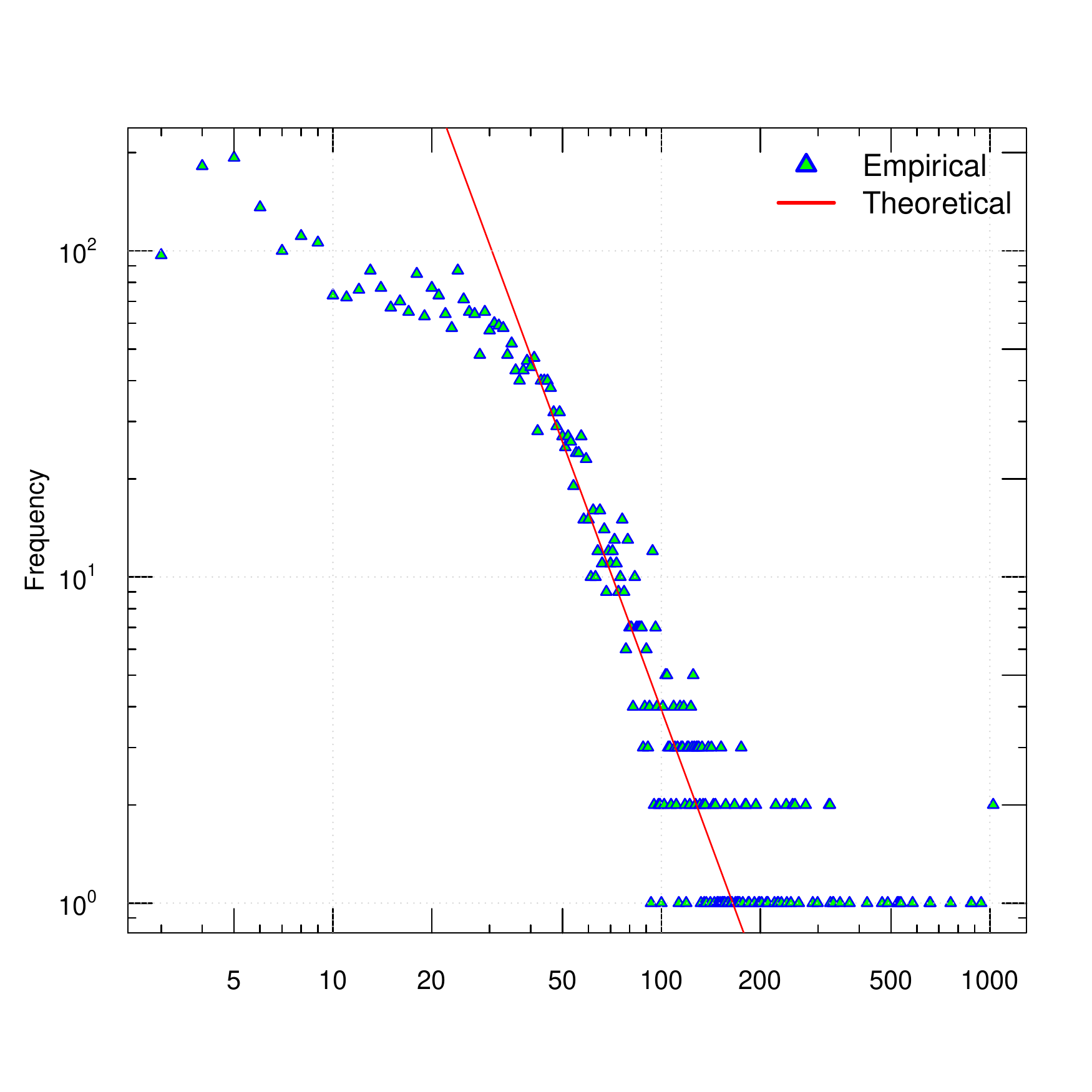}}
        \subfigure[MOSES]{\includegraphics[width=.121\textwidth]{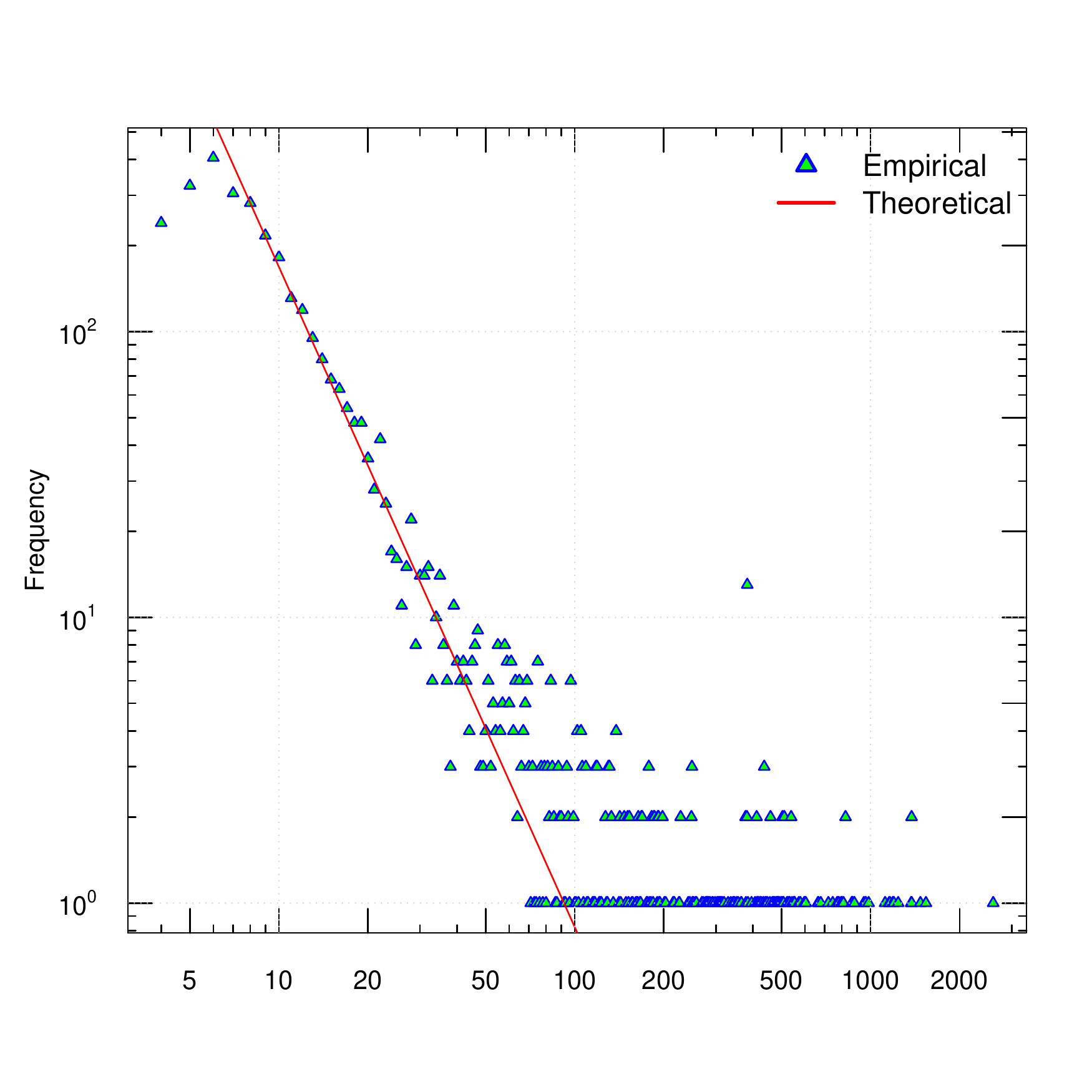}}
        \subfigure[SLPA]{\includegraphics[width=.121\textwidth]{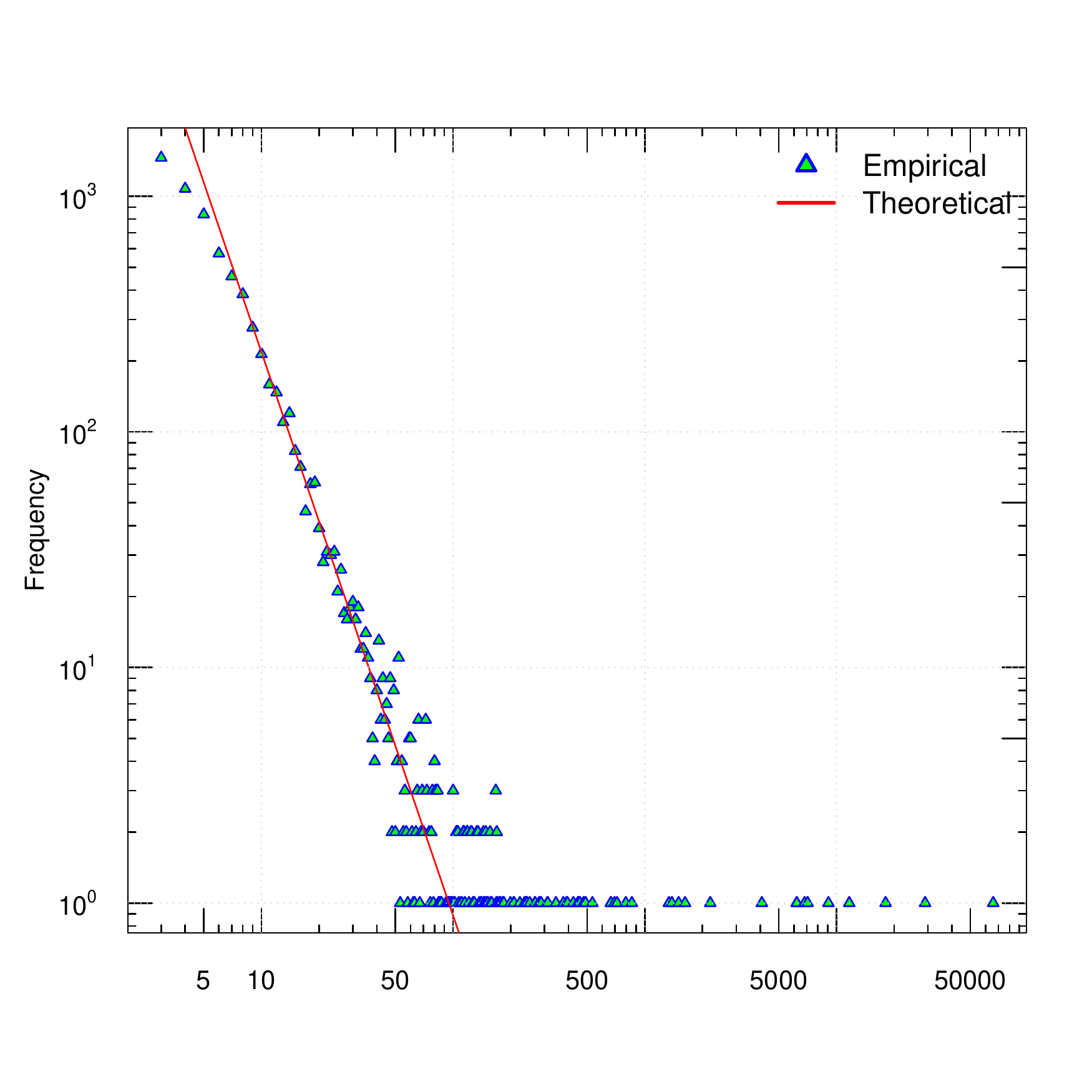}}
        \subfigure[DEMON]{\includegraphics[width=.121\textwidth]{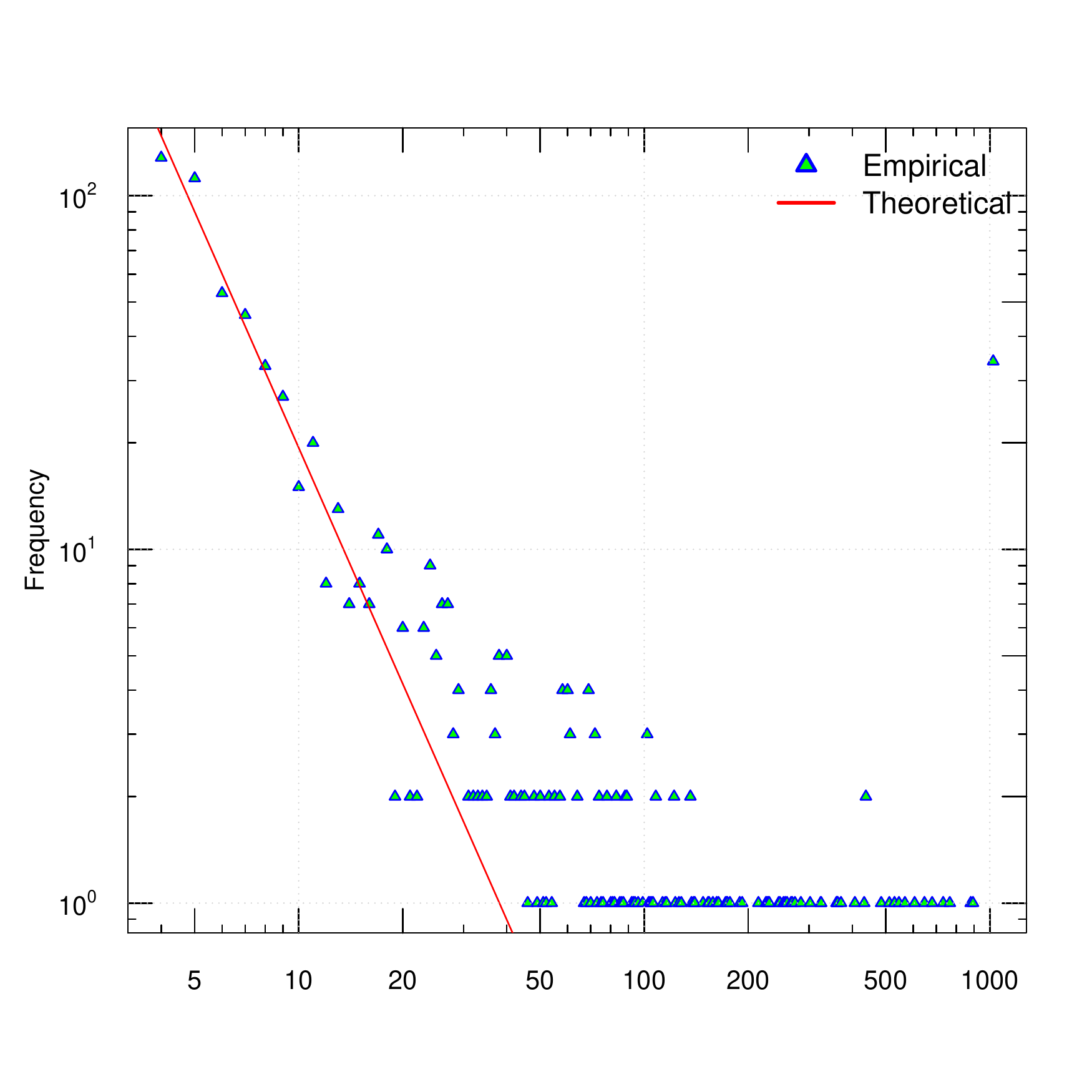}}

        \caption{\label{fig15}Log-log empirical Community size distribution (dots) and Power-Law estimate (line) of aNobii Ground-truth (a), LFM (b),  GCE (c), OSLOM (d), MOSES (e), SLPA (f), and DEMON (g)}
        \end{figure}

        \begin{figure}[!ht]
        \subfigure[Ground-truth]{\includegraphics[width=.121\textwidth]{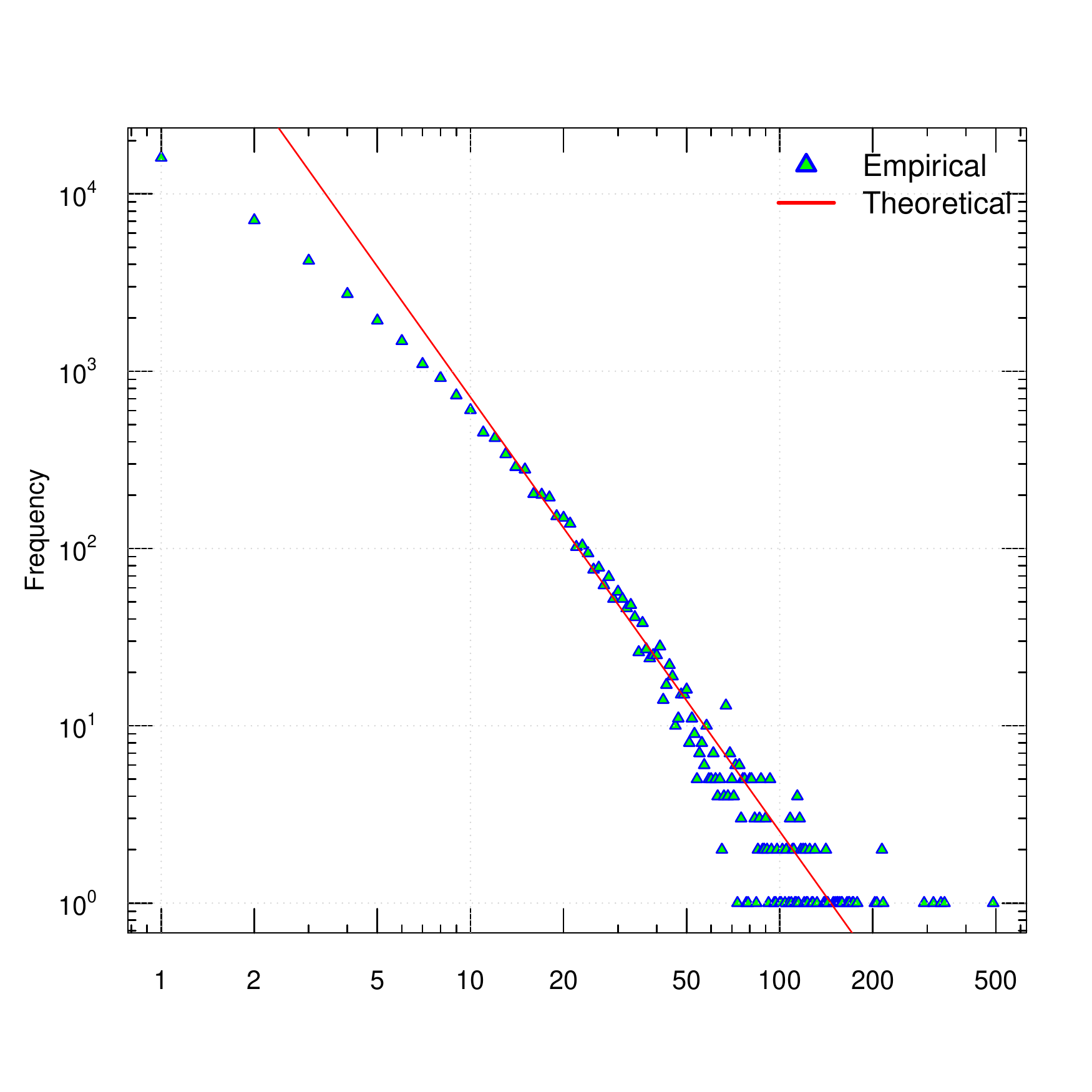}}
        \subfigure[LFM]{\includegraphics[width=.121\textwidth]{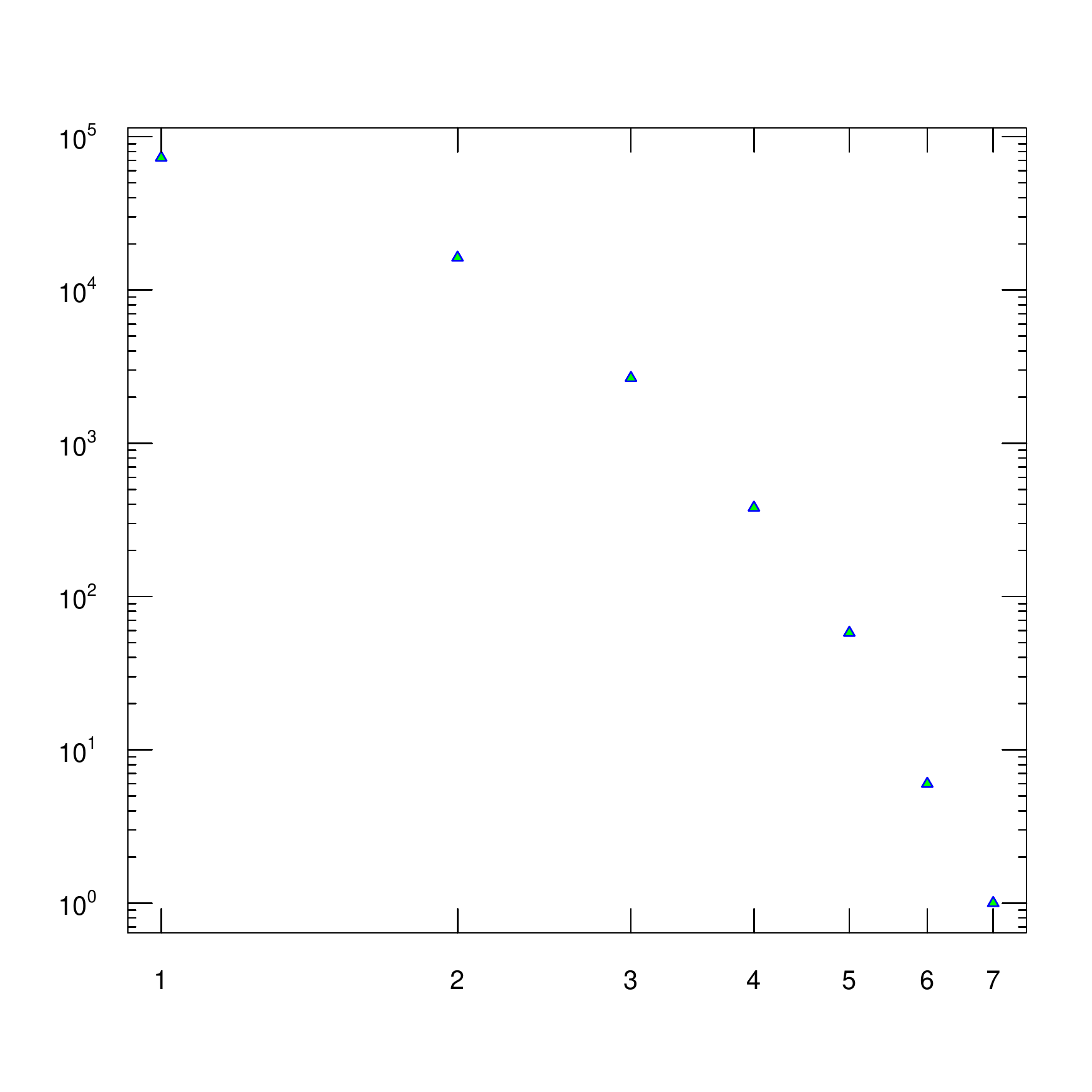}}
        \subfigure[GCE]{\includegraphics[width=.121\textwidth]{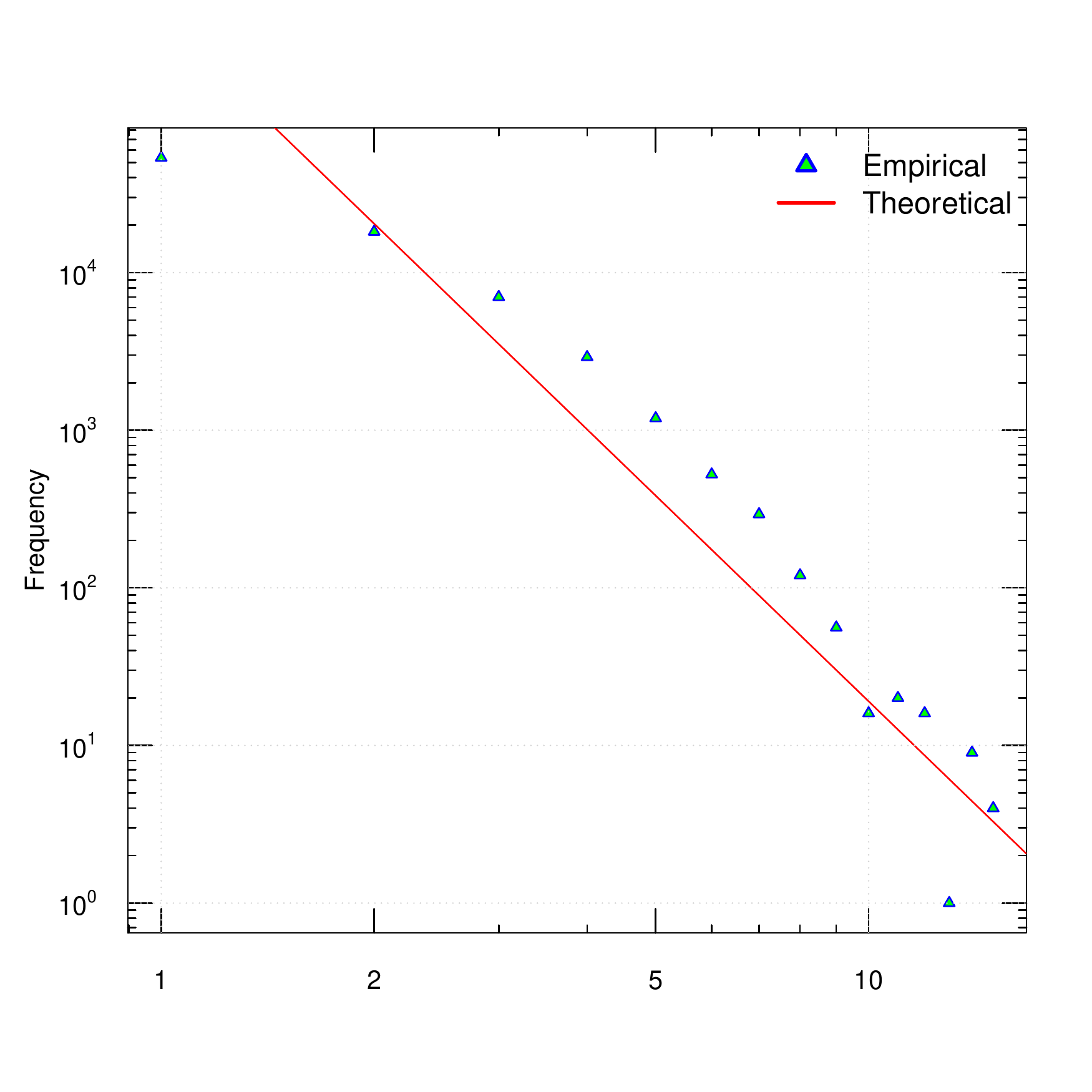}}
        \subfigure[OSLOM]{\includegraphics[width=.121\textwidth]{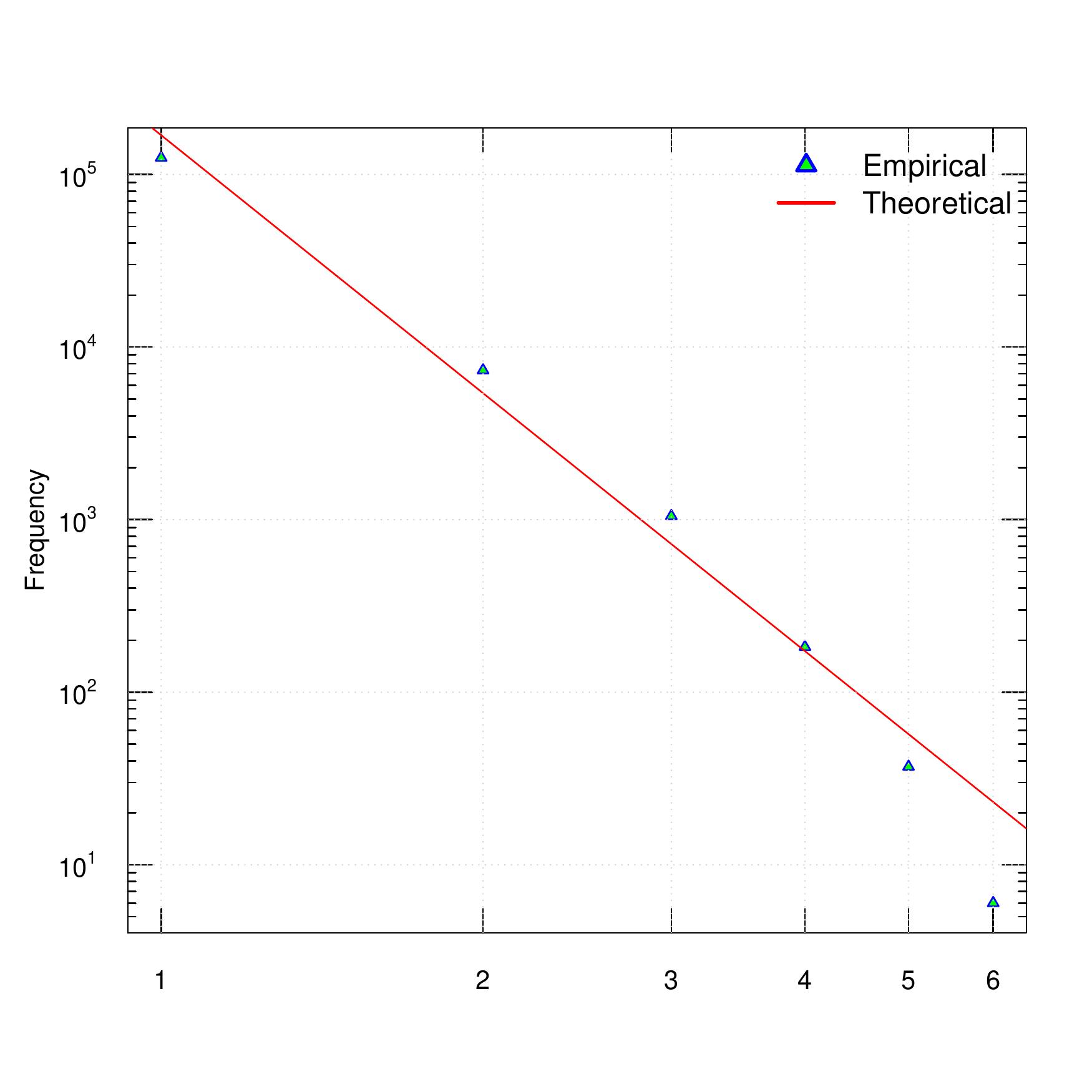}}
        \subfigure[MOSES]{\includegraphics[width=.121\textwidth]{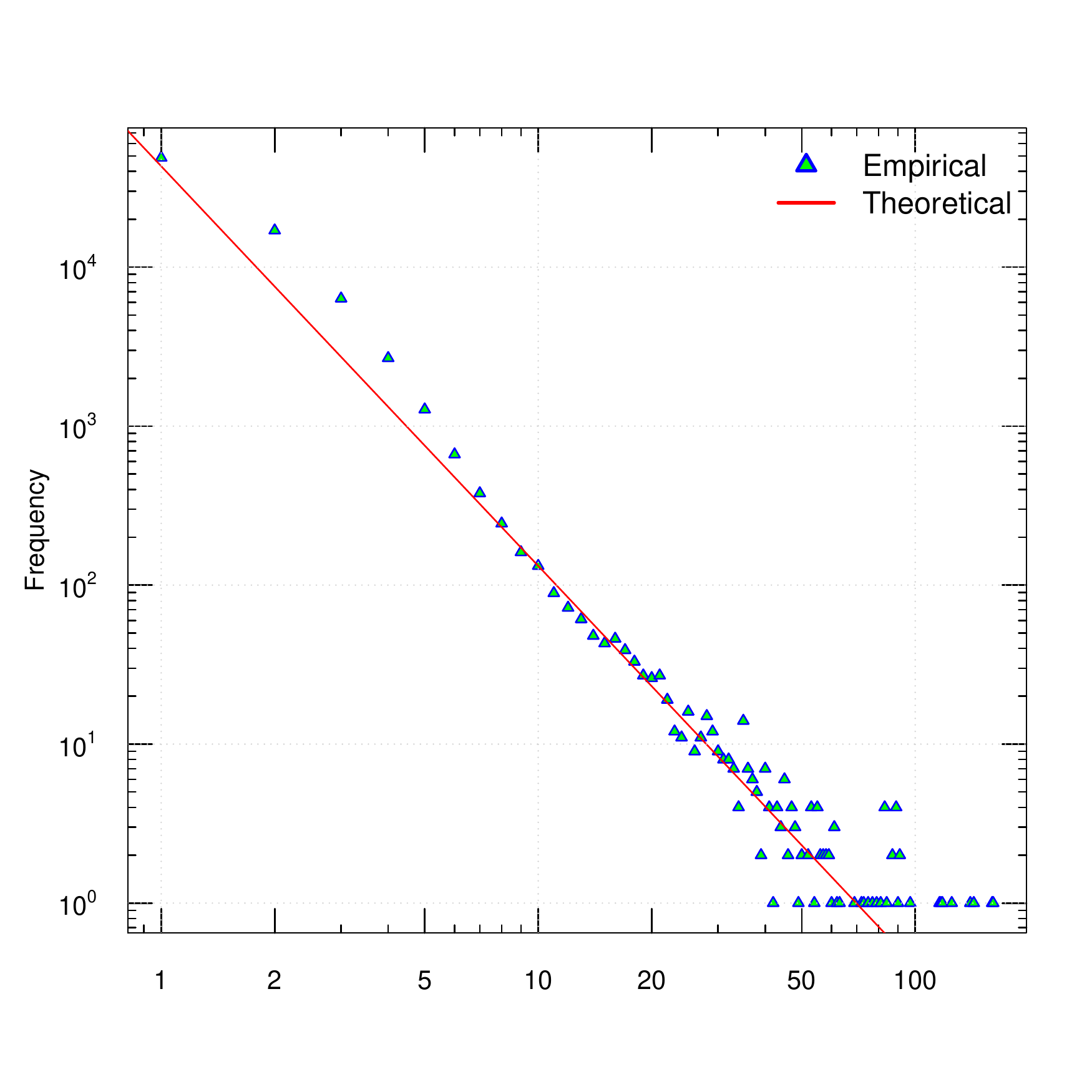}}
        \subfigure[SLPA]{\includegraphics[width=.121\textwidth]{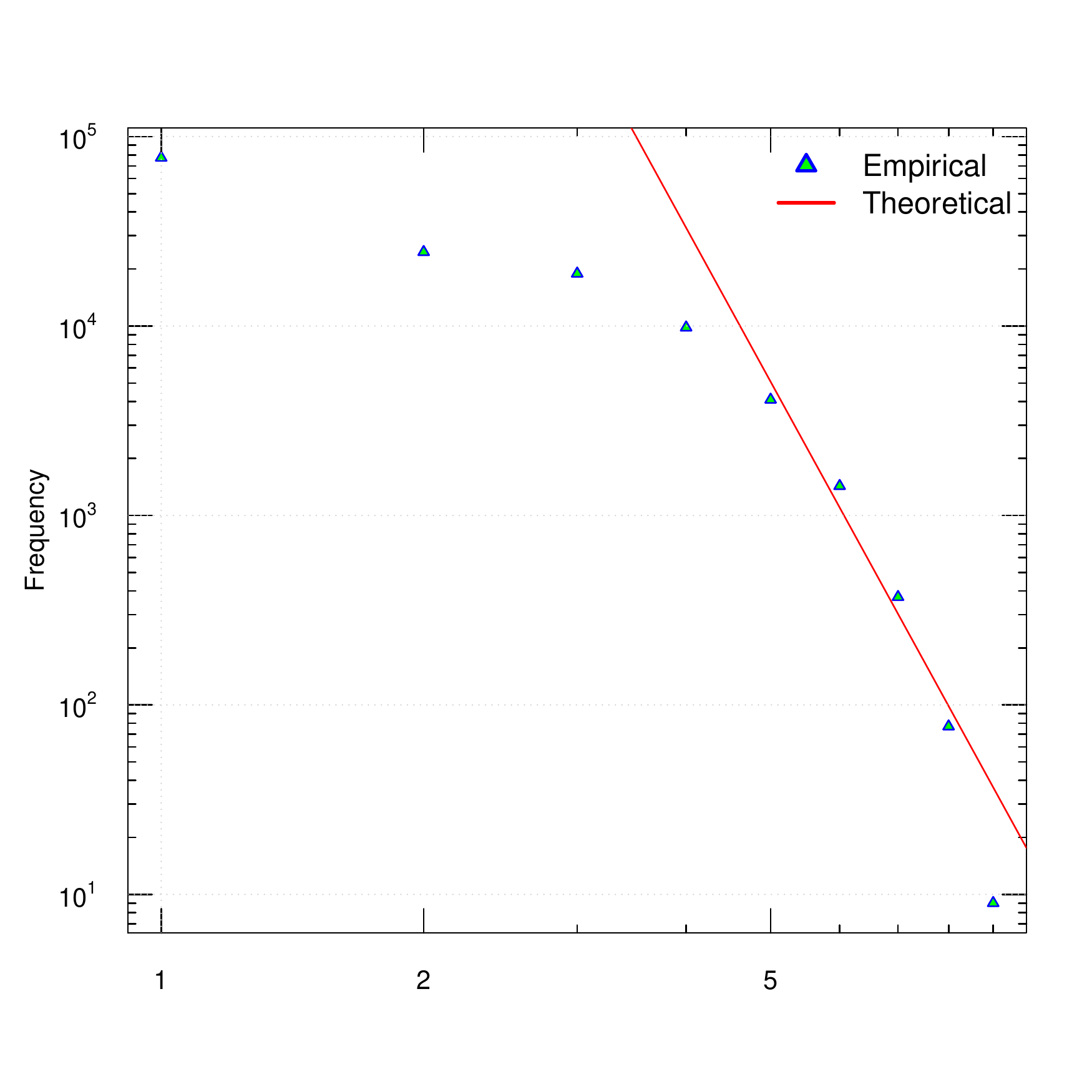}}
        \subfigure[DEMON]{\includegraphics[width=.121\textwidth]{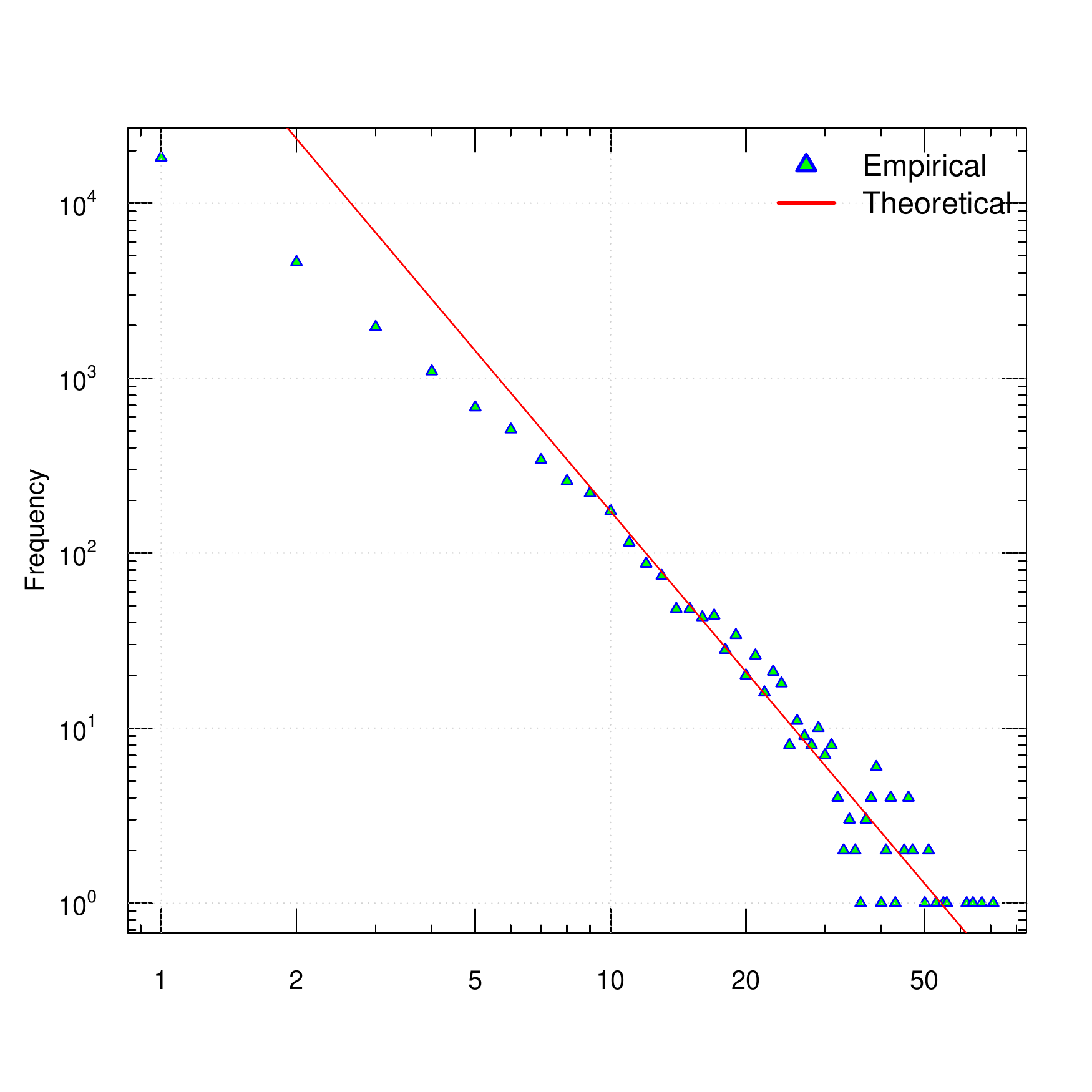}}

        \caption{\label{fig18}Log-log empirical Membership distribution (dot) and Power-Law estimate (line) of aNobii Ground-truth (a), LFM (b),  GCE (c), OSLOM (d), MOSES (e), SLPA (f), and DEMON (g)}
        \end{figure}

        \begin{figure}[!ht]
        \subfigure[Ground-truth]{\includegraphics[width=.121\textwidth]{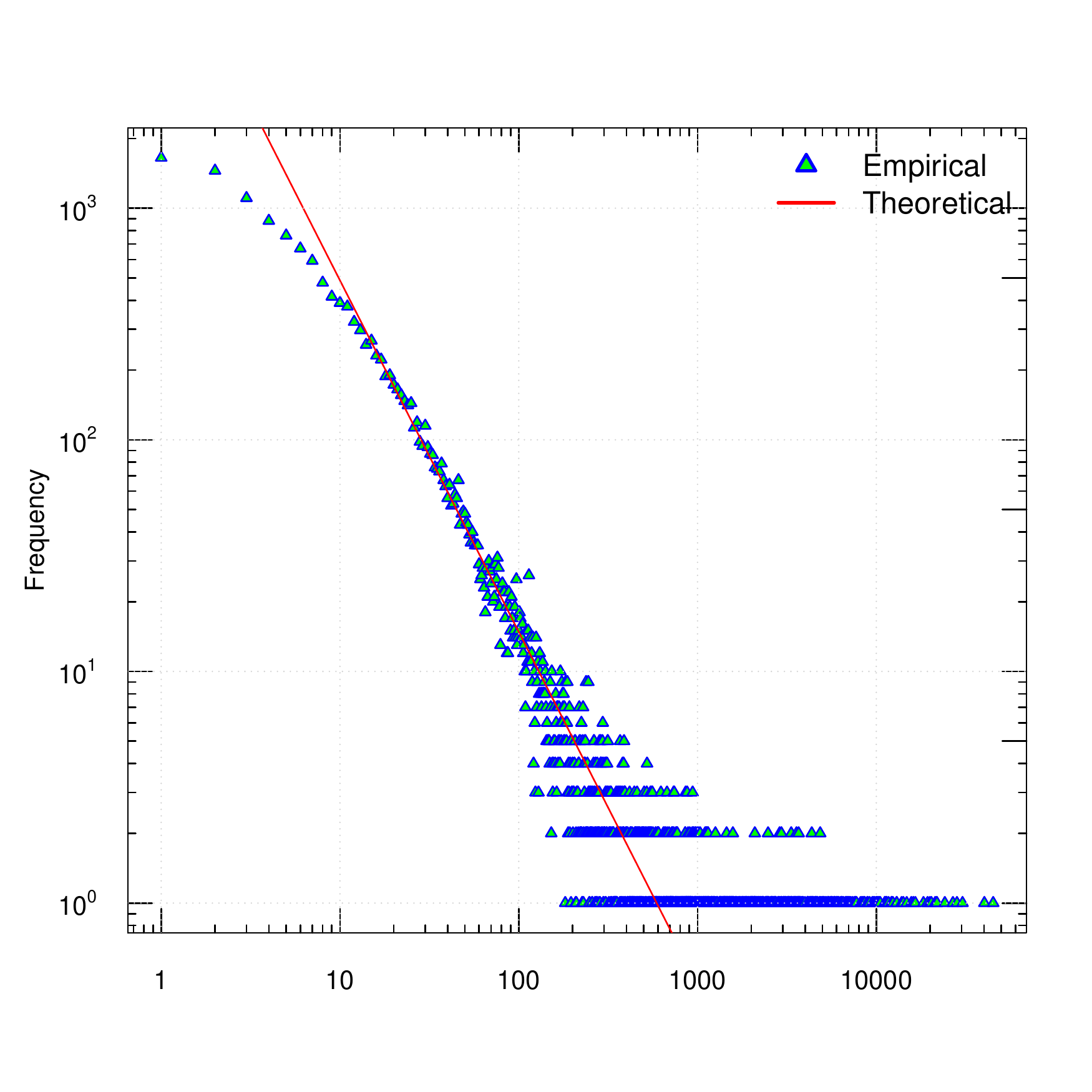}}
        \subfigure[LFM]{\includegraphics[width=.121\textwidth]{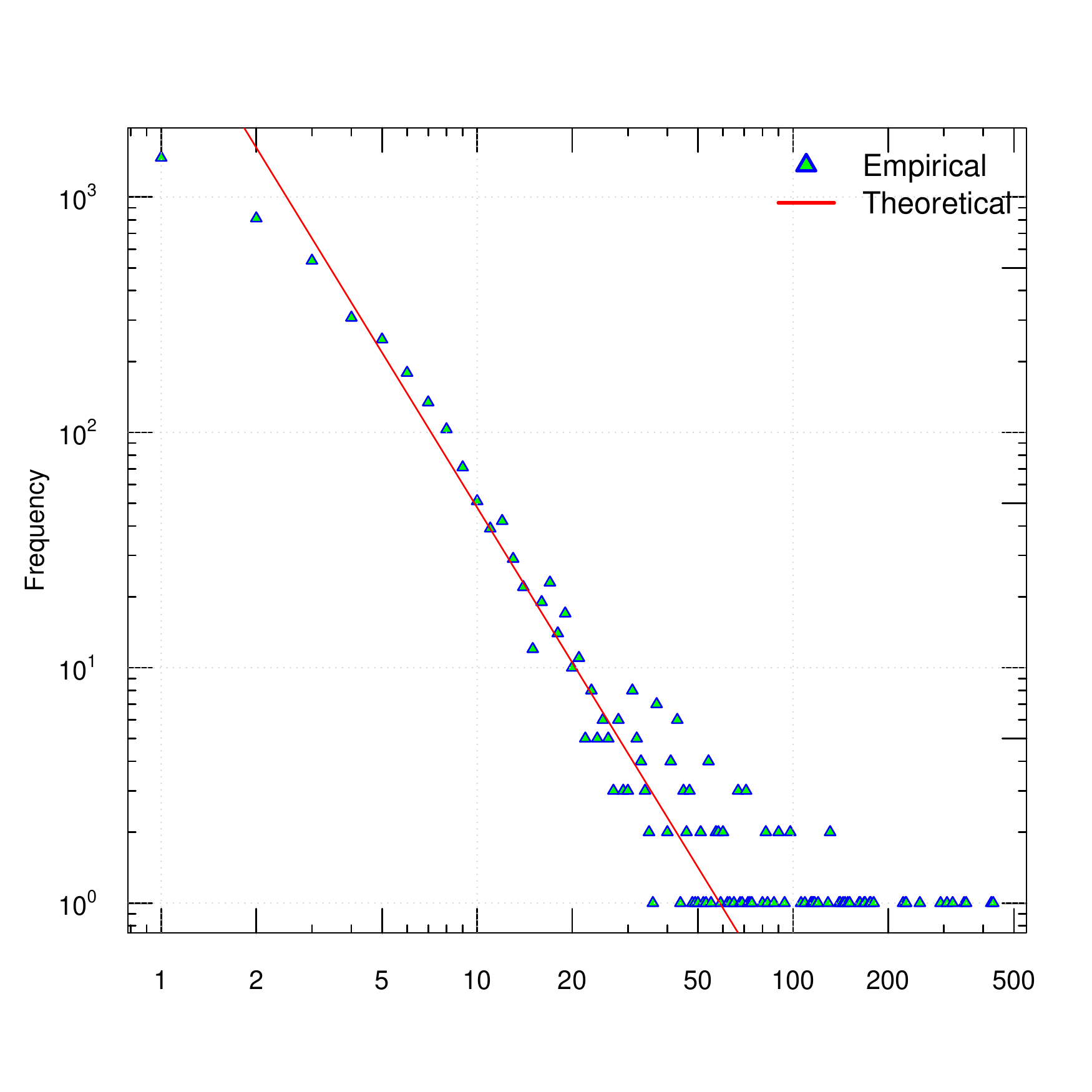}}
        \subfigure[GCE]{\includegraphics[width=.121\textwidth]{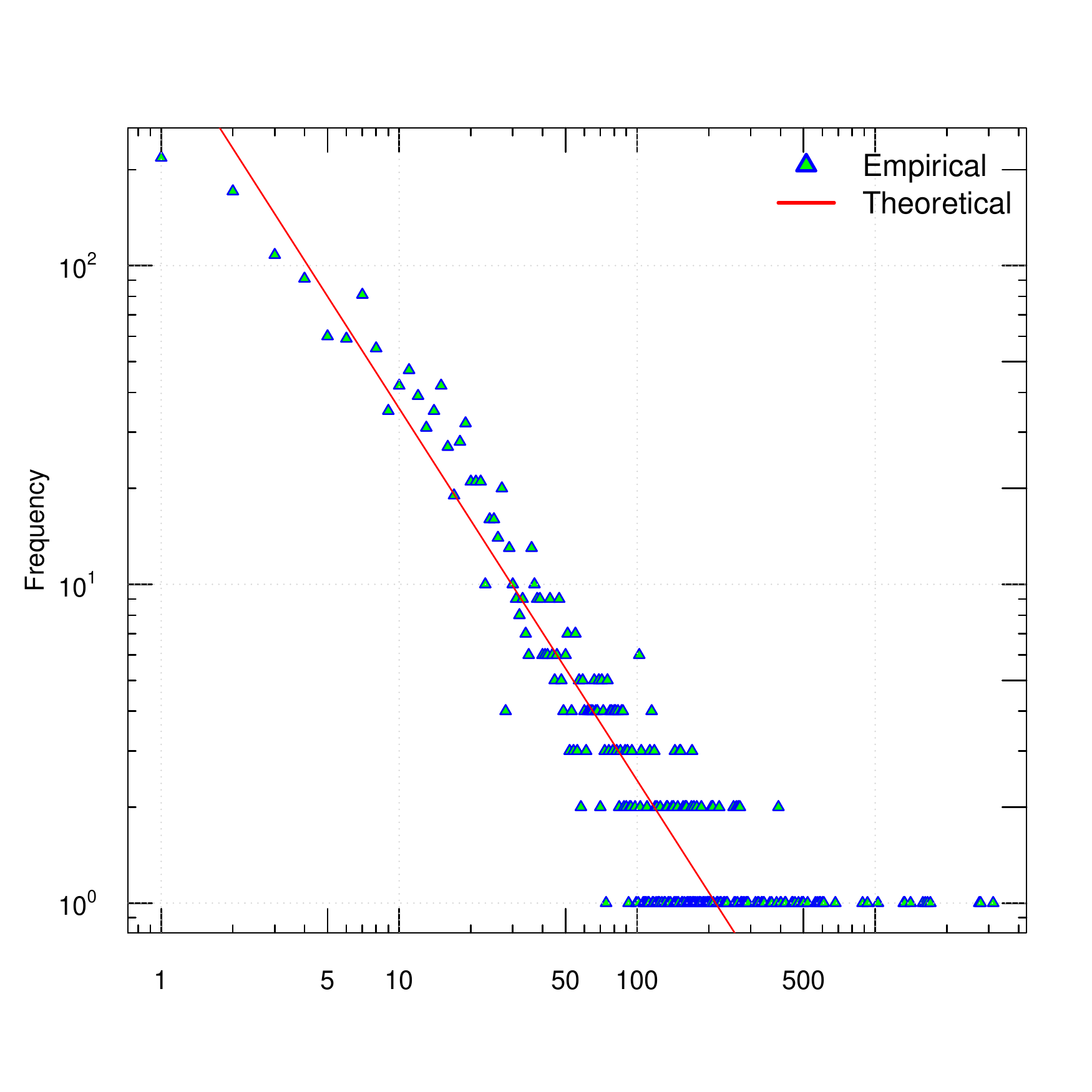}}
        \subfigure[OSLOM]{\includegraphics[width=.121\textwidth]{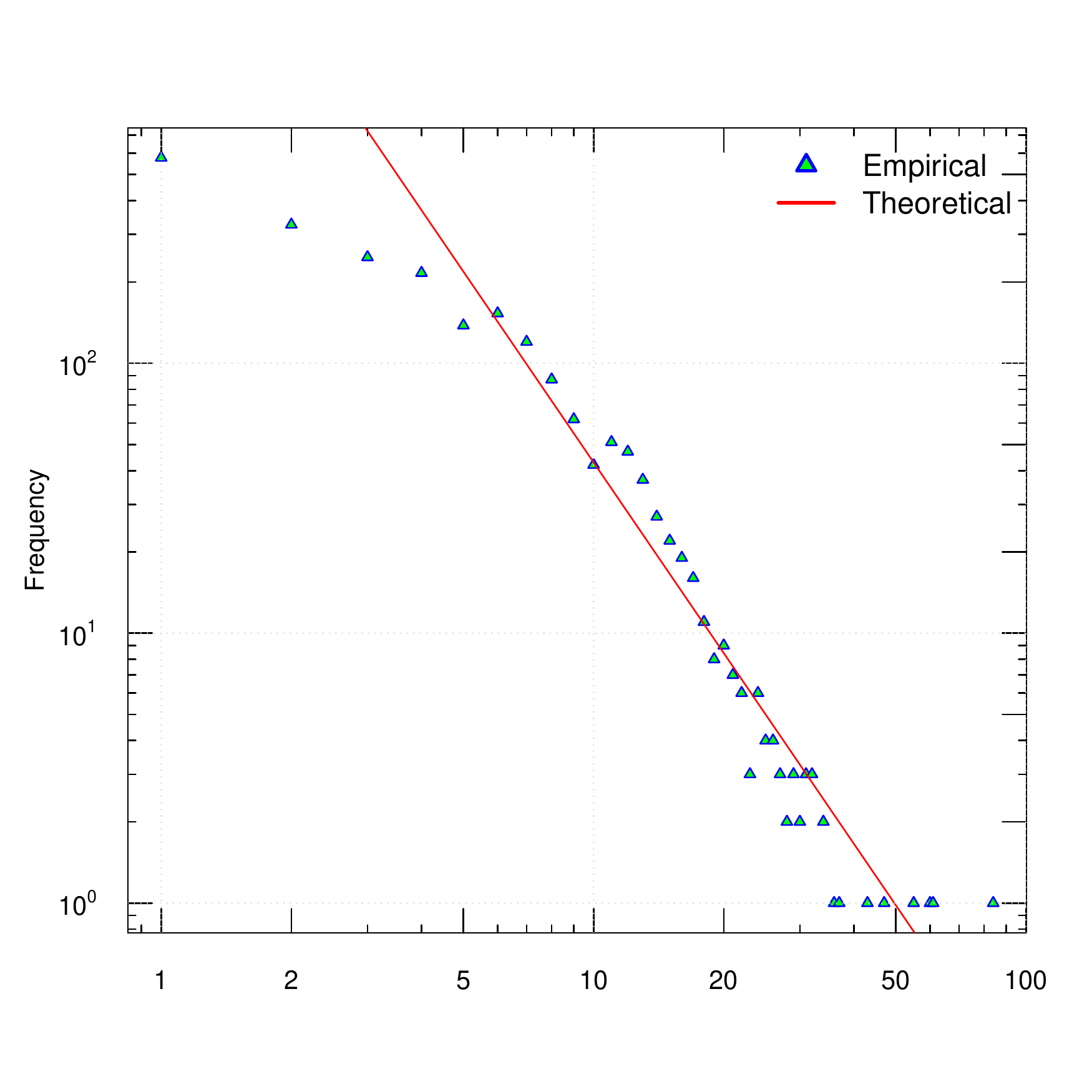}}
        \subfigure[SLPA]{\includegraphics[width=.121\textwidth]{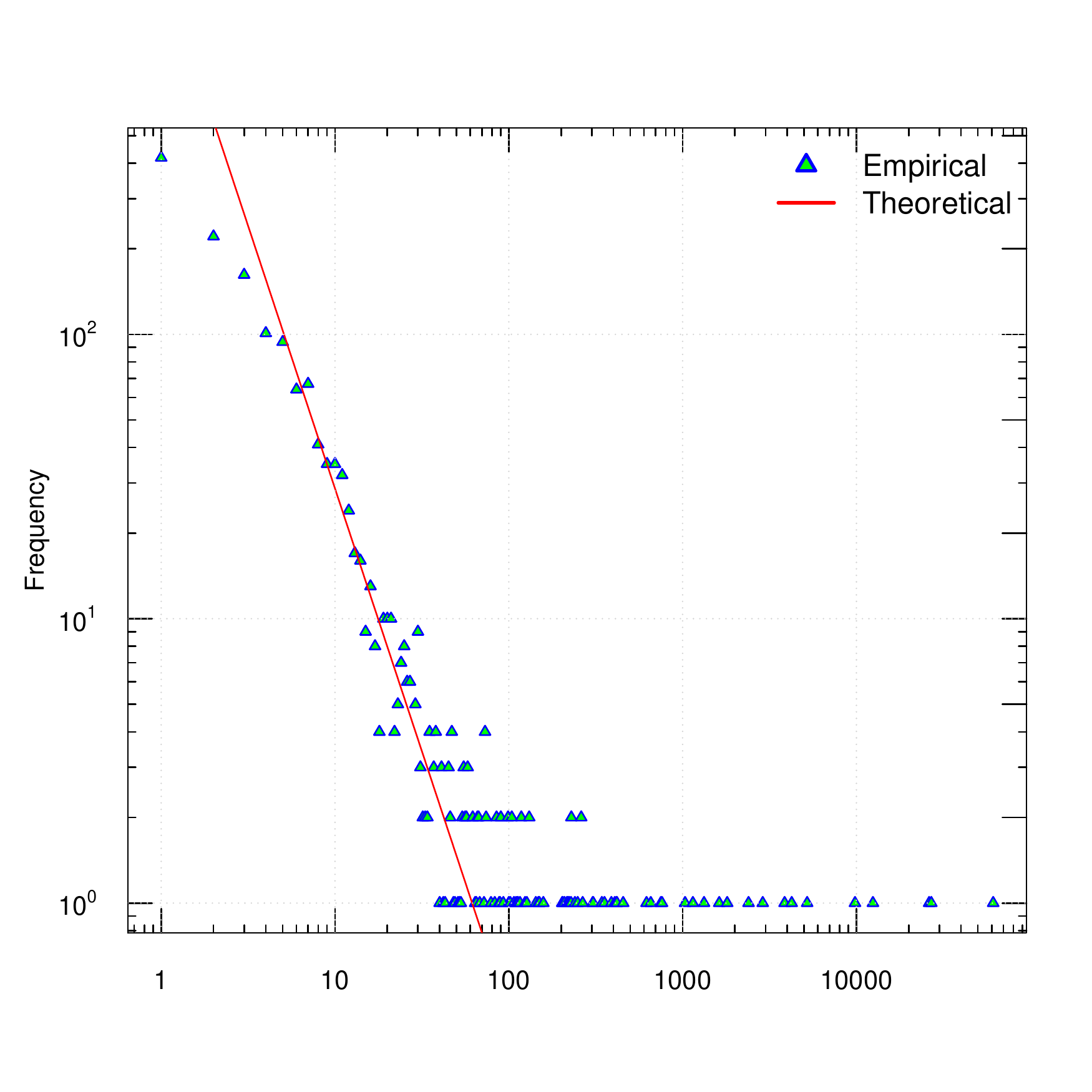}}
        \subfigure[DEMON]{\includegraphics[width=.121\textwidth]{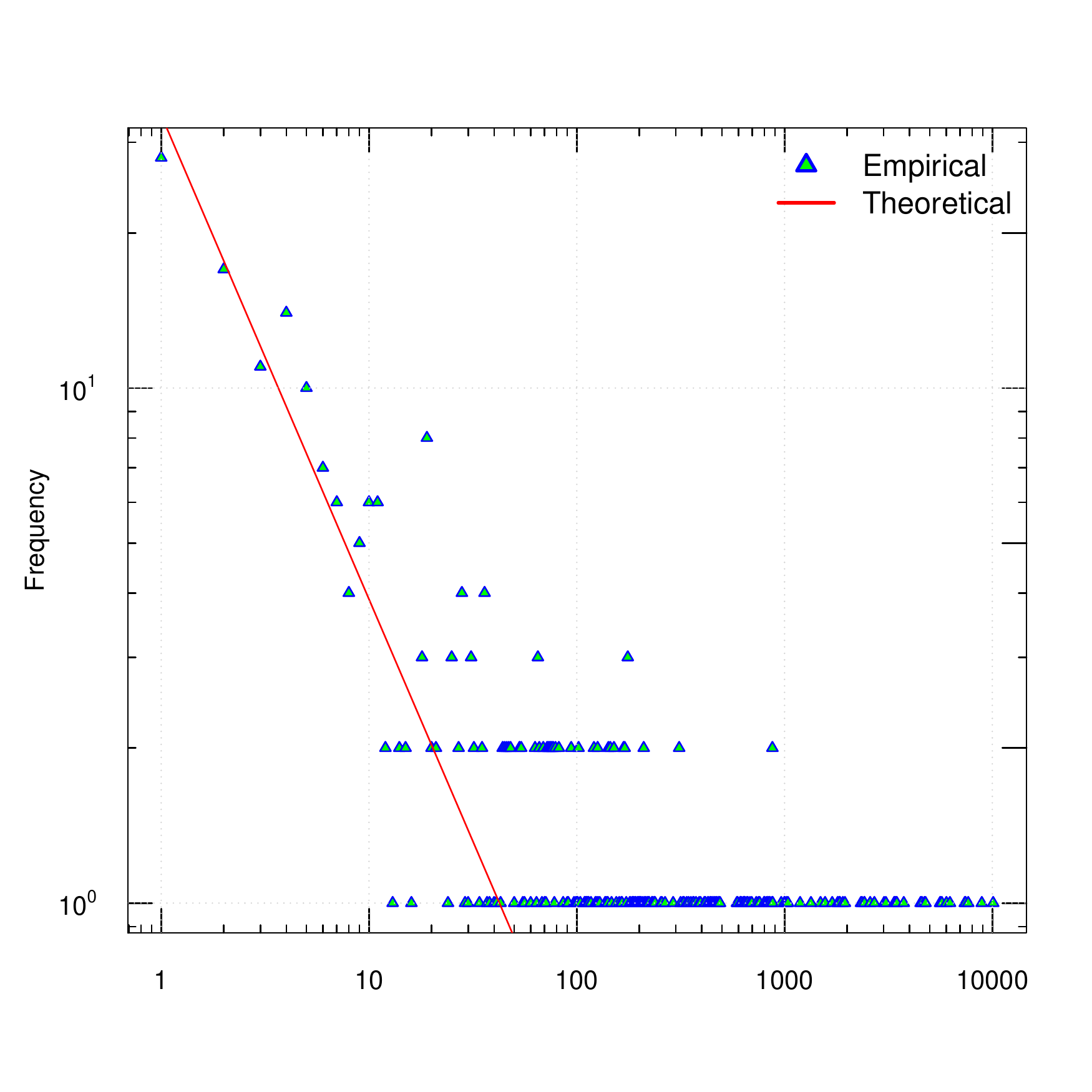}}
        \caption{\label{fig21}Log-log empirical Overlap Size distribution (dot) and Power-Law estimate (line) of aNobii Ground-truth (a), LFM (b),  GCE (c), OSLOM (d), MOSES (e), SLPA (f), and DEMON (g)}
        \end{figure}

        \begin{table}[!ht]
        \centering
        \caption{Basic properties of aNobii* and 'community-graph' of the overlapping community detection algorithms. The calculated properties are Number of nodes (V), Number of edges (E), Density ($\rho$), Diameter ($d$), Average shortest path ($l_{G}$), Average node degree ($\widetilde{deg}$), Max node degree ($\delta(G)$), Assortativity Coefficient ($\tau$), and Clustering Coefficient ($C$)}
        \label{table4}
        \begin{tabular}{lccccccccc}
        \hline
        &V&E&$\rho$&$d$&$l_{G}$&$\widetilde{deg}$&$\delta(G)$&$\tau$&$C$\\
        \hline
        aNobii* & 18970 & 823969 & 4.58E-03 & 8 & 5.53 & 2.04 & 7298 & -0.3 & 0.24 \\
        LFM* & 8159 & 15421 & 4.63E-04 & 12 & 8.14 & 2.51 & 212 & 0.01 & 0.06 \\
        GCE* & 2433 & 23246 & 7.86E-03 & 9 & 6.51 & 2.39 & 694 & -0.09 & 0.13 \\
        OSLOM* & 2492 & 8108 & 2.61E-03 & 11 & 8.13 & 2.91 & 92 & 0.07 & 0.09 \\
        MOSES* & 3396 & 107053 & 1.86E-02 & 6 & 5.11 & 2.03 & 1325 & -0.31 & 0.37 \\
        SLPA* & 5201 & 21413 & 1.58E-03 & 7 & 6.66 & 2.09 & 2576 & -0.31 & 0.02 \\
        DEMON* & 411 & 12655 & 1.50E-01 & 6 & 5.02 & 1.89 & 262 & -0.29 & 0.57 \\
        \hline
        \end{tabular}
        \end{table}

        \begin{table}[!ht]
        \centering
        \caption{KS-test values for the degree distribution with the aNobii dataset. The distribution under test are the Power-Law (PL), Beta (BE), Cauchy (CA), Exponential (E), Gamma (GM), Logistic (LO), Log-Normal (LN), Normal (N), Uniform (U), and Weibull (WB)}
        \label{table7}
        \begin{tabular}{lcccccccccc}
        \hline
         &  PL & BE & CA & E & GM & LO & LN & N & U & WB \\
        \hline
        aNobii* & 0.03 & 0.56 & 0.2 &	0.37 & 	0.52 & 	0.37 & 0.04 & 	0.39 & 	0.9 & 	0.21\\
		DEMON* & 	0.09 & 	0.11 & 	0.17 & 	0.19 & 	0.18 & 	0.15 & 	0.16 & 	0.17 & 	0.41 & 	0.09\\
        GCE* & 	0.03 & 	0.29 & 	0.25 & 	0.16 & 	0.26 & 	0.28 & 	0.06 & 	0.3 & 	0.86 & 	0.25\\
        LFM* & 	0.02 & 	0.42 & 	0.26 & 	0.36 & 	0.4 & 	0.36 & 	0.19 & 	0.38 & 	0.91 & 	0.2\\
        MOSES* & 0.07 & 	0.34 & 	0.22 & 	0.35 & 	0.26 & 	0.3 & 	0.07 & 	0.32 & 	0.77 & 	0.16\\
		OSLOM* & 0.05 & 	0.2 & 	0.25 & 	0.2 & 	0.2 & 	0.21 & 	0.11 & 	0.24 & 	0.76 & 	0.22\\
        SLPA* & 	0.03 & 	0.77 & 	0.16 & 	0.28 & 	0.76 & 	0.45 & 	0.11 & 	0.45 & 	0.97 & 	0.16\\
		\hline
        \end{tabular}
        \end{table}

        \begin{table}[!ht]
        \centering
        \caption{KS-test values for the Average clustering coefficient as a function of degree distribution with the aNobii dataset. The distribution under test are the Power-Law (PL), Beta (BE), Cauchy (CA), Exponential (E), Gamma (GM), Logistic (LO), Log-Normal (LN), Normal (N), Uniform (U), and Weibull (WB)}
        \label{table10}
        \begin{tabular}{lcccccccccc}
        \hline
         &  PL & BE & CA & E & GM & LO & LN & N & U & WB \\
        \hline
        aNobii* & 0.06 & 0.03 & 0.18 & 0.05 & 0.03 & 0.12 & 0.08 & 0.14 & 0.76 & 0.03 	\\
		DEMON* & 0.1 & 	0.06 & 	0.16 & 	0.09 & 	0.04 & 0.1 & 	0.1 & 	0.1 & 	0.32 & 	0.03 	\\
        GCE* & 	0.05 & 	0.08 & 	0.19 & 	0.04 & 	0.05 & 	0.16 & 	0.09 & 	0.18 & 	0.72 & 	0.03 	\\
        LFM* & 	0.07 & 	0.07 & 	0.22 & 	0.05 & 	0.05 & 	0.15 & 	0.09 & 	0.17 & 	0.55 & 	0.04 	\\
        MOSES* & 0.07 & 	0.05 & 	0.16 & 	0.07 & 	0.03 & 	0.14 & 	0.08 & 	0.14 & 	0.59 & 	0.04 	\\
        OSLOM* & 0.07 & 	0.04 & 	0.19 & 	0.09 & 	0.09 & 	0.12 & 	0.11 & 	0.11 & 	0.45 & 	0.06 	\\
		SLPA* & 	0.07 & 	0.48 & 	0.22 & 	0.09 & 	0.46 & 	0.33 & 	0.05 & 	0.35 & 	0.92 & 	0.06 	\\	
			\hline
        \end{tabular}
        \end{table}

        \begin{table}[!ht]
        \centering
        \caption{KS-test values for the Hop distance distribution with aNobii dataset. The distribution under test are the Power-Law (PL), Beta (BE), Cauchy (CA), Exponential (E), Gamma (GM), Logistic (LO), Log-Normal (LN), Normal (N), Uniform (U), and Weibull (WB)}
        \label{table15}
        \begin{tabular}{lcccccccccc}
        \hline
         &  PL & BE & CA & E & GM & LO & LN & N & U & WB \\
        \hline
        aNobii* & 0.37 & 0.05 & 0.34 & 0.15 & 0.68 & 0.96 & 0.93 & 0.03 & 0.72 & 	0.94\\
		DEMON* & 	0.44 & 	0.82 & 	0.4 & 	0.29 & 	0.71 & 	0.1 & 	0.66 & 	0.02 & 0.67 & 	0.56\\
		GCE* & 	0.95 & 	0.93 & 	0.09 & 	0.67 & 	0.33 & 	0.08 & 	0.97 & 	0.04 & 	0.25 & 	0.23\\
		LFM* & 	0.38 & 	0.04 & 	0.26 & 	0.16 & 	0.8 & 	0.78 & 	0.36 & 	0.03 & 	0.46 & 	0.76\\
		MOSES* & 	0.87 & 	0.24 & 	0.47 & 	0.31 & 	0.93 & 	0.87 & 	0.62 & 	0.05 & 	0.95 & 	0.88\\
		OSLOM* & 	0.23 & 	0.75 & 	0.86 & 	0.92 & 	0.18 & 	0.62 & 	0.46 & 	0.04 & 	0.99 & 	0.59\\
		SLPA* & 	0.87 & 	0.51 & 	0.42 & 	0.34 & 	0.81 & 	0.95 & 	0.16 & 	0.02 & 	0.67 & 	0.88\\
			\hline
        \end{tabular}
        \end{table}

        \begin{table}[!ht]
        \centering
        \caption{KS-test values for the Community size distribution with the aNobii dataset. The distribution under test are the Power-Law (PL), Beta (BE), Cauchy (CA), Exponential (E), Gamma (GM), Logistic (LO), Log-Normal (LN), Normal (N), Uniform (U), and Weibull (WB)}
        \label{table19}
        \begin{tabular}{lcccccccccc}
        \hline
         &  PL & BE & CA & E & GM & LO & LN & N & U & WB \\
        \hline
				Ground-truth & 0.04 & 0.62 & 0.24 & 0.68 & 0.62 & 0.45 & 0.23 & 0.46 & 0.96 & 0.17\\
				DEMON & 0.05 & 0.54 & 0.26 & 0.51 & 0.28 & 0.36 & 0.11 & 0.36 & 0.77 & 0.22\\
				GCE & 0.02 & 0.54 & 0.23 & 0.19 & 0.52 & 0.34 & 0.03 & 0.36 & 0.92 & 0.22\\
				LFM & 0.03 & 0.56 & 0.25 & 0.21 & 0.55 & 0.38 & 0.1 & 0.39 & 0.95 & 0.17\\
				MOSES & 0.04 & 0.6 & 0.17 & 0.46 & 0.55 & 0.38 & 0.11 & 0.39 & 0.89 & 0.18\\
				OSLOM & 0.01 & 0.25 & 0.23 & 0.07 & 0.22 & 0.26 & 0.09 & 0.28 & 0.86 & 0.26\\
				SLPA & 0.02 & 0.78 & 0.59 & 0.77 & 0.48 & 0.11 & 0.49 & 0.99 & 0.31 & 0.44\\
			\hline
        \end{tabular}
        \end{table}

        \begin{table}[!ht]
        \centering
        \caption{KS-test values for the Membership distribution with the aNobii dataset. The distribution under test are the Power-Law (PL), Beta (BE), Cauchy (CA), Exponential (E), Gamma (GM), Logistic (LO), Log-Normal (LN), Normal (N), Uniform (U), and Weibull (WB)}
        \label{table22}
        \begin{tabular}{lcccccccccc}
        \hline
        &  PL & BE & CA & E & GM & LO & LN & N & U & WB \\
        \hline
			    Ground-truth & 0.02 & 0.39 & 0.22 & 0.39 & 0.39 & 0.33 & 0.16 & 0.35 & 0.92 & 0.24\\
				DEMON & 0.03 & 0.63 & 0.26 & 0.63 & 0.63 & 0.33 & 0.25 & 0.35 & 0.87 & 0.24\\
				GCE & 0.02 & 0.64 & 0.31 & 0.64 & 0.64 & 0.38 & 0.37 & 0.35 & 0.8 & 0.3\\
				LFM & 0.03 & 0.31 & 0.21 & 0.41 & 0.58 & 0.17 & 0.21 & 0.97 & 0.82 & 0.28\\
				MOSES & 0.02 & 0.62 & 0.24 & 0.62 & 0.62 & 0.37 & 0.33 & 0.39 & 0.95 & 0.22\\
				OSLOM & 0.03 & 34 & 0.51 & 0.55 & 0.62 & 0.34 & 0.47 & 0.52 & 0.17 & 0.78\\
				SLPA & 0.03 & 0.57 & 0.24 & 0.57 & 0.57 & 0.35 & 0.27 & 0.33 & 0.63 & 0.23\\
		\hline
        \end{tabular}
        \end{table}

        \begin{table}[!ht]
        \centering
        \caption{KS-test values for the overlap size distribution with the aNobii dataset. The distribution under test are the Power-Law (PL), Beta (BE), Cauchy (CA), Exponential (E), Gamma (GM), Logistic (LO), Log-Normal (LN), Normal (N), Uniform (U), and Weibull (WB)}
        \label{table25}
        \begin{tabular}{lcccccccccc}
        \hline
         &  PL & BE & CA & E & GM & LO & LN & N & U & WB \\
        \hline
			Ground-truth & 0.02 & 0.76 & 0.24 & 0.55 & 0.74 & 0.44 & 0.07 & 0.44 & 0.95 & 0.21\\
			DEMON & 0.09 & 0.38 & 0.19 & 0.4 & 0.24 & 0.33 & 0.05 & 0.35 & 0.8 & 0.12\\
			GCE & 0.03 & 0.55 & 0.24 & 0.33 & 0.5 & 0.37 & 0.05 & 0.39 & 0.89 & 0.25\\
			LFM & 0.02 & 0.46 & 0.25 & 0.36 & 0.43 & 0.38 & 0.14 & 0.4 & 0.91 & 0.18\\
			OSLOM & 0.02 & 0.25 & 0.22 & 0.25 & 0.25 & 0.21 & 0.12 & 0.23 & 0.78 & 0.25\\
			SLPA & 0.03 & 0.71 & 0.24 & 0.73 & 0.7 & 0.47 & 0.13 & 0.48 & 0.98 & 0.31\\
		\hline
        \end{tabular}
        \end{table}

        \begin{table}[!ht]
        \centering
        \caption{Ranking of the algorithms based on basic properties with the aNobii  dataset. The calculated properties are Number of nodes (V), Number of edges (E), Density ($\rho$), Diameter ($d$), Average shortest path ($l_{G}$), Average node degree ($\widetilde{deg}$), Max node degree ($\delta(G)$), Assortativity Coefficient ($\tau$), and Clustering Coefficient ($C$). Kconsensus  and TOPSIS denotes respectively the final ranking using Kemeny consensus and TOPSIS.}
        \label{table57}
        \begin{tabular}{lccccccccccc}
        \hline
          &V&E&$\rho$&$d$&$l_{G}$&$\widetilde{deg}$&$\delta(G)$&$\tau$&$C$&Kconsensus&TOPSIS\\
        \hline
            LFM*&1&4&4&6&6&5&5&5&4&4&5\\
            GCE*&5&2&3&2&3&3&4&4&1&3&3\\
            OSLOM*&4&6&1&5&5&6&6&6&3&6&4\\
            MOSES*&3&1&5&3&1&2&1&2&2&1&1\\
            SLPA*&2&3&2&1&4&1&2&1&5&2&2\\
            DEMON*&6&5&6&3&2&4&3&3&6&5&6\\
        \hline
        \end{tabular}
        \end{table}

        \begin{table}[!ht]
          \centering
          \caption{Correlation of basic properties rankings for aNobii dataset. The calculated properties are Number of nodes (V), Number of edges (E), Density ($\rho$), Diameter ($d$), Average shortest path ($l_{G}$), Average node degree ($\widetilde{deg}$), Max node degree ($\delta(G)$), Assortativity Coefficient ($\tau$), and Clustering Coefficient ($C$)}
          \label{table58}
            \begin{tabular}{lccccccccc}
            \hline
               &V&E&$\rho$&$d$&$l_{G}$&$\widetilde{deg}$&$\delta(G)$&$\tau$&$C$ \\
            \hline
            V & 1 &   &   &   &   &   &   &   &  \\
            E & 0.2 & 1 &   &   &   &   &   &   &  \\
            $\rho$ & 0.26 & -0.26 & 1 &   &   &   &   &   &  \\
            $d$ & -0.29 & 0.52 & 0 & 1 &   &   &   &   &  \\
            $l_{G}$ & -0.54 & 0.54 & -0.6 & 0.57 & 1 &   &   &   &  \\
            $\widetilde{deg}$ & 0.14 & 0.77 & -0.14 & 0.86 & 0.54 & 1 &   &   &  \\
            $\delta(G)$ & 0.03 & 0.71 & -0.49 & 0.69 & 0.77 & 0.89 & 1 &   &  \\
            $\tau$ & 0.09 & 0.6 & -0.31 & 0.8 & 0.6 & 0.94 & 0.94 & 1 &  \\
            $C$  & -0.03 & 0.54 & 0.26 & 0 & 0.14 & 0.03 & -0.09 & -0.26 & 1 \\
            \hline
            \end{tabular}%
        \end{table}%

        \begin{table}[!ht]
          \centering
          \caption{Ranking of the algorithms based on microscopic properties with the aNobii dataset. The distribution under test are the Degree distribution (DD), the Average clustering coefficient as function of degree (Av), the Hop distance (HD). Kconsensus  and TOPSIS denotes respectively the final ranking using Kemeny consensus and TOPSIS.}
          \label{table580}
            \begin{tabular}{lccccc}
            \hline
              & \multicolumn{1}{l}{DD} & \multicolumn{1}{l}{Av} & \multicolumn{1}{l}{HD} & \multicolumn{1}{l}{Kconsensus} & \multicolumn{1}{l}{TOPSIS} \\
            \hline
            LFM* & 3 & 4 & 5 &  5 & 5 \\
            GCE* & 1 & 3 & 4 & 4  & 2 \\
            OSLOM* & 6 & 2 & 6 & 6  & 4 \\
            MOSES* & 2 & 1 & 2 &  2 & 1 \\
            SLPA* & 4 & 5 & 1 &  1 & 3 \\
            DEMON* & 5 & 6 & 3 &  3 & 6 \\
            \hline
            \end{tabular}%
        \end{table}%

        \begin{table}[!ht]
          \centering
          \caption{Correlation of the rankings of the microscopic properties with the aNobii dataset (Degree distribution (DD), the Average clustering coefficient as function of degree (Av), the Hop distance (HD))}
          \label{table64}
            \begin{tabular}{lccc}
            \hline
              & DD & Av & HD \\
              \hline
            DD & 1 &   &  \\
            Av & 0.31 & 1 &  \\
            HD & 0.25 & -0.25 & 1 \\
            \hline
            \end{tabular}%
        \end{table}%

        \begin{table}[!ht]
          \centering
          \caption{Mesoscopic properties ranking for aNobii dataset. The distribution under test are the Community size (CS), the Membership (M), the Overlap size (OS). Kconsensus  and TOPSIS denotes respectively the final ranking using Kemeny consensus and TOPSIS.}
          \label{table581}
            \begin{tabular}{lccccc}
            \hline
              & CS & M & OS & Kconsensus & TOPSIS \\
            \hline
            LFM & 3 & 2 & 6 & 6  & 3 \\
            GCE & 6 & 5 & 5 & 5  & 6 \\
            OSLOM & 5 & 6 & 3 & 3  & 5 \\
            MOSES & 1 & 1 & 2 & 2  & 1 \\
            SLPA & 4 & 3 & 4 &  4 & 4 \\
            DEMON & 2 & 4 & 1 & 1  & 2 \\
            \hline
            \end{tabular}%
        \end{table}%

        \begin{table}[!ht]
          \centering
          \caption{Correlation of the rankings of the microscopic properties for aNobii dataset (the Community size (CS), the Membership (M), the Overlap size (OS))}
          \label{table67}
            \begin{tabular}{lccc}
            \hline
              & CS & M & OS \\
              \hline
            CS & 1 &   &  \\
            M & 0.77 & 1 &  \\
            OS & 0.54 & -0.02 & 1 \\
            \hline
            \end{tabular}%
        \end{table}%

        \begin{table}[!ht]
          \centering
          \caption{Ranking of the algorithms based on all topological properties with the aNobii dataset. The calculated properties are Number of nodes (V), Number of edges (E), Density ($\rho$), Diameter ($d$), Average shortest path ($l_{G}$), Average node degree ($\widetilde{deg}$), Max node degree ($\delta(G)$), Assortativity Coefficient ($\tau$), Clustering Coefficient ($C$), the Degree distribution (DD), the Average clustering coefficient as function of degree (Av), the Hop distance (HD), the Community size (CS), the Membership (M), the Overlap size (OS).}
          \small
          \label{table582}
            \begin{tabular}{lccccccccccccccccc}
            \hline
              & \multicolumn{9}{c}{Basic properties} & \multicolumn{3}{c}{Microscopic properties} & \multicolumn{3}{c}{Mesoscopic} &  \multicolumn{2}{c}{MCDM Ranking}  \\
            \hline
              &V&E&$\rho$&$d$&$l_{G}$&$\widetilde{deg}$&$\delta(G)$&$\tau$&$C$& DD & Av & HD & CS & M & OS & Kconsensus & TOPSIS \\
            \hline
            LFM & 1 & 4 & 4 & 6 & 6 & 5 & 5 & 5 & 4 & 3 & 4 & 5 & 3 & 2 & 6 & 6  & 5 \\
            GCE & 5 & 2 & 3 & 2 & 3 & 3 & 4 & 4 & 1 & 1 & 3 & 4 & 6 & 5 & 5 &  5 & 3 \\
            OSLOM & 4 & 6 & 1 & 5 & 5 & 6 & 6 & 6 & 3 & 6 & 2 & 6 & 5 & 6 & 3 & 3  & 6 \\
            MOSES & 3 & 1 & 5 & 3 & 1 & 2 & 1 & 2 & 2 & 2 & 1 & 2 & 1 & 1 & 2 & 2  & 1 \\
            SLPA & 2 & 3 & 2 & 1 & 4 & 1 & 2 & 1 & 5 & 4 & 5 & 1 & 4 & 3 & 4 &  4 & 2 \\
            DEMON & 6 & 5 & 6 & 3 & 2 & 4 & 3 & 3 & 6 & 5 & 6 & 3 & 2 & 4 & 1 & 1  & 4 \\
            \hline
            \end{tabular}%
        \end{table}%

        \begin{table}[!ht]
          \centering
          \footnotesize
          \caption{Correlation of ranking of all topological properties with the aNobii dataset. The calculated properties are Number of nodes (V), Number of edges (E), Density ($\rho$), Diameter ($d$), Average shortest path ($l_{G}$), Average node degree ($\widetilde{deg}$), Max node degree ($\delta(G)$), Assortativity Coefficient ($\tau$), Clustering Coefficient ($C$), the Degree distribution (DD), the Average clustering coefficient as function of degree (Av), the Hop distance (HD), the Community size (CS), the Membership (M), the overlap size (OS).}
          \label{table70}
            \begin{tabular}{lccccccccccccccc}
            \hline
              & V & E & $\rho$ & $d$ & $l_{G}$ & $\widetilde{deg}$ & $\delta(G)$ & $\tau$ & $C$ & DD & Av & HD & CS & M & OS \\
              \hline
            V  & 1 &   &   &   &   &   &   &   &   &   &   &   &   &   &  \\
            E  & 0.2 & 1 &   &   &   &   &   &   &   &   &   &   &   &   &  \\
            $\rho$  & 0.26 & -0.26 & 1 &   &   &   &   &   &   &   &   &   &   &   &  \\
            $d$  & -0.29 & 0.52 & 0 & 1 &   &   &   &   &   &   &   &   &   &   &  \\
            $l_{G}$  & -0.54 & 0.54 & -0.6 & 0.57 & 1 &   &   &   &   &   &   &   &   &   &  \\
            $\widetilde{deg}$  & 0.14 & 0.77 & -0.14 & 0.86 & 0.54 & 1 &   &   &   &   &   &   &   &   &  \\
            $\delta(G)$  & 0.03 & 0.71 & -0.49 & 0.69 & 0.77 & 0.89 & 1 &   &   &   &   &   &   &   &  \\
            $\tau$  & 0.09 & 0.6 & -0.31 & 0.8 & 0.6 & 0.94 & 0.94 & 1 &   &   &   &   &   &   &  \\
            $C$  & -0.03 & 0.54 & 0.26 & 0 & 0.14 & 0.03 & -0.09 & -0.26 & 1 &   &   &   &   &   &  \\
            DD  & \textcolor[rgb]{ 1,  0,  0}{\textbf{0.14}} & \textcolor[rgb]{ 1,  0,  0}{\textbf{0.89}} & \textcolor[rgb]{ 1,  0,  0}{\textbf{-0.26}} & \textcolor[rgb]{ 1,  0,  0}{\textbf{0.29}} & \textcolor[rgb]{ 1,  0,  0}{\textbf{0.31}} & \textcolor[rgb]{ 1,  0,  0}{\textbf{0.49}} & \textcolor[rgb]{ 1,  0,  0}{\textbf{0.37}} & \textcolor[rgb]{ 1,  0,  0}{\textbf{0.26}} & \textcolor[rgb]{ 1,  0,  0}{\textbf{0.66}} & 1 &   &   &   &   &  \\
            Av  & \textcolor[rgb]{ 1,  0,  0}{\textbf{0.14}} & \textcolor[rgb]{ 1,  0,  0}{\textbf{0.37}} & \textcolor[rgb]{ 1,  0,  0}{\textbf{0.26}} & \textcolor[rgb]{ 1,  0,  0}{\textbf{-0.23}} & \textcolor[rgb]{ 1,  0,  0}{\textbf{0.14}} & \textcolor[rgb]{ 1,  0,  0}{\textbf{-0.09}} & \textcolor[rgb]{ 1,  0,  0}{\textbf{-0.03}} & \textcolor[rgb]{ 1,  0,  0}{\textbf{-0.26}} & \textcolor[rgb]{ 1,  0,  0}{\textbf{0.83}} & 0.31 & 1 &   &   &   &  \\
            HD  & \textcolor[rgb]{ 1,  0,  0}{\textbf{0.09}} & \textcolor[rgb]{ 1,  0,  0}{\textbf{0.6}} & \textcolor[rgb]{ 1,  0,  0}{\textbf{-0.31}} & \textcolor[rgb]{ 1,  0,  0}{\textbf{0.8}} & \textcolor[rgb]{ 1,  0,  0}{\textbf{0.6}} & \textcolor[rgb]{ 1,  0,  0}{\textbf{0.94}} & \textcolor[rgb]{ 1,  0,  0}{\textbf{0.94}} & \textcolor[rgb]{ 1,  0,  0}{\textbf{0.93}} & \textcolor[rgb]{ 1,  0,  0}{\textbf{-0.26}} & 0.26 & -0.26 & 1 &   &   &  \\
            CS  & \textcolor[rgb]{ 1,  0,  0}{\textbf{0.14}} & \textcolor[rgb]{ 1,  0,  0}{\textbf{0.2}} & \textcolor[rgb]{ 1,  0,  0}{\textbf{-0.77}} & \textcolor[rgb]{ 1,  0,  0}{\textbf{-0.11}} & \textcolor[rgb]{ 1,  0,  0}{\textbf{0.49}} & \textcolor[rgb]{ 1,  0,  0}{\textbf{0.2}} & \textcolor[rgb]{ 1,  0,  0}{\textbf{0.6}} & \textcolor[rgb]{ 1,  0,  0}{\textbf{0.43}} & \textcolor[rgb]{ 1,  0,  0}{\textbf{-0.37}} & \textcolor[rgb]{ 1,  0,  0}{\textbf{-0.03}} & \textcolor[rgb]{ 1,  0,  0}{\textbf{-0.03}} & \textcolor[rgb]{ 1,  0,  0}{\textbf{0.43}} & 1 &   &  \\
            M  & \textcolor[rgb]{ 1,  0,  0}{\textbf{0.6}} & \textcolor[rgb]{ 1,  0,  0}{\textbf{0.6}} & \textcolor[rgb]{ 1,  0,  0}{\textbf{-0.54}} & \textcolor[rgb]{ 1,  0,  0}{\textbf{0.01}} & \textcolor[rgb]{ 1,  0,  0}{\textbf{0.26}} & \textcolor[rgb]{ 1,  0,  0}{\textbf{0.49}} & \textcolor[rgb]{ 1,  0,  0}{\textbf{0.66}} & \textcolor[rgb]{ 1,  0,  0}{\textbf{0.54}} & \textcolor[rgb]{ 1,  0,  0}{\textbf{-0.09}} & \textcolor[rgb]{ 1,  0,  0}{\textbf{0.43}} & \textcolor[rgb]{ 1,  0,  0}{\textbf{0.09}} & \textcolor[rgb]{ 1,  0,  0}{\textbf{0.54}} & 0.77 & 1 &  \\
            OS  & \textcolor[rgb]{ 1,  0,  0}{\textbf{-0.6}} & \textcolor[rgb]{ 1,  0,  0}{\textbf{-0.14}} & \textcolor[rgb]{ 1,  0,  0}{\textbf{-0.43}} & \textcolor[rgb]{ 1,  0,  0}{\textbf{0.23}} & \textcolor[rgb]{ 1,  0,  0}{\textbf{0.71}} & \textcolor[rgb]{ 1,  0,  0}{\textbf{0.09}} & \textcolor[rgb]{ 1,  0,  0}{\textbf{0.43}} & \textcolor[rgb]{ 1,  0,  0}{\textbf{0.31}} & \textcolor[rgb]{ 1,  0,  0}{\textbf{-0.31}} & \textcolor[rgb]{ 1,  0,  0}{\textbf{-0.43}} & \textcolor[rgb]{ 1,  0,  0}{\textbf{-0.03}} & \textcolor[rgb]{ 1,  0,  0}{\textbf{0.31}} & 0.54 & -0.03 & 1 \\
            \hline
            \end{tabular}%
        \end{table}%

        \begin{table}[htbp!]
          \centering
          \caption{Correlation of the basic, microscopic, and mesoscopic rankings for aNobii dataset.}
          \label{table90}
            \begin{tabular}{lccc}
            \hline
              & Basic & Micro & Meso \\
              \hline
            Basic & 1 &   &  \\
            Micro & 0.6 & 1 &  \\
            Meso & -0.14 & 0.31 & 1 \\
            \hline
            \end{tabular}%
        \end{table}%

        \begin{table}[!ht]
          \centering
          \caption{Quality metrics values for aNobii ground-truth and the uncovered community structure. The calculated properties are Average Degree (AD), Average ODF (AO), Flake ODF (FO), Internal Density (ID), Max ODF (MO), and Overlapping Modularity (OM).}
          \label{table46}%
            \begin{tabular}{lcccccc}
            \hline
              & AD & AO & FO & ID & MO & OM \\
            \hline
            aNobii & 1.38 & 45.96 & 9.79 & 0.82 & 102.55 & 0.63 \\
            LFM&0.92&4.92&5.47&0.22&21.06&0.07\\
            GCE&3.37&4.16&16.76&0.22&68.29&0.17\\
            OSLOM&3.28&7.72&14.48&0.23&140.74&0.25\\
            MOSES&4.87&63.55&22.39&0.49&962.72&0.05\\
            SLPA&2.18&3.92&10.72&0.47&19.59&0.4\\
            DEMON&5.16&88.26&55.35&0.45&1393.01&0.04\\
            \hline
            \end{tabular}%
        \end{table}%

        \begin{table}[!ht]
          \centering
          \caption{Quality metrics ranking for overlapping community detection algorithms with the aNobii dataset. The calculated properties are Average Degree (AD), Average ODF (AO), Flake ODF (FO), Internal Density (ID), Max ODF (MO), and Overlapping Modularity (OM). Kconsensus denotes the quality metrics ranking using Kemeny consensus.}
          \label{table53}%
            \begin{tabular}{lcccccccc}
            \hline
              & AD & AO & FO & ID & MO & OM & Kconsensus & TOPSIS\\
            \hline
            LFM&1&3&2&5&3&4&3&4\\
            GCE&4&4&4&5&1&3&4&5\\
            OSLOM&3&2&3&4&2&2&2&3\\
            MOSES&5&1&5&1&5&5&1&1\\
            SLPA&2&5&1&2&4&1&5&2\\
            DEMON&6&6&6&3&6&6&6&6\\
            \hline
            \end{tabular}%
    \end{table}%

        \begin{table}[htbp!]
          \centering
            \caption{Correlation of the quality metrics ranking for aNobii dataset. The calculated properties are Average Degree (AD), Average ODF (AO), Flake ODF (FO), Internal Density (ID), Max ODF (MO), and Overlapping Modularity (OM). Kconsensus denotes the quality metrics ranking using Kemeny consensus.}
                  \label{table84}%
            \begin{tabular}{lcccccc}
            \hline
              & AD & AO & FO & ID & MO & OM \\
              \hline
            AD & 1 &   &   &   &   &  \\
            AO & 0.14 & 1 &   &   &   &  \\
            FO & 0.94 & 0.03 & 1 &   &   &  \\
            ID & -0.39 & 0.13 & -0.2 & 1 &   &  \\
            MO & 0.49 & 0.26 & 0.43 & -0.72 & 1 &  \\
            OM & 0.66 & 0.03 & 0.83 & -0.13 & 0.6 & 1 \\
            \hline
            \end{tabular}%
        \end{table}%

        \begin{table}[!ht]
          \centering
          \caption{Clustering metrics for aNobii ground-truth and the uncovered community structure by overlapping community detection algorithms. The calculated properties are NMI, Omega Index (OI) and F1-score.}
          \label{table28}
            \begin{tabular}{lccc}
            \hline
              &  NMI  &  OI  &  F1-score \\
            \hline
            LFM & 0.22 & 0.1 & 0.12 \\
            GCE & 0.14 & 0.22 & 0.12 \\
            OSLOM & 0.34 & 0.27 & 0.24 \\
            MOSES & 0.17 & 0.08 & 0.64 \\
            SLPA & 0.51 & 0.41 & 0.38 \\
            DEMON & 0.47 & 0.09 & 0.37 \\
            \hline
            \end{tabular}%
        \end{table}%

        \begin{table}[!ht]
          \centering
          \caption{Clustering metrics ranking for overlapping community detection algorithmes applied on aNobii. The calculated properties are NMI, Omega Index (OI) and F1-score. Kconsensus  and TOPSIS denotes respectively the final ranking using Kemeny consensus and TOPSIS.}
                  \label{table34}
            \begin{tabular}{lcccccc}
            \hline
              & NMI  & OI & F1-score & Kconsensus & TOPSIS\\
            \hline
            LFM & 4 & 4 & 5 & 4 & 6 \\
            GCE & 6 & 3 & 6 & 3 & 5 \\
            OSLOM & 3 & 2 & 4 & 2 & 3 \\
            MOSES & 5 & 6 & 1 & 6 & 2 \\
            SLPA & 1 & 1 & 2 & 1 & 1 \\
            DEMON & 2 & 5 & 3 & 5 & 4 \\
            \hline
            \end{tabular}%
        \end{table}%

        \begin{table}[htbp!]
          \centering
          \caption{Correlation of the clustering metrics ranking for overlapping community detection algorithmes applied on aNobii.}
                  \label{table87}
            \begin{tabular}{lccc}
            \hline
              & NMI & OI & F1-score \\
              \hline
            NMI & 1 &   &  \\
            OI & 0.42 & 1 &  \\
            F1-score & 0.42 & -0.25 & 1 \\

            \hline
            \end{tabular}%
        \end{table}%

        \begin{table}[ht!]
          \centering
          \caption{Correlation of the topological properties, the quality metrics and the clustering measures rankings using the Kconsensus strategy for aNobii dataset.}
          \label{table99}%
            \begin{tabular}{lccc}
            \hline
              & \multicolumn{1}{l}{Topo} & \multicolumn{1}{l}{Quality} & \multicolumn{1}{l}{Clustering} \\
            \hline
            Topo & 1 &   &  \\
            Quality & -0.16 & 1 &  \\
            clustering & 0.74& -0.31& 1 \\
            \hline
            \end{tabular}%
        \end{table}%

        \begin{table}[ht!]
          \centering
          \caption{Correlation of the topological properties, the quality metrics and the clustering measures rankings using the TOPSIS strategy for aNobii dataset.}
          \label{table990}%
            \begin{tabular}{lccc}
            \hline
              & \multicolumn{1}{l}{Topo} & \multicolumn{1}{l}{Quality} & \multicolumn{1}{l}{Clustering} \\
            \hline
            Topo & 1 &   &  \\
            Quality & 0.64 & 1 &  \\
            Clustering & 0.49 & 0.32 & 1 \\
            \hline
            \end{tabular}%
        \end{table}%

        \begin{table}[!ht]
          \centering
          \caption{Ranking of the algorithms based on all properties  with the aNobii dataset. The calculated properties are Number of nodes (V), Number of edges (E), Density ($\rho$), Diameter ($d$), Average shortest path ($l_{G}$), Average node degree ($\widetilde{deg}$), Max node degree ($\delta(G)$), Assortativity Coefficient ($\tau$), Clustering Coefficient ($C$), the Degree distribution (DD), the Average clustering coefficient as function of degree (Av), the Hop distance (HD), the Community size (CS), the membership (M), the Overlap size (OS), Average Degree (AD), Average ODF (AO), Flake ODF (FO), Internal Density (ID), Max ODF (MO), Overlapping Modularity (OM), NMI, Omega Index (OI) and F1-score. Kconsensus  and TOPSIS denotes respectively the final ranking using Kemeny consensus and TOPSIS.}
          \label{table61}
          \scriptsize
            \begin{tabular}{{p{.9cm}|}*{5}{p{.008cm}}*{1}{p{.1cm}}*{1}{p{.18cm}}*{1}{p{.008cm}}*{1}{p{.008cm}|}*{2}{p{.08cm}}*{1}{p{.08cm}|}*{2}{p{.08cm}}*{1}{p{.08cm}|}*{5}{p{.08cm}}*{1}{p{.18cm}|}*{1}{p{.18cm}}*{1}{p{.04cm}}*{1}{p{.9cm}|}*{2}{p{.7cm}}}
            \hline
              & \multicolumn{9}{c|}{Basic properties} & \multicolumn{3}{c|}{Microscopic} & \multicolumn{3}{c|}{Mesoscopic} & \multicolumn{6}{c|}{Clustering} & \multicolumn{3}{c|}{Quality} & \multicolumn{2}{c}{MCDM Ranking} \\
              \hline
              &V&E&$\rho$&$d$&$l_{G}$&$\widetilde{deg}$&$\delta(G)$&$\tau$&$C$& DD & Av & HD & CS & M & OS & AD & AO & FO & ID & MO & OM & NMI & OI & F1-score & Kconsensus & TOPSIS \\
            \hline
            LFM & 1 & 4 & 4 & 6 & 6 & 5 & 5 & 5 & 4 & 3 & 4 & 5 & 3 & 2 & 6 & 1 & 3 & 2 & 5 & 3 & 4 & 4 & 4 & 5 & 4 & 4 \\
            GCE & 5 & 2 & 3 & 2 & 3 & 3 & 4 & 4 & 1 & 1 & 3 & 4 & 6 & 5 & 5 & 4 & 4 & 4 & 5 & 1 & 3 & 6 & 3 & 6 & 3 & 3 \\
            OSLOM & 4 & 6 & 1 & 5 & 5 & 6 & 6 & 6 & 3 & 6 & 2 & 6 & 5 & 6 & 3 & 3 & 2 & 3 & 4 & 2 & 2 & 3 & 2 & 4 & 1 & 5 \\
            MOSES & 3 & 1 & 5 & 3 & 1 & 2 & 1 & 2 & 2 & 2 & 1 & 2 & 1 & 1 & 2 & 5 & 1 & 5 & 1 & 5 & 5 & 5 & 6 & 1 & 5 & 1 \\
            SLPA & 2 & 3 & 2 & 1 & 4 & 1 & 2 & 1 & 5 & 4 & 5 & 1 & 4 & 3 & 4 & 2 & 5 & 1 & 2 & 4 & 1 & 1 & 1 & 2 & 2 & 2 \\
            DEMON & 6 & 5 & 6 & 3 & 2 & 4 & 3 & 3 & 6 & 5 & 6 & 3 & 2 & 4 & 1 & 6 & 6 & 6 & 3 & 6 & 6 & 2 & 5 & 3 & 6 & 6 \\
            \hline
            \end{tabular}%
        \end{table}%

\clearpage
\section{Parameters for all datasets: PGP, AMAZON, and aNobii}
        \begin{table}[!ht]
        \centering
        \caption{Mean and Standard deviation of PGP, AMAZON and aNobii. The calculated properties are mean (M) and standard deviation (SD)}
        \label{table11}
        \begin{tabular}{lcccccc}
        \hline
        & \multicolumn{2}{c}{PGP} & \multicolumn{2}{c}{AMAZON} & \multicolumn{2}{c}{aNobii} \\
        \hline
        & M     & SD    & M     & SD &M     & SD      \\
        \hline
        Community-graph & 2.75 & 0.44 & 1.38  & 5.4& 0.84  & 3.2 \\
        CFINDER*&$\times$&$\times$&2.54&7.77&$\times$&$\times$\\
        LFM*&8.5&0.5&3.87&13.56&1.37&5.2\\
        GCE*&2.82&0.7&3.24&10.34&1.16&3.7\\
        OSLOM*&3.9&0.43&3.65&11.57&1.36&5.0\\
        LINKC*&5.1&0.51&2.8&11.6&$\times$&$\times$\\
        SVINET*&$\times$&$\times$&1.38&5.4&$\times$&$\times$\\
        MOSES*&$\times$&$\times$&2.03&7.91&0.88&3.1\\
        SLPA*&2.2&0.64&1.71&6.82&0.88&3.2\\
        DEMON*&2.81&0.69&1.47&5.65&0.89&2.6\\
        \hline
        \end{tabular}
        \end{table}

        \begin{table}[!ht]
        \centering
        \caption{Median, effective diameter and diameter of PGP, AMAZON and aNobii. The calculated properties are number of nodes Median (M), effective diameter (EM) and diameter (D)}
        \label{table12}
        \begin{tabular}{lccccccccc}
        \hline
        & \multicolumn{3}{c}{PGP}  & \multicolumn{3}{c}{AMAZON} & \multicolumn{3}{c}{aNobii}    \\
        \hline
           & M & EM & D & M & EM & D& M & EM & D\\
        \hline
        Community-graph &3.22&4.78 &11&4.86 &6.66 &12&2.73 & 3.79 &8\\
        CFINDER*&$\times$&$\times$&\multicolumn{1}{l}{$\times$}&7.27&10.51&20&$\times$&$\times$&\multicolumn{1}{l}{$\times$}\\
        LFM*&8.29&11.71&20&13.09&\multicolumn{1}{r}{18}&27&$\times$&$\times$&\multicolumn{1}{l}{$\times$}\\
        GCE*&3.35&4.73&9&9.79&\multicolumn{1}{r}{14}&22&3.28&4.73&9\\
        OSLOM*&3.28&4.55&9&11.14&15.75&24&$\times$&$\times$&\multicolumn{1}{l}{$\times$}\\
        LINKC*&2.3&4.55&10&$\times$&$\times$&$\times$&$\times$&$\times$&$\times$\\
        SVINET*&2.14&2.5&7&$\times$&$\times$&$\times$&$\times$&$\times$&$\times$\\
        MOSE*S&$\times$&$\times$&\multicolumn{1}{l}{$\times$}&7.39&9.96&17&2.62&3.7&6\\
        SLPA*&2.17&3.88&8&6.35&8.48&14&2.75&3.82&7\\
        DEMON*&2.24&3.27&6&5.2&7.03&12&2.18&3.22&6\\
        \hline
        \end{tabular}
        \end{table}

        \begin{table}[!ht]
        \centering
        \caption{The number of communities, communities maximum size, the communities average size and the Power-Law exponent for PGP, AMAZON and detected community structure. The calculated properties are the number of community (NC), the maximum size (MZ), the average size (AZ) and the Power-Law exponent (alpha)}
        \label{table16}
        \begin{tabular}{lcccccccccccc}
        \hline
        & \multicolumn{4}{c}{PGP}       & \multicolumn{4}{c}{AMAZON}   & \multicolumn{4}{c}{aNobii} \\
        \hline
                & NC    & MZ & AZ & alpha & NC & MZ & AZ & alpha& NC & MZ & AZ & alpha\\ \hline
        Ground-truth & 13712 & 24861 & 6.96 & 2.53 & 75149 & 53551 & 30.23 & 2.08 & 20387 & 3307 & 9.98 & 1.73\\
        CFINDER&\multicolumn{1}{l}{$\times$}&\multicolumn{1}{l}{$\times$}&\multicolumn{1}{l}{$\times$}&$\times$&\multicolumn{1}{r}{28402}&\multicolumn{1}{r}{1023}&10.16&2.55&$\times$&$\times$&$\times$&$\times$\\
        LFM&43558&1024&\multicolumn{1}{l}{6.73}&3.17&\multicolumn{1}{r}{21841}&\multicolumn{1}{r}{296}&6.84&3.98&\multicolumn{1}{r}{15781}&\multicolumn{1}{r}{1023}&7.31&2.69\\
        GCE&1187&7964&\multicolumn{1}{l}{55.38}&2.22&\multicolumn{1}{r}{17043}&\multicolumn{1}{r}{402}&16.32&4.09&\multicolumn{1}{r}{2827}&\multicolumn{1}{r}{3740}&48.05&2.6\\
        OSLOM&2577&430&\multicolumn{1}{l}{12.63}&2.21&\multicolumn{1}{r}{17007}&\multicolumn{1}{r}{325}&20.91&4.47&\multicolumn{1}{r}{3984}&\multicolumn{1}{r}{1024}&36.08&2.91\\
        LINKC&42443&3044&65,44&$\times$&$\times$&$\times$&$\times$&$\times$&$\times$&$\times$&$\times$&$\times$\\
        SVINET&\multicolumn{1}{l}{$\times$}&\multicolumn{1}{l}{$\times$}&\multicolumn{1}{l}{$\times$}&$\times$&\multicolumn{1}{r}{25302}&\multicolumn{1}{r}{1073}&19.51&2.86&$\times$&$\times$&$\times$&$\times$\\
        MOSES&\multicolumn{1}{l}{$\times$}&\multicolumn{1}{l}{$\times$}&\multicolumn{1}{l}{$\times$}&$\times$&\multicolumn{1}{r}{30240}&\multicolumn{1}{r}{151}&10.89&2.81&\multicolumn{1}{r}{3476}&\multicolumn{1}{r}{2598}&42.82&1.91\\
        SLPA&6658&15402&\multicolumn{1}{l}{14.81}&2.38&\multicolumn{1}{r}{33986}&\multicolumn{1}{r}{740}&13.26&3.22&\multicolumn{1}{r}{6803}&\multicolumn{1}{r}{5728}&37.44&2.22\\
        DEMON&1111&1023&\multicolumn{1}{l}{35.67}&1.8&\multicolumn{1}{r}{19839}&\multicolumn{1}{r}{572}&26.7&4.67&\multicolumn{1}{r}{754}&\multicolumn{1}{r}{1023}&87.06&1.58\\

		\hline
        \end{tabular}
        \end{table}
\bibliographystyle{elsarticle-harv}
\bibliography{biblio}


%
%
%
\end{document}